\documentstyle[prl,aps,psfig]{revtex}
\def\Bf#1{\mbox{\boldmath{$#1$}}}
\def\bF#1{\mbox{\scriptsize\boldmath{$#1$}}}
 
\def\IEq#1#2{{\mathstrut_{\displaystyle #1 }^{\displaystyle #2}}} 
\def\Ieq#1#2{{\mathstrut_{#1}^{#2}}}

\def\wt#1{\widetilde{#1}}
\def\wh#1{\widehat{#1}}
\def\ol#1{\overline{#1}}
\def\ul#1{\underline{\!#1\!}}
\def\Ol#1{\overline{\overline{#1}}}
\def\Bar#1{\bar{\bar{#1}}}
\def\BAr#1{\!\bar{\!\!\!\,\bar{#1}}}
\def\sh{\small\sharp}
\begin{document} 

\title{
\mbox{\boldmath{$Dynamical$}} correlation functions expressed
in terms of many-particle 
\mbox{\boldmath{$ground$}-\boldmath{$state$}}
wavefunction;
the 
\mbox{\boldmath{$dynamical$}} 
self-energy operator
\footnote{\tt To the memory of my mother.}
} 

\author{\sc Behnam Farid}

\address{Cavendish Laboratory, Department of Physics, 
University of Cambridge,\\
Madingley Road, Cambridge CB3 0HE, United Kingdom$^{\dagger}$\\
and Spinoza Institute, Department of Physics and Astronomy,
University of Utrecht,\\
Leuvenlaan 4, 3584 CE Utrecht, The Netherlands$^{\ddagger}$}

\date{Received 10 October 2001 and accepted 8 March 2002}

\maketitle

\begin{abstract}
\leftskip 54.8pt
\rightskip 54.8pt
We explicitly calculate the four leading-order terms of the formal 
asymptotic series for large $\vert\varepsilon\vert$ (where $\varepsilon$ 
denotes the external energy parameter) of the single-particle Green 
function $G_{\sigma}(\varepsilon)$, $\sigma \in \{ -{\sf s}, -{\sf s}+1, 
\dots, {\sf s}\}$, and the three leading-order terms of that of the 
self-energy operator $\Sigma_{\sigma}(\varepsilon)$, pertaining to systems 
of spin-${\sf s}$ fermions (${\sf s}=$ a half-integer) in $d$-dimensional 
spatial space, interacting through an arbitrary two-body potential 
$v({\Bf r}-{\Bf r}')$. These contributions, which are expressed in 
terms at most of a three-body static {\sl ground-state} correlation 
function, are amenable to accurate numerical calculation through 
employing correlated many-particle {\sl ground-state} wavefunctions 
such as determined within, e.g., the quantum Monte Carlo framework. Such 
calculations will provide indisputably reliable information with regard 
to the spectral function of the single-particle excitations at high 
energies of correlated systems, as well as energy moments of this 
spectral function, and thus help to assess the reliability of theoretical 
approaches that are applied in studying such systems. We give especial 
attention to $d=3$ and $v\equiv v_c$, the long-range Coulomb potential,
for which case we explicitly calculate the {\sl five} leading-order 
terms of the {\sl regularized} large-$\vert\varepsilon\vert$ asymptotic 
series of $\Sigma_{\sigma}(\varepsilon)$. Our considerations reveal 
some interesting aspects which are very specific to the behaviour 
of $v_c({\Bf r}-{\Bf r}')$ both at small {\sl and} large values of 
$\|{\Bf r}-{\Bf r}'\|$. In particular, we show that in these systems an 
inhomogeneity, even on the atomic scale, in the particle spin-polarization 
density gives rise to a pronounced effect, directly discernible in the 
inverse-photo-emission spectra; we show this effect to be absent in 
models with $v$ bounded at origin and those in which $v\equiv v_c$ but 
Umklapp processes are neglected, unless the ground state possesses 
long-range magnetic order. Our analyses shed light on the importance of 
the non-local part of the self-energy operator and disclose that some 
of strictly non-local contributions to this, transform into local ones 
upon replacing $v\not\equiv v_c$ by $v_c$, implying that a local 
approximation to $\Sigma_{\sigma}(\varepsilon)$ that for $v\not\equiv 
v_c$ proves accurate, is necessarily less accurate for $v\equiv v_c$. 
These findings establish a fundamental limitation of the so-called 
`dynamical mean-field' approximation to $\Sigma_{\sigma}(\varepsilon)$, 
which is {\sl strictly} local, specifically in applications where 
$v\equiv v_c$. We further explicitly establish some of the shortcomings 
of $\Sigma_{\sigma}(\varepsilon)$ as calculated within the framework of 
the many-body perturbation theory. In this context we demonstrate the 
empirically well-known inadequacy of the dynamically-screened exchange 
self-energy operator in particular for describing the photo-emission and 
inverse photo-emission spectra of interacting systems at intermediate 
and large transfer energies and put forward a workable scheme that rids 
this self-energy of its fundamental defects. We present ample explicit 
analyses of our results in terms of {\it un}correlated many-body 
wavefunctions.  
\end{abstract}

\vspace{2.25cm}
\noindent
\underline{\sf Preprint number: SPIN-2001/23 }
\vspace{0.5cm}

\noindent
\underline{\sf arXiv:cond-mat/0110481}\\
\underline{\sc Philosophical Magazine B, 2002, Vol.~82, 
No.~14, 1413-1610}
\vfill
\pagebreak
\widetext

\contentsline {section}
 {\numberline {I} Introduction}
{\pageref{s1}}
\contentsline {subsection}
 {\numberline {I.A} Background}
{\pageref{s2}}
\contentsline {subsection}
 {\numberline {I.B} Scope of the present work}
{\pageref{s3}}
\contentsline {subsection}
 {\numberline {I.C} Narrow-band systems; a brief survey}
{\pageref{s4}}
\contentsline {subsection}
 {\numberline {I.D} The organization of the work}
{\pageref{s5}}
\contentsline {section}
 {\numberline {II} Preliminaries}
{\pageref{s6}}
\contentsline {subsection}
 {\numberline {II.A} The system and its Hamiltonian}
{\pageref{s7}}
\contentsline {subsection}
 {\numberline {II.B} Some details concerning asymptotic series}
{\pageref{s8}}
\contentsline {section}
 {\numberline {III} Theory}
{\pageref{s9}}
\contentsline {subsection}
 {\numberline {III.A}{\hskip 3mm}Generalities}
{\pageref{s10}}
\contentsline {subsection}
 {\numberline {III.B}{\hskip 3mm}The single-particle spectral
function $A_{\sigma}({\Bf r},{\Bf r}';\varepsilon)$}
{\pageref{s11}}
\contentsline {subsection}
 {\numberline {III.C}{\hskip 3mm}Specific details}
{\pageref{s12}}
\contentsline {subsection}
 {\numberline {III.D}{\hskip 3mm}On the quasi-particle energies and 
wavefunctions}
{\pageref{s13}}
\contentsline {subsection}
 {\numberline {III.E}{\hskip 3mm}A simple example and some discussions}
{\pageref{s14}}
\contentsline {subsection}
 {\numberline {III.E.1}{\hskip 4mm} The Hamiltonian and some conventions} 
{\pageref{s15}}
\contentsline {subsection}
 {\numberline {III.E.2}{\hskip 4mm} Some notational conventions
(general)} 
{\pageref{s16}}
\contentsline {subsection}
 {\numberline {III.E.3}{\hskip 4mm} Some intermediate considerations} 
{\pageref{s17}}
\contentsline {subsection}
 {\numberline {III.E.4}{\hskip 4mm} The weak- and intermediate-coupling 
regimes} 
{\pageref{s18}}
\contentsline {subsection}
 {\numberline {III.E.5}{\hskip 4mm} The strong-coupling regime} 
{\pageref{s19}}
\contentsline {subsection}
 {\numberline {III.E.6}{\hskip 4mm} An illustrative example} 
{\pageref{s20}}
\contentsline {subsection}
 {\numberline {III.F}{\hskip 3mm}Evaluation of 
$G_{\sigma;\infty_2}({\Bf r},{\Bf r}')$}
{\pageref{s21}}
\contentsline {subsection} 
 {\numberline {III.F.1}{\hskip 4mm} Evaluation of 
$\Sigma_{\sigma;\infty_0}({\Bf r},{\Bf r}')$}
{\pageref{s22}}
\contentsline {subsection}
 {\numberline {III.F.2}{\hskip 4mm} The case of Coulomb-interacting 
fermions in the thermodynamic limit}
{\pageref{s23}}
\contentsline {subsection}
 {\numberline {III.G}{\hskip 3mm}Evaluation of 
$G_{\sigma;\infty_3}({\Bf r},{\Bf r}')$}
{\pageref{s24}}
\contentsline {subsection}
 {\numberline {III.G.1}{\hskip 4mm} Evaluation of
$\Sigma_{\sigma;\infty_1}({\Bf r},{\Bf r}')$}
{\pageref{s25}}
\contentsline {subsection}
 {\numberline {III.G.2}{\hskip 4mm} The case of Coulomb-interacting
fermions in the thermodynamic limit}
{\pageref{s26}}
\contentsline {subsection}
 {\numberline {III.H}{\hskip 3mm}Evaluation of 
$G_{\sigma;\infty_4}({\Bf r},{\Bf r}')$}
{\pageref{s27}}
\contentsline {subsection}
 {\numberline {III.H.1}{\hskip 4mm} Evaluation of
$\Sigma_{\sigma;\infty_2}({\Bf r},{\Bf r}')$}
{\pageref{s28}}
\contentsline {subsection}
 {\numberline {III.H.2}{\hskip 4mm} The case of Coulomb-interacting
fermions in the thermodynamic limit}
{\pageref{s29}}
\contentsline {subsection}
 {\numberline {III.I}{\hskip 3mm}The asymptotic series for the 
imaginary part of the self-energy for $\vert\varepsilon\vert\to\infty$}
{\pageref{s30}}
\contentsline {subsection}
 {\numberline {III.I.1}{\hskip 4mm} General considerations}
{\pageref{s31}}
\contentsline {subsection}
 {\numberline {III.I.2}{\hskip 4mm} Uniform isotropic systems; 
the single-particle spectral function
$\ol{A}_{\sigma}(k;\varepsilon)$}
{\pageref{s32}}
\contentsline {section}
 {\numberline {IV}{\hskip 2mm}Conventional perturbation theory and the 
case of dynamically-screened exchange self-energy operator
$\Sigma_{\sigma}^{\prime (1)}({\Bf r},{\Bf r}';\varepsilon)$}
{\pageref{s33}}
\contentsline {subsection}
 {\numberline {IV.A}{\hskip 3mm}Evaluation of 
$\Sigma_{\sigma;\infty_1}^{(1)}({\Bf r},{\Bf r}')$}
{\pageref{s34}}
\contentsline {subsection}
 {\numberline {IV.B}{\hskip 3mm}Evaluation of 
$\Sigma_{\sigma;\infty_2}^{(1)}({\Bf r},{\Bf r}')$}
{\pageref{s35}}
\contentsline {subsection}
 {\numberline {IV.C}{\hskip 3mm}Evaluation of 
$\Sigma_{\sigma;\infty_3}^{(1)}({\Bf r},{\Bf r}')$}
{\pageref{s36}}
\contentsline {subsection}
 {\numberline {IV.D}{\hskip 3mm}Correcting
$\Sigma_{\sigma}^{(1)}({\Bf r},{\Bf r}';\varepsilon) \equiv
\Sigma_{\sigma}^{\prime (1)}({\Bf r},{\Bf r}';\varepsilon)
+\hbar^{-1} v_H({\Bf r};[n])\, \delta({\Bf r}-{\Bf r}')$;
a workable scheme}
{\pageref{s37}}
\contentsline {section}
 {\numberline {V} Summary and concluding remarks}
{\pageref{s38}}
\contentsline {section}
 {\numberline {{}}\sc Acknowledgements}
{\pageref{s39}}
\contentsline {section}
 {\numberline {{}}\sc Note added in proof}
{\pageref{s39a}}
\contentsline {section}
 {\numberline {{}}\bf\sc Appendices}
{\pageref{s40}}
\contentsline {subsection}
 {\numberline {A}Non-orthogonality and over-completeness of
the set of Lehmann amplitudes}
{\pageref{s41}}
\contentsline {subsection}
 {\numberline {A.1}{\hskip 4mm}Basic considerations;
the overcompleteness}
{\pageref{s42}}
\contentsline {subsection}
 {\numberline {A.2}{\hskip 4mm}Quasi-particles revisited: an 
approximate treatment}
{\pageref{s43}}
\contentsline {subsection}
 {\numberline {B}On the density matrices $\Gamma^{(m)}$
and their association with $n_{\sigma}({\Bf r})$,
$\varrho_{\sigma}({\Bf r},{\Bf r}')$ and
${\sf g}_{\sigma,\sigma'}({\Bf r},{\Bf r}')$}
{\pageref{s44}}
\contentsline {subsection}
 {\numberline {C}The single-Slater-determinant approximation (SSDA)} 
{\pageref{s45}}
\contentsline {subsection}
 {\numberline {D}The time-reversal symmetry and the vanishing 
of ${\cal J}_{\sigma}({\Bf r},{\Bf r}')$}
{\pageref{s46}}
\contentsline {subsection}
 {\numberline {E}Symmetry of some correlation functions}
{\pageref{s47}}
\contentsline {subsection}
 {\numberline {E.1}{\hskip 1mm} Implicitly symmetric functions}
{\pageref{s48}}
\contentsline {subsection}
 {\numberline {E.2}{\hskip 1mm} Asymmetric functions
${\cal B}_{\sigma}({\Bf r},{\Bf r}')$ and
${\cal G}_{\sigma}({\Bf r},{\Bf r}')$ and their
symmetric combinations
${\cal D}_{\sigma}({\Bf r},{\Bf r}')$ and
${\cal F}_{\sigma}({\Bf r},{\Bf r}')$} 
{\pageref{s49}}
\contentsline {subsection}
 {\numberline {F}Regularization of some correlation functions 
pertaining to Coulomb-interacting fermion systems}
{\pageref{s50}}
\contentsline {subsection}
 {\numberline {F.1}{\hskip 1mm} ${\cal A}({\Bf r},{\Bf r}')$ and
its regularized form ${\cal A}'({\Bf r},{\Bf r}')$}
{\pageref{s51}}
\contentsline {subsection}
 {\numberline {F.1.a}{\hskip 3mm} Arbitrary systems; general arguments}
{\pageref{s52}}
\contentsline {subsection}
 {\numberline {F.1.b}{\hskip 3mm} Uniform and isotropic systems}
{\pageref{s53}}
\contentsline {subsection}
 {\numberline {F.1.c}{\hskip 3mm} On the behaviour of $\rho({\Bf r},
{\Bf r}')$ pertaining to uniform and isotropic ground states}
{\pageref{s54}}
\contentsline {subsection}
 {\numberline {F.1.d}{\hskip 3mm} On the van Hove pair correlation
function ${\sf g}_{\sigma,\sigma'}({\Bf r},{\Bf r}')$}
{\pageref{s55}}
\contentsline {subsection}
 {\numberline {F.2}{\hskip 1mm} ${\cal B}_{\sigma}({\Bf r},{\Bf r}')$ 
and its regularized forms ${\cal B}_{\sigma}'({\Bf r},{\Bf r}')$
and ${\cal B}_{\sigma}''({\Bf r},{\Bf r}')$}
{\pageref{s56}}
\contentsline {subsection}
 {\numberline {F.3}{\hskip 1mm} ${\cal G}_{\sigma}({\Bf r},{\Bf r}')$ 
and its regularized forms ${\cal G}_{\sigma}'({\Bf r},{\Bf r}')$
and ${\cal G}_{\sigma}''({\Bf r},{\Bf r}')$}
{\pageref{s57}}
\contentsline {subsection}
 {\numberline {F.4}{\hskip 1mm} ${\cal K}_{\sigma}({\Bf r},{\Bf r}')$ 
and its regularized forms ${\cal K}_{\sigma}'({\Bf r},{\Bf r}')$,
${\cal K}_{\sigma}''({\Bf r},{\Bf r}')$ 
and ${\cal K}_{\sigma}'''({\Bf r},{\Bf r}')$}
{\pageref{s58}}
\contentsline {subsection}
 {\numberline {F.5}{\hskip 1mm} ${\cal L}({\Bf r})$ 
and its regularized form}
{\pageref{s59}}
\contentsline {subsection}
 {\numberline {F.5.a}{\hskip 3mm} Basic considerations;
${\cal L}'({\Bf r})$, ${\cal L}''({\Bf r})$, ${\cal M}({\Bf r})$
and $\wt{\sf M}({\Bf r};z)$}
{\pageref{s60}}
\contentsline {subsection}
 {\numberline {F.5.b}{\hskip 3mm} The large-$\vert z\vert$ asymptotic 
series for $\wt{\sf M}({\Bf r};z)$; 
$\wt{\sf M}_{\infty_2}({\Bf r}\vert z)$,
${\sf M}_{\infty_2}^{\rm r}({\Bf r})$ and 
$\wt{\sf M}_{\infty_2}^{\rm s}({\Bf r}\| z)$}
{\pageref{s61}}
\contentsline {subsection}
 {\numberline {G}Regularization of ${\cal T}_{\sigma,\bar\sigma}
({\Bf r})$ and the large-$\vert z\vert$ asymptotic series for 
$\wt{\sf T}_{\sigma,\bar\sigma}({\Bf r};z)$
(${\sf T}^{\rm r}_{\sigma,\bar\sigma;\infty_2}({\Bf r})$, 
${\sf T}^{\rm s_b}_{\sigma,\bar\sigma;\infty_2}({\Bf r})$ and 
$\wt{\sf T}^{\rm s}_{\sigma,\bar\sigma;\infty_2}({\Bf r}\| z)$)}
{\pageref{s62}}
\contentsline {subsection}
 {\numberline {H} Regularization of the momentum representation of
$\Sigma_{\sigma;\infty_2}^{\rm s_b}$ and the double Fourier transform 
of $\wt{\Sigma}_{\sigma;\infty_2}^{\rm s_b}({\Bf r},{\Bf r}'\| z)$}
{\pageref{s63}}
\contentsline {subsection}
 {\numberline {I}Two basic integrals}
{\pageref{s64}}
\contentsline {subsection}
 {\numberline {J}Asymptotic behaviour of the density matrices
pertaining to uniform and isotropic Fermi liquids}
{\pageref{s65}}
\contentsline {subsection}
 {\numberline {K}On the differentiability property of the
ground-state partial number densities}
{\pageref{s66}}
\contentsline {section}
 {\numberline {{}}\sc Some frequently used notation}
{\pageref{s67}}
\contentsline {section}
 {\numberline {{}}\sc Abbreviations}
{\pageref{s68}}
\contentsline {section}
 {\numberline {{}}\sc References}
{\pageref{s69}}

\narrowtext
\twocolumn
\pagebreak
\section{Introduction}
\label{s1} 

\subsection{Background}
\label{s2} 

Calculation of properties of interacting many-particle systems is 
demanding for two reasons. Firstly, determination of the required 
eigenstate or ensemble of eigenstates is non-trivial and, secondly,
evaluation of expectation values of observables with respect to 
correlated many-particle wavefunctions entails evaluation of 
integrals over the configuration space of the system whose dimension 
is proportional to the number of particles in the system. For 
non-interacting {\sl many-particle} systems, both of these tasks 
are considerably simplified: as for the first task, eigenstates of 
these systems can be explicitly and directly constructed from those 
of a one-particle problem (as we shall see, because of the specific 
structure of these states, there is in fact {\sl no} need for having 
these in explicit form at one's disposal), and as for the second, 
following the fact that the expectation value of {\sl any} operator 
with respect to an uncorrelated many-body state $\vert {\Bf\xi}\rangle$ 
can be expressed in terms of integrals involving an associated 
single-particle Slater-Fock density matrix $\varrho_{{\Bf\xi};\sigma}
({\Bf r},{\Bf r}')$ (Appendices C and F), the dimension of the space 
over which explicit integration has to be carried out is determined 
by the nature of the operator (or observable) in question and thus 
does {\sl not} scale with the number of particles in the system. This 
is of course also {\sl almost} 
\footnote{\label{f1}
`Almost', because here in general the single-particle density matrix 
does {\sl not} suffice, but some higher-order correlation functions 
are required. }
the case for interacting systems; however, in contrast with the 
non-interacting case, where integrals over a large number of 
particle coordinates are fully determined by normalization, in the 
interacting case, these integrals need to be explicitly evaluated; 
owing to correlation, an integral over the configuration space, which 
is a {\sl product space}, does {\sl not} in general reduce into a 
product of low-dimensional integrals. This distinction between 
un-correlated and correlated states is explicit in the fact that in 
contrast with $\{ \varrho_{{\Bf\xi};\sigma} \}$, the single-particle 
density matrices pertaining to correlated states are {\sl not} 
idempotent. 

In view of our later considerations in this work, let us here briefly 
touch upon some of the crucial aspects that are involved in the 
calculation of the single-particle excitation energies of $N$-particle 
systems of fermions, as encoded in the single-particle Green function 
(GF) $G_{\sigma}({\Bf r},{\Bf r}';\varepsilon)$ (see \S~III.A). For 
systems with a small number of particles, the number of excited states 
whose energies are confined within a finite interval, is limited (here 
we disregard the case where such interval contains an accumulation 
point of the excitation spectrum),
\footnote{\label{f2}
The completeness of eigenstates of self-adjoint operators, following 
the Bolzano-Weierstrass theorem (Whittaker and Watson 1927, pp.~12 
and 13), implies that the eigenvalues of these must possess {\sl at 
least} one point of accumulation. }
so that for such systems, calculation of the single-particle excitation
energies over a finite range of energies is in principle feasible. In 
the thermodynamic limit, where the density of the single-particle 
excitation energies transforms into a continuum, 
\footnote{\label{f3}
The separation between the energy levels of a {\sl non}-interacting 
system scales like $L^{-2}$ (Landau and Lifshitz 1980, p.~14), where 
$L$ stands for the macroscopic length of the system (here we neglect 
the possibility of occurrence of gaps in the excitation spectra brought 
about through the presence of some possible external potential). 
Interaction amongst particles complicates this picture in that each 
single-particle excitation energy of an interacting system (that is 
$\varepsilon_{s;\sigma}$ as defined in Eq.~(\protect\ref{e19})) is 
some weighted average of a distribution of the `non-interacting' 
single-particle excitation energies (that is $\{\varepsilon_{\varsigma;
\sigma}^{(0)}\}$ as introduced in Eq.~(\protect\ref{e56})). This 
aspect, together with the over-completeness of the set $\{ f_{s;\sigma}
({\Bf r})\}$ (see Appendix A), can in principle give rise to a large 
degree of degeneracy in the single-particle energy levels. These 
mechanisms are equally operative in finite systems or those containing 
finite number of particles. }
calculation of these energies over a finite range amounts to 
calculation of a vast number of excited many-particle states and 
accordingly to that of a vast number of excited-states expectation 
values, which evidently {\sl cannot} be practicable.

It is very seldom the case that an interacting many-body Hamiltonian 
can be exactly diagonalized so that, in general, eigenstates of 
interacting systems must be obtained numerically. This task is greatly 
simplified if the ground state (GS) and possibly a small number of 
lowest-lying excited states are in demand, in which cases iterative 
methods, notably the Lanczos method (Golub and van Loan 1983), are very 
effective specifically because of the fact that their application 
does not require storage of the entire Hamiltonian matrix (in some 
appropriate representation), which in general is prohibitively 
large. With reference to our above remarks, it is evident that
for large systems, in particular those which are intended to mimic 
systems in the thermodynamic limit, the energies corresponding to a 
`small' number of lowest-lying eigenstates do not extend over an 
interval of any significant width, so that for these systems the 
numerical method of `exact diagonalization' is of severely limited 
potentiality as regards study of the excitation spectra (i.e. those 
corresponding to single-particle or multi-particle excitations) of 
interacting systems.

Insofar as GSs are concerned, the methods of quantum Monte Carlo 
(QMC) (for a recent review see Foulkes, {\sl et al.} (2001)) are of 
considerable interest. These provide both very accurate approximate 
GS wavefunctions for interacting systems and an effective method, 
based on the scheme of `importance sampling' (for example Negele and 
Orland (1988, chapter 8)), for evaluating the aforementioned 
multi-dimensional integrals that are encountered in the calculation 
of expectation values. In fact, making use of a generalization of 
the variational principle, it is possible within the QMC formalism 
to construct accurate variational wavefunctions for the excited 
states of interacting systems; however the prospects of this approach 
are limited by the fact that each variational excited state is 
required to be orthogonal onto the lower-lying {\sl exact} eigenstates 
(Peierls 1979; \S~3.5 herein). Further, here, as in the framework 
of the `exact-diagonalization' approach, the maximum range covered by 
the energies of the excited states that one may possibly be able to 
calculate, diminishes progressively for systems of increasing extent, 
so much so that, within this formalism {\sl rigorous} calculations 
of the excitation spectra over a finite energy range become impracticable 
for systems that are intended to mimic those in the thermodynamic limit.
In this connection we should like to emphasize that the difficulties
associated with the transformation of the distribution of the 
single-particle excitation energies into continuum, in consequence 
of effecting the thermodynamic limit, remain even in cases where in 
appropriate representations the {\sl non}-interacting single-particle 
energies are, in the thermodynamic limit, discrete and in general 
widely separated. For concreteness, consider homogeneous systems of 
fermions, with uniform GSs, and periodic solids. For vanishing coupling 
constant of interaction, in the former systems the single-particle 
excitations and their energies are uniquely specified in terms of the
wave-vector ${\Bf k}$ and spin index $\sigma$; in the latter systems, for 
each $({\Bf k},\sigma)$ there is a discrete set of single-particle 
states, each marked in addition to ${\Bf k}$ (now a {\sl reduced} 
wave-vector) and $\sigma$, by an integer-valued index $\ell$ (band index), 
whose corresponding energies are in general widely separated. In these 
systems, an interaction of even infinitesimally weak strength brings 
about a {\sl continuous} distribution of single-particle excitations whose 
energies envelope (not necessarily symmetrically) the well-defined and
well-separated (band) energies of the corresponding non-interacting 
systems (see Appendix A; see also footnote \ref{f2}). This phenomenon is 
precisely that which gives rise to the broadening of infinitely sharp 
peaks along the energy axis of the momentum representation of the 
single-particle spectral functions of non-interacting systems (here, 
the homogeneous systems with uniform GSs {\sl and} periodic solids) upon 
switching on of the inter-particle interaction and is interpreted as 
corresponding to the finite lifetimes of the single-particle excitations 
in interacting systems (see \S~III.D). It follows that, for the systems 
just considered, even by projecting the many-body Hamiltonian on to 
$(N\pm 1)$-particle Hilbert spaces characterized by the (reduced) 
wave-vector ${\Bf k}$, in general the number of $(N\pm 1)$-particle 
excited states whose energies (as measured with respect to the 
energy of the $N$-particle GS) cover a {\sl finite} interval of energy, 
remain prohibitively large for systems that are supposed to approximate 
those in the thermodynamic limit.

The above considerations make evident that in general determination
of the properties of the excited states of interacting systems is a 
demanding task, in particular when large systems with large number 
of particles are concerned. This aspect persists even when one's 
interest is narrowed to the calculation of the low-lying 
single-particle excitations spectra of these systems.

\subsection{Scope of the present work}
\label{s3}

In this work we put forward an approach for the determination of 
the single-particle GF $G_{\sigma}({\Bf r},{\Bf r}';\varepsilon)$ 
and the associated self-energy (SE) $\Sigma_{\sigma}({\Bf r},{\Bf r}';
\varepsilon)$ for {\sl large} values of $\vert\varepsilon\vert$ (see 
later, specifically \S\S~II.B and III.E), formulated solely in terms 
of the GS expectation values of {\sl static} operators, namely
the GS number densities $n_{\sigma}({\Bf r})$, $\sigma\in \{-{\sf s},
-{\sf s}+1,\dots,{\sf s}\}$, the GS density matrices $\varrho_{\sigma}
({\Bf r}',{\Bf r})$, etc. These functions can be relatively easily 
calculated within the framework of the existing QMC techniques {\sl or} 
that of the  {\sl local Ansatz} (Stollhoff and Fulde 1980, Stollhoff 
1996, 1998) which deals with construction of variational GS 
wavefunctions with especial emphasis on the atomic or `local' GS 
correlations. Knowledge of the asymptotic behaviour of $\Sigma_{\sigma}
({\Bf r},{\Bf r}';\varepsilon)$ for large $\vert\varepsilon\vert$ enables 
one to calculate directly the single-particle excitation energies of 
interacting systems at large energies. In this paper we explicitly 
calculate {\sl all} asymptotic contributions to $\Sigma_{\sigma}
({\Bf r},{\Bf r}';\varepsilon)$ decaying not faster than 
$1/\varepsilon^2$ for $\vert\varepsilon\vert\to\infty$. 

In order to illustrate the physical as well as practical relevance 
of the above-mentioned asymptotic contributions (see also \S~III.E), 
we point out that the leading asymptotic contribution to 
$\Sigma_{\sigma}({\Bf r},{\Bf r}';\varepsilon)$ for 
$\vert\varepsilon\vert\to\infty$ is equal to 
$\Sigma^{\sc hf}({\Bf r},{\Bf r}';[\varrho_{\sigma}])$, the 
Hartree-Fock exchange SE (Farid 1999a,c), 
\footnote{\label{f4}
The dependence on $\sigma$ of this contribution to the SE is solely 
through that of $\varrho_{\sigma}$ (see \S~III.F.1). }
as evaluated in terms of the {\sl exact} GS single-particle density matrix 
$\varrho_{\sigma}$; $\Sigma^{\sc hf}({\Bf r},{\Bf r}';[\varrho_{\sigma}])$ 
deviates from the Hartree plus exchange SE as determined within the 
framework of the (self-consistent) Hartree-Fock scheme by the fact that, 
in the latter, $\varrho_{\sigma}({\Bf r}',{\Bf r})$ is replaced by its 
Slater-Fock counterpart $\varrho_{{\rm s};\sigma}({\Bf r}',{\Bf r})$ 
which in strict contrast with $\varrho_{\sigma}({\Bf r}',{\Bf r})$, is 
idempotent (see \S~III.F). Considering the fact that the significance 
of the asymptotic contributions to $\Sigma_{\sigma}({\Bf r},{\Bf r}';
\varepsilon)-\Sigma^{\sc hf}({\Bf r},{\Bf r}';[\varrho_{\sigma}])$ for 
$\vert\varepsilon\vert\to\infty$ is controlled by the small parameter 
$e_0/\varepsilon$ (where $e_0$ stands for an energy scale characteristic 
of the system under consideration; see \S\S~II.B and III.E), while taking 
into account the relative accuracy of the Hartree-Fock single-particle 
excitation energies of specifically non-extensive interacting systems 
(a fact attested by the popularity of this scheme in quantum chemical 
applications (for example Szabo and Ostlund (1989))), it follows that 
the asymptotic contributions to $\Sigma_{\sigma}({\Bf r},{\Bf r}';
\varepsilon)$ for $\vert\varepsilon\vert\to\infty$ are physically 
significant. This aspect of these terms should be compared with that of 
the terms in the perturbation series for $\Sigma_{\sigma}({\Bf r},
{\Bf r}';\varepsilon)$ as determined within the framework of the 
conventional many-body perturbation theory (for example Fetter and
Walecka (1971)), where in contrast with $e_0/\varepsilon$, the 
strength of perturbation (or the dimensionless value of the coupling 
constant of the particle-particle interaction) is fixed and cannot 
be changed at will.
\footnote{\label{f5}
Since the {\sl exact} $\varrho_{\sigma}$ {\sl cannot} be expressed in 
terms of a finite-order perturbation series, it follows that already 
calculation of the {\sl exact} $\Sigma^{\sc hf}({\Bf r},{\Bf r}';
[\varrho_{\sigma}])$ amounts to a prohibitive many-body problem. }
In fact, as we explicitly demonstrate in this work (\S~IV), the 
leading {\sl perturbative} contribution to $\Sigma_{\sigma}({\Bf r},
{\Bf r}';\varepsilon)-\Sigma^{\sc hf}({\Bf r},{\Bf r}';[\varrho_{\sigma}])$, 
in terms of the dynamically screened interaction function $W({\Bf r},
{\Bf r}';\varepsilon)$ (Hubbard 1957) rather than the bare one, 
$v({\Bf r}-{\Bf r}')$, does {\sl not} fully reproduce even the leading 
asymptotic contribution to $\Sigma_{\sigma}({\Bf r},{\Bf r}';\varepsilon)
-\Sigma^{\sc hf}({\Bf r},{\Bf r}';[\varrho_{\sigma}])$ for 
$\vert\varepsilon\vert\to\infty$.

In addition to the above, the large-$\vert\varepsilon\vert$ asymptotic
series (AS) for the {\sl exact} $\Sigma_{\sigma}({\Bf r},{\Bf r}';
\varepsilon)$ has further significance in that its correct reproduction 
to some finite order by an approximate expression for $\Sigma_{\sigma}
({\Bf r},{\Bf r}';\varepsilon)$, guarantees that the energy moments (see 
Eqs.~(\ref{e37}) and (\ref{e38})) of the associated approximate 
single-particle spectral function $A_{\sigma}({\Bf r},{\Bf r}';
\varepsilon)$ (see Eq.~(\ref{e39}) for the definition) are to certain 
order identical with those of the {\sl exact} $A_{\sigma}({\Bf r},
{\Bf r}';\varepsilon)$ (see also \S\S~I.C and III.B). These energy 
moments are of {\sl direct} relevance to the energies of the 
single-particle excitations at large values of these energies. Conversely, 
{\sl none} of the $\varepsilon$ moments of $A_{\sigma}^{\rm Appr}({\Bf r},
{\Bf r}';\varepsilon)$ (with the exception of one which is determined 
by completeness, or closure; see Eq.~(\ref{e57})) is equal to the 
corresponding $\varepsilon$ moment pertaining to the exact $A_{\sigma}
({\Bf r},{\Bf r}';\varepsilon)$ when the {\sl approximate} SE, to which 
$A_{\sigma}^{\rm Appr}({\Bf r},{\Bf r}';\varepsilon)$ corresponds, fails 
to reproduce, to {\sl some} order, the terms in the large-$\vert
\varepsilon\vert$ AS pertaining to the {\sl exact} $\Sigma_{\sigma}
({\Bf r},{\Bf r}';\varepsilon)$. In this connection, we specifically 
analyse a well-known perturbative approximation to the SE operator, 
known as the $GW$ approximation (Hedin 1965) (\S~IV), and establish 
that its next-to-leading asymptotic term for $\vert\varepsilon\vert 
\to \infty$ consists solely of a {\sl local} contribution, in contrast 
with the exact result which in addition consists of a {\sl non}-local 
contribution. Given the fact that within the framework of the 
perturbation theory, the single-particle density matrix 
$\varrho_{\sigma}({\Bf r}',{\Bf r})$ that enters in the leading-order 
asymptotic contribution $\Sigma^{\sc hf}({\Bf r},{\Bf r}';
[\varrho_{\sigma}])$ to $\Sigma_{\sigma}({\Bf r},{\Bf r}';\varepsilon)$ 
is necessarily replaced by its Slater-Fock counterpart $\varrho_{{\rm s};
\sigma}({\Bf r}',{\Bf r})$, we observe that in practice {\sl none} of 
the terms in the large-$\vert\varepsilon\vert$ AS of this 
{\sl perturbative} approximation to $\Sigma_{\sigma}({\Bf r},{\Bf r}';
\varepsilon)$ coincides with its exact counterpart. These observations 
are in conformity with the well-known fact that $A_{\sigma}({\Bf r},
{\Bf r}';\varepsilon)$ as deduced from this approximate SE fails to 
reproduce the experimental photo-emission results (Hedin and Lundqvist 
1969, Almbladh and Hedin 1983), even in simple metals (Aryasetiawan, 
{\sl et al.} 1996) (see also Aryasetiawan and Gunnarsson (1995), 
Aryasetiawan and Karlsson (1996) and Farid (1997a, 1999a)). 

We point out that, although $A_{\sigma}({\Bf r},{\Bf r}';\varepsilon)$ 
pertaining to Hamiltonians with unbounded spectra, has an unbounded 
support and exhibits, in the thermodynamic limit, a power-law decay for 
$\vert\varepsilon\vert \to\infty$, implying thus {\sl unbounded} energy 
moments for $A_{\sigma}({\Bf r},{\Bf r}';\varepsilon)$ beyond some 
finite order,
\footnote{\label{f6}
For uniform and isotropic systems, to leading order in the AS, 
this function decays like $1/\vert\varepsilon\vert^3$ (see \S~III.I.2). 
Consequently, for these systems the integral of $\varepsilon^{m-1} \, 
A_{\sigma}({\Bf r},{\Bf r}';\varepsilon)$ over $(-E,E)$ is unbounded 
for $m \ge 3$ when $E\to\infty$. For definiteness, by the `order' of 
an $\varepsilon$ moment of $A_{\sigma}({\Bf r},{\Bf r}';\varepsilon)$, 
we refer to $(m-1)$ in the above expression. }
nonetheless by considering the moments as being defined in terms of 
$\varepsilon$ integrals over $(-E,E)$, the matrix elements of these 
with respect to some basis spanning the single-particle Hilbert space
of the problem (see \S~III.B) have, to {\sl any} finite order, 
well-defined and physically meaningful limits for $E\to\infty$. We 
consider this aspect in some detail and further expose an interplay 
between the {\sl over-completeness} (Appendix A) of the set of 
Lehmann (1954) amplitudes (see Eq.~(\ref{e18})) associated with the 
single-particle 
excitations in interacting systems and the nature of the distribution 
of the energies of these excitations (see Eq.~(\ref{e19})) along the 
energy axis (see \S\S~III.B and III.D).

Some elements of the formalism introduced in this work have been 
presented in (Farid 1999a), in the context of exposing the significance 
to the framework of many-body perturbation theory of a mean-field 
theory that is capable of reproducing the GS number densities 
$n_{\sigma}({\Bf r})$, $\sigma\in \{-{\sf s},-{\sf s}+1,\dots,{\sf s}\}$, 
pertaining to fully interacting systems.
\footnote{\label{f7}
As will become apparent later in this work, the contributions to the 
AS of $\Sigma_{\sigma}({\Bf r},{\Bf r}';\varepsilon)$ for 
$\vert\varepsilon\vert\to\infty$ overwhelmingly involve $n_{\sigma}
({\Bf r})$ and $\varrho_{\sigma}({\Bf r}',{\Bf r})$, the latter reducing 
to $n_{\sigma}({\Bf r})$ as ${\Bf r}'\to {\Bf r}$. Some of the arguments 
by Farid (1999a) are centred around the fact that a many-body perturbation 
theory based on a `non-interacting' Hamiltonian $\wh{H}_0$ whose GS 
number densities $n_{\sigma}({\Bf r})$, $\forall\sigma$, are identical 
with those pertaining to the GS of $\wh{H}$, has the evident advantage 
of {\sl exactly} reproducing (already in the zeroth order of 
perturbation theory) the pertinent contributions to the AS of 
$\Sigma_{\sigma}({\Bf r},{\Bf r}';\varepsilon)$ for $\vert\varepsilon\vert
\to\infty$. For a detailed discussion of this subject see (Farid 1999b).} 
In (Farid 1999a) we have merely considered the leading term in the 
AS of the {\sl exact}
\footnote{\label{f8}
In fact, the considerations in (Farid 1999a) are confined to systems 
of spin-less fermions or those of spin-$1/2$ fermions which are fully 
spin compensated. } 
$\Sigma_{\sigma}({\Bf r},{\Bf r}';\varepsilon)$, for $\vert\varepsilon
\vert\to\infty$, and paid {\sl no} attention whatever to the possible 
consequences of the specific features of the Coulomb potential 
$v_c({\Bf r}-{\Bf r}')$, namely its long range for $\|{\Bf r}-{\Bf r}'
\|\to\infty$ and its divergence for $\|{\Bf r}-{\Bf r}'\|\to 0$, to 
the behaviour of terms in this series. As we discuss in this work, these 
aspects of $v_c$ are of significant influence on the behaviour of 
the SE {\sl operator} $\Sigma_{\sigma}(\varepsilon)$ at large values 
of $\vert\varepsilon\vert$, both in the coordinate and in the momentum 
representation. In this connection it is appropriate to mention that, owing 
to the divergence of $v_c({\Bf r}-{\Bf r}')$ for $\|{\Bf r}-{\Bf r}'\| 
\to 0$, the nature of the AS for $\Sigma_{\sigma}(\varepsilon)$ is 
dependent upon the choice of representation; that is, depending on this, 
the contributions to the {\sl complete} AS need be combined differently 
in order to determine the appropriate terms in a {\sl finite}-order AS 
concerning $\vert\varepsilon\vert \to \infty$. In this paper, we 
primarily consider $\Sigma_{\sigma}(\varepsilon)$ in the coordinate 
representation; however, owing to the physical significance of 
$\Sigma_{\sigma}(\varepsilon)$ in the momentum representation, we in 
addition present many of the relevant expressions in this representation. 
We should emphasize that $v_c$ in $d=3$ is {\sl not} exceptional for 
its specific consequences to the behaviour of $\Sigma_{\sigma}
(\varepsilon)$ at large $\vert\varepsilon\vert$, as one can conceive 
of other two-body potentials, and in spatial dimensions not necessarily 
equal to $3$, for which the contributions to the AS for $\Sigma_{\sigma}
(\varepsilon)$ at large $\vert\varepsilon\vert$ are dissimilar to those 
corresponding to bounded and short-range interaction potentials. The 
distinctive aspect of $v_c$ in $d=3$ in the context of our present 
work is that, up to a multiplicative constant, it is the inverse of the 
single-particle kinetic-energy operator and this aspect directly results 
in some undefined contributions upon substitution of $v_c$ for $v$ in 
the expression for the coefficient of the $1/\varepsilon^2$ term in 
the formal large-$\vert\varepsilon\vert$ AS for $\Sigma_{\sigma}
({\Bf r},{\Bf r}';\varepsilon)$ (see Eq.~(\ref{e72})). Since 
$\Sigma_{\sigma}({\Bf r},{\Bf r}';\varepsilon)$ is a well-defined 
function, any fundamentally unbounded contribution in its 
large-$\vert\varepsilon\vert$ AS {\sl must} be compensated by equally 
fundamentally unbounded contributions in infinite number of subsequent 
terms in this series (see \S~II.B); appropriate summations over these 
contributions result in well-defined functions that involve 
transcendental functions of $\varepsilon$, such as 
$(-z/\varepsilon_0)^{1/2}/z^2$ and $\ln(-z/\varepsilon_0)/z^2$ where 
$z= \varepsilon\pm i\eta$, complementing the formal 
large-$\vert\varepsilon\vert$ AS for $\Sigma_{\sigma}({\Bf r},{\Bf r}';
\varepsilon)$ in terms of the asymptotic sequence $\{1,1/\varepsilon, 
1/\varepsilon^2,\dots\}$. Here $\varepsilon_0$ denotes an arbitrary 
positive energy so that $\varepsilon/\varepsilon_0$ is dimensionless.

As mentioned above, the large-$\vert\varepsilon\vert$ AS for 
$\Sigma_{\sigma}({\Bf r},{\Bf r}';\varepsilon)$ with which we deal in 
this work, is in terms of {\sl exact} {\it GS} correlation 
functions. Denoting the coefficient of $1/\varepsilon^m$ in this formal 
series by $\Sigma_{\sigma;\infty_m}$ (see Eq.~(\ref{e72})), we observe that 
$\Sigma_{\sigma;\infty_m}$ involves various $(m+1)$th-order combinations 
\footnote{\label{f9}
As we shall see in \S~III.C, from the Dyson equation it directly follows 
(see, e.g., footnote \protect\ref{f38} below) that $\{\Sigma_{\sigma;
\infty_p} \| p=0,1,\dots,m\}$ is the complete set necessary for
the evaluation of $G_{\sigma;\infty_{m+2}}$; conversely, determination 
of $\Sigma_{\sigma;\infty_m}$ is dependent upon the knowledge of 
$\{ G_{\sigma;\infty_p} \| p=0,1,\dots,m+2\}$. Through 
Eq.~(\protect\ref{e34}) below we observe that $G_{\sigma;\infty_{m+2}}$ 
involves $\wh{L}^{m+1}$ which implies $m+1$ {\sl nested} commutations 
with $\wh{H}$. This underlies our above statement, ``$(m+1)$th-order 
combinations of \dots''. See \S~II.B, the paragraph beginning with 
``Second, considering for the moment \dots''. } 
of the single-particle kinetic-energy operator $\tau$ (see Eq.~(\ref{e3})), 
the scalar external potential $u$ and the {\sl bare} particle-particle 
interaction function $v$. These aspects unequivocally establish 
that the {\sl exact} $\Sigma_{\sigma;\infty_m}$ would be identically 
reproduced by choosing as the starting point of the calculations, the 
{\sl complete} $(m+1)$th-order perturbation series expansion of the SE 
operator, that is the series {\sl up to and including} the $(m+1)$th-order 
contributions, in terms of the {\sl bare} particle-particle interaction 
function and the {\sl skeleton} SE diagrams (for the definition see 
Luttinger and Ward (1960)) which are determined in terms of the 
{\sl exact} single-particle GF pertaining to the {\sl interacting} 
system. In fact (see in particular Eq.~(\ref{e121}) and the associated 
text), the {\sl complete} AS for $\Sigma_{\sigma}({\Bf r},{\Bf r}';
\varepsilon)$ corresponding to $\vert\varepsilon\vert\to\infty$ is 
entirely equivalent with the {\sl complete} many-body perturbation 
series concerning the {\sl exact} $\Sigma_{\sigma}({\Bf r},{\Bf r}';
\varepsilon)$ for {\sl arbitrary} $\varepsilon$; the two series merely 
deviate by the fact that, whereas, in the perturbation series, 
contributions to $\Sigma_{\sigma}({\Bf r},{\Bf r}';\varepsilon)$ are 
arranged in accordance with the increasing powers of the coupling 
constant of $v$, in the AS of this function corresponding to 
$\vert\varepsilon\vert\to\infty$, the contributions are arranged in 
accordance with the increasing powers of $1/\varepsilon$. As we have 
indicated earlier in this Introduction (\S~I.A), the latter arrangement 
is superior to the former for sufficiently large values of 
$\vert\varepsilon\vert$ (see specifically \S~III.E).

Similar to perturbation series where unbounded contributions are
regularized through performing partial summations over {\sl infinite} 
number of pertinent unbounded contributions corresponding to 
higher-order terms in the perturbation series, here also unbounded 
terms in the expressions for the coefficient functions pertaining to 
the asymptotic sequence $\{ 1,1/\varepsilon, 1/\varepsilon^2, \dots\}$ 
are regularized through infinite summations over pertinent unbounded 
terms that occur in the higher-order terms as arranged in accordance 
with the powers of $1/\varepsilon$. For $v\equiv v_c$ and $d=3$, 
$\Sigma_{\sigma;\infty_2}({\Bf r},{\Bf r}')$ is unbounded and, performing
the latter infinite partial summations, we deduce that in this case, in 
a finite-order large-$\vert\varepsilon\vert$ AS for $\Sigma_{\sigma}
({\Bf r},{\Bf r}';\varepsilon)$, $\Sigma^{\sc hf}({\Bf r},{\Bf r}';
[\varrho_{\sigma}])+\Sigma_{\sigma;\infty_1}({\Bf r},
{\Bf r}')/\varepsilon$ is followed by contributions that decay like 
$1/\vert\varepsilon\vert^{3/2}$, $\ln(\vert\varepsilon/\varepsilon_0
\vert)/\varepsilon^2$, $1/\varepsilon^2$, $\dots$ rather than merely 
$1/\varepsilon^2$, $\dots$, as would be expected from the {\sl formal} 
structure of the AS for the SE corresponding to an unspecified $v$ 
(see Eq.~(\ref{e72})). In addition, we obtain that $\Sigma_{\sigma;
\infty_2}({\Bf r},{\Bf r}')$ involves a contribution proportional to 
$v^3({\Bf r}-{\Bf r}')$ which, although well defined, is {\sl not} 
integrable (in the sense of Riemann) in the case where $v\equiv v_c$ 
in $d=3$, thus demonstrating the momentum representation of 
$\Sigma_{\sigma;\infty_2}$ to be ill defined. This problem is removed 
through a further summation over an infinite number of similarly 
non-integrable contributions corresponding to $\Sigma_{\sigma;\infty_m}
({\Bf r},{\Bf r}')$ with $m > 2$, giving rise to an additional singular 
contribution of the form $\ln(\vert\varepsilon/\varepsilon_0
\vert)/\varepsilon^2$ to the AS of the momentum representation of 
$\Sigma_{\sigma}(\varepsilon)$ for $\vert\varepsilon\vert\to\infty$. 

To summarize, we observe that in the cases corresponding to $v\equiv 
v_c$ in $d=3$, the AS for $\Sigma_{\sigma}({\Bf r},{\Bf r}';\varepsilon)$, 
$\vert\varepsilon \vert\to\infty$, acquires {\sl singular} contributions, 
corresponding to the transcendental functions $\ln(-z/\varepsilon_0)$ 
and $(-z/\varepsilon_0)^{1/2}$; the latter function has its 
origin in the particular behaviour of $v_c({\Bf r}-{\Bf r}')$ for 
$\|{\Bf r}-{\Bf r}'\| \to 0$ and its appearance in the AS for 
$\Sigma_{\sigma}({\Bf r},{\Bf r}';\varepsilon)$ depends on the number 
density corresponding to particles of spin index $\sigma$ is in 
{\sl local} imbalance with the {\sl total} number density of particles 
whose spin indices are different from $\sigma$ (see Eqs.~(\ref{e213}) 
and (\ref{eg15})). Here, $z$ denotes a {\sl complex}-valued energy 
(see \S~II.B).

Although our considerations in this work are {\sl not} in the main
concerned with the behaviour of $\Sigma_{\sigma}({\Bf r},{\Bf r}';
\varepsilon)$ for $\varepsilon \approx \varepsilon_F$, 
\footnote{\label{f10}
We only very briefly touch upon $\varepsilon \approx \varepsilon_F$ 
in Appendix A. }
with $\varepsilon_F$ the Fermi energy (we assume for the moment 
metallic GSs), our above findings with regard to the existence of 
singular contributions corresponding to $(-z/\varepsilon_0)^{1/2}$ 
and $\ln(-z/\varepsilon_0)$ in the large-$\vert z\vert$ AS for 
$\wt{\Sigma}_{\sigma}(z)$ can be of significance for the nature of 
the terms to be expected in the AS for $\wt{\Sigma}_{\sigma}(z)$ 
corresponding to $\vert z -\varepsilon_F\vert\to 0$. 
\footnote{\label{f11}
In this paper we denote the analytic continuation of $f(\varepsilon)$, 
a function of the external energy parameter $\varepsilon$, into the 
physical Riemann sheet of the $z$ plane, that is the complex energy plane, 
by ${\tilde f}(z)$ (for details, see Farid (1999a,c)). See, for instance,
Eqs.~(protect\ref{e17}) and (\protect\ref{e24}) below. }
To clarify this statement, we first mention that $\wt{\Sigma}_{\sigma}(z)$ 
is analytic over the entire complex $z$ plane with the exception of the 
real axis (Luttinger 1961, Farid 1999a). Consequently, the branch cuts 
corresponding to the branch points at the point of infinity, associated 
with $(-z/\varepsilon_0)^{1/2}$, $\ln(-z/\varepsilon_0)$, etc., must be 
connected with further singular contributions to $\wt{\Sigma}_{\sigma}(z)$ 
in the {\sl finite} part of the complex energy plane, of the form 
$(z-\varepsilon_1)^{1/2}$, $\ln(z -\varepsilon_2)$, etc., respectively, 
where $\varepsilon_1$, $\varepsilon_2$, etc., are {\sl real} and 
{\sl finite}. By making the {\sl assumption} that $\varepsilon_1
=\varepsilon_2=\dots = \varepsilon_F$, the nature of the singular 
functions in the small-$\vert z-\varepsilon_F\vert$ AS of 
$\wt{\Sigma}_{\sigma}(z)$ are directly deduced from those in the 
large-$\vert z\vert$ AS of this function. Thus, the results presented 
above would imply that, under the conditions where 
$(-z/\varepsilon_0)^{1/2}/z^2$ appears in the large-$\vert z\vert$ 
AS of $\wt{\Sigma}_{\sigma}(z)$, one should expect appearance of 
$(z-\varepsilon_F)^{3/2}$ in the small-$\vert z -\varepsilon_F\vert$ 
AS of $\wt{\Sigma}_{\sigma}(z)$. Although a contribution of this form 
does {\sl not} imply breakdown of the Fermi-liquid picture for the 
corresponding metallic state (Farid 1999c), it nonetheless amounts 
to a considerable deviation from the behaviour of 
$\wt{\Sigma}_{\sigma}(z)$ corresponding to {\sl conventional} Fermi 
liquids in $d=3$. The above assumption {\sl cannot} in general be true;
\footnote{\label{f12}
This assumption is, however, easily shown to be true in the case of
{\sl conventional} Fermi liquids in $d=3$ in non-magnetic GSs. 
Nonetheless, our observations in \S~IV where we investigate the 
large-$\vert z\vert$ AS for $\wt{\Sigma}_{\sigma}^{(1)}(z)$, the 
first-order contribution to $\wt{\Sigma}_{\sigma}(z)$ according to 
the perturbation theory in terms of the dynamically screened 
particle-particle interaction function $W(\varepsilon)$ and the 
interacting GF, imply that this assumption {\sl cannot} be universally 
valid (see \S~V). In \S~IV we deduce that, {\sl independent} of the 
magnetic state of the GS, in the case of $v\equiv v_c$ and $d=3$ there 
is a contribution proportional to $(-z/\varepsilon_0)^{1/2}/z^2$ to the 
large-$\vert z\vert$ AS of $\wt{\Sigma}_{\sigma}^{(1)}(z)$ which 
survives substitution of $G_{\sigma}$ by its non-interacting counterpart 
$G_{0;\sigma}$, and of $W$ by its random-phase approximation as 
determined in terms of $G_{0;\sigma}$. Yet, in spite of these, there 
is {\sl no} evidence that, according to these widely used approximations, 
$\wt{\Sigma}_{\sigma}^{(1)}(z)$ would involve a term proportional to 
$(z-\varepsilon_F)^{3/2}$ in its AS for $\vert z -\varepsilon_F\vert \to 
0$. Further, one can show (B. Farid, 2001, unpublished) that for a system 
of Coulomb-interacting fermions confined to a plane (i.e. for $d=2$), 
the leading, $z$-independent, term in the large-$\vert z\vert$ AS of 
$\wt{\Sigma}_{\sigma}^{(1)}(z)$ is followed by a term proportional to 
$(-z/\varepsilon_0)^{1/2}/z$. The appearance of $(z-\varepsilon_0)^{1/2}$ 
in the small-$\vert z -\varepsilon_F\vert$ AS for 
$\wt{\Sigma}_{\sigma}^{(1)}(z)$ would not only imply a severe breakdown 
of the Fermi-liquid metallic state in $d=2$, but also that the value 
of the corresponding anomalous exponent would far exceed that of its 
counterpart pertaining to the {\sl one}-dimensional Luttinger (1963)
model (Mattis and Lieb 1965) (for a review see Voit (1994)). 
In spite of these arguments, it is possible that, from among many 
branch points at the point of infinity, {\sl some} of the associated 
points in $\{\varepsilon_j \| j=1,2,\dots \}$ may coincide with 
$\varepsilon_F$ (contrast with the extreme and unlikely proposition that
$\varepsilon_1 =\varepsilon_2=\dots =\varepsilon_F$). }
however, the fact remains that branch points of $\wt{\Sigma}_{\sigma}(z)$ 
at the point of infinity of the complex $z$ plane {\sl must necessarily} 
have their counterparts at finite values of energies along the real 
$\varepsilon$ axis. From this perspective, it is {\sl not} excluded that 
breakdown of Fermi-liquid metallic states in spatial dimensions $d$
greater than one, specifically in $d=2$, can be a consequence of the 
coincidence of one of the above-indicated energies $\{\varepsilon_j\}$ 
with $\varepsilon_F$. Accordingly, while attempting to establish the 
behaviour of $\wt{\Sigma}_{\sigma}(z)$ for $\vert z -\varepsilon_F\vert 
\to 0$, one may consider also to investigate the asymptotic behaviour 
of $\wt{\Sigma}_{\sigma}(z)$ for $\vert z\vert\to \infty$ as one of 
the preparatory steps.

\subsection{Narrow-band systems; a brief survey}
\label{s4}

Electronic systems with transition, rare-earth or actinide atoms, 
involving localized and open-shell $d$, $4f$ and $5f$ electrons 
respectively, manifest some of the most dramatic consequences of 
the electron-electron interaction (Fulde 1991, chapters 11, 12 
and 13). The strength of correlation in these systems is 
almost invariably quantified in terms of the value for the 
{\sl intra}-atomic energy $U$ ($I$ and $C$ in the earlier 
literature) in relation to the width $W$ of the tight-binding band 
associated with the overlap matrix element of an appropriate mean-field 
Hamiltonian with respect to the atomic (or Wannier) orbitals centred 
at the indicated atoms in neighbouring positions. Strong correlation 
in these systems is thus signified through the corresponding `large' 
values of the dimensionless constant $U/W$. Such a characterization 
has its parallel in uniform and isotropic systems (see \S~III.E), with 
the dimensionless parameter $r_s$ (see Eq.~(\ref{e94})) playing the role 
of $U/W$ where to leading order and up to a multiplicative constant 
of the order unity ($\approx -2.4$), $r_s$ is equal to the ratio of 
the GS expectation value of the interaction-energy operator $\wh{V}$ 
(see Eq.~(\ref{e2})) to that of the kinetic-energy operator $\wh{T}$, 
\footnote{\label{f13}
Measuring energy in Rydberg (see Eq.~(\protect\ref{e102}) below) and 
considering $v\equiv v_c$ in $d=3$, according to Gell-Mann and Brueckner 
(1957) (see also Pines and Nozi\`eres (1966)) for the total energy per 
particle $E_{N;0}/N$ of the paramagnetic GS of a uniform and isotropic 
electron system we have $E_{N;0}/N = 2.21/r_s^2 - 0.916/r_s + 0.062 
\ln(r_s) - 0.096 + \dots$, from which it is seen that the ratio of the 
second term (i.e. the {\sl bare} exchange energy) to the first one 
(i.e. the {\sl bare} kinetic energy) is indeed proportional to $r_s$. 
See text following Eq.~(\protect\ref{e101}) below. }
establishing a direct correspondence between large values of $r_s$ 
and strong correlation. Clearly, however, similar values of $U/W$ and 
$r_s$ do {\sl not} signify comparable physical consequences. Thus, 
whereas at the half-filling of the one-band Hubbard model in
$d=3$ (corresponding to one electron per orbital), $U/W \approx 1.15$ 
is the condition for the occurrence of metal-to-insulator transition 
(Hubbard 1964), not until $r_s \approx 10^2$ is this transition expected 
in uniform-electron systems (see Ceperley and Alder (1980), and Herman 
and March (1984) for a detailed analysis of the data by the former 
authors; for a more recent study see Ortiz, {\sl et al.} (1999)).

The low-energy physics of strongly correlated systems in which the 
{\sl intra-atomic} electron-electron interaction plays a prominent 
role is generally described by means of the Hubbard Hamiltonian 
(Anderson 1959, Ruijgrok 1962, Gutzwiller 1963, Izuyama, {\sl et al.} 
1963, Hubbard 1963, Kanamori 1963)
\footnote{\label{f14}
For a compilation of some of the relevant papers see Montorsi (1992). }
in which the number of orbitals per lattice cite can be as few as one, 
corresponding to the `one-band' Hubbard model. Such or comparable 
restrictions on the number of orbitals per cite implies that in contrast 
with the single-particle excitation spectrum pertaining to Hamiltonian
$\wh{H}$ in Eq.~(\ref{e1}), that pertaining to the Hubbard Hamiltonian 
(or any other Hamiltonian deduced from it) is {\sl bounded}. Consequently, 
$\wt{G}_{\sigma}({\Bf r},{\Bf r}';z)$ and $\wt{\Sigma}_{\sigma}({\Bf r},
{\Bf r}';z)$ pertaining to this model 
\footnote{\label{f15}
Here ${\Bf r}$ and ${\Bf r}'$ are vectors of a discrete set which 
thus may be denoted by ${\Bf r}_i$ and ${\Bf r}_j$ respectively. }
possess Laurent series expansions (see \S~II.B) for $\vert z\vert$ 
larger than some finite value which depends on the parameters of 
the Hamiltonian and $d$. Viewing these series as AS (see \S~II.B), we 
observe that, in dealing with the Hubbard Hamiltonian, the complications 
associated with the large-$\vert z\vert$ AS expansions of 
$\wt{G}_{\sigma}({\Bf r},{\Bf r}';z)$ and $\wt{\Sigma}_{\sigma}
({\Bf r},{\Bf r}';z)$ pertaining to the general Hamiltonian $\wh{H}$ 
in Eq.~(\ref{e1}) do {\sl not} arise. In other words, in the case of 
the Hubbard Hamiltonian, the status of $E$ as introduced in 
Eq.~(\ref{e38}) below is raised to that of a natural finite cut-off 
energy. Correspondingly, the energy moments integrals of the 
single-particle spectral function of this Hamiltonian (see 
Eqs.~(\ref{e37}) and (\ref{e38})) exist to any {\sl finite} order 
(see \S~I.B). With this property at hand, the problem concerning 
specification of the single-particle spectral function $A_{\sigma}
({\Bf r},{\Bf r}';\varepsilon)$ (see Eqs.~(\ref{e39}) and (\ref{e282}))
in terms of its $\varepsilon$-moments integrals is reduced to the 
classical ``problem of moments'' due to Tchebycheff (i.e. Chebyshev), 
Markoff (i.e. Markov), Stieltjes and others (for example Shohat and 
Tamarkin (1943)). To clarify this statement, we note that the 
$\varepsilon$ integrals in our work are Riemann integrals, specified 
by the Riemann measure ${\rm d}\varepsilon$, as opposed to Stieltjes 
integrals (Kreyszig 1978, pp.~225-227)
\footnote{\label{f16}
For excellent treatments of the theory of Stieltjes integrals see 
Titchmarsh (1939) and Hobson (1927). } 
specified by the measure ${\rm d}\xi(\varepsilon)$, where 
$\xi(\varepsilon)$ stands for a function of bounded variation over 
the relevant range of $\varepsilon$ so that, unless the single-particle 
spectral function $A_{\sigma}({\Bf r},{\Bf r}';\varepsilon)$ is one 
with exponentially decaying `tails' or with bounded support (such as 
is the case with the conventional Hubbard Hamiltonian), with the 
exception of a small number of low-order ones, the $\varepsilon$ 
moments of $A_{\sigma}({\Bf r},{\Bf r}';\varepsilon)$ pertaining to 
macroscopic systems do {\sl not} exist.
\footnote{\label{f17}
See \S~III.I.2 and recall that, in general, $A_{\sigma}({\Bf r},
{\Bf r}';\varepsilon)$ decays algebraically for $\vert\varepsilon\vert 
\to\infty$, as opposed to exponentially. }

According to Kubo and Tomita (1954), {\it ``the moment method''}, considered 
by the authors as {\it ``the most common and basic method''} used thus far 
{\it ``for the discussions of the line shapes of magnetic resonance 
absorption''}, was first employed in physics by Broer (1943), van Vleck 
(1948) and Pryce and Stevens (1951). To our knowledge, the application 
of the moments method in the form most akin to that discussed in \S~III.B 
of the present work is due to Harris and Lange (1967) and Kalashnikov and 
Fradkin (1969,1973). It also significantly features in the considerations 
by Nagaoka (1965,1966) and Brinkman and Rice (1970a).
\footnote{\label{f18}
For completeness, we also refer to work by Gordon (1968) which deals 
with the calculation of the thermodynamic properties of systems through 
employing the high-temperature expansion of the canonical partition 
function, which he formulates as a classical ``problem of moments''. }
In a series of 
articles by Nolting (1972, 1977, 1978), Nolting and Ole\'s (1979,1980,
1981,1987), Nolting and Borgie{\l} (1989), Eskes and Ole\'s (1994) and 
Eskes, {\sl et al.} (1994), the method has been extensively applied in 
conjunction with a two-pole {\sl Ansatz} for the single-particle spectral 
function as well as some appropriate auxiliary correlation functions 
(required for the purpose of achieving self-consistency in the pertinent 
calculations) pertaining to the Hubbard Hamiltonian. The significance of
correctly reproducing the $\varepsilon$ moments of the single-particle 
spectral function has been further accounted for in the construction of 
an approximate expression for the SE (Kajueter and Kotliar 1996) 
pertaining to the Hubbard Hamiltonian in $d=\infty$ (Metzner and 
Vollhardt 1989, Vollhardt 1993) 
\footnote{\label{f19}
In order to obtain a meaningful Hamiltonian for $d\to\infty$, the 
hopping term in the conventional Hubbard Hamiltonian is divided by 
$\sqrt{2 d}$ prior to taking the limit $d\to\infty$ (see Metzner and 
Vollhardt 1989). }
which is equivalent with the single-impurity Anderson (1961) 
model for the hybridized state of a single magnetic impurity and 
conduction electrons (Ohkawa 1991, Georges and Kotliar 1992).

The method of moments and the associated continued fraction expansions 
(for example, Gordon 1968) of the (single-particle) spectra of interacting 
systems {\sl within mean-field approximation frameworks} have been 
subjects of considerable interest in the past (see Haydock 1980). In 
the same way that application of the moments method in the case of 
interacting systems (see above) offers a way out of diagonalizing an 
interacting Hamiltonian (albeit in exchange for certain inaccuracies), it 
provides an alternative to bypassing exact diagonalization of mean-field 
Hamiltonians. Technically, however, particle-particle interaction renders 
calculation of the moments considerably difficult as it prevents these 
from being expressed in terms of products of matrices. Our considerations 
in the subsequent Sections of this paper clearly expose the complexity of 
the required calculations; these also show the prohibitive nature of the 
task of regularizing various contributions to the explicit expressions 
for the moments integrals in the cases where the interaction potential 
$v$ is not short range and bounded at zero separation of particles.

\subsection{The organization of the work}
\label{s5}

In \S~II.A we introduce the Hamiltonian on which our general 
considerations in this paper are based. In order to maintain 
flexibility, unless we indicate otherwise the dimension of the 
spatial space to which the system is confined is $d$ which is an 
{\sl integer}, however under specific circumstances can be made 
into a real or complex quantity (Wilson 1973).
\footnote{\label{f20}
For a comprehensive review of the Wilson-Fisher (Wilson 1973, 
Wilson and Fisher 1972) theory of dimensional continuation
see (Collins 1984, Chapter 4). For some specific details see (Farid 
2000b). The process of dimensional continuation in the style of Metzner 
and Vollhardt (1989) for the purpose of evaluating the limit $d\to
\infty$, requires some appropriate adjustments to the Hamiltonian 
of the system, so that it {\sl cannot} be identified with the
Wilson-Fisher process. }
Similarly, we leave the spin ${\sf s}$, corresponding to spin 
multiplicity $2 {\sf s}+1$, of the fermions unspecified.
\footnote{\label{f21}
Again, here we have in mind the possibility of employing such 
strategy as that of $1/{\sf s}$ expansion (see Auerbach 1993) at 
some future occasion. }
As our work is mainly directed towards determination of the behaviour 
of the single-particle GF $G_{\sigma}({\Bf r},{\Bf r}';\varepsilon)$ 
and the SE $\Sigma_{\sigma}({\Bf r},{\Bf r}';\varepsilon)$ at large
values of $\vert\varepsilon\vert$, in \S~II.B we present a detailed 
exposition of some specific aspects of the theory of AS of functions 
of complex variable $z$ for in particular large values of 
$\vert z\vert$. For two reasons we have considered inclusion of this 
exposition to be necessary: firstly, the available texts dealing with 
the theory of AS are {\sl not} sufficiently explicit in regard to the 
issues of primary relevance to our specific considerations in the 
present work;
\footnote{\label{f22}
We have, however, greatly benefited from specifically the text by 
Dingle (1973) on the subject matter. }
secondly, the main objects of our considerations are the Green and 
SE {\sl operators}, $G_{\sigma}(\varepsilon)$ and $\Sigma_{\sigma}
(\varepsilon)$ respectively, of which the above {\sl functions} are 
coordinate representations; it turns out that, in particular for systems 
of particles interacting through the Coulomb potential, these operators 
do {\sl not} admit of universal and meaningful {\sl finite-order} 
AS; rather such finite-order series crucially depend on the choice of 
the representation of these operators. In \S~II.B we give our full 
attention to this aspect, specifying the criteria for finite-order AS 
to be meaningful in the coordinate representation, and present the 
details concerning the unbounded contributions in the latter series as 
well as the principles underlying their regularization.

In \S~III we pursue the primary objective of the present work, namely 
determination of the general forms and explicit expressions for the 
terms in the formal AS corresponding to $G_{\sigma}({\Bf r},{\Bf r}';
\varepsilon)$ and $\Sigma_{\sigma}({\Bf r},{\Bf r}';\varepsilon)$ for 
$\vert\varepsilon\vert\to\infty$. We consider both the general case 
and the specific case concerning systems of fermions in $d=3$, 
interacting through the Coulomb potential $v_c$. We present the 
pertinent expressions in fully regularized form, amenable to direct 
numerical calculation; in Appendices we present the details 
underlying regularization of all unbounded contributions encountered 
in our work. In this Section we also consider some details concerning 
determination of the single-particle excitation energies in interacting 
systems, in particular in the regime of high excitation energies where 
our large-$\vert\varepsilon\vert$ asymptotic results for $\Sigma_{\sigma}
({\Bf r},{\Bf r}';\varepsilon)$ can be safely and fruitfully employed.
We devote the concluding parts of \S~III to a detailed consideration of 
a finite-order AS for ${\rm Im}\big[\Sigma_{\sigma}({\Bf r},{\Bf r}';
\varepsilon)\big]$ as $\vert\varepsilon\vert\to\infty$. Among others, 
these considerations clearly expose the ambiguities that can arise 
(and {\sl do} arise in the case of Coulomb interacting particles) as a 
result of exchanging the orders of infinite sums and integrals in 
the process of deducing finite-order AS for 
${\rm Im}\big[\Sigma_{\sigma}({\Bf r},{\Bf r}';\varepsilon)\big]$ 
corresponding to $\vert\varepsilon\vert\to\infty$.

In \S~IV we deal with the SE operator as determined within the
framework of the first-order perturbation theory in terms of the 
dynamically screened interaction function $W({\Bf r},{\Bf r}';
\varepsilon)$, as opposed to the bare interaction function 
$v({\Bf r}-{\Bf r}')$. In our general considerations we assume that 
the perturbation series is in terms of the {\sl exact} single-particle 
GF and the {\sl exact} screened interaction function. From the 
corresponding results we readily deduce those specific to the case 
in which the single-particle GF is identified with that pertaining to 
a mean-field Hamiltonian. We further explicitly demonstrate the 
significant deviation of the second leading term in the large-$\vert
\varepsilon\vert$ AS of the mentioned first-order result for 
$\Sigma_{\sigma}({\Bf r},{\Bf r}';\varepsilon)$ in comparison with 
the corresponding exact result, as deduced and presented in \S~III,
and put forward a practicable formalism that corrects for this and 
other similar deviations.

We conclude the main body of this work in \S~V by a summary and some 
remarks. In a relatively large number of Appendices (see Contents) 
we provide the details that either underly or are supplementary 
to our considerations in the main body of the paper. To facilitate 
use of this paper, in addition to the Contents, we provide two 
lists containing specifications of some of the frequently used symbols 
and abbreviations in this paper. Throughout this work we assume the 
value of the absolute temperature to be zero and the GS of the systems 
dealt with to be non-degenerate and {\sl normal} (as opposed to 
superfluid, for instance).

\section{Preliminaries}
\label{s6}

\subsection{The system and its Hamiltonian}
\label{s7}

In this work we deal with the following Hamiltonian
\begin{equation}
\label{e1}
\wh{H} = \wh{T} + \wh{U} + \wh{V},
\end{equation}
in which
\footnote{\label{f23}
Here the spatial integrals are over a $d$-dimensional space of 
volume $\Omega$. For macroscopic systems in $d$ spatial dimensions, 
we consider $\Omega = L^d$, where $L$ stands for the macroscopic 
length of the side of the $d$-dimensional hyper-cubic box into which 
the system is confined. }
\begin{eqnarray}
\label{e2}
\wh{T}&& {:=} \sum_{\sigma}\int {\rm d}^dr\;
\hat\psi_{\sigma}^{\dag}({\Bf r})\,\tau({\Bf r})\,
\hat\psi_{\sigma}({\Bf r}), \nonumber\\
\wh{U}&& {:=} \sum_{\sigma}\int {\rm d}^dr\;
\hat\psi_{\sigma}^{\dag}({\Bf r})\, u({\Bf r})\, 
\hat\psi_{\sigma}({\Bf r}),\\
\wh{V}&& {:=} \frac{1}{2} \sum_{\sigma,\sigma'}
\int {\rm d}^dr {\rm d}^dr'\;
\hat\psi_{\sigma}^{\dag}({\Bf r}) 
\hat\psi_{\sigma'}^{\dag}({\Bf r}') \nonumber\\
&&\;\;\;\;\;\;\;\;\;\;\;\;\;\;\;\;\;\;\;\;\;\;\;\;\;\;
\times v({\Bf r}-{\Bf r}')\,
\hat\psi_{\sigma'}({\Bf r}') 
\hat\psi_{\sigma}({\Bf r})\nonumber
\end{eqnarray}
are the second-quantization representations of the kinetic-energy
due to the single-particle kinetic-energy operator
\begin{equation}
\label{e3}
\tau({\Bf r}) {:=} \frac{-\hbar^2}{2 m_e} \, \nabla^2_{\Bf r},
\end{equation}
potential energy due to the {\sl local} electrostatic external potential 
$u({\Bf r})$ (e.g. ionic potential) and the particle-particle interaction 
energy, respectively. In Eq.~(\ref{e3}), $m_e$ denotes the mass of particles 
(e.g. electrons) and the subscript ${\Bf r}$ in $\nabla_{\Bf r}^2$ 
makes explicit that it operates on functions of ${\Bf r}$. In what 
follows we leave the dimension of the spatial space, $d$, unspecified. 
Further, in our general considerations we do not specify the two-body 
potential $v$; however, at regular intervals explicitly consider the case 
where $v$ is the long-range Coulomb potential in $d=3$, which we denote 
by $v_c$. Thus, whereas, in the following, $v$ is general and therefore 
may be identified with $v_c$, $v_c$ is specific and, unless explicitly 
indicated otherwise, strictly denotes the Coulomb potential in $d=3$. 
Further, unless we state to the contrary, in this paper, systems of 
fermions interacting through $v_c$ are implicitly assumed to be 
{\sl macroscopic}. Although our formalism is equally applicable to 
finite systems, independent of whether $v\equiv v_c$ or $v\not\equiv 
v_c$, some details of our considerations are redundant in regard to 
these systems.

For an appropriate treatment of the case corresponding to {\sl 
macroscopic} systems of particles interacting through the Coulomb 
potential
\footnote{\label{f24}
Since we define $v_c$ (see Eq.~(\protect\ref{e4})) in terms of $(-e)^2$, 
it has the dimension {\sl energy} and therefore is {\sl not} a 
{\sl potential} in the strict sense of the word. }
\begin{equation}
\label{e4}
v_c({\Bf r}-{\Bf r}') \equiv
\frac{e^2}{4\pi\epsilon_0} \frac{1}{\|{\Bf r}-{\Bf r}'\|},
\end{equation}
with $e^2$ the electron charge $-e$ squared and $\epsilon_0$ 
the vacuum permittivity, equal to $8.854\dots\times 10^{-12}$ F m$^{-1}$,
we need to supplement $\wh{H}$ in 
Eq.~(\ref{e1}) by $\wh{H}_{\kappa}$, where
\begin{equation}
\label{e5}
\wh{H}_{\kappa} {:=} -\varpi_{\kappa} \wh{N},\;\;\;
\varpi_{\kappa} {:=} \frac{e^2 n_0}{2\epsilon_0\kappa^2},
\end{equation}
and take the limit $\kappa\downarrow 0$ at the final stage of
calculations. In Eq.~(\ref{e5}) $\wh{N}$ stands for the {\sl total} 
number operator,
\begin{equation}
\label{e6}
\wh{N} \equiv \sum_{\sigma} \wh{N}_{\sigma},
\end{equation}
with 
\begin{equation}
\label{e7}
\wh{N}_{\sigma} {:=} \int {\rm d}^dr\; 
{\hat\psi}_{\sigma}^{\dag}({\Bf r})
{\hat\psi}_{\sigma}({\Bf r}),\;\;
\sigma\in \{-{\sf s},-{\sf s}+1,\dots,{\sf s}\},
\end{equation}
the {\sl partial} number operators, which mutually commute, i.e. 
\begin{equation}
\label{e8}
\big[ \wh{N}_{\sigma},\wh{N}_{\sigma'}\big]_- = 0,\;\;
\forall\, \sigma,\sigma'.
\end{equation}
Further, in Eq.~(\ref{e5})
\begin{equation}
\label{e9}
n_0 {:=} N/\Omega
\end{equation}
stands for the {\sl total} concentration of the fermions, with $N$ 
the {\sl total} number of fermions in the system (see Eq.~(\ref{e20}) 
below) and $\Omega$ the volume occupied by the system. For macroscopic
systems, that is those in the thermodynamic limit corresponding to
$N,\Omega\to\infty$ and $n_0$ a {\sl finite} constant, the Hamiltonian 
$\wh{H}_{\kappa}$ accounts for two contributions, namely the 
self-interaction energy of a positively charged uniform background 
of charge density $e n_0$, which is equal to $-\wh{H}_{\kappa}$, and 
the interaction energy of the negatively charged fermions with the 
above-mentioned background, which is equal to $2 \wh{H}_{\kappa}$ 
(Fetter and Walecka 1971, pp.~21-23). 
\footnote{\label{f25}
Our formulation slightly differs from that by Fetter and Walecka 
(1971) who do {\sl not} introduce $\wh{H}_{\kappa}$ involving the 
total number {\sl operator} $\wh{N}$; they rather introduce the 
$N$-particle GS expectation value of what we have denoted by 
$\wh{H}_{\kappa}$ which is sufficient so long as one does not need 
to deal with contributions to the SE operator beyond the 
first order in the particle-particle interaction. } 
The prefactor $2$ reflects the fact that here an energy of two 
mutually distinguishable systems of uniformly distributed particles 
is concerned; with each system containing $N$ particles, there are 
$N^2$ distinct pairs of interactions, which is to be contrasted with 
the self-interaction energy of a system of (uniformly distributed) 
$N$ particles for which there are $N (N-1)/2$ pairs of distinct 
interactions, which to the leading order in $N$ for $N\to\infty$ 
is equal to $N^2/2$.

In the above results, $\kappa$, which has the dimension of 
reciprocal metres, has its root in the `soft' cut-off function 
$\exp(-\kappa \|{\Bf r}-{\Bf r}'\|)$ with which $v_c({\Bf r}-{\Bf r})$ 
in Eq.~(\ref{e4}) has to be multiplied for the purpose of evaluating 
the electrostatic potential energies due to uniform charge 
distributions (see Eq.~(\ref{e13}) below). In this connection, we 
note that
\begin{equation}
\label{e10}
\int_{\Omega} {\rm d}^3r'\; 
\frac{\exp(-\kappa \|{\Bf r}-{\Bf r}'\|)}
{\|{\Bf r}-{\Bf r}'\|} \to \frac{4\pi}{\kappa^2}\;\;
\mbox{\rm for}\;\; \Omega\to \infty.
\end{equation}
In arriving at this result we have made the {\sl assumption} that $R 
\exp(-\kappa R)/\kappa \ll 1/\kappa^2$, where $R {:=} a \Omega^{1/3}$ 
with $a$ some finite positive constant (say, $a= (3/[4\pi])^{1/3}$), 
implying the condition $1/R \ll \kappa$. This signifies the fact that 
{\sl in using the result in Eq.~(\ref{e5}), $R\to\infty$ must be applied
{\it before} taking the limit $\kappa \downarrow 0$.} In view of our 
later considerations, we point out that, in cases where one encounters 
the `soft' cut-off $\exp(-\kappa \|{\Bf r}-{\Bf r}'\|)$, disregard for 
the condition $1/R \ll \kappa$ can lead to erroneous results, so that 
it is crucial to deal with integrals involving this cut-off function 
with sufficient care (see Appendix F, \S~1.b herein). It is further
important to realize that introduction of a cut-off function, such as 
$\exp(-\kappa \|{\Bf r}-{\Bf r}'\|)$, in conjunction with $v_c$, is 
{\sl not} a mere regularization device, but serves to account for the 
fact that the interaction potential $v_c$ mediates between two particles 
(of charge $-e$) described by square integrable wave functions (the 
totality of these functions constitutes the single-particle Hilbert 
space of the problem); this device thus enables one to introduce such 
gauge-non-invariant function as $\Sigma_{\sigma}({\Bf r},{\Bf r}'; 
\varepsilon)$ in complete isolation from the context in which it 
determines observable quantities (which are gauge invariant) where it 
is integrated with respect to ${\Bf r}$, ${\Bf r}'$ or ${\Bf r}$ 
{\sl and} ${\Bf r}'$ together with functions that are square integrable 
over an unbounded space (see, e.g., Eq.~(\ref{e84}) below).

In considering $\wh{H}_{\kappa} + \wh{H}$, with $\wh{H}$ as defined 
in Eq.~(\ref{e1}), we merely need to apply in Eq.~(\ref{e2}) the 
substitution
\begin{equation}
\label{e11}
u({\Bf r}) \rightharpoonup u({\Bf r}) - \varpi_{\kappa},
\end{equation}
on account of the definition for $\wh{N}$ in Eq.~(\ref{e6}) above. 
By decomposing, in {\sl macroscopic} systems, the {\sl total} number 
density $n({\Bf r})$ as follows
\begin{equation}
\label{e12}
n({\Bf r}) = n_0 + n'({\Bf r})\;\;
\mbox{\rm with}\;\;
\int {\rm d}^dr\; n'({\Bf r}) = 0,
\end{equation}
where $n_0$ is the total concentration of fermions, defined in 
Eq.~(\ref{e9}), with
\begin{eqnarray}
\label{e13}
v'({\Bf r}-{\Bf r}') {:=}
\left\{ \begin{array}{l}
v({\Bf r}-{\Bf r}'), \\ \\
v_c({\Bf r}-{\Bf r}')\,
{\rm e}^{-\kappa \|{\Bf r}-{\Bf r}'\|},\;\;\;
\kappa\downarrow 0,
\end{array} \right.
\end{eqnarray}
for the electrostatic Hartree
potential 
\begin{equation}
\label{e14}
v_H({\Bf r};[n]) {:=} \int {\rm d}^dr'\;
v'({\Bf r}-{\Bf r}')\, n({\Bf r}'),
\end{equation}
which is a {\sl linear} functional of $n$, in the case of $d=3$ 
and $v\equiv v_c$ we have
\begin{equation}
\label{e15}
v_H({\Bf r};[n]) \equiv \varpi_{\kappa} +
v_H({\Bf r};[n']).
\end{equation}
It is seen that $\varpi_{\kappa}$ cancels in the sum of the {\sl 
transformed} external potential, according to Eq.~(\ref{e11}), and 
$v_H({\Bf r};[n])$, according to Eq.~(\ref{e15}) so that, upon
adding the two potentials, $\kappa$ can be set equal to zero. For 
completeness, for finite systems of fermions (such as atoms and 
molecules) interacting through the Coulomb potential, $\kappa$ in 
Eq.~(\ref{e13}) can be set equal to zero from the outset. {\sl In the 
remaining part of this paper, unless we indicate or imply otherwise, 
$v({\Bf r}-{\Bf r}')$ will denote $v'({\Bf r}-{\Bf r}')$ as defined 
in Eq.~(\ref{e13}).}

As we shall see later in this work, the property
\begin{equation}
\label{e16}
\nabla_{\Bf r}^2\, v_c({\Bf r}-{\Bf r}')
= \frac{-e^2}{\epsilon_0}\, \delta({\Bf r}-{\Bf r}'),
\end{equation} 
together with $v_c({\Bf r}-{\Bf r}') \to\infty$ for $\|{\Bf r}-
{\Bf r}'\|\to 0$ have far-reaching consequences for systems of fermions 
interacting through the two-body potential $v_c$.  
\footnote{\label{f26}
From Eqs.~(\protect\ref{e14}) and (\protect\ref{e16}) one readily 
deduces $\nabla_{\Bf r}^2 v_H({\Bf r};[n]) = (-e^2/\epsilon_0)\, 
n'({\Bf r})$, with $n'({\Bf r})$ as defined in Eq.~(\protect\ref{e12}). 
The definiteness of this result depends on enforcing $\kappa=0$ 
{\sl subsequent to} evaluating $\nabla_{\Bf r}^2 v_H({\Bf r};[n])$ 
with $\kappa\not=0$. Without this, one would {\sl incorrectly} 
deduce $(-e^2/\epsilon_0)\, n({\Bf r})$ for the right-hand side
of the latter expression. }
This aspect is additional to that corresponding to the long range 
of $v_c({\Bf r} -{\Bf r}')$ for $\|{\Bf r}-{\Bf r}'\|\to \infty$ 
which is specific to macroscopic systems. 
\footnote{\label{f27}
It is important, however, to realize that the two properties are 
{\sl not} unrelated; both are associated with the fact that 
$v_c({\Bf r}-{\Bf r}')$ is a {\sl homogeneous} function (for example 
Ince 1927, p.~10) of degree minus unity, which is implicit in the 
expression in Eq.~(\protect\ref{e16}); with $\alpha > 0$, making use 
of the property $\delta(\alpha [{\Bf r}-{\Bf r}']) = \alpha^{-d} 
\delta({\Bf r}-{\Bf r}')$, $d=3$, from $\nabla_{\alpha\Bf r}^2 = 
\alpha^{-2} \nabla_{\Bf r}^2$ it follows that for $v_c$ satisfying 
Eq.~(\protect\ref{e16}), we must have $v_c(\alpha [{\Bf r}-{\Bf r}']) 
\equiv \alpha^{-1}\, v_c({\Bf r} -{\Bf r}')$. This is indeed satisfied 
by $v_c$ in Eq.~(\protect\ref{e4}). } 
To clarify this statement, suffice it for the moment to mention that 
in some of the relevant expressions (see Eqs.~(\ref{e202}) and 
(\ref{e203}) below), we will encounter terms involving $v({\Bf r}
-{\Bf r}') \tau({\Bf r}) v({\Bf r}-{\Bf r}')$, which, in view of 
Eqs.~(\ref{e3}), (\ref{e16}) and (\ref{e4}), is unbounded for 
$v\equiv v_c$. 

In the light of the above remarks, we should emphasize that the 
significance that we ascribe to the Coulomb potential in $d=3$ should 
{\sl not} be considered as implying that other two-body potentials 
than the Coulomb potential, and/or $d\not=3$ were devoid of such 
significance. It is in fact possible to bring about even very 
peculiar results for arbitrary $d$ by considering two-body 
potentials that are not necessarily physically feasible.

\subsection{Some details concerning asymptotic series}
\label{s8}

Since in this work our main attention is focussed on AS expansions 
(Whittaker and Watson 1927, chapter VIII, Copson 1965, Dingle 
1973, Murray 1974, Lauwerier 1977), here we briefly present some 
details concerning these. 

An {\sl asymptotic series} (AS) (in the sense of Poincar\'e
\footnote{\label{f28}
This restricted sense of AS (for details of the inherent 
restrictions of the Poincar\'e definition --- as well as the extended 
Poincar\'e-Watson definition --- see Dingle (1973)) is sufficient for 
our considerations in this paper. This follows from the fact that none 
of the functions of $\varepsilon$ dealt with in this paper involves 
exponentially-decaying components for $\vert\varepsilon\vert\to\infty$
(see footnote \protect\ref{f30} below).})
is based on an {\sl asymptotic sequence} (Lauwerier, 1977):
$\{\tilde\phi_m(z) \| m=0,1, \dots\}$ is an asymptotic sequence 
corresponding to the asymptotic region $z\to z_0$ provided that
$\tilde\phi_{m+1}(z)/\tilde\phi_m(z) \to 0$ for $z\to z_0$, which
property is formally denoted by $\tilde\phi_{m+1}(z) = 
o\big(\tilde\phi_m(z)\big)$, for $z\to z_0$. In our present work, 
$z_0$ is the point of infinity in the complex $z$ plane and $\{1,1/z,
1/z^2,\dots\}$ the asymptotic sequence of our choice; thus, unless 
we indicate otherwise, in what follows $z_0$ will be the point of 
infinity. 

For functions which are analytic outside a disc of finite radius 
centred around $z=0$, the series in terms of $\{1/z,1/z^2,\dots\}$ 
constitutes the {\sl principal part} (as opposed to the {\sl analytic 
part}) of the Laurent series of these functions; for functions which 
possess a pole of order $m$ at $z=0$ and are analytic elsewhere, the 
principal part of their corresponding Laurent series terminates at the 
$m$th term, corresponding to $1/z^m$; functions whose principal parts 
of their Laurent series do not terminate, must therefore possess a 
{\sl non}-isolated singularity at $z=0$. Through defining $\zeta {:=} 
1/z$ (which amounts to a conformal mapping), our AS would coincide 
with the Taylor series based upon $\zeta=0$ if the functions under 
consideration were analytic in (finite) neighbourhoods of the origin 
in the $\zeta$ plane. As a Taylor series is a power series, its 
region of convergence in the complex plane is a circular disk, inside 
which (that is, excluding the boundary of the disk) it converges 
absolutely and uniformly. In contrast, an AS may not converge.
\footnote{\label{f29}
In fact, according to Whittaker and Watson (1927), (Chapter VIII), 
divergence of a series is a prerequisite for this to be considered 
as an AS. }
Consequently, AS, in particular those based on the asymptotic 
sequence $\{1,1/z,1/z^2,\dots\}$, concern a wider class of functions 
than those capable of being described by means of Taylor or Laurent 
series.

For definiteness, consider the function ${\tilde f}(z)$ as representing 
the functions dealt with in the present work, such as the coordinate 
representations of the single-particle Green {\sl operator} 
$\wt{G}_{\sigma}(z)$ and of the associated SE {\sl operator} 
$\wt{\Sigma}_{\sigma}(z)$ (see footnote \ref{f11}). Further, consider 
the AS ${\tilde f}(z) \sim a_0 + a_1/z + a_2/z^2 + \dots$, with $a_0$, 
$a_1$, $\dots$, constants. By definition (according to Poincar\'e) we 
have $z^m \big({\tilde f}(z)-\sum_{j=0}^m a_j/z^j\big) \to 0$ as 
$\vert z\vert \to \infty$, for $m \ge 0$. 
\footnote{\label{f30}
The restriction of the Poincar\'e definition is evident from the fact that 
$\sum_{j=1}^m a_j/z^j$ is unique up to exponentially-decaying functions 
of $z$. In general, such exponential functions exist and play a crucial 
role in compensating for the Stokes discontinuities in asymptotic 
expansions of continuous functions on crossing the Stokes rays (see 
Dingle 1973); note in passing that these rays concern discontinuities 
in the {\sl form}, as opposed to those in the summed values, of these 
expansions (for example Morse and Feshbach (1953, p.~609)). }  
It is, however, possible that, for a given ${\tilde f}(z)$, the expression 
for $a_m$, for {\sl some} $m$, involves an unbounded contribution (see 
later), which we denote by $a_m^{\rm u}$ and thus write $a_m = a_m^{\rm u} 
+ a_m^{\rm b}$, where the bounded contribution $a_m^{\rm b}$ may or may 
not be vanishing. Considering the {\sl ordered} set $\{a_j\}$ as being 
defined in terms of integrals, $a_m^{\rm u}$ can be viewed as 
corresponding to a non-integrable integrand. Let $a_m$ be the first 
unbounded coefficient in the ordered set $\{a_j\}$. By {\sl assuming} 
${\tilde f}(z)$ to be a well-defined function, the {\sl unbounded} 
contribution $a_m^{\rm u}/z^m$, associated with $a_m/z^m$, in the 
AS ${\tilde f}(z) \sim a_0 + a_1/z + a_2/z^2 + \dots$ must be cancelled 
by equally unbounded contributions that {\sl must} necessarily exist 
in the series $a_{m+1}/z^{m+1} + a_{m+2}/z^{m+2} +\dots$, corresponding 
to unbounded coefficients $\{a_j^{\rm u} \| j > m\}$. The latter set 
{\sl cannot} be finite, for it is {\sl not} possible to maintain the 
balance between the unbounded contributions of the form 
$a_j^{\rm u}/z^j$, $j \ge m$, by means of a {\sl finite} set of 
coefficients for $z$ varying over an {\sl open} region of the complex 
$z$ plane, that is a neighbourhood of the point of infinity. Thus the 
series $\sum_{j=m}^{\infty} a_j^{\rm u}/z^j$, which for it to be 
meaningful must be viewed as representing an expression in which the 
summation over $j$ is carried out {\sl prior to} the explicit 
evaluation of each of $a_j^{\rm u}$, 
\footnote{\label{f31}
One may think of $a_j^{\rm u}$, $j \ge m$, as being defined in terms 
of integrals of bounded integrands, so that the unboundedness of these 
coefficients must be because these integrands are not integrable; the 
assumption with regard to the boundedness of $\sum_{j=m}^{\infty} 
a_m^{\rm u}/z^m$ is therefore a statement with regard to the 
integrability of the complete sum of the integrands associated with 
$a_j^{\rm u}/z^j$, $j\ge m$. Here the transcendental function 
$z^m \sum_{j=m}^{\infty} a_j^{\rm u}/z^j$ {\sl may} or {\sl may not}
involve a non-vanishing $z$-{\it in}dependent part; in cases it does, 
one could re-define $a_m^{\rm b}$ by absorbing this $z$-independent 
part into it so that {\sl by definition} $z^m \sum_{j=m}^{\infty} 
a_j^{\rm u}/z^j$ would be without a $z$-independent contribution. In 
this paper we do {\sl not} adopt such redefinition. See, for 
instance, Section 5.b in Appendix F. }
gives rise to a transcendental function of $z$ which for $\vert z\vert 
\to\infty$ is asymptotically {\sl more} dominant than $a_m^{\rm b}/z^m$. 
This follows from the fact that the possibility of the asymptotic 
dominance of $a_m^{\rm b}/z^m$ with respect to $\sum_{j=m}^{\infty} 
a_j^{\rm u}/z^j$ would be tantamount to $a_j^{\rm u}$ being vanishing, 
which is a contradiction. Following the same line of reasoning, one 
arrives at the conclusion that $\sum_{j=m}^{\infty} a_j^{\rm u}/z^j$ 
is asymptotically {\sl less} dominant than $a_{m-1}/z^{m-1}$ for 
$\vert z\vert\to\infty$ (assuming $m\ge 1$). In this work, we encounter 
a situation where $a_0$ and $a_1$ are bounded, but $a_2$ contains an 
unbounded contribution $a_2^{\rm u}$. The latter in turn consists of 
two contributions, $a_{2;1}^{\rm u}$ and $a_{2;2}^{\rm u}$ corresponding 
to two infinite sets $\{a_{2;1}^{\rm u}, a_{3;1}^{\rm u}, \dots\}$ and 
$\{a_{2;2}^{\rm u}, a_{3;2}^{\rm u},\dots\}$ for which $\sum_{j=2}^{\infty} 
a_{j;1}^{\rm u}/z^j \sim A_{2;1} \ln(-z/\varepsilon_0)/z^2$ and 
$\sum_{j=2}^{\infty} a_{j;2}^{\rm u}/z^j \sim A_{2;2} 
(-z/\varepsilon_0)^{1/2}/z^2$ hold, for $\vert z\vert \to\infty$. 
Note that, in the asymptotic region $\vert z\vert \to\infty$, both 
$\ln(-z/\varepsilon_0)/z^2$ and $(-z/\varepsilon_0)^{1/2}/z^2$ are 
indeed {\sl more} dominant
\footnote{\label{f32}
For more general AS than those according to the Poincar\'e definition, 
`more dominant' does {\sl not} necessarily imply `decaying less fast'. 
See the main text as well as footnote \protect\ref{f126}. }
than $1/z^2$ and {\sl less} dominant than $1/z$, in full conformity 
with our above general statement. 

The above discussions bring out one essential aspect related to 
{\sl finite}-order AS, namely that {\sl in general} deducing these 
through merely truncating an infinite-order (asymptotic) series  
(we refer to these as `{\it formal} finite-order AS'), gives rise 
to ill-defined functions and that derivation of meaningful 
finite-order AS may require partial summations of some infinite 
number of terms in the original infinite series. Unless implied by 
the context otherwise, our explicit references in this work to 
`finite-order asymptotic series' concern the well-defined (or 
regularized) AS consisting of finite number of terms.

A point of considerable importance to our investigations in this 
paper is the following. Consider 
\begin{eqnarray}
{\tilde f}({\Bf r};z) {:=} 
\int {\rm d}^dr'\; {\tilde g}({\Bf r},{\Bf r}';z)\nonumber
\end{eqnarray}
and suppose that 
\begin{eqnarray}
{\tilde g}({\Bf r},{\Bf r}';z) \sim \gamma_0({\Bf r},{\Bf r}') + 
\frac{\gamma_1({\Bf r},{\Bf r}')}{z} + \dots,\;\;
\mbox{\rm for}\;\; \vert z\vert\to\infty. \nonumber
\end{eqnarray}
From our above discussions it follows that with 
\begin{eqnarray}
\phi_j({\Bf r}) {:=} 
\int {\rm d}^dr'\; \gamma_j({\Bf r},{\Bf r}'), \nonumber
\end{eqnarray}
\begin{eqnarray}
{\tilde f}({\Bf r};z) 
\sim \sum_{j=0}^m \frac{\phi_j({\Bf r})}{z^j} \nonumber
\end{eqnarray}
is the AS with respect to the asymptotic sequence $\{1,1/z,\dots\}$ 
of ${\tilde f}({\Bf r};z)$, for $\vert z\vert \to\infty$, involving 
the complete $m+1$ {\sl leading} terms, provided that the integrals in 
terms of which the functions $\phi_j({\Bf r})$, $j=0,1,\dots, m$, are 
defined are bounded almost everywhere, that is bounded with the exception 
of a possible {\sl finite} set of ${\Bf r}$ points (see further; for 
periodic systems, the latter set of ${\Bf r}$ points concerns those in 
the primitive cell). Here it is important to note the {\sl sufficiency} 
of {\sl boundedness} of the integrals defining $\phi_j({\Bf r})$, 
$j=0,1,\dots, m$, for the validity of the latter statement. With 
reference to our above considerations, the possibility of 
{\sl non}-integrability of $\gamma_{m+1}({\Bf r},{\Bf r}')$ with 
respect to ${\Bf r}'$ (for almost {\sl all} ${\Bf r}$) implies that 
the most leading asymptotic term following $\phi_m({\Bf r})/z^m$ 
must decay either more {\sl slowly} than $1/\vert z\vert^{m+1}$ or as 
slowly as $1/\vert z\vert^{m}$, but with increasingly rapidly 
oscillatory behaviour as $\vert z\vert\to \infty$; for instance, with 
$z=\varepsilon$, one may encounter a function of the form 
$\cos(\varepsilon/\varepsilon_0 + \vartheta)/\varepsilon^{m}$, with 
$\varepsilon_0$ a constant energy and $\vartheta$ a constant phase. 
\footnote{\label{f33}
Recall that $\cos(z)$ is a transcendental function of $z$. }
Considerations involving AS of this nature lie outside the scope 
of the Poincar\'e formalism and further will not concern us in this 
paper (see footnote \ref{f28}) so that, in the context of our 
considerations in this paper, the term following $\phi_{m}({\Bf r})/z^m$ 
necessarily should decay more {\sl slowly} than $1/\vert z\vert^{m+1}$ 
when $\gamma_{m+1}({\Bf r},{\Bf r}')$ is non-integrable with respect to 
${\Bf r}'$ for almost {\sl all} ${\Bf r}$. 

In this work we shall be dealing with systems in which particles interact 
through, in principle, arbitrary potentials $v({\Bf r}-{\Bf r}')$ and 
$u({\Bf r})$, with each other and with a background respectively. As 
is the case with the Coulomb potential in $d=3$, $v({\Bf r}-{\Bf r}')$ 
can increase {\sl indefinitely} for $\|{\Bf r}-{\Bf r}'\|\to 0$, and
similarly can $\vert u({\Bf r})\vert$ increase indefinitely for $\|{\Bf r}
-{\Bf R}_j\|\to 0$, where $\{ {\Bf R}_j\}$ denotes the set of position 
vectors of the atomic nuclei in the system (see Appendix K). Both 
because of these and owing to the unboundedness of the kinetic-energy 
operator (for example Kreyszig (1978, chapter 10)),
\footnote{\label{f34}
In many practical calculations, through imposing ultraviolet (or 
large-momentum) cut-offs, the unbounded character of the kinetic-energy 
operator is suppressed. This is specifically the case in the conventional 
Hubbard model (for example Montorsi (1992), Gebhard (1997)) where the 
ultraviolet cut-off is implied by the finite number of (non-interacting) 
bands (or finite number of orbitals per lattice point) taken into 
account. }
we are confronted with a situation where in principle the problem has 
{\sl no} {\sl a priori} high-energy scale which is required for the 
notion `large $\vert\varepsilon\vert$' to have a precise meaning. 
\footnote{\label{f35} 
Later in this Section we specify how in practice an appropriate 
energy scale can be identified with respect to which $\vert\varepsilon
\vert$ can be considered as large. } 
This aspect is clearly illustrated by considering an example, taken 
from \S~III.H (see also Appendix H), where in what we call `the 
large-$\vert z\vert$ asymptotic series' for the SE operator in the 
coordinate representation, $\wt{\Sigma}_{\sigma}({\Bf r},{\Bf r}';z)$ 
(see Eqs.~(\ref{e62}) and (\ref{e72}) below), we encounter the sequence 
(see Eqs.~(\ref{e173}), (\ref{e185}) and (\ref{e199}); see also 
Eq.~(\ref{e212})) 
\begin{eqnarray}
\big\{ v^m({\Bf r}-{\Bf r}')\varrho_{\sigma}({\Bf r}',{\Bf r})/z^{m-1} 
\| m=1,2, \dots\big\}. \nonumber
\end{eqnarray}
Assuming $v$ to be the Coulomb potential $v_c$ in $d=3$, it is seen 
that whatever large finite value is assigned to $\vert z\vert$, by 
decreasing $\|{\Bf r}-{\Bf r}'\|$ beyond a certain $z$-dependent 
limit, the terms in this sequence, in the order presented, can be 
made to {\sl increase} in amplitude rather than to decrease, in evident 
contradiction with the perception that one may have of AS (see later). 
Irrespective of whether such understanding of AS is correct or not 
(it is {\sl not}, as we shall discuss below), this example brings out 
the mechanism by which in particular the short-range part of the 
Coulomb interaction potential stands in the way of introducing an 
{\sl a priori} high-energy scale in the context of the AS expansion 
of $\Sigma_{\sigma}({\Bf r},{\Bf r}';\varepsilon)$ for 
$\vert\varepsilon\vert\to\infty$.

The `perception' with regard to AS, to which we have referred above, 
is {\sl false}. This is directly established by realizing the fact 
that AS are named such by the very aspect of being a series 
{\sl in terms of} an {\sl asymptotic sequence} appropriate to some 
asymptotic region. In fact, if the terms in an AS had to be 
monotonically decreasing in magnitude, by the D'Alembert criterion all 
AS would have to be absolutely convergent in sufficiently small but 
non-vanishing neighbourhoods of the asymptotic point. As we have 
indicated at the outset of this Section (see footnote \ref{f29}), 
however, the authoritative text by Whittaker and Watson (1927, 
chapter VIII) explicitly reserves the designation `asymptotic series' 
for those which, {\sl in addition to} being asymptotic in the sense 
just indicated, are also divergent (see also Dingle (1973)).
\footnote{\label{f36}
In fact, in contrast with Taylor series and the principal parts of 
Laurent series which may be considered as being AS in terms of the 
asymptotic sequences $\{1,z,z^2,\dots\}$ and $\{1/z, 1/z^2,\dots\}$ 
appropriate to the asymptotic points $z=0$ and $1/z=0$ respectively, 
and whose convergence for a particular $z$ in a neighbourhood of 
these points, say $z=z_0$, implies their convergence for {\sl all} 
$z = \vert z_0\vert \exp(i\varphi)$, $\varphi\in [0,2\pi]$, 
{\sl asymptotic} series are in general associated with functions that
do {\sl not} admit such {\sl uniform} representations as Taylor 
and Laurent series. Consider for instance ${\tilde f}(z)$ which we 
assume to be bounded in a neighbourhood of $z=z_0$. The AS 
${\tilde f}(z) \sim f_0 + f_1 (z-z_0) + f_2 (z-z_0)^2 + \dots$, for 
$z\to z_0$, either converges when $\vert z - z_0 \vert < \eta$ for 
{\sl some} positive $\eta$ (in which case it converges uniformly), or 
it diverges for {\sl any} positive $\eta$. The latter divergence signals 
the non-analyticity of ${\tilde f}(z)$ at $z=z_0$; $z_0$ is for instance
a branch point of ${\tilde f}(z)$. Consequently, the neighbourhood of 
$z_0$ can be subdivided into a number of sectors in each of which 
${\tilde f}(z)$ admits an AS specific to that sector. Consider 
(Lauwerier 1977, p.~11) ${\tilde f}(z) {:=} \exp(z) + \exp(-z) 
\tanh(1/z)$ which has the following AS for $z\to 0$: ${\tilde f}(z) 
\sim 2 \cosh(z) \sim 2 + z^2 + \dots$, for ${\rm Re}(z) > 0$, and 
${\tilde f}(z) \sim 2 \sinh(z) \sim 2 z + z^3/3 + \dots$, for 
${\rm Re}(z) < 0$. }
For clarity, consider the series $\sum_{m=0}^{\infty} (-1)^m m! z^m$, 
which is divergent for {\sl any} non-vanishing $z$; from the D'Alembert 
criterion, it follows that, for this series to be convergent, it is 
necessary that $\vert z\vert \le 1/(m+1)$ as $m\to\infty$. This series 
can however be summed by means of the Borel, or Euler, summation 
technique (for example Lauwerier (1977)), resulting in (Farid 1999c) 
$\exp(1/z) {\rm E}_1(1/z)/z {=:} {\tilde f}(z)$, where ${\rm E}_1(\zeta)$ 
stands for the exponential-integral function (Abramowitz and Stegun 
1972, p.~228) (see also Appendix I for some relevant details); any 
finite-order truncation of the above series is an AS (with respect to 
the asymptotic sequence $\{1,z,z^2,\dots\}$) of ${\tilde f}(z)$ for 
$\vert z\vert\to 0$, with $-\pi < {\rm arg}(z) < \pi$. 

In spite of the above, it remains worth enquiring whether an AS for 
the SE operator corresponding to $\vert z\vert\to\infty$, possessing 
the peculiarities associated with $\|{\Bf r} -{\Bf r}'\|\to 0$ 
considered above, can be of {\sl physical} relevance. It is further 
of significance to know the energy scale beyond which a finite-order 
AS for the SE operator corresponding to $\vert z\vert \to\infty$ 
can be considered as being an accurate representation of the exact 
$\wt{\Sigma}_{\sigma}({\Bf r},{\Bf r}';z)$. Below we shall briefly 
address these two issues.

--- Firstly, the peculiarity concerning the short-distance behaviour 
of the coefficients in the large-$\vert z\vert$ AS for 
$\wt{\Sigma}_{\sigma}({\Bf r},{\Bf r}';z)$, pertaining to systems of
Coulomb-interacting particles, is {\sl partly} (see later) an aspect 
specific to the coordinate representation of $\wt{\Sigma}_{\sigma}(z)$, 
which gives, by its very nature, prominence to the behaviour of this 
{\sl operator} on {\sl all} length scales; such indiscriminate 
representation 
\footnote{\label{f37}
In this context it is important to recall that $\{ \vert {\Bf r}
\rangle\}$ are the eigenstates of the {\sl unbounded} 
${\hat {\Bf r}}$ operator which are $\delta$ function normalized, 
that is $\langle {\Bf r} \vert {\Bf r}'\rangle = \delta({\Bf r}
-{\Bf r}')$. } 
is physically non-essential for the SE operator which is not gauge 
invariant and therefore not an observable. Observable quantities, 
insofar as they are determined by $\wt{\Sigma}_{\sigma}(z)$, such 
as the single-particle excitation {\sl energies} (see \S~III.D), involve 
matrix elements of $\wt{\Sigma}_{\sigma}(z)$ with respect to a set 
of basis functions spanning the single-particle Hilbert space of the 
system under consideration (contrast this with the matrix elements of 
$\wt{\Sigma}_{\sigma}(z)$ with respect to $\vert {\Bf r}\rangle$ and 
$\vert {\Bf r}'\rangle$; see footnote \ref{f37}). Expressing the 
latter matrix elements in terms of integrals of $\wt{\Sigma}_{\sigma}
({\Bf r},{\Bf r}';z)$ and the coordinate representation of the pertinent 
basis functions ({\it cf}. Eq.~(\ref{e84}) below), one observes that a 
finite-order AS for $\wt{\Sigma}_{\sigma}({\Bf r},{\Bf r}';z)$, 
$\vert z\vert\to\infty$, is unequivocally well-defined provided its 
constituent functions the following hold true.

\vspace{0.2cm}
(A) They are bounded (more precisely, bounded almost everywhere).

\vspace{0.2cm}
(B) They are integrable (here, in the sense of Riemann) with 
respect to ${\Bf r}$ (or, equivalently, ${\Bf r}'$, since 
$\wt{\Sigma}_{\sigma}({\Bf r},{\Bf r}';z)$ is symmetric with 
respect to ${\Bf r} \rightleftharpoons {\Bf r}'$) over {\sl any} 
{\sl finite} region of the single-particle configuration space. 

\vspace{0.2cm}
(C) Also, for systems in the thermodynamic limit, they decay 
sufficiently rapidly at large distances from the origin, so that 
the pertinent integrals (see (B)) converge to finite values as 
the size of the system is made to approach infinity. 

\vspace{0.2cm}
From these requirements it follows that, for $v\equiv v_c$ in $d=3$, 
the terms $v_c^m({\Bf r}-{\Bf r}') \varrho_{\sigma}({\Bf r}',
{\Bf r})/z^{m-1}$, presented above, corresponding to $m=1,2$ are 
unreservedly well defined in a finite-order large-$\vert z\vert$ 
AS for $\wt{\Sigma}_{\sigma}({\Bf r},{\Bf r}';z)$, but further 
terms, corresponding to $m > 2$, are {\sl not}. We note in passing 
that this is the aspect to which we referred above through our use 
of the word `partly'; the non-integrability of terms $v_c^m({\Bf r}
-{\Bf r}') \varrho_{\sigma}({\Bf r}',{\Bf r})/z^{m-1}$ for $m > 2$, 
though an aspect of the coordinate representation of the pertinent 
operators, reflects the fact that, in the case at hand and in the 
momentum representation, the contributions to the SE operator 
immediately following those decaying like $1/z$, are {\sl not} those 
decaying like $1/z^2$, but some which decay more {\sl slowly} than this 
(but {\sl rapidly} than $1/\vert z\vert$). The {\sl total} contribution 
of the sequence of non-integrable terms  $\{ v_c^m({\Bf r}-{\Bf r}') 
\varrho_{\sigma}({\Bf r}',{\Bf r})/z^{m-1}\, \|\, m=3,4, \dots\}$ is 
directly seen to be equal to $z^{-2} v_c^3({\Bf r}-{\Bf r}') 
\varrho_{\sigma}({\Bf r}',{\Bf r})/[1-v_c({\Bf r}-{\Bf r}')/z]$, 
which is indeed an integrable function of ${\Bf r}$ and ${\Bf r}'$. 
As we demonstrate in Appendix H, the momentum representation of the 
latter function involves a term proportional to 
$\ln(-z/\varepsilon_0)/z^2$, where $\varepsilon_0$ stands for a 
constant energy, followed by one proportional to $1/z^2$.

--- Secondly, considering for the moment a bounded and short-range 
interaction potential $v$, a close inspection of the explicit 
expressions for the coefficient functions $\Sigma_{\sigma;\infty_m}
({\Bf r},{\Bf r}')$ corresponding to $m=0,1,2$ (see \S\S~III.F, III.G 
and III.H; see also Eq.~(\ref{e72}) below for the definition), reveals 
that $\Sigma_{\sigma;\infty_0}({\Bf r},{\Bf r}')$ and $\Sigma_{\sigma;
\infty_1}({\Bf r},{\Bf r}')$ are {\sl explicitly} proportional to 
the first and the second powers respectively of the coupling constant 
of the particle-particle interaction; {\sl higher-order coefficients 
$\Sigma_{\sigma;\infty_m}({\Bf r},{\Bf r}')$, corresponding to 
$m \ge 2$, are determined by a complex interplay between the 
interaction and the kinematics of the interacting particles, 
manifesting itself in a range of contributions which depend on all 
powers of the coupling constant of interaction from $2$ up to and 
including $m+1$} ({\it cf}. Eqs.~(\ref{e107})-(\ref{e109}) below). The 
latter aspect follows from the fact that the large-$\vert z\vert$ 
AS for $\wt{\Sigma}_{\sigma}({\Bf r},{\Bf r}';z)$ amounts to an 
ordering of interaction effects in accordance with powers of $1/z$ 
associated with them rather than powers of the coupling constant 
of interaction which corresponds to the ordering scheme of the 
many-body perturbation theory. To be explicit, a general term in 
the expression for $\Sigma_{\sigma;\infty_m}$, pertaining to a 
system described by the Hamiltonian in Eqs.~(\ref{e1}) and (\ref{e2}), 
involves $p_1$ times $\tau$, $p_2$ times $u$ and $p_3$ times $v$ (all 
in various orders), with $p_1$, $p_2$ and $p_3$ constrained by the 
condition $p_1 + p_2 + p_3 = m+1$; in the case of $m=0$, by necessity 
we have $p_1=p_2=0$, whereas since the first-order SE contributions 
in terms of $v$ are static and can therefore only contribute to 
$\Sigma_{\sigma;\infty_0}$, and since $v$ is a {\sl two}-particle 
interaction function, for $m\ge 1$, $p_3 \ge 2$ (see, for example, 
Eqs.~(\ref{e107}), (\ref{e108}) and (\ref{e109}) below). 

In the specific case where $v\equiv v_c$ in $d=3$ and for $m=0,1$, 
the dependences of $\Sigma_{\sigma;\infty_m}({\Bf r},{\Bf r}')$ on 
the coupling constant of interaction are the same as for bounded and
short-range interaction potentials $v$ discussed above; for $m \ge 2$ 
however, where $\Sigma_{\sigma;\infty_m}({\Bf r},{\Bf r}')$ involves 
{\sl fundamentally unbounded} contributions (it also involves
{\sl non-integrable} contributions whose regularization becomes
necessary for the determination of the momentum representation of 
$\Sigma_{\sigma;\infty_m}$; see conditions (A) - (C) introduced 
and discussed above), on regularizing these, the dependence on the 
coupling constant of interaction of the regularized 
$\Sigma_{\sigma;\infty_m}({\Bf r},{\Bf r}')$, which we denote by 
$\Sigma_{\sigma;\infty_m}({\Bf r},{\Bf r}'\vert\varepsilon)$ (see 
\S~III.E.2), is no longer polynomial, but transcendental; as we show 
in \S~III.H, $\Sigma_{\sigma;\infty_2}({\Bf r},{\Bf r}'\vert
\varepsilon)$ involves the logarithm of this coupling constant (see, 
e.g., Eq.~(\ref{e114}) below; we note in passing that the dependence 
of the right-hand side (RHS) of Eq.~(\ref{e113}) on the logarithm of 
the coupling constant of interaction has its origin in the regularization 
of the aforementioned {\sl non-integrable} contribution in 
$\Sigma_{\sigma;\infty_2}({\Bf r},{\Bf r}')$).

In metallic systems, the spherical average of the radius of the Fermi 
surface, $k_F$, defines the length-scale $\ell_F {:=} 1/k_F$ relevant 
to the low-temperature transport and thermodynamic properties of these 
systems. Consequently, for these systems, $\hbar\Sigma_{\sigma;\infty_m}
({\Bf r},{\Bf r}')$ should be proportional to $(g/\ell_F)^{m+1}$ for 
$m=0,1$, where $g$ stands for the coupling-constant of the particle-particle 
interaction; for $v\equiv v_c$ in $d=3$ we have $g = g_c\equiv
e^2/(4\pi\epsilon_0)$ so that, using $k_F = (3\pi^2 n_0)^{1/3}$, 
applicable to systems with simply-connected Fermi-surface geometries 
(here $n_0$ denotes the average total electron concentration; see 
Eq.~(\ref{e9}) above), we deduce the energy scale $e^2/(4\pi\epsilon_0 
\ell_F) {=:} e_0'$ to be on the order of $2$ Hartrees or less ($1$ 
Hartree $=2$ Rydberg $\approx 27.21$ eV); for Sodium, for instance, this 
scale is $\frac{1}{2}$ Hartree (for a precise formulation see \S~III.E 
and in particular \S~III.E.4). For these systems, on this scale of 
energies, $\hbar\Sigma_{\sigma;\infty_1}/\varepsilon$ should become 
comparable with $\Sigma_{\sigma; \infty_0}$. In \S~III.E where we 
introduce the formalism appropriate for dealing with uniform and 
isotropic GSs, we provide a quantitative analysis of the result given 
here and further determine the role played by contributions in the 
AS for the SE that decay more rapidly than $1/\varepsilon$ for 
$\vert\varepsilon\vert\to\infty$. However, irrespective of these 
quantitative results, it would be quite mistaken to assess the utility 
of a finite-order AS for $\wt{\Sigma}_{\sigma}(z)$ corresponding to
$\vert z\vert\to\infty$ solely in terms of the accuracy with which it 
describes the exact $\wt{\Sigma}_{\sigma}(z)$ at sufficiently large 
values of $\vert z\vert$. As we discuss in \S~III.B in some detail, 
such series, when exact to order $m$ in $1/z$, lead to the {\sl exact} 
large-$\vert z\vert$ AS of the single-particle GF $\wt{G}_{\sigma}(z)$ 
to all orders up to and including $m+2$ in $1/z$. 
\footnote{\label{f38}
From the Dyson equation one obtains $\wt{G}_{\sigma}(z) = \big( I - 
G_{0;\sigma}(z) \Sigma_{\sigma}(z)\big)^{-1} G_{0;\sigma}(z) = 
\wt{G}_{0;\sigma} + \wt{G}_{0;\sigma}(z) \wt{\Sigma}_{0;\sigma}(z) 
\wt{G}_{0;\sigma}(z) + \dots$. The fact that $\wt{\Sigma}_{\sigma}(z)$ 
is `sandwiched' between at least {\sl two} $\wt{G}_{0;\sigma}(z)$ 
functions, with $\wt{G}_{0;\sigma}(z) \sim \hbar I/z$ for $\vert 
z\vert\to \infty$ (see Eq.~(\protect\ref{e61})), underlies the $2$ 
in $m+2$. }
This in turn implies, via Eqs.~(\ref{e37}) and (\ref{e38}), that the 
energy moment integrals of the corresponding single-particle 
spectral function are exact to order $m+1$ (see footnote \ref{f6}). 
In other words, insofar as a particular finite-order energy moment 
integral of the single-particle spectral function is concerned (for 
definitions see Eqs.~(\ref{e37}) and (\ref{e38}) below), it is entirely 
{\sl immaterial} whether or not (if at all), in a finite-order AS for 
$\wt{\Sigma}_{\sigma}(z)$, the terms beyond a certain order in $1/z$ 
are properly taken into account. In \S~III.E.6, after having explicitly 
introduced some of the necessary ingredients that are essential to our 
present work, we clarify the above statements by means of a simple 
example concerning uniform and isotropic systems of fermions.

\section{Theory}
\label{s9}

\subsection{Generalities}
\label{s10}

Our starting point in this work is the Lehmann (1954) spectral 
representation for the single-particle GF corresponding to the 
$N$-particle GS (which we assume to be non-degenerate and 
{\sl normal}, i.e. with {\sl no} off-diagonal long-range order;
for definition see Reichl (1980)) of a system of spin-${\sf s}$ 
fermions, with ${\sf s}$ {\sl any} half-integer:
\begin{eqnarray}
\label{e17}
&&G_{\sigma}({\Bf r},{\Bf r}';\varepsilon)
= \hbar \sum_s f_{s;\sigma}({\Bf r}) 
f_{s;\sigma}^*({\Bf r}')\nonumber\\
&&\;\;\;\;\;\;\;\;\;
\times\left\{
 {{\Theta(\mu -\varepsilon_{s;\sigma})}\over 
\varepsilon - \varepsilon_{s;\sigma} - i\eta}
+{{\Theta(\varepsilon_{s;\sigma} -\mu)}\over 
\varepsilon - \varepsilon_{s;\sigma} + i\eta}
\right\}, \; \eta\downarrow 0,
\end{eqnarray}
where $\{ f_{s;\sigma}({\Bf r})\}$ are the `Lehmann amplitudes'
defined by
\begin{eqnarray}
\label{e18}
f_{s;\sigma}({\Bf r}) 
{:=} \left\{ \begin{array}{ll}
\langle\Psi_{N_{\sigma}-1,N_{\bar\sigma};s}\vert\hat\psi_{\sigma}
({\Bf r})\vert\Psi_{N;0}\rangle,
&\;\; \varepsilon_{s;\sigma} < \mu, \\ \\
\langle\Psi_{N;0}\vert\hat\psi_{\sigma}
({\Bf r})\vert\Psi_{N_{\sigma}+1,N_{\bar\sigma};s}\rangle,
&\;\; \varepsilon_{s;\sigma} > \mu,
\end{array} \right.
\end{eqnarray}
and
\begin{eqnarray}
\label{e19}
\varepsilon_{s;\sigma} {:=} \left\{ \begin{array}{ll}
E_{N;0} - E_{N_{\sigma}-1,N_{\bar\sigma};s},
&\;\;\; \varepsilon_{s;\sigma} < \mu, \\ \\
E_{N_{\sigma}+1,N_{\bar\sigma};s} - E_{N;0},
&\;\;\; \varepsilon_{s;\sigma} > \mu,
\end{array} \right.
\end{eqnarray}
are the single-particle excitation energies; here $\mu$ stands for 
the `chemical potential' (see text following Eq.~(\ref{e22}) below). 
With $\bar\sigma$ denoting the {\sl set} of $2 {\sf s}$ spin indices
complementary to $\sigma$, throughout this work we consider
\begin{equation}
\label{e20}
N = N_{\sigma} + N_{\bar\sigma},
\end{equation}
where 
\begin{equation}
\label{e21}
N_{\bar\sigma} {:=} \sum_{\sigma'\not=\sigma} N_{\sigma'}.
\end{equation}
In Eq.~(\ref{e18}), $\vert\Psi_{M_{\sigma},M_{\bar\sigma};s}\rangle$ 
denotes a simultaneous {\sl normalized} eigenstate of the many-body 
Hamiltonian $\wh{H}$, with $E_{M_{\sigma},M_{\bar\sigma};s}$ the 
corresponding eigen-energy,
\footnote{\label{f39}
We point out that our use of the word `eigen-energy', which in a 
broader context would have been `eigen-value', is suggestive that
$\wh{H}$ possesses a `point spectrum', which for systems with 
continuous degrees of freedom is incorrect. In general, the {\sl 
spectrum} of an operator is defined in terms of its associated 
{\sl resolvent operator} (here $\wt{G}_{\sigma}(z)$) and can be 
grouped into three {\sl disjoint} sets of {\sl a point spectrum} 
(i.e. {\sl eigenvalues}), {\sl a continuous spectrum} and 
{\sl a residual spectrum}. For a detailed discussion see the sections 
on the spectral theory in normed spaces in, for instance, Kreyszig 
(1978) and Debnath and Mikusi\'nski (1990). For discussions 
concerning the relevance of the resolvent operator and its definition 
in terms of the Stieltjes integral, see the book by von Neumann 
(1955). See also footnote \protect\ref{f2}. In our considerations 
in this work, the set $\{s\}$ is assumed to exhaust the complete 
spectral contents of $\wh{H}$. }
and the {\sl partial} number operators $\wh{N}_{\sigma}$, $\sigma= 
-{\sf s}, -{\sf s}+1, \dots, {\sf s}$, defined in Eq.~(\ref{e7}), 
with $\{M_{\sigma}\}$, the corresponding eigen-numbers; in 
Eqs.~(\ref{e18}) and (\ref{e19}), we have specifically employed 
the short-hand notations $\vert\Psi_{N;s}\rangle \equiv 
\vert\Psi_{N_{\sigma},N_{\bar\sigma};s}\rangle$ and $E_{N;s} 
\equiv E_{N_{\sigma}, N_{\bar\sigma};s}$. 

Above, $s$ is a compound variable which consists of all indices that 
characterize the pertinent eigenstates of $\wh{H}$ (see text following 
Eq.~(\ref{e44}) below and footnotes \ref{f50} and \ref{f109}), with 
$s=0$ reserved to {\sl symbolize} the non-degenerate GS of the system.
Further, $\mu$ denotes the ``chemical potential'', 
satisfying
\begin{equation}
\label{e22}
\mu_{N;\sigma}^- < \mu < \mu_{N;\sigma}^+,
\;\;\; \sigma\in \{-{\sf s},-{\sf s}+1,\dots,{\sf s}\},
\end{equation}
where
\begin{eqnarray}
\label{e23}
\mu_{N;\sigma}^- &{:=}& E_{N;0}-E_{N_{\sigma}-1,N_{\bar\sigma};0}, 
\nonumber\\
\mu_{N;\sigma}^+ &{:=}& E_{N_{\sigma}+1,N_{\bar\sigma};0}-E_{N;0}.
\end{eqnarray}
The existence of a $\mu$ which satisfies Eq.~(\ref{e22}) is implied 
by the requirement of stability of the GS of the system (see footnote 
\ref{f60}). In fact, $\mu_{N;\sigma}^{-}$ is independent of $\sigma$, 
that is it takes on the same value for {\sl all} $\sigma$ for which 
$N_{\sigma}\not=0$ holds. 

For our following considerations it is advantageous to introduce
the analytic continuation of $G_{\sigma}({\Bf r},{\Bf r}';\varepsilon)$
into the {\sl physical} Riemann sheet of the $z$ plane (see Farid 
1999c); by doing so, no need will arise to deal separately with 
retarded and advanced GFs. Denoting the analytically continued 
function by $\wt{G}_{\sigma}({\Bf r},{\Bf r}';z)$, for 
${\rm Im}\{z\}\not=0$ we have
\begin{equation}
\label{e24}
\wt{G}_{\sigma}({\Bf r},{\Bf r}';z)
= \hbar \sum_s \frac{f_{s;\sigma}({\Bf r}) 
f_{s;\sigma}^*({\Bf r}')}{z - \varepsilon_{s;\sigma}},
\end{equation}
from which the physical $G_{\sigma}({\Bf r},{\Bf r}';\varepsilon)$ 
as presented in Eq.~(\ref{e17}) is deduced according to 
\begin{equation}
\label{e25}
G_{\sigma}({\Bf r},{\Bf r}';\varepsilon) \equiv
\lim_{\eta\downarrow 0} 
\wt{G}_{\sigma}({\Bf r},{\Bf r}';\varepsilon\pm i\eta),
\;\;\;\; \varepsilon\, \IEq<> \,\mu.
\end{equation}
We note in passing that for ${\rm Im}(z)\not=0$, $\wt{G}_{\sigma}
({\Bf r},{\Bf r}';z)$ satisfies the reflection property (Luttinger 
1961, Farid 1999a,c, second paragraph of Appendix B)
\begin{equation}
\label{e26}
\wt{G}_{\sigma}({\Bf r},{\Bf r}';z^*) \equiv
\wt{G}_{\sigma}^*({\Bf r},{\Bf r}';z).
\end{equation}

The asymptotic expansion (see \S~II.B) of $\wt{G}_{\sigma}({\Bf r},
{\Bf r}';z)$ for $\vert z\vert\to\infty$ (in the Poincar\'e sense) 
in terms of the asymptotic sequence $\{1/z^m \| m=0,1,2,\dots\}$ 
is obtained by formally assuming an upper bound
\footnote{\label{f40}
This upper bound $E$ is made explicit in Eq.~(\protect\ref{e38}) below. }
to the absolute value of the single-particle energies 
$\{\varepsilon_{s;\sigma}\}$. For any $\vert z\vert$ greater than 
this upper bound, one obtains from Eq.~(\ref{e24}) the following 
uniformly convergent series
\begin{equation}
\label{e27}
\wt{G}_{\sigma}({\Bf r},{\Bf r}';z)
= \sum_{m=1}^{\infty} 
\frac{G_{\sigma;\infty_m}({\Bf r},{\Bf r}')}{z^m},
\end{equation}
where 
\begin{equation}
\label{e28}
G_{\sigma;\infty_m}({\Bf r},{\Bf r}') {:=} 
\hbar\sum_s\varepsilon_{s;\sigma}^{m-1}\, 
f_{s;\sigma}({\Bf r}) f^*_{s;\sigma}({\Bf r}'),\;\;
m=1,2,\dots.
\end{equation}
These are {\sl distributions} (Gelfand and Shilov 1964), as opposed to 
{\sl functions}, as can be deduced from the following consideration: 
\footnote{\label{f41}
These {\sl distributions} can viewed as being the limits of 
{\sl functions} $G_{\sigma;\infty_m}^{(E)}({\Bf r},{\Bf r}')$, 
defined in Eq.~(\protect\ref{e38}), for $E\to \infty$. }
making use of the definitions for the Lehmann amplitudes in 
Eq.~(\ref{e18}) and the canonical equal-time anticommutation 
(Fetter and Walecka 1971)
relations
\begin{eqnarray}
\label{e29}
\big[\hat\psi_{\sigma}^{\dag}({\Bf r}),
\hat\psi_{\sigma'}({\Bf r}')\big]_+
&&= \delta_{\sigma,\sigma'}\,\delta({\Bf r}-{\Bf r}'),
\nonumber\\
\big[\hat\psi_{\sigma}^{\dag}({\Bf r}),
\hat\psi_{\sigma'}^{\dag}({\Bf r}')\big]_+ &&= 
\big[\hat\psi_{\sigma}({\Bf r}),
\hat\psi_{\sigma'}({\Bf r}')\big]_+ = 0,
\end{eqnarray}
where $\big[ \wh{A},\wh{B}]_+ {:=} \wh{A} \wh{B} + \wh{B} \wh{A}$,
one readily obtains
\begin{eqnarray}
\label{e30}
&&G_{\sigma;\infty_1}({\Bf r},{\Bf r}') \equiv
\hbar\sum_s f_{s;\sigma}({\Bf r}) f_{s;\sigma}^*({\Bf r}')\nonumber\\
&&\;\;\;\;\;\;\;\;\;\;\;\;\;\;\;\;\;\;\;\;\,
=\hbar \sum_s \langle\Psi_{N;0}\vert \hat\psi_{\sigma}^{\dag}({\Bf r}')
\vert \Psi_{N_{\sigma}-1,N_{\bar\sigma};s}\rangle
\nonumber\\
&&\;\;\;\;\;\;\;\;\;\;\;\;\;\;\;\;\;\;\;\;\;\;\;\;\;\;\;\;\;
\times\langle \Psi_{N_{\sigma}-1,N_{\bar\sigma};s}\vert
\hat\psi_{\sigma}({\Bf r})\vert\Psi_{N;0}\rangle\nonumber\\
&&\;\;\;\;\;\;\;\;\;\;\;\;\;\;\;\;\;\;\;\;\,
+\hbar \sum_s \langle\Psi_{N;0}\vert \hat\psi_{\sigma}({\Bf r})
\vert \Psi_{N_{\sigma}+1,N_{\bar\sigma};s}\rangle\nonumber\\
&&\;\;\;\;\;\;\;\;\;\;\;\;\;\;\;\;\;\;\;\;\;\;\;\;\;\;\;\;\;
\times \langle \Psi_{N_{\sigma}+1,N_{\bar\sigma};s}\vert
\hat\psi_{\sigma}^{\dag}({\Bf r}')\vert\Psi_{N;0}\rangle
\nonumber\\
&&\;\;\;\;\;\;
=\hbar\langle\Psi_{N;0}\vert 
\hat\psi_{\sigma}^{\dag}({\Bf r}')
\hat\psi_{\sigma}({\Bf r}) +
\hat\psi_{\sigma}({\Bf r})
\hat\psi_{\sigma}^{\dag}({\Bf r}')\vert\Psi_{N;0}\rangle 
\nonumber\\
&&\;\;\;\;\;\;
\equiv \hbar\langle\Psi_{N;0}\vert 
\big[\hat\psi_{\sigma}({\Bf r}), 
\hat\psi_{\sigma}^{\dag}({\Bf r}')\big]_+ 
\vert\Psi_{N;0}\rangle \nonumber\\
&&\;\;\;\;\;\;
\equiv \hbar\delta({\Bf r}-{\Bf r}'),
\end{eqnarray}
which at the same time amounts to a statement concerning the 
completeness, or closure, 
\footnote{\label{f42}
As we shall discuss later, the set $\{f_{s;\sigma}({\Bf r})\}$ is 
{\sl overcomplete}, a fact manifested in the property
$\int {\rm d}^dr\, f_{s;\sigma}^*({\Bf r}) f_{s;\sigma}({\Bf r}) < 1$, 
for {\sl at least} one $s$, when $v\not\equiv 0$ (see Appendix A). }
of the Lehmann amplitudes. Thus already $G_{\sigma;\infty_1}({\Bf r},
{\Bf r}')$ is a distribution. In arriving at Eq.~(\ref{e30}) we have
made use of the completeness of $\{\vert \Psi_{N_{\sigma}\pm 1,
N_{\bar\sigma};s}\rangle \}$ in the Hilbert space consisting of
the direct product of the Hilbert spaces of $(N_{\sigma}\pm 1)$-
and $N_{\bar\sigma}$-particle states, that is
\begin{equation}
\label{e31}
\sum_s \vert\Psi_{N_{\sigma}\pm 1,N_{\bar\sigma};s}\rangle
\langle\Psi_{N_{\sigma}\pm 1,N_{\bar\sigma};s}\vert
= {\cal I}_{\sigma}^{\pm},
\end{equation}  
where ${\cal I}_{\sigma}^{\pm}$ stands for the unit operator in 
the above-mentioned Hilbert space.

We point out that $\{f_{s;\sigma}({\Bf r})\}$ is {\sl not} an 
orthogonal set, however reduces to an {\sl orthonormal} set as the 
coupling constant of the particle-particle interaction is set equal 
to zero (see Eqs.~(\ref{e46}) and (\ref{e47}) below; see also Appendix 
A). We further point out that $G_{\sigma;\infty_m}({\Bf r},{\Bf r}')$, 
$m=1,2,\dots$, are $c$-numbers. As we shall see in the subsequent 
sections, the explicit expressions for $G_{\sigma;\infty_m}({\Bf r},
{\Bf r}')$, in particular those concerning $m=2,3,4$, that are 
encountered in our explicit calculations involve gradient operators 
so that this knowledge with regard to the $c$-number nature of these 
distributions informs us that these differential operators act on 
the functions involved in the explicit expression for $G_{\sigma;
\infty_m}({\Bf r},{\Bf r}')$ and {\sl not} those exterior to it.

From Eq.~(\ref{e28}), making use of Eqs.~(\ref{e18}) and (\ref{e19}),
we have
\begin{eqnarray}
\label{e32}
G_{\sigma;\infty_2}({\Bf r},{\Bf r}') =
&&\hbar\sum_s (E_{N;0} -E_{N_{\sigma}-1,N_{\bar\sigma};s})  
\nonumber\\
\times\langle\Psi_{N;0}\vert \hat\psi_{\sigma}^{\dag}({\Bf r}')
\vert&&\Psi_{N_{\sigma}-1,N_{\bar\sigma};s}\rangle
\langle\Psi_{N_{\sigma}-1,N_{\bar\sigma};s}\vert
\hat\psi_{\sigma}({\Bf r})\vert\Psi_{N;0}\rangle\nonumber\\
+&&\hbar\sum_s (E_{N_{\sigma}+1,N_{\bar\sigma};s} - E_{N;0})  
\nonumber\\
\times\langle\Psi_{N;0}\vert \hat\psi_{\sigma}({\Bf r})
\vert&&\Psi_{N_{\sigma}+1,N_{\bar\sigma};s}\rangle
\langle\Psi_{N_{\sigma}+1,N_{\bar\sigma};s}\vert
\hat\psi_{\sigma}^{\dag}({\Bf r}')\vert\Psi_{N;0}\rangle.
\nonumber\\
\end{eqnarray}
Making use of $E_{M_{\sigma},M_{\bar\sigma};s}
\vert\Psi_{M_{\sigma},M_{\bar\sigma};s}\rangle = \wh{H} 
\vert\Psi_{M_{\sigma},M_{\bar\sigma};s}\rangle$ and employing
the completeness relation in Eq.~(\ref{e31}), we readily obtain
\begin{eqnarray}
\label{e33}
G_{\sigma;\infty_2}({\Bf r},{\Bf r}') =&& 
-\hbar\langle\Psi_{N;0}\vert \big[\wh{H},
\hat\psi_{\sigma}({\Bf r})\big]_- \hat\psi_{\sigma}^{\dag}
({\Bf r}')\vert\Psi_{N;0}\rangle \nonumber\\
&&-\hbar\langle\Psi_{N;0}\vert \hat\psi_{\sigma}^{\dag}({\Bf r}')
\big[\wh{H},\hat\psi_{\sigma}({\Bf r})\big]_- 
\vert\Psi_{N;0}\rangle\nonumber\\
\equiv 
\hbar \langle \Psi_{N;0}\vert&&
\Big[ \big[\hat\psi_{\sigma}({\Bf r}),\wh{H}\big]_-,
\hat\psi_{\sigma}^{\dag}({\Bf r}')\Big]_+
\vert\Psi_{N;0}\rangle,
\end{eqnarray}
in which $[\wh{A},\wh{B}]_- {:=} \wh{A}\wh{B} -\wh{B}\wh{A}$ 
denotes commutation. 

The expressions in Eqs.~(\ref{e30}) and (\ref{e33}) suggest that 
$G_{\sigma;\infty_m}({\Bf r},{\Bf r}')$ can be written as
\begin{eqnarray}
\label{e34}
G_{\sigma;\infty_m}({\Bf r},{\Bf r}') &\equiv& 
\hbar \langle \Psi_{N;0}\vert
\big[ \wh{L}^{m-1} \hat\psi_{\sigma}({\Bf r}), 
\hat\psi_{\sigma}^{\dag}({\Bf r}')\big]_+
\vert\Psi_{N;0}\rangle, \nonumber\\
& &\;\;\;\;\;\;\;\;\;\;\;\;\;\;\;\;\;\;\;\;\;\;\;\;\;\;\;\;\;
\;\;\;\;\;\;\;\;\;\;\;
m\ge 1,
\end{eqnarray}
where the Liouville super-operator $\wh{L} \equiv \wh{L}^1$ is
defined as
\footnote{\label{f43}
By definition, $\wh{L}^0 = I$, the unit operator in the space
of $\{ \hat\phi\}$, so that $\wh{L}^0 \hat\phi = \hat\phi$. }
\begin{equation}
\label{e35}
\wh{L} \hat\phi {:=} \big[\hat\phi,\wh{H}\big]_-,
\end{equation}
with $\hat\phi$ any operator in the second-quantization representation. 
The validity of Eq.~(\ref{e34}) is easily verified through substituting 
the RHS of Eq.~(\ref{e34}) in that of Eq.~(\ref{e27}) and formally 
carrying out the infinite sum $\sum_{m=1}^{\infty} (\wh{L}/z)^{m-1} 
=\big(I -\wh{L}/z\big)^{-1}$, thus obtaining
\begin{eqnarray}
\label{e36}
\wt{G}_{\sigma}({\Bf r},{\Bf r}';z) &&= 
\hbar \langle\Psi_{N;0}\vert \Big[ \big(z I - \wh{L}\big)^{-1}
\hat\psi_{\sigma}({\Bf r}), \hat\psi_{\sigma}^{\dag}({\Bf r}')
\Big]_+\vert\Psi_{N;0}\rangle,\nonumber\\
\end{eqnarray} 
which is the Mori-Zwanzig (Mori 1965, Zwanzig 1961) expression 
for $\wt{G}_{\sigma}({\Bf r},{\Bf r}';z)$ (see also Fulde (1991)). 
This result in addition demonstrates {\sl completeness} of the
series in Eq.~(\ref{e27}) for {\sl all} $z$.
\footnote{\label{f44}
The result in Eq.~(\protect\ref{e34}) is directly deduced by employing 
those in Eqs.~(\protect\ref{e37}) and (\protect\ref{e38}) below; making 
use of the time-integral representation of $\wt{G}_{\sigma}({\Bf r},
{\Bf r}';\varepsilon\pm i\eta)$ (that is the inverse of that presented 
in Eq.~(\protect\ref{ee9})), $G^{(E)}_{\sigma;\infty_m}({\Bf r},
{\Bf r}')$ in Eq.~(\protect\ref{e38}) is seen to be determined by 
${\rm d}^{m-1} G_{\sigma}({\Bf r}t,{\Bf r}'0)/{\rm d}t^{m-1}$ which 
following the defining expression in Eq.~(\protect\ref{ee7}) and the 
Heisenberg equation of motion (Fetter and Walecka 1971, Eq.~(6.29)) 
(see also Appendix E, and in particular Eq.~(\protect\ref{ee6})) 
$i\hbar\, {\rm d} \hat\psi_{\sigma}({\Bf r}t)/{\rm d}t = 
\wh{L}\hat\psi_{\sigma}({\Bf r}t)$, directly results in the 
expression on the RHS of Eq.~(\protect\ref{e34}). }

We note that for $m > 2$, the direct application of the expression 
in Eq.~(\ref{e34}), corresponding to the $\wh{H}$ in Eqs.~(\ref{e1}) 
and (\ref{e2}) leads to extremely complicated expressions. As we shall 
see in the following subsections, calculation of $G_{\sigma;\infty_m}
({\Bf r},{\Bf r}')$ is greatly facilitated through introduction of 
auxiliary fields and transposition of the $\wh{H}$ operators that are 
implied by $\wh{L}^{m-1}\hat\psi_{\sigma}({\Bf r})$ (in exchange for 
commutators of $\wh{H}$ with its adjacent field operators) to the 
left or right of all other operators;
\footnote{\label{f45}
For an {\sl odd} value of $m$, $\wh{L}^{m-1}\hat\psi_{\sigma}({\Bf r})$
in the expression for $G_{\sigma;\infty_m}({\Bf r},{\Bf r}')$ (see
Eq.~(\protect\ref{e34})) involves an {\sl even} number of $\wh{H}$, of 
which $(m-1)/2$ can be transposed to the left and $(m-1)/2$ to the right. 
This directly gives rise to an expression for $G_{\sigma;\infty_m}
({\Bf r},{\Bf r}')$ in terms of GS correlation functions which
{\sl explicitly} satisfies the symmetry requirement (see Appendix B; see 
also Eq.~(\protect\ref{e178}) below) $G_{\sigma;\infty_m}({\Bf r}',{\Bf r}) 
\equiv G_{\sigma;\infty_m}({\Bf r},{\Bf r}')$. For an {\sl even} value 
of $m$, on the other hand, this possibility cannot be exploited to the 
full (as the last $\wh{H}$ must be transposed either to the left or 
to the right), resulting in an expression for $G_{\sigma;\infty_m}
({\Bf r}, {\Bf r}')$ which, although {\sl implicitly} symmetric, is 
{\sl explicitly} highly {\sl asymmetric}. For this reason, evaluation of 
$G_{\sigma;\infty_m}({\Bf r},{\Bf r}')$ for {\sl even} values of $m$ 
is far more complicated than that for {\sl odd} values of $m$, 
specifically because it turns out to be extremely cumbersome to make 
explicit the symmetry of the directly evaluated expression for 
$G_{\sigma;\infty_m}({\Bf r},{\Bf r}')$. The latter task can be 
simplified by replacing an {\sl explicitly} asymmetric 
$G_{\sigma;\infty_m}({\Bf r},{\Bf r}')$ by $\frac{1}{2} 
[G_{\sigma;\infty_m}({\Bf r},{\Bf r}') + G_{\sigma;\infty_m}({\Bf r}',
{\Bf r})]$. This has, however, the severe disadvantage of masking the 
unintentional algebraic errors that may have incurred in the process 
of evaluating the original expression for $G_{\sigma;\infty_m}({\Bf r},
{\Bf r}')$. {\sl None} of the results presented in this work is based 
on such a symmetrization procedure. } 
only after this, should the explicit expression for $\wh{H}$ as presented 
in Eqs.~(\ref{e1}) and (\ref{e2}) be employed, followed by a process 
of {\sl normal ordering} of the field operators, transposing all the 
creation field operators to the left of the annihilation field operators. 
We should like to emphasize that neglect of this strategy results in such 
a vast number of terms that render a reliable calculation of {\sl even} 
$G_{\sigma;\infty_3}$ prohibitively difficult.

\subsection{The single-particle spectral functions 
$A_{\sigma}({\Bf r},{\Bf r}';\varepsilon)$}
\label{s11}

One can represent $G_{\sigma;\infty_m}({\Bf r},{\Bf r}')$ as 
follows:
\begin{equation}
\label{e37}
G_{\sigma;\infty_m}({\Bf r},{\Bf r}')
= \lim_{E\to\infty} G_{\sigma;\infty_m}^{(E)}({\Bf r},{\Bf r}')
\end{equation}
where
\begin{equation}
\label{e38}
G_{\sigma;\infty_m}^{(E)}({\Bf r},{\Bf r}')
{:=} \int_{-E}^{E} {\rm d}\varepsilon\;
\varepsilon^{m-1} 
A_{\sigma}({\Bf r},{\Bf r}';\varepsilon),\;\; m\ge 1,
\end{equation}
in which
\begin{eqnarray}
\label{e39}
A_{\sigma}({\Bf r},{\Bf r}';\varepsilon)
{:=} \frac{1}{2\pi i} \Big\{
\wt{G}_{\sigma}({\Bf r},{\Bf r}';\varepsilon-i\eta) -
\wt{G}_{\sigma}({\Bf r},&&{\Bf r}';\varepsilon+i\eta) \Big\},
\nonumber\\
&& \eta\downarrow 0,
\end{eqnarray} 
is the spectral function of the single-particle GF. Our use of 
$\lim_{E\to\infty} \int_{-E}^{E} {\rm d}\varepsilon\, (\dots)$ as
implied by Eqs.~(\ref{e37}) and (\ref{e38}), rather than simply 
$\int_{-\infty}^{\infty} {\rm d}\varepsilon\, (\dots)$, amounts to a 
representation of the {\sl distribution} $G_{\sigma;\infty_m}({\Bf r},
{\Bf r}')$ (see text following Eq.~(\ref{e28}) above) in terms of a 
sequence of {\sl functions} $G_{\sigma;\infty_m}^{(E)}({\Bf r},
{\Bf r}')$ for increasing values of $E$.

From Eq.~(\ref{e27}), making use of Eq.~(\ref{e37}) followed by 
substituting herein the RHS of Eq.~(\ref{e38}) for $G_{\sigma;
\infty_m}^{(E)}({\Bf r},{\Bf r'})$ and exchanging the order of 
the summation over $m$ with the integration over $\varepsilon$ and 
subsequently employing 
\footnote{\label{f46}
Formally, we assume that for a given {\sl finite} $E$, $\vert z\vert 
> E$. Compare with our statement preceding Eq.~(\protect\ref{e27}). }
$\frac{1}{z} \sum_{m=1}^{\infty} (\varepsilon/z)^{m-1}=1/(z-\varepsilon)$ 
and finally taking the limit $E\to\infty$, we obtain
\begin{equation}
\label{e40}
\wt{G}_{\sigma}({\Bf r},{\Bf r}';z) 
\equiv \int_{-\infty}^{\infty} {\rm d}\varepsilon\;
\frac{A_{\sigma}({\Bf r},{\Bf r}';\varepsilon)}{z -\varepsilon},
\end{equation}
which is the well-known integral representation of the single-particle 
GF in terms of its spectral function (for example Nozi\`eres 1964); 
making use of $1/(x\pm i\eta) = \wp(1/x) \mp i\pi \delta(x)$ for 
$\eta\downarrow 0$ (here $\wp$ stands for the Cauchy `principal value'), 
it is readily verified that this representation correctly reproduces 
the defining expression for $A_{\sigma}({\Bf r},{\Bf r}';\varepsilon)$ 
in Eq.~(\ref{e39}). The fact that the exact $\wt{G}_{\sigma}({\Bf r},
{\Bf r}';z)$ is recovered from the representation in Eq.~(\ref{e27}), 
as evidenced by both the result in Eqs.~(\ref{e36}) and (\ref{e40}), 
implies the completeness of this representation which in turn 
demonstrates the sufficiency in the context of the present work of the 
asymptotic sequence (see \S~II.B) $\{1,1/z,\dots\}$, corresponding 
to $\vert z\vert\to\infty$. 

Although $G_{\sigma;\infty_m}({\Bf r},{\Bf r}')$ pertaining to systems 
whose Hamiltonians have unbounded $(N\pm 1)$-particle spectra is 
{\sl not} bounded for $m \ge m_0$ (for a uniform and isotropic system,
specifically in $d=3$ when $v\equiv v_c$, it can be rigorously shown 
that $m_0=3$; see \S~III.I.2), 
\footnote{\label{f47}
The property that for the mentioned Hamiltonians, $G_{\sigma;\infty_m}
({\Bf r},{\Bf r}')$ is unbounded for $m \ge m_0$, with $m_0$ a 
{\sl finite} integer, implies that the point of infinity in the
$z$ plane is a singular point of $\wt{G}_{\sigma}({\Bf r},{\Bf r}';z)$.
The results in Eqs.~(\protect\ref{e73}) - (\protect\ref{e75}) below 
suggest that the point of infinity {\sl may} also be a singular 
point of $\wt{\Sigma}_{\sigma}({\Bf r},{\Bf r}';z)$, which our 
explicit calculations show indeed to be the case. }
the projection of $G_{\sigma;\infty_m}^{(E)}({\Bf r},{\Bf r}')$,
which we for simplicity assume not to involve any fundamentally 
unbounded contributions, on to the single-particle Hilbert space of 
the problem (see Appendix A) has a bounded limit for $E\to\infty$ 
for {\sl any} finite value of $m$ (in this context it is useful to 
consider our discussions in \S~II.B leading to conditions (A)-(C)). 
To demonstrate this, we first point out that, from the representation 
in Eq.~(\ref{e28}) one has
\begin{equation}
\label{e41}
G_{\sigma;\infty_m}^{(E)}({\Bf r},{\Bf r}')
\equiv \hbar\!\!\!\!\!
\sum_{s''\atop \vert \varepsilon_{s'';\sigma}\vert < E}
\varepsilon_{s'';\sigma}^{m-1}\,
f_{s'';\sigma}({\Bf r}) f_{s'';\sigma}^*({\Bf r}').
\end{equation}
Multiplying both sides of this expression, from left by 
$f_{s;\sigma}^*({\Bf r})$ and from right by $f_{s';\sigma}
({\Bf r}')$, and integrating the resulting expression 
with respect to ${\Bf r}$ and ${\Bf r}'$ we obtain
\begin{eqnarray}
\label{e42}
&&\int {\rm d}^dr {\rm d}^dr'\;
f_{s;\sigma}^*({\Bf r})\, 
G_{\sigma;\infty_m}^{(E)}({\Bf r},{\Bf r}')\,
f_{s';\sigma}({\Bf r}') \nonumber\\
&&\;\;\;
=\hbar \!\!\!\!\!
\sum_{s''\atop \vert \varepsilon_{s'';\sigma}\vert < E}
\varepsilon_{s'';\sigma}^{m-1}\,
\langle f_{s;\sigma}\vert f_{s'';\sigma}\rangle\,
\langle f_{s'';\sigma}\vert f_{s';\sigma}\rangle.
\end{eqnarray}
Let $\{\phi_{\varsigma}({\Bf r})\}$ be the complete {\sl orthonormal} 
set of eigenfunctions of the non-interacting single-particle 
Hamiltonian (see Eqs.~(\ref{e1}) and (\ref{e2})) 
\begin{equation}
\label{e43}
h_0({\Bf r}) {:=} \tau({\Bf r}) + u({\Bf r});
\end{equation}
we have
\footnote{\label{f48}
Our remarks in footnote \protect\ref{f39} apply equally here, that
is the `spectrum' of $h_0({\Bf r})$ is {\sl not} exhausted by its 
`point spectrum', etc. } 
\begin{equation}
\label{e44}
h_0({\Bf r})\, \phi_{\varsigma}({\Bf r})
= {\sf e}_{\varsigma}\, \phi_{\varsigma}({\Bf r}),
\end{equation}
where, as implied, $\{\varsigma\}$ fully characterizes the entire set 
of (single-particle) eigenstates of $h_0({\Bf r})$. We can therefore 
express the compound variable $s$ characterizing the Lehmann amplitude 
$f_{s;\sigma}({\Bf r})$ as follows
\footnote{\label{f49}
We note that $\varsigma$ is also a compound variable; consider the 
case of periodic crystals (i.e. $u({\Bf r})$ in Eq.~(\protect\ref{e43})
is periodic) where $\{ \phi_{\varsigma}({\Bf r})\}$ are Bloch 
functions with $\varsigma = ({\Bf k},\ell)$, ${\Bf k}$ denoting a 
vector in the first Brillouin zone of the underlying lattice and 
$\ell$ a band index. }
\begin{equation}
\label{e45}
s = (\varsigma,\alpha),
\end{equation}
where $\alpha$ is the so-called ``parameter of degeneracy'' (Klein 
and Prange 1958).
\footnote{\label{f50}
The designation ``parameter of degeneracy'', as employed by Klein and 
Prange (1958), is somewhat misleading. It is true that excited states 
of many-particle systems are largely degenerate (see footnote 
\protect\ref{f3} above); however, the necessity for the introduction of 
the ``parameter of degeneracy'' does {\sl not} arise from degeneracy 
of these states {\sl per se}, but from the fact that, for {\sl interacting} 
systems, the distinctive aspects of the $N\pm 1$-particle ground and 
excited states, in comparison with the $N$-particle GS, {\sl cannot} be 
regarded as being associated with properties of one or other 
single-particle wavefunction. Consequently, characterization of 
these states in relation to the $N$-particle GS {\sl cannot} be solely 
in terms of parameters describing a single-particle wavefunction. This 
viewpoint is consistent with the fact that although $\{ f_{s;\sigma}
({\Bf r})\}$ is complete, that is its constituent elements satisfy the 
{\sl closure} relation $\sum_s f_{s;\sigma}({\Bf r}) 
f_{s;\sigma}^*({\Bf r}') = \delta({\Bf r}-{\Bf r}')$ (see 
Eq.~(\protect\ref{e30})), we have $\int {\rm d}^dr\, f_{s;\sigma}^*
({\Bf r}) f_{s;\sigma}({\Bf r}) < 1$ for {\sl at least} one $s$ (unless 
$v\equiv 0$), which in combination with the former result, implies 
{\sl overcompleteness} of $\{f_{s;\sigma}({\Bf r})\}$ (see Appendix A). }
In general, we can write
\begin{equation}
\label{e46}
f_{s;\sigma}({\Bf r}) \equiv \phi_{\varsigma}({\Bf r})
+ \delta f_{s;\sigma}({\Bf r}),\;\;
\mbox{\rm with}\;\; s = (\varsigma,\alpha),
\end{equation}
where 
\begin{equation}
\label{e47}
\delta f_{s; \sigma}({\Bf r}) \to 0\;\;\;
\mbox{\rm for}\;\;\; v\to 0. 
\end{equation}
Since for increasing values of the magnitude of the single-particle 
excitation energies $\varepsilon_{s;\sigma}$, interaction effects 
diminish, the result in Eq.~(\ref{e47}) is equally applicable
to the cases where $\vert\varepsilon_{s;\sigma}\vert \to\infty$.
Under either of these two conditions (i.e. $v\to 0$ and/or
$\vert\varepsilon_{s;\sigma}\vert\to\infty$), which we designate 
as conditions of `weak correlation', making use of the orthonormality 
relation $\langle \phi_{\varsigma}\vert \phi_{\varsigma'}\rangle 
= \delta_{\varsigma,\varsigma'}$, we have  
\begin{equation}
\label{e48}
\langle f_{s;\sigma}\vert f_{s';\sigma}\rangle
\sim \delta_{s,s'} \equiv \delta_{\varsigma,\varsigma'}\;\;\;
\mbox{\rm (`weak correlation')},
\end{equation} 
where in the last equality we have made explicit that under the 
indicated condition, the `parameters of degeneracy' $\alpha$ and 
$\alpha'$, in $s=(\varsigma,\alpha)$ and $s' =(\varsigma',\alpha')$, 
are irrelevant to the leading order in the interaction. Note that 
Eq.~(\ref{e48}) applies even for $v\not\to 0$ when one or both of the 
conditions $\vert \varepsilon_{s;\sigma}\vert \to\infty$ and $\vert 
\varepsilon_{s';\sigma} \vert \to \infty$ (also identified with
`weak correlation' conditions) are satisfied.

From Eq.~(\ref{e42}) for sufficiently `weak correlation' (see above)
we obtain
\begin{eqnarray}
\label{e49}
\frac{1}{\hbar} \lim_{E\to\infty}
\int {\rm d}^dr {\rm d}^dr'\;
f_{s;\sigma}^*({\Bf r})&&G_{\sigma;\infty_m}^{(E)}
({\Bf r},{\Bf r}') f_{s';\sigma}({\Bf r}')\nonumber\\
&&\;\;\;\;\;\;\;\;\;
\sim\delta_{s,s'}\, \varepsilon_{s;\sigma}^{m-1},
\end{eqnarray}
which is indeed finite for any finite value of $m$. Since $G_{\sigma;
\infty_1}({\Bf r},{\Bf r}') = \hbar \delta({\Bf r}-{\Bf r}')$ (see 
Eq.~(\ref{e30}) above), for $m=1$, Eq.~(\ref{e49}) is seen exactly to 
reproduce Eq.~(\ref{e48}). This establishes the consistency of our 
above arguments. Making use of Eqs.~(\ref{e38}) and (\ref{e49}) we 
finally obtain 
\begin{eqnarray}
\label{e50}
\frac{1}{\hbar} \int_{-\infty}^{\infty} {\rm d}\varepsilon\;
\varepsilon^{m-1} \,
{\sf A}_{\sigma;s,s'}(\varepsilon)
&\sim& \delta_{s,s'}\, \varepsilon_{s;\sigma}^{m-1}\nonumber\\
& &\mbox{\rm (`weak correlation')}
\end{eqnarray}
where
\begin{equation}
\label{e51}
{\sf A}_{\sigma;s,s'}(\varepsilon) {:=}
\int {\rm d}^dr {\rm d}^dr'\;
f_{s;\sigma}^*({\Bf r}) A_{\sigma}({\Bf r},{\Bf r}';
\varepsilon) f_{s';\sigma}({\Bf r}'). 
\end{equation}
Considering uniform and isotropic systems, we have $\{\phi_{\varsigma}
({\Bf r})\} \equiv \{\Omega^{-1/2} \exp(i {\Bf k}\cdot {\Bf r})\}$ 
(implying identification of $\varsigma$ with ${\Bf k}$).
\footnote{\label{f51}
In choosing $\Omega^{-1/2} \exp(i {\Bf k}\cdot {\Bf r})$, we have
employed the `box' boundary condition, implying our further use of
$\Omega \delta_{{\Bf k},{\Bf k}'}/(2\pi)^d$ instead of
$\delta({\Bf k}-{\Bf k}')$. } 
Under the conditions of `weak correlation', from Eqs.~(\ref{e46}),
(\ref{e47}) and (\ref{e51}) we thus obtain
\begin{equation}
\label{e52}
{\sf A}_{\sigma;s,s'}(\varepsilon) \sim 
\ol{{\sf A}}_{\sigma;{\Bf k},{\Bf k}'}(\varepsilon)
\equiv \ol{A}_{\sigma}(\|{\Bf k}\|;\varepsilon)\,
\delta_{{\Bf k},{\Bf k}'}.
\end{equation}
From this and Eq.~(\ref{e50}) we deduce the following expression
\begin{equation}
\label{e53}
\frac{1}{\hbar} \int_{-\infty}^{\infty} {\rm d}\varepsilon\;
\varepsilon^{m-1}\, \ol{A}_{\sigma}(\|{\Bf k}\|;\varepsilon)
\sim \varepsilon_{{\Bf k};\sigma}^{m-1}.
\end{equation}
Here and in Eq.~(\ref{e52}), $\ol{A}_{\sigma}(\|{\Bf k}\|;\varepsilon)$ 
stands for the Fourier transform with respect to ${\Bf r}-{\Bf r}'$ of 
$A_{\sigma}({\Bf r},{\Bf r}';\varepsilon)$ defined in Eq.~(\ref{e39}) 
above (see \S~III.I.2). The asymptotic result in Eq.~(\ref{e53}) 
becomes the more accurate (i.e. it applies to cases with smaller 
values of $\vert\varepsilon_{{\Bf k};\sigma}\vert$, or of $\|{\Bf k}\|$) 
the weaker the strength of the particle-particle interaction. We shall 
encounter Eq.~(\ref{e53}) in \S~III.E.6 and further discuss its range of 
applicability. 

It is interesting to recall that the assumption with regard to the 
one-to-one correspondence between $\{\phi_{\varsigma}({\Bf r})\}$ and 
$\{ f_{s;\sigma}({\Bf r})\}$, or that between $\varsigma$ and $s$, 
which here is used under the provision of `weak correlation', is that 
which underlies the {\sl phenomenological} theory of Landau (1957) 
concerning the {\sl low}-lying single-particle excitations of interacting 
systems (for example Pines and Nozi\`eres (1966)). As we discuss in 
\S~III.C, the choice of $\wh{H}_0$, the `non-interacting' Hamiltonian, 
is not entirely arbitrary, as it is necessary that the GS of $\wh{H}_0$ 
be {\sl adiabatically connected} with that of $\wh{H}$ (see Farid (1997a) 
and the references herein); this aspect is of vital importance in the 
context of the Fermi-liquid theory. The significance of this requirement 
is easily illustrated by considering the case where for {\sl some} 
$\sigma$ the eigenvalue of the partial number operator $\wh{N}_{\sigma}$ 
corresponding to the GS of $\wh{H}_0$ is {\sl not} equal to that 
corresponding to the GS of $\wh{H}$. In such a case, {\sl no} 
number-preserving perturbation Hamiltonian is capable of adiabatically 
transforming the non-interacting GS into the interacting GS. This 
specific problem can be circumvented by considering (see Eq.~(\ref{e43})
above)
\begin{eqnarray}
\label{e54}
&&\wh{H}_0 = \sum_{\sigma} \int {\rm d}^dr\;
\hat\psi_{\sigma}^{\dag}({\Bf r}) h_{0;\sigma}({\Bf r})
\hat\psi_{\sigma}({\Bf r}),\\
\label{e55}
&&h_{0;\sigma}({\Bf r}) {:=} h_0({\Bf r}) + w_{\sigma}({\Bf r}),
\end{eqnarray}
where $w_{\sigma}({\Bf r})$ is {\sl some} spin-dependent local potential 
chosen in such a way that $\{N_{\sigma}\}$ is common to the GSs of 
$\wh{H}_0$ and $\wh{H}$. In fact, for the class of interacting number 
densities $\{n_{\sigma}({\Bf r}) \|\sigma = -{\sf s},-{\sf s}+1,
\dots,{\sf s} \}$ (see Eq.~(\ref{e163}) below), known as `pure-state 
non-interacting $v$-representable' densities, it is possible to 
choose $\{w_{\sigma}({\Bf r})\}$ in such a way that even 
$\{n_{0;\sigma}({\Bf r})\}$ is identical with its interacting 
counterpart $\{n_{\sigma}({\Bf r})\}$ (Farid 1997a,b, 1999b); this 
statement is most naturally described within the framework of the 
density-functional theory of Hohenberg and Kohn (1964) (for example
Dreizler and Gross (1990)). 

From the perspectives of the above considerations, it is therefore
more advantageous (see Appendix A) to relate the set of Lehmann 
amplitudes with the complete set of {\sl orthonormal} eigenstates 
of the following one-particle problem
\footnote{\label{f52}
The left-hand side of Eq.~(\protect\ref{e56}) may even accommodate
a {\sl non-local} contribution, of the form $\int {\rm d}^dr'\;
\Upsilon({\Bf r},{\Bf r}') \varphi_{\varsigma;\sigma}({\Bf r}')$,
where $\Upsilon_{\sigma}({\Bf r},{\Bf r}')$ may be chosen to be
$\Sigma_{\sigma;\infty_0}^{\sh}({\Bf r},{\Bf r}')$, introduced
in Eq.~(\protect\ref{e72}) below. }
\begin{equation}
\label{e56}
h_{0;\sigma}({\Bf r})\, \varphi_{\varsigma;\sigma}({\Bf r})
= \varepsilon_{\varsigma;\sigma}^{(0)}\,
\varphi_{\varsigma;\sigma}({\Bf r}).
\end{equation}
All our earlier comments concerning the set $\{\varsigma\}$
corresponding to the non-interacting eigenvalue problem in 
Eq.~(\ref{e44}) apply here. It is natural that $w_{\sigma}({\Bf r})$ 
be chosen in such a way that for $v\to 0$ it approaches zero or at 
most a constant, independent of ${\Bf r}$. Further, since for $\vert
\varepsilon_{\varsigma;\sigma}^{(0)}\vert\to\infty$, $\tau({\Bf r}) 
\varphi_{\varsigma;\sigma}({\Bf r}) \sim \varepsilon_{\varsigma;
\sigma}^{(0)}\, \varphi_{\varsigma;\sigma}({\Bf r})$, which also 
applies to $\phi_{\varsigma}({\Bf r})$ when $\vert {\sf e}_{\varsigma}
\vert\to\infty$, that is $\tau({\Bf r}) \phi_{\varsigma}({\Bf r}) \sim 
{\sf e}_{\varsigma}\phi_{\varsigma}({\Bf r})$, it follows that under 
the conditions of `weak correlation', specified in the text following 
Eq.~(\ref{e48}) above, the eigenvalue problem in Eq.~(\ref{e56}) is 
interchangeable with and, for the reason indicated, preferable to
that in Eq.~(\ref{e44}) (see Appendix A). It follows that 
Eqs.~(\ref{e48})-(\ref{e53}) retain their validity by choosing the 
`non-interacting' problem to be that corresponding to $h_{0;\sigma}
({\Bf r})$ in Eq.~(\ref{e55}) (see also footnote \ref{f52}) rather than 
that corresponding to $h_0({\Bf r})$ in Eq.~(\ref{e43}).

For completeness, we note that from Eqs.~(\ref{e30}) and (\ref{e38}) 
one has
\begin{equation}
\label{e57}
\frac{1}{\hbar} \int_{-\infty}^{\infty}
{\rm d}\varepsilon\;
A_{\sigma}({\Bf r},{\Bf r}';\varepsilon)
= \delta({\Bf r}-{\Bf r}'),
\end{equation}
the RHS of which is evidently {\sl in}dependent of the strength of
the particle-particle interaction, so that Eq.~(\ref{e57}) equally 
applies to {\sl non}-interacting systems. The result in Eq.~(\ref{e57}) 
is in fact nothing but a statement concerning the {\sl completeness} or 
{\sl closure} of both $\{ f_{s;\sigma}({\Bf r})\}$ (see Eq.~(\ref{e30})) 
and $\{\varphi_{\varsigma;\sigma}({\Bf r})\}$ in the single-particle 
Hilbert space of the system under consideration (see Appendix A). We 
point out that, under the conditions where Eq.~(\ref{e48}) as well as 
other related expressions apply, one can replace $s$, $s'$ by 
$\varsigma$ and $\varsigma'$ respectively. Thus, in contrast with 
the {\sl exact} case where the non-trivial dependence of 
$\varepsilon_{s;\sigma}$ on $\alpha$ through $s = (\varsigma,\alpha)$ 
implies that the single-particle excitation energies are {\sl not} 
sharply defined but are rather associated with peaks of finite widths 
in the single-particle spectral functions pertaining to interacting 
systems, in the cases pertinent to Eqs.~(\ref{e50}) and (\ref{e53}), 
the energies $\varepsilon_{\varsigma,\sigma}$ are sharply defined. 
Consequently, the latter energies should only be indicative of the 
{\sl average} locations of well-defined peaks of the single-particle 
spectral function along the $\varepsilon$ axis (see \S~III.D).

Before closing this Section, we mention that the series in
Eq.~(\ref{e27}) has an alternative of the form
\begin{equation}
\label{e58}
\wt{G}_{\sigma}({\Bf r},{\Bf r}';z)
= \sum_{m=1}^{\infty}
\frac{G_{\sigma;(1/\varepsilon_0)_m}({\Bf r},{\Bf r}')}
{(z- \varepsilon_0)^m};
\end{equation} 
here $\varepsilon_0$ is a {\sl real} constant energy, so that for
${\rm Im}(z)\not=0$, the expression in Eq.~(\ref{e58}) correctly
reproduces the exact reflection property in Eq.~(\ref{e26}).
As can be readily seen, the expression in Eq.~(\ref{e27}) corresponds 
to the case where $\varepsilon_0 = 0$; we have made this fact explicit 
by introducing the subscript $(1/\varepsilon_0)_m$ in the coefficient 
functions on the RHS of Eq.~(\ref{e58}). The expression in 
Eq.~(\ref{e58}) can be viewed as an infinite-order AS in terms of the 
asymptotic sequence $\{1,1/(z-\varepsilon_0),1/(z-\varepsilon_0)^2,
\dots\}$ appropriate for $\vert z-\varepsilon_0\vert\to \infty$. 

The relationship between $\{ G_{\sigma;\infty_m}\}$ and $\{ G_{\sigma;
(1/\varepsilon_0)_m} \}$ is direct and can be revealed as follows. By 
writing the denominator of the integrand on the RHS of Eq.~(\ref{e40})
as $(z-\varepsilon_0) [1 - (\varepsilon-\varepsilon_0)/(z
-\varepsilon_0)]$, making use of the geometric series for $1/[1
-(\varepsilon-\varepsilon_0)/(z-\varepsilon_0)]$ and subsequently
employing the binomial expansion 
\begin{eqnarray}
(\varepsilon-\varepsilon_0)^m
= \sum_{j=0}^{m} {m\choose j} 
(-\varepsilon_0)^{m-j} \varepsilon^j, \nonumber
\end{eqnarray}
from Eq.~(\ref{e58}) we immediately obtain the following results:
\footnote{\label{f53}
Here we define $(0)^{m-j} = \delta_{m,j}$. }
\begin{eqnarray}
\label{e59}
&&G_{\sigma;(1/\varepsilon_0)_m}({\Bf r},{\Bf r}')
\equiv \int  {\rm d}\varepsilon\;
(\varepsilon-\varepsilon_0)^{m-1}\,
A_{\sigma}({\Bf r},{\Bf r}';\varepsilon)\nonumber\\
&&\;\;\;
= \sum_{j=0}^{m-1} {m-1\choose j}\, 
(-\varepsilon_0)^{m-j-1}\, 
G_{\sigma;\infty_{j+1}}({\Bf r},{\Bf r}');
\end{eqnarray}
by restricting the $\varepsilon$ integral in Eq.~(\ref{e59}) to the 
interval $(-E,E)$, we obtain what we denote by $G_{\sigma;
(1/\varepsilon_0)_m}^{(E)}({\Bf r},{\Bf r}')$ in terms of $\{ G_{\sigma;
\infty_j}^{(E)}({\Bf r},{\Bf r}')\}$ defined in Eqs.~(\ref{e38}) and
(\ref{e41}). The first equality in Eq.~(\ref{e59}) implies that 
$\{G_{\sigma;(1/\varepsilon_0)_m}({\Bf r},{\Bf r}')\}$ are the 
{\sl central} $\varepsilon$ moments of the single-particle spectral 
function with respect to $\varepsilon_0$, while the second equality 
makes explicit the association between $G_{\sigma;(1/\varepsilon_0)_m}$ 
and the entire set $\{ G_{\sigma;\infty_1},\dots, G_{\sigma;
\infty_{m}}\}$. 

Making use of 
\begin{eqnarray}
\varepsilon^{m-1} &\equiv& (\varepsilon -\varepsilon_0 + 
\varepsilon_0)^{m-1} \nonumber\\
&\equiv& \sum_{j=0}^{m-1} {m-1\choose j} 
(\varepsilon_0)^{m-j-1} (\varepsilon-\varepsilon_0)^j \nonumber
\end{eqnarray} 
in the integral on the RHS of Eq.~(\ref{e38}) (identifying herein 
$E$ with $\infty$) and comparing the result with the first expression 
on the RHS of Eq.~(\ref{e59}), we readily obtain the inverse of 
the second expression on the RHS of Eq.~(\ref{e59}), namely
\begin{equation}
\label{e60}
G_{\sigma;\infty_m}
= \sum_{j=0}^{m-1} {m-1\choose j} (\varepsilon_0)^{m-j-1}\,
G_{\sigma;(1/\varepsilon_0)_{j+1}}.
\end{equation}
Note the change of $(-\varepsilon_0)$ in the RHS of Eq.~(\ref{e59})
into $(\varepsilon_0)$ on the RHS of Eq.~(\ref{e60}).

\subsection{Specific details}
\label{s12}

In what follows we deal with {\sl formal} truncated AS for 
$\wt{G}_{\sigma}({\Bf r},{\Bf r}';z)$ and $\wt{\Sigma}_{\sigma}
({\Bf r},{\Bf r}';z)$ for arbitrary interaction potentials $v$ and 
spatial dimensions $d$, {\sl without} considering whether the 
coefficients in these series are bounded or not ({\it cf}. $a_j^{\rm b}$ 
and $a_j^{\rm u}$ introduced in \S~II.B). Later, we shall in detail 
consider the issues related to the existence of the mentioned 
coefficients. It is important to realize that truncation of the infinite
series for $\wt{G}_{\sigma}({\Bf r},{\Bf r}';z)$ in Eq.~(\ref{e27})
(or Eq.~(\ref{e58})), as well as that for $\wt{\Sigma}_{\sigma}({\Bf r},
{\Bf r}';z)$ (see Eq.~(\ref{e72}) below), results in functions (which 
for the moment we assume to be well defined) that are analytic 
{\sl everywhere} in the complex $z$ plane except at the origin (or
at $z=\varepsilon_0$), where they possess a multiple {\sl pole} whose 
order depends on the number of terms incorporated in the pertinent 
finite-order AS. Although these series are appropriate for the region 
$\vert z\vert\to \infty$, the latter observation is relevant in that 
it shows that, in contrast with the exact functions $\wt{G}_{\sigma}
({\Bf r},{\Bf r}';z)$ and $\wt{\Sigma}_{\sigma}({\Bf r},{\Bf r}';z)$, 
the corresponding (finite-order) AS are single valued. 
\footnote{\label{f54}
We leave aside the fact that $\wt{\Sigma}_{\sigma}({\Bf r},{\Bf r}';z)$ 
is a {\sl bounded} function (almost everywhere in the ${\Bf r}$ and 
${\Bf r}'$ space) over the entire $z$ plane, including $z=0$ (or
$z=\varepsilon_0$). }
Since the coefficients in these formal finite-order series are 
real valued (see Appendix B; see also Eq.~(\ref{e178}) below), for 
cases where ${\rm Im}(z)=0$, these series are consequently also 
real-valued. In \S~III.I we show that any {\sl formal} finite-order 
AS for $\wt{\Sigma}_{\sigma}({\Bf r},{\Bf r}';z)$, as $\vert z\vert
\to\infty$, reduces to a {\sl formal} AS for ${\rm Re}[\Sigma_{\sigma}
({\Bf r},{\Bf r}';\varepsilon)]$, when $z\to \varepsilon \pm i\eta$, 
with $\eta\downarrow 0$ ({\it cf}. Eq.~(\ref{e65}) below); in this 
Section we also show how, from the latter, the large-$\vert\varepsilon
\vert$ AS for ${\rm Im}[\Sigma_{\sigma}({\Bf r},{\Bf r}';\varepsilon)]$ 
can be deduced and derive the {\sl explicit} expression for this 
series up to terms decreasing not more rapidly than $1/\varepsilon^2$. 
Making use of this as well as some other results that we obtain later 
in this paper, in \S~III.I.2 we calculate the explicit AS (up to and 
including the second leading term for the cases corresponding to 
bounded and short-range interaction functions and the third leading
term for the cases corresponding to $d=3$ and $v\equiv v_c$) for the 
Fourier transform of the single-particle spectral function pertaining 
to a uniform and isotropic system of fermions.

In view of our explicit calculations in the present paper, we 
restrict ourselves to considering the four leading terms of the
infinite series in Eq.~(\ref{e27}) and write
\begin{equation}
\label{e61}
\wt{G}_{\sigma}(z) \sim 
\frac{\hbar I}{z} +
\frac{G_{\sigma;\infty_2}}{z^2} +
\frac{G_{\sigma;\infty_3}}{z^3} +
\frac{G_{\sigma;\infty_4}}{z^4},\;\;\; 
\vert z \vert\to \infty,
\end{equation}
where we have employed a representation-free notation and used the 
result in Eq.~(\ref{e30}). 
\footnote{\label{f55}
Recall that for the unit operator $I$ in the single-particle Hilbert 
space we have $\langle {\Bf r}\vert I\vert {\Bf r}'\rangle 
= \delta({\Bf r}-{\Bf r}')$ (see text following Eq.~(\protect\ref{e81})
below). }
With reference to our considerations in \S~II.B, we emphasize that 
the series in Eq.~(\ref{e61}) is at this stage merely a formal device, 
since in general the coefficients $G_{\sigma;\infty_m}({\Bf r},{\Bf r}')$, 
$m=2,3,4$, do {\sl not} satisfy the equivalents of the criteria (A)-(C) 
in \S~II.B (which concern $\wt{\Sigma}_{\sigma}({\Bf r},{\Bf r}';z)$) 
for $\wt{G}_{\sigma}({\Bf r},{\Bf r}';z)$.

Before entering into details, we mention that, whereas $\wt{G}_{\sigma}(z)$ 
corresponding to $\wh{H}$ is well-specified, the freedom in the choice 
of the non-interacting Hamiltonian $\wh{H}_0$ implies that this is not 
the case for $\wt{G}_{0;\sigma}(z)$. Traditionally, the SE operator 
$\wt{\Sigma}_{\sigma}(z)$ is defined with respect to the $\wt{G}_{0;
\sigma}(z)$ pertaining to the truly non-interacting Hamiltonian, that 
is $\wh{H}_0 {:=} \wh{T} + \wh{U}$. However, as $\{\wt{G}_{0;\sigma}(z)\}$ 
is defined in terms of the GS of $\wh{H}_0$, it is possible that the 
$\{N_{\sigma}\}$ corresponding to this GS is different from that 
corresponding to the GS of $\wh{H}$, in which case $\wt{G}_{0;\sigma}
(z)$ and $\wt{G}_{\sigma}(z)$ {\sl cannot} be directly related to one 
another through $\wt{\Sigma}_{\sigma}(z)$. To circumvent this problem, 
in this work we assume $\wh{H}_0$ to be defined as in Eq.~(\ref{e54}) 
in terms of $h_{0;\sigma}({\Bf r})$ in Eq.~(\ref{e55}).

In the light of the above, in the following we denote the SE operator 
corresponding to $\wh{H}' {:=}\wh{H}-\wh{H}_0$, with $\wh{H}_0$ as 
defined in Eq.~(\ref{e54}), by $\wt{\Sigma}_{\sigma}^{\sh}(z)$, to 
be distinguished from $\wt{\Sigma}_{\sigma}(z)$, for which we have
\begin{equation}
\label{e62}
\wt{\Sigma}_{\sigma}({\Bf r},{\Bf r}';z) \equiv 
\wt{\Sigma}_{\sigma}^{\sh}({\Bf r},{\Bf r}';z) + 
\frac{1}{\hbar} w_{\sigma}({\Bf r}) \delta({\Bf r}-{\Bf r}').
\end{equation} 

From the Dyson equation 
\begin{equation}
\label{e63}
\wt{G}_{\sigma}(z) = \wt{G}_{0;\sigma}(z) +
\wt{G}_{0;\sigma}(z) \wt{\Sigma}_{\sigma}^{\sh}(z) 
\wt{G}_{\sigma}(z)
\end{equation}
we have (use of $\wt{\Sigma}_{\sigma}^{\sh}(z)$ in Eq.~(\ref{e63})
implies that $\wt{G}_{0;\sigma}(z)$ herein pertains to the 
single-particle Hamiltonian $h_{0;\sigma}({\Bf r})$ in Eq.~(\ref{e55}), 
to be distinguished from $h_0({\Bf r})$ in Eq.~(\ref{e43}))
\begin{equation}
\label{e64}
\wt{\Sigma}_{\sigma}^{\sh}(z) = \wt{G}_{0;\sigma}^{-1}(z) - 
\wt{G}_{\sigma}^{-1}(z).
\end{equation}
Above $\wt{\Sigma}_{\sigma}(z)$ and $\wt{\Sigma}_{\sigma}^{\sh}(z)$ 
stand for the analytic continuations into the physical Riemann sheet 
of the SE operators $\Sigma_{\sigma}(\varepsilon)$ and 
$\Sigma_{\sigma}^{\sh}(\varepsilon)$ respectively; the latter are 
recovered from the former according to ({\it cf}. Eq.~(\ref{e25}) 
above) 
\begin{eqnarray}
\label{e65}
\Sigma_{\sigma}(\varepsilon) &=& 
\lim_{\eta\downarrow 0}
\wt{\Sigma}_{\sigma}(\varepsilon\pm i \eta),\;\;\; 
\varepsilon\, \IEq<> \,\mu, \\
\label{e66}
\Sigma_{\sigma}^{\sh}(\varepsilon) &=& 
\lim_{\eta\downarrow 0}
\wt{\Sigma}_{\sigma}^{\sh}(\varepsilon\pm i \eta),\;\;\; 
\varepsilon\, \IEq<> \,\mu. 
\end{eqnarray}
We point out that, similar to $\wt{G}_{\sigma}({\Bf r},{\Bf r}';z)$,
$\wt{\Sigma}_{\sigma}({\Bf r},{\Bf r}';z)$ possesses the reflection 
property (see Eq.~(\ref{e26}) above)
\begin{equation}
\label{e67}
\wt{\Sigma}_{\sigma}({\Bf r},{\Bf r}';z^*) \equiv 
\wt{\Sigma}_{\sigma}^*({\Bf r},{\Bf r}';z)\;\;\;
\mbox{\rm when}\;\;\;
{\rm Im}(z)\not= 0. 
\end{equation}
The expression in Eq.~(\ref{e64}) serves to provide us with the 
AS of $\wt{\Sigma}_{\sigma}(z)$ for $\vert z\vert\to
\infty$ from those of $\wt{G}_{0;\sigma}(z)$ and $\wt{G}_{\sigma}(z)$. 
This is achieved through employing the following, which amounts to a 
generalization of a theorem from the calculus of asymptotic expansions
(Copson 1965, pp. 8 and 9): 
\footnote{\label{f56}
The generalization concerns extending a result applicable to 
{\sl scalar functions} to {\sl operators}, taking due account of the
proper ordering of the operators. } 
Let ${\tilde f}(z)$ possess the following AS with respect to the 
asymptotic sequence $\{1/z^m \| m=0,1,2,\dots\}$,
\begin{equation}
\label{e68}
{\tilde f}(z) \sim \frac{f_1}{z} + \frac{f_2}{z^2} + \dots,
\;\;\; \vert z\vert\to\infty.
\end{equation}
Then, provided $\det(f_1)\not=0$, one has
\begin{equation}
\label{e69}
{\tilde f}^{-1}(z) \sim z f_1^{-1} + f'_0 + \frac{f'_1}{z}
+ \frac{f'_2}{z^2} + \dots,
\;\;\; \vert z\vert\to \infty,
\end{equation}
where
\begin{eqnarray}
\label{e70}
f'_0 &\equiv& -f_1^{-1} f_2 f_1^{-1},\nonumber\\
f'_1 &\equiv& f_1^{-1} \big(f_2 f_1^{-1} f_2 - f_3\big) f_1^{-1},
\nonumber\\
f'_2 &\equiv& - f_1^{-1} \big( f_2 f_1^{-1} f_2 f_1^{-1} f_2
- f_2 f_1^{-1} f_3\nonumber\\
& &\;\;\;\;\;\;\;\;\;\;\;\;\;\;\;\;\;\;\;\;
- f_3 f_1^{-1} f_2 + f_4\big) f_1^{-1}. 
\end{eqnarray}
From the AS in Eq.~(\ref{e27}) (or Eq.~(\ref{e61})) we thus obtain
\begin{eqnarray}
\label{e71}
&&\wt{G}_{\sigma}^{-1}(z) \sim
\frac{z I}{\hbar} - \frac{G_{\sigma;\infty_2}}{\hbar^2}
+ \frac{G_{\sigma;\infty_2}^2 - 
\hbar G_{\sigma;\infty_3}}{\hbar^3 z}\nonumber\\
&&\;\;\;
- \frac{G_{\sigma;\infty_2}^3 -\hbar G_{\sigma;\infty_2}
G_{\sigma;\infty_3} - \hbar G_{\sigma;\infty_3} G_{\sigma;\infty_2}
+ \hbar^2 G_{\sigma;\infty_4}}{\hbar^4 z^2}\nonumber\\ 
&&\;\;\;
+ \dots.
\end{eqnarray}
The same expression applies to $\wt{G}_{0;\sigma}^{-1}(z)$, with 
the operators $G_{\sigma;\infty_p}$, $p=2,3,4,\dots$, on the RHS 
replaced by $G_{0;\sigma;\infty_p}$. Making use of these results 
and Eq.~(\ref{e64}), we readily obtain
\begin{equation}
\label{e72}
\wt{\Sigma}_{\sigma}^{\sh}(z) \sim \Sigma_{\sigma;\infty_0}^{\sh}
+ \frac{\Sigma_{\sigma;\infty_1}}{z} 
+ \frac{\Sigma_{\sigma;\infty_2}}{z^2} + \dots,\;\; \vert z\vert 
\to\infty,
\end{equation}
where
\begin{eqnarray}
\label{e73}
\Sigma_{\sigma;\infty_0}^{\sh} \equiv&&
\frac{-1}{\hbar^2} 
\big\{G_{0;\sigma;\infty_2} - G_{\sigma;\infty_2}\big\}, \\
\label{e74}
\Sigma_{\sigma;\infty_1} \equiv&&
\frac{1}{\hbar^3} 
\big\{G_{0;\sigma;\infty_2}^2 - G_{\sigma;\infty_2}^2\big\} 
\nonumber\\
- &&\frac{1}{\hbar^2} 
\big\{G_{0;\sigma;\infty_3} - G_{\sigma;\infty_3}\big\} \\
\label{e75}
\Sigma_{\sigma;\infty_2} \equiv&&
\frac{-1}{\hbar^4}
\big\{G_{0;\sigma;\infty_2}^3 - G_{\sigma;\infty_2}^3\big\}
\nonumber\\
+ &&\frac{1}{\hbar^3} \big\{
G_{0;\sigma;\infty_2} G_{0;\sigma;\infty_3} 
- G_{\sigma;\infty_2} G_{\sigma;\infty_3} \nonumber\\
&&\;\;\;\; + G_{0;\sigma;\infty_3} G_{0;\sigma;\infty_2} 
- G_{\sigma;\infty_3} G_{\sigma;\infty_2} \big\}
\nonumber\\
- &&\frac{1}{\hbar^2}
\big\{G_{0;\sigma;\infty_4} - G_{\sigma;\infty_4}\big\}.
\end{eqnarray}
These results clarify our restriction to the finite series in 
Eq.~(\ref{e61}). Note that the {\sl independence} from $z$ of the 
difference between $\wt{\Sigma}_{\sigma}^{\sh}(z)$ and 
$\wt{\Sigma}_{\sigma}(z)$ (see Eq.~(\ref{e62}) above) implies that 
{\sl only} the leading asymptotic term of these SEs are different; 
explicitly,
\begin{eqnarray}
\label{e76}
&&\Sigma_{\sigma;\infty_0}({\Bf r},{\Bf r}')
= \Sigma_{\sigma;\infty_0}^{\sh}({\Bf r},{\Bf r}') 
+ \frac{1}{\hbar} w_{\sigma}({\Bf r})\, 
\delta({\Bf r}-{\Bf r}'), \nonumber\\
&&\Sigma_{\sigma;\infty_m}({\Bf r},{\Bf r}') \equiv 
\Sigma_{\sigma;\infty_m}^{\sh}({\Bf r},{\Bf r}'),\;\;\;
m \ge 1. 
\end{eqnarray} 
This aspect, which clarifies our notation in Eq.~(\ref{e72}), implies 
that in the expressions for $\Sigma_{\sigma;\infty_m}$, $m=1,2$, in 
Eqs.~(\ref{e74}) and (\ref{e75}), the dependence upon $w_{\sigma}$ of 
$G_{0;\sigma;\infty_2}^2$ must cancel that of $G_{0;\sigma;\infty_3}$
and similarly for $G_{0;\sigma;\infty_2}^3$ and $G_{0;\sigma;\infty_4}$,
etc. (note that, since the Hamiltonian $\wh{H}$ is independent of 
$w_{\sigma}$, $G_{\sigma;\infty_m}$ and its various powers are naturally 
independent of $w_{\sigma}$); this we shall explicitly demonstrate in 
the following. On the basis of these considerations, the results in 
Eqs.~(\ref{e74}) and (\ref{e75}) reduce into the following
\begin{eqnarray}
\label{e77}
\Sigma_{\sigma;\infty_1} &=& \frac{1}{\hbar^3}
\big\{ \hbar G_{\sigma;\infty_3} - G_{\sigma;\infty_2}^2\big\}, \\
\label{e78}
\Sigma_{\sigma;\infty_2} &=&
\frac{1}{\hbar^4}\big\{
\hbar^2 G_{\sigma;\infty_4}
-\hbar G_{\sigma;\infty_2} G_{\sigma;\infty_3}\nonumber\\
& &\;\;\;\;\;
-\hbar G_{\sigma;\infty_3} G_{\sigma;\infty_2}
+ G_{\sigma;\infty_2}^3\big\}.
\end{eqnarray}

With reference to the alternative series expansion in Eq.~(\ref{e58}),
we note that the SE operator similarly admits of the
following series ({\it cf}. Eq.~(\ref{e72}))
\begin{eqnarray}
\label{e79}
\Sigma_{\sigma}^{\sh}(z) \sim
\Sigma_{\sigma;(1/\varepsilon_0)_0}^{\sh} +
\frac{\Sigma_{\sigma;(1/\varepsilon_0)_1}}{z-\varepsilon_0} &+&
\frac{\Sigma_{\sigma;(1/\varepsilon_0)_2}}{(z-\varepsilon_0)^2} + 
\dots,\nonumber\\
& &\;\; \vert z - \varepsilon_0\vert\to \infty,
\end{eqnarray}
where the coefficient operators 
$\Sigma_{\sigma;(1/\varepsilon_0)_0}^{\sh}$,
$\Sigma_{\sigma;(1/\varepsilon_0)_1}$ and $\Sigma_{\sigma;
(1/\varepsilon_0)_2}$ are obtained from the expressions in 
Eqs.~(\ref{e73}), (\ref{e74}) and (\ref{e75}) in which $G_{0;\sigma;
\infty_m}$ and $G_{\sigma;\infty_m}$ are replaced by $G_{0;\sigma;
(1/\varepsilon_0)_m}$ and $G_{\sigma;(1/\varepsilon_0)_m}$ 
respectively. The latter operators are obtained from the former ones 
through the second expression on the RHS of Eq.~(\ref{e59}). In this 
way one obtains the coefficients in the series in Eq.~(\ref{e79}) 
in terms of those in Eq.~(\ref{e72}) which we explicitly consider in 
this work. One can alternatively obtain the explicit expressions for 
$\{\Sigma_{\sigma;(1/\varepsilon_0)_m}\}$ in terms of 
$\{ \Sigma_{\sigma;\infty_m}\}$ through employing the RHS of
\begin{equation}
\label{e80}
\frac{1}{z^m} \equiv \frac{1}{(z-\varepsilon_0)^m}
\sum_{j=0}^{\infty} (-1)^j\,
\Big[\sum_{j'=1}^{m} {m\choose j'}\,
\Big(\frac{\varepsilon_0}{z-\varepsilon_0}\Big)^{j'} \Big]^{j}
\end{equation}
in that of Eq.~(\ref{e72}) and, in view of the expression on the RHS of 
Eq.~(\ref{e79}), identifying in the resulting expression the coefficient 
of $1/(z-\varepsilon_0)^m$ with $\Sigma_{\sigma;(1/\varepsilon_0)_m}$. 
In this way one readily obtains
\begin{eqnarray}
\label{e81}
\Sigma_{\sigma;(1/\varepsilon_0)_0}^{\sh} &\equiv& 
\Sigma_{\sigma;\infty_0}^{\sh}, \nonumber\\
\Sigma_{\sigma;(1/\varepsilon_0)_1} &\equiv& 
\Sigma_{\sigma;\infty_1}, \nonumber\\
\Sigma_{\sigma;(1/\varepsilon_0)_2} &\equiv& 
\Sigma_{\sigma;\infty_2} - \varepsilon_0\, 
\Sigma_{\sigma;\infty_1}.  
\end{eqnarray}
 
In the following Sections, we explicitly evaluate $G_{\sigma;\infty_m}$,
with $m=2,3,\dots$, and $\Sigma_{\sigma;\infty_m}$, with $m=0,1,\dots$, 
in the coordinate representation, where we have to do with $G_{\sigma;
\infty_m}({\Bf r},{\Bf r}') \equiv \langle {\Bf r}\vert G_{\sigma;
\infty_m}\vert {\Bf r}'\rangle$ and $\Sigma_{\sigma;\infty_m}({\Bf r},
{\Bf r}') \equiv \langle {\Bf r}\vert \Sigma_{\sigma;\infty_m}\vert 
{\Bf r}'\rangle$. Here $\vert {\Bf r}\rangle$ and $\vert {\Bf r}'\rangle$ 
stand for the eigenstates of the ${\hat {\Bf r}}$ operator, corresponding 
to eigenvalues ${\Bf r}$ and ${\Bf r}'$ respectively, which satisfy the 
normalization condition $\langle {\Bf r} \vert {\Bf r}'\rangle = 
\delta({\Bf r}-{\Bf r}')$. Consequently, for the $l$th power of an 
operator ${\sf O}$ (representing $G_{\sigma;\infty_m}$), making use of 
the `decomposition' of the unit operator in the single-particle Hilbert 
space, namely 
\begin{equation}
\label{e82}
I = \int {\rm d}^dr\; 
\vert {\Bf r}\rangle\langle {\Bf r}\vert, 
\end{equation}
one has
\begin{eqnarray}
\label{e83}
&&\langle {\Bf r}\vert {\sf O}^{2} \vert {\Bf r}'\rangle
= \int {\rm d}^dr_1 \;
\langle {\Bf r}\vert {\sf O}\vert {\Bf r}_1\rangle
\langle {\Bf r}_1\vert {\sf O}\vert {\Bf r}'\rangle,
\nonumber\\
&&\,\langle {\Bf r}\vert {\sf O}^{l} \vert {\Bf r}'\rangle
= \int {\rm d}^dr_1\;\dots {\rm d}^dr_{l-1}\;
\nonumber\\
&&\;\;\;\;\;\;
\times\langle {\Bf r}\vert {\sf O}\vert {\Bf r}_1\rangle
\langle {\Bf r}_1\vert {\sf O}\vert {\Bf r}_2\rangle \dots 
\langle {\Bf r}_{l-1}\vert {\sf O}\vert {\Bf r}'\rangle,
\;\;\; l > 2.
\end{eqnarray}

\subsection{On the quasi-particle energies and wavefunctions}
\label{s13}

It is well known that the `quasi-particle' wavefunctions 
$\{\psi_{\varsigma;\sigma}({\Bf r};\varepsilon)\}$ and energies 
$\{\varepsilon_{\varsigma;\sigma}\}$ are obtained through solving 
the following two equations (see Eqs.~(\ref{e44}) and (\ref{e56}) 
above)
\begin{eqnarray}
\label{e84}
h_{0;\sigma}({\Bf r})\, 
\psi_{\varsigma;\sigma}({\Bf r};\varepsilon)
&+& \hbar \int {\rm d}^dr'\;
\Sigma_{\sigma}^{\sh}({\Bf r},{\Bf r}';\varepsilon)\,
\psi_{\varsigma;\sigma}({\Bf r}';\varepsilon)\nonumber\\
& &\;\;\;\;\;\;\;\;\;\;\;
= {\cal E}_{\varsigma;\sigma}(\varepsilon)\,
\psi_{\varsigma;\sigma}({\Bf r};\varepsilon),\\
\label{e85}
{\cal E}_{\varsigma;\sigma}(\varepsilon_{\varsigma;\sigma}) 
&=& \varepsilon_{\varsigma;\sigma}.
\end{eqnarray}
For a given $(\varsigma;\sigma)$, however, Eq.~(\ref{e85}) may {\sl not} 
have a solution (Farid 1999a,c). In cases where Eq.~(\ref{e85}) does 
have a solution, this may {\sl not} be a {\sl regular} solution,  
characterized by the property of at least once continuous 
differentiability of ${\cal E}_{\varsigma;\sigma}(\varepsilon)$ with 
respect to $\varepsilon$ in a neighbourhood of the solution; such 
non-regular solution is exemplified by the Fermi energy $\varepsilon_F$ 
of metallic states of one-dimensional interacting systems (generically
described by the one-dimensional Luttinger (1963) model (Mattis and 
Lieb 1965) (for a review see Voit (1994))) when momentum (with which 
$\varsigma$ in our present notation should be identified) is set equal 
to one of the two Fermi momenta. 
\footnote{\label{f57}
As we have emphasized in (Farid 1999a,c), the possible solutions of  
Eqs.~(\protect\ref{e84}) and (\protect\ref{e85}) may {\sl not} correspond 
to the Landau quasi-particles (whence our above use of the quotation 
marks in denoting quasi-particles) for, even though such solutions 
correspond to single-particle excitations in the system, these excitations 
may {\sl not} be capable of being described (not even asymptotically) 
in such terms as are specific to single particles, characterized 
by a `smooth' energy dispersion (see Appendix A). Metallic interacting 
systems whose lowest-lying single-particle excitations can be described 
as such, constitute the Fermi-liquid ``universality'' class. } 
Furthermore, by the assumption of stability of the GS of the system 
under consideration, and the {\sl completeness} of the single-particle 
eigenfunctions $\{ \varphi_{\varsigma;\sigma}({\Bf r})\}$ of 
$h_{0;\sigma}({\Bf r})$ whose indices $\{\varsigma\}$ are inherited 
by the set of eigenfunctions $\{ \psi_{\varsigma;\sigma}({\Bf r};
\varepsilon) \}$ in Eq.~(\ref{e84}), 
\footnote{\label{f58}
It is here that the importance of the hypothesis concerning the 
{\sl adiabatic connection} (see Farid (1997a,b)) between the 
$N$-particle GSs of $\wh{H}_0$ and $\wh{H}$ becomes most clearly 
manifest. See \S~III.C where we briefly elaborate on the 
significance of $\Sigma_{\sigma}^{\sh}(\varepsilon)$ in comparison 
with $\Sigma_{\sigma}(\varepsilon)$. }
for a given $(\varsigma;\sigma)$, Eq.~(\ref{e85}) {\sl cannot} 
possess multiple solutions, for such solutions would imply 
incompleteness 
\footnote{\label{f59}
For the precise definition of {\sl completeness} of normed spaces,
and of Cauchy sequences, see Kreyszig (1978) and Debnath and 
Mikusi\'nski (1990). } 
of $\{ \varphi_{\varsigma;\sigma}({\Bf r}) \}$ as regards the Hilbert 
space of the single-particle excitations of the interacting system,
spanned by the overcomplete set of the Lehmann amplitudes
$\{ f_{s;\sigma}({\Bf r}) \}$ (see Appendix A). 

With reference to our discussions in \S~III.B concerning the 
distinction between the single-particle orbitals $\{\varphi_{\varsigma;
\sigma}({\Bf r})\}$ and the Lehmann amplitudes $\{ f_{s;\sigma}
({\Bf r})\}$, and specifically concerning that between indices $s$ and 
$\varsigma$ (see Eq.~(\ref{e45}) above), we point out that {\sl the 
possibility of non-existence of solutions to Eq.~(\ref{e85}) is 
{\it directly} related to the latter distinction}. To clarify this 
statement, it is important to recall that $\Sigma_{\sigma}(\varepsilon)$ 
is {\sl not} Hermitian for arbitrary values of $\varepsilon$. This 
implies that ${\cal E}_{\varsigma;\sigma}(\varepsilon) \equiv
\lim_{\eta\downarrow 0} \wt{\cal E}_{\varsigma;\sigma}(\varepsilon \pm 
i\eta)$, $\varepsilon\, \Ieq<> \,\mu$ ({\it cf}. Eq.~(\ref{e65}) above) 
is in general complex valued so that, unless ${\rm Im}[\Sigma_{\sigma}
({\Bf r},{\Bf r}';\varepsilon)] \equiv 0$ at $\varepsilon =
\varepsilon_0$, because Eq.~(\ref{e85}) is satisfied by $\varepsilon
=\varepsilon_0$, $\varepsilon_0$ must also be complex valued. Owing to 
the reflection property in Eq.~(\ref{e67}), such a complex-valued 
solution would imply satisfaction of Eq.~(\ref{e85}) also at $\varepsilon
=\varepsilon_0^*$. This possibility is ruled out by the assumption with 
regard to the stability of the (non-degenerate) $N$-particle GS of the 
system under consideration. We note in passing that the stability
of the GS implies ${\rm Im}[{\cal E}_{\varsigma;\sigma}(\varepsilon)]
\,\Ieq{\ge}{\le}\, 0$ for $\varepsilon \,\Ieq<> \, \mu$, $\forall
\varsigma, \sigma$ (Luttinger 1961, in particular Eqs.~(38), (49) 
and (50)). Given the fact that for extended systems, 
${\rm Im}[\Sigma_{\sigma}({\Bf r},{\Bf r}';\varepsilon)]$ is a function 
of $\varepsilon$ with unbounded support (see \S~III.I.2), we observe 
that from among the innumerably many single-particle excitations (see 
\S~III.B), Eq.~(\ref{e85}) {\sl may} in principle possess as few as 
{\sl two} solutions 
\footnote{\label{f60}
Previously (Farid 1999c) we have shown that in general there exist 
appropriate $\varsigma$ for which ${\cal E}_{\varsigma;\sigma}
(\mu_{N;\sigma}^{\mp}) =\mu_{N;\sigma}^{\mp}$, where 
$\mu_{N;\sigma}^{\mp}$ are defined in Eq.~(\protect\ref{e23}) above. 
In (extended) {\sl metallic} systems, $\mu_{N;\sigma}^{-} \equiv 
\varepsilon_F$, the Fermi energy of the interacting system. In 
general $\mu_{N;\sigma}^{-} < \mu < \mu_{N;\sigma}^{+}$ (see 
Eq.~(\protect\ref{e22}) above), where $\mu$ stands for the `chemical 
potential'. In the latter systems, $\mu_{N;\sigma}^{+} 
-\mu_{N;\sigma}^{-}$ is infinitesimally small (proportional to 
$1/N^{\alpha}$ with $\alpha > 0$) so that, by the constancy of 
$\mu$ with respect to $\{\sigma\}$, $\mu_{N;\sigma}^{\pm}$ must be 
up to an infinitesimal correction {\sl independent} of $\{\sigma\}$. 
The validity of this statement at finite temperatures is ascribed 
to the condition of thermodynamic equilibrium between particles, 
irrespective of their spin index. For completeness, we mention that 
for {\sl all} $\varepsilon \in [\mu_{N;\sigma}^-, \mu_{N;\sigma}^+]$ 
we have the {\sl exact} result ${\rm Im}[{\cal E}_{\varsigma;\sigma}
(\varepsilon)] \equiv 0$ (see Galitskii and Migdal 1958, and Luttinger 
1960, equations (6) and (94)). } 
which, as we have just indicated, must be real valued; complex-valued 
single-particle excitation energies, with the finite imaginary parts 
corresponding to the finite life-times of the associated single-particle 
excitations, can be deduced from a pair of equations similar to those 
in Eqs.~(\ref{e84}) and (\ref{e85}) corresponding to the analytic 
continuation of $\wt{\Sigma}_{\sigma}({\Bf r},{\Bf r}';z)$ into 
{\sl non}-physical Riemann sheets (for details see Farid (1999a,c)). The 
{\sl overcompleteness} of $\{ f_{s;\sigma}({\Bf r})\}$, which is related 
to $s = (\varsigma,\alpha)$ (see Eq.~(\ref{e45}) above and the associated 
text in \S~III.B), on the one hand, and the characterization of the 
eigenfunctions of Eq.~(\ref{e84}) by $\varsigma$, on the other hand, 
\footnote{\label{f61}
The eigenfunctions $\{\psi_{\varsigma;\sigma}({\Bf r};\varepsilon)\}$,
with $\psi_{\varsigma;\sigma}({\Bf r};\varepsilon) \equiv 
\lim_{\eta\downarrow 0} \wt{\psi}({\Bf r};\varepsilon\pm i\eta)$, 
$\varepsilon\, \Ieq{<}{>}\,\mu$, do {\sl not} form an orthonormal set, 
but a {\sl bi-orthonormal} set in combination with $\{ \psi'_{\varsigma;
\sigma}({\Bf r};\varepsilon) \}$ (Morse and Feshbach 1953, pp. 884-886, 
Layzer 1963), where $\psi'_{\varsigma;\sigma}({\Bf r};\varepsilon) 
{:=} \lim_{\eta\downarrow 0} {\tilde\psi}_{\varsigma;\sigma}({\Bf r};
\varepsilon \mp i\eta)$ for $\varepsilon \Ieq{<}{>} \mu$ (Farid 
1999a,c) ({\it cf}. Eq.~(\protect\ref{e66}) above and notice $\mp$ as 
compared with $\pm$). }
makes explicit that the role of the ``parameter of degeneracy'' $\alpha$ 
(see footnote \ref{f50}) is in a way taken over by the continuous 
energy parameter $\varepsilon$ in Eq.~(\ref{e84}) in combination with 
the general non-Hermiticity of $\Sigma_{\sigma}(\varepsilon)$; 
substituting $(\varsigma,\alpha)$ for $s$ in $\varepsilon_{s;\sigma}$ 
as defined in Eq.~(\ref{e19}), we observe that by the very token that 
for any given $\varsigma$ (which is capable of characterizing 
{\sl any} single-particle excitation of the system described by 
$\wh{H}_0$ (see Eqs.~(\ref{e54}) and (\ref{e55}) above)) there 
exists a continuous manifold of associated single-particle excitations 
of the {\sl interacting} system corresponding to the continuous variable 
$\alpha$, Eqs.~(\ref{e84}) and (\ref{e85}) {\sl must not} be capable 
of exhausting the spectrum of the single-particle excitations of the 
interacting system in the form of {\sl isolated} real-valued solutions. 
\footnote{\label{f62}
This aspect holds true also for finite systems. }
Conversely, in cases where, for a given $\varsigma$, Eqs.~(\ref{e84}) and 
(\ref{e85}) yield a real-valued solution, the `parameter of degeneracy' 
must be `quenched'; the range of variations in $\alpha$ in the 
neighbourhoods of its `quenched' values depends on the nature of the 
associated solutions (i.e. on whether these are isolated, regular 
or accumulation points).

The significance of the large-$\vert\varepsilon\vert$ AS of 
$\wt{\Sigma}_{\sigma}^{\sh}(z)$ in Eq.~(\ref{e72}) is made explicit 
through substituting this series for $\Sigma_{\sigma}^{\sh}({\Bf r},
{\Bf r}';\varepsilon)$ ({\it cf}. Eq.~(\ref{e66}) above) in 
Eq.~(\ref{e84}). We consider such substitution in the following 
Section where we explicitly deal with a uniform system in its uniform 
and isotropic GS. In \S~III.E.6 we discuss the consequences of neglecting 
the imaginary part of $\Sigma_{\sigma}({\Bf r},{\Bf r}';\varepsilon)$ 
by a {\sl finite-order} AS for this function.
\footnote{\label{f63}
We point out that the complex-valuedness of the {\sl regularized} 
$\Sigma_{\sigma;\infty_m}({\Bf r},{\Bf r}')$ for $m \ge 2$ (see
Eq.~(\protect\ref{e110}) below), originates from the {\sl infinite} partial 
summation of unbounded terms in the full large-$\vert\varepsilon\vert$ 
AS for $\Sigma_{\sigma}^{\sh}({\Bf r},{\Bf r}';\varepsilon)$. We 
further point out that, {\sl in principle}, it is possible explicitly 
to take into account ${\rm Im}[\Sigma_{\sigma}({\Bf r},{\Bf r}';
\varepsilon)]$ to any desired order in $1/\varepsilon$ (see \S~III.I) 
alongside the series for ${\rm Re}[\Sigma_{\sigma}({\Bf r},{\Bf r}';
\varepsilon)]$; in practice, however, calculation of the former is 
{\sl not} as direct as that of the latter (see 
Eq.~(\protect\ref{e227}) below). }
Therefore, suffice it to say for the moment that, although such a 
neglect by a finite-order series gives rise to real-valued solutions to 
Eq.~(\ref{e85}), such solutions have direct physical significance 
through their relevance to the $\varepsilon$-moments integrals of 
the single-particle spectral function (see \S~III.B). Further, as we 
discuss in \S~III.E.6, the contributions of the latter real-valued 
solutions to the single-particle spectral function, which show up as 
poles, can be broadened in a controlled fashion in anticipation of 
the condensation of these poles into branch cuts of $\Sigma_{\sigma}
({\Bf r},{\Bf r}';\varepsilon)$ in the limit where the complete 
infinite AS of this function is taken into account. Note in passing 
that Eqs.~(\ref{e72}) and (\ref{e76}) (see also Eq.~(\ref{e173}) 
below) clearly show the approach of the equation in Eq.~(\ref{e84}) 
to the {\sl exact} Hartree-Fock equation
\footnote{\label{f64}
The word `exact' here refers to the exact $\varrho_{\sigma}$, as 
opposed to $\varrho_{{\rm s};\sigma}$, which determines 
$\Sigma_{\sigma;\infty_0} \equiv \Sigma^{\sc hf}[\varrho_{\sigma}]$ 
(see Eq.~(\protect\ref{e173}) below). } 
for $\vert\varepsilon\vert\to\infty$ (see \S~I.B).

\subsection{A simple example and some discussions}
\label{s14}

Below we consider some quantitative aspects associated with
finite-order large-$\vert\varepsilon\vert$ AS for $\Sigma_{\sigma}
(\varepsilon)$. To this end, we specifically deal with a system of 
fermions in a uniform positively-charged background. For simplicity, 
we deal solely with the homogeneous and isotropic GS of this system.

\subsubsection{The Hamiltonian and some conventions} 
\label{s15}

The system with which we deal is described by the following 
Hamiltonian (for example Fetter and Walecka (1971, p.~25)):
\begin{eqnarray}
\label{e86}
&&\wh{H} = \sum_{\sigma'} \sum_{{\Bf k}'}\,
\frac{\hbar^2 {k'}^2}{2 m_e}\,
{\sf\hat a}_{{\Bf k}';\sigma'}^{\dag}
{\sf\hat a}_{{\Bf k}';\sigma'} \nonumber\\
&&\;\;
+\frac{1}{2\Omega}
\sum_{\sigma_1',\sigma_2'}\,
\sum'_{{\Bf k}',{\Bf p}',{\Bf q}'}\,
{\bar v}(q')\,
{\sf\hat a}_{{\Bf k}'+{\Bf q}';\sigma_1'}^{\dag}
{\sf\hat a}_{{\Bf p}'-{\Bf q}';\sigma_2'}^{\dag}
{\sf\hat a}_{{\Bf p}';\sigma_2'}
{\sf\hat a}_{{\Bf k}';\sigma_1'},\nonumber\\
\end{eqnarray}
where the prime over $\sum_{{\Bf q}'}'$ implies that the contribution 
corresponding to ${\Bf q}'={\bf 0}$ must be excluded. In Eq.~(\ref{e86}), 
${\bar v}(q')$, with $q' {:=} \|{\Bf q}'\|$, stands for the Fourier 
transform of the particle-particle interaction function $v({\Bf r}
-{\Bf r}')$; in what follows, we write
\begin{equation}
\label{e87}
v({\Bf r}-{\Bf r}') = g\, w({\Bf r}-{\Bf r}'),
\end{equation} 
where $g$ stands for the coupling constant of interaction and 
$w({\Bf r}-{\Bf r}')$ depends on $\| {\Bf r}-{\Bf r}'\|$; in the 
case of $v\equiv v_c$ in $d=3$, we have (see Eq.~(\ref{e13}) and
text following Eq.~(\ref{e15}) above)
\begin{equation}
\label{e88}
g_c = \frac{e^2}{4\pi\epsilon_0},\;\;\;
{\bar w}_c(q) = \frac{4\pi}{q^2 + \kappa^2},\;\;\; 
\kappa\downarrow 0,
\end{equation}
where, owing to the exclusion of ${\Bf q}'={\bf 0}$ form the 
${\Bf q}'$-sum on the RHS of Eq.~(\ref{e86}), it is permitted to 
identify $\kappa$ with zero without any further consequence. With 
reference to our considerations in \S~II.A, we further point out that, 
in dealing with the case corresponding to $v=v_c$ in $d=3$, the RHS 
of Eq.~(\ref{e86}) has to be supplemented by $2\wh{H}_{\kappa}$ with
$\wh{H}_{\kappa}$ as defined in Eq.~(\ref{e5}); here the prefactor
$2$ accounts for the fact that by excluding the term corresponding 
to ${\Bf q}'={\bf 0}$ in the sum on the RHS of Eq.~(\ref{e86}), we have 
{\sl explicitly} removed the SE of the positively-charged uniform 
background of charge density $e n_0$ (see text following Eq.~(\ref{e9}) 
above). Throughout this paper we assume
\begin{equation}
\label{e89}
g \equiv g_c,\;\;\; {\bar w}(q) \equiv 
{\sf\bar w}_{\{\lambda_1,\dots,\lambda_J\}}(q),
\end{equation}
with ${\sf\bar w}_{\{\lambda_1,\dots,\lambda_J\}}(q)$ possessing 
the scaling property
\begin{equation}
\label{e90}
{\sf\bar w}_{\{\alpha\lambda_1,\dots,\alpha\lambda_J\}}(\alpha q) 
= \alpha^{1-d}\, {\sf\bar w}_{\{\lambda_1,\dots,\lambda_J\}}(q),
\end{equation} 
that is ${\sf\bar w}_{\{\lambda_1,\dots,\lambda_J\}}(q)$ is a 
homogeneous function of $q$ and $\{\lambda_1,\dots,\lambda_J\}$ 
of degree $1-d$ (for example Ince (1927, p.~10)). Here, as in 
Eqs.~(\ref{e89}) and (\ref{e90}), $\{\lambda_1,\dots,\lambda_J\}$ 
stands for the set of $J$ parameters on which ${\bar w}(q)$ may 
depend; these parameters have the same dimension as $q$, that is 
inverse length. The property in Eq.~(\ref{e90}) is seen to be 
satisfied by ${\bar w}_c(q)$ in Eq.~(\ref{e88}) for which 
${\sf\bar w}_{\{\lambda_1,\dots,\lambda_J\}}(q)$ is equal to 
$4\pi/(q^2+\kappa^2)$, corresponding to $J=1$ and $\lambda_1=\kappa$, 
$\kappa\downarrow 0$; to give another example, for the Yukawa 
potential in $d=3$, ${\sf\bar w}_{\{\lambda_1,\dots,\lambda_J\}}(q)$ 
is equal to $4\pi/(q^2 + q_{\sc tf}^2)$, so that $J=1$ and 
$\lambda_1=q_{\sc tf}$, the Thomas-Fermi screening wavenumber.  

In Eq.~(\ref{e86}), ${\sf\hat a}_{{\Bf k};\sigma}$ stands for the 
annihilation operator corresponding to the particle with spin index
$\sigma$ in the single-particle state $\langle {\Bf r}\vert {\Bf k}
\rangle = \Omega^{-1/2} \exp(i {\Bf k}\cdot {\Bf r})$, characterized by 
wave-vector ${\Bf k}$; with the creation operator ${\sf\hat a}_{{\Bf k};
\sigma}^{\dag}$, which is the Hermitian conjugate of 
${\sf\hat a}_{{\Bf k};\sigma}$, we have the following 
anticommutation relations ({\it cf}. Eq.~(\ref{e29}) above)
\begin{eqnarray}
\label{e91}
&&\big[ {\sf\hat a}_{{\Bf k};\sigma}^{\dag},
{\sf\hat a}_{{\Bf k}';\sigma'} \big]_+ = \delta_{\sigma,\sigma'}
\delta_{{\Bf k},{\Bf k}'},\nonumber\\
&&\big[ {\sf\hat a}_{{\Bf k};\sigma}^{\dag},
{\sf\hat a}_{{\Bf k}';\sigma'}^{\dag} \big]_+ =
\big[ {\sf\hat a}_{{\Bf k};\sigma},
{\sf\hat a}_{{\Bf k}';\sigma'} \big]_+ = 0.
\end{eqnarray}

With
\begin{equation}
\label{e92}
a_0 {:=} \frac{ 4\pi \epsilon_0 \hbar^2}{m_e e^2}
\equiv \frac{\hbar^2}{m_e g_c}
\end{equation}
the Bohr radius 
\footnote{\label{f65}
In dealing with systems in which the fermions are not true elementary 
particles, but rather long-lived quasi-particles of a host material, 
the {\sl bare} fermion mass $m_e$ should be replaced by an appropriate 
effective mass $m_e^*$ and $\epsilon_0$ by $\epsilon_0^* {:=} \epsilon_0 
\epsilon_{\rm r}$, with $\epsilon_{\rm r}$ the relative dielectric 
constant of the host material. The effective Bohr radius $a_0^*$ thus 
obtained, is a more realistic measure (than the Bohr radius $a_0$) for 
the lengths relevant to the problem at hand. }
and $r_0$ the average distance between the particles (independent of 
their spin indices) in the GS, that is
\begin{equation}
\label{e93}
\Omega = \frac{2 \pi^{d/2}}{\Gamma(d/2) d}\, r_0^d\, N\;
\Longleftrightarrow\;
r_0 = \left(\frac{\Gamma(d/2) d}{2\pi^{d/2} n_0}\right)^{1/d},
\end{equation}
where $n_0$ stands for the total concentration of fermions, defined in 
Eq.~(\ref{e9}), we introduce the {\sl dimensionless} Wigner-Seitz radius
\begin{equation}
\label{e94}
r_s {:=} \frac{r_0}{a_0}.
\end{equation}
In Eq.~(\ref{e93}), $\Gamma(x)$ stands for the gamma function 
(Abramowitz and Stegun 1972, p.~255). 
\footnote{\label{f66}
Recall that $\Gamma(1/2)=\pi^{1/2}$ and that $\Gamma(1+x)=x\Gamma(x)$. }
Defining the dimension-less volume and wave-vectors according to
\begin{equation}
\label{e95}
{\bar\Omega} {:=} \frac{\Omega}{r_0^d},\;\;\;
{\bar {\Bf k}} {:=} r_0 {\Bf k},\;\;
\mbox{\rm etc.},
\end{equation}
the Hamiltonian in Eq.~(\ref{e86}) can be expressed as follows:
\begin{equation}
\label{e96}
\wh{H} = \frac{\hbar^2}{2 m_e a_0^2}\,
\frac{2}{r_s^2}\, \wh{\cal H},
\end{equation}
where
\begin{equation}
\label{e97}
\wh{\cal H} = \wh{\cal T} + \wh{\cal V},
\end{equation}
in which 
\begin{equation}
\label{e98}
\wh{\cal T} \equiv
\sum_{\sigma'} \sum_{{\bar {\Bf k}'}} 
\frac{1}{2} {\bar {k'}}^2\,
{\hat a}_{{\bar {\Bf k}'};\sigma'}^{\dag}
{\hat a}_{{\bar {\Bf k}'};\sigma'},
\end{equation}
and 
\begin{eqnarray}
\label{e99}
&&\wh{\cal V} \equiv
\frac{r_s}{2 {\bar\Omega}}
\sum_{\sigma_1',\sigma_2'}\,
\sum_{{\bar {\Bf k}'},{\bar {\Bf p}'},{\bar {\Bf q}'}}'\,
{\bar w}({\bar q}')\nonumber\\
&&\;\;\;\;\;\;\;\;\;\;\;\;\;\;\;\;\;\;\;\;\;
\times {\hat a}_{{\bar {\Bf k}'}+{\bar {\Bf q}'};\sigma_1'}^{\dag}
{\hat a}_{{\bar {\Bf p}'}-{\bar {\Bf q}'};\sigma_2'}^{\dag}
{\hat a}_{{\bar {\Bf p}'};\sigma_2'}
{\hat a}_{{\bar {\Bf k}'};\sigma_1'},
\end{eqnarray}
where ${\bar w}({\bar q}')$ stands for ${\bar{\sf w}}_{r_0\lambda_1,
\dots, r_0\lambda_J}(r_0\, q')$ (see Eq.~(\ref{e90}) above). Above we 
have introduced
\begin{equation}
\label{e100}
{\hat a}_{{\bar {\Bf k}};\sigma} {:=\,} 
{\sf\hat a}_{{\bar {\Bf k}}/r_0;\sigma} \equiv 
{\sf\hat a}_{{\Bf k};\sigma},\;\;\;\;
{\hat a}_{{\bar {\Bf k}};\sigma}^{\dag} {:=\,} 
{\sf\hat a}_{{\bar {\Bf k}}/r_0;\sigma}^{\dag} \equiv
{\sf\hat a}_{{\Bf k};\sigma}^{\dag}.
\end{equation}
From Eq.~(\ref{e91}) it trivially follows that
\begin{eqnarray}
\label{e101}
&&\big[ {\hat a}_{{\bar {\Bf k}};\sigma}^{\dag},
{\hat a}_{{\bar {\Bf k}}';\sigma'} \big]_+ = \delta_{\sigma,\sigma'}
\delta_{{\bar {\Bf k}},{\bar {\Bf k}}'},\nonumber\\
&&\big[ {\hat a}_{{\bar {\Bf k}};\sigma}^{\dag},
{\hat a}_{{\bar {\Bf k}}';\sigma'}^{\dag} \big]_+ =
\big[ {\hat a}_{{\bar {\Bf k}};\sigma},
{\hat a}_{{\bar {\Bf k}}';\sigma'} \big]_+ = 0.
\end{eqnarray}

The Hamiltonian $\wh{\cal H}$ is dimensionless and has the interesting 
property of making explicit that the ratio of the particle-particle
interaction energy to the kinetic energy in some eigenstate of the 
system, is to leading order and up to logarithmic corrections proportional 
to $r_s$ (whence $r_s\downarrow 0$ corresponds to `weak' correlation). 
We mention that
\footnote{\label{f67}
For $m_e$ we have here used the electron mass in vacuum, equal to 
$9.109~3897 \times 10^{-31}$~kg. With $a_0$ as defined in 
Eq.~(\protect\ref{e92}), our choice for $m_e$ implies that $a_0 = 
0.529~177~249 \times 10^{-10}$~m. }
\begin{equation}
\label{e102}
\frac{\hbar^2}{2 m_e a_0^2} = 1~\mbox{\rm Ry}
= 13.605~6981\; \mbox{\rm eV}
\end{equation}
so that, expressing energies in Rydbergs, $\wh{H}$ is equal to $2/r_s^2$ 
times $\wh{\cal H}$. We define
\begin{equation}
\label{e103}
e_0 {:=} \frac{\hbar^2}{m_e a_0^2 r_s^2} \equiv 
\frac{2}{r_s^2}\,\mbox{\rm Ry}.
\end{equation}
In what follows we express energies in units of $e_0$ and denote the 
normalized energies by their conventional symbols topped with a single 
bar, namely $\bar\varepsilon_k^{(0)} {:=} \varepsilon_k^{(0)}/e_0$; 
since in this paper we denote the diagonal elements of the Fourier 
transform with respect to ${\Bf r}$ and ${\Bf r}'$ of functions of 
$\|{\Bf r}-{\Bf r}'\|$ also through a {\sl single} bar placed on their 
well-established symbols, below we denote the normalized 
$\ol{\Sigma}_{\sigma}(k;\varepsilon)$ by topping this with an additional 
bar; thus
\begin{equation}
\label{e104}
\Ol{\Sigma}_{\sigma}({\bar k};\bar\varepsilon) \equiv
\frac{\hbar\ol{\Sigma}_{\sigma}(k;\varepsilon)}{e_0}. 
\end{equation}
Note that it would have been more appropriate to denote the left-hand
side (LHS) of Eq.~(\ref{e104}) by $\Ol{\hbar\Sigma}_{\sigma}({\bar k};
\bar\varepsilon)$, however we have refrained from doing so for
aesthetic reasons. Accordingly, we maintain to use the same
convention as in Eq.~(\ref{e104}) in the remaining part of this
paper.

In view of the expression in Eq.~(\ref{e85}), the (dimensionless)
single-particle energy $\bar\varepsilon_{{\bar k};\sigma}$ is the 
solution of the following equation (for details see Farid (1999a,c)):
\footnote{\label{f68}
In this equation, the LHS is simply the normalized 
${\cal E}_{\varsigma;\sigma}(\varepsilon)$, i.e. ${\cal E}_{\varsigma;
\sigma}(\varepsilon)/e_0$, as introduced in Eq.~(\protect\ref{e84}), in 
which $\varepsilon$ is accorded with the (normalized) subscripts of 
${\cal E}_{\varsigma;\sigma} \equiv {\cal E}_{k;\sigma}$, i.e. 
${\bar k}$ (see Eq.~(\protect\ref{e95})) and $\sigma$. }
\begin{equation}
\label{e105}
\bar\varepsilon_{{\bar k}}^{(0)} + 
\Ol{\Sigma}_{\sigma}
({\bar k};\bar\varepsilon_{{\bar k};\sigma}) = 
\bar\varepsilon_{{\bar k};\sigma},
\end{equation}
where (see Eq.~(\ref{ea35}) below)
\begin{equation}
\label{e106}
\bar\varepsilon_{{\bar k}}^{(0)} {:=}
\frac{\varepsilon_k^{(0)}}{e_0} \equiv
\frac{1}{2}\, {\bar k}^2.
\end{equation}

From the {\sl explicit} linear dependence on $r_s$ of $\wh{\cal V}$ in 
Eq.~(\ref{e99}), some general arguments (see footnote \ref{f9}) and 
inspection of the explicit results concerning $\{\Ol{\Sigma}_{\sigma;
\infty_m}({\bar k})\}$, we deduce the following general expressions
\begin{eqnarray}
\label{e107}
&&\Ol{\Sigma}_{\sigma;\infty_0}({\bar k})
\equiv r_s\, {\cal S}_{\sigma;\infty_0}^{(1)}({\bar k}),\\
\label{e108}
&&\Ol{\Sigma}_{\sigma;\infty_1}({\bar k})
\equiv r_s^2\, {\cal S}_{\sigma;\infty_1}^{(2)}({\bar k}),\\
\label{e109}
&&\Ol{\Sigma}_{\sigma;\infty_m}({\bar k})
\equiv r_s^2\, {\cal S}_{\sigma;\infty_m}^{(2)}({\bar k})
+\dots + r_s^{m+1}\,
{\cal S}_{\sigma;\infty_m}^{(m+1)}({\bar k}),\nonumber\\
&&\;\;\;\;\;\;\;\;\;\;\;\;\;\;\;\;\;\;\;\;\;\;\;\;\;\;\;\;\;\;\;\;
\;\;\;\;\;\;\;\;\;\;\;\;\;\;\;\;\;\;\;\;\;\;\;\;\;\;\;\;\;\;\;\; 
m\ge 2,
\end{eqnarray}
where ${\cal S}_{\sigma;\infty_j}^{(p)}({\bar k})$, $j=0,1,\dots, m$; 
$p=1,2,\dots, m+1$, are {\sl implicit} functions of $r_s$ which do 
{\sl not} explicitly depend on $r_s$; the expressions for these functions 
are readily deduced from the full expression for 
$\Ol{\Sigma}_{\sigma;\infty_j}({\bar k})$. The superscript $p$ 
in ${\cal S}_{\sigma;\infty_j}^{(p)}({\bar k})$ signifies $r_s^p$ 
by which it is multiplied. Diagrammatically, $r_s^p 
{\cal S}_{\sigma;\infty_j}^{(p)}({\bar k})$ is the total contribution 
to $\Ol{\Sigma}_{\sigma;\infty_j}({\bar k})$ due to all 
$p$th-order skeleton SE diagrams (for the definition see Luttinger 
and Ward (1960)) in terms of the {\sl bare} particle-particle interaction 
function $v$ and the {\sl exact} single-particle GF. 

The expression in Eq.~(\ref{e109}) does {\sl not} directly apply to 
$\Ol{\Sigma}_{\sigma;\infty_m}({\bar k})$ pertaining to systems 
of fermions interacting through specifically (but {\sl not} exclusively)
$v\equiv v_c$ in $d=3$. To deal with this case we need to introduce 
some new notation. Below we do this for general systems, that is including 
those with possibly inhomogeneous GSs; similar to our considerations 
in the earlier Sections, below we deal with the coordinate representation 
of the SE operator.

\subsubsection{Some notational conventions (general)}
\label{s16}

Consider the {\sl infinite}-order (asymptotic) series for 
$\wt{\Sigma}_{\sigma}({\Bf r},{\Bf r}';z)$ as presented in 
Eq.~(\ref{e72}), with $\Sigma_{\sigma;\infty_m}({\Bf r},{\Bf r}')$ 
the coefficient of $1/z^m$. Truncating this series (following the 
necessary regularizations (see \S~II.B)) at a finite order, we 
denote the coefficient of $1/z^m$ by $\wt{\Sigma}_{\sigma;\infty_m}
({\Bf r},{\Bf r}'\vert z)$ and introduce the following decomposition:
\footnote{\label{f69}
As we shall see, owing to the branch-cut discontinuities of the 
transcendental functions involved in the pertinent expressions, 
it is expedient to work with {\sl complex} $z$ and effect the 
substitution $z \rightharpoonup \varepsilon \pm i\eta$, $\eta
\downarrow 0$, only when required in a particular application. }
\begin{eqnarray}
\label{e110}
\wt{\Sigma}_{\sigma;\infty_m}({\Bf r},{\Bf r}'\vert z) &\equiv&
\Sigma_{\sigma;\infty_m}^{\rm r}({\Bf r},{\Bf r}')+
\Sigma_{\sigma;\infty_m}^{\rm s_b}({\Bf r},{\Bf r}')\nonumber\\
&+&\wt{\Sigma}_{\sigma;\infty_m}^{\rm s}({\Bf r},{\Bf r}'\| z).
\end{eqnarray}
Here $\Sigma_{\sigma;\infty_m}^{\rm r}({\Bf r},{\Bf r}')$, the 
`regular' part of $\wt{\Sigma}_{\sigma;\infty_m}({\Bf r},{\Bf r}' 
\vert z)$, consists of {\sl all} contributions to the general 
expression for $\Sigma_{\sigma;\infty_m}({\Bf r},{\Bf r}')$ that 
for the interaction function $v$ under consideration, specifically 
for $v\equiv v_c$, satisfy conditions (A)-(C) in \S~II.B; 
$\Sigma_{\sigma;\infty_m}^{\rm s_b}({\Bf r},{\Bf r}')$ denotes 
the totality of functions in $\Sigma_{\sigma;\infty_m}({\Bf r},
{\Bf r}')$ that satisfy the requirement (A) in \S~II.B (that is, 
they are bounded almost everywhere) but fail to satisfy either (B) 
or (C), or both ({\it cf}. Eq.~(\ref{e212}) below); finally, 
$\wt{\Sigma}_{\sigma;\infty_m}^{\rm s}({\Bf r},{\Bf r}'\| z)$ arises 
from the regularization of the unbounded terms $\Sigma_{\sigma;
\infty_p}^{\rm s}({\Bf r},{\Bf r}')$, $p=2,\dots,m$, in the expression 
for the original $\Sigma_{\sigma;\infty_p}({\Bf r},{\Bf r}')$, that 
is those which fail to satisfy requirement (A) in \S~II.B ({\it cf}. 
Eq.~(\ref{e214}) below). In the context of our considerations in 
this paper, the unbounded terms $\Sigma_{\sigma;\infty_p}^{\rm s}
({\Bf r},{\Bf r}')$ are those directly associated with the 
substitution $v\rightharpoonup v_c$ (see later); when not identically 
vanishing, the actual function $\wt{\Sigma}_{\sigma;\infty_m}^{\rm s}
({\Bf r},{\Bf r}'\| z)$ is required to satisfy the following 
{\sl three} conditions:
\footnote{\label{f70}
In these expressions are implicit the sufficiency of the Poincar\'e 
definition of AS in the context of our work in this paper. See \S~II.B. }
\begin{eqnarray}
\label{e111}
&&\frac{1}{\wt{\Sigma}_{\sigma;\infty_m}^{\rm s}
({\Bf r},{\Bf r}'\| z)} = o(1),\;\;
\frac{\wt{\Sigma}_{\sigma;\infty_m}^{\rm s}({\Bf r},
{\Bf r}'\| z)}{z} = o(1),\;\;
\mbox{\rm and}\nonumber\\
&&\;\;\;\wt{\Sigma}_{\sigma;\infty_m}^{\rm s}({\Bf r},
{\Bf r}'\| z)\;\;
\mbox{\rm contains {\it no} decaying parts},\;\;\; 
\nonumber\\
&&\;\;\;\;\;\;\;\;\;\;\;\;\;\;\;\;\;\;\;\;\;\;\;\;\;\;\;\;\;\;\;
\;\;\;\;\;\;\; \;\;\;\;\;\;\;\;\;\;\;\;\;\;\;\;\;\;\;\;
\mbox{\rm as}\;\; \vert z\vert\to\infty,
\end{eqnarray}
which imply that $\wt{\Sigma}_{\sigma;\infty_m}^{\rm s}({\Bf r},
{\Bf r}'\| z)$ may {\sl not} be the full contribution of the pertinent 
regularized functions (originating from $\Sigma_{\sigma;\infty_p}^{\rm s}
({\Bf r},{\Bf r}')$, $p=2,\dots,m$), but that contribution of these 
which upon dividing by $z^m$ is asymptotically more dominant than 
$1/z^{m+1}$ but less dominant than {\sl or} as dominant as (since 
$\wt{\Sigma}_{\sigma;\infty_m}^{\rm s}({\Bf r},{\Bf r}'\| z)$ 
{\sl may} consist of a non-vanishing $z$-independent part; see 
footnote \ref{f31}) $1/z^m$ for $\vert z\vert\to \infty$. In this 
paper we denote the {\sl full} contribution arising from the 
regularization of $\Sigma_{\sigma;\infty_p}^{\rm s}({\Bf r},
{\Bf r}')/z^p$ by $\wt{\Sigma}_{\sigma;\infty_p}^{\rm s}({\Bf r},
{\Bf r}';z)$ and that as arising from the regularization of 
$\Sigma_{\sigma;\infty_p}^{\rm s_b}({\Bf r},{\Bf r}')/z^p$ (so as 
to obtain an {\sl integrable} function with respect to ${\Bf r}$ and 
${\Bf r}'$) by $\wt{\Sigma}_{\sigma;\infty_p}^{\rm s_b}({\Bf r},
{\Bf r}';z)$. 

In order to demonstrate the possibility of dependence of 
$\wt{\Sigma}_{\sigma;\infty_m}^{\rm s}({\Bf r},{\Bf r}'\| z)$ on 
contributions arising from $\Sigma_{\sigma;\infty_p}^{\rm s}({\Bf r},
{\Bf r}')$ with $p \le m$ (but specifically with $p < m$), consider 
the case where, for instance, regularization of a component part of 
$\Sigma_{\sigma;\infty_p}^{\rm s}({\Bf r},{\Bf r}')$ would result in 
$(-z/\varepsilon_0)^{1/2} {\tilde g}({\Bf r},{\Bf r}';z)$, in which 
\begin{eqnarray}
{\tilde g}({\Bf r},{\Bf r}';z) \sim \sum_{j=p}^{\infty} 
\frac{g_{\infty_j}({\Bf r},{\Bf r}')}{z^j}, \nonumber
\end{eqnarray}
with $g_{\infty_p}({\Bf r},{\Bf r}')\not\equiv 0$ and $\{ g_{\infty_j}
({\Bf r},{\Bf r}') \|\, j \ge p\}$ bounded almost everywhere and 
integrable with respect to ${\Bf r}$ and ${\Bf r}'$. From this we 
observe that, although $(-z/\varepsilon_0)^{1/2} {\tilde g}({\Bf r},
{\Bf r}';z)$ has direct bearing on $\Sigma_{\sigma;\infty_p}^{\rm s}
({\Bf r},{\Bf r}')$ (or $\Sigma_{\sigma;\infty_p}({\Bf r},{\Bf r}')$), 
it nonetheless contributes to $\wt{\Sigma}_{\sigma;\infty_m}^{\rm s}
({\Bf r},{\Bf r}'\| z)$ with $m \ge p$; with reference to the conditions 
presented in Eq.~(\ref{e111}), it is readily observed that this 
contribution to $\wt{\Sigma}_{\sigma;\infty_m}^{\rm s}({\Bf r},{\Bf r}'
\| z)$ is equal to $(-z/\varepsilon_0)^{1/2} g_{\infty_{m}}({\Bf r},
{\Bf r}')$. 

We point out that the requirements in Eq.~(\ref{e111}) are necessary 
insofar as they guarantee that
\footnote{\label{f71}
Within the context of our considerations in this paper, 
$\Sigma_{\sigma;\infty_0}({\Bf r},{\Bf r}')$ and $\Sigma_{\sigma;
\infty_1}({\Bf r},{\Bf r}')$ are strictly regular, so that the
expression in Eq.~(\protect\ref{e111}) has relevance to $m \ge 2$, 
$m=2$ being specifically significant for $v_c$ in $d=3$. }
\begin{eqnarray}
\label{e112}
&&\wt{\Sigma}_{\sigma}({\Bf r},{\Bf r}';z)
\sim \Sigma_{\sigma;\infty_0}({\Bf r},{\Bf r}')
+\frac{\Sigma_{\sigma;\infty_1}({\Bf r},{\Bf r}')}{z}\nonumber\\
&&\;\;\;\;\;
+\frac{\wt{\Sigma}_{\sigma;\infty_2}({\Bf r},{\Bf r}'\vert z)}{z^2}
+\frac{\wt{\Sigma}_{\sigma;\infty_3}({\Bf r},{\Bf r}'\vert z)}{z^3}
+\dots
\end{eqnarray}
is a {\sl well-ordered} AS for $\wt{\Sigma}_{\sigma}({\Bf r},{\Bf r}';
z)$ as $\vert z\vert\to\infty$, that is the ratio of the $(m+1)$th term 
to the $m$th term approaches zero for $\vert z\vert \to \infty$ (see 
\S~II.B). We should emphasize that the possibility of the first 
appearance of a $z$-dependent coefficient function in the {\sl third} 
leading term on the RHS of Eq.~(\ref{e112}) is closely related with 
$v\equiv v_c$ {\sl and} $d=3$ that we explicitly consider in this 
paper; to compare, for $v\equiv v_c$ and $d=2$, the first $z$-dependent 
coefficient appears in the {\sl second} leading term of the 
large-$\vert z\vert$ AS for $\wt{\Sigma}_{\sigma}({\Bf r},{\Bf r}';z)$ 
(B. Farid, 2001, unpublished). 

As discussed in \S~II.B in detail, $\wt{\Sigma}_{\sigma;\infty_m}^{\rm s}
({\Bf r},{\Bf r}'\| z)$ necessarily depends on $z$, which dependence is 
{\sl not} polynomial (of finite order), but transcendental. Depending 
on whether or not the partial infinite summation corresponding to the 
process of regularization involves contributions with increasing 
`powers' of $v$, $\wt{\Sigma}_{\sigma;\infty_m}^{\rm s}({\Bf r},
{\Bf r}'\| z)$ is or is not a transcendental function of the coupling 
constant of interaction (or of $r_s$ in the systems with uniform and 
isotropic GSs, discussed above in this Section) respectively; in the 
latter case, it is a finite-order polynomial of the mentioned coupling 
constant (here we are solely considering the {\sl explicit} dependence 
of $\wt{\Sigma}_{\sigma;\infty_m}^{\rm s}({\Bf r},{\Bf r}'\| z)$ on the 
coupling constant of interaction and not its {\sl implicit} dependence). 
In any event, even if $\wt{\Sigma}_{\sigma;\infty_m}^{\rm s}({\Bf r},
{\Bf r}'\| z)$ should for a given $m$ depend {\sl strictly} polynomially 
on the interaction coupling constant, the mentioned non-holomorphic 
dependence of this function upon $z$ gives rise to a transcendental 
dependence of the single-particle energies on the coupling-constant 
of interaction.

\subsubsection{Some intermediate considerations}
\label{s17}

In the light of the above considerations, we observe that, when $d=3$ and 
$v\equiv v_c$, $\Ol{\Sigma}_{\sigma;\infty_m}^{\,\rm r}({\bar k})$ 
can be expressed {\sl exactly} as in Eq.~(\ref{e109}). Since by definition 
$\Sigma_{\sigma;\infty_m}^{\rm s_b}({\Bf r},{\Bf r'})$ fails to satisfy 
at least one of the conditions (B) and (C) in \S~II.B, it follows that 
the Fourier transform of this function does {\sl not} exist, so that 
calculation of the AS of the Fourier transform of $\wt{\Sigma}_{\sigma}
({\Bf r},{\Bf r}';z)$ involving terms of order $1/z^m$ has to be preceded 
by an infinite summation over a well-specified set of terms pertaining to 
$\{ \Sigma_{\sigma;\infty_p}^{\rm s_b}({\Bf r},{\Bf r}')/z^p\, \|\, p\ge 
m\}$, resulting in a function that satisfies conditions (B) and (C) in 
\S~II.B (for simplicity, but without loss of generality, here we have
assumed that $\Sigma_{\sigma;\infty_m}^{\rm s_b}({\Bf r},{\Bf r'})$ has 
not already been accounted for in the process of regularization of 
$\Sigma_{\sigma;\infty_p}^{\rm s_b}({\Bf r},{\Bf r'})$, with $p < m$;
see later); we denote this function, which in addition to ${\Bf r}$ and 
${\Bf r}'$ also depends on $z$, by $\wt{\Sigma}_{\sigma;
\infty_m}^{\rm s_b}({\Bf r},{\Bf r}'; z)$. Only {\sl after} the Fourier 
transformation of this function, can the contribution to the desired 
large-$\vert z\vert$ AS of the Fourier transform of $\wt{\Sigma}_{\sigma}
({\Bf r},{\Bf r}';z)$, as originating from $\Sigma_{\sigma;
\infty_m}^{\rm s_b}({\Bf r},{\Bf r}')$, be calculated. Precisely as is 
the case with the determination of $\wt{\Sigma}_{\sigma;\infty_m}^{\rm s}
({\Bf r},{\Bf r}'\| z)$, where account has to be taken of the asymptotic 
contributions due to functions arising from the regularization of 
unbounded contributions to $\Sigma_{\sigma;\infty_p}({\Bf r},{\Bf r}')$ 
with $p < m$ (see text succeeding Eq.~(\ref{e112}) above), here also the 
coefficient of $1/z^m$ in a finite-order AS for the Fourier transform of 
$\wt{\Sigma}_{\sigma}({\Bf r},{\Bf r}';z)$, $\vert z\vert \to\infty$, 
can in principle involve contributions from the large-$\vert z\vert$ AS 
of the Fourier transform of $\wt{\Sigma}_{\sigma;\infty_p}^{\rm s_b}
({\Bf r},{\Bf r}'; z)$ with $p < m$. Since in our considerations 
(specialized to $v\equiv v_c$ in $d=3$) $\Sigma_{\sigma;\infty_p}({\Bf r},
{\Bf r}')$ are fully regular (in the sense of \S~II.B) for $p=0,1$, it 
follows that in determining the coefficient of $1/z^2$ in the 
large-$\vert z\vert$ AS for the Fourier transform of $\wt{\Sigma}_{\sigma}
({\Bf r},{\Bf r}';z)$, there are {\sl no} contributions to be taken into 
account that would otherwise originate from the large-$\vert z\vert$ AS 
of the Fourier transform of $\wt{\Sigma}_{\sigma;\infty_p}^{\rm s_b}
({\Bf r},{\Bf r}';z)$ with $p < 2$. In Appendix H we determine both 
$\wt{\Sigma}_{\sigma;\infty_2}^{\rm s_b}({\Bf r},{\Bf r}';z)$ and the 
expressions for the relevant terms in the large-$\vert z\vert$ AS of the 
double Fourier transform of this function with respect to ${\Bf r}$ and 
${\Bf r}'$. For uniform systems of spin-$1/2$ fermions, we deduce the 
{\sl exact} leading-order term pertaining to the fully-interacting 
system, while for the same system we calculate the second term within 
the framework of the `single Slater-determinant' approximation (SSDA)
(see Appendix C).

Our considerations in Appendices F, G and H lead to the following
results:
\footnote{\label{f72}
In order to prevent confusion, below we replace the ${\bar k}$ 
argument of functions that do {\sl not} depend on ${\bar k}$ (i.e.
those that are naturally associated with {\sl local} operators)
by ``${\cdot}$''. We point out that since $\Sigma_{\sigma;
\infty_m}^{\rm s_b}({\Bf r},{\Bf r}')$ is {\sl non}-integrable 
(although it is bounded almost everywhere), for the `coefficient' of 
$1/z^m$ in the {\sl regularized} large-$\vert z\vert$ AS of the double 
Fourier transform of $\wt{\Sigma}_{\sigma}({\Bf r},{\Bf r}';z)$ with 
respect to ${\Bf r}$ and ${\Bf r}'$, that is $\wt{\Ol{\Sigma}}_{\sigma}
({\Bf q},{\Bf q}';z)$, which we denote by $\wt{\Ol{\Sigma}}_{\sigma;
\infty_m}({\Bf q},{\Bf q}'\vert z)$ ({\it cf}. Eq.~(\protect\ref{e112})), 
we have ({\it cf}. Eq.~(\protect\ref{e110})) $\wt{\Ol{\Sigma}}_{\sigma;
\infty_m} ({\Bf q},{\Bf q}'\vert z) = \Ol{\Sigma}_{\sigma;
\infty_m}^{\,\rm r}({\Bf q},{\Bf q}') + \wt{\Ol{\Sigma}}_{\sigma;
\infty_m}^{\rm s_b}({\Bf q},{\Bf q}'\| z) + \wt{\Ol{\Sigma}}_{\sigma;
\infty_m}^{\rm s}({\Bf q},{\Bf q}'\| z)$. Here both 
$\wt{\Ol{\Sigma}}_{\sigma;\infty_m}^{\rm s_b}({\Bf q},{\Bf q}'\| z)$ and 
$\wt{\Ol{\Sigma}}_{\sigma;\infty_m}^{\rm s}({\Bf q},{\Bf q}'\| z)$ are 
required to satisfy the three generic conditions specified in 
Eq.~(\protect\ref{e111}). The origin of the dependence on $z$ of 
$\wt{\Ol{\Sigma}}_{\sigma;\infty_m}^{\rm s_b}({\Bf q},{\Bf q}'\| z)$ 
should be evident (see \S~II.B). When appropriate, we denote the 
$z$-independent contribution to $\wt{\Ol{\Sigma}}_{\sigma;
\infty_m}^{\rm s_b}({\Bf q},{\Bf q}'\| z)$ by $\Ol{\Sigma}_{\sigma;
\infty_m}^{\rm s_b}({\Bf q},{\Bf q}')$ (see text following 
Eq.~(\protect\ref{eh11})). Thus, $r_s^3\, {\cal S}_{\sigma;
\infty_2}^{\rm s_b (3)}({\bar k})$ in Eq.~(\protect\ref{e113}) represents 
what we could have denoted as $\Ol{\Sigma}_{\sigma;\infty_2}^{\rm s_b}
({\bar k})$. We adopt a similar convention concerning 
$\wt{\Ol{\Sigma}}_{\sigma;\infty_m}^{\rm s}({\Bf q},{\Bf q}'\| z)$ 
({\it cf}. Eq.~(\protect\ref{e114})). }
\begin{equation}
\label{e113}
\wt{\Ol{\Sigma}}_{\sigma;\infty_2}^{\rm s_b}
({\bar k}\| {\bar z})
= r_s^3 
\big\{ -\frac{3}{2} \ln(-{\bar z}/r_s)
+ {\cal S}_{\sigma;\infty_2}^{\rm s_b (3)}({\bar k})\big\},
\end{equation}
\begin{eqnarray}
\label{e114}
\wt{\Ol{\Sigma}}_{\sigma;\infty_2}^{\rm s}({\cdot}\| {\bar z})
&=& \frac{3\, r_s^2}{\sqrt{2}}\, 
\big[ n_{0;\bar\sigma}/n_0 -n_{0;\sigma}/n_0\big]\,
(-{\bar z})^{1/2}\nonumber\\
& &+ r_s^3 \big\{ 3 \ln(-{\bar z}/r_s^2)
+ {\cal S}_{\sigma;\infty_2}^{{\rm s} (3)} \big\},
\end{eqnarray}
where
\footnote{\label{f73}
Here $\gamma$ stands for the Euler number (see Eq.~(\protect\ref{ef138})). }
\begin{eqnarray}
\label{e115}
&&{\cal S}_{\sigma;\infty_2}^{\rm s_b (3)}({\bar k}) =
\frac{3}{2} (\gamma-1)\nonumber\\
&&\;\;\;\;\;\;\;\;\;\;\;\;\;
+ 3 \int \frac{{\rm d}^3k'}{(2\pi)^3}\;
\frac{\ol{\varrho}_{\sigma}^{\rm h}(\|{\Bf k}'\|)}{n_0}\,
\ln\left(\|{\bar {\Bf k}} - r_0 {\Bf k}'\|\right);\\
\label{e116}
&&{\cal S}_{\sigma;\infty_2}^{\rm s (3)} = 
3 \ln(R/a_0) +
3 \int_0^{\infty}\!\! {\rm d}r\; \big[ \frac{\Theta(r-R)}{r} 
+ \frac{1}{g_c n_0}\, \Lambda^{\rm h}(r)\big],\nonumber\\ 
\end{eqnarray}
in which $a_0$ is the Bohr radius defined in Eq.~(\ref{e92}) and (see 
Eq.~(\ref{ef128}))
\begin{equation}
\label{e117}
R \gg \frac{e^2}{4\pi\epsilon_0\, \vert z\vert}.
\end{equation}
For $\ol{\varrho}_{\sigma}^{\rm h}$, $\Lambda^{\rm h}$ and $R$ see in 
particular Eqs.~(\ref{ej3}), (\ref{ef119}) and (\ref{ef127}) respectively; 
for the significance of superscript h see Eq.~(\ref{ef9}). Note that 
$\wt{\Ol{\Sigma}}_{\sigma;\infty_2}^{\rm s}({\cdot}\| {\bar z})$ in 
Eq.~(\ref{e114}) indeed satisfies the requirements in Eq.~(\ref{e111}) 
and that the first contribution on the RHS of Eq.~(\ref{e114}) is 
identically vanishing for $n_{\sigma}\equiv n_{\bar\sigma}$ (see Appendix 
G). Combining the results in Eqs.~(\ref{e113}) and (\ref{e114}), we have
\begin{eqnarray}
\label{e118}
\wt{\Ol{\Sigma}}_{\sigma;\infty_2}^{{\rm s_b}\oplus {\rm s}}
({\bar k}\| {\bar z}) &{:=}& 
\wt{\Ol{\Sigma}}_{\sigma;\infty_2}^{\rm s_b}
({\bar k}\| {\bar z}) +
\wt{\Ol{\Sigma}}_{\sigma;\infty_2}^{\rm s}
({\cdot}\| {\bar z}) \nonumber\\
&=& \frac{3\, r_s^2}{\sqrt{2}}\, 
\big[ n_{0;\bar\sigma}/n_0 -n_{0;\sigma}/n_0\big]\,
(-{\bar z})^{1/2}\nonumber\\
&+& r_s^3 \big\{
\frac{3}{2}\,\ln(-{\bar z}/r_s^3) 
+ {\cal S}_{\sigma;\infty_2}^{{\rm s_b}\oplus 
{\rm s} (3)}({\bar k})\big\}, 
\end{eqnarray}
where
\begin{equation}
\label{e119}
{\cal S}_{\sigma;\infty_2}^{{\rm s_b}\oplus {\rm s} (3)}({\bar k})
{:=} {\cal S}_{\sigma;\infty_2}^{\rm s_b (3)}({\bar k})
+ {\cal S}_{\sigma;\infty_2}^{\rm s (3)}.
\end{equation}
For completeness, we present the following expression (see Appendix
H), to be compared with that in Eq.~(\ref{e115}):
\begin{eqnarray}
\label{e120}
&&\left. 
{\cal S}_{\sigma;\infty_2}^{\rm s_b (3)}({\bar k})\right|_{\rm s}
= \frac{3}{2} (\gamma-1) + \frac{3}{16} 
\big[ ({\bar k}/{\bar k}_F)^2 - 17/3\big]\nonumber\\
&&\;\;\;
-\frac{3}{32} \frac{{\bar k}_F}{\bar k}\,
\Big[ \big(1 - {\bar k}/{\bar k}_F\big)^3
\big(3 + {\bar k}/{\bar k}_F\big)\, 
\ln(\vert {\bar k}_F - {\bar k}\vert)\nonumber\\
&&\;\;\;\;\;\;\;\;\;\;\;\;\;\;\;
-\big(1 + {\bar k}/{\bar k}_F \big)^3
\big( 3 - {\bar k}/{\bar k}_F \big)\,
\ln(\vert {\bar k}_F + {\bar k}\vert)\Big].
\end{eqnarray}

In what follows we mainly deal with the solution of the equation for 
single-particle energies, i.e. Eq.~(\ref{e105}). Below we shall for 
the large part deal with the general case where the complications 
such as those associated with $v\equiv v_c$ in $d=3$ are not present; 
at places, however, we shall make comments relevant to $v_c$ in $d=3$. 
We shall consider two regimes, corresponding to $r_s\Ieq{\sim}{<} 1$, 
and $r_s > 1$, the weak- and intermediate-coupling regime and 
strong-coupling regime respectively.

\subsubsection{The weak- and intermediate-coupling regimes} 
\label{s18}

Here we consider the solution of Eq.~(\ref{e105}) in the weak- and 
intermediate-coupling regimes, corresponding to $r_s < 1$ and
$r_s \approx 1$ respectively. To this end, we first deal with 
$\Ol{\Sigma}_{\sigma}({\bar k};\bar\varepsilon)$, corresponding 
to an {\sl unspecified} $\bar\varepsilon$, for which following 
Eqs.~(\ref{e72}), (\ref{e107}), (\ref{e108}) and (\ref{e109}), 
we have
\begin{equation}
\label{e121}
\Ol{\Sigma}_{\sigma}({\bar k};\bar\varepsilon)
= r_s {\cal S}_{\sigma;\infty_0}^{(1)}({\bar k})
+ \sum_{j=2}^{\infty} r_s^j \sum_{p=j-1}^{\infty}
\frac{{\cal S}_{\sigma;\infty_p}^{(j)}({\bar k})}{\bar\varepsilon^p}.
\end{equation}
We note that 
\begin{equation}
\label{e122}
r_s^j\,\sum_{p=j-1}^{\infty} 
\frac{{\cal S}_{\sigma;\infty_p}^{(j)}({\bar k})}
{\bar\varepsilon^p} \equiv 
\Ol{\Sigma}_{\sigma}^{(j)}({\bar k};\bar\varepsilon),
\end{equation} 
where $\Ol{\Sigma}_{\sigma}^{(j)}({\bar k};\bar\varepsilon)$ stands
for the total contribution of the $j$th-order SE terms of the {\sl 
exact} $\Ol{\Sigma}_{\sigma}({\bar k};\bar\varepsilon)$ within the 
framework of many-body perturbation theory in terms of the {\sl bare} 
particle-particle interaction function $v$ and the {\sl exact} 
single-particle GF (see \S~I.B and the text following Eq.~(\ref{e109}) 
above); diagrammatically, $\Ol{\Sigma}_{\sigma}^{(j)}({\bar k};
\bar\varepsilon)$ is described by the complete set of $j$th-order 
skeleton SE diagrams. Equation (\ref{e121}) makes explicit that in 
order for a perturbative approximation correctly to reproduce 
$\Sigma_{\sigma;\infty_m}$, it is necessary to employ the {\sl full} 
perturbation series to order $m+1$ (inclusive) (see \S~I.B), following 
the observation that the lower boundary of the sum on the LHS of 
Eq.~(\ref{e122}) is $j-1$. This rule is of course meaningful only when 
the pertinent finite-order perturbation series do {\sl not} involve 
unbounded contributions, thus excluding the cases corresponding to 
$v\equiv v_c$ in $d=3$ (except when only the first-order series is to 
be dealt with) for which this statement can be appropriately generalized.

The expression in Eq.~(\ref{e121}) is of considerable importance, for 
it makes explicit that for $\vert \bar\varepsilon\vert \gg 1$ we 
can write
\begin{equation}
\label{e123}
\sum_{p=j-1}^{\infty} 
\frac{{\cal S}_{\sigma;\infty_p}^{(j)}({\bar k})}
{\bar\varepsilon^p} \sim
\frac{{\cal S}_{\sigma;\infty_{j-1}}^{(j)}({\bar k})}
{\bar\varepsilon^{j-1}} + 
\frac{{\cal S}_{\sigma;\infty_j}^{(j)}({\bar k})}
{\bar\varepsilon^j} + \dots, 
\end{equation}
and consequently
\begin{eqnarray}
\label{e124}
\sum_{j=2}^{\infty} r_s^j \sum_{p=j-1}^{\infty}
\frac{{\cal S}_{\sigma;\infty_p}^{(j)}({\bar k})}{\bar\varepsilon^p}
&\sim& r_s \sum_{j=2}^{\infty}
\frac{{\cal S}_{\sigma;\infty_{j-1}}^{(j)}({\bar k})}
{(\bar\varepsilon/r_s)^{j-1}}\nonumber\\ 
& &\, +\sum_{j=2}^{\infty}
\frac{{\cal S}_{\sigma;\infty_j}^{(j)}({\bar k})}
{(\bar\varepsilon/r_s)^j} + \dots.
\end{eqnarray}
The RHS of this expression shows that a further simplification can be 
achieved when $\vert\bar\varepsilon\vert \gg r_s$, under which 
condition one can write
\begin{equation}
\label{e125}
r_s \sum_{j=2}^{\infty}
\frac{{\cal S}_{\sigma;\infty_{j-1}}^{(j)}({\bar k})}
{(\bar\varepsilon/r_s)^{j-1}} \sim 
r_s \frac{{\cal S}_{\sigma;\infty_1}^{(2)}({\bar k})}
{\bar\varepsilon/r_s} +
r_s \frac{{\cal S}_{\sigma;\infty_2}^{(3)}({\bar k})}
{(\bar\varepsilon/r_s)^2} + \dots,
\end{equation}
\begin{equation}
\label{e126}
\sum_{j=2}^{\infty}
\frac{{\cal S}_{\sigma;\infty_j}^{(j)}({\bar k})}
{(\bar\varepsilon/r_s)^j} \sim 
\frac{{\cal S}_{\sigma;\infty_2}^{(2)}({\bar k})}
{(\bar\varepsilon/r_s)^2} + \dots.
\end{equation}
With $\bar\varepsilon {:=} \varepsilon/e_0$, since (see 
Eq.~(\ref{e103}))
\begin{equation}
\label{e127}
\bar\varepsilon = \frac{ \varepsilon}{2\,\mbox{\rm Ry}}\, r_s^2
\;\iff\; 
\frac{\bar\varepsilon}{r_s} =
\frac{ \varepsilon}{2\,\mbox{\rm Ry}}\, r_s,
\end{equation}
we observe that there are a number different possibilities for 
satisfying both $\vert\bar\varepsilon\vert \gg 1$ and $\vert
\bar\varepsilon\vert/r_s \gg 1$. In the range of metallic densities, 
where $2 \Ieq{\sim}< r_s \Ieq{\sim}< 6$ (specifically, where $r_s > 1$, 
corresponding to the `intermediate-coupling' {\sl regime}), we observe 
that it is the condition $\vert\bar\varepsilon\vert/r_s \gg 1$ that 
determines the minimum $\vert \bar\varepsilon\vert$ beyond which both 
$\vert\bar\varepsilon\vert \gg 1$ and $\vert\bar\varepsilon\vert/r_s \gg 
1$ are satisfied. For Na with $r_s \approx 4$, $\vert\varepsilon\vert$ 
should exceed $\frac{1}{2}$~Ry in order for 
$\vert\bar\varepsilon\vert/r_s > 1$; for $\vert \varepsilon\vert 
\Ieq{\sim}> \frac{1}{2}$~Ry, however, we have $\vert\bar\varepsilon\vert 
\Ieq{\sim}> 4$. In the weak-coupling {\sl regime} where $r_s < 1$, on 
the other hand, it is the condition $\vert\bar\varepsilon\vert \gg 1$ 
that determines the minimum $\vert\bar\varepsilon\vert$ required for 
the satisfaction of both $\vert\bar\varepsilon\vert \gg 1$ and 
$\vert\bar\varepsilon\vert/r_s \gg 1$.

Assuming $\vert\bar\varepsilon\vert \gg \max(1,r_s)$, we can combine 
the results in Eqs.~(\ref{e121}), (\ref{e124}), (\ref{e125}) and 
(\ref{e126}) and obtain
\begin{eqnarray}
\label{e128}
&&\Ol{\Sigma}_{\sigma}({\bar k};\bar\varepsilon)
\sim r_s {\cal S}_{\sigma;\infty_0}^{(1)}({\bar k})
+ r_s^2\,\frac{{\cal S}_{\sigma;\infty_1}^{(2)}({\bar k})}
{\bar\varepsilon}\nonumber\\
&&\;\;\;\;\;\;\;\;
+ r_s^2\,\frac{{\cal S}_{\sigma;\infty_2}^{(2)}({\bar k})}
{\bar\varepsilon^2}
+ r_s^3\,\frac{{\cal S}_{\sigma;\infty_2}^{(3)}({\bar k})}
{\bar\varepsilon^2} + \dots.
\end{eqnarray}

With reference to the considerations leading to the result in 
Eq.~(\ref{e128}), one readily obtains the counterpart of the expression 
in Eq.~(\ref{e128}) for the case corresponding to $v\equiv v_c$ in $d=3$: 
since singular contributions to $\Ol{\Sigma}_{\sigma;\infty_m}({\bar k}
\vert z)$, as arising from $\wt{\Ol{\Sigma}}_{\sigma;\infty_m}^{\rm s}
({\bar k}\| {\bar z})$, exist only for $m\ge 2$ (here we are employing
the notational convention introduced in \S~III.E.2, for the diagonal 
components of the SE operator in the momentum representation; by 
symmetry, the off-diagonal components are identically vanishing), the 
first two terms on the RHS of Eq.~(\ref{e128}) remain unchanged
({\sl qua} form) in the case corresponding to $v\equiv v_c$ and $d=3$ 
({\it cf}. Eq.~(\ref{e112})). Denoting the counterparts of 
${\cal S}_{\sigma;\infty_2}^{(2)}({\bar k})$ and ${\cal S}_{\sigma;
\infty_2}^{(3)}({\bar k})$, as arising from $\Ol{\Sigma}_{\sigma;
\infty_2}^{\rm r}({\bar k})$ (the normalized diagonal component of the 
momentum representation of $\Sigma_{\sigma;\infty_2}^{\rm r}({\Bf r},
{\Bf r}')$ introduced in Eq.~(\ref{e110})), by ${\cal S}_{\sigma;
\infty_2}^{{\rm r}(2)}({\bar k})$ and ${\cal S}_{\sigma;
\infty_2}^{{\rm r}(3)}({\bar k})$ respectively, from Eqs.~(\ref{e110}), 
(\ref{e112}) and (\ref{e118}) for $\vert {\bar z}\vert \gg \max(1,r_s)$ 
we obtain ({\it cf}. Eq.~(\ref{e128}))
\begin{eqnarray}
\label{e129}
&&\wt{\Ol{\Sigma}}_{\sigma}({\bar k};{\bar z})
\sim r_s {\cal S}_{\sigma;\infty_0}^{(1)}({\bar k})
+ r_s^2\,\frac{{\cal S}_{\sigma;\infty_1}^{(2)}({\bar k})}
{{\bar z}}\nonumber\\
&&\;\;\;\;\;
+ r_s^2\,\frac{{\cal S}_{\sigma;\infty_2}^{{\rm r} (2)}({\bar k})
+ 3\big[ n_{0;\bar\sigma}/n_0 - n_{0;\sigma}/n_0\big]\, 
(-{\bar z}/2)^{1/2}}{{\bar z}^2}\nonumber\\
&&\;\;\;\;\;
+ r_s^3\,\frac{{\cal S}_{\sigma;\infty_2}^{{\rm r} (3)}({\bar k})
+ {\cal S}_{\sigma;\infty_2}^{{\rm s_b}\oplus {\rm s}}({\bar k})
+ (3/2)\, \ln(-{\bar z}/r_s^3)}{{\bar z}^2} + \dots.\nonumber\\
\end{eqnarray}
 
The results in Eqs.~(\ref{e128}) and (\ref{e129}) are evidently the 
appropriate expressions for the determination of the single-particle 
excitation energies $\{ \bar\varepsilon_{{\bar k};\sigma}\}$ in the 
`weak-coupling' {\sl regime}, that is $r_s < 1$ (note the increasing 
powers of $r_s$), at sufficiently large values of $\vert\bar\varepsilon
\vert =\vert\bar\varepsilon_{{\bar k};\sigma}\vert$. On the other hand, 
since in arriving at the results in Eqs.~(\ref{e128}) and (\ref{e129}), 
the basic assumption has been $\vert\bar\varepsilon\vert \gg \max(1,r_s)$, 
these are equally applicable in the intermediate-coupling regime (let 
us say, for $r_s$ in the range corresponding to metallic densities), 
and even strong-coupling regime, in exchange for larger values of 
$\vert\bar\varepsilon\vert =\vert\bar\varepsilon_{{\bar k};\sigma}\vert$, 
as implied by $\vert\bar\varepsilon\vert \gg \max(1,r_s)$. 

In the weak-coupling {\sl limit} corresponding to $r_s \ll 1$, the terms 
on the RHSs of Eqs.~(\ref{e128}) and (\ref{e129}) which are proportional 
to $r_s^3$ can be neglected in comparison with those proportional to 
$r_s^2$. In turn, the latter contributions are negligible in comparison 
with $r_s {\cal S}_{\sigma;\infty_0}^{(1)}({\bar k})$. The proportionality 
with $r_s$ of this term implies that Eq.~(\ref{e105}) in conjunction 
with Eqs.~(\ref{e128}) and (\ref{e129}) correctly yields 
$\bar\varepsilon_{{\bar k};\sigma} \to \bar\varepsilon_{\bar k}^{(0)}$ as 
$r_s\downarrow 0$. It is important to note that neglect of the 
contributions proportional to $r_s^3$ on the RHSs of Eqs.~(\ref{e128}) 
and (\ref{e129}), results in a notable simplification in the calculations, 
a fact that can be readily verified through inspecting the explicit 
expressions for $\Sigma_{\sigma;\infty_2}({\Bf r},{\Bf r}')$ and
$\Sigma_{\sigma;\infty_2}^{\rm r}({\Bf r},{\Bf r}')$ in Eqs.~(\ref{e199}) 
and (\ref{e211}) respectively, and eliminating all terms herein that 
involve the third power of the coupling constant of $v$.

Calculation of $\bar\varepsilon_{{\bar k};\sigma}$ from Eq.~(\ref{e105})
in which $\Ol{\Sigma}_{\sigma}({\bar k};\bar\varepsilon_{{\bar k};
\sigma})$ is replaced by a {\sl finite-order} series, such as that in 
Eq.~(\ref{e128}) or in Eq.~(\ref{e129}), involves the interesting aspect 
that unlike $\bar\varepsilon$, $\bar\varepsilon_{{\bar k};\sigma}$ is 
{\sl not} a free parameter so that in principle, in particular when $r_s$ 
is {\sl not} truly small, one does not have an {\sl a priori} reason for 
considering the solution as necessarily accurate. This can, however, be 
established through {\sl a posteriori} verifying whether the calculated 
$\bar\varepsilon_{{\bar k};\sigma}$, for a given ${\bar k}$, satisfies 
the underlying assumption $\vert\bar\varepsilon_{{\bar k};\sigma}\vert \gg 
\max(1,r_s)$. Since, however, $\bar\varepsilon_{{\bar k};\sigma} \sim 
\bar\varepsilon_{{\bar k}}^{(0)}$ for ${\bar k}\gg {\bar k}_F$, 
the sufficiency of Eqs.~(\ref{e128}) and (\ref{e129}) for ${\bar k} \gg 
{\bar k}_F$ is guaranteed by $\bar\varepsilon_{{\bar k}}^{(0)} 
\gg 1$.

Our above considerations concerning the accuracy of finite-order AS 
for $\Ol{\Sigma}_{\sigma}({\bar k};\bar\varepsilon)$ rely solely 
on the {\sl explicit} dependences on $r_s$ of various terms, disregarding 
the possibility that these terms, when stripped off of these dependences, 
can themselves be small or large or that they may exhibit considerable 
variation with respect to ${\bar k}$, so that accuracy of the above 
expressions may also depend on the value of ${\bar k}$ considered. In 
this connection it is instructive to consider $\Ol{\Sigma}_{\sigma;
\infty_0}({\bar k})$ within the framework of the SSDA of the GS 
wavefunction; for simplicity, we deal with a system of spin-$1/2$ fermions 
in the paramagnetic phase. For this we have (below the subscript {\rm `s'} 
is indicative of this approximation)
\footnote{\label{f74}
$(2/\pi) (9\pi/4)^{1/3} \approx 1.22$. }
\footnote{\label{f75}
Restricting the AS on the RHSs of Eqs.~(\protect\ref{e128}) and 
(\protect\ref{e129}) to their leading term $r_s {\cal S}_{\sigma;
\infty_0}^{(1)}({\bar k})$, while replacing ${\cal S}_{\sigma;
\infty_0}^{(1)}({\bar k})$ by the expression on the RHS of 
Eq.~(\protect\ref{e130}) below, one obtains $\Ol{\Sigma}_{\rm s}^{\sc hf}
({\bar k})$, the Hartree-Fock SE (within the framework of the SSDA). 
The single-particle excitation energy corresponding to this, following 
Eqs.~(\protect\ref{e105}) and (\protect\ref{e106}), is equal to 
${\bar k}^2/2 - (2/\pi) {\bar k}_F\, r_s\, {\sf F}({\bar k}/{\bar k}_F)$ 
which is exactly the single-particle energy dispersion according to the 
conventional Hartree-Fock scheme (for example Ashcroft and Mermin (1981, 
p. 334)). Here, ${\bar k}_F = (9\pi/4)^{1/3}$. }
\begin{equation}
\label{e130}
\left. {\cal S}_{\sigma;\infty_0}^{(1)}({\bar k})\right|_{\rm s}
= -\frac{2}{\pi}\,\left(\frac{9\pi}{4}\right)^{1/3}\, 
{\sf F}({\bar k}/{\bar k}_F),\;\; \sigma = \pm 1/2,
\end{equation}
where
\begin{eqnarray}
\label{e131}
{\sf F}(x) {:=} \frac{1}{2} + \frac{1-x^2}{4 x}\,
\ln\left|\frac{1+x}{1-x}\right|
&\sim& \frac{1}{3 x^2} + \frac{1}{15 x^4}+\dots,\nonumber\\
& &\;\;\;\;\;\;\;\;\;
\vert x\vert \to \infty.
\end{eqnarray}
With ${\sf F}(0)=1$ and ${\sf F}(1)=1/2$, one observes a significant 
dependence on ${\bar k}$ of ${\cal S}_{\sigma;\infty_0}^{(1)}
({\bar k})\vert_{\rm s}$ for ${\bar k} \in [0,{\bar k}_F]$. In 
Appendix F we calculate the contribution of the {\sl local} part 
of $\Sigma_{\sigma;\infty_1}$ to ${\cal S}_{\sigma;\infty_1}^{(2)}
({\bar k})$ within the framework of the SSDA, which amounts to 
$(2 {\sf a}_{\rm s}/\pi)\, (9\pi/4)^{2/3} \approx 1.32$ (see 
Eq.~(\ref{ef56})); the sign of this quantity is opposite to that 
of ${\cal S}_{\sigma;\infty_0}^{(1)}({\bar k})\vert_{\rm s}$ in 
Eq.~(\ref{e130}) and its value is almost equal to the absolute
value of ${\cal S}_{\sigma;\infty_0}^{(1)}({\bar k})\vert_{\rm s}$ 
for small values of ${\bar k}$.

With reference to the series in Eq.~(\ref{e121}), whether this be 
considered as an AS of $\Ol{\Sigma}_{\sigma}({\bar k};
\bar\varepsilon)$ for $\vert\bar\varepsilon\vert\to \infty$ or one 
for $r_s\to 0$, we observe that for $\bar\varepsilon > 0$ we {\sl may} 
have to do with an {\sl alternating} AS. 
\footnote{\label{f76}
For $\bar\varepsilon\to -\infty$, one should consider the AS at issue 
as consisting of two sub-series, one involving the {\sl even} powers 
of $\bar\varepsilon$ and one involving the {\sl odd} powers of 
$\bar\varepsilon$, each of which {\sl may} be alternating. }
If this turns out to be the case, then the {\sl theory of terminants} 
(Dingle 1973, chapter XXI) (see also Copson 1965 for an exposition 
of the method of Stieltjes, according to which the last incorporated
term in an alternating asymptotic series is simply multiplied by $1/2$) 
can be fruitfully employed for deducing very accurate results for 
$\Ol{\Sigma}_{\sigma}({\bar k};\bar\varepsilon)$ through 
appropriately terminating the AS.

\subsubsection{The strong-coupling regime} 
\label{s19}

Making use of Eqs.~(\ref{e107}) - (\ref{e109}), we can rewrite 
Eq.~(\ref{e105}) as follows:
\footnote{\label{f77}
Evidently, $\Ol{\Sigma}_{\sigma}({\bar k};\bar\varepsilon)$ $=r_s 
F_{1;\sigma;r_s}(\bar\varepsilon/r_s;{\bar k})$ $+\sum_{j=2}^{\infty} 
r_s^{-j+2}$ $F_{j;\sigma;r_s}(\bar\varepsilon/r_s;{\bar k})$ for arbitrary 
$\bar\varepsilon$ ({\sl cf}. Eq.~(\protect\ref{e121})). This expression 
is seen to be ideally suited to use when $\bar\varepsilon$ takes on a 
finite value and $r_s\to\infty$. Proceeding along the lines of \S~III.B, 
one can easily deduce a similar expression which is appropriate for the 
region $\bar\varepsilon\to \bar\varepsilon_F$. }
\begin{eqnarray}
\label{e132}
\bar\varepsilon_{{\bar k}}^{(0)}
&+& r_s F_{1;\sigma;r_s}(\bar\varepsilon_{{\bar k};\sigma}/r_s;{\bar k})
\nonumber\\
&+& \sum_{j=2}^{\infty} 
\frac{1}{r_s^{j-2}}\, 
F_{j;\sigma;r_s}(\bar\varepsilon_{{\bar k};\sigma}/r_s;{\bar k}) 
= \bar\varepsilon_{{\bar k};\sigma},
\end{eqnarray}
where
\begin{eqnarray}
\label{e133}
F_{1;\sigma;r_s}(x;{\bar k}) &{:=}& \sum_{p=0}^{\infty} 
\frac{ {\cal S}_{\sigma;\infty_p}^{(p+1)}({\bar k})}{x^p},\\
\label{e134}
F_{j;\sigma;r_s}(x;{\bar k}) &{:=}& \sum_{p=j}^{\infty}
\frac{{\cal S}_{\sigma;\infty_p}^{(p+2-j)}({\bar k})}{x^p},
\;\;\; j \ge 2.
\end{eqnarray}
The subscript $r_s$ in $F_{j;\sigma;r_s}(x;{\bar k})$, $j=1,2,\dots$, 
signifies the {\sl implicit} dependence on $r_s$ of $F_{j;\sigma;r_s}
(x;{\bar k})$. As long as the sums in Eqs.~(\ref{e132}) - (\ref{e134}) 
are {\sl not} truncated (or at most only partially truncated so 
that complete sets of unbounded contributions are accounted for), the 
result in Eq.~(\ref{e132}) is also applicable to the case corresponding 
to $v\equiv v_c$ in $d=3$. Note that Eq.~(\ref{e132}) is exact, so that 
it equally applies to small values of $r_s$; it is not difficult to 
verify that for $r_s\to 0$, the expression in Eq.~(\ref{e121}) is directly 
deduced from Eq.~(\ref{e132}) (and {\sl vice versa}); consider, for 
instance, the terms in Eq.~(\ref{e121}) up to and including the 
{\sl explicit} quadratic order in $r_s$; making use of the asymptotic 
results for $\vert x\vert\to\infty$ (corresponding to $r_s\to 0$)
\begin{eqnarray}
\label{e135}
&&F_{1;\sigma;r_s}(x;{\bar k}) \sim
{\cal S}_{\sigma;\infty_0}^{(1)}({\bar k})
+ \frac{1}{x}\, {\cal S}_{\sigma;\infty_1}^{(2)}({\bar k}),\\
\label{e136}
&&F_{j;\sigma;r_s}(x;{\bar k}) 
\sim \frac{1}{x^j}\, {\cal S}_{\sigma;\infty_j}^{(2)}({\bar k}),
\;\;\; j \ge 2,
\end{eqnarray}
which are directly deduced from the expressions in Eqs.~(\ref{e133}) 
and (\ref{e134}), Eq.~(\ref{e132}) is readily shown to transform into 
Eq.~(\ref{e121}) to the appropriate order.

From Eq.~(\ref{e132}) we observe that for $\bar\varepsilon_{{\bar k};
\sigma}$ to be bounded as $r_s\to\infty$ (the `strong-coupling' 
{\sl limit}), it is necessary that the following conditions be met:
\begin{eqnarray}
\label{e137}
&&\lim_{r_s\to\infty} r_s 
F_{1;\sigma;r_s}(\bar\varepsilon_{{\bar k};\sigma}/r_s;{\bar k}) = 
\phi_{1;\sigma}({\bar k}),\\
\label{e138}
&&\lim_{r_s\to\infty} r_s^{j-2} 
F_{j;\sigma;r_s}(\bar\varepsilon_{{\bar k};\sigma}/r_s;{\bar k}) = 
\phi_{j;\sigma}({\bar k}),\;\;\; j\ge 2,
\end{eqnarray}
where $\{\phi_{j;\sigma}({\bar k})\}$ stands for a set of bounded 
functions of ${\bar k}$; the dependence of these functions on ${\bar k}$ 
may be partly due to that of $\bar\varepsilon_{{\bar k};\sigma}$ on 
${\bar k}$. As can be readily inferred from the relationship between 
${\cal S}_{\sigma;\infty_m}^{(p)}$ and the contributions of the skeleton 
SE diagrams (see text following Eq.~(\ref{e122}) above), $\{F_{j;\sigma;
r_s}(x;{\bar k})\}$ are {\sl all} bounded functions of $r_s$ for 
{\sl all} $r_s$. Therefore, the most stringent condition as implied 
by the requirement that $\bar\varepsilon_{{\bar k};\sigma}$ be finite 
in the strong-coupling {\sl limit}, is that given in Eq.~(\ref{e137}), 
with $\phi_{1;\sigma}({\bar k})$ {\sl bounded}, implying that 
$F_{1;\sigma;r_s}(\bar\varepsilon_{{\bar k};\sigma}/r_s;{\bar k})$ 
should decay at the slowest like $1/r_s$ as $r_s\to\infty$; with 
$F_{1;\sigma;r_s}(\bar\varepsilon_{{\bar k};\sigma}/r_s;{\bar k})$ 
decaying in this limit like $1/r_s^{\eta}$ with $\eta > 1$, we have 
$\phi_{1;\sigma}({\bar k}) \equiv 0$. Thus, to the leading order in 
$1/r_s$ we have
\begin{equation}
\label{e139}
\bar\varepsilon_{{\bar k};\sigma} \sim 
\bar\varepsilon_{{\bar k}}^{(0)}
+ \phi_{1;\sigma}({\bar k}) 
+ \phi_{2;\sigma}({\bar k}),\;\;\; r_s\to\infty,
\end{equation} 
where $\phi_{1;\sigma}({\bar k})$ {\sl may} be identically vanishing. 
We note that from knowledge that, for sufficiently large value of 
$r_s$, the GS of the system under consideration is {\sl not} uniform 
but a Wigner crystal, it follows that the result in Eq.~(\ref{e139}) 
will be valid for $r_s$ less than a critical value, $r_s^{\rm c}$, 
which for spin-$1/2$ fermions in $d=3$ is of the order of $10^2$ 
(Ceperley and Alder 1980, Herman and March 1984, Ortiz, {\sl et al.} 
1999). We point out that by the Seitz (1940, pp. 343 and 344) theorem, 
\footnote{\label{f78}
With $E_0(n_0)$ denoting the GS total energy per particle of a 
uniform system and $n_0$ the total particle concentration (see 
Eq.~(\protect\ref{e9})), the Seitz theorem (Seitz 1940, pp. 343 and 
344) reads $\varepsilon_F = {\rm d}[n_0 E_0(n_0)]/{\rm d}n_0$. 
Assuming the system to be in the paramagnetic state 
(${\bar k}_{F;\sigma} = {\bar k}_{F;\bar\sigma}$), with $E_0(n_0) 
\equiv E_{\rm kin}(n_0) + E_{\rm xc}(n_0)$, where $E_{\rm kin}(n_0) 
= \frac{3}{5} \varepsilon_F^{(0)}$ and $E_{\rm xc}(n_0)$ the 
exchange-correlation energy per particle, we have $\varepsilon_F 
-\varepsilon_F^{(0)} {=:} \mu_{\rm xc}(r_s) \equiv 
{\rm d}[n_0 E_{\rm xc}(n_0)]/{\rm d}n_0$ ({\it cf}. 
Eq.~(\protect\ref{ea54})). We note in passing that $\varepsilon_F = 
\varepsilon_F^{(0)} + \mu_{\rm xc}(r_s)$ amounts to the well-known 
statement that the energy of the highest-occupied single-particle 
wavefunction of the Kohn-Sham (1965) equation, which features within 
the framework of the density-functional theory (Hohenberg and Kohn 
1964) (for example Dreizler and Gross (1990)), coincides with the 
Fermi energy of the interacting system. }
the value of $\bar\varepsilon_{{\bar k};\sigma}
-\bar\varepsilon_{{\bar k}}^{(0)}$ is fixed at ${\bar k}
={\bar k}_{F;\sigma}$, so that in the limit $r_s\to\infty$ (to be precise, 
$r_s\to r_s^{\rm c}$), knowledge of $\phi_{1;\sigma}({\bar k}_F)$ is 
sufficient to determine $\phi_{2;\sigma}({\bar k}_F)$, and vice versa.

The above results clearly demonstrate the complexity of calculation 
of $\bar\varepsilon_{{\bar k};\sigma}$ in the `strong-coupling' 
regime in comparison with that in the `weak-coupling' one. It is 
however possible that some scheme that interpolates the weak-coupling 
results with those of the strong-coupling {\sl limit}, may turn out 
to be accurate for use at all $r_s$. Consider {\sl for instance} the 
following {\sl interpolation} expressions:
\begin{eqnarray}
\label{e140}
&&F_{1;\sigma;r_s}^{\rm int.}(x;{\bar k}) = 
\frac{x\, {\cal S}_{\sigma;\infty_0}^{(1)}({\bar k})}{x + 1} + 
\frac{ x\,[ {\cal S}_{\sigma;\infty_0}^{(1)}({\bar k}) + 
 {\cal S}_{\sigma;\infty_1}^{(2)}({\bar k})]}{x^2 + 1},
\nonumber\\ \\
\label{e141}
&&F_{j;\sigma;r_s}^{\rm int.}(x;{\bar k}) = 
\frac{ {\cal S}_{\sigma;\infty_j}^{(2)}({\bar k})}{x^j + 
\gamma_{\sigma}(r_s)},
\;\;\; j \ge 2,
\end{eqnarray}
where the constant $\gamma_{\sigma}(r_s)$ is to be chosen such that 
$\bar\varepsilon_F - \bar\varepsilon_F^{(0)} = \mu_{\rm xc}(r_s)$,
or $\phi_{1;\sigma}({\bar k}_F) + \phi_{2;\sigma}({\bar k}_F) = 
\mu_{\rm xc}(r_s\to\infty)$, be satisfied (see footnote \ref{f78}). 
Note that for $\vert x\vert\to\infty$, the expressions in 
Eqs.~(\ref{e140}) and (\ref{e141}) satisfy those in Eqs.~(\ref{e135}) 
and (\ref{e136}) respectively and that, according to these interpolation 
formulae, 
\begin{eqnarray}
\label{e142}
&&\phi_{1;\sigma}({\bar k}) = \big\{ 
2 {\cal S}_{\sigma;\infty_0}^{(1)}({\bar k})
+ {\cal S}_{\sigma;\infty_1}^{(2)}({\bar k}) \big\}\,
\bar\varepsilon_{{\bar k};\sigma},\\
\label{e143}
&&\phi_{2;\sigma}({\bar k}) = \frac{1}{\gamma_{\sigma}(\infty)}\,
{\cal S}_{\sigma;\infty_2}^{(2)}({\bar k}),
\end{eqnarray}
which in combination with Eq.~(\ref{e139}) yield
\begin{equation}
\label{e144}
\bar\varepsilon_{{\bar k};\sigma} \sim
\frac{\bar\varepsilon_{{\bar k}}^{(0)}
+ {\cal S}_{\sigma;\infty_2}^{(2)}({\bar k})/\gamma_{\sigma}(\infty)}
{1 - [ 2 {\cal S}_{\sigma;\infty_0}^{(1)}({\bar k})
+ {\cal S}_{\sigma;\infty_1}^{(2)}({\bar k}) ]},\;\;\; 
r_s\to\infty.
\end{equation}
With reference to our considerations following Eq.~(\ref{e131}) above, 
one observes that, for small values of ${\bar k}$ and within the 
framework of the SSDA, the denominator of the expression on the RHS 
of Eq.~(\ref{e144}) is approximately equal to two.

Above we have dealt with uniform and isotropic systems of fermions, 
interacting through a bounded and short-range two-body potential $v$. 
As we have indicated at the outset of this Section, for $v\equiv v_c$ 
in $d=3$, Eq.~(\ref{e132}) retains its validity provided {\sl all} the 
summations, including those in Eqs.~(\ref{e133}) and (\ref{e134}), are 
fully or partially (but to some {\sl infinite} order; see above) 
taken into account. However, Eq.~(\ref{e109}) is no longer valid, in 
consequence of containing un-compensated unbounded contributions 
associated with $\Ol{\Sigma}_{\sigma;\infty_m}({\bar k})$, with $m\ge 2$. 
Following the regularization of the mentioned unbounded contributions, 
effected through carrying out appropriate partial infinite summations 
(see \S~II.B), we arrive at expressions in which for instance 
${\cal S}_{\sigma;\infty_2}^{(2)}({\bar k})$ of our above 
considerations (see Eq.~(\ref{e128})) is replaced by the {\sl regular} 
contribution to ${\cal S}_{\sigma;\infty_2}^{(2)}({\bar k})$, that is 
${\cal S}_{\sigma;\infty_2}^{{\rm r} (2)}({\bar k})$, supplemented by 
$3 [ n_{0;\bar\sigma}/n_0 - n_{0;\sigma}/n_0] (-{\bar z}/2)^{1/2}$ 
(see Eq.~(\ref{e129})) and ${\cal S}_{\sigma;\infty_2}^{(3)}({\bar k})$ 
(see Eq.~(\ref{e128})) by ${\cal S}_{\sigma;\infty_2}^{{\rm r} (3)}
({\bar k})$ supplemented by ${\cal S}_{\sigma;\infty_2}^{{\rm s_b}
\oplus {\rm s}}({\bar k}) + (3/2) \ln(-{\bar z}/r_s^3)$ (see 
Eq.~(\ref{e129})). In the concluding part of the following Section we 
shall very briefly discuss the consequences of these singular 
functions of $r_s$ and ${\bar z}$ for the energies of the 
single-particle excitations.

\subsubsection{An illustrative example} 
\label{s20}

Here we consider in some detail the nature of the single-particle 
excitation energies as obtained from Eq.~(\ref{e105}) in which
$\Ol{\Sigma}_{\sigma}({\bar k};\bar\varepsilon)$ is replaced by 
the following truncated series:
\footnote{\label{f79}
Rather than using the expression in Eq.~(\protect\ref{e121}), which 
has its root in the series in Eq.~(\protect\ref{e72}), one may use the 
equivalent of the expression in Eq.~(\protect\ref{e121}) corresponding 
to the alternative series in Eq.~(\protect\ref{e79}). The normalized 
energy parameter $\bar\varepsilon_0\equiv \varepsilon_0/e_0$ (see 
Eq.~(\protect\ref{e103}) above) in the alternative expression may be 
fixed either by the requirement $\vert \bar\varepsilon
-\bar\varepsilon_0\vert \gg 1$, where $\bar\varepsilon$ is close to the 
expected value of the sought-after solution $\bar\varepsilon_{{\bar k};
\sigma}$ of Eq.~(\protect\ref{e105}), or by the requirement that 
Eq.~(\protect\ref{e146}), with $\bar\varepsilon_{{\bar k};\sigma}$ in 
the denominator replaced by $\bar\varepsilon_{{\bar k};\sigma}
-\bar\varepsilon_0$ (see Eqs.~(\protect\ref{e79}) and (\protect\ref{e81})) 
be identically satisfied by $\bar\varepsilon = \bar\varepsilon_F$ for 
${\bar k}={\bar k}_{F;\sigma} \equiv {\bar k}_{F;\bar\sigma}$; 
$\bar\varepsilon_F =\bar\varepsilon_F^{(0)} + \bar\mu_{\rm xc}$ can be 
calculated through knowledge of the total energy of the system as 
a function of the total number density (see footnote \protect\ref{f78}). }
\begin{equation}
\label{e145}
\Ol{\Sigma}_{\sigma}({\bar k};\bar\varepsilon)
\sim \Ol{\Sigma}_{\sigma;\infty_0}({\bar k})
+ \frac{ \Ol{\Sigma}_{\sigma;\infty_1}({\bar k})}
{\bar\varepsilon}.
\end{equation}
We thus have
\begin{equation}
\label{e146}
\bar\varepsilon_{{\bar k}}^{(0)} +
r_s {\cal S}_{\sigma;\infty_0}^{(1)}({\bar k})
+ r_s^2 \frac{{\cal S}_{\sigma;\infty_1}^{(2)}({\bar k})}
{\bar\varepsilon_{{\bar k};\sigma}}
=\bar\varepsilon_{{\bar k};\sigma}.
\end{equation}
From this we have the equivalent equations
\begin{equation}
\label{e147}
x = \alpha + \frac{\beta}{x} 
\iff x^2 - \alpha x - \beta = 0,
\end{equation}
where
\begin{equation}
\label{e148}
x {:=} \bar\varepsilon_{{\bar k};\sigma},\;\;
\alpha {:=} \bar\varepsilon_{{\bar k}}^{(0)} +
r_s {\cal S}_{\sigma;\infty_0}^{(1)}({\bar k}),\;\;
\beta {:=} r_s^2 \,{\cal S}_{\sigma;\infty_1}^{(2)}({\bar k}).
\end{equation}

By {\sl assuming} $\vert \beta \vert \ll \vert \alpha\vert$ (the 
``weak-coupling'' limit --- see above), one observes that $x \approx 
\alpha$, or better $x \approx \alpha + \beta/\alpha$; by iteratively 
continuing this procedure of refining the solution, we obtain the 
following infinite continued-fraction expansion (Khinchin 1964) for 
the solution, which we denote by $x_{\rm w}$ (the subscript `w' 
indicates `weak coupling'):
\begin{equation}
\label{e149}
x_{\rm w} = \alpha + 
\frac{\beta}{\alpha +}\,
\frac{\beta}{\alpha +}\,
\frac{\beta}{\alpha +}\,\dots. 
\end{equation}
If on the other hand, $\vert \beta\vert \gg \vert\alpha\vert$ (the 
``strong-coupling'' limit; 
\footnote{\label{f80}
This ``strong-coupling'' limit does {\sl not} coincide with the true 
`strong-coupling' limit. This is evident from the fact that our point 
of departure in this Section, namely Eq.~(\protect\ref{e145}), is 
{\sl not} the appropriate one for considerations regarding the true 
`strong-coupling' regime (compare Eqs.~(\protect\ref{e121}) and 
(\protect\ref{e132})). Nonetheless, the direct implication of 
$\vert\beta\vert \gg \vert\alpha\vert$ for $r_s$ (see 
Eq.~(\protect\ref{e148})) renders identification of the latter condition 
with the `strong-coupling' limit meaningful for a {\sl finite} range of 
${\bar k}$ values. To appreciate the significance of the latter statement, 
note that, since $\bar\varepsilon_{{\bar k}}^{(0)}$ is a monotonically 
increasing function of ${\bar k}$ (see Eq.~(\protect\ref{e106})), it 
follows that independent of the value chosen for $r_s$, for a 
sufficiently large value of ${\bar k}$, $\vert\alpha\vert$ exceeds 
$\vert\beta\vert$ (see Eq.~(\protect\ref{e148})), so that the 
``strong-coupling'' limit as considered here only applies for a finite 
range of ${\bar k}$ values. }
we shall later discuss the appropriate sign of $\beta$ in this 
limit), the above iteration scheme will {\sl not} be useful as it 
converges towards the `weak-coupling' solution. 
\footnote{\label{f81}
This statement is easiest understood through a graphical representation 
of $\alpha + \beta/x$ in relation to line $y=x$ in the $x-y$ plane. }
Rather, with $f(x) {:=} \alpha + \beta/x$, one should solve the 
equivalent problem $f^{-1}(x) = x$, where $f^{-1}(x)$ denotes the 
inverse of $f(x)$, defined though $f^{-1}\big(f(x)\big) \equiv 
f\big(f^{-1}(x)\big) = x$. With $f^{-1}(x) \equiv \beta/(x-\alpha)$, 
assuming, for instance, $x\approx \beta^{1/2}$ (here we are assuming 
$\beta > 0$; see later), 
\footnote{\label{f82}
For $\vert\beta\vert \gg \vert\alpha\vert$, the equation on the LHS 
of Eq.~(\protect\ref{e147}) can be written as $x\approx \beta/x \iff 
x^2 \approx \beta$, whose solution, provided $\beta > 0$, is $x\approx 
\beta^{1/2}$. Consistency demands that $\beta/x\vert_{x=\beta^{1/2}}
=\beta^{1/2}$ be larger than $\vert\alpha\vert$. In other words, 
the ``strong-coupling'' solution applies so long as $\beta^{1/2} \gg 
\vert\alpha\vert$. }
an iterative treatment of the latter equation yields the solution 
which we denote by $x_{\rm s}$ (the subscript `s' indicates `strong 
coupling'):
\begin{equation}
\label{e150}
x_{\rm s} = 
-\frac{\beta}{\alpha +}\,
\frac{\beta}{\alpha +}\,
\frac{\beta}{\alpha +}\,\dots. 
\end{equation} 
From Eqs.~(\ref{e149}) and (\ref{e150}) we have $x_{\rm w} +
x_{\rm s} = \alpha$. 

For the quadratic equation on the RHS of Eq.~(\ref{e147}) we have
the following solutions:
\begin{equation}
\label{e151}
x_{\pm} {:=} \frac{1}{2} \big\{ \alpha \pm 
(\alpha^2 + 4\beta)^{1/2} \big\},
\end{equation}
from which one indeed has $x_+ + x_- = \alpha$. It is readily verified
that in fact $x_{\rm w} \equiv x_+$ and $x_{\rm s} \equiv x_-$, so that 
the RHS of Eq.~(\ref{e150}) is the continued-fraction expansion (Khinchin
1964) of $\frac{1}{2} \{\alpha - (\alpha^2 + 4\beta)^{1/2}\}$.

The exact solutions to Eq.~(\ref{e147}) reveal that in order for 
Eq.~(\ref{e147}) (or Eq.~(\ref{e146})) to possess real solutions, 
it is {\sl necessary} that $\alpha^2 \ge -4\beta$, which is satisfied 
in the `weak-coupling' limit, however, its satisfaction in the 
`strong-coupling' limit requires $\beta > 0$ (see above where we state 
that in this limit $x\approx \beta^{1/2}$ and that $\beta > 0$). We note 
in passing that $\alpha^2 > -4 \beta$ coincides with the condition for 
the convergence of the continued fractions on the RHSs of 
Eqs.~(\ref{e149}) and (\ref{e150}). With reference to our statements
in the last but one paragraph of \S~III.E.4, we point out that our 
explicit calculations show the {\sl local} contribution to $\Sigma_{\sigma;
\infty_1}(k)\vert_{\rm s}$ to be positive (see Appendix F) so that
the condition $\beta > 0$ is not ruled out ({\it cf}. Eqs.~(\ref{e187})
and (\ref{ef55})).

Approximating $\Sigma_{\sigma}(k;\varepsilon)$ by $\ol{\Sigma}_{\sigma;
\infty_0}(k) +\ol{\Sigma}_{\sigma;\infty_1}(k)/\varepsilon$, from the 
Dyson equation for the Fourier transform of the single-particle spectral 
function as defined in Eq.~(\ref{e39}) (see also Eqs.~(\ref{e52})
and (\ref{e237})) we 
readily obtain
\begin{equation}
\label{e152}
\ol{A}_{\sigma}(k;\varepsilon) = \hbar 
\Big\{ \frac{\vert \varepsilon_-\vert}{\vert
\varepsilon_+ - \varepsilon_- \vert}\,
\delta(\varepsilon-\varepsilon_-)
+\frac{\vert \varepsilon_+\vert}{\vert
\varepsilon_+ - \varepsilon_- \vert}\,
\delta(\varepsilon-\varepsilon_+)\Big\},
\end{equation} 
where (see Eqs.~(\ref{e148}) and (\ref{e103}) above) $\varepsilon_{\pm} 
\equiv e_0\, x_{\pm}$ and where we have assumed $x_+\not=x_-$. From 
Eq.~(\ref{e152}) we obtain ({\it cf}. Eqs.~(\ref{e38}) and (\ref{e53}))
\begin{eqnarray}
\label{e153}
&&\frac{1}{\hbar} \int_{-\infty}^{\infty} {\rm d}\varepsilon\;
\varepsilon^{m-1}\, 
\ol{A}_{\sigma}(k;\varepsilon)\nonumber\\
&&\;\;\;\;\;\;\;\;\;\;\;\;
= \frac{1}{\vert \varepsilon_+ - \varepsilon_-\vert}
\Big\{ \vert\varepsilon_-\vert \varepsilon_-^{m-1}
+ \vert\varepsilon_+\vert \varepsilon_+^{m-1}\Big\}.
\end{eqnarray}
In the `weak-coupling' limit we have $\varepsilon_- \approx 0$
and $\varepsilon_+ \approx e_0\alpha$ when $\alpha > 0$
and $\varepsilon_- \approx e_0\alpha$
and $\varepsilon_+ \approx 0$ when $\alpha < 0$. Employing these 
results, from Eq.~(\ref{e152}) we have 
\begin{equation}
\label{e154}
\ol{A}_{\sigma}(k;\varepsilon) 
\approx \hbar \delta(\varepsilon -e_0\alpha)
\end{equation}
(that is, we have only one single-particle excitation whose energy 
is equal to $e_0\alpha$), and from Eq.~(\ref{e153})
\begin{equation}
\label{e155}
\frac{1}{\hbar} \int_{-\infty}^{\infty} {\rm d}\varepsilon\;
\varepsilon^{m-1}\, 
\ol{A}_{\sigma}(k;\varepsilon) \approx (e_0\alpha)^{m-1},
\end{equation}
the RHS of which exactly coincides with that of Eq.~(\ref{e53}).

On the other hand, in the ``strong-coupling'' limit (see footnote 
\ref{f80} and note that {\sl here} this limit is meaningful only for
finite values of ${\bar k}$), $x_{\pm} \approx \pm \beta^{1/2}$ so 
that from Eq.~(\ref{e153}) we have
\footnote{\label{f83}
The principle underlying conversion of the dimensionless quantity 
$\beta$ into one with dimension energy, namely $e_0\beta^{1/2}$, 
is as follows. Starting from the `dimensionful' Dyson equation 
$\varepsilon_{k}^{(0)} + \sum_{m=0}^{\infty} \hbar
\ol{\Sigma}_{\sigma;\infty_m}(k)/\varepsilon_{k;\sigma}^m = 
\varepsilon_{k;\sigma}$, dividing both sides by $e_0$, with $e_0$
as defined in Eq.~(\protect\ref{e103}), and subsequently writing 
$\varepsilon_{k;\sigma}^m = e_0^m (\varepsilon_{k;\sigma}/e_0)^m 
\equiv e_0^m\, \bar\varepsilon_{{\bar k};\sigma}^m$, it immediately 
follows that $\Ol{\Sigma}_{\sigma;\infty_m}({\bar k}) \equiv 
\hbar\ol{\Sigma}_{\sigma;\infty_m}(k)/e_0^{m+1}$ (see text following
Eq.~(\protect\ref{e104})). }
\begin{equation}
\label{e156}
\frac{1}{\hbar} \int_{-\infty}^{\infty} {\rm d}\varepsilon\;
\varepsilon^{m-1}\, 
\ol{A}_{\sigma}(k;\varepsilon) \approx 
\frac{1}{2} \{1 + (-1)^{m-1}\}\,
\big(e_0\beta^{1/2}\big)^{m-1},
\end{equation}
which, with the exception of cases corresponding to {\sl even} 
values of $m$, similar to Eq.~(\ref{e155}) exactly reproduces the 
RHS of Eq.~(\ref{e53}) for $x_{\pm}$. It is of significance that both 
Eq.~(\ref{e155}) and Eq.~(\ref{e156}) yield the {\sl exact} result 
in Eq.~(\ref{e57}) corresponding to $m=1$ which concerns the 
interaction-independent normalization of the single-particle 
spectral function.

With reference to our discussions in \S~III.B, we point out that
in general (excluding the case corresponding to $m=1$) there is a 
{\sl fundamental} difference between the moments integral in 
Eq.~(\ref{e38}) (in conjunction with Eq.~(\ref{e37})) and moments 
integrals in Eqs.~(\ref{e53}), (\ref{e153}), (\ref{e155}) and 
(\ref{e156}) which involve the Fourier transform of the single-particle 
spectral function. As we have indicated earlier in this paper (see
also Appendix A), this follows from the fact that, for interacting 
systems, the Lehmann amplitudes $\{f_{s;\sigma}({\Bf r})\}$ pertaining 
to systems with uniform GSs are {\sl not} single plane waves 
$\{ \Omega^{-1/2} \exp(i {\Bf k}\cdot {\Bf r})\}$, so that 
{\sl non}-orthonormality that is a characteristic feature of the 
set of Lehmann amplitudes pertaining to {\sl interacting} systems 
(see Appendix A), is {\sl not} taken account of by Eqs.~(\ref{e53}), 
(\ref{e153}), (\ref{e155}) and (\ref{e156}) (see our remark following 
Eq.~(\ref{e53}) above). Since in the weak-coupling limit we have 
(symbolically) $\{ f_{s;\sigma}({\Bf r}) \} \approx \{ \Omega^{-1/2} 
\exp(i {\Bf k} \cdot {\Bf r})\}$ ({\it cf}. Eqs.~(\ref{e46}) -
(\ref{e48}) above), it follows that Eq.~(\ref{e155}) (which is 
the equivalent of Eq.~(\ref{e53}) in our present considerations) is a 
far more reliable representation of Eq.~(\ref{e42}) (in which $s'$ is 
set equal to $s$ and $E$ is identified with $\infty$) than 
Eq.~(\ref{e156}) is. Although the approximate SE in Eq.~(\ref{e145}) 
is undeniably too inaccurate in the strong-coupling regime (see 
\S~III.E.5 and footnote \ref{f80}), it is {\sl not} difficult to see 
that the deviation of Eq.~(\ref{e156}) for even values of $m$ from 
Eq.~(\ref{e53}) (more specifically, the fact that, unlike the RHS of 
Eq.~(\ref{e53}), that of Eq.~(\ref{e156}) is {\sl identically} vanishing 
for even values of $m$) is in reality almost entirely (insofar as 
small values of ${\bar k}$ are concerned; see footnote \ref{f80}
and later) attributable to the severe deviation of $\{ f_{s;\sigma}
({\Bf r})\}$ from $\{ \Omega^{-1/2} \exp(i {\Bf k} \cdot {\Bf r})\}$ 
in the strong-coupling regime: since $\ol{A}_{\sigma}(k;\varepsilon)$ is 
positive semi-definite, it follows that for {\sl any} $\ol{A}_{\sigma}
(k;\varepsilon)$ which is symmetric with respect to $\varepsilon$ (as 
$\ol{A}_{\sigma}(k;\varepsilon)$ in Eq.~(\ref{e152}) approximately is 
for $\varepsilon_{\pm} \approx \pm e_0\beta^{1/2}$), the integrals over 
$(-E,E)$, $E\to\infty$, of $\varepsilon^{m-1}\,\ol{A}_{\sigma}(k;
\varepsilon)$ must be identically vanishing for even values of $m$; what 
prevents this to be generally the case for the LHS of Eq.~(\ref{e50}), 
with $s=s'$ (where $s$ is justifiably identified with ${\Bf k}$ in the 
weak-coupling regime and whence Eq.~(\ref{e53})), is the fact that, 
depending on the $s$, and therefore $f_{s;\sigma}({\Bf r})$, under 
consideration, the weight in the spectral function ${\sf A}_{\sigma;s,s}
(\varepsilon)$, as distinct from $\ol{A}_{\sigma}(\|{\Bf k}\|;\varepsilon) 
\equiv \ol{A}_{\sigma;{\Bf k},{\Bf k}}(\varepsilon)$, can be considerably 
asymmetrically distributed with respect to $\varepsilon=0$; this 
possibility is {\sl not} capable of being fully realized in the 
strong-coupling regime when the ``parameter of degeneracy'' $\alpha$ in 
$s = (\varsigma,\alpha) \equiv ({\Bf k},\alpha)$ (see Eq.~(\ref{e45}) 
above) is {\sl not} accounted for and $s$ is reduced to ${\Bf k}$ (see 
Appendix A). Concerning the insufficiency of the expression in 
Eq.~(\ref{e145}) in the strong-coupling regime and its role in rendering 
the RHS of Eq.~(\ref{e156}) identically vanishing for even values of 
$m$ (see above where we ascribe the latter problem almost entirely to 
the severe deviation of $f_{s;\sigma}({\Bf r})$ from a single plane 
wave), we mention that a more accurate expression for the SE than that 
on the RHS of Eq.~(\ref{e145}) would in essence bring about some degree 
of asymmetry with respect to $\varepsilon=0$ in the corresponding 
$\ol{A}_{\sigma}(k;\varepsilon)$ (since the single-particle excitations 
with $\varepsilon_{s;\sigma} < \mu$ correspond to $(N_{\sigma} - 1 + 
N_{\bar\sigma})$-particle states whereas those with 
$\varepsilon_{s;\sigma} > \mu$ correspond to $(N_{\sigma} + 1 + 
N_{\bar\sigma})$-particle states; see Eq.~(\ref{e19}) above); however, 
this would {\sl not} change the fact that, in the strong-coupling 
regime, ${\sf A}_{\sigma;s,s}(\varepsilon)$ {\sl cannot} even
approximately be identified with $\ol{A}_{\sigma;{\Bf k},{\Bf k}}
(\varepsilon)$.

The above considerations clearly show how in the weak-coupling 
limit, the simple asymptotic expression for the SE, as presented 
in Eq.~(\ref{e145}), can reliably reproduce the exact results as
presented in Eq.~(\ref{e53}). In this connection we mention that we 
have performed numerical calculations within the framework of the 
SSDA (see Appendix C) and in doing so neglected the {\sl non}-local 
contribution to $\Ol{\Sigma}_{\sigma;\infty_1}({\bar k})$ in 
Eq.~(\ref{e145}) ({\it cf}. Eq.~(\ref{e188})); in conformity with 
our above theoretical findings, our calculated results for the 
bandwidths of ideal metals, over the entire range of the metallic 
densities, turn out to be qualitatively accurate and on physical 
grounds also quantitatively justifiable.

Now we briefly consider the cases where rather than the first two leading 
terms (as in Eq.~(\ref{e145}) above), account is taken of the first 
$p+1$, $p \ge 2$, leading terms in the large-$\vert\varepsilon\vert$ 
AS for the SE (see Eq.~(\ref{e72}) above). To keep our discussion
transparent, we first restrict our considerations to the case where 
$\{ \Sigma_{\sigma;\infty_j}({\Bf r},{\Bf r}')\| j=0,1, \dots,p\}$ are 
bounded (almost everywhere) and integrable so that the employed 
finite-order AS (both in the coordinate and in the momentum
representation) involves solely $\{1/\varepsilon^j \| j=0,1,\dots,p\}$ 
and {\sl no} transcendental functions of $\varepsilon$ (see \S~II.B 
--- see also Eq.~(\ref{e128}) and compare with Eq.~(\ref{e129})). 
Replacing $\ol{\Sigma}_{\sigma}(k;\varepsilon)$ in the Dyson equation 
(see Eq.~(\ref{e232}) below) by the latter series, one obtains a 
$(p+1)$th-order polynomial equation for the single-particle energies, 
to be contrasted with the second-order equation in Eq.~(\ref{e147}). The 
real-valuedness of the coefficients in the equation ({\it cf}. 
Eq.~(\ref{e148}) above; see also \S~III.I.2 and the second paragraph 
of Appendix B)
\footnote{\label{f84}
Here we neglect ${\rm Im}[\Sigma_{\sigma}(\varepsilon)]$ (see \S~III.I), 
but shall later comment on the relevance of this contribution. }
implies that in principle this equation has $p+1$ distinct real-valued 
solutions (it may {\sl in principle} also possess pairs of complex 
conjugate solutions, however these are pathological (see \S~III.D), so 
that provided that the constituent functions [i.e. the `coefficients'] 
of the polynomial equation are calculated accurately, we expect these 
solutions not to occur), to be compared with the two solutions 
$\varepsilon_{\pm}$ discussed above; the contributions of {\sl all} 
these solutions to the single-particle spectral function, which are 
poles, will be such that the integral of this function over 
$(-\infty,\infty)$ equals unity for {\sl all} values of $p$ ({\it cf}. 
Eq.~(\ref{e57}) above or, equivalently, Eq.~(\ref{e38}) in which $m$ 
is identified with $1$ and $E$ with $\infty$). In the limit $p\to\infty$, 
these poles condense into {\sl branch cuts} (for an explicit example 
see Farid (1999c, Appendix A)) and, unlike the function 
$\ol{A}_{\sigma}(k;\varepsilon)$ in Eq.~(\ref{e152}) which is 
unbounded at the locations of the solutions of Eq.~(\ref{e146}) (or 
Eq.~(\ref{e147})), the equivalent of this function corresponding to $p\to
\infty$ is bounded almost everywhere along the real $\varepsilon$-axis; 
the pronounced peaks of this function have thus in general finite 
widths, which are interpreted as corresponding to the finite lifetimes 
of the single-particle excitations in interacting systems. Considering 
the fact that the {\sl infinite}-order series for $\wt{G}_{\sigma}
({\Bf r},{\Bf r}';z)$ in Eq.~(\ref{e27}) and the associated 
{\sl infinite}-order series for $\wt{\Sigma}_{\sigma}({\Bf r},{\Bf r}';z)$ 
are exact representations of these functions (see \S~III.A), we observe 
that by replacing the $\delta$ functions, encountered in the 
single-particle spectral functions pertaining to finite values of $p$, 
by Lorentzians, that is by applying the substitution
\begin{equation}
\label{e157}
\delta(\varepsilon-\varepsilon_j) \rightharpoonup
\frac{\eta/\pi}{(\varepsilon-\varepsilon_j)^2 + \eta^2},
\end{equation}
with $\eta$ some appropriately chosen small positive value (decreasing 
for increasing values of $p$), an accelerated convergence towards the 
exact single-particle spectral function should be achieved as $p$ is 
increased towards $\infty$. In fact, by applying such a procedure in 
the case corresponding to $p=1$ which we have explicitly considered in 
this Section, one would view, for sufficiently small values of $\vert 
\varepsilon_+ - \varepsilon_-\vert$ (which condition precludes the case 
corresponding to the `strong-coupling' regime where $\vert\varepsilon_+ - 
\varepsilon_-\vert \approx 2 e_0 \beta^{1/2}$ with $\beta^{1/2} \gg 
\alpha$), the two solutions $\varepsilon_{\pm}$ as representing the
precursors to a single peak of finite width (of the order of 
$\vert\varepsilon_+ -\varepsilon_-\vert$; see Eq.~(\ref{e152})
above) in the exact $\ol{A}_{\sigma}(k;\varepsilon)$ corresponding to 
$p=\infty$.

As we have seen in Eq.~(\ref{e118}) above and shall encounter in 
\S~III.H, for some $v$ and $d$, specifically for $v\equiv v_c$ and
$d=3$, regularization of the unbounded contributions in $\Sigma_{\sigma;
\infty_2}({\Bf r},{\Bf r}')$, which amounts to performing sums over 
infinite number of related unbounded terms, results in transcendental 
functions of $z$ that possess both essential singularities at the point 
of infinity in the complex $z$ plane (as well as at $z=0$) {\sl and} 
finite imaginary parts for $z=\varepsilon \pm i\eta$, with $\varepsilon$ 
real and $\eta\downarrow 0$ (see Eqs.~(\ref{e113}) and (\ref{e114}) 
above), and similarly for the non-integrable contributions to 
$\Sigma_{\sigma;\infty_2}({\Bf r},{\Bf r}')$ which give rise to singular 
complex-valued parts in the regularized large-$\vert\varepsilon \vert$ 
AS for the momentum representation of the SE operator (see \S~III.E.3 
and Appendix H). In such cases, incorporating the latter imaginary 
contributions while, as in the above considerations, neglecting the 
contributions to ${\rm Im}[\Sigma_{\sigma}({\Bf r},{\Bf r}';\varepsilon)]$ 
which are associated with the regular, or bounded (and integrable), 
terms in the large-$\vert\varepsilon\vert$ AS for $\Sigma_{\sigma}
({\Bf r},{\Bf r}';\varepsilon)$ (see \S~III.I) gives rise to an 
equation for the single-particle energies that unlike that presented 
in Eq.~(\ref{e147}) is {\sl not} real (for real $\varepsilon$) and 
polynomial, but complex and transcendental. Consequently, we expect the 
solutions of these equations to be complex-valued, some of which may 
be pathological (possessing imaginary parts with incorrect sign, 
arising from the approximate nature of the calculations, based on a 
partial incorporation of the imaginary contributions to the SE), 
incorrectly implying the instability of the GS of the system under 
consideration (see \S~III.D). In the present case, where the 
quasi-particle equations involve transcendental functions of 
$\varepsilon$ (or $z$), the nature of solutions are not as 
straightforwardly classified as in the case where these equations are 
purely polynomial, which, following the fundamental theorem of algebra 
(Titchmarsh 1939, pp. 118 and 119), possess a well-specified class 
of solutions.

The full incorporation of the imaginary contributions of $\Sigma_{\sigma}
(\varepsilon)$ to the large-$\vert\varepsilon\vert$ AS of this function 
(see \S~III.I) does {\sl not} lead to any {\sl fundamental} change in the 
behaviour of the corresponding single-particle excitation energies in 
comparison with the behaviours of those just discussed, except that it 
gives rise to removal of the possible pathological solutions referred to 
above. In general, leaving aside the contributions of the aforementioned 
transcendental functions of $\varepsilon$ to the quasi-particle energies, 
the contribution of the regular imaginary terms may be effectively taken 
account of by means of a set of finite positive constants $\{\eta_j\}$ 
that through Eq.~(\ref{e157}) transform the $\delta$ functions contributing 
to the single-particle spectral function (i.e. the one deduced through
identifying {\sl all} imaginary terms in the polynomial equation with 
zero) into Lorentzians. This should also apply as a means for 
effectively taking account of the imaginary contributions originating 
from the aforementioned transcendental functions of $\varepsilon$ to 
the equation for the single-particle energies, provided that the 
real-valued solutions corresponding to the case where {\sl all} 
imaginary contributions to the SE are neglected are {\sl not} too 
close to the branch points of these transcendental functions (in our 
present considerations, these are at $z=0$ and at the point of infinity 
in the complex $z$ plane; use of Eq.~(\ref{e79}) as the starting point
of the calculations, rather than Eq.~(\ref{e72}), changes 
$z=0$ here into $z=\varepsilon_0$).

\subsection{Evaluation of $G_{\sigma;\infty_2}({\Bf r},{\Bf r}')$ }
\label{s21}

Before proceeding with the calculation of $G_{\sigma;\infty_2}({\Bf r},
{\Bf r}')$, we explicitly specify the direction along which in the 
following the kinetic-energy operator $\tau({\Bf r})$ in $h_0({\Bf r})$ 
(see Eq.~(\ref{e43}) above) is meant to act on functions of ${\Bf r}$. 
In spite of the fact that $\tau({\Bf r})$ (as well as $h_0({\Bf r})$) 
is Hermitian in the single-particle Hilbert space of the problem at 
hand, this specification is {\sl not} redundant. This follows from the 
fact that in our following considerations, the domain of operation of 
$\tau({\Bf r})$ (and of $h_0({\Bf r})$ for that matter) does {\sl not} 
extend to {\sl all} functions of ${\Bf r}$ to its right, a consequence 
of the specific way (namely, through {\sl local} (anti)commutation 
relations) in which $\tau({\Bf r})$ finds its entrance into the 
expressions to be considered below. In what follows, $\tau({\Bf r})$ 
(and $h_0({\Bf r})$) will {\sl strictly} act on all functions of 
${\Bf r}$ to its right that together with $\tau({\Bf r})$ are enclosed 
by appropriate brackets. When wishing to indicate operation of 
$\tau({\Bf r})$ ($h_0({\Bf r})$) to functions of ${\Bf r}$ to its left, 
we employ the notation $\tau^+({\Bf r})$ ($h^+_0({\Bf r})$), with the 
pertinent functions of ${\Bf r}$ enclosed by the same brackets that 
enclose $\tau^+({\Bf r})$ ($h^+({\Bf r})$). 

Using the canonical anticommutation relations in Eq.~(\ref{e29}), 
we deduce
\begin{equation}
\label{e158}
\wh{A}_{\sigma}({\Bf r}) {:=}
\big[\hat\psi_{\sigma}({\Bf r}),\wh{H}\big]_-
= \big\{ {\hat\alpha}({\Bf r})\, \hat\psi_{\sigma}({\Bf r})\big\},
\end{equation}
where
\begin{equation}
\label{e159}
{\hat\alpha}({\Bf r}) {:=} h_0({\Bf r}) 
+\sum_{\sigma'}\int {\rm d}^dr''\; v({\Bf r}-{\Bf r}'')
\hat\psi_{\sigma'}^{\dag}({\Bf r}'')\hat\psi_{\sigma'}({\Bf r}''). 
\end{equation}
By enclosing the RHS of Eq.~(\ref{e158}) within curly brackets, we 
have made explicit the fact that $h_0({\Bf r})$ that features in the 
defining expression for ${\hat\alpha}({\Bf r})$ in Eq.~(\ref{e159}), 
acts {\sl solely} upon $\hat\psi_{\sigma}({\Bf r})$ and {\sl no} other 
function to the right of the enclosed function (see above). Since 
$\wh{H}$ is Hermitian, by Hermitian conjugation of Eq.~(\ref{e158}) 
we obtain
\begin{equation}
\label{e160}
\wh{A}_{\sigma}^{\dag}({\Bf r}) 
\equiv -\big[\hat\psi_{\sigma}^{\dag}({\Bf r}),\wh{H}\big]_-
= \{ \hat\psi_{\sigma}^{\dag}({\Bf r})\, 
{\hat\alpha}^{\dag}({\Bf r}) \},
\end{equation}
where ${\hat\alpha}^{\dag}({\Bf r})$, according to Eq.~(\ref{e159}),
involves $h_0^+({\Bf r})$ and thus $\tau^+({\Bf r})$ which operates 
solely on $\hat\psi_{\sigma}^{\dag}({\Bf r})$ to its left, so that
we have $\{ \hat\psi_{\sigma}^{\dag}({\Bf r}) \tau^+ ({\Bf r}) \} 
\equiv \{ \tau({\Bf r}) \hat\psi_{\sigma}^{\dag}({\Bf r}) \}$. We 
should emphasize that, although our detailed specification of
$\wh{A}_{\sigma}({\Bf r})$ and of its Hermitian conjugate (as 
reflected in our introduction of curly braces in Eqs.~(\ref{e158}) and 
(\ref{e160})) seems excessive at this stage (and it is admittedly 
so as far as evaluation of $G_{\sigma;\infty_2}({\Bf r},{\Bf r}')$ 
is concerned), such a specification is of {\sl vital} significance 
for the correct determination of $G_{\sigma;\infty_m}({\Bf r},{\Bf r}')$, 
when $m > 2$. Further, our experience shows that without appropriate 
structuring of various contributions, as achieved through introducing 
$\wh{A}_{\sigma}({\Bf r})$, the explosive proliferation of terms with 
their various indices and arguments renders determination of the 
expression for $G_{\sigma;\infty_m}({\Bf r},{\Bf r}')$, with 
$m \ge 4$, well nigh impossible.

From Eqs.~(\ref{e34}) and (\ref{e158}) it follows that 
\begin{equation}
\label{e161}
G_{\sigma;\infty_2}({\Bf r},{\Bf r}') = 
\hbar \langle\Psi_{N;0}\vert [\wh{A}_{\sigma}
({\Bf r}),\hat\psi_{\sigma}^{\dag}({\Bf r}')]_+
\vert\Psi_{N;0}\rangle.
\end{equation}
Making use of the result on the RHS of Eq.~(\ref{e159}) and the
anticommutation relations in Eq.~(\ref{e29}), we arrive at
\footnote{\label{f85}
As we have indicated earlier (see \S~III.A), $G_{\sigma;\infty_2}
({\Bf r},{\Bf r}')$ is a $c$-number, so that $\nabla_{\Bf r}^2$
involved in $h_0({\Bf r})$ (see Eq.~(\protect\ref{e43}) above) on the 
RHS of Eq.~(\protect\ref{e162}) solely operates on $\delta({\Bf r}
-{\Bf r}')$ to its right and on {\sl no} function outside the 
curly brackets. }
\begin{eqnarray}
\label{e162}
G_{\sigma;\infty_2}({\Bf r},{\Bf r}') =
\hbar \Big\{ \big[ h_0({\Bf r})
&&+ v_H({\Bf r};[n])\big] \delta({\Bf r}-{\Bf r}')
\nonumber\\
&&-v({\Bf r}-{\Bf r}') \varrho_{\sigma}({\Bf r}',{\Bf r}) \Big\},
\end{eqnarray}
where the Hartree potential $v_H({\Bf r};[n])$ is defined in 
Eq.~(\ref{e14}) above, in which the {\sl total} number density 
$n$ is defined through (see Eq.~(\ref{ek1}))
\begin{eqnarray}
\label{e163}
n({\Bf r}) &&\equiv n_{\sigma}({\Bf r}) + 
n_{\bar\sigma}({\Bf r})\nonumber\\
&&\equiv \sum_{\sigma'} \langle\Psi_{N;0}\vert
\hat\psi_{\sigma'}^{\dag}({\Bf r})
\hat\psi_{\sigma'}({\Bf r})\vert\Psi_{N;0}\rangle;
\end{eqnarray}
since our considerations are {\sl not} restricted to systems of 
spin-$1/2$ fermions, we have the general expression
\begin{equation}
\label{e164}
n_{\bar\sigma}({\Bf r}) \equiv \sum_{\sigma'\not=\sigma}
n_{\sigma'}({\Bf r}),
\end{equation}
where the summation is over $2 {\sf s}$ contributions ({\it cf}.
Eq.~(\ref{e21}) above). Further, in Eq.~(\ref{e162}),
\begin{equation}
\label{e165}
\varrho_{\sigma}({\Bf r}',{\Bf r})
{:=} \langle\Psi_{N;0}\vert
\hat\psi_{\sigma}^{\dag}({\Bf r}')
\hat\psi_{\sigma}({\Bf r})
\vert\Psi_{N;0}\rangle
\end{equation}
stands for the GS partial density matrix pertaining to the interacting
system; in our present work, it is real and symmetric (see Appendix B). 
The non-orthonormality of the set $\{f_{s;\sigma}({\Bf r})\}$ pertaining 
to interacting systems (see Appendix A) implies that $\varrho_{\sigma}
({\Bf r},{\Bf r}')$ is {\sl strictly} non-idempotent, that is 
\begin{equation}
\label{e166}
\int {\rm d}^dr''\; \varrho_{\sigma}({\Bf r},{\Bf r}'')
\,\varrho_{\sigma}({\Bf r}'',{\Bf r}') \not\equiv
\varrho_{\sigma}({\Bf r},{\Bf r}').
\end{equation}
In the case of $v\equiv v_c$ in $d=3$, the sum $h_0({\Bf r}) + 
v_H({\Bf r};[n])$ on the RHS of Eq.~(\ref{e162}) is identical with 
$\tau({\Bf r}) + u({\Bf r}) +v_H({\Bf r};[n'])$, independent of 
$\varpi_{\kappa}$, so that it is well-defined for $\kappa\downarrow 0$ 
(see \S~II.A and Eqs.~(\ref{e11}) and (\ref{e15})).

For completeness, through making use of the completeness relation 
\begin{equation}
\label{e167}
\sum_{M_{\sigma},M_{\bar\sigma},s}
\vert\Psi_{M_{\sigma},M_{\bar\sigma};s}\rangle 
\langle\Psi_{M_{\sigma},M_{\bar\sigma};s}\vert = {\cal I},
\end{equation}
where ${\cal I}$ stands for the unit operator in the Fock space, 
taking into account the orthogonality of the eigenstates of 
$\{\wh{N}_{\sigma}\}$ corresponding to different eigenvalues, we 
obtain the familiar expression
\begin{equation}
\label{e168}
\varrho_{\sigma}({\Bf r},{\Bf r}')
= \sum_s^{<} f_{s;\sigma}^*({\Bf r}) f_{s;\sigma}({\Bf r}').
\end{equation} 
Here $\sum_s^{<}$ denotes the sum over all $s$ corresponding to
$\varepsilon_{s;\sigma} < \mu$. This specification is necessary 
through our use of functions from the set $\{ f_{s;\sigma}({\Bf r})\}$ 
(see Eq.~(\ref{e18}) above) which encompasses two fundamentally 
distinct classes of functions, 
\footnote{\label{f86}
Another way of stating this fact is that the compound variable $s$ 
corresponding to cases $\varepsilon_{s;\sigma} < \mu$ is inherently 
different from that corresponding to $\varepsilon_{s;\sigma} > \mu$. } 
namely those corresponding to $\varepsilon_{s;\sigma} < \mu$ and 
those corresponding to $\varepsilon_{s;\sigma} > \mu$; had we not 
employed symbols $f_{s;\sigma}^*({\Bf r})$ and $f_{s;\sigma}({\Bf r}')$, 
but instead $\langle \Psi_{N;0}\vert\hat\psi_{\sigma}^{\dag}({\Bf r})
\vert\Psi_{N_{\sigma}-1,N_{\bar\sigma};s}\rangle$ and 
$\langle\Psi_{N_{\sigma}-1,N_{\bar\sigma};s}\vert\hat\psi_{\sigma}
({\Bf r}')\vert\Psi_{N;0}\rangle$, respectively, the `symbolic' 
restriction, as implied by $<$ in $\sum_s^{<}$, would {\sl not} 
have been necessary. It is important to appreciate the significance 
of these aspects, since in spite of the {\sl restriction} that is 
implied by $<$ in $\sum_s^{<}$ (and similarly by $>$ in $\sum_s^{>}$, 
denoting a sum over all $s$ corresponding to $\varepsilon_{s;\sigma} 
> \mu$), nonetheless Eq.~(\ref{e31}) applies. In view of these 
statements, it is interesting to consider the similar expression as 
in Eq.~(\ref{e168}) for the Slater-Fock density matrix pertaining 
to the `non-interacting' Hamiltonian $\wh{H}_0$ in Eq.~(\ref{e54}). 
With reference to the eigenvalue problem in Eq.~(\ref{e56}) we have
\begin{equation}
\label{e169}
\varrho_{{\rm s};\sigma}({\Bf r},{\Bf r}') 
= \sum_{\varsigma}^{<} \varphi_{\varsigma;\sigma}^*({\Bf r})
\varphi_{\varsigma;\sigma}({\Bf r}'),
\end{equation} 
which in view of the expression in Eq.~(\ref{e46}) and the succeeding 
text (see also the text following Eq.~(\ref{e56}) above) indeed corresponds 
to $\varrho_{\sigma}({\Bf r},{\Bf r}')$ in Eq.~(\ref{e168}) for $v\to 0$. 
In Eq.~(\ref{e169}), $\sum_{\varsigma}^{<}$ denotes a sum over all 
$\varsigma$ corresponding to $\varepsilon_{\varsigma;\sigma}^{(0)} < 
\mu^{(0)}$ (see Eq.~(\ref{e168}) above and the subsequent text), where 
$\mu^{(0)}$ stands for the `chemical potential' of the `non-interacting' 
$N$-particle system. Taking into account the normalization to unity 
of the single-particle wavefunctions $\{\varphi_{\varsigma;\sigma}
({\Bf r})\}$, making use of $\varrho_{{\rm s};\sigma}({\Bf r},{\Bf r}) 
\equiv n_{0;\sigma}({\Bf r})$ and $\int {\rm d}^dr\; n_{0;\sigma}
({\Bf r})=N_{\sigma}$, from Eq.~(\ref{e169}) it follows that 
\begin{equation}
\label{e170}
\sum_{\varsigma} \Theta\big(\mu^{(0)} - 
\varepsilon_{\varsigma;\sigma}^{(0)} \big) = N_{\sigma};
\end{equation}
that is the number of terms involved in the summation on the RHS 
of Eq.~(\ref{e169}) is {\sl exactly} equal to $N_{\sigma}$. This is in 
stark contrast with the number of terms involved in the summation 
on the RHS of Eq.~(\ref{e168}) which, under the condition that
$N_{\sigma}\not=0$, is {\sl infinitely large} (as evidenced by
the completeness relation in Eq.~(\ref{e31})),
\footnote{\label{f87}
This fact underlies the property that, for systems in the thermodynamic 
limit, $\mu_{N;\sigma}^{\mp}$ (see Eq.~(\protect\ref{e23})) are 
{\sl accumulation} points of the set of single-particle excitations in 
these systems; correspondingly, $z=\mu_{N;\sigma}^{\mp}$ are singular 
points of both $\wt{G}_{\sigma}(z)$ and $\wt{\Sigma}_{\sigma}(z)$ (Farid 
1999a,c). Note that, for metallic systems, $\mu_{N;\sigma}^{-} \equiv 
\varepsilon_F$ (independent of $\sigma$ so long as $N_{\sigma}\not=0$). 
See footnote \protect\ref{f2}. }
even though naturally $\int {\rm d}^dr\; \varrho_{\sigma}({\Bf r},
{\Bf r}) = N_{\sigma}$ holds (see our considerations leading to 
Eqs.~(\ref{e48}) and (\ref{e50}) in \S~III.B). Such a distinctive
contrast between $\varrho_{\sigma}({\Bf r},{\Bf r}')$ and 
$\varrho_{{\rm s};\sigma}({\Bf r},{\Bf r}')$ reveals one of the
most dramatic consequences of interaction and thereby of the
{\sl overcompleteness} of the set of the Lehmann amplitudes 
$\{ f_{s;\sigma}({\Bf r})\}$ (see Appendix A). 

The expression for $G_{0;\sigma;\infty_2}({\Bf r},{\Bf r}')$ is 
readily deduced from that in Eq.~(\ref{e162}) through setting the 
coupling constant of the particle-particle interaction to zero and 
noting that, in $\wh{H}_0$, the external potential $u({\Bf r})$ is 
replaced by $u({\Bf r}) + w_{\sigma}({\Bf r})$ ({\it cf.} 
Eqs.~(\ref{e55}) and (\ref{e62})); thus one obtains
\begin{equation}
\label{e171}
G_{0;\sigma;\infty_2}({\Bf r},{\Bf r}') = \hbar \Big\{
\big[ h_0({\Bf r}) + w_{\sigma}
({\Bf r})\big] \delta({\Bf r}-{\Bf r}') \Big\},
\end{equation}
which similar to $G_{\sigma;\infty_2}({\Bf r},{\Bf r}')$ in
Eq.~(\ref{e162}) is a $c$-number. As in Eq.~(\ref{e162}), we have 
emphasized this fact by means of curly brackets on the RHS of 
Eq.~(\ref{e171}); in the following we shall {\sl not} employ
this convention when there is no likelihood for confusion.

Before closing this Section, it is relevant to point out that,
although in the case of when $v\equiv v_c$ we need to effect the 
transformation in Eq.~(\ref{e11}), which through Eq.~(\ref{e43}) 
results in the corresponding $h_0({\Bf r})$ undergoing the 
transformation (see Eq.~(\ref{e5}))
\begin{equation}
\label{e172}
h_0({\Bf r}) \rightharpoonup h_0({\Bf r}) 
-\varpi_{\kappa},
\end{equation}
such a transformation must {\sl not} be effected insofar as the 
$h_0({\Bf r})$ in Eq.~(\ref{e171}) is concerned. This follows from 
the fact that $\varpi_{\kappa}$, which according to Eq.~(\ref{e5})
is proportional to the coupling constant of the particle-particle
interaction (i.e., $e^2$), is identically vanishing for the 
{\sl non}-interacting system.

\subsubsection{Evaluation of 
$\Sigma_{\sigma;\infty_0}({\Bf r},{\Bf r}')$ }
\label{s22}

From Eqs.~(\ref{e73}), (\ref{e76}), (\ref{e162}) and (\ref{e171}) 
we obtain
\begin{eqnarray}
\label{e173}
&&\Sigma_{\sigma;\infty_0}({\Bf r},{\Bf r}')
= \frac{1}{\hbar} \Big\{
v_H({\Bf r};[n]) \delta({\Bf r}-{\Bf r}')\nonumber\\
&&\;\;\;\;\;\;\;\;
-v({\Bf r}-{\Bf r}') 
\varrho_{\sigma}({\Bf r}',{\Bf r})\Big\} \equiv 
\Sigma^{\sc hf}({\Bf r},{\Bf r}';[\varrho_{\sigma}]),
\end{eqnarray} 
where $\Sigma^{\sc hf}({\Bf r},{\Bf r}';[\varrho_{\sigma}])$ stands
for the non-local SE within the Hartree-Fock framework. Clearly, 
$\Sigma_{\sigma;\infty_0}({\Bf r},{\Bf r}')$ does not {\sl explicitly} 
depend on the single-particle spin index $\sigma$. We point out, however, 
that this SE involves both the {\sl exact} number density $n$ and the 
{\sl exact} single-particle density matrix $\varrho_{\sigma}$, so that 
$\Sigma^{\sc hf}({\Bf r},{\Bf r}';[\varrho_{\sigma}])$ should be 
distinguished from that which is determined within the self-consistent 
Hartree-Fock framework where $\varrho_{\sigma}$ is replaced 
by the Slater-Fock density matrix $\varrho_{{\rm s};\sigma}$.

\subsubsection{The case of Coulomb-interacting fermions in
the thermodynamic limit }
\label{s23}

In the case where $v\equiv v_c$, $v_H({\Bf r};[n])$ on the RHS of 
Eq.~(\ref{e173}) must be replaced by $v_H({\Bf r};[n'])$ (see 
Eq.~(\ref{e15}) above). This follows from the fact that, as we have 
pointed out in the closing part of the previous Section, in this case 
$h_0({\Bf r})$ in Eq.~(\ref{e171}) must {\sl not} be subjected to the 
transformation in Eq.~(\ref{e172}) so that $\Sigma_{\sigma;\infty_0}
({\Bf r},{\Bf r}')$ `inherits' $v_H({\Bf r};[n'])$ from the expression 
for $G_{\sigma;\infty_2}({\Bf r},{\Bf r}')$ in Eq.~(\ref{e162}) (see 
our remark in the paragraph directly following Eq.~(\ref{e171}) above). 
We thus have
\begin{eqnarray}
\label{e174}
\Sigma_{\sigma;\infty_0}({\Bf r},{\Bf r}')
&=&\frac{1}{\hbar} \Big\{
v_H({\Bf r};[n']) \delta({\Bf r}-{\Bf r}')\nonumber\\
& &\;\;\;\;\;\;\;\;
-v_c({\Bf r}-{\Bf r}') \varrho_{\sigma}({\Bf r}',{\Bf r})\Big\}.
\end{eqnarray}

\subsection{Evaluation of 
$G_{\sigma;\infty_3}({\Bf r},{\Bf r}')$}
\label{s24}

Making use of Eq.~(\ref{e158}), while employing the definition
in Eq.~(\ref{e34}), we have
\begin{eqnarray}
\label{e175}
G_{\sigma;\infty_3}({\Bf r},{\Bf r}') &&=
\hbar \langle \Psi_{N;0}\vert \Big[
\big[\wh{A}_{\sigma}({\Bf r}),
\wh{H}\big]_-,\hat\psi_{\sigma}^{\dag}({\Bf r}')\Big]_+
\vert\Psi_{N;0}\rangle.
\nonumber\\
\end{eqnarray} 
As we have indicated earlier, direct evaluation of the anticommutator 
and subsequently the commutator in this expression leads to a vast
number of terms and therefore should not be attempted. On the other
hand, by employing the identity
\begin{eqnarray}
\label{e176}
\Big[ \big[\wh{A}_{\sigma}({\Bf r}),&&
\wh{H}\big]_-,\hat\psi_{\sigma}^{\dag}({\Bf r}')\Big]_+
\equiv \wh{A}_{\sigma}({\Bf r}) \big[\wh{H},
\hat\psi_{\sigma}^{\dag}({\Bf r}')\big]_- \nonumber\\
&&+\wh{A}_{\sigma}({\Bf r}) \hat\psi_{\sigma}^{\dag}({\Bf r}') \wh{H} 
-\wh{H} \wh{A}_{\sigma}({\Bf r}) \psi_{\sigma}^{\dag}({\Bf r}')
\nonumber\\
&&+\hat\psi_{\sigma}^{\dag}({\Bf r}') \wh{A}_{\sigma}({\Bf r}) \wh{H} 
-\big[\hat\psi_{\sigma}^{\dag}({\Bf r}'), \wh{H}\big]_-
\wh{A}_{\sigma}({\Bf r}) \nonumber\\
&&-\wh{H} \hat\psi_{\sigma}^{\dag}({\Bf r}')
\wh{A}_{\sigma}({\Bf r}),
\end{eqnarray}
one achieves considerable simplification by the fact that in calculating 
$G_{\sigma;\infty_3}({\Bf r},{\Bf r}')$ according to the expression in 
Eq.~(\ref{e175}), $\wh{H}$ in the second, third, fourth and sixth terms 
on the RHS of Eq.~(\ref{e176}) can be replaced by the $c$-number 
$E_{N;0}$. Moreover, the commutation of $\wh{H}$ with 
$\hat\psi_{\sigma}^{\dag}({\Bf r}')$, which according to Eq.~(\ref{e160}) 
is equal to $\wh{A}_{\sigma}^{\dag}({\Bf r}')$, results in the desirable 
situation in which in comparison with $\wh{H}$, the number of field 
operators is decreased by one (see Eqs.~(\ref{e158}) and (\ref{e159})). 
Making use of the result in Eq.~(\ref{e176}), from Eq.~(\ref{e175}) one 
readily obtains ({\it cf}. Eq.~(\ref{e34}) in which $m$ is identified 
with $1$)
\begin{eqnarray}
\label{e177}
G_{\sigma;\infty_3}({\Bf r},{\Bf r}') 
= \hbar \langle\Psi_{N;0}\vert \big[\wh{A}_{\sigma}({\Bf r}),
\wh{A}_{\sigma}^{\dag}({\Bf r}')\big]_+\vert
\Psi_{N;0}\rangle.
\end{eqnarray}
This expression makes explicit the {\sl general} property $G_{\sigma;
\infty_m}({\Bf r}',{\Bf r}) \equiv G_{\sigma;\infty_m}^*({\Bf r},
{\Bf r}')$; since in this work we choose the GS wavefunction to be 
real valued, we have, however, $G_{\sigma;\infty_m}^*({\Bf r},{\Bf r}') 
\equiv G_{\sigma;\infty_m}({\Bf r},{\Bf r}')$ (see Appendix B) so that
\begin{equation}
\label{e178}
G_{\sigma;\infty_m}({\Bf r}',{\Bf r}) \equiv
G_{\sigma;\infty_m}({\Bf r},{\Bf r}'),\;\; \forall\, m. 
\end{equation}
Despite the validity of this symmetry property for {\sl all} $m$, it 
turns out that only for {\sl odd} values of $m$, use of identities 
similar to that in Eq.~(\ref{e176}) results in an {\sl explicitly} 
symmetric $G_{\sigma;\infty_m}({\Bf r},{\Bf r}')$ (see footnote 
\ref{f45}).

Making use of the expression for $\wh{A}_{\sigma}({\Bf r})$ in
Eq.~(\ref{e158}) and the anticommutation relations in Eq.~(\ref{e29})
followed by a process of normal-ordering of the field operators, 
we arrive at
\begin{eqnarray}
\label{e179}
&&G_{\sigma;\infty_3}({\Bf r},{\Bf r}') = \hbar \Big\{ \Big[
h_0({\Bf r}) h_0({\Bf r}') \nonumber\\
&&\;\;\;\;\;
+ h_0({\Bf r}) v_H({\Bf r}';[n])
+ h_0({\Bf r}') v_H({\Bf r};[n]) 
\nonumber\\
&&\;\;\;\;\;
+ {\cal A}({\Bf r},{\Bf r})
+ \int {\rm d}^dr''\; v^2({\Bf r}-{\Bf r}'') n({\Bf r}'') \Big] 
\,\delta({\Bf r}-{\Bf r}') \nonumber\\
&&\;\;\;\;\;
- \big(h_0({\Bf r})+h_0({\Bf r}')\big) v({\Bf r}-{\Bf r}')
\varrho_{\sigma}({\Bf r}',{\Bf r}) \nonumber\\
&&\;\;\;\;\;
- v^2({\Bf r}-{\Bf r}') \varrho_{\sigma}({\Bf r}',{\Bf r})\nonumber\\
&&\;\;\;\;\;
+ v({\Bf r}-{\Bf r}') \big({\cal B}_{\sigma}({\Bf r},{\Bf r}')
+ {\cal B}_{\sigma}({\Bf r}',{\Bf r})\big)\Big\},
\end{eqnarray}
where the functions ${\cal A}({\Bf r},{\Bf r}')$ and ${\cal B}_{\sigma}
({\Bf r},{\Bf r}')$ are defined in Appendix B (see Eqs.~(\ref{eb28})
and (\ref{eb29}) respectively). We note in passing that ${\cal A}
({\Bf r}',{\Bf r}) \equiv {\cal A}({\Bf r},{\Bf r}')$, but 
${\cal B}_{\sigma}({\Bf r}',{\Bf r}) \not\equiv {\cal B}_{\sigma}
({\Bf r},{\Bf r}')$, and further that
\begin{eqnarray}
\label{e180}
&&E_{N;0} \equiv \langle\Psi_{N;0}\vert \wh{H}\vert
\Psi_{N;0}\rangle\nonumber\\
&&\;\;\;
= \sum_{\sigma}\int {\rm d}^dr\;
\{ \lim_{{\Bf r}'\to {\Bf r}} h_0({\Bf r}) \varrho_{\sigma}
({\Bf r}',{\Bf r}) -\frac{1}{2} {\cal B}_{\sigma}({\Bf r},{\Bf r}) \},
\end{eqnarray}
where the limit ${\Bf r}'\to {\Bf r}$ is necessary owing to
$\nabla_{\Bf r}^2$ in the definition for $h_0({\Bf r})$ (see
Eq.~(\ref{e43}) above). Making use of the fact that
\begin{eqnarray}
\label{e181}
&&-\frac{1}{2} \int {\rm d}^dr\; \sum_{\sigma} {\cal B}_{\sigma}
({\Bf r},{\Bf r})
\equiv \frac{N (N-1)}{2} \int {\rm d}^dr\;\nonumber\\
&&\;\;\;\;\;\;\;\;\;\;\;\;
\times\int {\rm d}^dr'\;
v({\Bf r}-{\Bf r}') \sum_{\sigma,\sigma'}
g_{\sigma,\sigma'}({\Bf r},{\Bf r}'),
\end{eqnarray}
with $g_{\sigma,\sigma'}({\Bf r},{\Bf r}')$ the van Hove pair-correlation
function (see Eqs.~(\ref{eb17}) and (\ref{ef88})), the expression in 
Eq.~(\ref{e180}) transforms into the standard expression for $E_{N;0}$ 
(for example March, {\sl et al.} (1967, p.~10)). In the 
thermodynamic limit we have $N (N-1)\, g_{\sigma,\sigma'}({\Bf r},{\Bf r}') 
= n_0^2\, {\sf g}_{\sigma,\sigma'}({\Bf r},{\Bf r}')$, where $n_0$ is 
the total concentration of the fermions (see Eq.~(\ref{e9}) above) and 
${\sf g}_{\sigma,\sigma'}({\Bf r},{\Bf r}')$ the {\sl normalized} van 
Hove pair-correlation function (see Eq.~(\ref{eb20})). We note in passing 
that following Galitskii and Migdal (1958) we have $E_{N;0}= \hbar^{-1} 
\int {\rm d}^dr\; \lim_{{\Bf r}'\to {\Bf r}} \int_{-\infty}^{\mu} 
{\rm d}\varepsilon\; [\varepsilon + h_0({\Bf r})] \sum_{\sigma} 
A_{\sigma}({\Bf r},{\Bf r}';\varepsilon)$ (see Eqs.~(\ref{e22}), 
(\ref{e39}) and (\ref{e43}) above).

The expression for $G_{0;\sigma;\infty_3}({\Bf r},{\Bf r}')$ is readily 
obtained from Eq.~(\ref{e179}) through equating the coupling constant of 
the particle-particle interaction in Eq.~(\ref{e179}) with zero and 
replacing $h_0$ herein by $h_{0;\sigma} \equiv h_0 + w_{\sigma}$ defined 
in Eq.~(\ref{e55}) above. Thus,
\begin{eqnarray}
\label{e182}
G_{0;\sigma;\infty_3}({\Bf r},{\Bf r}')
=\hbar\, h_{0;\sigma}({\Bf r}) h_{0;\sigma}({\Bf r}')\, 
\delta({\Bf r}-{\Bf r}').
\end{eqnarray}

\subsubsection{Evaluation of 
$\Sigma_{\sigma;\infty_1}({\Bf r},{\Bf r}')$ }
\label{s25}

From the expression in Eq.~(\ref{e83}), making use of the result in 
Eq.~(\ref{e162}), one readily obtains (see Eq.~(\ref{e77}) above)
\begin{eqnarray}
\label{e183}
&&\langle {\Bf r}\vert G_{\sigma;\infty_2}^2\vert {\Bf r}'\rangle
= \hbar^2 \Big\{
\Big[\big(h_0({\Bf r}) + v_H({\Bf r};[n])\big)\nonumber\\
&&\;\;\;\;\;\;\;\;\;\;\;\;\;\;\;\;\;\;\;\;\;\;\;\,
\times\big(h_0({\Bf r}') + v_H({\Bf r}';[n])\big)\Big] 
\delta({\Bf r}-{\Bf r}') \nonumber\\
&&
-\Big[ h_0({\Bf r}) + h_0({\Bf r}') 
+ v_H({\Bf r};[n]) + v_H({\Bf r}';[n]) \Big]\nonumber\\
&&\;\;\;\;\;\;\;\;\;\;\;\;\;\;\;\;\;\;\;\;\;\;\;\;\;\;\;\;
\;\;\;\;\;\;\;\;\;\;\;\;\;\;\;\;\;\,
\times v({\Bf r}-{\Bf r}') \varrho_{\sigma}({\Bf r}',{\Bf r})
\nonumber\\
&&
+\int {\rm d}^dr''\;
v({\Bf r}-{\Bf r}'') \varrho_{\sigma}({\Bf r}'',{\Bf r})
\varrho_{\sigma}({\Bf r}',{\Bf r}'') 
v({\Bf r}''-{\Bf r}') \Big\}.
\end{eqnarray}
From this one immediately deduces that
\begin{eqnarray}
\label{e184}
\langle {\Bf r}\vert G_{0;\sigma;\infty_2}^2\vert {\Bf r}'\rangle
= \hbar^2 h_{0;\sigma}({\Bf r}) h_{0;\sigma}({\Bf r}')\,
\delta({\Bf r}-{\Bf r}').
\end{eqnarray}
It follows that $\hbar^{-3} G_{0;\sigma;\infty_2}$ and $\hbar^{-2} 
G_{0;\sigma;\infty_3}$ in the expression for $\Sigma_{\sigma;\infty_1}$ 
in Eq.~(\ref{e74}) exactly cancel, leading to the expected result that 
$\Sigma_{\sigma;\infty_m}$ must {\sl not} depend on $w_{\sigma}$ for 
$m \ge 1$ (see the text following Eqs.~(\ref{e75}) and (\ref{e76})). 
From Eqs.~(\ref{e74}) (or Eq.~(\ref{e77})), (\ref{e179}), (\ref{e182}), 
(\ref{e183}) and (\ref{e184}) we eventually obtain
\begin{eqnarray}
\label{e185}
&&\Sigma_{\sigma;\infty_1}({\Bf r},{\Bf r}') = \frac{1}{\hbar}
\Big\{
\Big[{\cal A}({\Bf r},{\Bf r}) - v_H^2({\Bf r};[n]) 
\nonumber\\
&&\;\;\;
\;\;\;\;\;\;\;\;\;\;\;\;\;\;\;\;\;\;\;\;\;\;\;
+ \int {\rm d}^dr''\; v^2({\Bf r}-{\Bf r}'') n({\Bf r}'') \Big] 
\delta({\Bf r}-{\Bf r}')
\nonumber\\
&&\;\;\;
- v^2({\Bf r}-{\Bf r}') \varrho_{\sigma}({\Bf r}',{\Bf r})
\nonumber\\
&&\;\;\;
+ v({\Bf r}-{\Bf r}') \big[{\cal B}_{\sigma}({\Bf r},{\Bf r}')
+ {\cal B}_{\sigma}({\Bf r}',{\Bf r})\big]
\nonumber\\
&&\;\;\;
+ \big[v_H({\Bf r};[n]) + v_H({\Bf r}';[n]) \big]
v({\Bf r}-{\Bf r}') \varrho_{\sigma}({\Bf r}',{\Bf r})
\nonumber\\
&&\;\;\;
-\int {\rm d}^dr''\;
v({\Bf r}-{\Bf r}'') 
v({\Bf r}'-{\Bf r}'') 
\varrho_{\sigma}({\Bf r},{\Bf r}'')
\varrho_{\sigma}({\Bf r}'',{\Bf r}') \Big\}.\nonumber\\
\end{eqnarray}
It is interesting to note that similar to $\Sigma_{\sigma;\infty_0}
({\Bf r},{\Bf r}')$, $\Sigma_{\sigma;\infty_1}({\Bf r},{\Bf r}')$ 
consists of both {\sl local} and {\sl non-local} contributions and 
that, analogous to $\Sigma_{\sigma;\infty_0}$ (see Eq.~(\ref{e173})), 
{\sl only} the non-local contribution to $\Sigma_{\sigma;\infty_1}$ 
{\sl explicitly} depends on the spin index $\sigma$. This aspect 
directly exposes the significance of {\sl non}-local contributions to 
the SE operator in representing the {\sl direct} influence of the spin 
state of a single-particle excitation on its behaviour. This result in
addition unequivocally demonstrates that the SE operator necessarily 
{\sl explicitly} depends on $\sigma$, and further that any {\sl local} 
approximation to the $\varepsilon$-dependent part of $\Sigma_{\sigma}
({\Bf r},{\Bf r}';\varepsilon)$ neglects this explicit dependence at 
least to order $1/\varepsilon$ for large $\vert\varepsilon\vert$; in 
fact, our considerations in \S\S~III.H.1,2 establish that unless 
$v\equiv v_c$ (and in this case, unless $n_{\sigma}({\Bf r}) \not\equiv 
n_{\bar\sigma}({\Bf r})$), the local part of $\Sigma_{\sigma;\infty_2}$ 
is also explicitly {\sl independent} of $\sigma$ and the explicit 
dependence on $\sigma$ of $\Sigma_{\sigma;\infty_2}$ is associated 
with its non-local part.
\footnote{\label{f88}
Thus, leaving aside the case corresponding to $v\equiv v_c$, any 
approximation to $\Sigma_{\sigma}({\Bf r},{\Bf r}';\varepsilon)$, 
the local part of whose $\varepsilon$-dependent term to second 
order in $1/\varepsilon$ {\sl explicitly} depends on $\sigma$, 
is necessarily erroneous. } 
Our considerations in \S~III.H.2 reveal that some explicitly 
$\sigma$-dependent terms in $\Sigma_{\sigma;\infty_2}$, which ordinarily 
are {\sl non-local}, transmute into {\sl local} contributions as 
$v\rightharpoonup v_c$, $d\rightharpoonup 3$, so that for the 
Coulomb-interacting fermions in $d=3$, also the local contribution 
to $\Sigma_{\sigma;\infty_2}$ {\sl explicitly} depends on $\sigma$. 
Consequently, a {\sl local} approximation to the $\varepsilon$-dependent 
part of $\Sigma_{\sigma}({\Bf r},{\Bf r}';\varepsilon)$ that may prove 
reliable (for a certain range of $\varepsilon$) in applications 
corresponding to short-range and bounded $v$, should necessarily be
less satisfactory for $v\equiv v_c$ in $d=3$.

\subsubsection{The case of Coulomb-interacting fermions
in the thermodynamic limit }
\label{s26}

We now exclusively discuss the case of systems of fermions interacting 
through $v_c$ in $d=3$. One immediately observes that the expression 
in Eq.~(\ref{e185}) is {\sl not} directly amenable to explicit (numerical) 
calculation, for in this case several of the terms contain unbounded 
contributions for $\kappa \downarrow 0$. We shall now present a 
re-formulation of the expression in Eq.~(\ref{e185}) which is free 
from such contributions.

In Appendix~F we show that, although the individual terms on the RHS 
of Eq.~(\ref{e185}) enclosed by square brackets and post-multiplied 
by $\delta({\Bf r}-{\Bf r}')$ are unbounded for the case $v\equiv v_c$ 
and $d=3$, their {\sl total} contribution, namely (see Eq.~(\ref{ef2}))
\begin{eqnarray}
\label{e186}
{\cal A}'({\Bf r},{\Bf r}) &{:=}&
{\cal A}({\Bf r},{\Bf r}) - v_H^2({\Bf r};[n])
\nonumber\\
& &\;\;
+\int {\rm d}^dr''\; v^2({\Bf r}-{\Bf r}'') 
n({\Bf r}''),
\end{eqnarray}
is a bounded function for this case. Thus, for the {\sl local} part of 
$\Sigma_{\sigma;\infty_1}({\Bf r},{\Bf r}')$ we have the well-defined 
expression
\begin{equation}
\label{e187}
\Sigma_{\sigma;\infty_1}^{\rm l}({\Bf r},{\Bf r}')
= \frac{1}{\hbar} {\cal A}'({\Bf r},{\Bf r})\, 
\delta({\Bf r}-{\Bf r}').
\end{equation}
In a similar manner, making use of Eqs.~(\ref{ef95}) and 
(\ref{ef98}), for the {\sl non}-local part of $\Sigma_{\sigma;
\infty_1}({\Bf r},{\Bf r}')$ we obtain
\begin{eqnarray}
\label{e188}
&&\Sigma_{\sigma;\infty_1}^{\rm nl}({\Bf r},{\Bf r}')
= \frac{1}{\hbar}\Big\{
-v^2({\Bf r}-{\Bf r}') \varrho_{\sigma}({\Bf r}',{\Bf r})
\nonumber\\
&&\;\;\;\;\;
+ v({\Bf r}-{\Bf r}') \big[
{\cal B}''_{\sigma}({\Bf r},{\Bf r}') + 
{\cal B}''_{\sigma}({\Bf r}',{\Bf r})\big]\nonumber\\
&&\;\;\;\;\;
-\int {\rm d}^dr''\; v({\Bf r}-{\Bf r}'') 
v({\Bf r}'-{\Bf r}'') \nonumber\\
&&\;\;\;\;\;\;\;\;\;\;\;\;\;\;\;\;\;\;\;\;\;\;\;\;\;\;\;
\times \varrho_{\sigma}({\Bf r}',{\Bf r}'')
\varrho_{\sigma}({\Bf r}'',{\Bf r})\Big\},
\end{eqnarray}
where we have defined ${\cal B}''_{\sigma}$ in Eq.~(\ref{ef98}). 
We point out that our use of $d$ and $v$ here (as opposed to $d=3$ 
and $v\equiv v_c$) is meant to convey the fact that the results in 
Eqs.~(\ref{e186}) and (\ref{e188}) are general and {\sl not} specific 
to $d=3$ and $v\equiv v_c$. 

As can be immediately observed, the individual contributions on the 
RHS of Eq.~(\ref{e188}) are {\sl not} bounded for $v\equiv v_c$ in 
$d=3$ as $\|{\Bf r}-{\Bf r}'\|\to 0$, the most singular contribution 
being the first term, which for $\|{\Bf r}-{\Bf r}'\| \to 0$ diverges 
like $1/\|{\Bf r}-{\Bf r}'\|^2$ (in this limit, $\varrho_{\sigma}
({\Bf r}',{\Bf r})\to n_{\sigma}({\Bf r})$). This divergence is 
integrable in $d=3$, so that $\Sigma_{\sigma;\infty_1}^{\rm nl}
({\Bf r},{\Bf r}')$ can be Fourier transformed with respect to 
${\Bf r}$ and ${\Bf r}'$ (see conditions (B) and (C) in \S~II.B). As we 
shall see in \S~III.H.1, $\Sigma_{\sigma;\infty_2}({\Bf r},{\Bf r}')$ 
involves the contribution $-\hbar^{-1} v^3({\Bf r}-{\Bf r}') 
\varrho_{\sigma}({\Bf r}',{\Bf r})$ (see Eq.~(\ref{e212}) below) which 
for $v\equiv v_c$ in $d=3$ is {\sl not} integrable. We shall however 
show (see Appendix H) that a re-summation of the infinite series of 
functions $-v_c^m({\Bf r}-{\Bf r}') \varrho_{\sigma}({\Bf r}',
{\Bf r})/z^{m-1}$, $m \ge 3$, each of which is non-integrable (see 
condition (B) in \S~II.B) for $m \ge 3$, gives rise to an integrable 
contribution which is amenable to Fourier transformation (for some 
relevant details see \S~II.B).

\subsection{Evaluation of $G_{\sigma;\infty_4}({\Bf r},{\Bf r}')$ }
\label{s27}

From Eq.~(\ref{e34}), along the lines of \S~III.F we obtain
\begin{eqnarray}
\label{e189}
G_{\sigma;\infty_4}({\Bf r},{\Bf r}')
= \hbar \langle\Psi_{N;0}\vert
\big[ \wh{L} \wh{A}_{\sigma}({\Bf r}),\wh{A}_{\sigma}^{\dag}
({\Bf r}')\big]_+ \vert\Psi_{N;0}\rangle.
\end{eqnarray} 
It is interesting to compare $G_{\sigma;\infty_1}({\Bf r},{\Bf r}')$ 
and $G_{\sigma;\infty_2}({\Bf r},{\Bf r}')$ in Eqs.~(\ref{e30}) and 
(\ref{e161}), on the one hand, and $G_{\sigma;\infty_3}({\Bf r},
{\Bf r}')$ and $G_{\sigma;\infty_4}({\Bf r},{\Bf r}')$ in 
Eqs.~(\ref{e177}) and (\ref{e189}), on the other hand, from which 
one observes that, in the case of the latter two functions (to be 
precise, {\sl distributions}), the operators $\wh{A}_{\sigma}
({\Bf r})$ and $\wh{A}_{\sigma}^{\dag}({\Bf r}')$ have taken over 
the role played by $\hat\psi_{\sigma}({\Bf r})$ and 
$\hat\psi_{\sigma}^{\dag}({\Bf r}')$, respectively. This similarity 
obtains for {\sl all} pairs $G_{\sigma;\infty_{2m-1}}({\Bf r},
{\Bf r}')$ and $G_{\sigma;\infty_{2m}}({\Bf r},{\Bf r}')$, $m\ge 1$
where, at each level, new operators replace those at lower levels of 
the hierarchy (analogous to $\wh{A}_{\sigma}({\Bf r})$ replacing 
$\hat\psi_{\sigma}({\Bf r})$). It should, however, be noted that, 
contrary to $\hat\psi_{\sigma}({\Bf r})$ and $\hat\psi_{\sigma}^{\dag}
({\Bf r}')$, the subsequent operators are {\sl not} canonical, that 
is they do {\sl not} satisfy the set of anticommutation relations 
as given in Eq.~(\ref{e29}). 

By some straightforward algebra, the expression in Eq.~(\ref{e189}) 
can be rearranged into the following form: 
\begin{eqnarray}
\label{e190}
&&G_{\sigma;\infty_4}({\Bf r},{\Bf r}') 
= h_0({\Bf r}')\, G_{\sigma;\infty_3}({\Bf r},{\Bf r}') \nonumber\\
&&\;\;\;
+\hbar \sum_{\sigma'} \int {\rm d}^dr''\; 
v({\Bf r}'-{\Bf r}'') \nonumber\\
&&\;\;\;\;\;\;\;\;\;
\times\langle\Psi_{N;0}\vert 
\Big[\big[{\hat A}_{\sigma}({\Bf r}),\wh{H}\big]_{-},
{\hat\xi}_{\sigma,\sigma'}({\Bf r}',{\Bf r}'')\Big]_{+}
\vert\Psi_{N;0}\rangle, 
\end{eqnarray}
where
\begin{equation}
\label{e191}
{\hat\xi}_{\sigma,\sigma'}({\Bf r}',{\Bf r}'')
{:=} \hat\psi_{\sigma}^{\dag}({\Bf r}') 
\hat\psi_{\sigma'}^{\dag}({\Bf r}'')
\hat\psi_{\sigma'}({\Bf r}'').
\end{equation}
A detailed investigation reveals that it is technically advantageous 
to employ the following identity for the purpose of further simplifying
the result in Eq.~(\ref{e190}),
\begin{eqnarray}
\label{e192}
&&\langle\Psi_{N;0}\vert \Big[
\big[{\hat A}_{\sigma}, \wh{H}\big]_{-},
{\hat\xi}_{\sigma,\sigma'}\Big]_{+}
\vert\Psi_{N;0}\rangle 
\nonumber\\
&&\;\;\;\;\;\;\;\;\;\;\;\;
= \langle\Psi_{N;0}\vert \Big[
\big[ \wh{H}, 
{\hat\xi}_{\sigma,\sigma'} \big]_{-},
{\hat A}_{\sigma} \Big]_{+}
\vert\Psi_{N;0}\rangle.
\end{eqnarray}

After some algebra, we obtain
\begin{eqnarray}
\label{e193}
&&\big[\wh{H},
{\hat\xi}_{\sigma,\sigma'}({\Bf r}',{\Bf r}'')\big]_{-}
= h_0({\Bf r}')\, {\hat\xi}_{\sigma,\sigma'}({\Bf r}',{\Bf r}'')
\nonumber\\
&&\;\;\;
+\lim_{{\tilde {\Bf r}}''\to {\Bf r}''}
\tau({\Bf r}'') \big[
\hat\psi_{\sigma}^{\dag}({\Bf r}')
\hat\psi_{\sigma'}^{\dag}({\Bf r}'')
\hat\psi_{\sigma'}({\tilde {\Bf r}}'')\nonumber\\
&&\;\;\;\;\;\;\;\;\;
\;\;\;\;\;\;\;\;\;\;\;\;\;\;\;\;
-\hat\psi_{\sigma}^{\dag}({\Bf r}')
\hat\psi_{\sigma'}^{\dag}({\tilde {\Bf r}}'')
\hat\psi_{\sigma'}({\Bf r}'')\big]\nonumber\\
&&\;\;\;
+v({\Bf r}'-{\Bf r}'')\,
\hat\psi_{\sigma}^{\dag}({\Bf r}')
\hat\psi_{\sigma'}^{\dag}({\Bf r}'')
\hat\psi_{\sigma'}({\Bf r}'')\nonumber\\
&&\;\;\;
+\sum_{\sigma_1'}\int {\rm d}^dr_1''\;
v({\Bf r}'-{\Bf r}_1'')\nonumber\\
&&\;\;\;\;\;\;\;\;\times
\hat\psi_{\sigma}^{\dag}({\Bf r}')
\hat\psi_{\sigma_1'}^{\dag}({\Bf r}_1'')
\hat\psi_{\sigma'}^{\dag}({\Bf r}'')
\hat\psi_{\sigma'}({\Bf r}'')
\hat\psi_{\sigma_1'}({\Bf r}_1'').
\end{eqnarray}
In evaluating the anticommutation of $\big[\wh{H}, 
{\hat\xi}_{\sigma,\sigma'}\big]_{-}$ with ${\hat A}_{\sigma}$, we make 
use of the expression in Eq.~(\ref{e158}). After some lengthy but 
otherwise straightforward algebra, we arrive at
\begin{eqnarray}
\label{e194}
&&G_{\sigma;\infty_4}({\Bf r},{\Bf r}') = \hbar
\Big\{ \Big[ h_0({\Bf r}) h_0({\Bf r}) h_0({\Bf r})\nonumber\\
&&+\big\{ h_0({\Bf r}) h_0({\Bf r})
+ h_0({\Bf r}') h_0({\Bf r}') + h_0({\Bf r}) h_0({\Bf r}') \big\}\,
v_H({\Bf r};[n]) \nonumber\\ 
&&+\big\{ {\cal A}({\Bf r},{\Bf r}) h_0({\Bf r}')
+ {\cal A}({\Bf r}',{\Bf r}') h_0({\Bf r})
+ {\cal A}({\Bf r},{\Bf r}') h_0({\Bf r}) \big\}\nonumber\\
&& + \int {\rm d}^dr''\; \big\{ v({\Bf r}-{\Bf r}'')
v({\Bf r}-{\Bf r}'')\nonumber\\
&&\;\;\;\;\;\;\;\;\;\;\;\;\;\;\;
+ v({\Bf r}'-{\Bf r}'')
v({\Bf r}'-{\Bf r}'')\big\}\, n({\Bf r}'')\,
h_0({\Bf r})\nonumber\\
&&+\int {\rm d}^dr''\; v({\Bf r}'-{\Bf r}'')
v({\Bf r}-{\Bf r}'') n({\Bf r}'')\,
h_0({\Bf r}') \nonumber\\
&&+\int {\rm d}^dr''\; v({\Bf r}'-{\Bf r}'')
\lim_{{\tilde {\Bf r}}''\to {\Bf r}''} \tau({\Bf r}'')\,
v({\Bf r}-{\Bf r}'')\, 
\varrho({\Bf r}'',{\tilde {\Bf r}}'')\nonumber\\
&&-\int {\rm d}^dr''\; v({\Bf r}'-{\Bf r}'')
v({\Bf r}-{\Bf r}'')
\lim_{{\tilde {\Bf r}}''\to {\Bf r}''} \tau({\Bf r}'')\,
\varrho({\tilde {\Bf r}}'',{\Bf r}'')\nonumber\\
&&+ \int {\rm d}^dr''\; v({\Bf r}'-{\Bf r}'')
v({\Bf r}-{\Bf r}'') v({\Bf r}-{\Bf r}'') 
n({\Bf r}'')\nonumber\\
&&+3 \int {\rm d}^dr_1'' {\rm d}^dr_2''\;
v({\Bf r}'-{\Bf r}_1'') v({\Bf r}-{\Bf r}_1'')
v({\Bf r}-{\Bf r}_2'')\nonumber\\
&&\;\;\;\;\;\;\;\;\;\;\;\;\;\;\;
\times \sum_{\sigma_1',\sigma_2'}
\Gamma^{(2)}({\Bf r}_1''\sigma_1',{\Bf r}_2''\sigma_2';
{\Bf r}_1''\sigma_1',{\Bf r}_2''\sigma_2')\nonumber\\
&&+\int {\rm d}^dr_1'' {\rm d}^dr_2'' {\rm d}^dr_3''\;
v({\Bf r}'-{\Bf r}_1'')
v({\Bf r}'-{\Bf r}_2'')
v({\Bf r}-{\Bf r}_3'')\nonumber\\
&&\;\;\;\;\;
\times\sum_{\sigma_1',\sigma_2',\sigma_3'}
\Gamma^{(3)}({\Bf r}_1''\sigma_1',{\Bf r}_2''\sigma_2',
{\Bf r}_3''\sigma_3'';
{\Bf r}_1''\sigma_1',{\Bf r}_2''\sigma_2',
{\Bf r}_3''\sigma_3'')\Big]\nonumber\\
&&\;\;\;\;\;\;\;\;\;\;\;\;\;\;\;\;\;\;\;\;\;\;\;\;\;\;\;\;\;\;\;\;
\;\;\;\;\;\;\;\;\;\;\;\;\;\;\;\;\;\;\;\;\;\;\;\;\;\;\;\;\;\;\;\;\;
\times \delta({\Bf r}-{\Bf r}') \nonumber\\
&&-\Big[ h_0({\Bf r}) h_0({\Bf r}) 
+ h_0({\Bf r}') h_0({\Bf r}') 
+ h_0({\Bf r}') h_0({\Bf r}) \nonumber\\
&&\;\;\;\;\;\;
+\big\{ h_0({\Bf r}) + h_0({\Bf r}')\big\} v({\Bf r}-{\Bf r}')
\nonumber\\
&&\;\;\;\;\;\;
+v({\Bf r}-{\Bf r}') \big\{ h_0({\Bf r}) + h_0({\Bf r}')\big\}
\nonumber\\
&&\;\;\;\;\;\;
+v({\Bf r}-{\Bf r}') v({\Bf r}-{\Bf r}')
\Big] v({\Bf r}-{\Bf r}')
\varrho_{\sigma}({\Bf r}',{\Bf r})\nonumber\\
&&
+v({\Bf r}-{\Bf r}') h_0({\Bf r}) h_0({\Bf r}')
\varrho_{\sigma}({\Bf r}',{\Bf r})\nonumber\\ 
&&+\Big[ h_0({\Bf r}) + h_0({\Bf r}') +
2 v({\Bf r}-{\Bf r}') \Big]
v({\Bf r}-{\Bf r}') \nonumber\\
&&\;\;\;\;\;\;\;\;\;\;\;\;\;\;\;\;\;
\;\;\;\;\;\;\;\;\;\;\;\;\;\;\;\;\;\;
\times\big\{ {\cal B}_{\sigma}({\Bf r},{\Bf r}')
+ {\cal B}_{\sigma}({\Bf r}',{\Bf r})\big\}\nonumber\\
&&-v({\Bf r}-{\Bf r}') \big\{ 
\tau({\Bf r}) {\cal B}_{\sigma}({\Bf r},{\Bf r}')+\tau({\Bf r}') 
{\cal B}_{\sigma}({\Bf r}',{\Bf r})\big\} \nonumber\\
&&+v({\Bf r}-{\Bf r}') v({\Bf r}-{\Bf r}')
{\cal D}_{\sigma}({\Bf r},{\Bf r}')-v({\Bf r}-{\Bf r}') 
{\cal F}_{\sigma}({\Bf r},{\Bf r}')\nonumber\\
&&-\int {\rm d}^dr''\; v({\Bf r}'-{\Bf r}'')
\tau({\Bf r}') v({\Bf r}-{\Bf r}')\nonumber\\
&&\;\;\;\;\;\;\;\;\;\;\;\;\;\;\;\;\;\;\;\;\;\;\;\;\;\;\;\;\;
\times
\sum_{\sigma'} \Gamma^{(2)}({\Bf r}'\sigma,
{\Bf r}''\sigma';{\Bf r}\sigma,{\Bf r}''\sigma')\nonumber\\
&&-\int {\rm d}^dr''\; v({\Bf r}-{\Bf r}'')
\tau({\Bf r}) v({\Bf r}-{\Bf r}')\nonumber\\
&&\;\;\;\;\;\;\;\;\;\;\;\;\;\;\;\;\;\;\;\;\;\;\;\;\;\;\;\;\;
\times
\sum_{\sigma'} \Gamma^{(2)}({\Bf r}\sigma,
{\Bf r}''\sigma';{\Bf r}'\sigma,{\Bf r}''\sigma')\nonumber\\
&&+\int {\rm d}^dr''\; v({\Bf r}-{\Bf r}'')
v({\Bf r}'-{\Bf r}'') \lim_{{\tilde {\Bf r}}''\to {\Bf r}''}
\tau({\Bf r}'') \nonumber\\
&&\;\;\;\;\;\;\;\;\;\;\;\;\;\;\;\;\;\;\;\;\;\;\,
\times \big\{ \sum_{\sigma'}
\Gamma^{(2)}({\Bf r}'\sigma,{\tilde {\Bf r}}''\sigma';
{\Bf r}\sigma,{\Bf r}''\sigma')\nonumber\\
&&\;\;\;\;\;\;\;\;\;\;\;\;\;\;\;\;\;\;\;\;\;\;\;\;\;
+\sum_{\sigma'}
\Gamma^{(2)}({\Bf r}'\sigma,{\Bf r}''\sigma';
{\Bf r}\sigma,{\tilde {\Bf r}}''\sigma') \big\} \nonumber\\
&&-2 v({\Bf r}-{\Bf r}') \int {\rm d}^dr''\;
v({\Bf r}-{\Bf r}'') v({\Bf r}'-{\Bf r}'')\nonumber\\
&&\;\;\;\;\;\;\;\;\;\;\;\;\;\;\;\;\;\;\;\;\;\;\;\;\;\;\;\;\;
\times
\sum_{\sigma'} \Gamma^{(2)}({\Bf r}'\sigma,
{\Bf r}''\sigma';{\Bf r}\sigma,{\Bf r}''\sigma')\nonumber\\
&&-\int {\rm d}^dr''\; v({\Bf r}'-{\Bf r}'')
\lim_{{\tilde {\Bf r}}''\to {\Bf r}''} \tau({\Bf r}'')\,
v({\Bf r}-{\Bf r}'') \nonumber\\
&&\;\;\;\;\;\;\;\;\;\;\;\;\;\;\;\;\;\;\;\;\;\;\,
\times \big\{ \sum_{\sigma'}
\Gamma^{(2)}({\Bf r}'\sigma,{\Bf r}''\sigma';
{\Bf r}\sigma,{\tilde {\Bf r}}''\sigma')\nonumber\\
&&\;\;\;\;\;\;\;\;\;\;\;\;\;\;\;\;\;\;\;\;\;\;\;\;\;
+\sum_{\sigma'}
\Gamma^{(2)}({\Bf r}'\sigma,{\tilde {\Bf r}}''\sigma';
{\Bf r}\sigma,{\Bf r}''\sigma') \big\} \nonumber\\
&&+{\cal J}_{\sigma}({\Bf r},{\Bf r}') \nonumber\\
&&-2 v({\Bf r}-{\Bf r}') \int {\rm d}^dr_1'' {\rm d}^dr_2''\;
v({\Bf r}'-{\Bf r}_1'')
v({\Bf r}-{\Bf r}_2'')\nonumber\\
&&\;\;\;\;\;\;\;
\times \sum_{\sigma_1',\sigma_2'}
\Gamma^{(3)}({\Bf r}'\sigma,{\Bf r}_1''\sigma_1',
{\Bf r}_2''\sigma_2';
{\Bf r}\sigma,{\Bf r}_1''\sigma_1',{\Bf r}_2''\sigma_2')
\Big\}.
\end{eqnarray}
For the symmetric functions ${\cal D}_{\sigma}({\Bf r},{\Bf r}')$ and 
${\cal F}_{\sigma}({\Bf r},{\Bf r}')$ see Appendix~E. In Eq.~(\ref{e194}) 
we have arranged the terms in such a way that the expected symmetry of 
$G_{\sigma;\infty_4}({\Bf r},{\Bf r}')$ with respect to the exchange 
${\Bf r} \rightleftharpoons {\Bf r}'$ (see Eq.~(\ref{e178}) above; see 
also Appendix B) is made maximally explicit. In Appendix E we demonstrate 
the implicit symmetry of the contributions in Eq.~(\ref{e194}) that are 
{\sl not} manifestly symmetric with respect to ${\Bf r} \rightleftharpoons 
{\Bf r}'$. In Appendix D we prove that the explicitly asymmetric 
function ${\cal J}_{\sigma}({\Bf r},{\Bf r}')$ on the RHS of 
Eq.~(\ref{e194}) (for the defining expression see Eq.~(\ref{ed1})) is 
{\sl identically} vanishing due to the assumed time-reversal symmetry of 
the GS. We have retained ${\cal J}_{\sigma}({\Bf r},{\Bf r}')$ in 
Eq.~(\ref{e194}) in order to preserve the most complete expression for 
$G_{\sigma;\infty_4}({\Bf r},{\Bf r}')$ as deduced through carrying 
out a series of unbiased algebraic manipulations. Finally, from 
Eq.~(\ref{e194}) one readily obtains $G_{0;\sigma;\infty_4}({\Bf r},
{\Bf r}') = \hbar\, h_{0;\sigma}({\Bf r}) h_{0;\sigma}({\Bf r}) 
h_{0;\sigma}({\Bf r})\, \delta({\Bf r}-{\Bf r}')$ ({\it cf}.
Eqs.~(\ref{e171}) and (\ref{e182})).

\subsubsection{Evaluation of 
$\Sigma_{\sigma;\infty_2}({\Bf r},{\Bf r}')$ }
\label{s28}

Making use of the expression in Eq.~(\ref{e83}) and employing 
the results in Eqs.~(\ref{e162}) and (\ref{e179}), we 
straightforwardly obtain
\begin{eqnarray}
\label{e195}
&&\langle {\Bf r}\vert G_{\sigma;\infty_2} 
G_{\sigma;\infty_3}\vert {\Bf r}'\rangle
= \hbar^2 \Big\{
\big[ h_0({\Bf r}) + v_H({\Bf r};[n]) \big]
\nonumber\\
&&\times \big[ h_0({\Bf r}) h_0({\Bf r}')
+ h_0({\Bf r}) v_H({\Bf r}';[n])
+ h_0({\Bf r}') v_H({\Bf r};[n])\nonumber\\
&&\;\;\;\;
+ {\cal A}({\Bf r},{\Bf r}) + \int {\rm d}^dr''\; 
v^2({\Bf r}-{\Bf r}'') n({\Bf r}'') \big]\,
\delta({\Bf r}-{\Bf r}') \nonumber\\
&&+\big[ h_0({\Bf r}) + v_H({\Bf r};[n]) \big]\nonumber\\
&&\;\;\;
\times\big[ - \big\{ h_0({\Bf r}) + h_0({\Bf r}')\big\}
v({\Bf r}-{\Bf r}') \varrho_{\sigma}({\Bf r}',{\Bf r}) \nonumber\\
&&\;\;\;\;\;\;\;
-v^2({\Bf r}-{\Bf r}')\, 
\varrho_{\sigma}({\Bf r}',{\Bf r})\nonumber\\
&&\;\;\;\;\;\;\;
+v({\Bf r}-{\Bf r}') \big\{
{\cal B}_{\sigma}({\Bf r},{\Bf r}') +
{\cal B}_{\sigma}({\Bf r}',{\Bf r})\big\}\big]\nonumber\\ 
&&-\big[ h_0({\Bf r}') h_0({\Bf r}') 
+v_H({\Bf r}';[n]) h_0({\Bf r}') \nonumber\\
&&\;\;\;\;\;
+h_0({\Bf r}') v_H({\Bf r}';[n]) 
+ {\cal A}({\Bf r}',{\Bf r}')\nonumber\\
&&\;\;\;\;\;
+ \int {\rm d}^dr''\;
v({\Bf r}'-{\Bf r}'') n({\Bf r}'') v({\Bf r}''-{\Bf r}) \big] 
\nonumber\\
&&\;\;\;\;\;\;\;\;\;\;\;\;\;\;
\;\;\;\;\;\;\;\;\;\;\;\;\;\;\;\;\;\;\;\;\;\;\;\;\;\;\;\;
\times v({\Bf r}-{\Bf r}') 
\varrho_{\sigma}({\Bf r}',{\Bf r}) \nonumber\\
&&+\int {\rm d}^dr''\;
v({\Bf r}-{\Bf r}'') \varrho_{\sigma}({\Bf r}'',{\Bf r})
\nonumber\\
&&\;\;\;
\times \big[\big\{ h_0({\Bf r}'') + h_0({\Bf r}')\big\}
v({\Bf r}''-{\Bf r}') \varrho_{\sigma}({\Bf r}',{\Bf r}'')
\nonumber\\
&&\;\;\;\;\;
+v({\Bf r}''-{\Bf r}') \varrho_{\sigma}({\Bf r}',{\Bf r}'')
v({\Bf r}'-{\Bf r}'')\nonumber\\
&&\;\;\;\;\;
-v({\Bf r}''-{\Bf r}') \big\{
{\cal B}_{\sigma}({\Bf r}'',{\Bf r}')
+ {\cal B}_{\sigma}({\Bf r}',{\Bf r}'')\big\}\big] \Big\},
\end{eqnarray} 
from which we readily deduce that (see text preceding Eq.~(\ref{e171})
above)
\begin{eqnarray}
\label{e196}
&&\langle {\Bf r}\vert G_{0;\sigma;\infty_2} 
G_{0;\sigma;\infty_3}\vert {\Bf r}'\rangle
= \hbar^2 h_{0;\sigma}({\Bf r}) h_{0;\sigma}({\Bf r}) 
h_{0;\sigma}({\Bf r})\, \delta ({\Bf r}-{\Bf r}').\nonumber\\
\end{eqnarray}
Further, in a similar manner to the above, we obtain
\begin{eqnarray}
\label{e197}
&&\langle {\Bf r}\vert G_{\sigma;\infty_3} 
G_{\sigma;\infty_2}\vert {\Bf r}'\rangle
= \hbar^2 \Big\{ \big[ h_0({\Bf r}) h_0({\Bf r}) \nonumber\\
&&\;\;\;
+h_0({\Bf r}) v_H({\Bf r};[n]) 
+v_H({\Bf r};[n]) h_0({\Bf r}) \nonumber\\
&&\;\;\;
+ {\cal A}({\Bf r},{\Bf r})
+\int {\rm d}^dr''\; v^2({\Bf r}-{\Bf r}'') n({\Bf r}'')
\big]\nonumber\\
&&\;
\times \Big( \big[ h_0({\Bf r}) + v_H({\Bf r};[n])\big]\,
\delta({\Bf r}-{\Bf r}')
-v({\Bf r}-{\Bf r}') 
\varrho_{\sigma}({\Bf r}',{\Bf r})\Big) \nonumber\\
&&-\big[ h_0({\Bf r}') + v_H({\Bf r}';[n])\big]
\big[ h_0({\Bf r}) + h_0({\Bf r}')\big]
v({\Bf r}-{\Bf r}') 
\varrho_{\sigma}({\Bf r}',{\Bf r}) \nonumber\\ 
&&+\big[ h_0({\Bf r}') + v_H({\Bf r}';[n])\big]
v({\Bf r}-{\Bf r}') \nonumber\\
&&\;\;\;\;\;\;\;\;\;\;\;
\times \big[ {\cal B}_{\sigma}({\Bf r},{\Bf r}') + 
{\cal B}_{\sigma}({\Bf r}',{\Bf r})
-v({\Bf r}-{\Bf r}')\varrho_{\sigma}({\Bf r}',{\Bf r}) \big]
\nonumber\\
&&+\int {\rm d}^dr''\; 
\Big( \big[ h_0({\Bf r}) + h_0({\Bf r}'') 
+ v({\Bf r}-{\Bf r}'') \big]
\varrho_{\sigma}({\Bf r}'',{\Bf r}) \nonumber\\
&&\;\;\;\;\;\;\;\;\;\;\;\;\;\;\;\;\;\;\;\;\;\;\;\;\;\;\;\;\;
- {\cal B}_{\sigma}({\Bf r},{\Bf r}'') 
- {\cal B}_{\sigma}({\Bf r}'',{\Bf r}) \Big)
v({\Bf r}-{\Bf r}'') \nonumber\\
&&\;\;\;\;\;\;\;\;\;\;\;\;\;\;\;\;\;\;\;\;\;\;\;\;\;\;\;
\;\;\;\;\;\;\;\;\;
\times v({\Bf r}'-{\Bf r}'') 
\varrho_{\sigma}({\Bf r}',{\Bf r}'') \Big\},
\end{eqnarray}
from which we infer that
\begin{eqnarray}
\label{e198}
&&\langle {\Bf r}\vert G_{0;\sigma;\infty_3} 
G_{0;\sigma;\infty_2}\vert {\Bf r}'\rangle
= \hbar^2 h_{0;\sigma}({\Bf r}) h_{0;\sigma}({\Bf r}) 
h_{0;\sigma}({\Bf r})\, \delta ({\Bf r}-{\Bf r}').\nonumber\\
\end{eqnarray}

Combining the above results in accordance with the expression in 
Eq.~(\ref{e75}) (or Eq.~(\ref{e78})), we finally arrive at (below 
$\varrho \equiv \sum_{\sigma'} \varrho_{\sigma'}$ and $n \equiv 
\sum_{\sigma'} n_{\sigma'}$)
\begin{eqnarray}
\label{e199}
&&\Sigma_{\sigma;\infty_2}({\Bf r},{\Bf r}') = \frac{1}{\hbar}
\Big\{ \Big[ 
v_H^3({\Bf r};[n]) 
\nonumber\\
&&- v_H({\Bf r};[n]) v_H({\Bf r}';[n]) h_0({\Bf r})
-2 v_H({\Bf r};[n]) {\cal A}({\Bf r},{\Bf r}) \nonumber\\
&&-2 v_H({\Bf r};[n]) \int {\rm d}^dr''\;
v^2({\Bf r}-{\Bf r}'') n({\Bf r}'') \nonumber\\
&&+ {\cal A}({\Bf r},{\Bf r}') h_0({\Bf r})\nonumber\\
&&+\int {\rm d}^dr''\; 
v({\Bf r}-{\Bf r}'') v({\Bf r}'-{\Bf r}'') n({\Bf r}'')
h_0({\Bf r}') \nonumber\\
&&+\int {\rm d}^dr''\; v({\Bf r}-{\Bf r}'')
\lim_{{\tilde {\Bf r}}''\to {\Bf r}''}
\tau({\Bf r}'') v({\Bf r}-{\Bf r}'') \varrho({\Bf r}'',
{\tilde {\Bf r}}'')\nonumber\\
&&-\int {\rm d}^dr''\;
v^2({\Bf r}-{\Bf r}'') \lim_{{\tilde {\Bf r}}''\to {\Bf r}''}
\tau({\Bf r}'') \varrho({\tilde {\Bf r}}'',{\Bf r}'')\nonumber\\
&&+\int {\rm d}^dr''\; v^3({\Bf r}-{\Bf r}'') n({\Bf r}'')
\nonumber\\
&&+3 \int {\rm d}^dr_1'' {\rm d}^dr_2''\;
v^2({\Bf r}-{\Bf r}_1'') v({\Bf r}-{\Bf r}_2'') \nonumber\\
&&\;\;\;\;\;\;\;\;\;\;\;\;\;\;\;\;\;\;\;\;\;\;\;
\times \sum_{\sigma_1',\sigma_2'} 
\Gamma^{(2)}({\Bf r}_1''\sigma_1',{\Bf r}_2''\sigma_2';
{\Bf r}_1''\sigma_1',{\Bf r}_2''\sigma_2')\nonumber\\
&&+ \int {\rm d}^dr_1'' {\rm d}^dr_2'' {\rm d}^dr_3''\;
v({\Bf r}-{\Bf r}_1'') v({\Bf r}-{\Bf r}_2'')
v({\Bf r}-{\Bf r}_3'') \nonumber\\
&&\;
\times \sum_{\sigma_1',\sigma_2',\sigma_3'}
\Gamma^{(3)}({\Bf r}_1''\sigma_1',
{\Bf r}_2''\sigma_2',{\Bf r}_3''\sigma_3';
{\Bf r}_1''\sigma_1', {\Bf r}_2''\sigma_2',{\Bf r}_3''\sigma_3')
\Big]\nonumber\\
&&\;\;\;\;\;\;\;\;\;\;\;\;\;\;\;\;\;\;\;\;\;\;\;\;\;\;\;\;\;\;\;\;
\;\;\;\;\;\;\;\;\;\;\;\;\;\;\;\;\;\;\;\;\;\;\;\;\;\;\;\;\;\;\;
\times\delta({\Bf r}-{\Bf r}')\nonumber\\
&&-\Big[
v_H^2({\Bf r};[n])+v_H^2({\Bf r}';[n]) 
+v_H({\Bf r};[n]) v_H({\Bf r}';[n]) \nonumber\\
&&\;\;\;
-v_H({\Bf r};[n]) v({\Bf r}-{\Bf r}')
-v_H({\Bf r}';[n]) v({\Bf r}-{\Bf r}')\nonumber\\
&&\;\;\;
-v_H({\Bf r};[n]) h_0({\Bf r})
-v_H({\Bf r}';[n]) h_0({\Bf r}')\nonumber\\ 
&&\;\;\;
-{\cal A}({\Bf r},{\Bf r})
-{\cal A}({\Bf r}',{\Bf r}')\nonumber\\
&&\;\;\;
-\int {\rm d}^dr''\; \big\{ v^2({\Bf r}-{\Bf r}'') 
+ v^2({\Bf r}'-{\Bf r}'') \big\}\,
n({\Bf r}'') \nonumber\\
&&\;\;\;
+ v({\Bf r}-{\Bf r}') 
\big\{ h_0({\Bf r}) + h_0({\Bf r}')\big\}
+ v^2({\Bf r}-{\Bf r}') \Big]
\nonumber\\
&&\;\;\;\;\;\;\;\;\;\;\;\;\;\;\;\;\;\;\;\;\;\;\;\;\;\;\;
\;\;\;\;\;\;\;\;\;\;\;\;\;\;\;
\times v({\Bf r}-{\Bf r}') 
\varrho_{\sigma}({\Bf r}',{\Bf r})\nonumber\\
&&+\frac{1}{2} v^2({\Bf r}-{\Bf r}')
\big\{ h_0({\Bf r}) + h_0({\Bf r}')\big\}
\varrho_{\sigma}({\Bf r}',{\Bf r})\nonumber\\
&&+\big\{ v_H({\Bf r};[n]) + v_H({\Bf r}';[n])\big\}\nonumber\\
&&\;\;\;
\times \big\{ \int {\rm d}^dr''\; v({\Bf r}-{\Bf r}'')
v({\Bf r}'-{\Bf r}'') \varrho_{\sigma}({\Bf r},{\Bf r}'')
\varrho_{\sigma}({\Bf r}'',{\Bf r}')\nonumber\\
&&\;\;\;\;\;\;
- v({\Bf r}-{\Bf r}') 
\big[ {\cal B}_{\sigma}({\Bf r},{\Bf r}')
+ {\cal B}_{\sigma}({\Bf r}',{\Bf r})\big] \big\}\nonumber\\
&&+\int {\rm d}^dr''\; v({\Bf r}-{\Bf r}'')
v({\Bf r}'-{\Bf r}'') v_H({\Bf r}'';[n])\nonumber\\
&&\;\;\;\;\;\;\;\;\;\;\;\;\;\;\;\;\;\;\;\;\;\;\;\;\;\;\;
\;\;\;\;\;\;\;\;\;\;\;\;\;\;\;
\times\varrho_{\sigma}({\Bf r},{\Bf r}'')
\varrho_{\sigma}({\Bf r}'',{\Bf r}')\nonumber\\
&&-\int {\rm d}^dr_1'' {\rm d}^dr_2'' \;
v({\Bf r}-{\Bf r}_1'') v({\Bf r}'-{\Bf r}_2'')
v({\Bf r}_1''-{\Bf r}_2'')\nonumber\\
&&\;\;\;\;\;\;\;\;\;\;\;\;\;\;\;\;\;\;\;\;\;\;\;\;\;\;\;
\times\varrho_{\sigma}({\Bf r},{\Bf r}_1'')
\varrho_{\sigma}({\Bf r}_1'',{\Bf r}_2'')
\varrho_{\sigma}({\Bf r}_2'',{\Bf r}')
\nonumber\\
&&-\int {\rm d}^dr''\;
v({\Bf r}'-{\Bf r}'') \varrho_{\sigma}({\Bf r}'',{\Bf r}')
h_0({\Bf r}'') 
\nonumber\\
&&\;\;\;\;\;\;\;\;\;\;\;\;\;\;\;\;\;\;\;\;\;\;\;\;\;\;\;
\;\;\;\;\;\;\;\;\;\;\;\;\;\;\;
\times v({\Bf r}''-{\Bf r}) 
\varrho_{\sigma}({\Bf r},{\Bf r}'')\nonumber\\
&&-\int {\rm d}^dr''\;
v({\Bf r}-{\Bf r}'') v^2({\Bf r}'-{\Bf r}'')
\varrho_{\sigma}({\Bf r},{\Bf r}'')\,
\varrho_{\sigma}({\Bf r}'',{\Bf r}') \nonumber\\
&&-\int {\rm d}^dr''\;
v^2({\Bf r}-{\Bf r}'') v({\Bf r}'-{\Bf r}'')
\varrho_{\sigma}({\Bf r}',{\Bf r}'')\, 
\varrho_{\sigma}({\Bf r}'',{\Bf r}) \nonumber\\
&&+\int {\rm d}^dr''\;
v({\Bf r}-{\Bf r}'') v({\Bf r}'-{\Bf r}'')
\varrho_{\sigma}({\Bf r}'',{\Bf r})\nonumber\\
&&\;\;\;\;\;\;\;\;\;\;\;\;\;\;\;\;\;\;\;\;\;\;\;\;\;\;\;\;\;
\times \big\{ {\cal B}_{\sigma}({\Bf r}'',{\Bf r}')
+ {\cal B}_{\sigma}({\Bf r}',{\Bf r}'')\big\}\nonumber\\
&&+\int {\rm d}^dr''\;
v({\Bf r}'-{\Bf r}'') v({\Bf r}-{\Bf r}'')
\varrho_{\sigma}({\Bf r}'',{\Bf r}')\nonumber\\
&&\;\;\;\;\;\;\;\;\;\;\;\;\;\;\;\;\;\;\;\;\;\;\;\;\;\;\;\;\;
\times \big\{ {\cal B}_{\sigma}({\Bf r}'',{\Bf r})
+ {\cal B}_{\sigma}({\Bf r},{\Bf r}'')\big\}\nonumber\\
&&+\frac{3}{2} v^2({\Bf r}-{\Bf r}')
\big\{ {\cal B}_{\sigma}({\Bf r},{\Bf r}')
+ {\cal B}_{\sigma}({\Bf r}',{\Bf r})\big\} \nonumber\\
&&-\frac{1}{2} v({\Bf r}-{\Bf r}')
\big\{ {\cal G}_{\sigma}({\Bf r},{\Bf r}')
+ {\cal G}_{\sigma}({\Bf r}',{\Bf r})\big\}\nonumber\\
&&+\frac{1}{2} v({\Bf r}-{\Bf r}')
\big\{ h_0({\Bf r}) + h_0({\Bf r}') \big\}
\big\{ {\cal B}_{\sigma}({\Bf r},{\Bf r}')
+ {\cal B}_{\sigma}({\Bf r}',{\Bf r})\big\}\nonumber\\
&&- v({\Bf r}-{\Bf r}') \big\{
\tau({\Bf r}) {\cal B}_{\sigma}({\Bf r},{\Bf r}')
+ \tau({\Bf r}') 
{\cal B}_{\sigma}({\Bf r}',{\Bf r})\big\}\nonumber\\
&&-\int {\rm d}^dr''\;
v({\Bf r}'-{\Bf r}'') \tau({\Bf r}') v({\Bf r}-{\Bf r}')
\nonumber\\
&&\;\;\;\;\;\;\;\;\;\;\;\;\;\;\;\;\;\;\;\;\;\;\;\;\;\;\;
\times \sum_{\sigma'} \Gamma^{(2)}({\Bf r}'\sigma,{\Bf r}''\sigma';
{\Bf r}\sigma,{\Bf r}''\sigma')\nonumber\\
&&-\int {\rm d}^dr''\;
v({\Bf r}-{\Bf r}'') \tau({\Bf r}) v({\Bf r}-{\Bf r}')
\nonumber\\
&&\;\;\;\;\;\;\;\;\;\;\;\;\;\;\;\;\;\;\;\;\;\;\;\;\;\;\;
\times \sum_{\sigma'} \Gamma^{(2)}({\Bf r}\sigma,{\Bf r}''\sigma';
{\Bf r}'\sigma,{\Bf r}''\sigma')\nonumber\\
&&-2 v({\Bf r}-{\Bf r}')
\int {\rm d}^dr''\;
v({\Bf r}-{\Bf r}'') v({\Bf r}'-{\Bf r}'')\nonumber\\
&&\;\;\;
\;\;\;\;\;\;\;\;\;\;\;\;\;\;\;\;\;\;\;\;\;\;\;\;\;\;
\times\sum_{\sigma'} \Gamma^{(2)}({\Bf r}'\sigma,{\Bf r}''\sigma';
{\Bf r}\sigma,{\Bf r}''\sigma')\nonumber\\
&&+\int {\rm d}^dr''\; v({\Bf r}-{\Bf r}'')
v({\Bf r}'-{\Bf r}'') \lim_{{\tilde {\Bf r}}''\to {\Bf r}''}
\tau({\Bf r}'')\nonumber\\
&&\;\;\;
\;\;\;\;\;\;\;\;\;\;\;\;\;\;\;\;\;\;\;\;\;\;\;\;\;\;
\times \big\{
\sum_{\sigma'}\Gamma^{(2)}({\Bf r}'\sigma,{\tilde {\Bf r}}''\sigma';
{\Bf r}\sigma,{\Bf r}''\sigma')\nonumber\\
&&\;\;\;\;\;\;
\;\;\;\;\;\;\;\;\;\;\;\;\;\;\;\;\;\;\;\;\;\;\;\;\;\;
+\sum_{\sigma'}\Gamma^{(2)}({\Bf r}'\sigma,{\Bf r}''\sigma';
{\Bf r}\sigma,{\tilde {\Bf r}}''\sigma')\big\} \nonumber\\
&&-\int {\rm d}^dr''\; v({\Bf r}'-{\Bf r}'')
\lim_{{\tilde {\Bf r}}''\to {\Bf r}''} \tau({\Bf r}'')
v({\Bf r}-{\Bf r}'') \nonumber\\
&&\;\;\;
\;\;\;\;\;\;\;\;\;\;\;\;\;\;\;\;\;\;\;\;\;\;\;\;\;\;
\times \big\{
\sum_{\sigma'}\Gamma^{(2)}({\Bf r}'\sigma,{\tilde {\Bf r}}''\sigma';
{\Bf r}\sigma,{\Bf r}''\sigma')\nonumber\\
&&\;\;\;\;\;\;
\;\;\;\;\;\;\;\;\;\;\;\;\;\;\;\;\;\;\;\;\;\;\;\;\;\;
+\sum_{\sigma'}\Gamma^{(2)}({\Bf r}'\sigma,{\Bf r}''\sigma';
{\Bf r}\sigma,{\tilde {\Bf r}}''\sigma')\big\} \nonumber\\
&&+ {\cal J}_{\sigma}({\Bf r},{\Bf r}')\nonumber\\
&&-2 v({\Bf r}-{\Bf r}') \int {\rm d}^dr_1'' {\rm d}^dr_2''\;
v({\Bf r}-{\Bf r}_1'') v({\Bf r}'-{\Bf r}_2'')\nonumber\\
&&\;\;\;\;\;\;
\times \sum_{\sigma_1',\sigma_2'}
\Gamma^{(3)}({\Bf r}'\sigma,{\Bf r}_1''\sigma_1',
{\Bf r}_2''\sigma_2';
{\Bf r}\sigma,{\Bf r}_1''\sigma_1',{\Bf r}_2''\sigma_2') \Big\}.
\end{eqnarray}
For the {\sl asymmetric} function ${\cal G}_{\sigma}({\Bf r},
{\Bf r}')$ see Eq.~(\ref{ee18}). It is readily verified that the above 
expression appropriately satisfies the required symmetry property 
(see Appendix B) $\Sigma_{\sigma;\infty_2}({\Bf r},{\Bf r}') \equiv 
\Sigma_{\sigma;\infty_2}({\Bf r}',{\Bf r})$. As in the case of 
$G_{\sigma;\infty_4}({\Bf r},{\Bf r}')$ in Eq.~(\ref{e194}), we have 
retained the {\sl identically} vanishing function ${\cal J}_{\sigma}
({\Bf r},{\Bf r}')$ (see Appendix D; see also text following 
Eq.~(\ref{e194})) in the above expression for $\Sigma_{\sigma;
\infty_2}({\Bf r},{\Bf r}')$ so as to preserve the most complete 
expression for this function as obtained through performing a 
sequence of unbiased calculations.

\subsubsection{The case of Coulomb-interacting fermions in the 
thermodynamic limit}
\label{s29}

Here we consider the case in which $v\equiv v_c$ and $d=3$. Similar
to $\Sigma_{\sigma;\infty_0}({\Bf r},{\Bf r}')$ and $\Sigma_{\sigma;
\infty_1}({\Bf r},{\Bf r}')$, whose expressions as presented in 
Eqs.~(\ref{e173}) and (\ref{e185}) respectively involve integrals 
that owing to the long range of $v_c$ are unbounded for $v\equiv v_c$, 
$\Sigma_{\sigma;\infty_2}({\Bf r},{\Bf r}')$ as expressed in 
Eq.~(\ref{e199}) is determined in terms of integrals that owing to the 
latter property of $v_c$ are unbounded when $v$ herein is identified 
with $v_c$ (and $d$ with $3$). Consequently, the expression in 
Eq.~(\ref{e199}) must be reformulated prior to effecting the substitution
$v\rightharpoonup v_c$ (see \S~II.B). A second aspect associated with 
$v_c({\Bf r}-{\Bf r}')$ that gains prominence in the expression for 
$\Sigma_{\sigma;\infty_2}({\Bf r},{\Bf r}')$ and remains prominent in 
the expressions for $\Sigma_{\sigma;\infty_m}({\Bf r},{\Bf r}')$, for 
{\sl all} $m > 2$, is its divergence like $1/\|{\Bf r}-{\Bf r}'\|$ for
$\|{\Bf r}-{\Bf r}'\|\to 0$ (see however footnote \ref{f27}). 
\footnote{\label{f89}
We point out that $\Sigma_{\sigma;\infty_2}({\Bf r},{\Bf r}')$
stands on the `boundary' where the behaviours of $v_c({\Bf r}
-{\Bf r}')$ at both small and large $\|{\Bf r}-{\Bf r}'\|$ play 
equally significant roles; concerning $\Sigma_{\sigma;\infty_m}
({\Bf r},{\Bf r}')$, the behaviour of $v_c({\Bf r}-{\Bf r}')$ 
at large $\|{\Bf r}-{\Bf r}'\|$ is dominant for $m=0,1$, while 
this dominance is shared by the behaviour of $v_c({\Bf r}-{\Bf r}')$ 
at small $\|{\Bf r}-{\Bf r}'\|$ for $m \ge 2$. }
 
A systematic analysis of the expression on the RHS of Eq.~(\ref{e199}) 
reveals that (for some details see Appendices F and G) $\Sigma_{\sigma;
\infty_2}({\Bf r},{\Bf r}')$ is fundamentally {\sl unbounded}, that 
is in contrast with $\Sigma_{\sigma;\infty_0}({\Bf r},{\Bf r}')$ and 
$\Sigma_{\sigma;\infty_1}({\Bf r},{\Bf r}')$ the unbounded contributions 
to the constituent terms of $\Sigma_{\sigma;\infty_2}({\Bf r},{\Bf r}')$ 
do {\sl not} fully cancel. As the SE $\wt{\Sigma}_{\sigma}({\Bf r},
{\Bf r}';z)$ is itself a bounded function almost everywhere (see 
footnote \ref{f54}), it follows that these contributions to 
$\Sigma_{\sigma;\infty_2}({\Bf r},{\Bf r}')$ {\sl must} have their 
compensating counterparts in the expressions for $\Sigma_{\sigma;
\infty_m}({\Bf r},{\Bf r}')$ corresponding to $m > 2$ ({\it cf}. 
\S~II.B). A thorough inspection of the expression on the RHS of 
Eq.~(\ref{e199}) further reveals that the {\sl explicitly}-presented 
functions $h_0({\Bf r})$ and $v_H({\Bf r};[n])$ in this expression 
{\sl cannot} be brought into such combinations as $h_0({\Bf r}) + 
v_H({\Bf r};[n])$, $h_0({\Bf r})-h_0({\Bf r}')$ or $v_H({\Bf r};[n]) 
-v_H({\Bf r}';[n])$ where the divergent contribution $\varpi_{\kappa}$, 
corresponding to $\kappa\downarrow 0$, is fully cancelled.
\footnote{\label{f90}
Here, as in other places in the present work, $v_H({\Bf r};[n])$ is 
defined according to Eq.~(\protect\ref{e14}) with $v'$ as defined in 
Eq.~(\protect\ref{e13}); neglect in Eq.~(\protect\ref{e13}) of the 
cut-off function $\exp(-\kappa \|{\Bf r}-{\Bf r}'\|)$, with 
$\kappa\downarrow 0$, can lead to {\sl ambiguity} in the expressions, 
even at the formal level where the possible boundedness of certain 
integrals is {\sl not} at issue. }
This observation implies that some unbounded contributions involving 
$\varpi_{\kappa}$ must necessarily be {\sl implicit} in the expression 
on the RHS of Eq.~(\ref{e199}), which detailed considerations show 
indeed to be the case. These considerations have led us to the complete 
identification of the set of unbounded and non-integrable functions 
in the expression on the RHS of Eq.~(\ref{e199}) (see criteria (A) - (C) 
in \S~II.B) together with their compensating counterparts (see \S~II.B) 
in the formal expressions for $\Sigma_{\sigma;\infty_m}({\Bf r},{\Bf r}')$ 
with $m > 2$.

Our analyses demonstrate that $\Sigma_{\sigma;\infty_2}({\Bf r},
{\Bf r}')$ contains {\sl four} fundamentally {\sl unbounded} 
contributions, of which three originate from the {\sl local} part 
of $\Sigma_{\sigma;\infty_2}({\Bf r},{\Bf r}')$ (that is that part 
which in Eq.~(\ref{e199}) is post-multiplied by 
$\delta({\Bf r}-{\Bf r}')$), and the fourth originates from what 
in the case of $v\not\equiv v_c$ is a (bounded) {\sl non}-local 
contribution to $\Sigma_{\sigma;\infty_2}({\Bf r},{\Bf r}')$; 
as $v\rightharpoonup v_c$, this is shown to transform into a 
{\sl local} unbounded contribution
\footnote{\label{f91}
By `local unbounded' contribution we refer to a two-point function 
of the form $f({\Bf r},{\Bf r}')\, \delta({\Bf r}-{\Bf r}')$, with
$f({\Bf r},{\Bf r}')$ an unbounded function for $\|{\Bf r}-{\Bf r}'
\|\to 0$. }
(see text following Eq.~(\ref{e185}) above). It follows that all unbounded 
contributions to $\Sigma_{\sigma;\infty_2}({\Bf r},{\Bf r}')$, arising 
from the identification of $v$ with $v_c$ and $d$ with $3$, are local. 
It is interesting to mention that the latter fourth contribution (i.e.
${\cal I}_4$ in Eq.~(\ref{e203}) below) turns out to belong to the same 
category as one of the former three local unbounded contributions (i.e. 
${\cal I}_3$ in Eq.~(\ref{e202}) below) and its effect is one of 
rendering $\Sigma_{\sigma;\infty_2}({\Bf r},{\Bf r}')$ {\sl explicitly} 
dependent upon $\bar\sigma$, the totality of the spin components 
complementary to $\sigma$ (see the definition in the text following 
Eq.~(\ref{e19}) above; see also Eq.~(\ref{e209}) below). More importantly, 
the regularized expression (see Appendix G) corresponding to these two 
contributions, that is ${\cal I}_3$ and ${\cal I}_4$, which is shown to 
be proportional to $\sum_{\sigma'\not=\sigma} n_{\sigma'}({\Bf r})
-n_{\sigma}({\Bf r}) \equiv n_{\bar\sigma}({\Bf r}) - n_{\sigma}
({\Bf r})$ (for the case of spin-$1/2$ fermions, this amounts to 
$n_{\downarrow}({\Bf r})-n_{\uparrow}({\Bf r})$ when $\sigma=\uparrow$), 
has a distinctive asymptotic behaviour for $\vert z\vert\to\infty$: it 
decays in magnitude like $1/\vert z\vert^{3/2}$, to be compared with 
the decay of other contributions to the large-$\vert z\vert$ AS for 
$\wt{\Sigma}_{\sigma}({\Bf r},{\Bf r}';z)$ following $\Sigma_{\sigma;
\infty_1}/z$, which depend on $\ln(-z/\varepsilon_0)/z^2$, $1/z^2$,
etc. (here $\varepsilon_0$ denotes an arbitrary constant energy). The 
dependence of the mentioned contribution on $n_{\bar\sigma}({\Bf r})
-n_{\sigma}({\Bf r})$ provides a direct means by which to establish the 
existence and to determine the magnitude of this possible local 
imbalance in the distribution of the particles with spin $\sigma$ with 
respect to that of the remaining particles. Aside from the interest of 
this result to the interpretation of the experimental 
{\sl inverse} photo-emission data (see Eqs.~(\ref{e234}) and 
(\ref{e239}) below), it unequivocally demonstrates that models that 
through neglect of Umklapp processes do not account for the possibility 
of {\sl local} imbalance in the distribution of particles with 
different spins (to be distinguished from the long-range ordering of 
spins, corresponding to, e.g., ferromagnetism or anti-ferromagnetism) 
are {\sl not} capable of manifesting the latter asymptotic behaviour 
at large valued of $\vert z\vert$. Further, since this particular 
behaviour owes its existence to the $1/\|{\Bf r}-{\Bf r}'\|$ singularity 
of $v_c({\Bf r}-{\Bf r}')$ at ${\Bf r}={\Bf r}'$, it is equally not 
accounted for in theoretical considerations based on models that 
involve bounded interaction potentials, such as is the case in the 
Hubbard model (for example Montorsi (1992), Gebhard (1997)) involving 
a finite {\sl intra}-atomic interaction potential.

For concreteness, we now present the aforementioned {\sl four} 
fundamentally unbounded contributions in the expression on the RHS of 
Eq.~(\ref{e199}) specific to $v\equiv v_c$ in $d=3$. An immediate and 
unequivocal recognition of these contributions in Eq.~(\ref{e199}) is 
{\sl not} possible, since a substantial reorganization of terms in 
Eq.~(\ref{e199}) will have to be effected before the following 
contributions are deduced (see Appendices F and G for some details). 
We have
\begin{eqnarray}
\label{e200}
&&{\cal I}_1 {:=} \int {\rm d}^dr''\;
v^3({\Bf r}-{\Bf r}'')\, n({\Bf r}'')\,
\delta({\Bf r}-{\Bf r}');\\
\label{e201}
&&{\cal I}_2 {:=} \int {\rm d}^dr_1'' {\rm d}^dr_2''\;
v^2({\Bf r}-{\Bf r}_1'') v({\Bf r}-{\Bf r}_2'') \nonumber\\
&&\;\times \sum_{\sigma_1',\sigma_2'}
\Big[\Gamma^{(2)}({\Bf r}_1''\sigma_1',{\Bf r}_2''\sigma_2';
{\Bf r}_1''\sigma_1',{\Bf r}_2''\sigma_2')
- n_{\sigma_1'}({\Bf r}_1'') n_{\sigma_2'}({\Bf r}_2'')\Big]
\nonumber\\
&&\;\;\;\;\;\;\;\;\;\;\;\;\;\;\;\;\;\;\;\;\;\;\;\;\;\;\;\;\;\;
\;\;\;\;\;\;\;\;\;\;\;\;\;\;\;\;\;\;\;\;\;\;\;\;\;\;\;\;
\times\delta({\Bf r}-{\Bf r}');\\
\label{e202}
&&{\cal I}_3 {:=} \int {\rm d}^dr''\;
v({\Bf r}-{\Bf r}'') n({\Bf r}'') 
\big(\tau({\Bf r}'') v({\Bf r}-{\Bf r}'')\big)
\,\delta({\Bf r}-{\Bf r}');\nonumber\\ \\
\label{e203}
&&{\cal I}_4 {:=} -2 v({\Bf r}-{\Bf r}')
\varrho_{\sigma}({\Bf r}',{\Bf r})
\big(\tau({\Bf r}) v({\Bf r}-{\Bf r}')\big). 
\end{eqnarray}
It can be easily verified that for $v\equiv v_c$ in $d=3$, ${\cal I}_1$ 
is unbounded for two reasons: firstly, the singularity of $v_c^3({\Bf r}
-{\Bf r}'')$ at ${\Bf r}'' = {\Bf r}$ is {\sl not} integrable, and 
secondly, excluding a finite neighbourhood of ${\Bf r}''={\Bf r}$, 
${\cal I}_1$ further diverges as the volume of the system, $\Omega$, 
tends to infinity.
\footnote{\label{f92}
To be precise, by employing the $v'$ corresponding to $v_c$ in 
Eq.~(\protect\ref{e13}), the calculated ${\cal I}_1$ diverges 
(logarithmically) as $\kappa\downarrow 0$. }
It can also be shown (see Appendix F) that similarly ${\cal I}_2$
diverges as $\Omega\to\infty$. However, since the leading term in 
the large-$\|{\Bf r}_1''\|$ AS of the integral over 
${\Bf r}_2''$ in Eq.~(\ref{e201}) is equal to $-v_c({\Bf r}-{\Bf r}_1'') 
n({\Bf r}_1'')$ (see Eq.~(\ref{ef125}) and footnote \ref{f134}), the 
sum (see Eq.~(\ref{ef118})) 
\begin{equation} 
\label{e204}
{\cal I}_1 + {\cal I}_2 {=:} {\cal M}({\Bf r})\,
\delta({\Bf r}-{\Bf r}')
\end{equation}
does {\sl not} suffer from the latter problem of diverging in 
consequence of $\Omega\to\infty$; the integrand corresponding 
to ${\cal M}({\Bf r})$, as originating from ${\cal I}_1$, is 
non-integrable, however, owing to the indicated singularity of 
$v_c^3({\Bf r}-{\Bf r}'')$ at ${\Bf r}''={\Bf r}$. This can be 
remedied through a summation over an infinite number of 
non-integrable counterparts pertaining to 
$\{\Sigma_{\sigma;\infty_p}({\Bf r},{\Bf r}')\, \| \, p > 2\}$; 
we present the details of this renormalization procedure in 
Appendix F. 

Following Eq.~(\ref{e16}), we have (see Eq.~(\ref{e3}))
\begin{equation}
\label{e205}
\tau({\Bf r}) v_c({\Bf r}-{\Bf r}')
= \frac{\hbar^2 e^2}{2 m_e \epsilon_0}\, 
\delta({\Bf r}-{\Bf r}'),
\end{equation}
which upon substitution into the expressions on the RHSs of
Eqs.~(\ref{e202}) and (\ref{e203}), with $v\equiv v_c$, results in 
\begin{eqnarray}
\label{e206}
&&{\cal I}_3 \rightharpoonup 
\frac{\hbar^2 e^2}{2 m_e \epsilon_0}
v_c({\Bf r}-{\Bf r}') n({\Bf r})\, \delta({\Bf r}-{\Bf r}'),\\
\label{e207}
&&{\cal I}_4 \rightharpoonup 
-\frac{\hbar^2 e^2}{m_e \epsilon_0}
v_c({\Bf r}-{\Bf r}') n_{\sigma}({\Bf r})\, 
\delta({\Bf r}-{\Bf r}'),
\end{eqnarray}
both of which are unbounded as a consequence of the combination 
of $v_c({\Bf r}-{\Bf r}')$ with $\delta({\Bf r}-{\Bf r}')$ (see
footnote \ref{f91} and the text citing it). In Appendix G we consider 
the regularization of ${\cal T}_{\sigma,\bar\sigma}({\Bf r})$ in 
\begin{equation}
\label{e208}
{\cal I}_3 + {\cal I}_4 {=:} {\cal T}_{\sigma,\bar\sigma}({\Bf r})\,
\delta({\Bf r}-{\Bf r}'),
\end{equation}
where
\begin{eqnarray}
\label{e209}
{\cal T}_{\sigma,\bar\sigma}({\Bf r}) &{:=}&
\int {\rm d}^3r''\; v_c({\Bf r}-{\Bf r}'')
\big[ n_{\bar\sigma}({\Bf r}'') -
n_{\sigma}({\Bf r}'')\big]\nonumber\\
&&\;\;\;\;\;\;\;\;\;\;\;\;\;\;\;\;\;\;\;
\times \big( \tau({\Bf r}'')
v_c({\Bf r}-{\Bf r}'')\big).
\end{eqnarray}

Following the above considerations, we now present the expression 
for $\wt{\Sigma}_{\sigma;\infty_2}({\Bf r},{\Bf r}'\vert z)$ (see 
Eq.~(\ref{e110}) above) pertaining to a macroscopic system for which we 
use the decomposition of $n({\Bf r})$ as presented in Eq.~(\ref{e12}). 
In the expressions that we provide below, the contribution of 
$v_H({\Bf r};[n_0])$, which may or may not be unbounded (as nowhere
do we require identity of $v$ with $v_c$), is {\sl naturally} cancelled 
against a counterpart in $h_0({\Bf r})$ (see Eq.~(\ref{e43}) above); 
with reference to Eqs.~(\ref{e15}) and (\ref{e172}), we should bear mind 
that the explicit expression for $\varpi_{\kappa}$ in Eq.~(\ref{e5}) 
is specific to Coulomb interacting fermion systems in $d=3$. The 
expression for $\wt{\Sigma}_{\sigma;\infty_2}({\Bf r},{\Bf r}'\vert z)$ 
that we present below is therefore specifically, but {\sl not} 
exclusively, suitable for the cases corresponding to $v\equiv v_c$ 
in $d=3$.

With reference to Eq.~(\ref{e110}) (see the considerations in 
\S~III.E.2), we write
\begin{eqnarray}
\label{e210}
\wt{\Sigma}_{\sigma;\infty_2}({\Bf r},{\Bf r}'\vert z) \equiv 
\Sigma_{\sigma;\infty_2}^{\rm r}({\Bf r},{\Bf r}') &+& 
\Sigma_{\sigma;\infty_2}^{\rm s_b}({\Bf r},{\Bf r}')\nonumber\\
&+& \wt{\Sigma}_{\sigma;\infty_2}^{\rm s}({\Bf r},{\Bf r}'\| z),
\end{eqnarray}
where (see Eq.~(\ref{e199}) above; below we suppress the identically 
vanishing function ${\cal J}_{\sigma}({\Bf r},{\Bf r}')$ that is 
encountered in Eq.~(\ref{e199}); see Appendix D)
\begin{eqnarray}
\label{e211}
&&\Sigma_{\sigma;\infty_2}^{\rm r}({\Bf r},{\Bf r}')
= \frac{1}{\hbar} \Big\{ \Big( {\cal A}'({\Bf r},{\Bf r}')
\big( \tau({\Bf r}) + u({\Bf r}) + 
v_H({\Bf r};[n'])\big) \nonumber\\
&&
-\frac{\hbar^2}{2 m_e}
\int {\rm d}^dr''\; v({\Bf r}-{\Bf r}'') 
[{\Bf\nabla}_{{\Bf r}''} v({\Bf r}-{\Bf r}'')]
\cdot [{\Bf\nabla}_{{\Bf r}''} n({\Bf r}'')]\nonumber\\
&&\;\;\;\;\;\;\;\;\;\;\;\;\;\;\;\;\;\;\;\;\;\;\;\;\;\;\;\;\;\;
\;\;\;\;\;\;\;\;\;\;\;\;\;\;\;\;\;\;\;\;\;\;\;
+{\cal L}''({\Bf r}) \Big)\,
\delta({\Bf r}-{\Bf r}')\nonumber\\
&&
+\big( {\cal B}''_{\sigma}({\Bf r},{\Bf r}')
+ {\cal B}''_{\sigma}({\Bf r}',{\Bf r}) \big)
\big(\tau({\Bf r}) v({\Bf r}-{\Bf r}')\big)\nonumber\\
&&
+ \big( {\cal A}'({\Bf r},{\Bf r}) 
+ {\cal A}'({\Bf r}',{\Bf r}') \big)
v({\Bf r}-{\Bf r}') \varrho_{\sigma}({\Bf r}',{\Bf r})
\nonumber\\
&&
+\frac{1}{2} v({\Bf r}-{\Bf r}')
\big( v_H({\Bf r};[n'])-v_H({\Bf r}';[n']) \big) 
\nonumber\\
&&\;\;\;\;\;\;\;\;\;\;\;
\times \big(\tau({\Bf r}) - \tau({\Bf r}')
+ u({\Bf r}) - u({\Bf r}') \nonumber\\
&&\;\;\;\;\;\;\;\;\;\;\;\;\;\;\;\;\;\,
+ v_H({\Bf r};[n']) - v_H({\Bf r}';[n'])  \big)
\varrho_{\sigma}({\Bf r}',{\Bf r})\nonumber\\
&&
+\frac{1}{2} v({\Bf r}-{\Bf r}')
\varrho_{\sigma}({\Bf r}',{\Bf r})
\big( \tau({\Bf r}) v_H({\Bf r};[n'])
+ \tau({\Bf r}') v_H({\Bf r}';[n']) \big)\nonumber\\
&&
-\frac{1}{2} v^2({\Bf r}-{\Bf r}') 
\big( \tau({\Bf r}) + \tau({\Bf r}')
+ u({\Bf r}) + u({\Bf r}') \nonumber\\
&&\;\;\;\;\;\;\;\;\;\;\;\;\;\;\;\;\;\;\;\;\;\;\;\;\;
+ v_H({\Bf r};[n']) + v_H({\Bf r}';[n']) \big)
\varrho_{\sigma}({\Bf r}',{\Bf r}) \nonumber\\
&&
+\frac{1}{2} v({\Bf r}-{\Bf r}')
\big( \tau({\Bf r}) + \tau({\Bf r}')
+ u({\Bf r}) + u({\Bf r}') \nonumber\\
&&\;\;\;\;\;\;\;\;\;\;\;\;\;\;\;\;\;\;\;\;
+v({\Bf r}-{\Bf r}')\big) 
\big({\cal B}''_{\sigma}({\Bf r},{\Bf r}')
+ {\cal B}''_{\sigma}({\Bf r}',{\Bf r}) \big)\nonumber\\
&&
+ v({\Bf r}-{\Bf r}') \big( v_H({\Bf r};[n']) 
{\cal B}''_{\sigma}({\Bf r}',{\Bf r})
+ v_H({\Bf r}';[n']) 
{\cal B}''_{\sigma}({\Bf r},{\Bf r}')\big)\nonumber\\
&&
-\frac{1}{2} v({\Bf r}-{\Bf r}') 
\big( {\cal G}''_{\sigma}({\Bf r},{\Bf r}')
+ {\cal G}''_{\sigma}({\Bf r}',{\Bf r}) \big) \nonumber\\
&&
-2 v({\Bf r}-{\Bf r}') 
{\cal K}'''_{\sigma}({\Bf r},{\Bf r}')\nonumber\\
&&
+\frac{\hbar^2}{m_e} v({\Bf r}-{\Bf r}')
\big( [{\Bf\nabla}_{{\Bf r}} v({\Bf r}-{\Bf r}')]
\cdot [{\Bf\nabla}_{{\Bf r}} 
\varrho_{\sigma}({\Bf r}',{\Bf r})] \nonumber\\
&&\;\;\;\;\;\;\;\;\;\;\;\;\;\;\;\;\;\;\;\;\;\;\;
+ [{\Bf\nabla}_{{\Bf r}'} v({\Bf r}-{\Bf r}')]
\cdot [{\Bf\nabla}_{{\Bf r}'} 
\varrho_{\sigma}({\Bf r}',{\Bf r})] \big) \nonumber\\
&&
-\frac{\hbar^2}{2 m_e} v({\Bf r}-{\Bf r}')
\big( [{\Bf\nabla}_{{\Bf r}} v_H({\Bf r};[n'])]
\cdot [{\Bf\nabla}_{{\Bf r}} 
\varrho_{\sigma}({\Bf r}',{\Bf r})] \nonumber\\
&&\;\;\;\;\;\;\;\;\;\;\;\;\;\;\;\;\;\;\;\;\;\;\;\;\,
+ [{\Bf\nabla}_{{\Bf r}'} v_H({\Bf r}';[n'])]
\cdot [{\Bf\nabla}_{{\Bf r}'} 
\varrho_{\sigma}({\Bf r}',{\Bf r})] \big) \nonumber\\
&&
+v({\Bf r}-{\Bf r}') \int {\rm d}^dr''\; 
\sum_{\sigma'} \big( \Gamma^{(2)}({\Bf r}'\sigma,
{\Bf r}''\sigma';{\Bf r}\sigma,{\Bf r}''\sigma')\nonumber\\
&&\;\;\;\;\;\;\;\;\;\;\;\;\;\;\;\;\;\;\;\;\;\;\;\;\;\;
\;\;\;\;\;\;\;\;\;\;\;\;\;\;\;\;\;\;\;\;\;\;
-n_{\sigma'}({\Bf r}'')
\varrho_{\sigma}({\Bf r}',{\Bf r}) \big)\nonumber\\
&&\;\;\;\;\;\;\;\;\;\;\;
\times \big( \tau({\Bf r}) v({\Bf r}-{\Bf r}'')
+\tau({\Bf r}') v({\Bf r}'-{\Bf r}'') \big) \nonumber\\
&&
-\frac{\hbar^2}{m_e}
\int {\rm d}^dr''\;
\big( v({\Bf r}-{\Bf r}')
[{\Bf\nabla}_{{\Bf r}} v({\Bf r}-{\Bf r}'')]\nonumber\\
&&\;\;\;\;\;\;\;\;\;\;\;\;\;\;\;\;\;\;\;\;
-v({\Bf r}-{\Bf r}'') [{\Bf\nabla}_{{\Bf r}} 
v({\Bf r}-{\Bf r}')]\big)\nonumber\\
&&\;
\cdot \big[ {\Bf\nabla}_{{\Bf r}} \sum_{\sigma'}
\big( \Gamma^{(2)}({\Bf r}'\sigma,
{\Bf r}''\sigma';{\Bf r}\sigma,{\Bf r}''\sigma')
- n_{\sigma'}({\Bf r}'') 
\varrho_{\sigma}({\Bf r}',{\Bf r})\big) \big]\nonumber\\
&&
-\frac{\hbar^2}{m_e}
\int {\rm d}^dr''\;
\big( v({\Bf r}-{\Bf r}')
[{\Bf\nabla}_{{\Bf r}'} v({\Bf r}'-{\Bf r}'')]\nonumber\\
&&\;\;\;\;\;\;\;\;\;\;\;\;\;\;\;\;\;\;\;\;
-v({\Bf r}'-{\Bf r}'') [{\Bf\nabla}_{{\Bf r}'} 
v({\Bf r}-{\Bf r}')]\big)\nonumber\\
&&\;
\cdot \big[ {\Bf\nabla}_{{\Bf r}'} \sum_{\sigma'}
\big( \Gamma^{(2)}({\Bf r}\sigma,
{\Bf r}''\sigma';{\Bf r}'\sigma,{\Bf r}''\sigma')
- n_{\sigma'}({\Bf r}'') 
\varrho_{\sigma}({\Bf r}',{\Bf r})\big) \big]\nonumber\\
&&
-\frac{\hbar^2}{m_e}
\int {\rm d}^dr''\; [{\Bf\nabla}_{{\Bf r}''}
v({\Bf r}-{\Bf r}'')]\cdot
[{\Bf\nabla}_{{\Bf r}''}
v({\Bf r}'-{\Bf r}'')]\nonumber\\
&&\;\;\;\;\;\;\;\;\;\;\;\;\;\;\;\;\;\;\;\;\;\;\;\;\;
\;\;\;\;\;
\times\sum_{\sigma'} 
\Gamma^{(2)}({\Bf r}'\sigma,{\Bf r}''\sigma';
{\Bf r}\sigma,{\Bf r}''\sigma')\nonumber\\
&&
-\frac{\hbar^2}{2 m_e} \int {\rm d}^dr''\;
[{\Bf\nabla}_{{\Bf r}''} v({\Bf r}-{\Bf r}'')
\varrho_{\sigma}({\Bf r},{\Bf r}'')]\nonumber\\
&&\;\;\;\;\;\;\;\;\;\;\;\;\;\;\;\;\;\;\;\;\;\;\,
\cdot [{\Bf\nabla}_{{\Bf r}''} v({\Bf r}'-{\Bf r}'')
\varrho_{\sigma}({\Bf r}',{\Bf r}'')]\nonumber\\
&&
-\int {\rm d}^dr''\; v({\Bf r}-{\Bf r}'')
v({\Bf r}'-{\Bf r}'') \nonumber\\
&&\;\;\;
\times \big( u({\Bf r}'') + v_H({\Bf r}'';[n'])
-v({\Bf r}-{\Bf r}'') - v({\Bf r}'-{\Bf r}'') \big)
\nonumber\\
&&\;\;\;\;\;\;\;\;\;\;\;\;\;\;\;\;\;\;\;\;\;\;\;\;\;\;\;\;\;
\;\;\;\;\;\;\;\;\;\;\;\;\;\;
\times \varrho_{\sigma}({\Bf r}',{\Bf r}'')
\varrho_{\sigma}({\Bf r}'',{\Bf r})\nonumber\\
&&
-\int {\rm d}^dr_1'' {\rm d}^dr_2''\;
v({\Bf r}-{\Bf r}_1'') v({\Bf r}'-{\Bf r}_2'')
v({\Bf r}_1''-{\Bf r}_2'')\nonumber\\
&&\;\;\;\;\;\;\;\;\;\;\;\;\;\;\;\;\;\;\;\;\;\;\;\;\;
\;\;\,
\times \varrho_{\sigma}({\Bf r},{\Bf r}_1'')
\varrho_{\sigma}({\Bf r}_1'',{\Bf r}_2'')
\varrho_{\sigma}({\Bf r}_2'',{\Bf r}') \Big\},
\end{eqnarray}
\begin{equation}
\label{e212}
\Sigma_{\sigma;\infty_2}^{\rm s_b}({\Bf r},{\Bf r}')
= -\frac{1}{\hbar}\,
v^3({\Bf r}-{\Bf r}') \varrho_{\sigma}({\Bf r}',{\Bf r}),
\end{equation}
\begin{eqnarray}
\label{e213}
\wt{\Sigma}_{\sigma;\infty_2}^{\rm s}({\Bf r},{\Bf r}'\| z)
&=& \frac{1}{\hbar} \Big\{
{\sf M}^{\rm r}_{\infty_2}({\Bf r}) 
+ \wt{\sf M}^{\rm s}_{\infty_2}({\Bf r}\| z)\nonumber\\
& &\;\;\;\;\;\;\;
+\wt{\sf T}_{\sigma,\bar\sigma;\infty_2}^{\rm s}({\Bf r}\| z) 
\Big\}\, \delta({\Bf r}-{\Bf r}').
\end{eqnarray}
For functions ${\cal A}'$, ${\cal B}_{\sigma}''$, ${\cal G}_{\sigma}''$ 
and ${\cal K}_{\sigma}'''$, their definitions and significant properties,
see Appendix F (see specifically Eqs.~(\ref{ef3}), (\ref{ef98}),
(\ref{ef104}) and (\ref{ef112}) respectively). For the explicit
expressions for ${\sf M}^{\rm r}_{\infty_2}({\Bf r})$, 
$\wt{\sf M}^{\rm s}_{\infty_2}({\Bf r}\| z)$ and 
$\wt{\sf T}^{\rm s}_{\sigma,\bar\sigma;\infty_2}({\Bf r}\| z)$ see 
Eqs.~(\ref{ef144}), (\ref{ef145}) and (\ref{eg15}) respectively. The 
`singular' contribution arising from $\Sigma_{\sigma;\infty_2}({\Bf r},
{\Bf r}')$ that initializes the infinite series (i.e. that underlying
the required regularization process) of unbounded terms whose relevant 
asymptotic contribution to $\wt{\Sigma}_{\sigma;\infty_2}({\Bf r},
{\Bf r}'\vert z)$ for $\vert z\vert\to\infty$ is $\wt{\Sigma}_{\sigma;
\infty_2}^{\rm s}({\Bf r},{\Bf r}'\| z)$ on the RHS of Eq.~(\ref{e213}), 
is the following (see Eqs.~(\ref{e204}) and (\ref{e208}) above)
\begin{equation}
\label{e214}
\Sigma_{\sigma;\infty_2}^{\rm s}({\Bf r},{\Bf r}')
\equiv \frac{1}{\hbar} \big\{ {\cal M}({\Bf r})
+ {\cal T}_{\sigma,\bar\sigma}({\Bf r}) \big\}\, 
\delta({\Bf r}-{\Bf r}'),
\end{equation}
where ${\cal M}({\Bf r})$ is defined in Eq.~(\ref{ef118}) below and 
${\cal T}_{\sigma,\bar\sigma}({\Bf r})$ in Eq.~(\ref{e209}) above 
(see Appendix G). We note that $\Sigma_{\sigma;\infty_2}({\Bf r},
{\Bf r}')\equiv\Sigma_{\sigma;\infty_2}^{\rm r}({\Bf r},{\Bf r}') + 
\Sigma_{\sigma;\infty_2}^{\rm s_b}({\Bf r},{\Bf r}') +
\Sigma_{\sigma;\infty_2}^{\rm s}({\Bf r},{\Bf r}')$, with the 
constituent parts being those presented in Eqs.~(\ref{e211}), 
(\ref{e212}) and (\ref{e214}) respectively, {\sl is} fully identical with 
$\Sigma_{\sigma;\infty_2}({\Bf r},{\Bf r}')$ in Eq.~(\ref{e199}). This 
follows from the fact that {\sl no} specific aspect of $v_c$ has been 
assumed in arriving at the expressions in Eqs.~(\ref{e211}), (\ref{e212}) 
and (\ref{e214}). In contrast, the expression for $\wt{\Sigma}_{\sigma;
\infty_2}^{\rm s}({\Bf r},{\Bf r}'\| z)$ in Eq.~(\ref{e213}), to be 
contrasted with $\Sigma_{\sigma;\infty_2}^{\rm s}({\Bf r},{\Bf r}')$ 
in Eq.~(\ref{e214}), {\sl is} specific to $v\equiv v_c$ in $d=3$ and 
has its origin in an infinite number of terms in the {\sl formal} 
expressions pertaining to $\{ \Sigma_{\sigma;\infty_p}^{\rm s}({\Bf r},
{\Bf r}')\, \|\, p\ge 2\}$. We note that $\wt{\sf T}_{\sigma,\bar\sigma;
\infty_2}^{\rm s}({\Bf r}\| z) \equiv 0$ for $n_{\sigma}({\Bf r}) 
\equiv n_{\bar\sigma}({\Bf r})$ (see Eq.~(\ref{eg15})).

While employing the expression in Eq.~(\ref{e211}) for $v\equiv v_c$ 
in $d=3$, the following simplified results (see Eq.~(\ref{e205}) above
and footnote \ref{f26}) are at one's disposal:
\begin{eqnarray}
\label{e215}
&&\big( {\cal B}''_{\sigma}({\Bf r},{\Bf r}')
+ {\cal B}''_{\sigma}({\Bf r}',{\Bf r}) \big)
\tau({\Bf r}) v_c({\Bf r}-{\Bf r}') \nonumber\\
&&\;\;\;\;\;\;\;\;\;\;\;\;\;\;\;\;\;\;\;\;\;\;\;\;\;
\;\;\;\;\;\;\;\;\;
= \frac{\hbar^2 e^2}{m_e \epsilon_0}\, 
{\cal B}''_{\sigma}({\Bf r},{\Bf r})\,
\delta({\Bf r}-{\Bf r}')
\end{eqnarray}
and
\begin{eqnarray}
\label{e216}
&&\tau({\Bf r}) v_H({\Bf r};[n']) + \tau({\Bf r}')
v_H({\Bf r}';[n']) \nonumber\\
&&\;\;\;\;\;\;\;\;\;\;\;\;\;\;\;\;\;\;\;\;\;\;\;\;\;
\;\;\;\;\;\;\;\;\;
= \frac{\hbar^2 e^2}{2 m_e \epsilon_0}\,
\big( n'({\Bf r}) + n'({\Bf r}')\big).
\end{eqnarray}
As can be easily verified from Eq.~(\ref{ef97}) (see also 
Eq.~(\ref{ef98})), in contrast with ${\cal B}'_{\sigma}({\Bf r},
{\Bf r})$, ${\cal B}''_{\sigma}({\Bf r},{\Bf r})$ is {\sl bounded} 
and well-defined. For the cases corresponding to $v\not\equiv v_c$, 
the LHS of Eq.~(\ref{e215}) accounts for a {\sl non}-local contribution 
to $\Sigma_{\sigma;\infty_2}^{\rm r}({\Bf r},{\Bf r}')$ which, as the 
RHS shows, transforms into a {\sl local} contribution to 
$\Sigma_{\sigma;\infty_2}^{\rm r}({\Bf r},{\Bf r}')$ in the cases 
corresponding to $v\equiv v_c$. This aspect is significant in that 
it demonstrates that {\sl any} local approximation to 
$\wt{\Sigma}_{\sigma}({\Bf r},{\Bf r}';z)$, designed for an 
{\sl arbitrary} two-body potential $v$, will fail correctly to 
reproduce the local part of $\wt{\Sigma}_{\sigma}({\Bf r},{\Bf r}';z)$ 
when applied to cases where $v\equiv v_c$ (and $d=3$). In this 
connection we recall that a crucial contribution to 
${\cal T}_{\sigma,\bar\sigma}({\Bf r})$, namely ${\cal I}_4$ in 
Eq.~(\ref{e203}), and consequently 
$\wt{\sf T}^{\rm s}_{\sigma,\bar\sigma;\infty_2}({\Bf r}\| z)$ 
(see Eq.~(\ref{eg15})),
\footnote{\label{f93}
The first term (corresponding to $m=1$) in the series representation 
of $\wt{\sf T}_{\sigma,\bar\sigma}({\Bf r};z)$ in Eq.~(\protect\ref{eg1}), 
is simply $z^{-2}\, {\cal T}_{\sigma,\bar\sigma}({\Bf r})$ where 
${\cal T}_{\sigma,\bar\sigma}({\Bf r})$ is introduced in 
Eq.~(\protect\ref{e209}). } 
has its origin in what for a general $v$ amounts to a {\sl non}-local 
term in the expression for $\Sigma_{\sigma;\infty_2}({\Bf r},{\Bf r}')$. 
As we explicitly demonstrate in \S~IV, in the first-order perturbation 
series for $\wt{\Sigma}_{\sigma}({\Bf r},{\Bf r}';z)$ in terms of the 
dynamical screened interaction function $W(\varepsilon)$ (Hubbard 1957), 
as opposed to the static bare interaction function $v$ (here, $v_c$), 
{\sl non}-local terms are missing altogether in the corresponding 
expression for $\Sigma_{\sigma;\infty_1}({\Bf r},{\Bf r}')$ (which we 
denote by $\Sigma_{\sigma;\infty_1}^{(1)}({\Bf r},{\Bf r}')$; see
Eq.~(\ref{e257}) below); further, the absence of a crucial non-local 
contribution (namely, ${\cal I}_4$ introduced in Eq.~(\ref{e203}) 
above) in the first-order counterpart of $\Sigma_{\sigma;\infty_2}
({\Bf r},{\Bf r}')$, that is $\Sigma_{\sigma;\infty_2}^{(1)}({\Bf r},
{\Bf r}')$ (see Eq.~({\ref{e267}) below), results in $\wt{\sf T}^{(1)}
({\Bf r};z)$ (see Eq.~(\ref{e277}) below), to be contrasted with
$\wt{\sf T}_{\sigma,\bar\sigma}({\Bf r};z)$ (see Eq.~(\ref{e209}) 
above) in the exact treatment. 

For completeness, as can be seen from Eqs.~(\ref{e210}), (\ref{e213})
and (\ref{eg15}), in general, in the cases corresponding to $v\equiv 
v_c$, $\hbar^{-1} \wt{\sf T}_{\sigma,\bar\sigma}^{\rm s}({\Bf r}\| z)\, 
\delta({\Bf r}-{\Bf r}')$ is a dominant contribution to the local part 
of the {\sl regularized} $\Sigma_{\sigma;\infty_2}({\Bf r},{\Bf r}')$, 
that is $\wt{\Sigma}_{\sigma;\infty_2}({\Bf r},{\Bf r}'\vert z)$; for 
increasing values of $\vert z\vert$, this contribution decays {\sl 
most slowly} in comparison with {\sl all} other asymptotic contributions 
to $\wt{\Sigma}_{\sigma}({\Bf r},{\Bf r}';z)$ that follow $\Sigma_{\sigma;
\infty_1}({\Bf r},{\Bf r}')/z$ (see Eq.~(\ref{e112}) above). This again 
makes evident that {\sl non}-local contributions to $\Sigma_{\sigma}
({\Bf r},{\Bf r}';z)$ can acquire considerable significance when 
$v\equiv v_c$. 

We close this Section by pointing out the important fact that, 
{\sl independent} of whether the GS of the system under consideration 
is metallic or insulating, $\Sigma_{\sigma;\infty_2}({\Bf r},{\Bf r}')$ 
involves both unbounded and non-integrable contributions (see 
Eqs.~(\ref{e214}) and (\ref{e212}) respectively) in cases where 
$v \equiv v_c$ in $d=3$ ({\sl not} exclusively, however). Since
regularization of the unbounded contributions to $\Sigma_{\sigma;
\infty_2}^{\rm s}({\Bf r},{\Bf r}')$ in Eq.~(\ref{e214}) requires 
a summation over an {\sl infinite} number of terms pertaining to 
$\Sigma_{\sigma;\infty_p}({\Bf r},{\Bf r}')$, $p > 2$, in view of the 
direct relationship between $\Sigma_{\sigma;\infty_p}({\Bf r},{\Bf r}')$ 
and the many-body perturbation series for $\wt{\Sigma}_{\sigma}({\Bf r},
{\Bf r}';z)$ in terms of the bare $v$ and the {\sl exact} $G_{\sigma}$ 
(see \S~I.B (the paragraph in which footnote \ref{f9} is cited for the
first time) as well as Eqs.~(\ref{e107}), (\ref{e108}), (\ref{e109}) 
and (\ref{e122}) above), it follows that {\sl no} finite-order perturbation 
theory (beyond the first order) for $\wt{\Sigma}_{\sigma}({\Bf r},
{\Bf r}';z)$ in terms of the {\sl bare} $v$ can suffice in cases where 
$v\equiv v_c$ and $d=3$. To be explicit, depending on whether
$n_{\bar\sigma}({\Bf r})-n_{\sigma}({\Bf r}) \not\equiv 0$ or
$n_{\bar\sigma}({\Bf r})-n_{\sigma}({\Bf r}) \equiv 0$, ({\it cf}. 
Eq.~(\ref{eg15})), such perturbation series involves unbounded 
contributions already at respectively the second or third order in the 
perturbation theory (see Eq.~(\ref{e129}) above) when $v\equiv v_c$ in 
$d=3$. To appreciate this point, suppose that, in contradiction to our
assertion, a $p$th-order perturbation theory of the type mentioned
would suffice, with $p > 2$ (let us say, for a system with an insulating 
GS). In such an event, $\Sigma_{\sigma;\infty_{p-1}}({\Bf r},{\Bf r}')$ 
would have to be integrable and bounded almost everywhere ({\it cf.} 
conditions (A), (B) and (C) in \S~II.B), which our considerations in 
this Section show not to be the case when $p=3$. 

To complete our above arguments, it remains to consider the following 
observation: the many-body perturbation theory for $\wt{\Sigma}_{\sigma}
({\Bf r},{\Bf r}';z)$ to which we have referred above, is one in terms 
of the {\sl exact} single-particle GF $G_{\sigma}$ and the {\sl bare}
particle-particle interaction function $v$, diagrammatically represented 
by means of skeleton diagrams (Luttinger and Ward 1960). Since 
$G_{\sigma}$ is an implicit functional of $v$, depending on $v$ to 
infinite order, it follows that the contribution of any single skeleton 
diagram to $\wt{\Sigma}_{\sigma}({\Bf r},{\Bf r}';z)$ is in fact 
non-perturbative. Consequently, the question may arise concerning 
relevance, if not validity, of our above assertion, in its general 
form, with regard to {\sl finite}-order perturbation series for 
$\wt{\Sigma}_{\sigma}({\Bf r},{\Bf r}';z)$ in cases corresponding to 
$v\equiv v_c$ and $d=3$. This question is readily dealt with, 
unequivocally establishing the validity of our above assertion, by 
considering the fact that the unbounded and non-integrable contributions 
to $\Sigma_{\sigma;\infty_m}({\Bf r},{\Bf r}')$, with $m\ge 2$, remain 
(a fact that can be explicitly verified in the case when $m=2$) by 
replacing the exact functions $n_{\sigma}$, $\varrho_{\sigma}$, 
$\Gamma^{(m)}$, in the pertinent expressions by their counterparts 
within the framework of the SSDA. In doing so it is important that the 
underlying single Slater determinant (SSD) (see Appendix C), namely 
$\vert\Phi_{N;0}\rangle$, is the GS of an $\wh{H}_0$ with which the 
exact GS $\vert\Psi_{N;0}\rangle$ is adiabatically connected (see the 
paragraph preceding that containing Eq.~(\ref{e62})). To appreciate 
this aspect, consider the case where $n_{\sigma}({\Bf r}) \not\equiv 
n_{\bar\sigma}({\Bf r})$ so that $\wt{\sf T}^{\rm s}_{\sigma,\bar\sigma;
\infty_2}({\Bf r}\| z) \not\equiv 0$ (see Eq.~(\ref{eg15})); evidently, 
to an inappropriate $\vert\Phi_{N;0}\rangle$, for which $n_{{\rm s};
\sigma}({\Bf r}) \equiv n_{{\rm s};\bar\sigma}({\Bf r})$, corresponds 
an identically-vanishing $\wt{\sf T}^{\rm s}_{\sigma,\bar\sigma;
\infty_2}({\Bf r}\| z)\vert_{\rm s}$.

\subsection{The asymptotic series of the imaginary part of the 
self-energy for $\vert\varepsilon\vert\to \infty$}
\label{s30}

\subsubsection{General considerations}
\label{s31}

In the earlier Sections we have dealt with the AS of the {\sl full} SE 
for large values of $\vert\varepsilon\vert$. As we have emphasized (see 
\S~III.C), unless the entire series, or at least an {\sl infinite} 
subset of its terms, is summed, a finite-order AS for $\Sigma_{\sigma}
({\Bf r},{\Bf r}';\varepsilon)$ {\sl cannot} but be real valued (see 
also the second paragraph in Appendix B). Here we deduce a finite-order 
large-$\vert\varepsilon\vert$ AS for ${\rm Im}\Sigma_{\sigma}({\Bf r},
{\Bf r}';\varepsilon)$ which in conjunction with a finite-order AS 
for ${\rm Re}\Sigma_{\sigma}({\Bf r},{\Bf r}';\varepsilon)$, coinciding 
with that for $\Sigma_{\sigma}({\Bf r},{\Bf r}';\varepsilon)$, provide 
to some finite order in $1/\varepsilon$ the {\sl complete} asymptotic 
behaviour of $\Sigma_{\sigma}({\Bf r},{\Bf r}';\varepsilon)$ for 
$\vert\varepsilon\vert\to\infty$. For a concise notation, in this 
Sections we denote $\Sigma_{\sigma}({\Bf r},{\Bf r}';\varepsilon)$ and 
related {\sl functions}, such as $\Sigma_{\sigma;\infty_m}({\Bf r},
{\Bf r}')$, by $\Sigma_{\sigma}(\varepsilon)$ and $\Sigma_{\sigma;
\infty_m}$ respectively; thus below $\Sigma_{\sigma}(\varepsilon)$ (and 
$\Sigma_{\sigma;\infty_m}$, etc.) should {\sl not} be identified with 
the SE {\sl operator}, but with its coordinate representation, which 
is a {\sl function}. 

The imaginary part of $\Sigma_{\sigma}(\varepsilon) \equiv
\wt{\Sigma}_{\sigma}(\varepsilon \pm i\eta)$, $\varepsilon\,
\Ieq<> \,\mu$, $\eta\downarrow 0$, is obtained in terms of its 
real part from the following Kramers-Kr\"onig relation
\begin{equation}
\label{e217}
{\rm Im}\wt{\Sigma}_{\sigma}(\varepsilon \pm i\eta)
= \frac{\pm 1}{\pi}\, {\rm sgn}(\mu -\varepsilon)\,
{\it S}_{\sigma}(\varepsilon),\;\;\; \eta\downarrow 0,
\end{equation}
where 
\begin{eqnarray}
\label{e218}
{\it S}_{\sigma}(\varepsilon) {:=}
\wp\!\int_{-\infty}^{\infty} {\rm d}\varepsilon'\;
\frac{{\rm Re}\Sigma_{\sigma}(\varepsilon') 
-\Sigma_{\sigma;\infty_0}}{\varepsilon' 
-\varepsilon}.
\end{eqnarray}
In Eq.~(\ref{e217}), $\mu$ stands for the `chemical potential' introduced 
in Eq.~(\ref{e22}) above. Separating the interval of $\varepsilon'$ 
integration into $(-\infty,0]$ and $[0,\infty)$ and employing the variable 
transformation $\varepsilon' \rightharpoonup -\varepsilon'$ in the 
first integral followed by the variable transformation $\varepsilon' 
\rightharpoonup 1/\varepsilon'$ in the integral thus obtained as 
well as the second integral, we arrive at
\begin{eqnarray}
\label{e219}
&&{\it S}_{\sigma}(\varepsilon) = \varepsilon\,
\wp\!\int_0^{\infty} {\rm d}\varepsilon'\;
\Big[(1+1/[\varepsilon\varepsilon'])\,
{\rm Re}\Sigma_{\sigma}(1/\varepsilon') \nonumber\\
&&\;\;
+(1-1/[\varepsilon\varepsilon'])\,
{\rm Re}\Sigma_{\sigma}(-1/\varepsilon')
-2 \Sigma_{\sigma;\infty_0}\Big]/
\big(1-{\varepsilon'}^2\varepsilon^2\big).\nonumber\\
\end{eqnarray}
Now we decompose the interval of the $\varepsilon'$ integration in
Eq.~(\ref{e219}) into $[0,\Delta]$ and $[\Delta,\infty)$, where 
$\Delta$ is some positive small number; later we shall assume 
$\Delta > 1/\vert\varepsilon \vert$ when $\vert\varepsilon\vert \to 
\infty$; we denote the former integral by ${\it S}_{\sigma}^{(1)}
(\varepsilon)$ and the latter by ${\it S}_{\sigma}^{(2)}(\varepsilon)$. 
In dealing with the integral over $[0,\Delta]$, we employ ({\it cf}. 
Eq.~(\ref{e72}) and see Eq.~(\ref{e62}) above)
\begin{equation}
\label{e220}
\Sigma_{\sigma}(\pm 1/\varepsilon') = \sum_{m=0}^{\infty}
(\pm 1)^m\, \Sigma_{\sigma;\infty_m}\, {\varepsilon'}^m,
\end{equation}
from which, after some algebra, we obtain
\begin{eqnarray}
\label{e221}
&&(1+1/[\varepsilon\varepsilon']) \Sigma_{\sigma}(1/\varepsilon')+
(1-1/[\varepsilon\varepsilon']) \Sigma_{\sigma}(-1/\varepsilon')
-2\Sigma_{\sigma;\infty_0} \nonumber\\
&&\;\;\;
= \frac{2}{\varepsilon} \Sigma_{\sigma;\infty_1}
+ 2 \sum_{m=1}^{\infty}
\big[ \Sigma_{\sigma;\infty_{2m}} + \frac{1}{\varepsilon}
\Sigma_{\sigma;\infty_{2m+1}} \big] {\varepsilon'}^{2m}.
\end{eqnarray}
{\sl Formally}, the RHS of Eq.~(\ref{e221}) is real valued for real 
values of $\varepsilon'$.  Substituting this expression in the integral 
on the RHS of Eq.~(\ref{e219}) with the interval of the $\varepsilon'$ 
integration restricted to $[0,\Delta]$, we encounter the following 
two standard integrals
\begin{eqnarray}
\label{e222}
\wp\!\int_0^{\Delta} {\rm d}\varepsilon'\;
\frac{1}{1 - {\varepsilon'}^2 \varepsilon^2}
&=& \frac{1}{\varepsilon} \tanh^{-1}(1/[\Delta \varepsilon])
\nonumber\\
&=& \frac{1}{\Delta\, \varepsilon^2} +
{\cal O}(1/\varepsilon^4),
\end{eqnarray}
\begin{eqnarray}
\label{e223}
&&\wp\!\int_0^{\Delta} {\rm d}\varepsilon'\;
\frac{{\varepsilon'}^{2m}}{1 - {\varepsilon'}^2 \varepsilon^2}
= \frac{-\Delta^{2m-1}}{(2m-1) \varepsilon^2}\nonumber\\
&&\;\;\;\;\times
{}_2F_1\big(1/2-m,1;3/2-m;1/[\Delta^2 \varepsilon^2]\big)\nonumber\\
&&\;\;\;\;
= -\frac{\Delta^{2m-1}}{(2m-1) \varepsilon^2}
+ {\cal O}(1/\varepsilon^4),
\end{eqnarray}
where $\tanh^{-1}(x) \equiv {\rm arc}\!\tanh(x)$ stands for the 
inverse of $\tanh(x)$ and ${}_2F_1(a,b;c;x)$ for the generalized
Gauss hypergeometric function (Abramowitz and Stegun 1972, p.~556). 
From the above results it immediately follows that
\begin{eqnarray}
\label{e224}
&&{\it S}_{\sigma}^{(1)}(\varepsilon) =
\big[-2 \sum_{m=1}^{\infty} \Sigma_{\sigma;\infty_{2m}}
\,\frac{\Delta^{2m-1}}{2m - 1}\big]\,
\frac{1}{\varepsilon}\nonumber\\
\;\;&&+\big[\frac{2 \Sigma_{\sigma;\infty_1} }{\Delta} 
-2 \sum_{m=1}^{\infty} \Sigma_{\sigma;\infty_{2m+1}}\,
\frac{\Delta^{2m-1}}{2m-1}\big]\,
\frac{1}{\varepsilon^2} + {\cal O}(\frac{1}{\varepsilon^3}).
\nonumber\\
\end{eqnarray}
We note in passing that $\sum_{m=1}^{\infty} \Delta^{2m-1}/(2m-1)
= {\rm arc}\!\tanh(\Delta)$. 

Further, it is easily verified that 
\begin{eqnarray}
\label{e225}
& &{\it S}_{\sigma}^{(2)}(\varepsilon)\nonumber\\
& &\;\;
= -\frac{1}{\varepsilon}\,
\wp\!\int_{\Delta}^{\infty}
\frac{{\rm d}\varepsilon'}{{\varepsilon'}^2}\;
\frac{{\rm Re}\Sigma_{\sigma}(1/\varepsilon')
+{\rm Re}\Sigma_{\sigma}(-1/\varepsilon')
-2\Sigma_{\sigma;\infty_0}}{1-1/[{\varepsilon'}^2
\varepsilon^2]}\nonumber\\
& &\;\;\;\;\;
-\frac{1}{\varepsilon^2}\,
\wp\!\int_{\Delta}^{\infty}
\frac{{\rm d}\varepsilon'}{{\varepsilon'}^3}\;
\frac{{\rm Re}\Sigma_{\sigma}(1/\varepsilon')
-{\rm Re}\Sigma_{\sigma}(-1/\varepsilon')
}{1-1/[{\varepsilon'}^2 \varepsilon^2]}.
\end{eqnarray}
For $\vert\varepsilon\vert > 1/\Delta$, taking into account that, in 
Eq.~(\ref{e225}) $\varepsilon' \ge \Delta$, we can employ in the RHS 
of Eq.~(\ref{e225}) the uniformly convergent series $1/(1-
1/[{\varepsilon'}^2 \varepsilon^2]) = \sum_{m=0}^{\infty} 
(1/[{\varepsilon'}^2\varepsilon^2])^m$ and thus obtain a uniformly 
convergent series in powers of $1/\varepsilon$, which amounts to 
the large-$\vert\varepsilon\vert$ AS for ${\it S}_{\sigma}^{(2)}
(\varepsilon)$. Combining the result thus obtained for 
${\it S}_{\sigma}^{(2)}(\varepsilon)$ and that for 
${\it S}_{\sigma}^{(1)}(\varepsilon)$ in Eq.~(\ref{e224}), we arrive at 
\begin{eqnarray}
\label{e226}
&&{\it S}_{\sigma}(\varepsilon) =
\Big[-2 \sum_{m=1}^{\infty} \Sigma_{\sigma;\infty_{2m}}\,
\frac{\Delta^{2m-1}}{2m-1}\nonumber\\
&& -\wp\!\int_{\Delta}^{\infty} 
\frac{{\rm d}\varepsilon'}{{\varepsilon'}^2}\;
\big( {\rm Re}\Sigma_{\sigma}(1/\varepsilon')
+ {\rm Re}\Sigma_{\sigma}(-1/\varepsilon')
-2\Sigma_{\sigma;\infty_0}\big)\Big]\,\frac{1}{\varepsilon}
\nonumber\\
&& +\Big[ \frac{2\Sigma_{\sigma;\infty_1}}{\Delta}
- 2 \sum_{m=1}^{\infty} \Sigma_{\sigma;\infty_{2m+1}}\,
\frac{\Delta^{2m-1}}{2m-1}\nonumber\\
&&\;\;\;
-\wp\!\int_{\Delta}^{\infty}
\frac{{\rm d}\varepsilon'}{{\varepsilon'}^3}\;
\big( {\rm Re}\Sigma_{\sigma}(1/\varepsilon')
-{\rm Re}\Sigma_{\sigma}(-1/\varepsilon')\big) \Big]
\frac{1}{\varepsilon^2}\nonumber\\
&&\;\;\;\;\;\;\;\;\;\;\;\;\;\;\;\;\;\;\;\;\;\;\;\;\;\;\;\;\;\;\;\;\;
+ {\cal O}(\frac{1}{\varepsilon^3}),
\;\;\;\mbox{\rm for}\;\ \vert\varepsilon\vert > 1/\Delta.
\end{eqnarray}
One can readily verify that the derivatives with respect to $\Delta$ 
of the coefficients of the $1/\varepsilon$ and $1/\varepsilon^2$ 
terms on the RHS of Eq.~(\ref{e226}) are {\sl identically} vanishing, 
appropriately reflecting the fact that $\Delta$ does not appear in 
the defining expression for ${\it S}_{\sigma}(\varepsilon)$. We note 
that the {\sl exact} $\Delta$ independence of the coefficients in 
the series in Eq.~(\ref{e226}) is a direct consequence of taking {\sl full}  
account of the series involving $\{\Sigma_{\sigma;\infty_{2m}}\}$ and
$\{\Sigma_{\sigma;\infty_{2m+1}}\}$; truncation of these series at some 
finite order will result in the violation of this exact property. The 
magnitude of the error thus introduced can be made arbitrary 
small by reducing $\Delta$, leading, however, through the condition
$\vert\varepsilon\vert > 1/\Delta$ to the requirement that the 
corresponding expression for ${\it S}_{\sigma}(\varepsilon)$ be taken 
as accurate for increasingly larger values of $\vert\varepsilon\vert$. 

We point out that, according to the asymptotic results obtained in 
the present work for ${\rm Re}\Sigma_{\sigma}(\varepsilon)$, we have 
${\rm Re}\Sigma_{\sigma}(1/\varepsilon') + {\rm Re}\Sigma_{\sigma}
(-1/\varepsilon') - 2\Sigma_{\sigma;\infty_0} = o(\varepsilon')$ 
for $\varepsilon'\to 0$, where $f(x) = o\big(g(x)\big)$ denotes the 
property $f(x)/g(x) \to 0$ for $x\to 0$. As a result of this fact, we 
observe that the first integral over $\varepsilon'$ on the RHS of 
Eq.~(\ref{e226}) is bounded for $\Delta\downarrow 0$. Consequently, 
insofar as the coefficient of the $1/\varepsilon$ on the RHS of 
Eq.~(\ref{e226}) is concerned, we can set herein $\Delta$ equal to 
zero, which leads to the disappearance of the infinite sum in the 
expression for this coefficient. We cannot, however, {\sl directly} 
follow this procedure of setting $\Delta$ equal to zero in the 
coefficient of $1/\varepsilon^2$ on the RHS of Eq.~(\ref{e226}). 

From Eqs.~(\ref{e217}) and (\ref{e226}), setting $\Delta$ in the
expression for the coefficient of $1/\varepsilon$ on the RHS of 
Eq.~(\ref{e226}) equal to zero, we finally obtain (see the second 
part of the {\it notes added in proof})
\begin{eqnarray}
\label{e227}
&&{\rm Im}\wt{\Sigma}_{\sigma}(\varepsilon \pm i\eta)
= \frac{\pm {\rm sgn}(\mu - \varepsilon)}{\pi} \nonumber\\
&&\;\;\;\;\;
\times\Big\{-\wp\!\int_{0}^{\infty} 
\frac{{\rm d}\varepsilon'}{{\varepsilon'}^2}\;
\big( {\rm Re}\Sigma_{\sigma}(1/\varepsilon') \nonumber\\
&&\;\;\;\;\;\;\;\;\;\;\;\;\;\;\;\;\;\;\;\;\;\;\;\;\;\;\;
+ {\rm Re}\Sigma_{\sigma}(-1/\varepsilon')
-2\Sigma_{\sigma;\infty_0}\big)\,\frac{1}{\varepsilon}
\nonumber\\
&& +\Big[ \frac{2\Sigma_{\sigma;\infty_1}}{\Delta}
- 2 \sum_{m=1}^{\infty} \Sigma_{\sigma;\infty_{2m+1}}\,
\frac{\Delta^{2m-1}}{2m-1}\nonumber\\
&&\;\;\;\;\;
-\wp\!\int_{\Delta}^{\infty}
\frac{{\rm d}\varepsilon'}{{\varepsilon'}^3}\;
\big( {\rm Re}\Sigma_{\sigma}(1/\varepsilon')
-{\rm Re}\Sigma_{\sigma}(-1/\varepsilon')\big) \Big]
\frac{1}{\varepsilon^2} \nonumber\\
&&\;\;\;\;\;\;\;\;\;\;\;\;\;\;\;\;\;\;\;\;\;\;\;\;\;\;\;\;\;\;\;\;
+{\cal O}(\frac{1}{\varepsilon^3}) \Big\}, 
\;\;\;\mbox{\rm for}\;\ \vert\varepsilon\vert > 1/\Delta.
\end{eqnarray}

Now we discuss an important limitation of the expression in 
Eq.~(\ref{e227}) in specific cases which we shall indicate below. As 
should be evident, the validity of the result in Eq.~(\ref{e227})
vitally depends on the assumption concerning admissibility of 
exchanging orders of summation and integration, which assumption 
has been basic in enabling us {\sl term by term} to integrate the 
series representation $\sum_{m=0}^{\infty} \Sigma_{\sigma;
\infty_m}/\varepsilon^m$ for $\Sigma_{\sigma}(\varepsilon)$; the 
derivation of the expression for ${\it S}_{\sigma}^{(1)}(\varepsilon)$ 
in Eq.~(\ref{e224}), which has directly resulted in the expression for 
${\rm Im}\wt{\Sigma}_{\sigma}(\varepsilon \pm i\eta)$ in 
Eq.~(\ref{e227}), crucially depends on the integral in Eq.~(\ref{e223}). 
However, as we have observed in \S~III.H, {\sl not} all terms 
contributing to $\Sigma_{\sigma;\infty_m}({\Bf r},{\Bf r}')$, with 
$m \ge 2$, are well defined,
\footnote{\label{f94}
We point out that here $m=2$ is specific to systems in $d=3$ with 
particles interacting through the Coulomb potential $v_c$. Thus $m=2$ 
is {\sl not} a universal number in the context of our present work. }
and that regularization of such contributions requires {\sl infinite} 
summations over specific category of equally ill-defined contributions 
pertaining to $\Sigma_{\sigma;\infty_p}({\Bf r},{\Bf r}')$ with 
$p > m$ (see \S~II.B). It is exactly owing to these ill-defined 
contributions that the aforementioned procedure of term-by-term 
integration fails, necessitating us to resort to such procedure as that 
employed in Appendices F and G, where we evaluate the regularized 
contributions explicitly for strictly complex $z$ (which contributions 
we denote by $\wt{\Sigma}_{\sigma;\infty_m}^{\rm s}({\Bf r},{\Bf r}'\|
z)$; see Eq.~(\ref{e110}) above) and subsequently deduce the real and 
imaginary parts of $\Sigma_{\sigma;\infty_m}^{\rm s}({\Bf r},{\Bf r}'\| 
\varepsilon)$ through the substitution $z \rightharpoonup \varepsilon
\pm i\eta$, with $\eta\downarrow 0$ ({\it cf}. Eq.~(\ref{e65}) above).
 
We can summarize the above considerations as follows. In deducing 
a finite-order large-$\vert\varepsilon\vert$ AS for 
${\rm Im}\Sigma_{\sigma}(\varepsilon)$, one should consider the 
following decomposition of the SE into `regular' and `singular' 
contributions ({\it cf}. Eq.~(\ref{e110}) above; see also conditions 
(A)-(C) in \S~II.B)
\begin{eqnarray}
\label{e228}
\wt{\Sigma}_{\sigma}({\Bf r},{\Bf r}';z) \equiv
\wt{\Sigma}_{\sigma}^{{\rm r}\oplus {\rm s_b}}({\Bf r},{\Bf r}';z) + 
\wt{\Sigma}_{\sigma}^{\rm s}({\Bf r},{\Bf r}';z). 
\end{eqnarray}
The `regular' together with `singular but bounded (almost everywhere)' 
contribution $\wt{\Sigma}_{\sigma}^{{\rm r}\oplus {\rm s_b}}({\Bf r},
{\Bf r}';z)$ is defined as being one that possesses the following 
(asymptotic) series expansion
\begin{equation}
\label{e229}
\wt{\Sigma}_{\sigma}^{{\rm r}\oplus {\rm s_b}}({\Bf r},{\Bf r}';z)
\equiv \sum_{m=0}^{\infty}
\frac{\Sigma_{\sigma;\infty_m}^{\rm r}({\Bf r},{\Bf r}') +
\Sigma_{\sigma;\infty_m}^{\rm s_b}({\Bf r},{\Bf r}')}{z^m},
\end{equation}
in which the constituent coefficient functions $\{\Sigma_{\sigma;
\infty_m}^{\rm r}\}$ and $\{\Sigma_{\sigma;\infty_m}^{\rm s_b}\}$ are 
defined in \S~III.E.2; the infinite series in Eq.~(\ref{e229}) is 
characterized by the property that its truncation to a finite number 
of terms results in a well-defined AS for $\wt{\Sigma}_{\sigma}^{{\rm r}
\oplus {\rm s_b}}({\Bf r},{\Bf r}';z)$ corresponding to $\vert 
z\vert\to\infty$ (although, unless $\Sigma_{\sigma;\infty_m}^{\rm s_b}
({\Bf r},{\Bf r}') \equiv 0$ for {\sl all} $m$ taken into account, 
the resulting series is {\sl not} necessarily suitable for calculating 
an associated finite-order AS for the momentum representation of 
$\wt{\Sigma}_{\sigma}(z)$ for $\vert z\vert\to\infty$). The `singular' 
contribution $\wt{\Sigma}_{\sigma}^{\rm s}({\Bf r},{\Bf r}';z)$, on 
the other hand, is defined as one that is neither `regular' nor 
`singular but bounded (almost everywhere)'; we have (see Eq.~(\ref{e112}) 
above and footnote \ref{f71})
\begin{equation}
\label{e230}
\wt{\Sigma}_{\sigma}^{\rm s}({\Bf r},{\Bf r}';z) 
\equiv \sum_{m=2}^{\infty} 
\frac{\wt{\Sigma}_{\sigma;\infty_m}^{\rm s}
({\Bf r},{\Bf r}'\| z)}{z^m}.
\end{equation}
The AS in Eq.~(\ref{e227}) applies to ({\it cf}.
Eq.~(\ref{e65}) above)
\begin{equation}
\label{e231}
\Sigma_{\sigma}^{{\rm r}\oplus {\rm s_b}}(\varepsilon) \equiv
\lim_{\eta\downarrow 0} 
\wt{\Sigma}_{\sigma}^{{\rm r}\oplus {\rm s_b}}(\varepsilon \pm i\eta), 
\;\;\;\varepsilon\, \IEq{<}{>} \,\mu, 
\end{equation}
with the provision that {\sl all} the functions $\Sigma_{\sigma}$ and 
$\Sigma_{\sigma;\infty_m}$ in Eq.~(\ref{e227}) be accordingly replaced 
by $\Sigma_{\sigma}^{{\rm r}\oplus {\rm s_b}}$ and $\Sigma_{\sigma;
\infty_m}^{\rm r}+\Sigma_{\sigma;\infty_m}^{\rm s_b}$ respectively. 
In this connection, it is most relevant to realize that within the 
confines of our considerations in this paper, where we specifically 
focus on the peculiarities of $v_c$ in $d=3$ (see footnote \ref{f71}), 
the leading-order contribution to the large-$\vert\varepsilon\vert$
AS for ${\rm Im}\Sigma_{\sigma}^{\rm s}({\Bf r},{\Bf r}';\varepsilon)$ 
is {\sl sub-dominant} in comparison with $1/\varepsilon$, so that the 
$\Sigma_{\sigma}$ functions in the coefficient of the $1/\varepsilon$ 
on the RHS of Eq.~(\ref{e227}), remain the {\sl complete} functions, as 
opposed to their `regular' and `singular but bounded' parts.
\footnote{\label{f95}
This aspect, which is reflected in the lower bound of the summation 
on the RHS of Eq.~(\protect\ref{e230}), is implicit in the considerations 
preceding Eq.~(\protect\ref{e227}) which led to the possibility of 
identifying $\Delta$ in the first term on the RHS of 
Eq.~(\protect\ref{e226}) with zero. In this connection we point out 
that our above conclusion with regard to the leading-order contribution 
in the large-$\vert\varepsilon\vert$ AS for ${\rm Im}\Sigma_{\sigma}
(\varepsilon)$ is fully borne out by our explicit calculations in 
\S~III.I.1. Following our brief references in footnote \protect\ref{f12} 
and in the text following Eq.~(\protect\ref{e112}) to systems of 
fermions confined to $d=2$ and interacting through $v\equiv v_c$, we 
note that the leading-order contribution to the 
large-$\vert\varepsilon\vert$ AS for ${\rm Im}\Sigma_{\sigma}^{\rm s}
(\varepsilon)$ pertaining to these systems is asymptotically more 
dominant than $1/\varepsilon$ (B. Farid, 2001, unpublished). }

Determination of ${\rm Im}\Sigma_{\sigma}^{\rm s}({\Bf r},{\Bf r}';
\varepsilon)$, on the other hand, should either proceed from 
Eqs.~(\ref{e217}) and (\ref{e218}), with the provision that {\sl all} 
the $\Sigma_{\sigma}$ functions herein be replaced by 
$\Sigma_{\sigma}^{\rm s}$, or from an explicitly evaluated 
$\wt{\Sigma}_{\sigma}^{\rm s}(z)$ by means of an appropriate analytic 
continuation of the latter function on to the real $\varepsilon$-axis 
(see Appendices F, G and H; see also Eqs.~(\ref{ef146}), (\ref{eg17}) 
and (\ref{eh21})). Note that, since in all cases $\Sigma_{\sigma;
\infty_0}$ is regular, so that we have $\Sigma_{\sigma;\infty_0}^{\rm r} 
\equiv \Sigma_{\sigma;\infty_0}$, it follows that $\Sigma_{\sigma;
\infty_0}^{\rm s_b} \equiv 0$ and $\wt{\Sigma}_{\sigma;\infty_0}^{\rm s}
\equiv 0$. Thus, in employing Eqs.~(\ref{e217}) and (\ref{e218}), 
while dealing with $\Sigma_{\sigma}^{\rm s}(\varepsilon)$, we have 
{\sl no} $\varepsilon$-{\it in}dependent contribution on the RHS of 
Eq.~(\ref{e218}). Note further that $\wt{\Sigma}_{\sigma}^{\rm s}(z)$ 
may be identically vanishing, as is the case for systems of fermions 
interacting through bounded and short-range $v$.

Finally, we point out that the relevance of the expression in 
Eq.~(\ref{e227}) is largely conceptual rather than necessarily
practical; as is evident from the $\varepsilon'$ integrals on 
the RHS of Eq.~(\ref{e227}), which together cover the interval 
$[0, \infty)$, the mere knowledge of a finite-order AS for 
${\rm Re}\Sigma_{\sigma}(\varepsilon)$ (read 
${\rm Re}\Sigma_{\sigma}^{{\rm r}\oplus {\rm s_b}}(\varepsilon)$) 
corresponding to $\vert\varepsilon\vert\to\infty$ is {\sl not} 
sufficient for the purpose of determining ${\rm Im}\Sigma_{\sigma}
(\varepsilon)$ (read ${\rm Im}\Sigma_{\sigma}^{{\rm r}\oplus {\rm s_b}}
(\varepsilon)$) at large values of $\vert\varepsilon\vert$. Below 
we shall encounter an instance where the conceptual relevance of 
the result in Eq.~(\ref{e227}) becomes fully apparent.

\subsubsection{Uniform isotropic systems; the single-particle spectral 
function $\ol{A}_{\sigma}(k;\varepsilon)$}
\label{s32}

Here we determine some leading-order terms in the large-$\vert
\varepsilon\vert$ AS of the single-particle spectral function 
$A_{\sigma}({\Bf r},{\Bf r}';\varepsilon)$ pertaining to a uniform 
and isotropic system. Consequently, here as in \S~III.E we employ 
the notation $k {:=} \|{\Bf k}\|$. The uniformity of the system 
entails that the Dyson equation, which is an integral equation in 
the coordinate representation, reduces to an algebraic equation 
in the momentum representation; we have 
\begin{equation}
\label{e232}
\wt{\ol{G}}_{\sigma}(k;z) =
\wt{\ol{G}}_{0;\sigma}(k;z) +
\wt{\ol{G}}_{0;\sigma}(k;z)
\wt{\ol{\Sigma}}_{\sigma}^{\sh}(k;z)
\wt{\ol{G}}_{\sigma}(k;z),
\end{equation} 
where $\wt{\Sigma}_{\sigma}^{\sh}$ is defined in Eq.~(\ref{e62}); since 
the constant external potential to which fermions are exposed is 
independent of $\sigma$, it is the dependence on $\sigma$ of the GF 
pertaining to the `non-interacting' system, i.e. $\wt{\ol{G}}_{0;
\sigma}(k;z)$, that has necessitated introduction in Eq.~(\ref{e232}) 
of $\wt{\Sigma}_{\sigma}^{\sh}(z)$ in place of $\wt{\Sigma}_{\sigma}(z)$ 
(see \S~III.C). 
\footnote{\label{f96}
By choosing the `non-interacting' GF to be that pertaining to 
$h_{0;\sigma}({\Bf r})$ (see Eq.~(\protect\ref{e55})) (with $u({\Bf r})$ 
a constant), rather than $h_0({\Bf r})$ (see Eq.~(\protect\ref{e43})), 
we are capable of dealing with (homogeneous) magnetic GSs. } 
The isotropy of the system together with our specific choice of gauge 
(see Appendix B) imply that the real and imaginary parts of the functions 
in Eq.~(\ref{e232}) are the Fourier transforms of the real and imaginary 
parts respectively of their real-space counterparts. In other words, the 
result in Eq.~(\ref{e227}) which, as we have clearly stated at the outset 
of \S~III.I.1, is applicable to ${\rm Im}[\Sigma_{\sigma}({\Bf r},
{\Bf r}';\varepsilon)]$, also applies to ${\rm Im}[\ol{\Sigma}_{\sigma}(k;
\varepsilon)] \equiv {\rm Im}[\ol{\Sigma}_{\sigma}^{\sh}(k;\varepsilon)]$.
The same is the case concerning ${\rm Re}[\ol{\Sigma}_{\sigma}(k;
\varepsilon)]$, that is we can employ the result in Eq.~(\ref{e112}), 
with $\Sigma_{\sigma;\infty_0}({\Bf r},{\Bf r}')$, $\Sigma_{\sigma;
\infty_1}({\Bf r},{\Bf r}')$, $\dots$ herein replaced by 
$\ol{\Sigma}_{\sigma;\infty_0}(k)$, $\ol{\Sigma}_{\sigma;\infty_1}(k)$, 
$\dots$. For our discussions that follow, for $\vert\varepsilon\vert
\to\infty$ we thus write
\begin{eqnarray}
\label{e233}
&&{\rm Re} \ol{\Sigma}_{\sigma}(k;\varepsilon)
\sim \ol{\Sigma}_{\sigma;\infty_0}(k) + 
\frac{ \ol{\Sigma}_{\sigma;\infty_1}(k)}{\varepsilon},\\ 
\label{e234}
&&{\rm Im} \wt{\ol{\Sigma}}_{\sigma}(k;\varepsilon \pm i\eta)
\sim \frac{\pm {\rm sgn}(\mu-\varepsilon)}{\pi}\,\nonumber\\
&&
\times\left\{ \begin{array}{ll}
\displaystyle \frac{\ol{\Xi}_{\sigma}(k)}{\varepsilon}
+ \frac{\ol{\Pi}_{\sigma}(k)}{\varepsilon^2},&
\mbox{\rm (I)}\\ \\
\displaystyle \frac{\ol{\Xi}_{\sigma}(k)}{\varepsilon}
-4\pi^2 \big(\frac{e^2}{4\pi\epsilon_0}\big)^2\,
\frac{(n_{0;\bar\sigma} - n_{0;\sigma})}{\sqrt{2 m_e}}\,
\frac{\Theta(\varepsilon-\mu)}{\varepsilon^{3/2}}
\nonumber\\ \nonumber \\
+\displaystyle\frac{\ol{\Pi}_{\sigma}(k)
-4 \pi^3 \hbar^{-1} (e^2/[4\pi\epsilon_0])^3 
n_{0;\bar\sigma}\,
\Theta(\varepsilon-\mu)}{\varepsilon^2},&
\mbox{\rm (II)}
\end{array} \right.\\
\end{eqnarray}
where $\ol{\Xi}_{\sigma}(k)$ and $\ol{\Pi}_{\sigma}(k)$ stand for 
the diagonal elements of the spatial Fourier transforms with respect 
to ${\Bf r}$ and ${\Bf r}'$ of the coefficients of $1/\varepsilon$ 
and $1/\varepsilon^2$ on the RHS of Eq.~(\ref{e227}). In Eq.~(\ref{e234}), 
(I) corresponds to $v\not\equiv v_c$ and (II) to $v\equiv v_c$ in $d=3$. 
Note that depending on whether $v\not\equiv v_c$ or $v\equiv v_c$,
$\Xi_{\sigma}({\Bf r},{\Bf r}')$ and $\Pi_{\sigma}({\Bf r},{\Bf r}')$, 
and consequently $\ol{\Xi}_{\sigma}(k)$ and $\ol{\Pi}_{\sigma}(k)$,
are, as clearly specified in the text following Eq.~(\ref{e227}),
slightly different functions. From Eq.~(\ref{e234}) we observe that for 
$\varepsilon \to -\infty$ (`photo-emission'), 
${\rm Im}\ol{\Sigma}_{\sigma}(k;\varepsilon)$ corresponding to the 
cases (I) and (II) have identical functional forms and that, for 
$\varepsilon\to +\infty$ (`inverse-photo-emission'), 
${\rm Im}\ol{\Sigma}_{\sigma}(k;\varepsilon)$ corresponding to cases
(I) and (II) are fundamentally different especially when 
$n_{0;\bar\sigma}\not= n_{0;\sigma}$. For clarity, $n_{0;\bar\sigma}$
(note the $\bar\sigma$, the complement of $\sigma$) in the last term on 
the RHS of Eq.~(\ref{e234}), entry (II), has its origin in the combination 
of two contributions: one being $\hbar^{-1} {\sf M}^{\rm s}_{\infty_2}
({\Bf r}\| z)\,\delta({\Bf r}-{\Bf r}')$ (see Eqs.~(\ref{e213}), 
(\ref{ef145}) and (\ref{ef146})) which depends on $n$ (the {\sl total} 
number density), and the other $\wt{\Sigma}_{\sigma;\infty_2}^{\rm s_b}
({\Bf r},{\Bf r}';z)$ (see Eqs.~(\ref{e212}), (\ref{eh2}) and (\ref{eh21})) 
which depends on $n_{\sigma}$. We note in passing that the requirement 
for the stability of the GS (see \S~III.D) implies that $\ol{\Xi}_{\sigma}
(k) \ge 0$, $\forall\, k,\sigma$. 
 
From Eq.~(\ref{e55}), making use of 
\begin{equation}
\label{e235}
\wt{\ol{G}}_{0;\sigma}(k;z)
= \frac{\hbar}{z - {\bar h}_{0;\sigma}(k) },
\end{equation} 
with ${\bar h}_{0;\sigma}(k)$ the Fourier transform of $h_{0;\sigma}
({\Bf r})$ as defined in Eq.~(\ref{e55}) above, one immediately obtains
\begin{eqnarray}
\label{e236}
\wt{\ol{G}}_{\sigma}(k;z)
= \frac{\hbar}{z - [ {\bar h}_0(k) + \hbar 
\wt{\ol{\Sigma}}_{\sigma}(k;z) ]},
\end{eqnarray}
where ${\bar h}_0(k)$ denotes the Fourier transform of $h_0({\Bf r})$ 
as defined in Eq.~(\ref{e43}) above. From Eq.~(\ref{e236}) and the 
definition in Eq.~(\ref{e39}), we have
\begin{eqnarray}
\label{e237}
&&\ol{A}_{\sigma}(k;\varepsilon)
= \frac{\hbar}{\pi}\,\nonumber\\
&& \times \frac{\hbar \vert 
{\rm Im}\ol{\Sigma}_{\sigma}(k;\varepsilon)\vert }
{\big( \varepsilon - [{\bar h}_0(k) + \hbar 
{\rm Re}\ol{\Sigma}_{\sigma}(k;\varepsilon)]\big)^2 + 
\big( \hbar {\rm Im}\ol{\Sigma}_{\sigma}(k;\varepsilon) \big)^2}.
\end{eqnarray}
In arriving at this result, we have made use of the fact that
\begin{equation}
\label{e238}
\wt{\ol{\Sigma}}_{\sigma}(k;\varepsilon\pm i\eta) 
= {\rm Re} \ol{\Sigma}_{\sigma}(k;\varepsilon) 
\mp i \vert {\rm Im}\ol{\Sigma}_{\sigma}(k;\varepsilon)\vert,\;\;
\eta \downarrow 0,
\end{equation}
where $\ol{\Sigma}_{\sigma}(k;\varepsilon)$ stands for the 
{\sl physical} SE (see Eq.~(\ref{e65}) above). Making use of the 
results in Eqs.~(\ref{e233}) and (\ref{e234}), from Eq.~(\ref{e237}) 
for $\vert\varepsilon\vert\to\infty$ we immediately deduce that
\begin{eqnarray}
\label{e239}
&&\ol{A}_{\sigma}(k;\varepsilon) \sim
\frac{\hbar^2}{\pi^2}\,
\frac{ {\rm sgn}(\varepsilon-\mu)}{\varepsilon^3}\nonumber\\
&&\;
\times\left\{ \begin{array}{ll}
\displaystyle \ol{\Xi}_{\sigma}(k)
+ \frac{2 \big({\bar h}_0(k) +\hbar 
\ol{\Sigma}_{\sigma;\infty_0}(k) \big)
\,\ol{\Xi}_{\sigma}(k)
+\ol{\Pi}_{\sigma}(k)}{\varepsilon}, \!\!\!\!\!\!\! \\  
&\mbox{\rm (I)}\nonumber\\ \nonumber\\
\displaystyle \ol{\Xi}_{\sigma}(k) - 
4\pi^2 \big(\frac{e^2}{4\pi\epsilon_0}\big)^2\,
\frac{(n_{0;\bar\sigma} - n_{0;\sigma})}{\sqrt{2 m_e}} 
\frac{\Theta(\varepsilon-\mu)}{\varepsilon^{1/2}}\\ \\
\;\;
+ \big[ 2 \big({\bar h}_0(k) +\hbar 
\ol{\Sigma}_{\sigma;\infty_0}(k) \big)
\,\ol{\Xi}_{\sigma}(k) +
\ol{\Pi}_{\sigma}(k) \\ \\
\;\;\;\;\;
-4\pi^3 \hbar^{-1} (e^2/[4\pi\epsilon_0])^3\, 
n_{0;\bar\sigma}\,
\Theta(\varepsilon-\mu)\big]/\varepsilon.&
\mbox{\rm (II)}
\end{array} \right.\\
\end{eqnarray}
By Fourier transforming both sides of Eq.~(\ref{e239}) back to the 
$({\Bf r},{\Bf r}')$-space, one arrives at the result that, for the 
uniform isotropic system under consideration, the large-$\vert
\varepsilon\vert$ asymptotic behaviour of $A_{\sigma} ({\Bf r},
{\Bf r}';\varepsilon)$ (more precisely, $A_{\sigma}^{\rm h}(\|{\Bf r}
-{\Bf r}'\|;\varepsilon)$; see Eq.~(\ref{ef9})) is similar to that 
of $\ol{A}_{\sigma}(k;\varepsilon)$ in Eq.~(\ref{e239}). The results
in Eq.~(\ref{e239}) are specifically significant in revealing that
for the uniform and isotropic system under consideration, the energy 
moments of $\ol{A}_{\sigma}(k;\varepsilon)$, or of $A_{\sigma}({\Bf r},
{\Bf r}';\varepsilon)$, are {\sl not} bounded for $m \ge 3$ (see 
Eqs.~(\ref{e37}) and (\ref{e38}) above as well as footnote \ref{f6}). 
It is important to point out that, following the specification in 
Eq.~(\ref{e37}), $\ol{G}_{\sigma;\infty_m}(k)$ is bounded for $m=3$;
whereas the leading term in the AS for the integrand of the 
$\varepsilon$ integral on the RHS of Eq.~(\ref{e38}) decays like 
$1/\varepsilon$ for large values of $\vert\varepsilon\vert$, the 
integral over $(-E,E)$ has a well-defined limit for $E\to\infty$, 
following the fact that $1/\varepsilon$ is an {\sl odd} function of 
$\varepsilon$. 

The expression in Eq.~(\ref{e239}) in particular exposes some significant 
aspects of the (angle-resolved) photo-emission and inverse-photo-emission 
data and the way in which the latter are considerably affected through an 
imbalance between $n_{0;\bar\sigma}$ and $n_{0;\sigma}$. We should 
emphasize that the result in Eq.~(\ref{e239}) is {\sl strictly} only 
valid for uniform and isotropic systems and that, for other systems, the 
behaviour of the single-particle spectral function is {\sl not} as 
simple and transparent as that in Eq.~(\ref{e239}). We postpone 
considerations concerning $A_{\sigma}({\Bf r},{\Bf r}';\varepsilon)$ 
and $\ol{\ol{A}}_{\sigma}({\Bf k},{\Bf k}';\varepsilon)$ as appropriate 
to inhomogeneous systems to a future publication. Suffice it for the 
moment to mention that even a small amount of inhomogeneity is capable 
of bringing about {\sl considerable} change in the behaviour of the 
single-particle spectral function in comparison with that in the 
expression in Eq.~(\ref{e239}).

Finally, we point out that the dependence on $\varepsilon$ of 
$\ol{A}_{\sigma}(k;\varepsilon)$ pertaining to uniform and 
isotropic systems of Coulomb-interacting fermions confined to 
$d$-dimensional space with $d\not=3$, in particular with $d=2$ 
(B. Farid, 2001, unpublished) (see footnote \ref{f12}), is so 
distinctly different from that of $\ol{A}_{\sigma}(k;\varepsilon)$ 
in Eq.~(\ref{e239}) (corresponding to $d=3$) that it suggests the 
feasibility of an experimental determination of the (effective) 
dimension of the space to which fermions are confined, through 
investigating the $\varepsilon$ dependence of the corresponding 
single-particle spectral functions.

\section{Conventional perturbation theory and the case of 
dynamically screened exchange self-energy operator
$\Sigma_{\sigma}^{\prime (1)}(\lowercase{\Bf r},\lowercase{\Bf r}';
\varepsilon)$ }
\label{s33}

The dynamically screened exchange SE operator is the first-order
contribution to (see Eq.~(\ref{e62}) above)
\begin{eqnarray}
\label{e240}
&&\Sigma_{\sigma}'({\Bf r},{\Bf r}';\varepsilon) 
{:=} \Sigma_{\sigma}({\Bf r},{\Bf r}';\varepsilon)
-\frac{1}{\hbar} v_H({\Bf r};[n])\,\delta({\Bf r}-{\Bf r}') \nonumber\\
&&\;\;
\equiv \Sigma_{\sigma}^{\sh}({\Bf r},{\Bf r}';\varepsilon) 
+\frac{1}{\hbar} \big(w_{\sigma}({\Bf r}) - v_H({\Bf r};[n]) \big) 
\,\delta({\Bf r}-{\Bf r}')
\end{eqnarray}
in the perturbation series expansion of this function (see \S~III.C) in 
terms of the dynamically-screened particle-particle interaction function 
$W({\Bf r},{\Bf r}';\varepsilon)$ (Hubbard 1957) which in contrast 
with the bare particle-particle interaction $v({\Bf r}-{\Bf r}')$ is 
a function of both ${\Bf r}$ and ${\Bf r}'$, rather than merely 
$\|{\Bf r}-{\Bf r}'\|$ (leaving aside the case of uniform and isotropic 
systems), and a function of the external energy parameter $\varepsilon$; 
in Eq.~(\ref{e240}), $n$ in the argument of $v_H({\Bf r};[n])$ (see 
Eq.~(\ref{e14}) above) is the total GS number density pertaining to 
the interacting system. 

It is well known, in particular for systems with metallic GSs (for
example Mattuck (1992, \S~10.4)), that the perturbation series for 
$\Sigma_{\sigma}(\varepsilon)$ in terms of $v$ involves unbounded 
contributions when (not exclusively) $v\equiv v_c$ in $d=3$, associated 
with integrals that owing to the long range of the Coulomb potential are 
infrared divergent. These unbounded contributions can be traced back 
as arising from the polarization insertions that collectively constitute 
the random-phase approximation (RPA) to the polarization function; the 
{\sl total} contribution of the RPA series to the polarization function 
modifies the behaviour of the corresponding $W({\Bf r},{\Bf r}';
\varepsilon=0)$ for $\|{\Bf r}-{\Bf r}'\|\to \infty$, in comparison 
with that of $v_c({\Bf r}-{\Bf r}')$, in such a way 
\footnote{\label{f97}
Previously (Farid 1999a, footnote 48) we have shown that for a uniform 
and isotropic system of spin-$1/2$ fermions in the paramagnetic state, 
for $\|{\Bf r}\|\to \infty$ one has $W^{\sc rpa}(\|{\Bf r}\|;0) 
\sim A \cos(2 k_F \|{\Bf r}\|)/\|{\Bf r}\|^3$, where $A$ stands for 
a constant and $k_F$ for the Fermi wavenumber. } 
that substitution of $W^{\sc rpa}(\varepsilon)$ (or in a further 
approximation, of $W^{\sc rpa}(\varepsilon=0)$) for $W(\varepsilon)$ 
in the perturbation series for $\Sigma_{\sigma}(\varepsilon)$ in 
terms of this function, results in a series that is free from the 
aforementioned unbounded contributions. The conventional wisdom 
concerning this subject (for example Mattuck (1992, \S~10.4)) is that 
for non-metallic systems (i.e. those whose low-lying single-particle 
excitation spectrum is fully gapped) the above-mentioned unbounded 
contributions do {\sl not} occur, so that for these systems the 
many-body perturbation series for $\Sigma_{\sigma}(\varepsilon)$ in 
terms of $v_c$ does {\sl not} contain unbounded contributions. Our 
considerations in \S~III.H.2 have unequivocally shown that this 
viewpoint is {\sl not} correct (see in particular the last two 
paragraphs in \S~III.H.2).
\footnote{\label{f98}
Note that ${\cal M}({\Bf r})$ is a third-order contribution, whereas 
${\cal T}_{\sigma,\bar\sigma}({\Bf r})$ is a second-order one; this is 
particularly evident from the expression for ${\sf M}^{\rm s}_{\infty_2}
({\Bf r}\| z)$ in Eq.~(\protect\ref{ef145}) where one observes 
$(e^2/[4\pi\epsilon_0])^3$ on the RHS and from the expression for 
$\wt{\sf T}^{\rm s}_{\sigma,\bar\sigma;\infty_2}({\Bf r}\| z)$ in 
Eq.~(\protect\ref{eg15}) where one observes $(e^2/[4\pi\epsilon_0])^2$ 
on the RHS. It is interesting to note that whereas the analysis as 
presented in Mattuck (1992) shows breakdown of perturbation theory in 
terms of $v_c$ at the second order for metallic GSs, the second-order
contribution ${\cal T}_{\sigma,\bar\sigma}({\Bf r})$ is identically
vanishing for $n_{\sigma}({\Bf r}) \equiv n_{\bar\sigma}({\Bf r})$
(see Eq.~(\protect\ref{e129}), from which it is evident that, for 
$n_{0;\sigma}=n_{0;\bar\sigma}$, the first transcendental function of 
$r_s$, namely $\ln(-{\bar z}/r_s^3)$, is pre-multiplied by $r_s^3$). 
This is a remarkable result in that the counterpart of 
${\cal T}_{\sigma,\bar\sigma}({\Bf r})$ within the framework of the 
first-order perturbation theory in terms of the dynamically screened 
interaction $W(\varepsilon)$, namely ${\cal T}^{(1)}({\Bf r})$ 
introduced in Eq.~(\protect\ref{e278}) below (an unbounded function in the 
case of $v\equiv v_c$ and $d=3$), is {\sl never} vanishing. For clarity, 
the stark contrast between ${\cal T}_{\sigma,\bar\sigma}({\Bf r})$ and 
${\cal T}^{(1)}({\Bf r})$ has its origin in the {\sl incorrect} 
description of the non-local part of $\Sigma_{\sigma;\infty_p}({\Bf r},
{\Bf r}')$ by $\Sigma_{\sigma;\infty_p}^{(1)}({\Bf r},{\Bf r}')$ for 
$p\ge 1$ (see the last two paragraphs in \S~III.H.2); for a correct 
reproduction of ${\cal T}_{\sigma,\bar\sigma}({\Bf r})$ it is necessary 
also to consider the second-order skeleton SE diagrams in terms of $v_c$ 
and the exact single-particle GF. }
This observation makes evident that, when choosing to employ a many-body 
perturbation series for $\Sigma_{\sigma}(\varepsilon)$, the series 
must be, in the cases corresponding to $v\equiv v_c$ in $d=3$ (not 
exclusively), in terms of $W(\varepsilon)$ rather than $v_c$, independent 
of whether the GS of the system is metallic or otherwise.

Below we consider the expression for the first-order contribution to
the perturbation series of $\Sigma_{\sigma}'(\varepsilon)$ in terms 
of $W(\varepsilon)$, as deduced from that of $\Sigma_{\sigma}^{\sh}
(\varepsilon)$, and assume both $G_{\sigma}(\varepsilon)$ and 
$W(\varepsilon)$ herein to coincide with those pertaining to the 
interacting system; we denote this first-order term, or the 
`screened-exchange' SE (also called the $GW$ SE), by 
$\Sigma_{\sigma}^{\prime (1)}(\varepsilon)$. At places we also comment 
on the consequences of replacing $G_{\sigma}(\varepsilon)$ by 
$G_{0;\sigma}(\varepsilon)$ (pertaining to the `non-interacting'
system corresponding to $\wh{H}_0$ in Eq.~(\ref{e54})) and 
$W(\varepsilon)$ by $W^{\sc rpa}(\varepsilon)$; the latter function 
can in turn be one evaluated in terms of either $G_{\sigma}(\varepsilon)$ 
or $G_{0;\sigma}(\varepsilon)$.

For the first-order, or the screened-exchange, SE we have 
\begin{eqnarray}
\label{e241}
\Sigma_{\sigma}^{\prime (1)}({\Bf r},{\Bf r}';\varepsilon)
= \frac{i}{\hbar} \int_{-\infty}^{\infty}
\frac{{\rm d}\varepsilon'}{2\pi\hbar}\;
G_{\sigma}({\Bf r},{\Bf r}';\varepsilon-\varepsilon')\nonumber\\
\times W({\Bf r},{\Bf r}';\varepsilon')\,
\exp(-i\varepsilon' \eta/\hbar),\;\; \eta\downarrow 0.
\end{eqnarray}
Later in this Section, we denote the SE $\Sigma_{\sigma}({\Bf r},
{\Bf r}';\varepsilon)$ associated through Eq.~(\ref{e240}) with 
$\Sigma_{\sigma}^{\prime (1)}({\Bf r},{\Bf r}';\varepsilon)$, by
$\Sigma_{\sigma}^{(1)}({\Bf r},{\Bf r}';\varepsilon)$. The screened 
interaction function can be written as
\begin{equation}
\label{e242}
W({\Bf r},{\Bf r}';\varepsilon) = v({\Bf r}-{\Bf r}')
+ W'({\Bf r},{\Bf r}';\varepsilon),
\end{equation}
where $W'(\varepsilon)$ has the following Lehmann-type 
spectral representation 
\begin{eqnarray}
\label{e243}
&&W'({\Bf r},{\Bf r}';\varepsilon) =
\sum_{s,\sigma}\, w_{s;\sigma}({\Bf r}) 
w_{s;\sigma}^*({\Bf r}')\nonumber\\
&&\;\;\;\;\;\;\;\;\;\;\;\;\;\;\;\;\;
\times\Big\{ \frac{1}{\varepsilon - e_{s} + i\eta}
-\frac{1}{\varepsilon + e_{s} - i\eta} \Big\},\;\;\;
\eta \downarrow 0,
\end{eqnarray} 
where 
\begin{eqnarray}
\label{e244}
w_{s;\sigma}({\Bf r}) {:=} \int {\rm d}^dr'\;
v({\Bf r}-{\Bf r}')\, \rho_{s;\sigma}({\Bf r}'),
\end{eqnarray}
with
\begin{eqnarray}
\label{e245}
\rho_{s;\sigma}({\Bf r}) {:=} 
\langle\Psi_{N;0}\vert \hat\psi_{\sigma}^{\dag}({\Bf r})
\hat\psi_{\sigma}({\Bf r}) - n_{\sigma}({\Bf r}) 
\vert \Psi_{N;s}\rangle,
\end{eqnarray}
so that $\rho_{0;\sigma}({\Bf r}) \equiv 0$, and
\begin{equation}
\label{e246}
e_{s} {:=} E_{N,s} - E_{N,0} \geq 0,
\end{equation}
which are the energies of the {\sl neutral} excitations of the 
$N$-particle system; the assumed non-degeneracy of the GS implies that 
$e_s > 0$ for {\sl all} $s\not=0$; as should be evident from $E_{N;0}$, 
here $s=0$ symbolically denotes the $N$-particle GS of $\wh{H}$. We point 
out that the compound variable $s$, $s\not=0$, in the above expressions 
characterizes the excited $(N_{\sigma}+N_{\bar\sigma})$-particle 
eigenstates of $\wh{H}$, so that for {\sl interacting} systems it is 
a fundamentally different compound variable from that encountered in 
the Lehmann representation for $G_{\sigma}(\varepsilon)$ in 
Eq.~(\ref{e17}) (and as subscript of the associated functions), where 
$s$ characterizes the ground and excited $(N_{\sigma} \pm 1
+N_{\bar\sigma})$-particle eigenstates of $\wh{H}$. When the two types 
of variable are encountered in the same expression (as in 
Eq.~(\ref{e251}) below), each is uniquely identified through its 
association with functions that are known to have their origins in 
$G_{\sigma}(\varepsilon)$ (such as $\{f_{s;\sigma}({\Bf r})\}$) or 
$W(\varepsilon)$ (such as $\{ w_{s;\sigma}({\Bf r}) \}$). 

Further, the apparent independence of $\{e_s\}$ on $\sigma$ originates 
from the fact that $\big\{\hat\psi_{\sigma}^{\dag}({\Bf r}) 
\hat\psi_{\sigma}({\Bf r}) \big\}$ commutes with $\big\{ 
\wh{N}_{\sigma}\big\}$, implying that $\hat\psi_{\sigma}^{\dag}({\Bf r}) 
\hat\psi_{\sigma}({\Bf r})$, $\forall\sigma$, has {\sl vanishing} 
amplitude with respect to the ground and {\sl any} excited $N$-particle 
state of $\wh{H}$ whose $N_{\sigma}$, satisfying $N_{\sigma}
+N_{\bar\sigma}=N$, differs from that pertaining to the GS. 
\footnote{\label{f99}
The following two points are clarifying. First, the fact that to 
{\sl all} orders of perturbation theory we merely encounter
$W'(\varepsilon') \equiv \sum_{\sigma'} W'_{\sigma',\sigma'}
(\varepsilon')$ (see Eqs.~(\protect\ref{e241}) and (\protect\ref{e242})) 
in the expression for $\Sigma_{\sigma}(\varepsilon)$, rather than 
$W'_{\sigma',\sigma''}(\varepsilon')$ followed by appropriate trace 
operations involving other functions of $\sigma'$ and $\sigma''$, is to 
do with the assumed spin {\sl independence} of $v$ (for completeness, 
$W'_{\sigma',\sigma''}({\Bf r},{\Bf r}';\varepsilon)$ is defined in 
terms of $[\hat\psi_{\sigma'}^{\dag}({\Bf r}) \hat\psi_{\sigma'}({\Bf r}) 
- n_{\sigma'}({\Bf r})]$ and $[\hat\psi_{\sigma''}^{\dag}({\Bf r}') 
\hat\psi_{\sigma''}({\Bf r}') - n_{\sigma''}({\Bf r}')]$). Second, 
although $e_s$ is independent of $\sigma$, its actual participation in 
the expression for $W'_{\sigma,\sigma}(\varepsilon)$ is determined by 
$w_{s;\sigma}$, or $\rho_{s;\sigma}$ (see Eq.~(\protect\ref{e245})); 
thus although $s$ is {\sl independent} of $\sigma$, through the 
dependence on both $s$ {\sl and} $\sigma$ of $w_{s;\sigma}$, $s$ to 
the right of $\sum_s$ on the RHS of Eq.~(\protect\ref{e243}) can be 
considered as an {\sl implicit} function of $\sigma$ (one can of course 
equally consider $\sigma$ in Eq.~(\protect\ref{e243}) to be an 
{\sl implicit} function of $s$). This aspect is made evident through 
replacing the interacting states on the RHS of Eq.~(\protect\ref{e245}) 
by their `non-interacting' counterparts (i.e. SSDs), whereon 
$W'({\Bf r},{\Bf r}';\varepsilon)$ in Eq.~(\protect\ref{e243}) 
transforms into the leading-order term in the RPA series for this 
function; comparing this with the well-known RPA result for 
`non-interacting' fermions, for these systems, one directly establishes 
the mentioned {\sl implicit} dependence on $\sigma$ of $s$. } 
This may be compared with the dependence upon $\sigma$ of the 
single-particle excitation energies $\{\varepsilon_{s;\sigma}\}$, 
defined in Eq.~(\ref{e19}), which has its root in the fact that 
$\big\{\hat\psi_{\sigma}({\Bf r})\big\}$ and 
$\big\{\hat\psi_{\sigma}^{\dag}({\Bf r})\big\}$ do {\sl not} commute 
with $\big\{\wh{N}_{\sigma}\big\}$. With reference to our considerations 
in \S~III.F (see in particular footnote \ref{f87}), we mention that, 
for exactly the same reasons that $\varepsilon=\mu_{N;\sigma}^{\mp}$ 
are accumulation points of the set of single-particle excitations 
energies, $\varepsilon=\lambda_N$, with $\lambda_N {:=} \min\{e_s \|\, 
s\not=0\}$, is an accumulation point of the set of neutral excitations 
energies of the $N$-particle system. Accordingly, $z=\pm \lambda_N$ 
are {\sl non}-isolated singularities of $\wt{W}(z)$, analogous to 
$z=\mu_{N;\sigma}^{\mp}$ which are {\sl non}-isolated singularities 
of $\wt{G}_{\sigma}(z)$ and $\wt{\Sigma}_{\sigma}(z)$. Further, similar 
to $[\mu_{N;\sigma}^{-}, \mu_{N;\sigma}^{+}]$ (see Eq.~(\ref{e22})), 
$[-\lambda_N,\lambda_N]$ can be infinitesimally small for systems in 
the thermodynamic limit, but it is never vanishing (see Eq.~(\ref{e246}) 
above and the subsequent text).

For our later considerations, we introduce the analytic continuation
into the physical Riemann sheet of $W'({\Bf r},{\Bf r}';\varepsilon)$,
namely ({\it cf}. Eq.~(\ref{e24}) above and see footnote \ref{f11})
\begin{equation}
\label{e247}
\wt{W}'({\Bf r},{\Bf r}';z) = 2\sum_{s,\sigma}\, e_s\,
\frac{w_{s;\sigma}({\Bf r}) 
w_{s;\sigma}^*({\Bf r}')}{z^2 - e_{s}^2},
\end{equation} 
which is directly deduced from the representation in Eq.~(\ref{e243}). 
The `physical' $W'(\varepsilon)$ is obtained from $\wt{W}'(z)$ according 
to ({\it cf}. Eq.~(\ref{e25}) above)
\begin{equation}
\label{e248}
W'(\varepsilon) = \lim_{\eta\downarrow 0} 
\wt{W}'(\varepsilon\pm i\eta),\;\;\;
\varepsilon\, \IEq<> \, 0.
\end{equation} 

From the expression in Eq.~(\ref{e247}) we obtain the 
following series:
\begin{equation}
\label{e249}
\wt{W}'({\Bf r},{\Bf r}';z) = \sum_{m=1}^{\infty}
\frac{W'_{\infty_{2m}}({\Bf r},{\Bf r}')}{z^{2m}},
\end{equation}
where
\begin{equation}
\label{e250}
W'_{\infty_{m}}({\Bf r},{\Bf r}')
{:=} 2 \sum_{s,\sigma} e_{s}^{m-1}
w_{s;\sigma}({\Bf r}) w_{s;\sigma}^*({\Bf r}').
\end{equation}
The series in Eq.~(\ref{e249}) is in all essential respects similar to 
that for $\wt{G}_{\sigma}({\Bf r},{\Bf r}';z)$ presented in Eq.~(\ref{e27}). 
Thus, for instance, by truncating this series at some finite order, one 
obtains a {\sl formal} finite-order AS (see \S~II.B) for $\wt{W}'({\Bf r},
{\Bf r}';z)$ corresponding to $\vert z\vert\to\infty$. Our explicit 
calculations of $W'_{\infty_{m}}({\Bf r},{\Bf r}')$ for $m=1, 2, 3$ 
(see Eqs.~(\ref{e256}), (\ref{e264}) and (\ref{e273}) respectively; 
note that, of these three functions, only the one corresponding to 
$m=2$ is relevant to the large-$\vert z\vert$ AS for $\wt{W}'({\Bf r},
{\Bf r}';z)$) show that in the cases corresponding $v\equiv v_c$ in 
$d=3$, these functions are {\sl bounded} for ${\Bf r} \not={\Bf r}'$;
however, already for $m=2$, $W'_{\infty_{m}}({\Bf r},{\Bf r}')$ diverges 
as $\|{\Bf r}-{\Bf r}'\|\to 0$. Since, in the expression for 
$\Sigma_{\sigma;\infty_2}^{(1)}({\Bf r},{\Bf r}')$ to be presented in 
Eq.~(\ref{e253}) below, $W'_{\infty_{2}}({\Bf r},{\Bf r}')$ is 
multiplied by $\delta({\Bf r}-{\Bf r}')$, the mentioned divergence of 
$W'_{\infty_{2}}({\Bf r},{\Bf r}')$ for ${\Bf r}={\Bf r}'$ implies 
that, similar to the $\Sigma_{\sigma;\infty_2}({\Bf r},{\Bf r}')$, 
$\Sigma_{\sigma;\infty_2}^{(1)}({\Bf r},{\Bf r}')$ is fundamentally 
unbounded in the case of $v\equiv v_c$ in $d=3$. With reference to 
the discussions in \S\S~II.B and III.H, it follows that, in the correct 
finite-order AS of $\wt{\Sigma}_{\sigma}^{(1)}({\Bf r},{\Bf r}';z)$ 
for $\vert z\vert\to\infty$, the term decaying like $1/\vert z\vert$ 
is {\sl not} directly followed by one decaying like $1/\vert z^2\vert$, 
rather by one or several asymptotically more dominant terms (see
specifically Eqs.~(\ref{e277}) and (\ref{e278}) below and the related 
discussions). In this work we do {\sl not} discuss the details of 
these terms in much depth, in particular we do {\sl not} give special 
attention to the case corresponding to $v\equiv v_c$ in $d=3$.
 
Before proceeding, we note in passing that, in contrast with
$\wt{G}_{\sigma}({\Bf r},{\Bf r}';z)$, which, according to the 
expression presented in Eq.~(\ref{e27}), is determined by {\sl all} 
$G_{\sigma;\infty_m}({\Bf r},{\Bf r}')$, $\wt{W}'({\Bf r},{\Bf r}';z)$, 
according to the expression in Eq.~(\ref{e249}), is determined by a 
{\sl proper} subset of all $W'_{\infty_m}({\Bf r},{\Bf r}')$. 
Interestingly, as we shall see later (see Eqs.~(\ref{e253}) -
(\ref{e255}) below), the {\sl entire} set $\big\{W'_{\infty_m}
({\Bf r},{\Bf r}')\big\}$ contributes to the large-$\vert z\vert$ 
AS for $\wt{\Sigma}_{\sigma}^{(1)}({\Bf r},{\Bf r}';z)$. We shall 
clarify the mechanism underlying this counter-intuitive and interesting 
phenomenon later in this Section.

From the Lehmann representation for $G_{\sigma}({\Bf r},{\Bf r}';
\varepsilon)$ in Eq.~(\ref{e17}), making use of the residue theorem 
and subsequently analytically continuing the resulting expression 
for (see Eq.~(\ref{e240}) and compare with Eq.~(\ref{e62}) above)
\begin{eqnarray}
\Sigma_{\sigma}^{(1)}({\Bf r},{\Bf r}';\varepsilon) \equiv 
\Sigma_{\sigma}^{\sh (1)}({\Bf r},{\Bf r}';\varepsilon)
+\frac{1}{\hbar} w_{\sigma}({\Bf r})\, 
\delta({\Bf r}-{\Bf r}') \nonumber
\end{eqnarray}
into the physical Riemann sheet, from the defining expression in 
Eq.~(\ref{e241}) and from Eqs.~(\ref{e240}) and (\ref{e173}) 
we obtain
\begin{eqnarray}
\label{e251}
&&\wt{\Sigma}_{\sigma}^{(1)}({\Bf r},{\Bf r}';z)
= \Sigma^{\sc hf}({\Bf r},{\Bf r}';[\varrho_{\sigma}]) 
\nonumber\\
&&\;\;\;
+\frac{1}{\hbar} \sum_{s,s'} \frac{f_{s;\sigma}({\Bf r}) 
f_{s;\sigma}^*({\Bf r}')\, \sum_{\sigma'} w_{s';\sigma'}({\Bf r}) 
w_{s';\sigma'}^*({\Bf r}')}{z - \varepsilon_{s;\sigma}
- {\rm sgn}(\varepsilon_{s;\sigma}-\mu)\, e_{s'}},
\end{eqnarray}
from which $\Sigma_{\sigma}^{(1)}(\varepsilon)$ follows according 
to $\Sigma_{\sigma}^{(1)}(\varepsilon) = \lim_{\eta\downarrow 0}
\wt{\Sigma}_{\sigma}^{(1)}(\varepsilon \pm i\eta)$, $\varepsilon 
\Ieq<> \mu$ ({\it cf}. Eq.~(\ref{e65})). From the expression in 
Eq.~(\ref{e251}), one straightforwardly infers 
\begin{eqnarray}
\label{e252}
&&\wt{\Sigma}_{\sigma}^{(1)}({\Bf r},{\Bf r}';z) \sim
\Sigma^{\sc hf}({\Bf r},{\Bf r}';[\varrho_{\sigma}])
+ \frac{\Sigma_{\sigma;\infty_1}^{(1)}({\Bf r},{\Bf r}')}{z}
\nonumber\\
&&\;\;\;\;\;\;\;\;\;\;\;\;
+ \frac{\Sigma_{\sigma;\infty_2}^{(1)}({\Bf r},{\Bf r}')}{z^2}
+\frac{\Sigma_{\sigma;\infty_3}^{(1)}({\Bf r},{\Bf r}')}{z^3},
\;\;\vert z\vert\to\infty.
\end{eqnarray}
Here we deal with one more asymptotic term than in the case of the 
exact $\wt{\Sigma}_{\sigma}({\Bf r},{\Bf r}';z)$ (see Eq.~(\ref{e72})
above); this will be helpful in identifying some characteristic features 
of $\wt{\Sigma}_{\sigma}({\Bf r},{\Bf r}';z)$ which this function has 
in common with $\wt{\Sigma}_{\sigma}^{(1)}({\Bf r},{\Bf r}';z)$. A 
comparison of the RHS of Eq.~(\ref{e252}) with that of Eq.~(\ref{e72}) 
(in combination with Eq.~(\ref{e62})) reveals that the leading term 
in the AS of $\Sigma_{\sigma}^{(1)}(\varepsilon)$ for $\vert\varepsilon
\vert \to\infty$ coincides with that of the {\sl exact} $\Sigma_{\sigma}
(\varepsilon)$. 

Employing the binomial series expansions of $\big(\varepsilon_{s;
\sigma} + {\rm sgn}(\varepsilon_{s;\sigma}-\mu) e_{s'}\big)^m$, 
$m=0,1,2,\dots$, in the RHS of Eq.~(\ref{e251}), after some algebraic 
manipulations, for the functions introduced in Eq.~(\ref{e252}) 
we obtain
\begin{eqnarray}
\label{e253}
\Sigma_{\sigma;\infty_1}^{(1)}({\Bf r},{\Bf r}') 
&&\equiv \frac{1}{2 \hbar^2}
G_{\sigma;\infty_1}({\Bf r},{\Bf r}') 
W'_{\infty_1}({\Bf r},{\Bf r}')
\nonumber\\
&&\equiv \frac{1}{\hbar} 
\big\{\sum_{s',\sigma'} \vert w_{s';\sigma'}({\Bf r})
\vert^2\big\}\, \delta({\Bf r}-{\Bf r}'),
\end{eqnarray}
\begin{eqnarray}
\label{e254}
\Sigma_{\sigma;\infty_2}^{(1)}({\Bf r},{\Bf r}')
&\equiv& \frac{1}{2 \hbar^2}
G_{\sigma;\infty_2}({\Bf r},{\Bf r}')
W'_{\infty_1}({\Bf r},{\Bf r}') \nonumber\\
&+&\frac{1}{2 \hbar^2} G_{\sigma;\infty_1}({\Bf r},{\Bf r}')
W'_{\infty_2}({\Bf r},{\Bf r}') \nonumber\\
&-&\frac{1}{\hbar} \varrho_{\sigma}({\Bf r}',{\Bf r})
W'_{\infty_2}({\Bf r},{\Bf r}'),
\end{eqnarray}
\begin{eqnarray}
\label{e255}
\Sigma_{\sigma;\infty_3}^{(1)}({\Bf r},{\Bf r}')
&\equiv& \frac{1}{2 \hbar^2}
G_{\sigma;\infty_3}({\Bf r},{\Bf r}')
W'_{\infty_1}({\Bf r},{\Bf r}')\nonumber\\
&+&\frac{1}{\hbar^2} G_{\sigma;\infty_2}({\Bf r},{\Bf r}')
W'_{\infty_2}({\Bf r},{\Bf r}')\nonumber\\
&+&\frac{1}{2 \hbar^2} G_{\sigma;\infty_1}({\Bf r},{\Bf r}')
W'_{\infty_3}({\Bf r},{\Bf r}') \nonumber\\
&-&\frac{2}{\hbar} {\cal D}_{\sigma}({\Bf r},{\Bf r}')
W'_{\infty_2}({\Bf r},{\Bf r}').
\end{eqnarray}
From these expressions, one readily observes that, in agreement with 
our earlier statement, whereas $\wt{W}'(z)$ is determined by 
$W'_{\infty_{m}}$, $m=2,4,\dots$, the terms in the large-$\vert z\vert$
AS for $\wt{\Sigma}_{\sigma}^{(1)}(z)$ are determined 
by $\wt{W}'_{\infty_{m}}(z)$, $m=1,2,\dots$. This interesting aspect
can be understood by realizing the fact that although the representation 
for $\wt{W}(z)$ in Eq.~(\ref{e249}) similar to that for $\wt{G}_{\sigma}
(z)$ in Eq.~(\ref{e27}) is {\sl exact}, {\sl no} finite-order truncation 
of the sum on the RHS of Eq.~(\ref{e249}) is capable of correctly 
reproducing $\wt{W}(z)$ in a neighbourhood of $z=0$; in other words, 
in evaluating the $\varepsilon'$ integral on the RHS of Eq.~(\ref{e241}), 
the summation over $s$ corresponding to $W({\Bf r},{\Bf r}';\varepsilon')$ 
(see Eqs.~(\ref{e242}) and (\ref{e249})) has to be carried out first 
(for otherwise one has to do with integrands that are non-integrably 
singular at $\varepsilon'=0$), thus resulting in an expression involving 
$W'_{\infty_{m}}$ for {\sl all} $m$. 

It can be shown that an approximation scheme that is capable of 
reproducing the exact $W'_{\infty_2}$ (see later) yields a $\wt{W}'(z)$ 
that satisfies the $f$-sum rule; however, unless such an approximation 
yield the exact $W'_{\infty_1}$, already $\Sigma_{\sigma;\infty_1}^{(1)}$ 
($\equiv \Sigma_{\sigma;\infty_1}^{\prime (1)}$) deviates from the 
expected $\Sigma_{\sigma;\infty_1}^{(1)}$ (Engel, {\sl et al.} 1991). 
As we demonstrate below, the latter function itself deviates from that 
according to the exact theory, that is $\Sigma_{\sigma;\infty_1}({\Bf r},
{\Bf r}')$ as presented in Eq.~(\ref{e185}) above. It follows that, in 
certain applications, an approximately-evaluated $\Sigma_{\sigma}^{(1)}
({\Bf r},{\Bf r}';\varepsilon)$ may yield more accurate results than a 
$\Sigma_{\sigma}^{(1)}({\Bf r},{\Bf r}';\varepsilon)$ that is evaluated 
more accurately. This observation is clearly borne out by the calculations 
reported by Farid (1997a).

Before proceeding with the evaluation of the contributions on the RHSs 
of Eqs.~(\ref{e253}) - (\ref{e255}) in terms of the GS correlation 
functions $\{ \Gamma^{(m)}\}$, we indicate three noteworthy observations 
that are directly made from the expressions in Eqs.~(\ref{e252}) and 
(\ref{e253}). 

--- Firstly, formally the leading term in the AS of 
$\wt{\Sigma}_{\sigma}^{(1)}(z)$, for $\vert z\vert\to\infty$, that 
is $\Sigma_{\sigma;\infty_0}^{(1)}$, coincides with that of the 
{\sl exact} $\wt{\Sigma}_{\sigma}(z)$ (see Eq.~(\ref{e173}) above); in 
practice, however, since the $\Sigma^{\sc hf}({\Bf r},{\Bf r}';
[\varrho_{\sigma}])$ on the RHS of Eq.~(\ref{e252}) is {\sl necessarily} 
evaluated in terms of a `non-interacting', or Slater-Fock, density matrix 
$\varrho_{{\rm s};\sigma}({\Bf r},{\Bf r}')$ (which, among others, in 
contrast with the interacting $\varrho_{\sigma}({\Bf r},{\Bf r}')$ 
{\sl is} idempotent; see Appendix C), this contribution deviates 
from the exact $\Sigma^{\sc hf}({\Bf r},{\Bf r}';[\varrho_{\sigma}])$. 

--- Secondly, the next-to-leading asymptotic term pertaining to 
$\wt{\Sigma}_{\sigma}^{(1)}({\Bf r},{\Bf r}';z)$, for $\vert z\vert 
\to\infty$, that is $\Sigma_{\sigma;\infty_1}^{(1)}$, is purely 
{\sl local}, in evident contrast with the exact result presented in 
Eq.~(\ref{e185}) above (see Eqs.~(\ref{e187}) and (\ref{e188})). 

--- Thirdly, the latter, purely local asymptotic contribution does 
{\sl not} explicitly depend on the spin index $\sigma$, which amounts 
to a manifest shortcoming (see specifically the discussions in \S~V).

Below we evaluate the expressions for the various functions that 
feature on the RHSs of Eqs.~(\ref{e253}) - (\ref{e255}), in terms 
of GS correlation functions $\{\Gamma^{(m)}\}$ (see Appendix B). 
In view of our extensive considerations in the earlier parts of this 
work (as well as in some related Appendices) in regard to {\sl exact} 
functions $\Sigma_{\sigma;\infty_m}({\Bf r},{\Bf r}')$, $m=0,1,2$, below 
we shall {\sl not} be exhaustive and thus mainly concentrate on some 
specific aspects of the indicated expressions.

\subsection{Evaluation of 
$\Sigma_{\sigma;\infty_1}^{(1)}({\Bf r},{\Bf r}')$ }
\label{s34}

From the definitions in Eqs.~(\ref{e250}), (\ref{e245}) and 
(\ref{e246}), making use of the completeness relation for $\big\{ 
\vert\Psi_{N;s}\rangle\big\}$ in the Hilbert space of $(N_{\sigma}
+N_{\bar\sigma})$-particle states, that is $I = \sum_s 
\vert\Psi_{N_{\sigma},N_{\bar\sigma};s}\rangle \langle\Psi_{N_{\sigma},
N_{\bar\sigma};s}\vert$, and subsequently the normal ordering of the 
involved field operators, we obtain
\begin{eqnarray}
\label{e256}
&&W'_{\infty_1}({\Bf r},{\Bf r}')
= 2 \int {\rm d}^dr''\; v({\Bf r}-{\Bf r}'') 
v({\Bf r}'-{\Bf r}'') n({\Bf r}'')\nonumber\\ 
&&\;\;\;\;
-2\sum_{\sigma} v_H({\Bf r};[n_{\sigma}]) 
v_H({\Bf r}';[n_{\sigma}]) + 2 {\cal C}({\Bf r},{\Bf r}'),
\end{eqnarray}
where ${\cal C}({\Bf r},{\Bf r}')$ is defined in Eq.~(\ref{eb30}) (see 
Appendix B). From Eqs.~(\ref{e253}) and (\ref{e256}) we therefore have
\begin{eqnarray}
\label{e257}
&&\Sigma_{\sigma;\infty_1}^{(1)}({\Bf r},{\Bf r}')
= \frac{1}{\hbar} \big[ {\cal C}({\Bf r},{\Bf r})
- \sum_{\sigma'} v_H^2({\Bf r};[n_{\sigma'}])
\nonumber\\
&&\;\;\;\;\;\;\;\;
+\int {\rm d}^dr''\; v^2({\Bf r}-{\Bf r}'') 
n({\Bf r}'') \big] \delta({\Bf r}-{\Bf r}'),
\end{eqnarray}
to be compared with $\Sigma_{\sigma;\infty_1}({\Bf r},{\Bf r}')$ in 
Eq.~(\ref{e185}) (see also Eqs.~(\ref{e187}) and (\ref{e188})). We 
point out that for systems in which $n_{\sigma}({\Bf r}) \equiv 
\frac{1}{2{\sf s}+1} n({\Bf r})$ holds for {\sl all} $\sigma$ 
(the `paramagnetic' state of systems of spin-${\sf s}$ fermions), 
invoking the linearity of the Hartree potential in regard to its 
dependence on the number density, we have
\begin{eqnarray}
&&\sum_{\sigma'} v_H({\Bf r};[n_{\sigma'}]) 
v_H({\Bf r}';[n_{\sigma'}])\nonumber\\
&&\;\;\;\;\;\;\;\;\;\;\;\;\;\;\;\;\;\;\;\;\;\;\;\; 
=\frac{1}{2 {\sf s}+1}\, v_H({\Bf r};[n]) v_H({\Bf r}';[n]).\nonumber
\end{eqnarray}
This clearly brings out the most significant {\sl quantitative} 
difference between $v_H({\Bf r};[n]) v_H({\Bf r};[n])$ on the RHS of 
Eq.~(\ref{e185}) and $\sum_{\sigma'} v_H({\Bf r};[n_{\sigma'}]) 
v_H({\Bf r}';[n_{\sigma'}])$ on the RHS of Eq.~(\ref{e257}).

Within the framework in which the $N$-particle GS wavefunction is 
approximated by a SSD (see Appendix C), one has (below, as elsewhere 
in this work, we indicate application of this approximation by either 
`$\vert_{\rm s}$' or a subscript `s', to be distinguished from `$s$') 
\begin{eqnarray}
\label{e258}
&&\left. \Sigma_{\sigma;\infty_1}^{(1)}
({\Bf r},{\Bf r}')\right|_{\rm s}
= {\cal A}_{\rm s}'({\Bf r},{\Bf r})\, 
\delta({\Bf r}-{\Bf r}').
\end{eqnarray}
We present the expression for ${\cal A}_{\rm s}'({\Bf r},{\Bf r}')$, 
the SSDA to ${\cal A}'({\Bf r},{\Bf r}')$, defined in Eq.~(\ref{ef3}), 
in Eq.~(\ref{ef59}) (see also Eqs.~(\ref{ef55}) and (\ref{ef56})). 
The RHS of Eq.~(\ref{e258}) coincides with the {\sl local} part of the 
{\sl exact} $\Sigma_{\sigma;\infty_1}({\Bf r},{\Bf r}')$ with the 
$N$-particle GS similarly replaced by a SSD, that is (see 
Eq.~(\ref{e187}) above)
\begin{equation}
\label{e259}
\left. \Sigma_{\sigma;\infty_1}^{\rm l}
({\Bf r},{\Bf r}')\right|_{\rm s} \equiv 
\left. \Sigma_{\sigma;\infty_1}^{(1)}
({\Bf r},{\Bf r}')\right|_{\rm s}.
\end{equation}

In view of the fact that $\Sigma_{\sigma;\infty_1}^{(1)}({\Bf r},
{\Bf r}')$ is purely {\sl local} and in view of Eq.~(\ref{e259}),
it is of interest to mention that, within the SSDA, for the 
{\sl non-local} (`nl') part of the {\sl exact} $\Sigma_{\sigma;
\infty_1}({\Bf r},{\Bf r}')$ one has
\begin{eqnarray}
\label{e260}
&&\left.\Sigma_{\sigma;\infty_1}^{\rm nl}
({\Bf r},{\Bf r}')\right|_{\rm s}
= - v^2({\Bf r}-{\Bf r}') 
\varrho_{\sigma}({\Bf r}',{\Bf r}) \nonumber\\
&&\;\;\;
+\int {\rm d}^dr''\;
\big[ v({\Bf r}-{\Bf r}') v({\Bf r}-{\Bf r}'')
+ v({\Bf r}-{\Bf r}') v({\Bf r}'-{\Bf r}'')
\nonumber\\
&&\;\;\;\;\;\;
-v({\Bf r}-{\Bf r}'') v({\Bf r}'-{\Bf r}'')\big]
\varrho_{{\rm s};\sigma}({\Bf r}',{\Bf r}'') 
\varrho_{{\rm s};\sigma}({\Bf r}'',{\Bf r}).
\end{eqnarray}
The non-triviality of this result makes evident that indeed
$\wt{\Sigma}_{\sigma}^{(1)}({\Bf r},{\Bf r}';z)$ neglects a category of 
correlation effects already at order $1/z$. This is readily understood 
by considering the fact (see \S\S~III.E.1,3) that the {\sl exact} 
$\Sigma_{\sigma;\infty_1}({\Bf r},{\Bf r}')$ is, in its {\sl explicit} 
dependence on $v$, a {\sl quadratic} functional of $v$ and thus must 
not be capable of being correctly described within the framework of a 
{\sl first-order} perturbation theory. In general, and disregarding 
the possibility that, for arbitrary $m$, $\Sigma_{\sigma;\infty_m}
({\Bf r},{\Bf r}')$ can involve fundamentally unbounded contributions 
(see \S\S II.B and III), this function {\sl explicitly} depends to 
{\sl all} orders from the second (assuming $m\ge 1$) up to and 
including the $(m+1)$th order on $v$ (see \S~II.B, the paragraph 
starting with `--- Secondly, considering for the moment \dots'; see 
also Eqs.~(\ref{e107}) - (\ref{e109}) above), so that analogously 
$\Sigma_{\sigma;\infty_m}({\Bf r},{\Bf r}')$ {\sl cannot} be capable 
of being correctly described within the framework of a $p$th-order 
perturbation theory when $p \le m$. In this connection we point out 
that the difference between the exact $\Sigma_{\sigma;\infty_m}
({\Bf r},{\Bf r}')$ and its counterpart within the framework of the 
SSDA lies in the {\sl implicit} dependence of these functions on $v$, 
with the former depending to {\sl all} orders on $v$. 

Considering the intimate relationship between the set of $\Sigma_{\sigma;
\infty_m}$, $m=0,1,\dots$, and the $\varepsilon$ moments of the 
single-particle spectral function (see \S\S~III.B and III.E.6 as well 
as \S~III.I.2), one directly observes how SE operators evaluated at 
{\sl finite} orders in the perturbation theory are fundamentally incapable 
of reproducing the {\sl exact} single-particle excitation spectra. In 
this connection it is important to bear in mind how a change by 
$\pm 1$ in the order of the perturbation series for $\Sigma_{\sigma}
({\Bf r},{\Bf r}';\varepsilon)$, can give rise to appearance or 
disappearance of contributions (which in specific contexts may or may 
not be significant) to the calculated SE; in the case considered above, 
the second-order contribution at issue turns out to be nothing less 
than the {\sl entire} non-local contribution to $\wt{\Sigma}_{\sigma}
({\Bf r},{\Bf r}';z)$ at order $1/z$ for $\vert z\vert\to\infty$.

\subsection{Evaluation of 
$\Sigma_{\sigma;\infty_2}^{(1)}({\Bf r},{\Bf r}')$ }
\label{s35}
 
From Eq.~(\ref{e254}) we observe that, with the exception of
$W'_{\infty_2}({\Bf r},{\Bf r}')$, in earlier Sections we have 
already determined all functions that contribute to the expression 
for $\Sigma_{\sigma;\infty_2}^{(1)}({\Bf r},{\Bf r}')$. 

From Eqs.~(\ref{e250}), (\ref{e245}) and (\ref{e246}) we have
\begin{eqnarray}
\label{e261}
&&W'_{\infty_2}({\Bf r},{\Bf r}') \equiv
2 \sum_{s',\sigma'} e_{s'} w_{s',\sigma'}({\Bf r})
w_{s',\sigma'}^*({\Bf r}') \nonumber\\
&&= 2 \int {\rm d}^dr_1'' {\rm d}^dr_2''\;
v({\Bf r}-{\Bf r}_1'') v({\Bf r}'-{\Bf r}_2'')
\Upsilon_1({\Bf r}_1'',{\Bf r}_2''),
\end{eqnarray}
in which
\begin{eqnarray}
\label{e262}
&&\Upsilon_1({\Bf r}_1,{\Bf r}_2) {:=}
\sum_{s',\sigma'} e_{s'} \rho_{s';\sigma'}({\Bf r}_1)
\rho_{s';\sigma'}({\Bf r}_2)\nonumber\\
&&\;\;
= \sum_{\sigma'} \langle\Psi_{N;0}\vert
\Big\{ \hat\psi_{\sigma'}^{\dag}({\Bf r}_1)
\wh{A}_{\sigma'}({\Bf r}_1) -
\wh{A}_{\sigma'}^{\dag}({\Bf r}_1) 
\hat\psi_{\sigma'}({\Bf r}_1) \Big\}\nonumber\\
&&\;\;\;\;\;\;\;\;\;\;\;\;\;\;\;\;\;\;\;\;\;\;\;\;\;\;\;\;\;
\;\;\;
\times \hat\psi_{\sigma'}^{\dag}({\Bf r}_2)
\hat\psi_{\sigma'}({\Bf r}_2)\vert\Psi_{N;0}\rangle,
\end{eqnarray}
where $\wh{A}_{\sigma}({\Bf r})$ is defined in Eq.~(\ref{e158}).
Making use of the anticommutation relations in Eq.~(\ref{e29})
and some algebra, we deduce
\begin{eqnarray}
\label{e263}
\Upsilon_1({\Bf r}_1,{\Bf r}_2) &=&
\lim_{{\tilde {\Bf r}}_1\to {\Bf r}_1}
\tau({\Bf r}_1) \Big\{
\delta({\Bf r}_1-{\Bf r}_2)\,
\sum_{\sigma'} \varrho_{\sigma'}({\tilde {\Bf r}}_1,{\Bf r}_1)
\nonumber\\
&&\;\;\;\;
-\delta({\tilde {\Bf r}}_1-{\Bf r}_2)\,
\sum_{\sigma'} \varrho_{\sigma'}({\Bf r}_1,{\Bf r}_2)\Big\},
\end{eqnarray}
which upon substitution in Eq.~(\ref{e261}) results in 
\begin{eqnarray}
\label{e264}
W'_{\infty_2}({\Bf r},{\Bf r}') 
&=& \frac{-\hbar^2}{m_e} \int {\rm d}^dr''\;
v({\Bf r}-{\Bf r}'')\nonumber\\
& &\;\;\;
\times {\Bf\nabla}_{{\Bf r}''} \cdot
\big( n({\Bf r}'') {\Bf\nabla}_{{\Bf r}''}
v({\Bf r}'-{\Bf r}'')\big).
\end{eqnarray} 
In obtaining this result, we have employed
\begin{equation}
\label{e265}
\lim_{{\tilde {\Bf r}}\to {\Bf r}}
{\Bf\nabla}_{\Bf r}
\varrho_{\sigma}({\tilde {\Bf r}},{\Bf r})
= \frac{1}{2} {\Bf\nabla}_{{\Bf r}} n_{\sigma}({\Bf r}),
\end{equation}
which follows from the symmetry property $\varrho_{\sigma}({\Bf r}',
{\Bf r}) \equiv \varrho_{\sigma}({\Bf r},{\Bf r}')$ (see Appendix B). 
We draw attention to the fact that, were $W'({\Bf r},{\Bf r}';
\varepsilon)$ evaluated in terms of the non-interacting state 
$\vert\Phi_{N;0}\rangle$ (to be contrasted with $\vert\Psi_{N;0}
\rangle$ on the RHS of Eq.~(\ref{e245})), the total number density 
$n({\Bf r}'')$ on the RHS of Eq.~(\ref{e264}) would have to be 
$n_{\rm s}({\Bf r}'')$ corresponding to $\vert\Phi_{N;0}\rangle$ (Engel 
and Farid 1993). For $\vert\Phi_{N;0}\rangle$ coinciding with the GS of 
the `non-interacting' Hamiltonian due to Kohn and Sham (1965) within 
the framework of the GS density-functional theory, the corresponding 
$n_s({\Bf r}'')$ by construction is identical with $n({\Bf r})$ (for 
this to be possible, it is, however, required that $n({\Bf r})$ be 
so-called `pure-state non-interacting $v$-representable'; for details 
see Dreizler and Gross 1990; see also Farid (1998)). Thus, for instance, 
the $W^{\prime\, {\sc rpa}}_{\infty_2}({\Bf r},{\Bf r}')$ corresponding 
to $W^{\prime\, {\sc rpa}}({\Bf r},{\Bf r}';\varepsilon)$ as evaluated 
in terms of the single-particle GF of the Kohn-Sham Hamiltonian 
associated with the GS of $\wh{H}$, is {\sl identical} with the 
{\sl exact} $W'_{\infty_2}({\Bf r},{\Bf r}')$ (Farid 1999b);
see footnote \ref{f7}.

Making use of Eq.~(\ref{e168}) which in conjunction with the closure 
relation for the Lehmann amplitudes in Eq.~(\ref{e30}) yields
\begin{eqnarray}
\label{e266}
\sum_s^{>} f_{s;\sigma}({\Bf r}) f_{s;\sigma}^*({\Bf r}')
= \delta({\Bf r}-{\Bf r}') 
-\varrho_{\sigma}({\Bf r}',{\Bf r}), 
\end{eqnarray}
with $\sum_s^{>}$ denoting a sum over all $s$ corresponding to
$\varepsilon_{s;\sigma} > \mu$ (see text following Eq.~(\ref{e168})
above), from Eqs.~(\ref{e254}), (\ref{e162}), (\ref{e256}), 
(\ref{e30}) and (\ref{e264}) we obtain
\begin{eqnarray}
\label{e267}
&&\Sigma_{\sigma;\infty_2}^{(1)}({\Bf r},{\Bf r}')
= \frac{1}{\hbar} \big\{ \int {\rm d}^dr''\; v({\Bf r}-{\Bf r}'')
v({\Bf r}'-{\Bf r}'') n({\Bf r}'')\nonumber\\
&&\;\;\;\;\;\;\;
-\sum_{\sigma'} v_H({\Bf r};[n_{\sigma'}])
v_H({\Bf r}';[n_{\sigma'}]) + {\cal C}({\Bf r},{\Bf r}')\big\}
\nonumber\\
&&\;\;\;\;\;\;\;\;\;\;\;\;\;\;\;
\times \frac{1}{\hbar} G_{\sigma;\infty_2}({\Bf r},{\Bf r}')\nonumber\\
&&\;\;\;\;\;\;\;\;\;\;\;\;\;\;\;\;\;\;\;\;\;
+\frac{1}{\hbar}
\big\{ \delta({\Bf r}-{\Bf r}') - 2 \varrho_{\sigma}({\Bf r}',
{\Bf r}) \big\}\nonumber\\
&&\times \frac{-\hbar^2}{2 m_e}
\int {\rm d}^dr''\; v({\Bf r}-{\Bf r}'')
{\Bf\nabla}_{{\Bf r}''}\cdot \big( n({\Bf r}'')
{\Bf\nabla}_{{\Bf r}''} v({\Bf r}'-{\Bf r}'')\big).
\nonumber\\ 
\end{eqnarray}
We have {\sl not} attempted to present the above expression in such a 
way as to make explicit the {\sl local} and {\sl non}-local 
contributions to $\Sigma_{\sigma;\infty_2}^{(1)}({\Bf r},{\Bf r}')$.
The expression in Eq.~(\ref{e267}) is of interest to us, firstly, because 
a cursory comparison of it with its {\sl exact} counterpart (see 
Eqs.~(\ref{e185}), (\ref{e187}) and (\ref{e188}) above) immediately 
reveals the extent to which $\Sigma_{\sigma;\infty_2}^{(1)}({\Bf r},
{\Bf r}')$ neglects interaction effects, 
\footnote{\label{f100}
Note for instance that the approximation in Eq.~(\protect\ref{e267}) 
does {\sl not} involve the contribution $-v^3({\Bf r}-{\Bf r}') 
\varrho_{\sigma}({\Bf r}',{\Bf r})$, which is explicit in the 
expression for the {\sl exact} $\Sigma_{\sigma;\infty_2}({\Bf r},
{\Bf r}')$ on the RHS of Eq.~(\protect\ref{e199}) (see also 
Eq.~(\protect\ref{e212})); as we discuss in Appendix H, this 
contribution, which is {\sl non}-integrable in the case of $v\equiv v_c$ 
in $d=3$, following an infinite-order summation of related non-integrable 
functions in $\{ \Sigma_{\sigma;\infty_m} \| m \ge 2\}$, gives rise to 
a {\sl singular} contribution to the large-$\vert z\vert$ AS pertaining 
to the momentum representation of $\wt{\Sigma}_{\sigma}(z)$. }
and secondly, because it exposes a contribution which in the case of 
$v\equiv v_c$ in $d=3$ is fundamentally unbounded and which is directly 
associated with an equally fundamentally unbounded contribution in the 
formal expression for $\Sigma_{\sigma;\infty_3}^{(1)}({\Bf r},{\Bf r}')$ 
that we shall consider later. In all essential aspects, these unbounded 
contributions conform with our deductions on the basis of the expressions 
for the {\sl exact} $\Sigma_{\sigma;\infty_m}({\Bf r},{\Bf r}')$ 
corresponding to $m \le 2$ as regards the {\sl complete} infinite series 
of unbounded contributions whose summation (see \S~III.H.2) accounts for 
a singular contribution to the finite-order AS for $\wt{\Sigma}_{\sigma}
({\Bf r},{\Bf r}';z)$ decaying like $1/\vert z\vert^{3/2}$ as 
$\vert z\vert\to\infty$ (see Eqs.~(\ref{e210}), (\ref{e213}) and 
(\ref{eg11}) - (\ref{eg15})). Before entering into details, we mention 
that the required symmetry of $\Sigma_{\sigma;\infty_2}^{(1)}({\Bf r},
{\Bf r}')$ with respect to the exchange ${\Bf r}\rightleftharpoons 
{\Bf r}'$, which is {\sl not} explicit in the term arising from the 
product of $\varrho_{\sigma}({\Bf r}',{\Bf r})$ with the last integral 
on the RHS of Eq.~(\ref{e267}), is made explicit through applying the 
divergence theorem (or its generalizations in $d\not=3$) which gives 
rise to the symmetric integrand $[{\Bf\nabla}_{{\Bf r}''} v({\Bf r}
-{\Bf r}'')] \cdot [{\Bf\nabla}_{{\Bf r}''} v({\Bf r}'-{\Bf r}'')] 
n({\Bf r}'')$.

Now we consider the following expression whose LHS originates from 
the RHS of Eq.~(\ref{e267}),
\begin{eqnarray}
\label{e268}
&&\frac{-\hbar^2}{2 m_e} \int {\rm d}^dr''\;
v({\Bf r}-{\Bf r}'') {\Bf\nabla}_{{\Bf r}''}\cdot 
\big( n({\Bf r}'') {\Bf\nabla}_{{\Bf r}''} 
v({\Bf r}'-{\Bf r}'')\big) \nonumber\\
&&\equiv \frac{-\hbar^2}{2 m_e} \int {\rm d}^dr''\;
v({\Bf r}-{\Bf r}'') [{\Bf\nabla}_{{\Bf r}''} v({\Bf r}'
-{\Bf r}'')]\cdot [{\Bf\nabla}_{{\Bf r}''} 
n({\Bf r}'')] \nonumber\\
&&\;\;
+\int {\rm d}^dr''\; v({\Bf r}-{\Bf r}'') n({\Bf r}'')\,
\big(\tau({\Bf r}'') v({\Bf r}'-{\Bf r}'')\big).
\end{eqnarray}
When multiplied by $\frac{1}{\hbar} \delta({\Bf r}-{\Bf r}')$, the last 
term on the RHS of Eq.~(\ref{e268}) is seen to be identical with a term 
in the expression for the {\sl exact} $\Sigma_{\sigma;\infty_2}({\Bf r},
{\Bf r}')$ in Eq.~(\ref{e199}) (see ${\cal I}_3$ as defined in 
Eq.~(\ref{e202}) above). The identity of the two terms implies that in 
the expression for $\Sigma_{\sigma;\infty_3}^{(1)}({\Bf r},{\Bf r}')$, to 
be discussed in the following Section, we have to find a term identical 
with the term that we infer in Appendix G from the general structure 
of the {\sl exact} $\{ \Sigma_{\sigma;\infty_m} \| m \le 2\}$. This 
indeed turns out to be the case.

\subsection{Evaluation of 
$\Sigma_{\sigma;\infty_3}^{(1)}({\Bf r},{\Bf r}')$ }
\label{s36}

From Eqs.~(\ref{e250}), (\ref{e245}) and (\ref{e246}), making 
repeated use of the anticommutation relations in Eq.~(\ref{e29}), 
we obtain  
\begin{eqnarray}
\label{e269}
&&W'_{\infty_3}({\Bf r},{\Bf r}') \equiv
2 \sum_{s',\sigma'} e_{s'}^2\, w_{s',\sigma'}({\Bf r})
w_{s',\sigma'}^*({\Bf r}') \nonumber\\
&&= 2 \int {\rm d}^dr_1'' {\rm d}^dr_2''\;
v({\Bf r}-{\Bf r}_1'') v({\Bf r}'-{\Bf r}_2'')
\Upsilon_2({\Bf r}_1'',{\Bf r}_2''),
\end{eqnarray}
in which
\begin{eqnarray}
\label{e270}
&&\Upsilon_2({\Bf r}_1,{\Bf r}_2) {:=}
\sum_{s',\sigma'} e_{s'}^2\, \rho_{s';\sigma'}({\Bf r}_1)
\rho_{s';\sigma'}({\Bf r}_2)\nonumber\\
&&\;\;
= \sum_{\sigma'} \langle\Psi_{N;0}\vert
\Big\{ \wh{B}_{\sigma'}({\Bf r}_1) + 
\wh{B}_{\sigma'}^{\dag}({\Bf r}_1) 
-2 \wh{A}_{\sigma'}^{\dag}({\Bf r}_1)
\wh{A}_{\sigma'}({\Bf r}_1)\Big\}\nonumber\\
&&\;\;\;\;\;\;\;\;\;\;\;\;\;\;\;\;\;\;\;\;\;\;\;\;\;\;\;\;\;
\;\;\;
\times \hat\psi_{\sigma'}^{\dag}({\Bf r}_2)
\hat\psi_{\sigma'}({\Bf r}_2)\vert\Psi_{N;0}\rangle,
\end{eqnarray}
where ${\hat A}_{\sigma}({\Bf r})$ is defined in Eq.~(\ref{e158}) and
\begin{eqnarray}
\label{e271}
&&\wh{B}_{\sigma}({\Bf r}) {:=}
\hat\psi_{\sigma}^{\dag}({\Bf r})
\big[ \wh{A}_{\sigma}({\Bf r}),\wh{H}\big]_-
\nonumber\\
&&\;\;
=\hat\psi_{\sigma}^{\dag}({\Bf r}) 
\wh{A}_{\sigma}({\Bf r}) \wh{H} 
+ \wh{A}_{\sigma}^{\dag}({\Bf r}) \wh{A}_{\sigma}({\Bf r})
-\wh{H} \hat\psi_{\sigma}^{\dag}({\Bf r})
\wh{A}_{\sigma}({\Bf r}).
\end{eqnarray}
After some algebra, we deduce
\begin{eqnarray}
\label{e272}
&&\Upsilon_2({\Bf r}_1,{\Bf r}_2)
= \sum_{\sigma'} \varrho_{\sigma'}({\Bf r}_1,{\Bf r}_2)
\tau({\Bf r}_1) \tau({\Bf r}_2) \delta({\Bf r}_1
-{\Bf r}_2)\nonumber\\
&&\;\;\;
-\big[\tau({\Bf r}_2) \sum_{\sigma'} 
\varrho_{\sigma'}({\Bf r}_1,{\Bf r}_2)\big]
\tau({\Bf r}_1) \delta({\Bf r}_1-{\Bf r}_2)
\nonumber\\
&&\;\;\;
-\big[\tau({\Bf r}_1)\sum_{\sigma'}
\varrho_{\sigma'}({\Bf r}_1,
{\Bf r}_2)\big] \tau({\Bf r}_2) \delta({\Bf r}_1-
{\Bf r}_2)\nonumber\\
&&\;\;\;
+\big[\tau({\Bf r}_1) \tau({\Bf r}_2)
\sum_{\sigma'} \varrho_{\sigma'}({\Bf r}_1,{\Bf r}_2)\big]
\delta({\Bf r}_1-{\Bf r}_2)\nonumber\\
&&\;\;\;
+2 \lim_{{\tilde {\Bf r}}_1\to {\Bf r}_1\atop
{\tilde {\Bf r}}_2\to {\Bf r}_2}
\tau({\Bf r}_1) \tau({\Bf r}_2)\nonumber\\
&&\;\;\;\;\;\;\;\;\;\;\;\;\;\;\;
\times \sum_{\sigma'}
\Big\{ \Gamma^{(2)}({\tilde {\Bf r}}_1\sigma',
{\Bf r}_2\sigma';{\Bf r}_1\sigma',
{\tilde {\Bf r}}_2\sigma')\nonumber\\
&&\;\;\;\;\;\;\;\;\;\;\;\;\;\;\;\;\;\;\;\;\;\;\;\;
-\Gamma^{(2)}({\tilde {\Bf r}}_1\sigma',
{\tilde {\Bf r}}_2\sigma';
{\Bf r}_1\sigma',{\Bf r}_2\sigma')\Big\},
\end{eqnarray}
which on substitution in the expression on the RHS of 
Eq.~(\ref{e269}) results in
\begin{eqnarray}
\label{e273}
&&W'_{\infty_3}({\Bf r},{\Bf r}') =
2\int {\rm d}^dr''\; v({\Bf r}-{\Bf r}'')\nonumber\\
&&\;\;\;\;\;\;\;\;\;\;\;\;
\times \lim_{{\tilde {\Bf r}}''\to {\Bf r}''}
\tau^2({\Bf r}'') 
v({\Bf r}'-{\Bf r}'') \sum_{\sigma'} 
\varrho_{\sigma'}({\tilde {\Bf r}}'',{\Bf r}'')\nonumber\\
&&\;\;\;
-2\int {\rm d}^dr''\;
v({\Bf r}-{\Bf r}'') 
\lim_{{\tilde {\Bf r}}''\to {\Bf r}''}
\tau({\Bf r}'') v({\Bf r}'-{\Bf r}'')\nonumber\\
&&\;\;\;\;\;\;\;\;\;\;\;\;
\times\big[\tau({\Bf r}'') \sum_{\sigma'} 
\varrho_{\sigma'}({\tilde {\Bf r}}'',{\Bf r}'')\big]\nonumber\\
&&\;\;\;
-2\int {\rm d}^dr''\;
v({\Bf r}'-{\Bf r}'') 
\lim_{{\tilde {\Bf r}}''\to {\Bf r}''}
\tau({\Bf r}'') v({\Bf r}-{\Bf r}'')\nonumber\\
&&\;\;\;\;\;\;\;\;\;\;\;\;
\times\big[\tau({\Bf r}'') \sum_{\sigma'} 
\varrho_{\sigma'}({\Bf r}'',{\tilde {\Bf r}}'')\big]
\nonumber\\
&&\;\;\;
+2\int {\rm d}^dr''\;
v({\Bf r}-{\Bf r}'') v({\Bf r}'-{\Bf r}'')\nonumber\\
&&\;\;\;\;\;\;\;\;\;\;\;\;
\times \lim_{{\tilde {\Bf r}}''\to {\Bf r}''}
\tau({\Bf r}'') \tau({\tilde {\Bf r}}'')
\sum_{\sigma'} \varrho_{\sigma'}({\Bf r}'',
{\tilde {\Bf r}}'')\nonumber\\
&&\;\;\;
+4\int {\rm d}^dr_1'' {\rm d}^dr_2''\;
v({\Bf r}-{\Bf r}_1'') v({\Bf r}'-{\Bf r}_2'')
\nonumber\\
&&\;\;\;\;\;\;\;\;\;\;
\times \lim_{{\tilde {\Bf r}}_1''\to {\Bf r}_1''\atop
{\tilde {\Bf r}}_2''\to {\Bf r}_2''}
\tau({\Bf r}_1'') \tau({\Bf r}_2'')\nonumber\\
&&\;\;\;\;\;\;\;\;\;\;\;\;\;\;\;
\times
\sum_{\sigma'}
\Big\{ \Gamma^{(2)}({\tilde {\Bf r}}_1''\sigma',
{\Bf r}_2''\sigma';{\Bf r}_1''\sigma',
{\tilde {\Bf r}}_2''\sigma')\nonumber\\
&&\;\;\;\;\;\;\;\;\;\;\;\;\;\;\;\;\;\;\;\;\;\;\;\;
-\Gamma^{(2)}({\tilde {\Bf r}}_1''\sigma',
{\tilde {\Bf r}}_2''\sigma';
{\Bf r}_1''\sigma',{\Bf r}_2''\sigma')\Big\}.
\end{eqnarray}
Making use of the general result 
\begin{equation}
\label{e274}
\tau^2 f g = [\tau^2 f] g + f [\tau^2 g] + \dots,
\end{equation}  
we have, for the first integral on the RHS of Eq.~(\ref{e273}),
\begin{eqnarray}
\label{e275}
&&\int {\rm d}^dr''\; v({\Bf r}-{\Bf r}'')
\lim_{{\tilde {\Bf r}}''\to {\Bf r}''}
\tau^2({\Bf r}'') 
v({\Bf r}'-{\Bf r}'') \nonumber\\
&&\;\;\;\;\;\;\;\;\;\;\;\;\;\;\;\;\;\;\;\;\;\;\;\;\;
\;\;\;\;\;\;\;\;\;\;\;\;\;\;\;\;\;\;
\times\sum_{\sigma'} 
\varrho_{\sigma'}({\tilde {\Bf r}}'',{\Bf r}'')\nonumber\\
&&=\int {\rm d}^dr''\; v({\Bf r}-{\Bf r}'') n({\Bf r}'')
\big(\tau^2({\Bf r}'') v({\Bf r}'-{\Bf r}'')\big) +\dots,
\nonumber\\
\end{eqnarray}
which gives rise to the following result for the first term on the
RHS of Eq.~(\ref{e255}) concerning $\Sigma_{\sigma;\infty_3}^{(1)}
({\Bf r},{\Bf r}')$ (see Eq.~(\ref{e30}) above):
\begin{eqnarray}
\label{e276}
&&\frac{1}{2\hbar^2} G_{\sigma;\infty_1}({\Bf r},{\Bf r}')
W'_{\infty_3}({\Bf r},{\Bf r}')
= \frac{1}{\hbar} \int {\rm d}^dr''\; v({\Bf r}-{\Bf r}'')
n({\Bf r}'') \nonumber\\
&&\;\;\;\;\;\;\;\;\;\;\;\;\;\;\;
\times \big(\tau^2({\Bf r}'') v({\Bf r}-{\Bf r}'') \big)
\,\delta({\Bf r}-{\Bf r}') + \dots.
\end{eqnarray}
The (local) contribution which we have explicitly presented on the RHS 
of Eq.~(\ref{e276}), {\sl exactly} coincides with what is inferred 
from the expressions for the {\sl exact} $\Sigma_{\sigma;\infty_m}
({\Bf r},{\Bf r}')$, with $m=1,2$, in Appendix G. However, since 
$\Sigma_{\sigma;\infty_1}^{(1)}({\Bf r},{\Bf r}')$ does {\sl not}
contain a {\sl non}-local contribution, {\sl nor} does the 
{\sl non}-local contribution pertaining to $\Sigma_{\sigma;
\infty_2}^{(1)}({\Bf r},{\Bf r'})$ involve a term similar to that on 
the RHS of Eq.~(\ref{e203}) (see also Eq.~(\ref{e207}) above), from 
our above considerations it follows that, in regard to 
$\wt{\Sigma}_{\sigma}^{(1)}({\Bf r},{\Bf r}';z)$, the equivalent of 
$\wt{\sf T}_{\sigma,\bar\sigma}({\Bf r};z)$, defined in 
Eq.~(\ref{eg1}), is the following: 
\begin{eqnarray}
\label{e277}
&&\wt{\sf T}^{(1)}({\Bf r};z)
{:=} \frac{1}{z}
\sum_{m=1}^{\infty}
\int {\rm d}^3r''\; v_c({\Bf r}-{\Bf r}'') n({\Bf r}'') \nonumber\\
&&\;\;\;\;\;\;\;\;\;\;\;\;
\times \big[ \frac{1}{z^m} \tau^m({\Bf r}'')
v_c({\Bf r}-{\Bf r}'')\big]\nonumber\\
&&= \frac{1}{z^2} \int {\rm d}^3r''\;
v_c({\Bf r}-{\Bf r}'') n({\Bf r}'') \nonumber\\
&&\;\;\;\;\;\;\;\;\;\;\;\;
\times \tau({\Bf r}'') \big(1 - \frac{1}{z} \tau({\Bf r}'')
\big)^{-1} v_c({\Bf r}-{\Bf r}''),
\end{eqnarray}
which in contrast with $\wt{\sf T}_{\sigma,\bar\sigma}({\Bf r};z)$ does 
{\sl not} explicitly depend on $\sigma$. All the results concerning 
$\wt{\sf T}_{\sigma,\bar\sigma}({\Bf r};z)$, determined in Appendix G, 
are directly applicable to $\wt{\sf T}^{(1)}({\Bf r};z)$, with the sole
requirement that $[n_{\bar\sigma}({\Bf r}) - n_{\sigma}({\Bf r})]$ 
in the expressions associated with $\wt{\sf T}_{\sigma,\bar\sigma}
({\Bf r};z)$ be replaced by $n({\Bf r})$. The considerable significance 
of $\wt{\sf T}_{\sigma,\bar\sigma}({\Bf r};z)$ in the case $v\equiv v_c$ 
in $d=3$ to the behaviour of $\wt{\Sigma}_{\sigma}({\Bf r},{\Bf r}';z)$ 
at large $\vert z\vert$ on the one hand (see \S~III.H.2), and the 
significant difference between $\wt{\sf T}_{\sigma,\bar\sigma}({\Bf r};
z)$ and $\wt{\sf T}^{(1)}({\Bf r};z)$ on the other hand, unequivocally 
show a fundamental shortcoming of $\wt{\Sigma}_{\sigma}^{(1)}
({\Bf r},{\Bf r}';z)$ (see \S~IV). For later reference, in analogy with 
${\cal T}_{\sigma,\bar\sigma}({\Bf r})$ in Eq.~(\ref{e209}) we define
\begin{eqnarray}
\label{e278}
{\cal T}^{(1)}({\Bf r}) &{:=}&
\int {\rm d}^3r''\; v_c({\Bf r}-{\Bf r}'')\, n({\Bf r}'')\nonumber\\
&&\;\;\;\;\;\;\;\;\;\;\;\;\;\;\;\;
\times \big( \tau({\Bf r}'')
v_c({\Bf r}-{\Bf r}'')\big).
\end{eqnarray}

Finally, because $G_{\sigma;\infty_1}({\Bf r},{\Bf r}') \propto 
\delta({\Bf r}-{\Bf r}')$ (see Eq.~(\ref{e30}) above) and in view 
of Eq.~(\ref{e255}), the contributions to $\Sigma_{\sigma;
\infty_3}^{(1)}({\Bf r},{\Bf r}')$ as arising from the second and 
third integrals on the RHS of Eq.~(\ref{e273}), are for $v\equiv 
v_c$ in $d=3$ identical and fundamentally unbounded; these 
contributions are {\sl partly} cancelled by the contribution from 
the first term on the RHS of Eq.~(\ref{e273}), associated with a 
term on the RHS of Eq.~(\ref{e274}) whose explicit form we have 
not presented. The existence of such unbounded contributions to 
$\Sigma_{\sigma;\infty_3}^{(1)}({\Bf r},{\Bf r}')$ which are 
{\sl not} related to any unbounded counterparts in the expression 
for $\Sigma_{\sigma;\infty_2}^{(1)}({\Bf r},{\Bf r}')$, implies that 
(see \S~II.B) the $1/z^2$ term in the large-$\vert z\vert$ AS for 
$\wt{\Sigma}_{\sigma}^{(1)}({\Bf r},{\Bf r}';z)$ is followed by a 
term, or terms, more dominant than $1/z^3$.

\subsection{Correcting 
$\Sigma_{\sigma}^{(1)}({\Bf r},{\Bf r}';\varepsilon) \equiv
\Sigma_{\sigma}^{\prime (1)}({\Bf r},{\Bf r}';\varepsilon)
+ \hbar^{-1} v_H({\Bf r};[n])\, \delta({\Bf r}-{\Bf r}')$; 
a workable scheme }
\label{s37}

The considerations in the above three Sections, A, B and C have 
exposed the shortcomings of $\Sigma_{\sigma}^{(1)}({\Bf r},{\Bf r}';
\varepsilon)$ as reflected in the behaviour of this function for 
`large' values of $\vert\varepsilon\vert$. In view of our discussions 
in \S~III.B, it is evident that the deviation of $\Sigma_{\sigma}^{(1)}
({\Bf r}, {\Bf r}';\varepsilon)$ from its exact counterpart 
$\Sigma_{\sigma}({\Bf r},{\Bf r}';\varepsilon)$ (see Eq.~(\ref{e240})) 
in the regime of large $\vert\varepsilon\vert$ implies deviation of 
the $\varepsilon$ moments integrals of the associated single-particle 
spectral function from those of the exact $A_{\sigma}({\Bf r},{\Bf r}';
\varepsilon)$. With our knowledge of the first few terms in the 
large-$\vert\varepsilon\vert$ AS of the {\sl exact} $\Sigma_{\sigma}
({\Bf r},{\Bf r}';\varepsilon)$ (as well as those of 
$\Sigma_{\sigma}^{(1)}({\Bf r},{\Bf r}';\varepsilon)$), fully expressed 
in terms of GS correlation functions, we are in a position to correct 
for the above-mentioned shortcomings of $\Sigma_{\sigma}^{(1)}
({\Bf r},{\Bf r}';\varepsilon)$, while retaining a function applicable 
for {\sl all} values of $\varepsilon$. For simplicity of notation, 
in the following we employ ${\sf f}(\varepsilon)$ as denoting 
\footnote{\label{f101}
Owing to $\Sigma_{\sigma;\infty_2}^{\rm s}({\Bf r},{\Bf r}')$, 
analogous considerations as below but concerning $\Ol{\Sigma}_{\sigma}
({\Bf q},{\Bf q}';\varepsilon)$, the momentum representation of 
$\Sigma_{\sigma}(\varepsilon)$, require some non-essential alteration 
of the following in the case where, for example, $v\equiv v_c$ and 
$d=3$; see footnote \protect\ref{f72}. Insofar as homogeneous systems 
with uniform and isotropic GSs are concerned, the counterparts of the 
expression in Eq.~(\protect\ref{e279}) for the diagonal elements of 
the SE operator in the momentum representation are those in 
Eqs.~(\protect\ref{e128}) and (\protect\ref{e129}) (by symmetry, the 
off-diagonal elements are identically vanishing). }
$\Sigma_{\sigma} ({\Bf r},{\Bf r}';\varepsilon)$ for which we consider 
the following large-$\vert\varepsilon \vert$ AS
\begin{eqnarray}
\label{e279}
&&{\sf f}(\varepsilon) \sim {\sf f}_0 + 
\frac{{\sf f}_1' + i\, {\sf f}_1''}{\varepsilon}
+ \frac{[\Theta(\mu-\varepsilon) - i \Theta(\varepsilon-\mu)]\, 
{\sf f}_{3/2}}{\vert\varepsilon\vert^{3/2} } \nonumber\\
&&\;\;\;\;
+ \frac{ \ln\vert\varepsilon/\varepsilon_0\vert\, 
{\sf f}_{\rm ln} }{\varepsilon^2}
+ \frac{{\sf f}_2' + i\, [{\sf f}_2'' - \pi \Theta(\varepsilon-\mu)\,
{\sf f}_{\rm ln}] }{\varepsilon^2},\;\;\; 
\vert\varepsilon\vert \to \infty, \nonumber\\
\end{eqnarray}
where the coefficient functions ${\sf f}_0$, ${\sf f}_1'$, ${\sf f}_1''$, 
etc., are real valued (see Appendix B). In writing the expression in 
Eq.~(\ref{e279}), we have assumed ${\sf f}_1 \equiv {\sf f}_1' + i\, 
{\sf f}_1''$ to be well defined so that, in cases where $v\equiv v_c$, 
it is required that $d=3$ (see footnote \ref{f12} and \S~1.a in Appendix 
F). Further, in the cases corresponding to bounded and short-range 
interaction functions $v$, we have ${\sf f}_{3/2} \equiv {\sf f}_{\rm ln} 
\equiv 0$. With reference to our considerations in \S~III.E.2 (see in 
particular Eq.~(\ref{e112}); see also Eq.~(\ref{e229})), one readily 
identifies ${\sf f}_0$ with $\Sigma_{\sigma;\infty_0}({\Bf r},{\Bf r}')$, 
${\sf f}_1'$ with $\Sigma_{\sigma;\infty_1}({\Bf r},{\Bf r}')$ and 
${\sf f}_2'$ with $\Sigma_{\sigma;\infty_2}^{\rm r}({\Bf r},{\Bf r}') + 
\Sigma_{\sigma;\infty_2}^{\rm s_b}({\Bf r},{\Bf r}')$; the imaginary 
parts ${\sf f}_1''$ and ${\sf f}_2''$ are the coefficients of 
$1/\varepsilon$ and $1/\varepsilon^2$ on the RHS of Eq.~(\ref{e227}) 
(in identifying ${\sf f}_1''$ and ${\sf f}_2''$ with the latter
coefficients, account has to be taken of the provisions indicated in 
the paragraphs subsequent to Eq.~(\ref{e227})); we point out, however, 
that, for the sake of simplicity of presentation, here we have 
exceptionally deviated from our general convention (see footnote 
\ref{f31} and text following Eq.~(\ref{e111}) above) and incorporated 
$\hbar^{-1} {\sf M}^{\rm r}_{\infty_2}({\Bf r})\, \delta({\Bf r}
-{\Bf r}')$, which is a component part of $\wt{\Sigma}_{\sigma;
\infty_2}^{\rm s}({\Bf r},{\Bf r}'\| z)$ (see Eqs.~(\ref{e213}) and 
(\ref{ef144})), into ${\sf f}_2'$. Further, ${\sf f}_{3/2}$ and 
${\sf f}_{\rm ln}$ directly {\sl correspond to} $\hbar^{-1} 
\wt{\sf T}^{\rm s}_{\sigma,\bar\sigma;\infty_2}({\Bf r}\|z)\, 
\delta({\Bf r}-{\Bf r}')$ (see Eqs.~(\ref{eg15}) and (\ref{eg17})) and 
$\hbar^{-1} \wt{\sf M}_{\infty_2}^{\rm s}({\Bf r}\| z)\, \delta({\Bf r}
-{\Bf r}')$ (see Eqs.~(\ref{e213}) and (\ref{ef145})) respectively; 
since we explicitly account for the $\varepsilon$-dependent parts of 
the latter functions in the expression on the RHS of Eq.~(\ref{e279}), 
these parts are left out of ${\sf f}_{3/2}$ and ${\sf f}_{\rm ln}$. 

It is interesting to note that for $v\equiv v_c$ and $d=3$, 
${\sf f}_{3/2}$ is {\sl positive} for $n_{\bar\sigma}({\Bf r}) > 
n_{\sigma}({\Bf r})$, {\sl zero} for $n_{\bar\sigma}({\Bf r}) = 
n_{\sigma}({\Bf r})$, and {\sl negative} for $n_{\bar\sigma}({\Bf r}) < 
n_{\sigma}({\Bf r})$ (see Eqs.~(\ref{eg15}), and (\ref{eg17})), and 
further that ${\sf f}_{\rm ln}$ is {\sl strictly positive} (see 
Eq.~(\ref{ef145})). Considering the fact that stability of the GS 
demands ${\rm Im}[{\cal E}_{\varsigma;\sigma}(\varepsilon)] \, 
\Ieq{\ge}{\le}\, 0$ for $\varepsilon \,\Ieq<> \, \mu$, $\forall
\varsigma,\sigma$ (see \S~III.D, and in particular Eqs.~(\ref{e84}) 
and (\ref{e85})), the expression on the RHS of Eq.~(\ref{e279}) is 
seen to expose a crucial balance that should exist between various 
imaginary contributions in the large-$\vert\varepsilon\vert$ AS for 
$\Sigma_{\sigma}({\Bf r},{\Bf r}';\varepsilon)$ in order for an assumed 
(normal) GS, in terms of which these contributions are calculated, to 
be the true and stable GS of the system under consideration. In this 
light, one specifically clearly observes the significance of the 
interaction potentials $v({\Bf r}-{\Bf r}')$ that for $\| {\Bf r}
-{\Bf r}'\| \to 0$ behave like the Coulomb potential in $d=3$ in 
bringing about particular GSs that otherwise would not be feasible; 
in this connection, recall that, for bounded interaction potentials, 
${\sf f}_{3/2}$ is identically vanishing. 

We are now in a position to introduce the scheme to which we have
referred above; this scheme almost effortlessly removes the shortcomings 
of $\Sigma_{\sigma}^{(1)}({\Bf r},{\Bf r}';\varepsilon)$ as discussed 
in \S\S~IV.A, B, C. To this end, we consider an expression for the 
large-$\vert\varepsilon\vert$ AS of $\Sigma_{\sigma}^{(1)}({\Bf r},
{\Bf r}';\varepsilon)$ which {\sl functionally} is identical with that 
in Eq.~(\ref{e279}), with ${\sf f}(\varepsilon)$, ${\sf f}_0$, 
${\sf f}_1'$, ${\sf f}_1''$, ${\sf f}_{3/2}$, \dots herein replaced by 
${\sf f}^{(1)}(\varepsilon)$, ${\sf f}_0^{(1)}$, ${\sf f}_1^{\prime (1)}$, 
${\sf f}_1^{\prime\prime (1)}$, ${\sf f}_{3/2}^{(1)}$, \dots respectively. 
Consider now the functions $\Phi_{\alpha}(\varepsilon,\mu;\Delta)$, 
$\alpha =0, 1, 3/2, 2$, defined as follows:
\begin{eqnarray}
\label{e280}
&&\Phi_0(\varepsilon,\mu;\Delta) {:=}
\frac{\varepsilon^4 -\mu^4}{\varepsilon^4 + \Delta^4}
\sim 1 - \frac{\mu^4+\Delta^2}{\varepsilon^4}\;\;\;
\mbox{\rm for}\;\;\; \vert \varepsilon\vert \to\infty,\nonumber\\
&&\Phi_1(\varepsilon,\mu;\Delta) {:=}
\frac{\varepsilon^5 -\mu^5}{\varepsilon^6 + \Delta^6}
\sim \frac{1}{\varepsilon} - \frac{\mu^5}{\varepsilon^6}\;\;\;
\mbox{\rm for}\;\;\; \vert \varepsilon\vert \to\infty,\nonumber\\
&&\Phi_{3/2}(\varepsilon,\mu;\Delta) {:=}
\frac{\varepsilon^6 -\mu^6}
{\vert\varepsilon\vert^{15/2} + \Delta^{15/2}}
\sim \frac{1}{\vert\varepsilon\vert^{3/2}} - 
\frac{\mu^6}{\vert\varepsilon\vert^{15/2}} \nonumber\\
&&\;\;\;\;\;\;\;\;\;\;\;\;\;\;\;\;\;\;\;\;\;\;\;\;\;\;\;\;\;\;\;\;
\;\;\;\;\;\;\;\;\;\;\;\;\;\;\;\;\;\;\;\;\;\;\;\;\;\;\;
\mbox{\rm for}\;\;\; \vert \varepsilon\vert \to\infty,\nonumber\\
&&\Phi_2(\varepsilon,\mu;\Delta) {:=}
\frac{\varepsilon^6-\mu^6}{\varepsilon^8 + \Delta^8}
\sim \frac{1}{\varepsilon^2} - \frac{\mu^6}{\varepsilon^8}\;\;\;
\mbox{\rm for}\;\;\; \vert \varepsilon\vert \to\infty.
\end{eqnarray}
The characteristic aspects of these functions are that for $\varepsilon 
\to \mu$ they {\sl smoothly} approach zero
\footnote{\label{f102}
Specifically, they are continuously differentiable functions of 
$\varepsilon$ in a neighbourhood of $\varepsilon=\mu$ so that, for
the cases where the GS of the system under consideration is metallic, 
the modified SE, to be introduced in Eq.~(\protect\ref{e281}) below, 
is {\sl not} by construction made to correspond to a non-Fermi-liquid 
metallic state (Farid 1999c); see Appendix A, the paragraph including 
Eq.~(\protect\ref{ea61}). }
and for $\vert\varepsilon\vert \to \infty$, not only are their 
leading-order terms $1$, $1/\varepsilon$, 
$1/\vert\varepsilon\vert^{3/2}$ and $1/\varepsilon^2$ respectively 
({\it cf}. Eq.~(\ref{e279})), but also their next-to leading-order 
terms {\sl all} decay {\sl more rapidly} than $1/\varepsilon^2$.

Let now ${\sf f}^{(1)}_{\rm m}(\varepsilon)$ denote the {\sl modified} 
SE, deduced from the first-order SE ${\sf f}^{(1)}(\varepsilon)$, 
which we define as follows:
\begin{eqnarray}
\label{e281}
&&{\sf f}^{(1)}_{\rm m}(\varepsilon) {:=} {\sf f}^{(1)}(\varepsilon)
+\big[ {\sf f}_0 - {\sf f}_0^{(1)} \big]\,
\Phi_0(\varepsilon,\mu;\Delta_0) \nonumber\\
&&\;\;
+ \big[ {\sf f}_1' + i {\sf f}_1'' - {\sf f}_1^{\prime (1)}
- i {\sf f}_1^{\prime\prime (1)} \big]\,
\Phi_1(\varepsilon,\mu;\Delta_1)\nonumber\\
&&\;\;
+\big[\Theta(\mu-\varepsilon) - i \Theta(\varepsilon-\mu)\big]
\big[ {\sf f}_{3/2} - {\sf f}_{3/2}^{(1)} \big]\,
\Phi_{3/2}(\varepsilon,\mu;\Delta_{3/2})\nonumber\\
&&\;\;
+\Big\{ 
\ln\left|\frac{\varepsilon-\mu}{\varepsilon_0}\right|\,
\big[ {\sf f}_{\rm ln} - {\sf f}_{\rm ln}^{(1)} \big] 
+ {\sf f}_2' + i [ {\sf f}_2'' - \pi \Theta(\varepsilon-\mu)\,
{\sf f}_{\rm ln} ] \nonumber\\
&&\;\;\;\;\;\;\;
- {\sf f}_2^{\prime (1)} - i 
[ {\sf f}_2^{\prime\prime (1)} - \pi \Theta(\varepsilon-\mu)\,
{\sf f}_{\rm ln}^{(1)} ] \Big\}\,
\Phi_2(\varepsilon,\mu;\Delta_2),\nonumber\\
\end{eqnarray} 
where $\Delta_0$, $\Delta_1$, $\Delta_{3/2}$ and $\Delta_2$ are finite 
constants (about which more later) and $\mu$ stands for the `chemical 
potential' associated with ${\sf f}^{(1)}(\varepsilon)$, that is 
$\Sigma_{\sigma}^{(1)}({\Bf r},{\Bf r}';\varepsilon)$ (see Eqs.~(\ref{e22}) 
and (\ref{e23}) above). Assuming the latter function to have been 
evaluated in terms of the {\sl exact} $G_{\sigma}(\varepsilon)$, from 
the expression on the RHS of Eq.~(\ref{e251}) it can be readily deduced 
that the $\mu$ associated with ${\sf f}^{(1)}(\varepsilon)$ coincides 
with that associated with the {\sl exact} ${\sf f}(\varepsilon)$. We 
should emphasize that for {\sl non-metallic} GSs, where $\mu_{N;\sigma}^+ 
- \mu_{N;\sigma}^-$ (see Eqs.~(\ref{e22}) and (\ref{e23}) above) is 
finite, the statement with regard to the equality of the two `chemical 
potentials', associated with ${\sf f}(\varepsilon)$ and ${\sf f}^{(1)}
(\varepsilon)$, is rather meaningless, as for such GSs $\mu$ is {\sl not} 
uniquely defined (the absolute temperature is assumed to be zero in our 
considerations); however, for metallic GSs, for which $\mu_{N;\sigma}^+ 
- \mu_{N;\sigma}^- \sim 1/N^{\alpha}$, with $\alpha > 0$ (see footnote 
\ref{f60}), the above statement concerning the two $\mu$ values is both 
meaningful and non-trivial. We note in passing, that the $\mu$ associated 
with the ${\sf f}^{(1)}(\varepsilon) \equiv \Sigma_{\sigma}^{(1)}({\Bf r},
{\Bf r}';\varepsilon)$ (pertaining to a metallic GS; see above) 
evaluated in terms of the single-particle GF $G_{0;\sigma}(\varepsilon)$ 
pertaining to the {\sl exact} `non-interacting' Kohn-Sham (Kohn and Sham 
1965) Hamiltonian associated with the GS of $\wh{H}$ (Farid 1997a,b) 
equally coincides with the {\sl exact} $\mu$. For completeness, our 
considerations in Appendix A (\S~2 herein) indicate a way out also
{\sl explicitly} to correct the behaviour of ${\sf f}^{(1)}
(\varepsilon)$ for $\varepsilon$ close to $\mu$ (note that, there are 
some {\sl implicit} `corrections' to the behaviour of ${\sf f}^{(1)}
(\varepsilon)$ for $\varepsilon\to\mu$, brought about by the global 
conditions to be satisfied by ${\sf f}_{\rm m}^{(1)}(\varepsilon)$, 
enforced through the choice of the parameters $\Delta_0$, $\Delta_1$, 
$\dots$ in Eq.~(\ref{e281}); see later).

With reference to the appearance of $\ln\vert (\varepsilon
-\mu)/\varepsilon_0\vert$ on the RHS of Eq.~(\ref{e281}), to be 
contrasted with $\ln\vert\varepsilon/\varepsilon_0\vert$ on the 
RHS of Eq.~(\ref{e279}), we point out that, because
\begin{eqnarray}
\ln\left| \frac{\varepsilon-\mu}{\varepsilon_0}\right|
\sim \ln\left|\frac{\varepsilon}{\varepsilon_0}\right| 
-\frac{\mu}{\varepsilon}\;\;\;
\mbox{\rm for}\;\;\; \vert\varepsilon\vert\to\infty, \nonumber
\end{eqnarray}
the deviation of $\ln\vert (\varepsilon-\mu)/\varepsilon_0\vert$ from 
$\ln\vert\varepsilon/\varepsilon_0\vert$ gives rise to a contribution 
to the RHS of Eq.~(\ref{e281}) that for $\vert\varepsilon\vert\to\infty$
decays like $1/\varepsilon^3$. This is by one power of $1/\varepsilon$ 
less significant than the least significant term on the RHS of 
Eq.~(\ref{e279}) above. From this and in view of the 
large-$\vert\varepsilon\vert$ asymptotic results on the RHSs of the 
expressions in Eq.~(\ref{e280}), it follows that for $\vert\varepsilon
\vert\to\infty$, ${\sf f}^{(1)}_{\rm m}(\varepsilon)$ by construction 
exactly reproduces the asymptotic expression corresponding to the 
{\sl exact} ${\sf f}(\varepsilon)$ as presented in Eq.~(\ref{e279}).
In this connection, it is important to realize that taking account of 
{\sl any} number of correction terms (and thus neglecting the remaining 
terms) on the RHS of Eq.~(\ref{e281}) amounts to an improvement with 
respect to ${\sf f}^{(1)}(\varepsilon)$.

Concerning $\Delta_0$, $\Delta_1$, $\dots$ in the expression for 
${\sf f}_{\rm m}^{(1)}(\varepsilon)$ on the RHS of Eq.~(\ref{e281}), 
these free parameters do {\sl not} need to be all different and 
thus may in principle be identified with one parameter, $\Delta$ 
(see, however, later where we elaborate on the conditions to be 
satisfied by ${\sf f}_{\rm m}^{(1)}(\varepsilon)$ which signify the 
importance of sufficient flexibility in the expression for 
${\sf f}_{\rm m}^{(1)}(\varepsilon)$ and thus of $\Delta_0$, 
$\Delta_1$, $\dots$). Similarly, one can consider an alternative 
expression for ${\sf f}_{\rm m}^{(1)}(\varepsilon)$ where $[{\sf f}_0 
-{\sf f}_0^{(1)}] \Phi_0(\varepsilon,\mu;\Delta_0)$, etc., is replaced 
by ${\sf f}_0 \Phi_0(\varepsilon,\mu;\Delta_0')-{\sf f}_0^{(1)} 
\Phi_0(\varepsilon,\mu;\Delta_0'')$, etc., thus taking into account the 
possibility that ${\sf f}(\varepsilon)$, which ${\sf f}_{\rm m}^{(1)}
(\varepsilon)$ is supposed reliably to approximate, may differently 
approach its large-$\vert\varepsilon\vert$ asymptotic behaviour than 
${\sf f}^{(1)}(\varepsilon)$. Further, although ${\sf f}_{\rm m}^{(1)}
(\varepsilon)$ depends on $\Delta_0$, $\Delta_1$, $\dots$ so long as 
these are identified with {\sl finite} values, ${\sf f}_{\rm m}^{(1)}
(\varepsilon)$ exactly reproduces the RHS of Eq.~(\ref{e279}) for 
sufficiently large values of $\vert\varepsilon\vert$. In spite of 
this, the following aspects have to be carefully considered in choosing 
the values for $\Delta_0$, $\Delta_1$, $\dots$. 

--- Firstly, as can be obtained from Eqs.~(\ref{e39}) and (\ref{e24}),
\begin{equation}
\label{e282}
A_{\sigma}({\Bf r},{\Bf r}';\varepsilon) 
= \hbar\sum_{s} f_{s;\sigma}({\Bf r}) f_{s;\sigma}^*({\Bf r}')\,
\delta(\varepsilon-\varepsilon_{s;\sigma}),
\end{equation}
from which one deduces not only that $A_{\sigma}({\Bf r},{\Bf r};
\varepsilon)\ge 0$, $\forall\varepsilon$, but also that $A_{\sigma}
(\varepsilon)$, of which $A_{\sigma}({\Bf r},{\Bf r}';\varepsilon)$ 
is the coordinate representation, is a positive semi-definite 
operator for {\sl all} $\varepsilon$. This essential property must 
be preserved.

--- Secondly, by the requirement of conservation of particles of 
spin index $\sigma$, $\forall\sigma$, it is necessary that the 
single-particle spectral functions associated with the modified 
SEs, corresponding to all $\sigma$, satisfy the following equation
\begin{equation}
\label{e283}
\frac{1}{\hbar} \int_{-\infty}^{\mu} {\rm d}\varepsilon\;
\int {\rm d}^dr\; A_{\sigma}({\Bf r},{\Bf r};\varepsilon) 
= N_{\sigma},\;\; \forall\sigma,
\end{equation}
where $\{ N_{\sigma} \}$ is common to both the interacting and 
non-interaction GSs (see the paragraph containing Eqs.~(\ref{e54}) 
and (\ref{e55}) above). 

--- Thirdly, too small values for $\Delta_0$, $\Delta_1$, $\dots$ 
gives rise to a relatively rapid variation of ${\sf f}_{\rm m}^{(1)}
(\varepsilon)$ for $\varepsilon$ departing from $\mu$, while too large 
values for these parameters entail that, for a relatively wide range 
of $\varepsilon$ values away from $\varepsilon=\mu$, the behaviour of 
${\sf f}_{\rm m}^{(1)}(\varepsilon)$ is unduly dominated by that of 
${\sf f}^{(1)}(\varepsilon)$. In view of the considerations in 
\S~III.E, it is natural to expect that for any interacting system 
there exists an energy, similar to $e_0$ defined in Eq.~(\ref{e103}), 
with respect to which $\vert\varepsilon\vert$ can be considered as 
being `small' or `large' (for the specific system dealt with in 
\S~III.E, $\vert\varepsilon/e_0\vert$ is shown to be `large' provided 
it is larger than $\max(1,r_s)$) and which sets the scale over which 
a change in $\varepsilon$ produces a comparable change in 
$\hbar {\sf f}(\varepsilon)$. In this light we expect the parameters 
$\Delta_0$, $\Delta_1$, $\dots$ to be all on the order of the latter 
energy scale. Since ${\sf f}^{(1)}(\varepsilon)$ can be relatively 
easily calculated within the framework of the SSDA, one may deduce 
very good `estimates' (see later) for $\Delta_0$, $\Delta_1$, $\dots$ 
by means of fitting 
\begin{eqnarray}
&&{\sf f}_0^{(1)} \Phi_0(\varepsilon,\mu;\Delta_0)
+ \big[ {\sf f}_1^{\prime (1)} 
+i {\sf f}_1^{\prime\prime (1)} \big]\,
\Phi_1(\varepsilon,\mu;\Delta_1) \nonumber\\
&&\;\;\;
+\big[\Theta(\mu-\varepsilon) - i \Theta(\varepsilon-\mu)\big]
{\sf f}_{3/2}^{(1)}\, \Phi_{3/2}(\varepsilon,\mu;\Delta_{3/2})
\nonumber\\
&&\;\;\;
+\Big\{ \ln\left|\frac{\varepsilon-\mu}{\varepsilon_0}\right|\,
{\sf f}_{\rm ln}^{(1)} + {\sf f}_2^{\prime (1)} \nonumber\\
&&\;\;\;\;\;\;\;\;\;
+i [ {\sf f}_2^{\prime\prime (1)} 
-\pi\Theta(\varepsilon-\mu)\,
{\sf f}_{\rm ln}^{(1)} ] \Big\}\, 
\Phi_2(\varepsilon,\mu;\Delta_2)\nonumber
\end{eqnarray}
to the directly calculated ${\sf f}^{(1)}(\varepsilon)$ at `large' 
values of $\vert\varepsilon\vert$, taking into account the requirements
indicated above, in particular those implied by Eq.~(\ref{e283}). We note 
in passing that in practice it will be more convenient to obtain (at 
least) some of the coefficients ${\sf f}_0^{(1)}$, ${\sf f}_1^{\prime (1)}$, 
${\sf f}_1^{\prime\prime (1)}$, $\dots$ from the directly calculated 
${\sf f}^{(1)}(\varepsilon)$ (by means of some appropriate fitting 
procedure), than to determine these through evaluation of the pertinent 
expressions presented in \S\S~IV.A, B, C. Consequently, direct 
calculation of ${\sf f}^{(1)}(\varepsilon)$ at `large' values of 
$\vert\varepsilon\vert$ not only helps estimate the appropriate
values for $\Delta_0$, $\Delta_1$, $\dots$ but also facilitates
calculation of the sought-after coefficients ${\sf f}_0^{(1)}$, 
${\sf f}_1^{\prime (1)}$, ${\sf f}_1^{\prime\prime (1)}$, $\dots$.

--- Fourthly, attention should be paid to the fact that since 
${\sf f}_{\rm m}^{(1)}(\varepsilon)-{\sf f}^{(1)}(\varepsilon)$ is 
{\sl not} directly associated with a set of SE diagrams, 
${\sf f}_{\rm m}^{(1)}(\varepsilon)$ and the associated GF should 
{\sl not} in general identically satisfy the Luttinger-Ward (1960) 
identity presented in Eq.~(\ref{ea68}) (see the paragraph containing the 
indicated identity in Appendix A), so that this identity may be utilized
as a subsidiary condition (complimentary to the non-negativity condition
for $A_{\sigma}(\varepsilon)$ and those in Eq.~(\ref{e283}) above) to 
be fulfilled in determining the parameters $\Delta_0$, $\Delta_1$, 
$\dots$. 

\vspace{0.2cm}
We point out that, since determination of ${\sf f}_1''$ and 
${\sf f}_2''$ requires the knowledge of the real part of the 
{\sl exact} SE for $\varepsilon\in (-\infty, \infty)$ (see 
Eq.~(\ref{e227}) above), in practice it is necessary first to consider 
${\rm Re}[{\sf f}_{\rm m}^{(1)}(\varepsilon)]$ (whose expression does 
{\sl not} involve, amongst others, ${\sf f}_1''$ and ${\sf f}_2''$) 
and subsequent to choosing some appropriate values for the parameters 
$\Delta_0$, $\Delta_1$, $\dots$, determine ${\rm Im}[{\sf f}_{\rm m}^{(1)}
(\varepsilon)]$ from the Kramers-Kr\"onig relation in Eqs.~(\ref{e217}) 
and (\ref{e218}). We should emphasize that, since all the above conditions 
concern ${\sf f}_{\rm m}^{(1)}(\varepsilon)$, rather that 
${\rm Re}[{\sf f}_{\rm m}^{(1)}(\varepsilon)]$ and
${\rm Im}[{\sf f}_{\rm m}^{(1)}(\varepsilon)]$ separately, the 
aforementioned `appropriate' values for $\Delta_0$, $\Delta_1$, 
$\dots$ will have to be determined self-consistently. 

Finally, since the modified SE ${\sf f}_{\rm m}^{(1)}(\varepsilon)$
as evaluated within the framework of the SSDA may prove to be an 
accurate approximation to the exact ${\sf f}(\varepsilon)$, one may 
attempt to determine the `non-interacting' Hamiltonian underlying 
this approximation in a self-consistent manner along the lines of 
(Farid 1997a,b, 1999b).

\section{Summary and concluding remarks}
\label{s38}

In this work we have determined and presented the four leading 
terms in the {\sl formal} large-$\vert\varepsilon\vert$ AS of the 
single-particle GF $G_{\sigma}(\varepsilon)$ and the three leading 
terms in that of the SE operator $\Sigma_{\sigma}(\varepsilon)$ 
pertaining to a system of interacting fermions; for the first-order 
contribution to the SE in a perturbation series in terms of the 
{\sl exact} single-particle GF $G_{\sigma}(\varepsilon)$ pertaining 
to the {\sl interacting} system and the {\sl exact} dynamically 
screened interaction function $W(\varepsilon)$, as opposed to the 
static bare interaction function $v$, we have determined the 
{\sl four} leading asymptotic terms for $\Sigma_{\sigma}(\varepsilon)$. 
All the indicated terms, which are expressed in terms of {\sl static} 
{\sl GS} correlation functions, are amenable to numerical evaluation 
in terms of GS wavefunctions that are determined within such 
frameworks as the quantum Monte Carlo method. Each term in the 
large-$\vert\varepsilon\vert$ AS for $\Sigma_{\sigma}(\varepsilon)$ 
can be shown to correspond to well-specified contributions in the 
perturbation series of $\Sigma_{\sigma}(\varepsilon)$ in terms of $v$ 
and the {\sl exact} single-particle GF of the {\sl interacting} 
system, diagrammatically represented by means of skeleton SE diagrams; 
thus $\Sigma_{\sigma;\infty_0}$, the leading-order term in the 
large-$\vert\varepsilon\vert$ AS for $\Sigma_{\sigma}(\varepsilon)$, 
is the {\sl full} contribution of the first-order Hartree and Fock 
diagrams (which constitute the only first-order skeleton SE diagrams) 
and $\Sigma_{\sigma;\infty_m}/\varepsilon^m$ (see Eq.~(\ref{e72})), 
with $m \ge 1$, corresponds to the collection of skeleton diagrams 
of order $p$, with $p \in \{ 2,\dots, m+1\}$ (see specifically 
\S\S~III.E.4, 5). On replacing the exact GS correlation functions 
that feature in the expressions for the coefficients $\Sigma_{\sigma;
\infty_m}$, $m=0, 1, \dots$ (which in the case of the {\sl exact} 
SE we have presented in explicit form for $m=0,1$ and $2$) by those 
pertaining to a SSDA of the GS wavefunction of the interacting system, 
one arrives at highly non-trivial expressions for the mentioned 
coefficients that qualitatively (and sometimes even quantitatively) 
reproduce the most essential characteristics of the exact coefficients. 
This observation unequivocally establishes the crucial significance 
of skeleton SE diagrams, in comparison with the non-skeleton diagrams, 
within the framework of many-body perturbation theory. For clarity, 
with a SSDA for the GS wavefunction is associated a `non-interacting', 
or mean-field, many-body Hamiltonian $\wh{H}_0$ (see Eqs.~(\ref{e54}) 
and (\ref{e55})), of which the SSD is the GS and which defines the 
perturbation $\wh{H}_1 {:=} \wh{H} -\wh{H}_0$; consequently, the 
deviation of the SSDA to $\Sigma_{\sigma;\infty_m}$, with $m \ge 0$, 
from the exact $\Sigma_{\sigma;\infty_m}$ is attributable to the 
contributions of an {\sl infinite} set of {\sl non}-skeleton proper 
SE diagrams in terms of $G_{0;\sigma}$ (pertaining to the 
aforementioned Hamiltonian $\wh{H}_0$) whose `skeletons' are the 
skeleton diagrams of order one when $m=0$ and order $p$, with 
$p\in\{2,\dots,m+1\}$, when $m\ge 1$.
\footnote{\label{f103}
One should of course also take into account that $\Sigma_{\sigma;
\infty_0}^{\sh}$ (see Eqs.~(\protect\ref{e62}) and (\protect\ref{e72})) 
is directly dependent on the choice of $\wh{H}_0$. }
The influence of these neglected contributions is already apparent 
from the leading-order term $\Sigma_{\sigma;\infty_0}$ in the 
large-$\vert\varepsilon\vert$ AS of the SE operator in which one 
encounters the exact GS single-particle density matrix 
$\varrho_{\sigma}$ in explicit form (see Eq.~(\ref{e173})); the latter, 
in contrast with its SSDA counterpart $\varrho_{{\rm s};\sigma}$, is 
strictly {\sl non}-idempotent. That is, whereas $\varrho_{{\rm s};
\sigma} \varrho_{{\rm s};\sigma} \equiv \varrho_{{\rm s};\sigma}$, 
unless $v\equiv 0$, $\varrho_{\sigma} \varrho_{\sigma} \not\equiv 
\varrho_{\sigma}$. 

The asymptotic results for the SE operator can be directly employed 
in order to obtain accurate and well-controlled approximate results 
for the energies of the single-particle excitations and the associated 
wavefunctions (\S~III.D) in the regime of `large' excitation energies. 
For uniform and isotropic GSs we have {\sl explicitly} established an 
interplay between the single-particle excitation energies and the value 
for the average inter-particle distance in the GS and delineated the 
region where the latter energies can be considered as being `large'
(\S~III.E). In this connection it is important to note that, within the 
framework of the (self-consistent) Hartree-Fock approximation, the 
single-particle excitation energies are calculated in terms of 
$\Sigma^{\sc hf}({\Bf r},{\Bf r}';[\varrho_{{\rm s};\sigma}])$ over 
the {\sl entire} range of these energies; leaving aside the difference 
between $\varrho_{{\rm s};\sigma}$ and $\varrho_{\sigma}$ (see above), 
the fundamental shortcoming of the Hartree-Fock, or indeed {\sl any} 
static scheme, is already apparent from the absence of such 
{\sl non-trivial} contribution as $\Sigma_{\sigma;\infty_1}/z$ (see 
Eq.~(\ref{e72}); see also Eqs.~(\ref{e185}), (\ref{e187}) and 
(\ref{e188})) in the Hartree-Fock SE. We should here emphasize that our 
{\sl general} considerations in this work are based on the assumption 
that the interaction potential is such that, for an arbitrary finite
value of $m$, $m \in \{1,2, \dots\}$, $\Sigma_{\sigma;\infty_m}({\Bf r},
{\Bf r}')$ is well-defined and bounded for almost {\sl all} ${\Bf r}$ 
and ${\Bf r}'$. We have, however, explicitly considered the case where 
$d=3$ and $v\equiv v_c$, the Coulomb potential, for which we have 
established $\Sigma_{\sigma;\infty_m}({\Bf r},{\Bf r}')$ to be 
fundamentally unbounded and in need of being regularized (see \S~II.B) 
for $m\ge 2$; this regularization is effected through infinite partial 
summations over specific unbounded terms pertaining to $\Sigma_{\sigma;
\infty_p}({\Bf r},{\Bf r}')$ with $p > m$ (note that an unbounded 
contribution to $\Sigma_{\sigma;\infty_p}({\Bf r},{\Bf r}')$ can have 
already been taken account of in the process of regularizing an unbounded 
term in the expression for $\Sigma_{\sigma;\infty_{m'}}({\Bf r},
{\Bf r}')$ with $m' < m$). It is important to bear in mind the condition 
$d=3$, since for instance in the case of $d=2$ and Coulomb-interacting 
fermions, already $\Sigma_{\sigma;\infty_1}({\Bf r},{\Bf r}')$ is 
fundamentally unbounded and its regularization (see above) gives rise 
to a term immediately subsequent to $\Sigma_{\sigma;\infty_0}({\Bf r},
{\Bf r}')\equiv \Sigma^{\sc hf}({\Bf r},{\Bf r}';[\varrho_{\sigma}])$ 
in a finite-order large-$\vert\varepsilon\vert$ AS expansion of 
$\Sigma_{\sigma}({\Bf r},{\Bf r}';\varepsilon)$ that {\sl decays} like 
$1/\vert\varepsilon\vert^{1/2}$ (B. Farid, 2001, unpublished). Evidently, 
therefore, our statements with regard to Coulomb-interacting fermions both 
in this Section and in other parts of this paper should {\sl not} be taken 
as applying to all $d$; from this perspective it is significant also 
to recall that the Coulomb potential is not unique in bringing about 
fundamentally unbounded contributions associated with $\Sigma_{\sigma;
\infty_m}({\Bf r},{\Bf r}')$ and needful of regularization. Be it as it 
may, the appearance of the singular function $(-z/\varepsilon_0)^{1/2}/z$, 
to be contrasted with $(-z/\varepsilon_0)^{1/2}/z^2$ which corresponds 
to $d=3$, in the large-$\vert z\vert$ AS of $\wt{\Sigma}_{\sigma}(z)$ 
pertaining to systems of Coulomb-interacting fermions in $d=2$, 
demonstrates a fundamental hazard inherent in adopting formalisms that 
are specific to $d=\infty$ and {\sl bounded} $v$, to applications 
corresponding to systems of particles interacting through, e.g., 
$v\equiv v_c$ (see the following paragraph) and finite values of $d$, 
specifically $d=2$ and $3$. For the case of $v\equiv v_c$ and $d=3$ we 
explicitly calculate the first {\sl five} leading-order terms in the 
large-$\vert z\vert$ AS for $\wt{\Sigma}_{\sigma}(z)$. 

Our investigations concerning the SE operator within the framework 
of the first-order perturbation theory in terms of the dynamically 
screened interaction function $W(\varepsilon)$, as opposed to the 
bare interaction function $v$, have revealed that, even though this 
approximation to the SE were to be evaluated in terms of the {\sl exact} 
single-particle GF $G_{\sigma}(\varepsilon)$ and the {\sl exact} 
$W(\varepsilon)$, the resulting expression for $\Sigma_{\sigma;
\infty_1}$, the coefficient of the next-to-leading order term in the 
large-$\vert\varepsilon\vert$ AS for $\Sigma_{\sigma}^{(1)}
(\varepsilon)$, would solely consist of a {\sl local} contribution 
(see Eq.~(\ref{e257}) as well as Eq.~(\ref{e259})), in contrast with 
the {\sl exact} $\Sigma_{\sigma;\infty_1}$ which in addition consists 
of a non-trivial {\sl non-local} contribution (see Eqs.~(\ref{e185}), 
(\ref{e187}) and (\ref{e188})). This would not be as severe a 
shortcoming, were it not that the latter non-local contribution not 
only implicitly but also {\sl explicitly} depends on the spin state 
of the single-particle excitations. Further, as our direct calculations 
show, for systems of fermions in $d=3$ interacting through the Coulomb 
potential $v_c$, the next-to-leading term in the 
large-$\vert\varepsilon\vert$ AS for $\Sigma_{\sigma}({\Bf r},{\Bf r}';
\varepsilon)$ is followed by a {\sl local} term decaying like 
$1/\vert\varepsilon \vert^{3/2}$ for $\vert\varepsilon\vert\to\infty$ 
(to be contrasted with the term decaying like $1/\varepsilon^2$ in the 
cases where interaction potential is not as singular as the Coulomb 
potential $v_c({\Bf r}-{\Bf r}')$ at zero distance 
\footnote{\label{f104}
The long range of $v_c({\Bf r}-{\Bf r}')$ for $\|{\Bf r}-{\Bf r}'\|
\to\infty$ brings about a contribution proportional to 
$\ln(-z/\varepsilon_0)/z^2$, with $z=\varepsilon \pm i\eta$, $\eta
\downarrow 0$, for $\varepsilon\to \pm\infty$ (see 
Eqs.~(\protect\ref{e213}) and (\protect\ref{ef145})). }
and/or cases where the spectrum of the single-particle excitations
is bounded from above) which is fundamentally incorrectly described 
by the SE according to the first-order perturbation theory: compare 
the expressions in Eqs.~(\ref{e209}) and (\ref{e278}) (compare also 
Eq.~(\ref{eg1}) with Eq.~(\ref{e277})). This shortcoming arises
because, for $v \rightharpoonup v_c$ in $d=3$, some of the 
{\sl non}-local contributions to $\Sigma_{\sigma;\infty_2}$ transform 
into {\sl local} contributions; the first-order perturbation theory 
taking merely an {\sl incomplete} account of the non-local part of 
$\Sigma_{\sigma;\infty_2}$ (as we have indicated above, it takes 
{\sl no} account of the non-local part of $\Sigma_{\sigma;\infty_1}$), 
it fails to reproduce, even approximately, the local contributions as 
arising from non-local ones in $\Sigma_{\sigma;\infty_2}$ in consequence 
of the transformation $v\rightharpoonup v_c$. In addition to showing 
a main shortcoming of $\Sigma_{\sigma}^{\prime (1)}(\varepsilon)$ 
(see Eq.~(\ref{e241})), this observation unequivocally demonstrates 
that a dynamical local approximation to the SE operator that proves 
{\sl accurate} for systems of particles interacting through 
$v\not\equiv v_c$, is necessarily {\sl less accurate} when $v$ is 
identified with $v_c$.
This finding is specifically relevant in connection with the so-called 
`dynamical mean-field approximation' (for a review see Georges, 
{\sl et al.} 1996) to the SE operator which has found application in 
{\sl ab initio} calculations in real materials and is referred to as 
`LDA$^{++}$' (Lichtenstein and Katsnelson 1998, Katsnelson and 
Lichtenstein 2000); although the SE employed in this approach is 
$\varepsilon$-dependent, it is purely {\sl local}; it has its origin 
in the solution of the single-impurity Anderson (1961) problem and is 
related to the SE of the Hubbard model through the equivalence (Ohkawa 
1991, Georges and Kotliar 1992) of the former with the latter in $d=
\infty$ (Metzner and Vollhardt 1989). Our considerations unambiguously
show that, whereas the `dynamical mean-field approximation' to the SE 
operator may be accurate while dealing with the Hubbard Hamiltonian 
corresponding to a finite $d$ (in particular to $d=2,3$), it necessarily 
provides a less accurate description of the exact SE operator in systems 
where the inter-particle interaction is the Coulomb interaction. In 
this connection note that, in dealing with the Hubbard Hamiltonian, one 
has to deal with an intra-atomic particle-particle interaction potential 
that in stark contrast with the Coulomb potential $v_c$ is both bounded 
and localized. For a discussion of the limitations of the `dynamical 
mean-field approximation' from the standpoint of other authors (ours is 
{\sl exclusively} based on our observations in the present paper) 
see the review article by Georges, {\sl et al.} (1996, \S~IX). 

We have put forward a practicable formalism (see \S~IV.D) that
overcomes the shortcomings of the first-order SE operator 
(in terms of $W$) at large values of $\vert\varepsilon\vert$. The 
modified SE (see Eq.~(\ref{e281})) is applicable for {\sl all} 
values of $\varepsilon$ and, in view of its built-in correct 
asymptotic behaviour for $\vert\varepsilon\vert\to\infty$, is related 
to a single-particle spectral function $A_{\sigma}(\varepsilon)$ whose 
$\varepsilon$ moments integrals (to some finite order, depending on 
the number of terms in the large-$\vert\varepsilon\vert$ AS of the 
exact SE the modified SE is designed exactly to reproduce; see 
Eq.~(\ref{e281})) are {\sl identical} with those of the {\sl exact} 
$A_{\sigma}(\varepsilon)$ (see \S~III.B).

One of the consequences of the aforementioned misrepresentation of the 
SE within the framework of the first-order perturbation theory is that, 
according to this approximate theory, magnetism in its various forms 
is, roughly speaking, driven by `secondary' effects, as the SE in this 
framework does {\sl not} give due prominence to the {\sl explicit} 
dependence of the exact SE on the spin indices of the single-particle 
excitations. The relevance of this observation can be appreciated by 
recalling the early discussions concerning the occurrence of 
ferromagnetism in transition metal compounds and the corresponding 
deriving mechanisms (Slater 1953, Wohlfarth 1953, van Vleck 1953), 
while bearing in mind that the {\sl explicit} dependence of 
$\Sigma^{\sc hf}[\varrho_{\sigma}]\equiv \Sigma_{\sigma;\infty_0}
[\varrho_{\sigma}]$ (see Eq.~(\ref{e173})) upon $\sigma$ is solely
through $\varrho_{\sigma}$ (in practice, through $\varrho_{{\rm s};
\sigma}$ calculated self-consistently) and that, according to the 
current understanding of the subject (see also the following paragraph), 
`correlation' (as distinct from `exchange') is an essential aspect of 
magnetism in the indicated compounds (for example Fulde (1991)).

Concerning the relevance of the {\sl explicit} dependence on $\sigma$ 
of the exact $\Sigma_{\sigma}(\varepsilon)$, Harris and Lange (1967) 
observed that {\it ``in addition to band narrowing, one should include 
the spin-dependent shifts in the band energy which favor ferromagnetism 
when discussing questions of magnetic stability.''} Harris and Lange 
(1967) further observed that {\it ``the electron excitations have a 
longer lifetime in the paramagnetic phase than in the antiferromagnetic 
phase''}, which is in full conformity with our finding that the 
coefficient of the term in the large-$\vert\varepsilon\vert$ AS of 
$\Sigma_{\sigma}(\varepsilon)$ that decays like $1/\vert\varepsilon
\vert^{3/2}$ is {\sl identically} vanishing when the distribution of the 
spin-$\sigma$ (read `spin-$\uparrow$' for electrons) particles is in 
{\sl local} balance with that of the spin-$\bar\sigma$ (read 
`spin-$\downarrow$' for electrons) particles (see Eqs.~(\ref{eg7}) 
and (\ref{eg15})). This aspect is appreciated by considering the fact 
that the appearance of a term in the large-$\vert\varepsilon\vert$ 
AS for $\Sigma_{\sigma}(\varepsilon)$ decaying like $1/\vert\varepsilon
\vert^{3/2}$, there where a term decaying like $1/\varepsilon^2$, 
{\sl or} like $\ln(\vert\varepsilon/\varepsilon_0\vert)/\varepsilon^2$, 
would be expected, amounts to a broadening of the single-particle 
spectral function (compare cases (I) and (II) in Eq.~(\ref{e239})).
We point out that the term decaying like $1/\vert\varepsilon
\vert^{3/2}$ owes its existence to the short-range part of the Coulomb 
interaction function (see \S~III.H.2). It is important to realize 
that this observation and that by Harris and Lange (1967) based on 
investigations corresponding to the Hubbard Hamiltonian (involving a 
{\sl bounded} intra-atomic interaction), quoted above, do {\sl not} 
imply contradiction, for in our considerations we have dealt with 
systems with unbounded single-particle spectra, while in the case of 
the Hubbard model considered by Harris and Lange (1967), the 
single-particle spectrum is bounded. Consequently (see \S~I.C), in 
the former case the single-particle spectral function $A_{\sigma}
(\varepsilon)$ has an unbounded support, whereas in the latter case 
this function has a bounded support for which a {\sl finite}-order 
AS representation in terms of the sequence $\{1, 1/\varepsilon, 
1/\varepsilon^2,\dots\}$, similar to that presented in Eq.~(\ref{e239}), 
{\sl cannot} suffice. In this connection, recall that (see \S~II.B) the 
sequence $\{1, 1/\varepsilon,1/\varepsilon^2,\dots\}$ is an asymptotic 
one in the region $\vert\varepsilon\vert\to\infty$.

While dealing with the large-$\vert\varepsilon\vert$ asymptotic behaviour 
of the single-particle spectral function $\ol{A}_{\sigma}(k;\varepsilon)$ 
pertaining to uniform and isotropic GSs (see \S~III.I), we have emphasized 
the significance of inhomogeneity in GSs and the non-trivial influence 
of this (through, e.g., the Umklapp processes in periodic crystals) on 
the relationship between the SE operator and the single-particle spectral 
function. In particular, in such systems it is {\sl not} possible to 
deduce the diagonal elements of the self-energy operator in the momentum 
representation solely from those of the single-particle spectral function 
(and vice versa) as measured by means of angle-resolved photo-emission 
and inverse photo-emission spectroscopy, {\sl not even} in the regime 
of large transfer energies. Consequently, any attempt to model the 
diagonal elements of the self-energy operator in the momentum space, 
entirely on the basis of the measured diagonal elements of the 
single-particle spectral function of inhomogeneous systems, is 
{\sl a priori} doomed to be inaccurate, not only quantitatively, but 
also qualitatively. 

Our considerations have led us to a further conclusion which to our 
knowledge has never earlier been drawn in other works. In order to 
present this conclusion, we need first to provide some background 
detail. To this end consider the many-body perturbation series for 
$\Sigma_{\sigma}({\Bf r},{\Bf r}';\varepsilon)$ in terms of the 
{\sl bare} particle-particle interaction function $v$ and the skeleton 
SE diagrams. For definiteness consider $d=3$. It is well-known that 
for metallic GSs and $v\equiv v_c$, this series involves {\sl unbounded} 
contributions, arising from the long range of $v_c$ (for example 
Mattuck (1992, \S~10.4)); 
\footnote{\label{f105}
These unbounded contributions correspond to the set of polarization 
diagrams that constitute the RPA to the polarization function. }
these unbounded contributions are eliminated through expressing the 
perturbation series in terms of the {\sl screened} interaction function 
$W({\Bf r},{\Bf r}';\varepsilon)$ (see Eq.~(\ref{e242})) which, in 
contrast with the bare interaction function $v_c$, is dependent on 
$\varepsilon$. On the other hand, to our knowledge for systems whose 
GSs are non-metallic, {\sl no} unbounded contributions in the 
above-mentioned perturbation series had been expected thus far, 
so that for these systems the series in terms of $v_c$ is universally 
believed to be well-defined. 
\footnote{\label{f106}
A `well-defined' perturbation series is {\sl not} necessarily 
convergent. }
The interesting conclusion to which we have just referred is that the 
mentioned perturbation series, in terms of the bare $v_c$, is in fact
ill-defined {\sl even} for the latter systems (see the last two 
paragraphs of \S~III.H.2). We have arrived at this conclusion through 
the observation that for $v\equiv v_c$ and $d=3$, the coefficient of 
$1/\varepsilon^2$, i.e. $\Sigma_{\sigma;\infty_2}$, involves unbounded 
contributions arising from {\sl both} the long range of $v_c$ {\sl and} 
its specific form of divergence at the origin; 
\footnote{\label{f107}
Our calculations show that the divergence of $v_c({\Bf r}-{\Bf r}')$ for 
$\|{\Bf r}-{\Bf r}'\|\to 0$ plays a role in cases where $n_{\sigma}
({\Bf r}) \not\equiv n_{\bar\sigma}({\Bf r})$, where $n_{\bar\sigma}
({\Bf r})$ stands for the {\sl total} number density of fermions with 
the exclusion of those associated with the spin index $\sigma$ (see 
Eqs.~(\protect\ref{e163}) and (\protect\ref{e164})); in the case of 
spin-$1/2$ fermions, ${\bar\sigma} = -\sigma$ so that $n_{\bar\sigma}
({\Bf r}) \equiv n_{-\sigma}({\Bf r})$. }
in order to retain a well-defined finite-order AS for $\Sigma_{\sigma}
({\Bf r},{\Bf r}';\varepsilon)$ corresponding to $\vert\varepsilon\vert
\to\infty$, it is {\sl necessary} to perform partial summations over 
{\sl infinite} number of specific contributions pertaining to 
$\Sigma_{\sigma;\infty_m}({\Bf r},{\Bf r}')$, $m=2,3,4, \dots$. The 
necessity for performing such summations (here particularly owing to 
the specific behaviour of $v_c({\Bf r}-{\Bf r}')$ for $\|{\Bf r}
-{\Bf r}'\|\to 0$), irrespective of the nature of the GS, whether 
metallic or otherwise, makes evident that a {\sl finite}-order 
perturbation series in terms of the bare $v$ does {\sl not} suffice in 
cases where $v\equiv v_c$ and $d=3$, even for systems with non-metallic 
GSs (\S~II.B).

Performing, for $v\equiv v_c$ and $d=3$, the above-mentioned infinite 
partial summations, we have obtained that, in the large-$\vert\varepsilon
\vert$ AS for $\Sigma_{\sigma}({\Bf r},{\Bf r}';\varepsilon)$, 
$\Sigma^{\sc hf}({\Bf r},{\Bf r}';[\varrho_{\sigma}]) +\Sigma_{\sigma;
\infty_1}({\Bf r},{\Bf r}')/\varepsilon$ is followed by contributions 
that decay like $1/\vert\varepsilon\vert^{3/2}$, 
$\ln(\vert\varepsilon/\varepsilon_0\vert)/\varepsilon^2$, 
$1/\varepsilon^2$, $\dots$, rather than merely $1/\varepsilon^2$, 
$\dots$ as would be expected from the {\sl formal} structure of the 
AS for the SE corresponding to short-range and bounded interaction 
functions $v$ (see Eq.~(\ref{e72})). In addition, we have deduced that 
$\Sigma_{\sigma;\infty_2}({\Bf r},{\Bf r}')$ involves a contribution 
proportional to $v^3({\Bf r}-{\Bf r}')$ which, although well-defined 
in the case of $v\equiv v_c$ in $d=3$, is {\sl not} integrable, 
implying (see \S~II.B) an associated fundamentally unbounded contribution 
in the momentum representation of $\Sigma_{\sigma;\infty_2}$. We have 
resolved this problem through a further summation over an infinite 
number of pertinent non-integrable contributions pertaining to 
$\Sigma_{\sigma;\infty_m}({\Bf r},{\Bf r}')$, with $m \ge 2$, and 
deduced an {\sl additional} singular contribution of the form 
$\ln(\vert\varepsilon/\varepsilon_0\vert)/\varepsilon^2$ to the 
AS of the momentum representation of $\Sigma_{\sigma}(\varepsilon)$ 
for $\vert\varepsilon\vert\to\infty$. We observe that the most 
dominant singular contribution to a finite-order AS of 
$\wt{\Sigma}_{\sigma}(z)$ for $\vert z\vert\to\infty$ in the cases 
corresponding to $v\equiv v_c$ in $d=3$, both in the coordinate 
representation and in the momentum representation, is proportional to 
$(-z/\varepsilon_0)^{1/2}/z^2$ which has its origin in the specific 
behaviour of $v_c({\Bf r}-{\Bf r}')$ for $\|{\Bf r}-{\Bf r}'\| \to 0$
(the existence of this contribution also vitally depends on the
single-particle excitation spectra being unbounded from above). 
Importantly, this singular contribution in a finite-order 
large-$\vert z\vert$ AS for $\wt{\Sigma}_{\sigma}(z)$ owes its 
existence to the number density of particles with spin index $\sigma$ 
being in at least local {\sl imbalance} with the total number density 
of the remaining particles in the system (see footnote \ref{f107}). 

The appearance of functions with branch-cut singularity in the 
AS expansion of $\wt{\Sigma}_{\sigma}(z)$ for $\vert z\vert\to\infty$, 
such as $(-z/\varepsilon_0)^{1/2}/z^2$ and $\ln(-z/\varepsilon_0)/z^2$ 
(see \S~III.H), implies not only that $\wt{\Sigma}_{\sigma}(z)$ 
possesses a manifold of branch-point singularities at the point of 
infinity in the complex $z$ plane, but also that $\wt{\Sigma}_{\sigma}
(z)$ must possess associated branch-point singularities in the 
{\sl finite} part of the $z$ plane. Since with the exclusion of the 
real energy axis, $\wt{\Sigma}_{\sigma}(z)$ is analytic everywhere 
on the complex $z$ plane, it follows that the latter branch-point 
singularities must be located on the finite part of the real energy 
axis. Knowledge of the location of each of these points, say 
$\varepsilon_j$, together with that of the associated singular 
function in the AS of $\wt{\Sigma}_{\sigma}(z)$, for $\vert z\vert
\to\infty$, is sufficient partially to uncover the behaviour of 
$\wt{\Sigma}_{\sigma}(z)$ for $\vert z-\varepsilon_j \vert\to 0$. 
This aspect is of particular significance in the case of systems 
with metallic GSs where the coincidence of $\varepsilon_j$ (and 
possibly other singular points) with the Fermi energy $\varepsilon_F$ 
of the system provides one with information of potentially considerable 
significance (this depending on the singular function at issue) 
concerning the nature of the underlying metallic state (see \S~I.B). 
In this connection it is important to realize that determination of 
the terms in the AS of $\wt{\Sigma}_{\sigma}(z)$ pertinent to $\vert 
z\vert\to\infty$ is significantly less demanding than that of the 
terms in the AS of $\wt{\Sigma}_{\sigma}(z)$ appropriate to 
$\vert z-\varepsilon_F\vert\to 0$. Consequently, although to our 
knowledge there exists no {\sl a priori} reason for the coincidence 
of the Fermi energy of systems with metallic GSs with one or some of the 
counterparts of the singularities of $\wt{\Sigma}_{\sigma}(z)$ at the 
point of infinity of the $z$ plane (see footnote \ref{f12}; although 
there {\sl may} be conditions, as yet to be established, under which 
such coincidence would become unavoidable), it is nonetheless 
advantageous to establish the singular behaviour of 
$\wt{\Sigma}_{\sigma}(z)$ for $\vert z\vert\to \infty$ in considerations 
that primarily are concerned with the behaviour of $\wt{\Sigma}_{\sigma}
(z)$ for $\vert z-\varepsilon_F\vert\to 0$; knowledge of the behaviour
of $\wt{\Sigma}_{\sigma}(z)$ for $\vert z\vert \to \infty$ provides one 
with much valuable information not only concerning the single-particle 
excitations at high excitation energies (\S~III.E) but also, through 
the interplay between the latter and the energy moments integrals of 
the single-particle spectral function $A_{\sigma}(\varepsilon)$ (see 
\S~III.B), concerning the positions, widths and other aspects of the 
prominent peaks in $A_{\sigma}(\varepsilon)$.

Finally, although our attention in the present work has been mainly 
centred around the single-particle GF and the SE operator, much of 
our considerations are directly applicable to other dynamical 
correlation functions pertaining to interacting systems. Of these,
to name but two, one is the dynamical density-density response 
function (see Engel and Farid 1993), which has featured in our 
investigations through the dynamically screened interaction function 
$W(\varepsilon)$ (see \S~IV), and the dynamical spin-spin correlation 
function, which although it is related to the latter response function 
(see footnote \ref{f99}), has {\sl not} appeared as prominently as
$W(\varepsilon)$ in the foreground of our discussions. In particular 
we should like to emphasize that the scheme proposed in \S~IV.D for 
removing the shortcomings of an approximation to the SE operator, can 
be trivially extended for improving the approximate expressions for 
any of the above-mentioned dynamical correlation functions; one may 
think of these approximate expressions as being those according to 
the RPA of these functions.

\section*{Acknowledgements}
\label{s39}

It is a pleasure for me to thank Professor Peter Littlewood for his 
kind hospitality at Cavendish Laboratory where the major part of the 
work presented here was carried out and Professor Philip Stamp for 
his kind hospitality at Spinoza Institute where this work was completed. 
My special thanks are due to Dr Richard Needs whose encouragement
has been essential for my undertaking of the work reported here. 
I dedicate this work to the memory of my mother, Afagh 
Ghardashem-Farid (1936-2000).

\vspace{0.5cm}
\noindent 
{\it Note added in proof.} 
\label{s39a}
It has very recently come to our attention that Deisz, {\sl et al.}
(1997) have explicitly demonstrated that for large 
$\vert\varepsilon\vert$, the exact $1/\varepsilon$ dependence of 
the self-energy pertaining to the single-band Hubbard Hamiltonian 
is entirely determined by the second-order skeleton self-energy 
diagram evaluated in terms of the exact single-particle Green 
function. Our findings (see \S\S~I.B, III.E, III.H and V) not 
only conform with this result, but they reveal an intimate
relationship between $\Sigma_{\sigma;\infty_m}$ (the coefficient 
corresponding to the $1/\varepsilon^m$ term, $\forall m$, in the 
large-$\vert\varepsilon\vert$ asymptotic series of $\Sigma_{\sigma}
(\varepsilon)$ pertaining to a {\sl general} Hamiltonian) and 
specific sets of skeleton self-energy diagrams.

Further, the coefficient of $1/\varepsilon^2$ on the RHS of 
Eq.~(\ref{e227}) can be {\sl simplified} by absorbing the 
term $2\Sigma_{\sigma;\infty_1}/\Delta$ in the $\varepsilon'$ 
integral. Upon this, through the same line of reasoning as 
presented in the paragraph preceding Eq.~(\ref{e227}), for the 
coefficient of $1/\varepsilon^2$ on the RHS of Eq.~(\ref{e227}), 
that is $\Pi_{\sigma}$ (see the text following equation 
Eq.~(\ref{e234})), we obtain
\begin{eqnarray}
&&\Pi_{\sigma} = -\wp\!\int_0^{\infty}
\frac{{\rm d}\varepsilon'}{{\varepsilon'}^3}\;
\Big\{ {\rm Re}\Sigma_{\sigma}(1/\varepsilon') \nonumber\\
&&\;\;\;\;\;\;\;\;\;\;\;\;\;\;\;\;\;\;\;\;
-{\rm Re}\Sigma_{\sigma}(-1/\varepsilon')
-2\varepsilon'\,\Sigma_{\sigma;\infty_1}\Big\}.
\nonumber
\end{eqnarray}
\hfill $\Box$

\section*{Appendices}
\label{s40}

\begin{appendix}
\section{Non-orthogonality and over-completeness of the set of
Lehmann amplitudes }
\label{s41}

Here we demonstrate that, for systems of interacting fermions, the 
Lehmann amplitudes $\{ f_{s;\sigma}({\Bf r})\}$ as defined in 
Eq.~(\ref{e18}) do {\sl not} form an orthonormal set. This, together 
with the fact that $\{f_{s;\sigma}({\Bf r})\}$ satisfies the closure 
relation (see Eq.~(\ref{e30})), establishes the {\sl over-completeness} 
of the set of the Lehmann amplitudes. This aspect is important for a 
proper understanding of the nature of the single-particle excitations 
in interacting systems in particular it sheds light on the significance 
of Eq.~(\ref{e42}). In \S~A.2, by employing a number of the general 
results of \S~A.1, we deduce a simple approximate expression for the 
dispersion of the single-particle excitation energies in uniform 
systems. This provides us with an opportunity to view some of the 
abstract concepts of \S~A.1 in a more applied context.

\subsection{Basic considerations; the overcompleteness} 
\label{s42}

Consider the `non-interacting' Hamiltonian $\wh{H}_0$ in Eq.~(\ref{e54}) 
with the single-particle Hamiltonian $h_{0;\sigma}({\Bf r})$ defined 
in Eq.~(\ref{e55}). Let $\{\varphi_{\varsigma;\sigma}({\Bf r}) \}$ be 
the {\sl complete} set of orthonormal eigenfunctions of $h_{0;\sigma}
({\Bf r})$ (see Eq.~(\ref{e56})). 
\footnote{\label{f108}
The considerations of this Appendix could have been based on any 
{\sl complete} and orthonormal set of one-particle orbitals (note 
that `completeness' implies, among other things, compatibility with 
the single-particle Hilbert space of the problem at hand), notably 
$\{\phi_{\varsigma}({\Bf r})\}$, the orthonormal set of eigenfunctions 
of the non-interacting single-particle Hamiltonian $h_0({\Bf r})$ 
defined in Eq.~(\protect\ref{e43}) (see Eq.~(\protect\ref{e44})). The 
drawback of employing any other complete set than $\{\varphi_{\varsigma;
\sigma}({\Bf r})\}$ would be to deprive ourselves of the ability to relate 
our results to a `non-interacting' system and thereby of the possibility 
to attribute certain properties to `interaction'. Although viewed from 
this perspective, it would appear that the choice of $\{\phi_{\varsigma}
({\Bf r})\}$ would be more appropriate, and although in certain cases the 
choice between $\{\phi_{\varsigma}({\Bf r})\}$ and $\{\varphi_{\varsigma;
\sigma}({\Bf r})\}$ can be immaterial, {\sl in general} the GS of the 
truly non-interacting Hamiltonian $\wh{H}_0 = \wh{T}+\wh{U}$ (see 
Eq.~(\protect\ref{e1})), corresponding to $h_0({\Bf r})$, is {\sl not} 
adiabatically connected with that of the fully interacting Hamiltonian 
(see \S~III.C, in particular the paragraph following that containing 
Eq.~(\protect\ref{e61})). Consequently, in such cases any reference to 
`interaction effects' would be spurious, for the adiabatic evolution 
of the GS of the $\wh{H}_0$ through the adiabatic `switching on' of 
$\wh{V}$ (see Eq.~(\protect\ref{e1})) would not lead to the true GS of 
$\wh{H}$ but to some other eigenstate. It follows that the choice of 
$\{\varphi_{\varsigma;\sigma}({\Bf r})\}$ is far superior than that 
of $\{\phi_{\varsigma}({\Bf r})\}$. It is significant to realize that 
(see Eq.~(\protect\ref{e56}) and the subsequent discussions) by reducing 
the value of the coupling constant of the particle-particle interaction, 
the set $\{\varphi_{\varsigma;\sigma}({\Bf r})\}$ naturally reduces to 
$\{\phi_{\varsigma}({\Bf r})\}$ so that our results in this Appendix 
also retain a natural connection with the truly non-interacting system 
described by $\wh{T}+\wh{U}$. } 
Completeness implies the closure relation
\begin{equation}
\label{ea1}
\sum_{\varsigma} \varphi_{\varsigma;\sigma}^*({\Bf r})
\varphi_{\varsigma;\sigma}({\Bf r}') = \delta({\Bf r}-{\Bf r}').
\end{equation}
With ${\sf\hat a}_{\varsigma;\sigma}$ the fermion annihilation 
operator corresponding to $\varphi_{\varsigma;\sigma}({\Bf r})$, 
satisfying ({\it cf.} Eq.~(\ref{e91}))
\begin{eqnarray}
\label{ea2}
\big[ {\sf\hat a}_{\varsigma;\sigma}^{\dag}, 
{\sf\hat a}_{\varsigma';\sigma'} \big]_+ &=& \delta_{\sigma,\sigma'}\,
\delta_{\varsigma,\varsigma'},\nonumber\\
\big[ {\sf\hat a}_{\varsigma;\sigma}^{\dag}, 
{\sf\hat a}_{\varsigma';\sigma'}^{\dag} \big]_+ &=& 
\big[ {\sf\hat a}_{\varsigma;\sigma}, 
{\sf\hat a}_{\varsigma';\sigma'} \big]_+ = 0, 
\end{eqnarray}
we have the following representation for the annihilation field 
operator $\hat\psi_{\sigma}({\Bf r})$:
\begin{equation}
\label{ea3}
\hat\psi_{\sigma}({\Bf r}) = \sum_{\varsigma}
{\sf\hat a}_{\varsigma;\sigma}\,\varphi_{\varsigma;\sigma}({\Bf r}).
\end{equation}
From the definition for the Lehmann amplitudes in Eq.~(\ref{e18})
we thus deduce that
\begin{equation}
\label{ea4}
f_{s;\sigma}({\Bf r}) = \sum_{\varsigma} 
\chi_{s;\sigma}(\varsigma)\,
\varphi_{\varsigma;\sigma}({\Bf r}),
\end{equation}
where
\begin{eqnarray}
\label{ea5}
\chi_{s;\sigma}(\varsigma) {:=} \left\{ \begin{array}{ll}
\langle\Psi_{N_{\sigma}-1,N_{\bar\sigma};s} \vert
{\sf\hat a}_{\varsigma;\sigma}
\vert \Psi_{N;0}\rangle,
&\;\;\; \varepsilon_{s;\sigma} < \mu, \\ \\
\langle\Psi_{N;0} \vert {\sf\hat a}_{\varsigma;\sigma}
\vert \Psi_{N_{\sigma}+1,N_{\bar\sigma};s}\rangle,
&\;\;\; \varepsilon_{s;\sigma} > \mu.
\end{array} \right.
\end{eqnarray}
Thus we have
\begin{equation}
\label{ea6}
S_{s,s'} {:=} \int {\rm d}^dr\; f_{s;\sigma}^*({\Bf r})
f_{s';\sigma}({\Bf r}) = \sum_{\varsigma} 
\chi_{s;\sigma}^*(\varsigma)
\chi_{s';\sigma}(\varsigma).
\end{equation}
Let now ${\sf n}_{\sigma}(\varsigma)$ be the GS distribution function 
corresponding to the particles associated with $\{{\sf\hat a}_{\varsigma;
\sigma}\}$, defined as ({\it cf}. Eq.~(\ref{ej2}))
\begin{equation}
\label{ea7}
{\sf n}_{\sigma}(\varsigma) {:=}
\langle \Psi_{N;0}\vert 
{\sf\hat a}_{\varsigma;\sigma}^{\dag} 
{\sf\hat a}_{\varsigma;\sigma} 
\vert \Psi_{N;0}\rangle.
\end{equation}
With
\begin{eqnarray}
\label{ea8}
\nu_{\sigma}(\varsigma)
{:=} \left\{ \begin{array}{ll}
{\sf n}_{\sigma}(\varsigma), 
&\;\;\; \varepsilon_{s;\sigma} < \mu, \\
1-{\sf n}_{\sigma}(\varsigma),
&\;\;\; \varepsilon_{s;\sigma} > \mu,
\end{array} \right.
\end{eqnarray}
for $\nu_{\sigma}(\varsigma) \not= 0$, we define
\begin{eqnarray}
\label{ea9}
&&\vert\Psi^{0}_{N_{\sigma}-1,N_{\bar\sigma};\varsigma}\rangle
{:=} \frac{1}{\sqrt{\nu_{\sigma}(\varsigma)}}\,
{\sf\hat a}_{\varsigma;\sigma} \vert\Psi_{N;0}\rangle,\\
\label{ea10}
&&\vert\Psi^{0}_{N_{\sigma}+1,N_{\bar\sigma};\varsigma}\rangle
{:=} \frac{1}{\sqrt{\nu_{\sigma}(\varsigma)}}\,
{\sf\hat a}_{\varsigma;\sigma}^{\dag} \vert\Psi_{N;0}\rangle.
\end{eqnarray}
It can be readily verified that the state vectors defined in 
Eqs.~(\ref{ea9}) and (\ref{ea10}) are normalized to unity and 
moreover are eigenstates of the {\sl partial} number operators 
$\wh{N}_{\sigma}$, $\sigma = -{\sf s}, -{\sf s}+1, \dots, 
{\sf s}$, corresponding respectively to $N_{\sigma}-1$ and 
$N_{\sigma}+1$ particles with spin $\sigma$.

Making use of the completeness of $\{\vert\Psi_{N_{\sigma}-1,
N_{\bar\sigma};s}\rangle \}$ and $\{\vert\Psi_{N_{\sigma}+1,
N_{\bar\sigma};s}\rangle \}$ in the $(N_{\sigma}-1
+N_{\bar\sigma})$- and $(N_{\sigma}+1+N_{\bar\sigma})$-particle 
subspaces respectively of the Fock space (see Eq.~(\ref{e31})), 
one can write
\begin{equation}
\label{ea11}
\vert\Psi^{0}_{N_{\sigma}\pm 1,N_{\bar\sigma};\varsigma}\rangle
= \sum_s \gamma^{\pm}_{s;\sigma}(\varsigma) 
\vert\Psi_{N_{\sigma}\pm 1,N_{\bar\sigma};s}\rangle,
\end{equation}
from which, following the normalization to unity of the state 
vectors on both sides of Eq.~(\ref{ea11}), one obtains
\begin{equation}
\label{ea12}
\sum_s \vert \gamma^{\pm}_{s;\sigma}(\varsigma) \vert^2 \equiv 1,
\;\;\; \forall \, \varsigma.
\end{equation}
Rewriting Eq.~(\ref{ea9}) as 
\begin{eqnarray}
{\sf\hat a}_{\varsigma;\sigma} \vert\Psi_{N;0}\rangle
= \sqrt{\nu_{\sigma}(\varsigma)}\,
\vert\Psi^{0}_{N_{\sigma}-1,N_{\bar\sigma};\varsigma}\rangle
\nonumber
\end{eqnarray}
and Eq.~(\ref{ea10}) as
\begin{eqnarray}
\langle\Psi_{N;0}\vert {\sf\hat a}_{\varsigma;\sigma} 
= \sqrt{\nu_{\sigma}(\varsigma)}\,
\langle\Psi^{0}_{N_{\sigma}+1,N_{\bar\sigma};\varsigma}\vert,
\nonumber
\end{eqnarray}
from Eq.~(\ref{ea5}), making use of the representation in 
Eq.~(\ref{ea11}) together with the orthonormality property 
$\langle\Psi_{N_{\sigma}\pm 1,N_{\bar\sigma};s}\vert 
\Psi_{N_{\sigma}\pm 1,N_{\bar\sigma};s'}\rangle = \delta_{s,s'}$, 
it readily follows that
\begin{eqnarray}
\label{ea13}
\chi_{s;\sigma}(\varsigma) = \left\{ \begin{array}{ll}
\sqrt{\nu_{\sigma}(\varsigma)}\, 
\gamma^{-}_{s;\sigma}(\varsigma),
&\;\;\; \varepsilon_{s;\sigma} < \mu, \\ \\
\sqrt{\nu_{\sigma}(\varsigma)}\, 
\gamma^{+*}_{s;\sigma}(\varsigma),
&\;\;\; \varepsilon_{s;\sigma} > \mu,
\end{array} \right.
\end{eqnarray}
from which we deduce that
\begin{equation}
\label{ea14}
\sum_{\varsigma} \chi_{s;\sigma}^*(\varsigma)
\chi_{s;\sigma}(\varsigma)
= \sum_{\varsigma} \nu_{\sigma}(\varsigma)\,
\vert\gamma^{\mp}_{s;\sigma}(\varsigma)\vert^2,\;\;
\varepsilon_{s;\sigma}\, \IEq>< \, \mu.
\end{equation}
The case corresponding to $s\not=s'$, involving {\sl four} 
possibilities associated with the four combinations 
$(\varepsilon_{s;\sigma} < \mu, \varepsilon_{s';\sigma} < \mu)$, 
$(\varepsilon_{s;\sigma} < \mu, \varepsilon_{s';\sigma} > \mu)$, 
etc., can be easily dealt with; however, we shall not consider these
in this Appendix.

From $\langle\Psi_{N_{\sigma}\pm 1,N_{\bar\sigma};s}\vert 
\Psi_{N_{\sigma}\pm 1,N_{\bar\sigma};s'}\rangle = \delta_{s,s'}$ and
Eq.~(\ref{ea11}) we have ({\it cf.} Eq.~(\ref{ea12}))
\begin{eqnarray}
\label{ea15}
&&\sum_{\varsigma} \vert \gamma^{\pm}_{s;\sigma}(\varsigma) \vert^2 
= \langle\Psi_{N_{\sigma}\pm 1,N_{\bar\sigma};s}\vert
\nonumber\\
&&\;\;\;
\times\Big( \sum_{\varsigma} \vert\Psi^{0}_{N_{\sigma}\pm 1,
N_{\bar\sigma};\varsigma}\rangle
\langle \Psi^{0}_{N_{\sigma}\pm 1,
N_{\bar\sigma};\varsigma}\vert\Big)
\vert\Psi_{N_{\sigma}\pm 1,N_{\bar\sigma};s}\rangle = 1,
\nonumber\\
&&\;\;\;\;\;\;\;\;\;\;\;\;\;\;\;\;\;\;\;\;\;\;\;\;\;\;\;\;\;
\;\;\;\;\;\;\;\;\;\;\;\;\;\;\;\;\;\;\;\;\;\;\;\;\;\;\;\;\;\;
\;\;\;\;\;\;\;\;\;
\forall\, s,
\end{eqnarray}
where we have made use of the fact that the projection operator on 
the RHS of Eq.~(\ref{ea15}), enclosed by large parentheses, can be 
replaced by the unit operator $I$ (and that $\vert\Psi_{N_{\sigma}\pm 1,
N_{\bar\sigma};s}\rangle$ are normalized to unity). It is important 
to note that although in the case of interacting systems the compound 
variable $\varsigma$ associated with the states defined in 
Eqs.~(\ref{ea9}) and (\ref{ea10}) belongs to a {\sl proper} subset 
of the set to which the compound variable $s$ belongs,
\footnote{\label{f109}
Following Klein and Prange (1958) (see footnote 14 and the text 
following Eq.~(51) herein), we may write $s=(\varsigma,\alpha)$ where
$\alpha$ stands for a ``parameter of degeneracy'', distinguishing 
various states corresponding to a given $\varsigma$. It is this 
extension of the parameter space required for marking the 
$(N\pm 1)$-particle eigenstates of $\wh{H}$ (with the $N$-particle
GS as the reference state), in comparison with that required for 
marking a complete basis set for the single-particle Hilbert state of 
the problem, that gives rise to the over-completeness of 
$\{ f_{s;\sigma}({\Bf r})\}$. As we have pointed out in footnote 
\protect\ref{f50}, the designation ``parameter of degeneracy'' does 
not appropriately reflect the true significance of $\alpha$. }
implying that in general 
\begin{equation}
\label{ea16}
\langle \Psi^{0}_{N_{\sigma}\pm 1,N_{\bar\sigma};\varsigma}\vert 
\wh{H} \vert \Psi^{0}_{N_{\sigma}\pm 1,N_{\bar\sigma};\varsigma'}
\rangle \not= 0,\;\; \mbox{\rm for}\;\; \varsigma \not=\varsigma',
\end{equation}
nonetheless, $\{ \vert\Psi^{0}_{N_{\sigma}\pm 1,N_{\bar\sigma};
\varsigma}\rangle\}$ is {\sl complete} within the Hilbert spaces of 
the $(N_{\sigma}\pm 1+N_{\bar\sigma})$-particle states; the indicated 
difference between the nature of the variables $s$ and $\varsigma$ 
reflects the {\sl over-completeness} of the set 
$\{ \vert\Psi_{N_{\sigma}\pm 1, N_{\bar\sigma};s}\rangle\}$ within 
the above-mentioned Hilbert spaces. This over-completeness implies 
that the overlap matrix ${\Bf S}$, whose $(s,s')$ component is defined 
in Eq.~(\ref{ea6}), is defective, that is $\det[{\Bf S}]=0$. 

For {\sl `non-interacting'} systems,
\footnote{\label{f110}
In this case, all state vectors that in our above considerations 
have involved $\Psi$ (such as $\vert\Psi_{N_{\sigma}\pm 1,
N_{\bar\sigma};s}\rangle$, etc.), should be viewed as having been
replaced by their non-interacting counterparts involving $\Phi$ 
(such as $\vert\Phi_{N_{\sigma}\pm 1,N_{\bar\sigma};\varsigma}\rangle$, 
etc.). See footnote \protect\ref{f108} for the basic property required 
from `non-interacting' systems.  }
we have ({\it cf.} Eq. (\ref{ea11}))
\begin{equation}
\label{ea17}
\gamma^{\mp}_{s;\sigma}(\varsigma) = \delta_{s,\varsigma},\;\;\;
\nu_{\sigma}(\varsigma) = 1,
\end{equation}
from which, in conjunction with Eqs.~(\ref{ea13}) and (\ref{ea6}), 
it follows that $\int {\rm d}^dr\, f_{s;\sigma}^*({\Bf r}) f_{s;
\sigma}({\Bf r})=1$ where $f_{s;\sigma}({\Bf r}) \equiv 
\varphi_{\varsigma;\sigma}({\Bf r})$ (note that, as our notation 
involving $\delta_{s,\varsigma}$ makes explicit, here the variables $s$ 
and $\varsigma$ are elements of the same set; see text following 
Eq.~(\ref{e48})). 
 
For {\sl interacting} systems, we have 
\footnote{\label{f111}
Consider uniform and isotropic systems of interacting spin-$1/2$ 
fermions in $d=3$. The fact that for these systems $0 < {\sf n}_{\sigma}
(k) < 1$, $\forall k$, so that $0 < \nu_{\sigma}(k) < 1$, $\forall k$, 
implies that for the case at hand {\sl all} the following uses of the 
word `some' are in fact to be replaced by the word `all'. Most 
importantly, for these systems $\int {\rm d}^dr\; f_{s;\sigma}^*({\Bf r}) 
f_{s;\sigma}({\Bf r}) < 1$ for {\sl all} $s$ ({\it cf}. 
Eq.~(\protect\ref{ea20})). It has to be noted, however, that the rapid 
decay of ${\sf n}(k)$ for $k\to\infty$ as presented in 
Eq.~(\protect\ref{ej4}) (see also the text following this equation) 
and consequently the rapid approach of the corresponding $\nu(k) \equiv 
\sum_{\sigma=\uparrow,\downarrow} \nu_{\sigma}(k)$ towards unity (see 
Eq.~(\protect\ref{ea8})) as $k\to\infty$, is indicative of the fact 
that, for $\varepsilon_{s;\sigma}$ larger than a small multiple of 
$\varepsilon_F$ (say, larger than twice to three times $\varepsilon_F$), 
to a very good approximation $\int {\rm d}^dr\; f_{s;\sigma}^*({\Bf r}) 
f_{s;\sigma}({\Bf r})$ can be identified with unity ({\it cf}. 
Eq.~(\protect\ref{e48}) above). This in turn specifies the range of 
$\varepsilon$ above which ${\rm Im}[\ol{\Sigma}_{\sigma}(k;\varepsilon)]$ 
can be considered as being negligible. }
\begin{equation}
\label{ea18}
0 \le \nu_{\sigma}(\varsigma) \le 1
\;\;\;\;\mbox{\rm and}\;\;\;\; \exists \varsigma
\;\;\mbox{\rm for which}\;\;
\nu_{\sigma}(\varsigma) \not= 1,
\end{equation}
from which and Eq.~(\ref{ea15}) it follows that
\begin{eqnarray}
\label{ea19}
\sum_{\varsigma} \nu_{\sigma}(\varsigma)
\vert\gamma^{\mp}_{s;\sigma}(\varsigma)\vert^2 
&<& \sum_{\varsigma} 
\vert\gamma^{\mp}_{s;\sigma}(\varsigma)\vert^2 = 1,\nonumber\\
& &\;\;\;\;\;\;\;\;\;\;\;\;\;\mbox{\rm for {\sl some}}\;\; s,
\end{eqnarray}
which implies, through Eqs.~(\ref{ea6}) and (\ref{ea14}), that for 
{\sl interacting} systems we must have
\begin{equation}
\label{ea20}
\int {\rm d}^dr\;
f_{s;\sigma}^*({\Bf r})\,
f_{s;\sigma}({\Bf r}) < 1,
\;\;\mbox{\rm for {\sl some}}\;\; s.
\end{equation}
This demonstrates that, for interacting systems, $\{ f_{s;\sigma}
({\Bf r})\}$ is {\sl not} normalized to unity for {\sl some}, if
not all (see footnote \ref{f111}), $s$. With reference to our 
qualification `some', we point out that, if for a {\sl given} $s$, 
say $s_0$, $\gamma^{\mp}_{s_0;\sigma}(\varsigma)$ happens to be 
vanishing {\sl exclusively} for a non-empty set $S_0$ of $\varsigma$
values for which $\nu_{\sigma}(\varsigma) \not= 1$, then $\sum_{\varsigma}
\nu_{\sigma}(\varsigma)\vert\gamma^{\mp}_{s_0;\sigma}(\varsigma)\vert^2
= 1$ obtains and thus $\int {\rm d}^dr\, f_{s_0;\sigma}^*({\Bf r}) 
f_{s_0;\sigma}({\Bf r}) = 1$; evidently, however, since 
$\gamma^{\mp}_{s;\sigma}(\varsigma) \vert_{\varsigma\in S_0} = 0$ 
{\sl cannot} be valid for {\sl all} $s$, for otherwise Eq.~(\ref{ea12}) 
would be violated, it follows that $f_{s;\sigma} ({\Bf r})$ {\sl cannot} 
be of unit norm for {\sl all} $s$. In contrast, the independence of 
$\nu_{\sigma}(\varsigma)$ from $s$ implies that the possibility of
$\int {\rm d}^dr\, f_{s;\sigma}^*({\Bf r}) f_{s;\sigma}({\Bf r}) 
= 1$ for some $s$ must be a matter of accident, so that in general 
and so long as $v\not\equiv 0$, the word `{\sl some}' in Eq.~(\ref{ea20}) 
should be interchangeable with the word `{\sl all}'.

Along the same lines as above, one can demonstrate that 
$\{ f_{s;\sigma}({\Bf r})\}$ is {\sl not} orthogonal for interacting 
systems. It is important to realize that the observation that for 
{\sl interacting} systems the norm of $f_{s;\sigma}({\Bf r})$ is 
{\sl less} than unity (for {\sl some}, if not {\sl all}, $s$)
conforms with the fact that, while for interacting systems the 
compound variable $\varsigma$ belongs to a {\sl proper} subset of 
the set to which the compound variable $s$ belongs, nonetheless both 
$\{ f_{s;\sigma}({\Bf r})\}$ and $\{\varphi_{\varsigma;\sigma}
({\Bf r})\}$ satisfy the closure relation as presented in 
Eqs.~(\ref{e30}) and (\ref{ea1}) respectively.

To close our above discussion, we proceed by briefly elaborating 
on the over-completeness of $\{ f_{s;\sigma}({\Bf r})\}$ in the 
case of $v\not\equiv 0$. For definiteness, consider a case where 
$h_{0;\sigma}({\Bf r})$ or, what is the same, $u({\Bf r}) + 
w_{\sigma}({\Bf r})$, is invariant under the operations of some 
(discrete or continuous) symmetry group. From this it follows that 
$\{\varphi_{\varsigma;\sigma}({\Bf r})\}$ can be arranged to form 
the basis for the unitary irreducible representations of the 
underlying symmetry group. By doing this, a specific index $\varsigma$ 
identifies $\varphi_{\varsigma;\sigma}({\Bf r})$ with one specific 
basis function pertaining to one specific unitary irreducible 
representation of a general representation of the symmetry group of 
$h_{0;\sigma}({\Bf r})$ (for example Cornwell (1984, pp. 81-83)). Thus 
for instance, by assuming $u({\Bf r}) + w_{\sigma}({\Bf r})$ to be 
invariant under a discrete translation group, we can write $\varsigma 
= ({\Bf k},\ell)$, where ${\Bf k}$ stands for a point inside the first 
Brillouin zone corresponding to the underlying Bravais lattice and 
$\ell$, an integer, for a `band index'. For clarity, here $\ell$ 
singles out one of the (one-dimensional) irreducible representations 
of the discrete translation group, characterized by ${\Bf k}$. The 
function $\varphi_{({\Bf k},\ell);\sigma}({\Bf r})$ is thus a Bloch 
function. Now consider the system of {\sl interacting} fermions. From 
Eqs.~(\ref{ea4}) and (\ref{ea13}), and in view of the fact that 
Eq.~(\ref{ea17}) strictly {\sl only} applies to `non-interacting' 
systems, we observe that here, in contrast with the `non-interacting' 
case, $\{f_{s;\sigma}({\Bf r})\}$ {\sl cannot} be a basis for some 
{\sl unitary} irreducible representation of the symmetry group of 
$u({\Bf r}) + w_{\sigma}({\Bf r})$; note that already the result in 
Eq.~(\ref{ea20}) defies the notion of unitarity. Returning to our 
above example where $\varsigma = ({\Bf k},\ell)$, it follows that in 
general $s$ {\sl cannot} coincide with one particular $({\Bf k},\ell)$, 
but a collection of these, that is all those $({\Bf k},\ell)$ 
associated with non-vanishing contributions to the sum on the RHS of 
Eq.~(\ref{ea4}). In general we have 
\begin{equation}
\label{ea21}
s =\left. \cup_{\varsigma}\, 
\varsigma\right|_{\chi_{s;\sigma}(\varsigma)\not= 0}. 
\end{equation}
In words, considering $s$ as a set, it consists of the union of 
{\sl all} $\varsigma$ (also a {\sl set} of parameters) for which 
$\chi_{s;\sigma}(\varsigma) \not=0$ holds; owing to the dependence
of $\chi_{s;\sigma}(\varsigma)$ on $s$, Eq.~(\ref{ea21}) amounts to
an implicit equation, an intricacy that has its origin in the 
particle-particle interaction. This repeated incorporation of various 
elements of the set $\{\varsigma\}$, which is already associated with a 
{\sl complete} set of basis functions, into the elements of $\{ s\}$, 
implies that the set $\{ f_{s;\sigma}({\Bf r})\}$ must in fact be 
{\sl over-complete}. 

Finally, from the above observations it is evident that ``degeneracy'' 
as in ``parameter of degeneracy'' (Klein and Prange 1958), should be 
viewed in a somewhat different light than may be implied by the 
authors (see footnote \ref{f50}). This is important since, on the one 
hand, Klein and Prange (1958) deal with $({\Bf p},\alpha)$ ({\it cf}. 
Eq.~(\ref{e45}) above), where ${\Bf p}$ stands for the momentum vector 
associated with the single-particle excitations of a translational 
invariant system and $\alpha$ denotes the {\it ``parameter of 
degeneracy distinguishing states of given momentum''} while, on the 
other hand, our above discussion makes evident that, even for a 
translational invariant system (here an attribute of the {\sl GS}, 
but by no means that of the excited states of such system), 
$f_{s;\sigma}({\Bf r})$ is {\sl not} in general capable of being 
characterized by a {\sl single} momentum, as opposed to a distribution 
of momenta.

\subsection{Quasi-particles revisited: an approximate treatment}
\label{s43}

Here we derive a simple expression for the dispersion of the
single-particle energies of an interacting uniform system based on 
an {\sl Ansatz} concerning the $N\pm 1$-particle eigenstates of the 
interacting Hamiltonian $\wh{H}$ as presented in Eq.~(\ref{e1}). 
This energy dispersion is determined in terms of the single- and 
two-particle {\sl static} correlation functions $\Gamma^{(1)}$ and 
$\Gamma^{(2)}$ pertaining to the GS of this system. As we shall 
explicitly show, our {\sl Ansatz} with regard to the indicated eigenstates 
of the system provides an accurate expression for the single-particle 
excitation energies $\bar\varepsilon_{{\bar {\Bf k}};\sigma}$ for 
$\|{\Bf k}-{\Bf k}_{F;\sigma}\|\to 0$ in the weak-coupling regime. 
In our general considerations that follow we assume the Hamiltonian 
to be that pertaining to uniform systems, however, do {\sl not} impose 
this property on the GS of the system. 

Consider the Hamiltonian in Eq.~(\ref{e86}). Making use of the 
expression in Eq.~(\ref{e96}), for the GS energy of the corresponding 
system we have (see Eq.~(\ref{e103}))
\begin{equation}
\label{ea22}
E_{N;0} = e_0\, \bar{E}_{N;0},\;\;\;\;
\bar{E}_{N;0} = \langle\Psi_{N;0}\vert \wh{\cal H}
\vert\Psi_{N;0}\rangle.
\end{equation}
We {\sl assume} the $(N_{\sigma}\pm 1+N_{\bar\sigma})$-particle 
eigenstates of the interacting system to be characterized by a single 
wave-vector ${\bar {\Bf k}}$ (here we consider the case where the 
change in the number of particles is brought about through a change 
in that of particles with spin $\sigma$); thus we denote these by 
$\vert\Psi^{0}_{N_{\sigma} \pm 1,N_{\bar\sigma};{\bar {\Bf k}}}\rangle$. 
With 
\begin{equation}
\label{ea23}
\bar\mu {:=} \frac{\mu}{e_0} 
\end{equation}
the normalized `chemical potential' (see \S~III.E.1), we have the 
following for the {\sl single-particle} excitation energies 
({\it cf}. Eq.~(\ref{e19})):
\begin{eqnarray}
\label{ea24}
\bar\varepsilon_{{\bar {\Bf k}};\sigma} {:=} \left\{ \begin{array}{ll}
\bar{E}_{N;0} - \bar{E}_{N_{\sigma}-1,N_{\bar\sigma};{\bar {\Bf k}}},
&\;\;\; \bar\varepsilon_{{\bar {\Bf k}};\sigma} < \bar\mu, \\ \\
\bar{E}_{N_{\sigma}+1,N_{\bar\sigma};{\bar {\Bf k}}} - \bar{E}_{N;0},
&\;\;\; \bar\varepsilon_{{\bar {\Bf k}};\sigma} > \bar\mu.
\end{array} \right.
\end{eqnarray}

With (see Eq.~(\ref{ea8}) above)
\begin{eqnarray}
\label{ea25}
\nu_{\sigma}({\bar {\Bf k}})
{:=} \left\{ \begin{array}{ll}
{\sf n}_{\sigma}({\bar {\Bf k}}/r_0),
&\;\;\; \bar\varepsilon_{{\bar {\Bf k}};\sigma} < \bar\mu, \\
1-{\sf n}_{\sigma}({\bar {\Bf k}}/r_0),
&\;\;\; \bar\varepsilon_{{\bar {\Bf k}};\sigma} > \bar\mu,
\end{array} \right.
\end{eqnarray}
where ${\sf n}_{\sigma}({\Bf k})$ denotes the GS momentum distribution
function corresponding to spin-$\sigma$ particles, defined in 
Eq.~(\ref{ej2}), for $\nu_{\sigma}({\bar {\Bf k}})\not=0$ we make the 
following {\sl Ansatz} for the $(N_{\sigma}\mp 1+N_{\bar\sigma})$-particle 
eigenstates of the interacting system
\footnote{\label{f112}
These {\sl Ans\"atze} are similar in spirit to that by Anderson (1959) 
(see also Anderson (1987)). } 
\begin{eqnarray}
\label{ea26}
&&\vert\Psi^{0}_{N_{\sigma}-1,N_{\bar\sigma};{\bar {\Bf k}}}\rangle
{:=} \frac{1}{\sqrt{\nu_{\sigma}\big({\bar {\Bf k}}\big)}}\,
{\hat a}_{{\bar {\Bf k}};\sigma} \vert\Psi_{N;0}\rangle,\;\;
\bar\varepsilon_{{\bar {\Bf k}};\sigma} < \bar\mu,\\
\label{ea27}
&&\vert\Psi^{0}_{N_{\sigma}+1,N_{\bar\sigma};{\bar {\Bf k}}}\rangle
{:=} \frac{1}{\sqrt{\nu_{\sigma}\big({\bar {\Bf k}}\big)}}\,
{\hat a}_{{\bar {\Bf k}};\sigma}^{\dag} \vert\Psi_{N;0}\rangle,\;\;
\bar\varepsilon_{{\bar {\Bf k}};\sigma} > \bar\mu.
\end{eqnarray}
It is instructive to examine the properties of these states for the 
case where the system is non-interacting so that $\nu_{\sigma}
({\bar {\Bf k}}) = 1$ for $\bar\varepsilon_{{\bar {\Bf k}};\sigma} 
\,\Ieq><\,\bar\mu$. 

It can be readily shown that the states in Eqs.~(\ref{ea26}) and 
(\ref{ea27}) are normalized to unity and, moreover, are eigenstates 
of the {\sl partial} number operators, in particular of 
$\wh{N}_{\sigma}$ corresponding to eigenvalues $(N_{\sigma}\mp 1)$
respectively (assuming $N_{\sigma} \ge 1$).

Following the {\sl Ans\"atze} in Eqs.~(\ref{ea26}) and (\ref{ea27}), for 
\begin{eqnarray}
\bar{E}_{N_{\sigma}\mp 1,N_{\bar\sigma};{\bar {\Bf k}}} {:=} 
\langle\Psi^{0}_{N_{\sigma}\mp 1,N_{\bar\sigma};
{\bar {\Bf k}}}\vert \wh{\cal H}
\vert\Psi^{0}_{N_{\sigma}\mp 1,N_{\bar\sigma};
{\bar {\Bf k}}}\rangle \nonumber
\end{eqnarray}
we have
\begin{eqnarray}
\label{ea28}
\bar{E}_{N_{\sigma}-1,N_{\bar\sigma};{\bar {\Bf k}}}
&=& \frac{1}{\nu_{\sigma}({\bar {\Bf k}})}\,
\langle \Psi_{N;0}\vert {\hat a}_{{\bar {\Bf k}};\sigma}^{\dag}
\wh{\cal H}\, {\hat a}_{{\bar {\Bf k}};\sigma}
\vert \Psi_{N;0}\rangle, \nonumber\\
& &\;\;\;\;\;\;\;\;\;\;\;\;\;\;\;\;\;\;\;\;\;\;\;\;\;\;\;\;\;\;\;\;\;
\bar\varepsilon_{{\bar {\Bf k}};\sigma} < \bar\mu,\\
\label{ea29}
{\bar E}_{N_{\sigma}+1,N_{\bar\sigma};{\bar {\Bf k}}}
&=& \frac{1}{ \nu_{\sigma}({\bar {\Bf k}})}\,
\langle \Psi_{N;0}\vert {\hat a}_{{\bar {\Bf k}};\sigma}
\wh{\cal H}\, {\hat a}_{{\bar {\Bf k}};\sigma}^{\dag}
\vert \Psi_{N;0}\rangle, \nonumber\\
& &\;\;\;\;\;\;\;\;\;\;\;\;\;\;\;\;\;\;\;\;\;\;\;\;\;\;\;\;\;\;\;\;\;
\bar\varepsilon_{{\bar {\Bf k}};\sigma} > \bar\mu.
\end{eqnarray}
Making use of the identity
\begin{equation}
\label{ea30}
\wh{\cal H}\, {\hat a}_{{\bar {\Bf k}};\sigma}
\equiv \big[ \wh{\cal H}, {\hat a}_{{\bar {\Bf k}};\sigma} \big]_-
+ {\hat a}_{{\bar {\Bf k}};\sigma}\, \wh{\cal H},
\end{equation}
from Eqs.~(\ref{ea24}), (\ref{ea28}) and (\ref{ea29}) we readily 
deduce that
\begin{eqnarray}
\label{ea31}
&&\bar\varepsilon_{{\bar {\Bf k}};\sigma} \equiv 
\left\{ \begin{array}{ll}
-\langle\Psi_{N;0}\vert
{\hat a}_{{\bar {\Bf k}};\sigma}^{\dag}
\big[\wh{\cal H}, {\hat a}_{{\bar {\Bf k}};\sigma}\big]_-
\vert\Psi_{N;0}\rangle/\nu_{\sigma}({\bar {\Bf k}}), \\
\;\;\;\;\;\;\;\;\;\;\;\;\;\;\;\;\;\;\;\;\;\;\;\;\;\;\;\; 
\;\;\;\;\;\;\;\;\;\;\;\;\;\;\;\;\;\;\;\;\;\;\;
\bar\varepsilon_{{\bar {\Bf k}};\sigma} < \bar\mu, \\ \\
-\langle\Psi_{N;0}\vert
\big[ \wh{\cal H}, {\hat a}_{{\bar {\Bf k}};\sigma}\big]_-
{\hat a}_{{\bar {\Bf k}};\sigma}^{\dag}
\vert\Psi_{N;0}\rangle/\nu_{\sigma}({\bar {\Bf k}}), &\\
\;\;\;\;\;\;\;\;\;\;\;\;\;\;\;\;\;\;\;\;\;\;\;\;\;\;\;\; 
\;\;\;\;\;\;\;\;\;\;\;\;\;\;\;\;\;\;\;\;\;\;\;
\bar\varepsilon_{{\bar {\Bf k}};\sigma} > \bar\mu. &\\
\end{array} \right.\nonumber\\
\end{eqnarray}
It is interesting to note that, according to Eq.~(\ref{ea31}),
$\bar\varepsilon_{{\bar {\Bf k}};\sigma}$ is obtained from the 
knowledge of the GS; in this connection note that $\nu_{\sigma}
({\bar {\Bf k}})$ is also a GS property (see Eqs.~(\ref{ej2}) 
and (\ref{ea25})).

From the linearity of $[\, ,\,]_-$ we have
\begin{equation}
\label{ea32}
\big[ \wh{\cal H},{\hat a}_{{\bar {\Bf k}};\sigma}\big]_-
= \big[ \wh{\cal T},{\hat a}_{{\bar {\Bf k}};\sigma}\big]_-
+\big[ \wh{\cal V},{\hat a}_{{\bar {\Bf k}};\sigma}\big]_-;
\end{equation}
in the case of $d=3$ and $v\equiv v_c$, the RHS of Eq.~(\ref{ea32})
has to be supplemented by $\big[2\wh{\cal H}_{\bar\kappa},
{\hat a}_{{\bar {\Bf k}};\sigma}\big]_-$ where ({\it cf}. 
Eq.~(\ref{e5}))
\begin{equation}
\label{ea33}
\wh{\cal H}_{\bar\kappa} {:=} \frac{1}{e_0}\, \wh{H}_{\kappa}
\equiv -\bar\varpi_{\bar\kappa}\,\wh{N};\;\;\;
\bar\varpi_{\bar\kappa} {:=} \frac{3}{2 r_s\,\bar\kappa^2},\;\;
\bar\kappa {:=} r_0 \kappa.
\end{equation}
For the reason underlying the prefactor $2$ in 
$2\wh{\cal H}_{\bar\kappa}$ see the text following Eq.~(\ref{e9}). 

Making use of the anticommutation relations in Eq.~(\ref{e101}),
we readily obtain 
\begin{equation}
\label{ea34}
\big[ \wh{\cal T},{\hat a}_{{\bar {\Bf k}};\sigma}\big]_-
= -\frac{1}{2}\, {\bar k}^2\, {\hat a}_{{\bar {\Bf k}};\sigma},
\end{equation}
We point out that, by identifying $\wh{\cal V}$ with zero, requiring, in 
the case of $v\equiv v_c$, identification of $\wh{\cal H}_{\bar\kappa}$ 
with zero, from Eqs.~(\ref{ea31}) and (\ref{ea34}) we immediately 
obtain the expected result for the non-interacting single-particle 
energy dispersion
\begin{equation}
\label{ea35}
\bar\varepsilon_{{\bar {\Bf k}};\sigma}
\rightharpoonup
\bar\varepsilon_{{\bar {\Bf k}}}^{(0)}
{:=} \frac{1}{2}\, {\bar k}^2,\;\;\;
\bar\varepsilon_{{\bar {\Bf k}}}^{(0)}\, \IEq>< \, 
\bar\mu.
\end{equation}
Thus by writing
\begin{equation}
\label{ea36}
\bar\varepsilon_{{\bar {\Bf k}};\sigma}
\equiv \bar\varepsilon_{{\bar {\Bf k}}}^{(0)}
+ \Delta \bar\varepsilon_{{\bar {\Bf k}};\sigma},
\end{equation}
the contribution $\Delta\bar\varepsilon_{{\bar {\Bf k}};\sigma}$ is 
obtained from the expression on the RHS of Eq.~(\ref{ea31}) through 
replacing $\wh{\cal H}$ herein by $\wh{\cal V}$ (by $2 
\wh{\cal H}_{\bar\kappa} +\wh{\cal V}$ when $v\equiv v_c$) as defined 
in Eq.~(\ref{e99}). Consequently, the {\sl explicit} dependence on 
$r_s$ of $\Delta\bar\varepsilon_{{\bar {\Bf k}};\sigma}$ is linear; 
as is evident from Eq.~(\ref{ea31}), the {\sl implicit} dependence 
on $r_s$ of this function has its root in the dependences on $r_s$ of 
$\nu_{\sigma}({\bar {\Bf k}})$ and $\vert\Psi_{N;0}\rangle$. 
We draw attention to the fact that in view of Eqs.~(\ref{e105}) and 
(\ref{ea36}), $\Delta\bar\varepsilon_{{\bar {\Bf k}};\sigma}$ is to 
be identified with $\Ol{\Sigma}_{\sigma}({\bar {\Bf k}};
\bar\varepsilon_{{\bar {\Bf k}};\sigma})$, the on-the-mass-shell SE.         

Now we proceed with the determination of the expression for
$\Delta\bar\varepsilon_{{\bar {\Bf k}};\sigma}$ in terms of 
practically calculable GS correlation functions. Making use of the 
expression on the RHS of Eq.~(\ref{e99}) and the anticommutation
relations in Eq.~(\ref{e101}), after some straightforward algebra 
we obtain
\footnote{\label{f113}
Making use of Eqs.~(\protect\ref{ea30}), (\protect\ref{ea34}) and
(\protect\ref{ea37}) and the fact that $\wh{\cal H} \vert\Psi_{N;0}\rangle 
= {\bar E}_{N;0} \vert\Psi_{N;0}\rangle$, one can directly establish 
the deviation of the states in Eqs.~(\protect\ref{ea26}) and 
(\protect\ref{ea27}) from the true eigenstates of $\wh{\cal H}$. This 
consideration also paves the way for constructing improved states in 
comparison with those in Eqs.~(\protect\ref{ea26}) and 
(\protect\ref{ea27}). }
\begin{equation}
\label{ea37}
\big[ \wh{\cal V}, {\hat a}_{{\bar {\Bf k}};\sigma} \big]_-
= \frac{r_s}{{\bar\Omega}} \sum_{\sigma'}
\sum_{{\bar {\Bf k}}',{\bar {\Bf q}}'}'\,
{\bar w}({\bar q}')\,
{\hat a}_{{\bar {\Bf k}}'+{\bar {\Bf q}}';\sigma'}^{\dag}
{\hat a}_{{\bar {\Bf k}}+{\bar {\Bf q}}';\sigma}
{\hat a}_{{\bar {\Bf k}}';\sigma'}.
\end{equation}
From this expression, one can directly calculate 
$\Delta\bar\varepsilon_{{\bar {\Bf k}};\sigma}$ corresponding to the 
case $\bar\varepsilon_{{\bar {\Bf k}};\sigma} < \bar\mu$ in terms of 
the two-particle GS correlation function $\Gamma^{(2)}({\Bf r}_1\sigma_1,
{\Bf r}_2\sigma_2;{\Bf r}_1'\sigma_1',{\Bf r}_2'\sigma_2')$, defined 
in Appendix B (see Eq.~(\ref{eb8})). In order to obtain a similar 
expression for $\Delta\bar\varepsilon_{{\bar {\Bf k}};\sigma}$ 
corresponding to the case $\bar\varepsilon_{{\bar {\Bf k}};\sigma} 
> \bar\mu$, we need first to bring the pertinent operators into normal 
order. Making use of Eq.~(\ref{e101}) we obtain
\begin{eqnarray}
\label{ea38}
&&\sum_{\sigma'} \sum_{{\bar {\Bf k}}',{\bar {\Bf q}}'}'
{\bar w}({\bar q}')\,
{\hat a}_{{\bar {\Bf k}}'+{\bar {\Bf q}}';\sigma'}^{\dag}
{\hat a}_{{\bar {\Bf k}}+{\bar {\Bf q}}';\sigma}
{\hat a}_{{\bar {\Bf k}}';\sigma'}
{\hat a}_{{\bar {\Bf k}};\sigma}^{\dag}\nonumber\\
&&\;\;\;\;\;
=\sum_{{\bar {\Bf q}}'}' {\bar w}({\bar q}')\,
{\hat a}_{{\bar {\Bf k}}+{\bar {\Bf q}}';\sigma}^{\dag}
{\hat a}_{{\bar {\Bf k}}+{\bar {\Bf q}}';\sigma} \nonumber\\
&&\;\;\;\;\;
- \sum_{\sigma'} \sum_{{\bar {\Bf k}}',{\bar {\Bf q}}'}'
{\bar w}({\bar q}')\,
{\hat a}_{{\bar {\Bf k}};\sigma}^{\dag}
{\hat a}_{{\bar {\Bf k}}'+{\bar {\Bf q}}';\sigma'}^{\dag}
{\hat a}_{{\bar {\Bf k}}+{\bar {\Bf q}}';\sigma}
{\hat a}_{{\bar {\Bf k}}';\sigma'}.
\end{eqnarray}
For later use, we present the following expression (see Eq.~(\ref{ea33}) 
above)
\begin{equation}
\label{ea39}
\big[\wh{\cal H}_{\bar\kappa}, 
{\hat a}_{{\bar {\Bf k}};\sigma}\big]_-
=\bar\varpi_{\bar\kappa}\, {\hat a}_{{\bar {\Bf k}};\sigma}.
\end{equation}

Before casting the expression for $\Delta\bar\varepsilon_{{\bar {\Bf k}};
\sigma}$ into a form which is suitable for direct (numerical)
calculations, we note that the following pairs of Fourier transforms 
apply
\begin{eqnarray}
\label{ea40}
&&{\sf\hat a}_{{\Bf k};\sigma}
= \frac{1}{\Omega^{1/2}}
\int_{\Omega} {\rm d}^dr\;
{\rm e}^{-i {\Bf k}\cdot {\Bf r} }\,
\hat\psi_{\sigma}({\Bf r}),\\
\label{ea41}
&&\hat\psi_{\sigma}({\Bf r})
= \frac{1}{\Omega^{1/2}} \sum_{\Bf k}\,
{\rm e}^{+i {\Bf k}\cdot {\Bf r} }\,
{\sf\hat a}_{{\Bf k};\sigma}.
\end{eqnarray}
Our use of {\sl summation} over wave-vectors (both here and in the 
earlier expressions) signifies that we employ a {\sl box} boundary 
condition; equivalent expressions for the case where wave-vectors are 
in continuum, are obtained through the substitutions
\begin{equation}
\label{ea42}
\sum_{\Bf k} (\dots) \rightharpoonup
\frac{\Omega}{(2\pi)^d} \int {\rm d}^dk\; (\dots),\;\;
\delta_{{\Bf k},{\Bf k}'} \rightharpoonup
\frac{(2\pi)^d}{\Omega}\,
\delta({\Bf k}-{\Bf k}').
\end{equation}

Making use of Eq.~(\ref{ea40}), we obtain
\begin{eqnarray}
\label{ea43}
&&\sum_{\sigma'}\langle\Psi_{N;0}\vert
{\hat a}_{{\bar {\Bf k}};\sigma}^{\dag}
{\hat a}_{{\bar {\Bf k}}'+{\bar {\Bf q}}';\sigma'}^{\dag}
{\hat a}_{{\bar {\Bf k}}+{\bar {\Bf q}}';\sigma}
{\hat a}_{{\bar {\Bf k}}';\sigma'}\vert\Psi_{N;0}\rangle
\nonumber\\
&&\;\;\;
=\frac{-1}{\Omega^2}
\int_{\Omega} \Pi_{j=1}^4 {\rm d}^dr_j\;
{\rm e}^{i {\Bf k} \cdot ({\Bf r}_1 - {\Bf r}_3)}\,
{\rm e}^{i {\Bf k}'\cdot 
({\Bf r}_2 - {\Bf r}_4)}\,
{\rm e}^{i {\Bf q}'\cdot 
({\Bf r}_2 - {\Bf r}_3)}\nonumber\\
&&\;\;\;\;\;\;\;\;\;\;\;\;\;\;\;\;\;
\times \sum_{\sigma'}
\Gamma^{(2)}({\Bf r}_1\sigma,{\Bf r}_2\sigma';
{\Bf r}_3\sigma,{\Bf r}_4\sigma').
\end{eqnarray}
Consequently, through employing (see Eq.~(\ref{e89}))
\begin{equation}
\label{ea44}
w(\|{\Bf r}\|) = 
\int \frac{{\rm d}^dq'}{(2\pi)^d}\;
{\bar w}(\|{\Bf q}'\|)\, {\rm e}^{i {\Bf q}'\cdot {\Bf r} },
\end{equation}
we arrive at
\begin{eqnarray}
\label{ea45}
\beta_{{\bar {\Bf k}};\sigma}^{-} &{:=}& \frac{1}{r_s}
\langle\Psi_{N;0}\vert
{\hat a}_{{\bar {\Bf k}};\sigma}^{\dag} \big[ \wh{\cal V},
{\hat a}_{{\bar {\Bf k}};\sigma}\big]_- \vert\Psi_{N;0}\rangle
\nonumber\\
&=& -r_0\int {\rm d}^dr_1 \;
{\rm e}^{i {\Bf k}\cdot {\Bf r}_1}
\int {\rm d}^dr_2\; w(\|{\Bf r}_2\|)
\nonumber\\
&&\;\;\times
\sum_{\sigma'} \Gamma^{(2)}\big({\Bf r}_1\sigma,
{\Bf r}_2\sigma';{\bf 0}\sigma,
{\Bf r}_2\sigma'\big),\;\;\;\mbox{\rm (I)}\\
\label{ea46}
\beta_{{\bar {\Bf k}};\sigma}^{-} &{:=}& \frac{1}{r_s}
\langle\Psi_{N;0}\vert
{\hat a}_{{\bar {\Bf k}};\sigma}^{\dag} 
\big[ \wh{\cal V}+2\wh{\cal H}_{\bar\kappa},
{\hat a}_{{\bar {\Bf k}};\sigma}\big]_- \vert\Psi_{N;0}\rangle
\nonumber\\
&=& -r_0\int {\rm d}^3r_1 \;
{\rm e}^{i {\Bf k}\cdot {\Bf r}_1}
\int {\rm d}^3r_2\; w_c(\|{\Bf r}_2\|)
\nonumber\\
&&\;\;
\times \sum_{\sigma'} 
\Big\{\Gamma^{(2)}\big({\Bf r}_1\sigma,
{\Bf r}_2\sigma';{\bf 0}\sigma,
{\Bf r}_2\sigma'\big)\nonumber\\
& &\;\;\;\;\;\;\;\;\;\;\;\;\;\;\;\;\;\;\;\;\;\;\;\;\;\;\;
-n_{0;\sigma'}
\varrho_{\sigma}({\Bf r}_1,{\bf 0})\Big\},\;\;\;\mbox{\rm (II)}
\end{eqnarray}
where we have used Eqs.~(\ref{ea37}) and (\ref{ea43}) as well as 
the shift property in Eq.~(\ref{eb14}) specific to uniform GSs. For 
completeness, the expressions in Eqs.~(\ref{ea45}) and (\ref{ea46}) 
have been deduced from ones involving 
\begin{eqnarray}
\frac{1}{\Omega} \int {\rm d}^dr_3\;
\Gamma^{(2)}\big(({\Bf r}_1+{\Bf r}_3)
\sigma,({\Bf r}_2+{\Bf r}_3)\sigma';
{\Bf r}_3\sigma,({\Bf r}_2+{\Bf r}_3)\sigma'),
\nonumber
\end{eqnarray}
which in consequence of the mentioned shift property together with 
$\Omega^{-1} \int {\rm d}^dr_3 = 1$ has reduced into $\Gamma^{(2)}
\big({\Bf r}_1\sigma,{\Bf r}_2\sigma';{\bf 0}\sigma,{\Bf r}_2\sigma'\big)$. 
In Eq.~(\ref{ea45}), (I) indicates the case corresponding to short-range 
interaction potentials, and in Eq.~(\ref{ea46}), (II) the case 
corresponding to $v\equiv v_c$ in $d=3$.

Making use of Eqs.~(\ref{ea37}), (\ref{ea38}) and (\ref{ea40}), 
in an analogous manner as above, for
\begin{eqnarray}
\label{ea47}
&&\beta_{{\bar {\Bf k}};\sigma}^{+}
{:=} \frac{1}{r_s}\, 
\left\{ \begin{array}{ll}
\langle\Psi_{N;0}\vert \big[\wh{\cal V},
{\hat a}_{{\bar {\Bf k}};\sigma}\big]_-
{\hat a}_{{\bar {\Bf k}};\sigma}^{\dag}
\vert\Psi_{N;0}\rangle,&
\mbox{\rm (I)}\\ \\
\langle\Psi_{N;0}\vert \big[\wh{\cal V}+2\wh{\cal H}_{\bar\kappa},
{\hat a}_{{\bar {\Bf k}};\sigma}\big]_-
{\hat a}_{{\bar {\Bf k}};\sigma}^{\dag}
\vert\Psi_{N;0}\rangle&
\mbox{\rm (II)}\\
\end{array} \right.\nonumber\\
\end{eqnarray}
we obtain
\begin{equation}
\label{ea48}
\beta_{{\bar {\Bf k}};\sigma}^{+}
=r_0 \int {\rm d}^dr_1\;
{\rm e}^{i {\Bf k}\cdot {\Bf r}_1}\,
w(\|{\Bf r}_1\|)\,
\Gamma^{(1)}\big({\Bf r}_1\sigma;{\bf 0}\sigma\big)
-\beta_{{\bar {\Bf k}};\sigma}^{-}. 
\end{equation}
In Eq.~(\ref{ea47}), (I) and (II) have the same significance as 
in Eqs.~(\ref{ea45}) and (\ref{ea46}). For completeness, we point 
out that
\begin{eqnarray}
\label{ea49}
&& \int {\rm d}^dr_1\;
{\rm e}^{i {\Bf k}\cdot {\Bf r}_1} \, w(\|{\Bf r}_1\|)\,
\Gamma^{(1)}\big({\Bf r}_1\sigma;{\bf 0}\sigma\big)\nonumber\\
&&\;\;\;\;\;\;\;\;\;\;\;\;\;\;\;\;\;\;\;\;\;\;\;
\equiv \int \frac{{\rm d}^dq'}{(2\pi)^d}\;
{\bar w}(q')\,
{\sf n}_{\sigma}({\Bf k}+{\Bf q}').
\end{eqnarray}
Since (unless ${\sf n}_{\sigma}({\Bf k}) \equiv 0$, corresponding to 
a GS with {\sl no} particles of spin $\sigma$, i.e. $N_{\sigma}=0$),
${\sf n}_{\sigma}({\Bf k})$ is positive, from Eq.~(\ref{ea49}) and 
the assumption that ${\bar w}(q') \ge 0$ for all $q'$ (a non-attractive
interaction potential), it follows that the first term on the RHS of 
Eq.~(\ref{ea48}) is in general positive definite.

Combining the above results, from Eq.~(\ref{ea36}) we obtain
\begin{equation}
\label{ea50}
\bar\varepsilon_{{\bar {\Bf k}};\sigma} =
\bar\varepsilon_{{\bar {\Bf k}}}^{(0)}
- r_s\, \frac{\beta_{{\bar {\Bf k}};\sigma}^{\mp}}
{\nu_{\sigma}({\bar {\Bf k}})},\;\;
\bar\varepsilon_{{\bar {\Bf k}};\sigma}\,
\IEq>< \,\bar\mu.
\end{equation}
Note how effects of interaction on the single-particle excitation
energies are accounted for by this expression; in particular the 
occurrence of $\nu_{\sigma}({\bar {\Bf k}})$ in denominator clearly 
indicates the non-perturbative nature of this expression. On the other 
hand, since according to the expression in Eq.~(\ref{ea50}), the 
single-particle excitation energy is a well-defined real-valued 
quantity for {\sl all} ${\bar {\Bf k}}$, this expression should be 
necessarily of limited validity; as we have discussed in \S~III.D, 
the equation for the single-particle excitation energies (see 
Eq.~(\ref{e105})) does {\sl not} have any (real-valued) solution 
in regions along the $\bar\varepsilon$ axis where 
${\rm Im}[\Ol{\Sigma}_{\sigma}({\bar {\Bf k}};\bar\varepsilon)] 
\not\equiv 0$; real-valued $\bar\varepsilon_{{\bar {\Bf k}};\sigma}$ 
for {\sl all} ${\bar {\Bf k}}$ is only possible when 
${\rm Im}[\Ol{\Sigma}_{\sigma}({\bar {\Bf k}};\bar\varepsilon)]\equiv 
0$, $\forall\varepsilon$, which is {\sl exclusively} the case for 
non-interacting systems, or within mean-field frameworks. 
\footnote{\label{f114}
We should emphasize that here the real-valuedness of the single-particle 
excitation energies $\varepsilon_{s;\sigma}$ as defined in 
Eq.~(\protect\ref{e19}) is {\sl not} at issue; at issue is the 
real-valuedness of $\varepsilon_{{\Bf k};\sigma}$, characterized solely 
by ${\Bf k}$ rather than $s=({\Bf k},\alpha)$, with $\alpha$ the 
``parameter of degeneracy'' introduced and discussed in \S~III.B (see 
Eq.~(\protect\ref{e45})). In this context, one can think of a 
generalization of the {\sl Ans\"atze} in Eqs.~(\protect\ref{ea26}) and 
(\protect\ref{ea27}) according to which the pertinent states are 
characterized not only by ${\Bf k}$ but also by a set of additional 
parameters that render the employed $(N_{\sigma}\pm 1 
+N_{\bar\sigma})$-particle states better variational eigenstates of 
the interacting Hamiltonian than those presented in 
Eqs.~(\protect\ref{ea26}) and (\protect\ref{ea27}). These additional 
parameters bring about the possibility that, for a given ${\Bf k}$, the 
optimal parameters are {\sl not} unique, that is a multiplicity of 
different values for these parameters give rise to variational states 
corresponding to different $(N_{\sigma}\pm 1 +N_{\bar\sigma})$-particle 
eigenstates of the interacting Hamiltonian. In consequence of this 
mechanism, on taking full or sufficiently detailed account of $\alpha$, 
one establishes that to a given ${\Bf k}$ corresponds not one eigenstate, 
but a continuous {\sl distribution} of eigenstates whose corresponding 
energies, as measured with respect to the energy of the $(N_{\sigma}
+N_{\bar\sigma})$-particle GS of the interacting Hamiltonian, show up 
as peaks of finite widths in the depiction of the single-particle 
spectral function $A_{\sigma}({\Bf k};\varepsilon)$ along the 
$\varepsilon$ axis, which naturally are {\sl not} capable of being 
described in terms of real-valued energies 
$\varepsilon_{{\Bf k};\sigma}$. }
On the other hand, since ${\rm Im}[\Ol{\Sigma}_{\sigma}({\bar {\Bf k}};
\bar\varepsilon_F)]\equiv 0$, $\forall {\bar {\Bf k}}$ (Galitskii and 
Migdal 1958, Luttinger 1960, equations (6) and (94)), it is {\sl not} 
possible {\sl a priori} to decide on the possible inaccuracy of the 
expression in Eq.~(\ref{ea50}) for $\|{\bar {\Bf k}}
-{\bar {\Bf k}}_{F;\sigma}\|\to 0$ on the grounds that it is real valued 
for {\sl all} ${\bar {\Bf k}}$. Later in this Section we demonstrate 
that $\bar\varepsilon_{{\bar {\Bf k}};\sigma}$ in Eq.~(\ref{ea50}) 
coincides to at least linear order in $r_s$ with the expected energy 
dispersion in the weak-coupling regime. This implies that, through 
substitution of $\bar\varepsilon_{{\bar k};\sigma}$ on both sides of 
Eq.~(\ref{e146}) by the expression on the RHS of Eq.~(\ref{ea50}), one 
can solve for $\beta_{{\bar k};\sigma}^{\mp}$ in the weak-coupling 
regime and thus bypass the direct evaluation of $\beta_{{\bar k};
\sigma}^{\mp}$ according to the expressions in Eqs.~(\ref{ea45}), 
(\ref{ea46}) and (\ref{ea48}).

For {\sl metals} we must have
\begin{equation}
\label{ea51}
\lim_{{\bar {\Bf k}} \to {\bar {\Bf k}}_{F;\sigma}-{\bf 0}^+}
\bar\varepsilon_{{\bar {\Bf k}};\sigma} =
\lim_{{\bar {\Bf k}}\to {\bar {\Bf k}}_{F;\sigma}+{\bf 0}^+}
\bar\varepsilon_{{\bar {\Bf k}};\sigma},
\end{equation}
which through Eq.~(\ref{ea50}) implies
\begin{equation}
\label{ea52}
\frac{\beta_{{\bar {\Bf k}};\sigma}^{-}}
{\nu_{\sigma}({\bar {\Bf k}}-{\bf 0}^+)}
=\frac{\beta_{{\bar {\Bf k}};\sigma}^{+}}
{\nu_{\sigma}({\bar {\Bf k}}+{\bf 0}^+)},
\;\;\mbox{\rm for}\;\;
{\bar {\Bf k}}={\bar {\Bf k}}_{F;\sigma}; 
\end{equation}
our use of $\pm {\bf 0}^+$ here is dictated by the fact that 
${\sf n}_{\sigma}({\bar {\Bf k}}/r_0)$, and thus $\nu_{\sigma}
({\bar {\Bf k}})$ (see Eq.~(\ref{ea25})), is non-analytic at 
${\bar {\Bf k}}={\bar {\Bf k}}_{F;\sigma}$; for metallic GSs 
corresponding to continuously differentiable $\Ol{\Sigma}_{\sigma}
({\bar k}_{F;\sigma};\bar\varepsilon)$ with respect to $\bar\varepsilon$ 
in a neighbourhood of $\bar\varepsilon =\bar\varepsilon_F$, 
${\sf n}_{\sigma}({\bar k}/r_0)$ is {\sl discontinuous} at ${\bar k} 
={\bar k}_{F;\sigma}$ (Farid 1999c). Note that, since $\bar\varepsilon_F$ 
is {\sl independent} of $\sigma$, Eqs.~(\ref{ea51}) and (\ref{ea52}) 
still apply on replacing $\sigma$ on their RHSs by $\sigma'\not=\sigma$ 
so long as $N_{\sigma'}\ge 1$ (see footnote \ref{f87}). With reference 
to our considerations in \S\S~III.E.5,6, on account of the Seitz (1940, 
pp. 343 and 344) theorem, within the framework of our {\sl Ansatz} 
for (see footnote \ref{f78}) 
\begin{eqnarray}
\label{ea53}
\bar\mu_{{\rm xc}} {:=} 
\frac{{\rm d} [n {\bar E}_{\rm xc}(n)]}
{{\rm d} n} &\equiv&
\left. (\bar\varepsilon_{{\bar {\Bf k}};\sigma} - 
\bar\varepsilon_{{\bar {\Bf k}}}^{(0)})\right|_{{\bar {\Bf k}}
={\bar {\Bf k}}_{F;\sigma}} \nonumber \\
&\equiv& \bar\varepsilon_F - \bar\varepsilon_F^{(0)},
\end{eqnarray}
we have
\begin{equation}
\label{ea54}
\bar\mu_{{\rm xc}} = \left.
\frac{-r_s\, \beta_{{\bar {\Bf k}};\sigma}^{\mp}}
{\nu_{\sigma}({\bar {\Bf k}}\mp 
{\bf 0}^+)}\right|_{{\bar {\Bf k}}={\bar {\Bf k}}_{F;\sigma}},
\;\;\; \forall\,\sigma\;\;\mbox{\rm with}\;\; N_{\sigma} \not= 0.
\end{equation}
This expression can serve as a basis for determining the accuracy 
of the {\sl Ans\"atze} from which it is deduced, by comparing
$\bar\mu_{{\rm xc}}$ (for some finite range of $r_s$ values) 
as calculated according to Eq.~(\ref{ea54}) with that deduced from 
the total energy pertaining to the correlated $N$-particle GS
$\vert\Psi_{N;0}\rangle$ in Eqs.~(\ref{ea26}) and (\ref{ea27}) (for 
details see footnote \ref{f78}). 

In the remaining part of this Section we specialize to systems with 
uniform {\sl and} isotropic metallic GSs. We further assume that the 
Fermi sea pertaining to particles with spin $\sigma$ is simply 
connected so that it is characterized by a single Fermi wavenumber 
${\bar k}_{F;\sigma}$.

Within the SSDA, where
\begin{eqnarray}
\label{ea55}
&&\left. \sum_{\sigma'}\Big\{
\Gamma^{(2)}({\Bf r}_1\sigma,{\Bf r}_2\sigma';
{\bf 0}\sigma,{\Bf r}_2\sigma') - n_{0;\sigma'}\, 
\varrho_{\sigma}({\Bf r}_1,{\bf 0})\Big\}\right|_{\rm s} \nonumber\\
&&\;\;\;\;\;\;\;\;\;\;\;\;\;\;\;\;\;\;\;\;\;
= - \varrho_{{\rm s};\sigma}^{\rm h}(\|{\Bf r}_2\|)\,
\varrho_{{\rm s};\sigma}^{\rm h}(\|{\Bf r}_1-{\Bf r}_2\|),
\end{eqnarray}
from Eq.~(\ref{ea46}) for $v\equiv v_c$ in $d=3$ we readily obtain
\begin{equation}
\label{ea56}
\left. \beta_{{\bar k};\sigma}^{-}\right|_{\rm s}
= \frac{2}{\pi}\, {\bar k}_{F;\sigma}
{\sf F}({\bar k}/{\bar k}_{F;\sigma})\,
{\sf n}_{{\rm s};\sigma}({\bar k}),
\end{equation}
where ${\sf F}(x)$ is defined in Eq.~(\ref{e131}) and 
\begin{equation}
\label{ea57}
{\sf n}_{{\rm s};\sigma}({\bar k}) \equiv 
\Theta({\bar k}_{F;\sigma}-{\bar k})
\end{equation}
is the GS momentum distribution function pertaining to the system of 
non-interacting fermions. In obtaining the result in Eq.~(\ref{ea56}) we 
have made use of the expression for $\varrho_{{\rm s};\sigma}^{\rm h}
(\|{\Bf r}\|)$ as presented in Eq.~(\ref{ef24}). With ({\it cf}.
Eq.~(\ref{eb16}))
\footnote{\label{f115}
Here we have $\Gamma^{(1)}_{\rm s}({\Bf r}_1\sigma;{\bf 0}\sigma) 
\equiv \varrho_{{\rm s};\sigma}^{\rm h}(\|{\Bf r}_1\|)$ (see 
Eqs.~(\protect\ref{eb16}) and (\protect\ref{ef9})). Consequently, 
the result in Eq.~(\protect\ref{ea58}) is relevant to that in 
Eq.~(\protect\ref{e130}). }
\begin{eqnarray}
\label{ea58}
&&r_0 \int {\rm d}^3r_1\; {\rm e}^{i {\Bf k}\cdot {\Bf r}_1}\,
w_c(\|{\Bf r}_1\|)\, \Gamma^{(1)}_{\rm s}({\Bf r}_1\sigma;
{\bf 0}\sigma) \nonumber\\
&&\;\;\;\;\;\;\;\;\;\;\;\;\;\;\;\;\;\;\;\;\;\;\;\;\;\;\;\;\;
\;\;\;\;\;\;
= \frac{2}{\pi}\, {\bar k}_{F;\sigma} 
{\sf F}({\bar k}/{\bar k}_{F;\sigma}),
\end{eqnarray}
from Eqs.~(\ref{ea48}) and (\ref{ea56}) we obtain
\begin{equation}
\label{ea59}
\left. \beta_{{\bar k};\sigma}^{+}\right|_{\rm s}
= \frac{2}{\pi} {\bar k}_{F;\sigma}
{\sf F}({\bar k}/{\bar k}_{F;\sigma})\,
\big(1- {\sf n}_{{\rm s};\sigma}({\bar k}) \big).
\end{equation}
From Eqs.~(\ref{ea50}), (\ref{ea56}) and (\ref{ea59}) we thus have
\begin{equation}
\label{ea60}
\left. \bar\varepsilon_{{\bar k};\sigma}\right|_{\rm s} = 
\bar\varepsilon_{{\bar k}}^{(0)}
- \frac{2}{\pi}\, {\bar k}_{F;\sigma}\,
r_s\, {\sf F}({\bar k}/{\bar k}_{F;\sigma}),
\end{equation}
which for systems of spin-$1/2$ fermions in the paramagnetic state, 
for which ${\bar k}_{F;\sigma} \equiv {\bar k}_F = (9\pi/4)^{1/3}
\approx 1.919$, exactly reproduces the conventional Hartree-Fock 
result for the single-particle excitation energies (see footnote 
\ref{f75}). It follows that the expression in Eq.~(\ref{ea50}) is exact 
to at least the linear order in $r_s$ in the weak-coupling regime. 
In the light of our considerations subsequent to Eq.~(\ref{ea50}) 
above, the latter observation is mainly of significance for 
$\| {\bar {\Bf k}}-{\bar {\Bf k}}_{F;\sigma}\|\to 0$. 

Now we proceed by assuming that in addition to being uniform and
isotropic, the GS of the system is also a Fermi-liquid metallic 
state. Consequently (for example Farid (1999c)), in a neighbourhood 
of $k_{F;\sigma}$, we have the following relation (unless stated
otherwise, below energies and momenta are {\sl not} normalized)
\begin{equation}
\label{ea61}
\varepsilon_{k;\sigma} = 
\varepsilon_F + \hbar v_{F;\sigma} (k - k_{F;\sigma})
+ o(k-k_{F;\sigma}),
\end{equation}
where 
\begin{equation}
\label{ea62}
v_{F;\sigma} {:=} \frac{1}{\hbar}
\left.\frac{{\rm d} 
\varepsilon_{k;\sigma} }{{\rm d}k}\right|_{k=k_{F;\sigma}}
\equiv \frac{\hbar k_{F;\sigma}}{m_{\sigma}^{\star}}
\end{equation}
is the Fermi velocity in terms of which the effective mass
$m_{\sigma}^{\star}$ for quasi-particles with spin index $\sigma$
is defined. From the quasi-particle equation in Eq.~(\ref{e105}) 
we have, however, 
\footnote{\label{f116}
Below $\ol{\Sigma}(k;\varepsilon)$ denotes the Fourier representation 
of the SE operator and {\sl not} the normalized SE as in 
Eq.~(\protect\ref{e104}). }
\begin{equation}
\label{ea63}
\frac{1}{\hbar} \left.\frac{{\rm d} \varepsilon_{k;\sigma}}
{{\rm d} k}\right|_{k=k_{F;\sigma}} = Z_{F;\sigma} 
\Big\{ v_{F;\sigma}^{(0)} +
\left.\frac{ {\rm d}\ol{\Sigma}_{\sigma}(k;\varepsilon_{F})}
{{\rm d} k}\right|_{k=k_{F;\sigma}} \Big\},
\end{equation}
where $v_{F;\sigma}^{(0)} {:=} \hbar^{-1} 
{\rm d}\varepsilon_{k}^{(0)}/{\rm d} k\vert_{k=k_{F;\sigma}}$ denotes 
the Fermi velocity of the non-interacting particles (note that the 
dependence on $\sigma$ of $v_{F;\sigma}^{(0)}$ is entirely due to 
that of $k_{F;\sigma}$ on $\sigma$). With the amount of the jump 
$Z_{F;\sigma}$ in the momentum-distribution function ${\sf n}_{\sigma}
(k)$ (see Eq.~(\ref{ej7})) at $k=k_{F;\sigma}$ being determined from 
\begin{equation}
\label{ea64}
Z_{F;\sigma} = \Big( 1 -\hbar
\left.\frac{{\rm d}\ol{\Sigma}_{\sigma}(k_{F;\sigma};\varepsilon)}
{{\rm d}\varepsilon}\right|_{\varepsilon
=\varepsilon_{F}}\Big)^{-1},
\end{equation}
we observe that the knowledge of $Z_{F;\sigma}$, $v_{F;\sigma}$ (or 
$m_{\sigma}^{\star}$) and $\varepsilon_{F}$ is sufficient to describe 
fully the leading-order asymptotic behaviour of $\ol{\Sigma}_{\sigma}
(k;\varepsilon)$ corresponding to a Fermi-liquid metallic state for $k\to 
k_{F;\sigma}$ and $\varepsilon\to\varepsilon_{F}$. This follows from 
the fact that $\ol{\Sigma}_{\sigma}(k;\varepsilon)$ pertaining to a 
Fermi-liquid metallic state is uniquely characterized by the following 
two properties (Farid 1999c). 

\vspace{0.2cm}
(A) $\ol{\Sigma}_{\sigma}(k_{F;\sigma};\varepsilon)$ is continuously 
differentiable with respect to $\varepsilon$ in a neighbourhood of 
$\varepsilon=\varepsilon_{F}$. 

\vspace{0.2cm}
(B) $\ol{\Sigma}_{\sigma}(k;\varepsilon_{F})$ is continuously 
differentiable with respect to $k$ in a neighbourhood of 
$k=k_{F;\sigma}$. 

\vspace{0.2cm}
\noindent
Consequently (and this constitutes the foundation of Eq.~(\ref{ea61}) 
above), for $\wt{\ol{\Sigma}}_{\sigma}(k;z)$ pertaining to a 
Fermi-liquid metallic state we can write
\begin{eqnarray}
\label{ea65}
\wt{\ol{\Sigma}}_{\sigma}(k;z) 
&=& (\varepsilon_{F} - \varepsilon_{F;\sigma}^{(0)})
+ {\sf A}_{\sigma} (k - k_{F;\sigma}) 
+ {\sf B}_{\sigma} (z - \varepsilon_{F})\nonumber\\ 
& &\;\;\;\;\;\;\;\;\;\;
+ o(k-k_{F;\sigma}) + o(z-\varepsilon_{F}),
\end{eqnarray}
where $\varepsilon_{F;\sigma}^{(0)} {:=}
\varepsilon_{k}^{(0)}\vert_{k = k_{F;\sigma}}$,
\begin{equation}
\label{ea66}
{\sf A}_{\sigma} {:=} 
\left.\frac{ {\rm d}\ol{\Sigma}_{\sigma}(k;\varepsilon_{F})}
{{\rm d} k}\right|_{k=k_{F;\sigma}} \equiv
\frac{v_{F;\sigma}}{Z_{F;\sigma}} - v_{F;\sigma}^{(0)},
\end{equation}
\begin{equation}
\label{ea67}
{\sf B}_{\sigma} {:=}
\left.\frac{{\rm d}\ol{\Sigma}_{\sigma}(k_{F;\sigma};\varepsilon)}
{{\rm d}\varepsilon}\right|_{\varepsilon
=\varepsilon_{F}} \equiv
\frac{1}{\hbar} \Big(1 - \frac{1}{Z_{F;\sigma}}\Big).
\end{equation}
We point out that, to the order in which $\wt{\ol{\Sigma}}_{\sigma}
(k;z)$ is presented in Eq.~(\ref{ea65}), it is real-valued for $z$ 
identified with a real-valued energy parameter $\varepsilon$. The 
first term on the RHS of Eq.~(\ref{ea65}) can be directly determined 
from the expression in Eq.~(\ref{ea50}); it can also be determined 
from knowledge of the total energy of the interacting system as 
function of the constant particle density $n_0$ through application 
of the Seitz (1940, pp. 343 and 344) theorem (see Eqs.~(\ref{ea53}) 
and (\ref{ea54}); see also footnote \ref{f78}). As is evident from 
Eqs.~(\ref{ea48}) and (\ref{ea49}), calculation of 
$\beta_{{\bar {\Bf k}};\sigma}^{+}$ requires knowledge of 
${\sf n}_{\sigma}({\Bf k})$ from which $Z_{F;\sigma}$ is deduced, 
through $Z_{F;\sigma} = {\sf n}_{\sigma}(k_{F;\sigma}-0^+) - 
{\sf n}_{\sigma}(k_{F;\sigma}+0^+)$ (Migdal 1957, Luttinger 1960) 
(see Eq.~(\ref{ej7})). With the knowledge of $Z_{F;\sigma}$ and
$v_{F;\sigma}$ (or $m_{\sigma}^{\star}$; see Eq.~(\ref{ea62})
above), ${\sf A}_{\sigma}$ is readily obtained through employing 
the expression in Eq.~(\ref{ea66}); ${\sf B}_{\sigma}$, on the other 
hand, is fully determined by $Z_{F;\sigma}$.

In \S~IV.D we present a {\sl constrained} interpolation of the SE 
operator $\Sigma_{\sigma}(\varepsilon)$, interpolating between the 
first-order term of the perturbation series of this operator in 
terms of the dynamically-screened particle-particle interaction 
functions $W$ for $\varepsilon$ close to the `chemical potential' 
$\mu$, and the exact asymptotic expression (as described in terms 
of a finite number of leading-order terms of the 
large-$\vert\varepsilon\vert$ AS for $\Sigma_{\sigma}(\varepsilon)$) 
for `large' values of $\vert\varepsilon\vert$. Although the 
constraints imposed on the global behaviour of the mentioned 
interpolation expression, which we in \S~IV.D denote by 
${\sf f}_{\rm m}^{(1)}(\varepsilon)$, necessarily bring about change 
in the behaviour of the interpolated $\Sigma_{\sigma}(\varepsilon)$ for 
$\varepsilon$ close to $\mu$, in comparison with that of the first-order 
approximation to the SE operator (for one such constraint see the next 
paragraph), such change is {\sl not} explicitly controlled. This 
limitation can be removed by means of extending the scheme in \S~IV.D 
in such a way that it admits {\sl direct} adjustment for $\varepsilon$ 
in the vicinity of $\mu$ and imposing the expected behaviour on the 
interpolating function in this regime. In this context, the expression 
in Eq.~(\ref{ea65}), combined with the large-$\vert z\vert$ asymptotic 
expressions for $\wt{\ol{\Sigma}}_{\sigma}(k;z)$ in Eqs.~(\ref{e128}) 
and (\ref{e129}), is of particular relevance to uniform and isotropic 
metallic states that can be classified as Fermi liquids.

With reference to an observation by Kajueter and Kotliar (1996) in their 
treatment of the Anderson impurity model, we point out that imposition 
of the Luttinger-Ward (1960) {\sl identity} 
\footnote{\label{f117}
See Eq.~(63) in the paper by Luttinger and Ward (1960). See also 
Eqs.~(12) and (13) in the paper by Langer and Ambegaokar (1961) and 
Eq.~(51) in the paper by Farid (1999a); the expressions in the latter 
two works are less restrictive than the expression in the original 
work by Luttinger and Ward (1960) which pertains to uniform and 
isotropic systems and moreover involves a trace over momenta. A 
careful analysis of the details underlying the work by the latter 
workers reveals that these restrictions are {\sl not} essential. We 
note in passing that the two expressions in Eq.~(\protect\ref{e68}) 
are {\sl not} independent; one is obtained from the other through 
integration by parts, which in consequence of the expressions in 
Eqs.~(\protect\ref{e61}) and (\protect\ref{e72}) does {\sl not} lead 
to finite contributions corresponding to the end-points of the 
integration interval. For completeness, let ${\tilde f}_{\sigma}(z) 
{:=} \wt{G}_{\sigma}(z) {\rm d}\wt{\Sigma}_{\sigma}(z)/{\rm d}z$ and 
${\tilde g}_{\sigma}(z) {:=} \wt{\Sigma}_{\sigma}(z) 
{\rm d}\wt{G}_{\sigma}(z)/{\rm d}z$. With (see Eqs.~(\protect\ref{e25}), 
(\protect\ref{e65}), (\protect\ref{e26}) and (\protect\ref{e67}) and 
compare with Eq.~(\protect\ref{e39})) ${\sf f}_{\sigma}(\varepsilon) 
{:=} \lim_{\eta\downarrow 0} [{\tilde f}_{\sigma}(\varepsilon-i\eta) - 
{\tilde f}_{\sigma}(\varepsilon+i\eta)]/(2\pi i)$ and ${\sf g}_{\sigma}
(\varepsilon) {:=} \lim_{\eta\downarrow 0} [{\tilde g}_{\sigma}
(\varepsilon-i\eta) - {\tilde g}_{\sigma}(\varepsilon+i\eta)]/(2\pi i)$, 
Eq.~(\protect\ref{ea68}) can be shown to be equivalent to
$\int_{-\infty}^{\mu} {\rm d}\varepsilon\; {\sf f}_{\sigma}
(\varepsilon)\equiv -\int_{-\infty}^{\mu} {\rm d}\varepsilon\; 
{\sf g}_{\sigma}(\varepsilon)\equiv 0$. }
\begin{eqnarray}
\label{ea68}
&&\int_{\mu-i\infty}^{\mu+i\infty} {\rm d}z\; \wt{G}_{\sigma}(z)\, 
\frac{{\rm d}\wt{\Sigma}_{\sigma}(z)}{{\rm d}z} \nonumber\\
&&\;\;\;\;\;\;\;\;\;
\equiv -\int_{\mu-i\infty}^{\mu+i\infty} {\rm d}z\;
\wt{\Sigma}_{\sigma}(z)\,
\frac{{\rm d}\wt{G}_{\sigma}(z)}{{\rm d}z} \equiv 0
\end{eqnarray} 
on an approximate expression for $\wt{\ol{\Sigma}}_{\sigma}(k;z)$ 
can considerably enhance the accuracy of the approximation. For 
completeness we note that inspection of the work by Luttinger and 
Ward (1960) (see specifically the text following Eq.~(36) in the
latter work) reveals that the identity in Eq.~(\ref{ea68}) is a direct 
manifestation of the conservation of energy at the interaction 
vertices of the diagrammatic representations of $\wt{G}_{\sigma}(z)$ 
and $\wt{\Sigma}_{\sigma}(z)$. Thus Eq.~(\ref{ea68}) amounts to an 
identity not only for the {\sl exact} $\wt{G}_{\sigma}(z)$ and 
$\wt{\Sigma}_{\sigma}(z)$, but also for approximate 
$\wt{\Sigma}_{\sigma}(z)$ and $\wt{G}_{\sigma}(z)$ that in particular 
are calculated within the framework of the many-body perturbation 
theory; Eq.~(\ref{ea68}) may however be violated when the expressions 
for $\wt{\Sigma}_{\sigma}(z)$ and $\wt{G}_{\sigma}(z)$ are postulated. 
Consequently, only in these cases can Eq.~(\ref{e68}) amount to a 
nontrivial condition.

Finally, consider ({\it cf}. Eq.~(\ref{ea62}) above)
\begin{equation}
\label{ea69}
{\bar v}_{{\bar k};\sigma} {:=} 
\frac{{\rm d}\bar\varepsilon_{{\bar k};\sigma}}
{{\rm d}{\bar k}}.
\end{equation}
From the expression in Eq.~(\ref{ea50}) we obtain
\begin{eqnarray}
\label{ea70}
&&{\bar v}_{{\bar k};\sigma} = {\bar v}_{{\bar k}}^{(0)}
- \frac{r_s}{\nu_{\sigma}({\bar k})}\, \nonumber\\
&&\;\;\;\;\;\;\times
\Big( \frac{{\rm d}}{{\rm d}{\bar k}}\, 
\beta_{{\bar k};\sigma}^{\mp}
- \beta_{{\bar k};\sigma}^{\mp}\, \frac{{\rm d}}{{\rm d}{\bar k}}
\ln\big(\nu_{\sigma}({\bar k})\big) \Big),\;\; 
{\bar k}\, \IEq{>}{<}\, {\bar k}_{F;\sigma},
\end{eqnarray}
where ${\bar v}_{{\bar k}}^{(0)} {:=} 
{\rm d}\bar\varepsilon_{\bar k}^{(0)}/{\rm d}{\bar k}$.
The expression in Eq.~(\ref{ea70}) is of interest to us particularly 
for ${\bar k} \uparrow {\bar k}_{F;\sigma}$ and ${\bar k}\downarrow 
{\bar k}_{F;\sigma}$. With reference to Eq.~(\ref{ea25}), from this, 
one would expect that divergent derivatives with respect to ${\bar k}$ 
of $\beta_{{\bar k};\sigma}^{\mp}$ and $\nu_{\sigma}({\bar k})$ to 
the left and right of ${\bar k}_{F;\sigma}$ respectively would 
necessarily imply a vanishing $m_{\sigma}^{\star}$ (see Eq.~(\ref{ea62})
above) and consequently breakdown of Fermi-liquid metallic state. As 
we have discussed in (Farid 1999c), 
\footnote{\label{f118}
See the paragraph preceding that containing Eq.~(33) in the paper
by Farid (1999c). }
for Fermi liquids ${\rm d} {\sf n}_{\sigma}({\bar k})/{\rm d}{\bar k}$
can in principle diverge for ${\bar k}\uparrow {\bar k}_{F;\sigma}$. In 
fact, for a uniform and isotropic system of fermions interacting through a 
short-range potential, Belyakov (1961) (see also Sartor and Mahaux (1980)) 
calculated the GS ${\sf n}_{\sigma}({\bar k})$ of which the left and 
the right derivatives with respect to ${\bar k}$ are logarithmically 
divergent at ${\bar k}_{F;\sigma}$ (see Eq.~(\ref{ej9}) and the 
subsequent text). It is seen from Eq.~(\ref{ea70}) that {\sl cancellation} 
of the possible divergent contributions to the terms enclosed by the 
large parentheses allows for a {\sl finite} renormalization of 
${\bar v}_{F;\sigma}$ with respect to ${\bar v}_{F;\sigma}^{(0)}$, or 
a non-vanishing $m_{\sigma}^{\star}$, in spite of the possible 
divergence of ${\rm d} {\sf n}_{\sigma}({\bar k})/{\rm d}{\bar k}$ to 
the left and/or to the right of ${\bar k}={\bar k}_{F;\sigma}$ (see 
Appendix J, in particular the last paragraph herein).

From Eq.~(\ref{ea70}) we further observe the possibility that
${\bar v}_{{\bar k};\sigma}$ {\sl may} vanish for ${\bar k}
={\bar k}_{F;\sigma}$, implying, according to Eq.~(\ref{ea62}),
a divergent $m_{\sigma}^{\star}$. This possibility is {\sl not}, 
however, directly tied with the condition $Z_{F;\sigma} \to 0$, where 
$Z_{F;\sigma}$ denotes the amount of jump in ${\sf n}_{\sigma}
({\bar k})$ at ${\bar k}={\bar k}_{F;\sigma}$. This is in contrast with 
the expected behaviour according to the Gutzwiller (1963, 1964, 1965) 
{\sl Ansatz} for the GS wavefunction of the Hubbard Hamiltonian in
conjunction with the Gutzwiller {\sl approximation} (for example 
Gebhard (1997, \S~3.4)), where $m^{\star}/m_e = 1/Z_F$ (see Brinkman 
and Rice (1970b, Eq.~(7)) and note that Brinkman and Rice denote 
$Z_F$ by $q$). For completeness, we mention that in the considerations 
by Brinkman and Rice (1970b) the contribution to the effective mass 
due to the momentum dependence of the SE has {\sl not} been taken into 
account; this amounts to neglect of the second term enclosed by large 
parentheses on the RHS of Eq.~(\ref{ea63}). We emphasize that the 
expression in Eq.~(\ref{ea50}), and thus that in Eq.~(\ref{ea70}), 
does {\sl not} involve any approximation associated with the correlated 
GS of the $N$-particle system so that $\Gamma^{(1)}$ and $\Gamma^{(2)}$ 
which determine $\beta_{{\bar k};\sigma}^{\mp}$ (see Eqs.~(\ref{ea45}), 
(\ref{ea46}) and (\ref{ea48})) are the {\sl exact} GS correlations 
functions. Consequently, the approximate nature of the energy expression 
in Eq.~(\ref{ea50}) is {\sl wholly} attributable to our assumptions in 
Eqs.~(\ref{ea26}) and (\ref{ea27}) concerning the ground and excited 
states of the $(N_{\sigma}\mp 1 + N_{\bar\sigma})$-particle system. 
In the considerations by Brinkman and Rice (1970b), on the other hand, 
the underlying approximation (aside from that, which is unimportant in 
the present context, just mentioned above) concerns the GS of the 
$N$-particle system under consideration, taken to be the Gutzwiller 
(1963, 1964, 1965) {\sl Ansatz} and dealt with according to the 
Gutzwiller (1963, 1964, 1965) {\sl approximation}.

\hfill $\Box$

\section{On the density matrices $\Gamma^{(\lowercase{m})}$
and their association with 
$\lowercase{n}_{\sigma}(\lowercase{\Bf r})$,
$\varrho_{\sigma}(\lowercase{\Bf r},\lowercase{\Bf r}')$ and
$\lowercase{\sf g}_{\sigma,\sigma'}
(\lowercase{\Bf r},\lowercase{\Bf r}')$}
\label{s44}

Here we introduce a hierarchy of {\sl static} correlation functions, 
denoted by $\Gamma^{(m)}(x_1,\dots,x_m;x_1',\dots,x_m')$, $0\le m\le N$, 
with $x_i \equiv {\Bf r}_i\sigma_i$, pertaining to the $N$-particle 
GSs of interacting systems of spin-${\sf s}$ fermions described by the 
many-body Hamiltonian $\wh{H}$ in Eq.~(\ref{e1}). These functions are 
defined as the GS expectation values of the normal-ordered products of 
$m$ {\sl pairs} of creation and annihilation field operators in the 
Schr\"odinger picture (for this and other `pictures', see, for example
Fetter and Walecka (1971, pp. 53-59)). In this Appendix we deduce the
expression for $\Gamma^{(m)}$ in terms of a configuration-space integral 
of the GS wavefunction. These expressions are specifically useful for 
the purpose of numerical calculations; in these calculations, under the 
condition $N - m \Ieq{\sim}> 4$, the mentioned configuration-space 
integrals are most efficiently evaluated by means of the Monte Carlo 
importance sampling technique (Negele and Orland 1988, chapter 8). 

Since the first-quantized counterpart of $\wh{H}$ in Eq.~(\ref{e1}) is 
real, unless otherwise stated or implied, in the following as well as in 
other parts of this work we assume the coordinate representation of the 
$N$-particle GS of $\wh{H}$ to be real-valued; with this assumption we 
fix the global gauge of the problem at hand, that is, once the 
zero of the external potential $u({\Bf r})$ in Eq.~(\ref{e2}) has been 
fixed, it is no longer permitted to subject $u({\Bf r})$ to a constant 
(i.e. {\sl global}) arbitrary shift, which otherwise is of no consequence 
to observable quantities. We further assume the GS of $\wh{H}$ to be 
{\sl normal} and {\sl non}-degenerate. 
 
We define (for example March, {\sl et al.} (1967 chapter 1))
\begin{eqnarray}
\label{eb1}
&&\Gamma^{(m)}({\Bf r}_1\sigma_1,\dots,{\Bf r}_m\sigma_m;
{\Bf r}_1'\sigma_1',\dots,{\Bf r}_m'\sigma_m')\nonumber\\
&&\;\;\;\;\;\;\;\;\;\;\;\;\,
{:=} \langle\Psi_{N;0}\vert
\hat\psi_{\sigma_1}^{\dag}({\Bf r}_1)\dots
\hat\psi_{\sigma_m}^{\dag}({\Bf r}_m)\nonumber\\
&&\;\;\;\;\;\;\;\;\;\;\;\;\;\;\;\;\;\;\;\;\;\;\;\;
\times\hat\psi_{\sigma_m'}({\Bf r}_m')\dots
\hat\psi_{\sigma_1'}({\Bf r}_1')\vert\Psi_{N;0}\rangle.
\end{eqnarray}
Orthogonality of $N$-particle states corresponding to different 
$N_{\sigma}$, $\forall\sigma$, or, what is
the same, conservation of spin implies that
\begin{eqnarray}
\label{eb2}
&&\Gamma^{(m)}({\Bf r}_1\sigma_1,
\dots,{\Bf r}_m\sigma_m;{\Bf r}_1'\sigma_1',
\dots,{\Bf r}_m'\sigma_m') \equiv 0,\nonumber\\
&&\;\;\;\;\;\;\;\;\;\;\;\;\;\;\;\;
\neg\exists\, {\cal P}^{(m)}\;\;
\mbox{\rm for which}\;\;
\sigma_i = \sigma'_{{\cal P}^{(m)}i},
\;\forall i,
\end{eqnarray}
where ${\cal P}^{(m)}i$ stands for a permutation of $i$ over
the set $\{1,\dots,m\}$.

The prescription with regard to the coordinate representation of 
the $(N-1)$-particle state resulting from the operation of 
$\hat\psi_{\sigma}({\Bf r}) \equiv \hat\psi(x)$, with $x\equiv 
{\Bf r}\sigma$, on an $N$-particle state, such as $\vert\Psi_{N;0}
\rangle$ whose coordinate representation we denote by $\Psi_{N;0}
(x_1,\dots,x_N)$, is as follows (for example McWeeny 1992, 
pp.~460-464):
\begin{eqnarray}
\label{eb3}
\hat\psi(x) \vert\Psi_{N;0}\rangle 
\rightharpoonup N^{1/2} \Psi_{N;0}(x_1,\dots,x_{N-1},x),
\end{eqnarray}
that is, the coordinate representation of the state corresponding 
to $\hat\psi(x)\vert\Psi_{N;0}\rangle$ is equal to $N^{1/2}$ times 
the coordinate representation of $\vert\Psi_{N;0}\rangle$ in which
the $N$th spin-orbit coordinate $x_N$ has been replaced by $x \equiv
{\Bf r}\sigma$; the state thus obtained, is an $(N-1)$-particle state 
(it is, however, not an eigenstate 
\footnote{\label{f119}
From Eq.~(\protect\ref{e158}), one observes, however, that for systems 
composed of finite number of particles and localized in a finite region 
of space around the origin, such as atoms and molecules, 
$\hat\psi_{\sigma}({\Bf r}) \vert\Psi_{N;0}\rangle$ up to normalization 
asymptotically approaches an $(N-1)$-particle {\sl eigenstate} of $\wh{H}$ 
for $\|{\Bf r}\|\to\infty$; in contrast with what the pertinent equations 
might suggest at the first glance, the eigenvalue corresponding to this 
$(N-1)$-particle eigenstate {\sl is} larger than $E_{N;0}$, following 
the fact that $E_{N-1;s} \ge E_{N-1;0}$ and that the first ionization 
potential $I_1 {:=} E_{N-1;0}-E_{N;0}$ is positive (as an aside we note 
that the smallest $I_1$ for elements in the periodic table amounts to 
$3.89$~eV, pertaining to the element Cs (Perdew, {\sl et al.} 1982)). }
of $\wh{H}$ nor is it normalized to unity) in which $x$ plays the role 
of an {\sl external} parameter. From Eq.~(\ref{eb3}) and the fact that
$\hat\psi(x)\vert\Psi_{N;0}\rangle$ is an $(N-1)$-particle state, it 
follows that 
\begin{eqnarray}
\label{eb4}
\hat\psi(x')\, \hat\psi(x) \vert\Psi_{N;0}\rangle \rightharpoonup 
(N-1)^{1/2} N^{1/2}\nonumber\\ 
\times\Psi_{N;0}(x_1,\dots,x_{N-2},x',x).
\end{eqnarray}
Since $\hat\psi^{\dag}(x) \equiv \hat\psi_{\sigma}^{\dag}({\Bf r})$ 
is the Hermitian conjugate of $\hat\psi(x)$, from Eq.~(\ref{eb3})
we readily obtain
\begin{eqnarray}
\label{eb5}
\langle\Psi_{N;0}\vert \hat\psi^{\dag}(x)
\rightharpoonup N^{1/2} \Psi_{N;0}^*(x_1,\dots,x_{N-1},x),
\end{eqnarray}
and similarly, from Eq.~(\ref{eb4}),
\begin{eqnarray}
\label{eb6}
\langle\Psi_{N;0}\vert \hat\psi^{\dag}(x)\,\hat\psi^{\dag}(x')
\rightharpoonup (N-1)^{1/2} N^{1/2} \nonumber\\
\times\Psi_{N;0}^*(x_1,\dots,x_{N-2},x',x).
\end{eqnarray}
The results in Eqs.~(\ref{eb3}) and (\ref{eb4}) can be easily
generalized for products of an arbitrary number of field
operators, taking into account however that 
\begin{eqnarray}
\label{eb7}
&&\hat\psi(x_M) \hat\psi(x_{M-1})\dots \hat\psi(x_1)
\vert\Psi_{N_{\sigma},N_{\bar\sigma};0}\rangle 
\equiv 0 \nonumber\\
&&\;\;\;\;\;\;\;\;\;\;\;\;\;\;\;\;\;\;\;\;\;\;\;\;\;\;
\mbox{\rm when}\;\;\; 
\sum_{i=1}^M \delta_{\sigma_i,\sigma} > N_{\sigma},\; 
\forall\sigma,
\end{eqnarray}
which reflects the fact that the partial number operators 
$\big\{\wh{N}_{\sigma}\big\}$ are positive {\sl semi}-definite.

From the above prescriptions and Eq.~(\ref{eb1}) we immediately 
obtain
\begin{eqnarray}
\label{eb8}
\Gamma^{(m)}(&&{\Bf r}_1\sigma_1,\dots,{\Bf r}_m\sigma_m;
{\Bf r}_1'\sigma_1',\dots,{\Bf r}_m'\sigma_m')\nonumber\\
\equiv&&\frac{N!}{(N-m)!}
\int {\rm d}x_{m+1}\dots {\rm d}x_N\;\nonumber\\
&&\;\;\;\times\Psi_{N;0}^*({\Bf r}_1\sigma_1,\dots,{\Bf r}_m\sigma_m,
x_{m+1},\dots,x_N)\nonumber\\
&&\;\;\;\times\Psi_{N;0}({\Bf r}_1'\sigma_1',\dots,{\Bf r}_m'\sigma_m',
x_{m+1},\dots,x_N),
\end{eqnarray}
where we have permutated the coordinates of the wavefunctions and
moreover introduced the following short-hand notation:
\begin{equation}
\label{eb9}
x_i {:=} {\Bf r}_i\,\sigma_i,\;\;\;\;
\int {\rm d}x_i\; (\dots) {:=} 
\sum_{\sigma_i} \int {\rm d}^dr_i\; (\dots).
\end{equation}
In the two cases corresponding to $m=N-1$ and $m=N$, the integral 
on the RHS of Eq.~(\ref{eb8}) should be understood as signifying the 
following: for the case $m=N-1$, $\int {\rm d}x_{m+1}\dots {\rm d}x_N 
\Rightarrow \int {\rm d}x_N$, and for the case $m=N$, $\int 
{\rm d}x_{m+1}\dots {\rm d}x_N \Rightarrow 1$ (i.e. {\sl no} 
integration at all). From Eq.~(\ref{eb8}) we can consistently define
\begin{equation}
\label{eb10}
\Gamma^{(0)} = 1\;\;\;\; 
(\Gamma^{(0)}\;\mbox{\rm has {\sl no} arguments}).
\end{equation}
From Eq.~(\ref{eb8}) we further deduce the following rule of
contraction  
\begin{eqnarray}
\label{eb11}
&&\int {\rm d}x_m\;
\Gamma^{(m)}(x_1,\dots,x_m;x_1',\dots,x_m)\nonumber\\
&&\;\;\;\;
= (N-m+1)\,
\Gamma^{(m-1)}(x_1,\dots,x_{m-1};x_1',\dots,x_{m-1}'),
\nonumber\\
&&\;\;\;\;\;\;\;\;\;\;\;\;\;\;\;\;\;\;\;\;\;\;\;\;\;\;\;\;\;\;\;\;\;
\;\;\;\;\;\;\;\;\;\;\;\;\;\;\;\;\;\;\;\;\;\;\;\;\;\;\;\;\; m \ge 1.
\end{eqnarray}

For fermion wavefunctions $\Psi_{N;0}(x_1,\dots,x_N)$, the odd-parity 
(even-parity) permutations of $\{x_1,\dots, x_N\}$ (do not) change the 
sign of $\Psi_{N;0}(x_1,\dots,x_N)$, from which, making use of the
expression in Eq.~(\ref{eb8}), the following relationships are 
immediately deduced:
\begin{eqnarray}
\label{eb12}
&&\Gamma^{(m)}(x_1,\dots,x_j,\dots,x_i,\dots,x_m;
x_1',\dots,x_m')\nonumber\\
=\mp &&\Gamma^{(m)}(x_1,\dots,x_i,\dots,x_j,\dots,x_m;
x_1',\dots,x_m'),\nonumber\\ 
&&\Gamma^{(m)}(x_1,\dots,x_m;
x_1',\dots,x_j',\dots,x_i',\dots,x_m')\nonumber\\
= \mp &&\Gamma^{(m)}(x_1,\dots,x_m;
x_1',\dots,x_i',\dots,x_j',\dots,x_m'),
\end{eqnarray}
depending on whether $x_i \rightleftharpoons x_j$ and $x_i' 
\rightleftharpoons x_j'$ correspond to odd-parity (upper signs) or 
even-parity (lower signs) permutations.

Since by assumption $\Psi_{N;0}(x_1,\dots,x_N)$ is real-valued (see 
above), from Eq.~(\ref{eb8}) we observe that $\Gamma^{(m)}(\{x_i\};
\{x_i'\})$, which in general is complex-valued and transforms into 
the conjugate value of the latter function upon the exchange $\{x_i\} 
\rightleftharpoons \{x_i'\}$, is in our considerations real-valued 
and thus satisfies 
\begin{eqnarray}
\label{eb13}
&&\Gamma^{(m)}(x_1',\dots,x_m';x_1,\dots,x_m)\nonumber\\
\equiv &&\Gamma^{(m)}(x_1,\dots,x_m;x_1',\dots,x_m').
\end{eqnarray}

It can be easily verified that, for the {\sl uniform} GSs of homogeneous 
systems, the following shift property holds:
\begin{eqnarray}
\label{eb14}
&&\Gamma^{(m)}({\tilde {\Bf r}}_1\sigma_1,\dots,
{\tilde {\Bf r}}_m\sigma_m;
{\tilde {\Bf r}}'_1\sigma_1',\dots,
{\tilde {\Bf r}}'_m\sigma_m')\nonumber\\
&&\;\;\;\;\;
\equiv\Gamma^{(m)}({\Bf r}_1\sigma_1,\dots,{\Bf r}_m\sigma_m;
{\Bf r}'_1\sigma_1',\dots,{\Bf r}'_m\sigma_m'),
\end{eqnarray}
where
\begin{equation}
\label{eb15}
{\tilde {\Bf r}}_i {:=} {\Bf r}_i + {\Bf r}_0,\;\;\;\;
{\tilde {\Bf r}}'_i {:=} {\Bf r}'_i + {\Bf r}_0,\;\;
\forall\, i,
\end{equation}
in which ${\Bf r}_0$ stands for an arbitrary constant vector.

For some $(m, \{x_i\}, \{x_i'\})$, $\Gamma^{(m)}(\{x_i\};\{x_i'\})$ 
coincides with well-known static correlation functions. Three of these 
that feature in our present work are as follows.

\vspace{0.2cm}
(i) The partial density matrix
\begin{equation}
\label{eb16}
\varrho_{\sigma}({\Bf r},{\Bf r}') {:=}
\langle\Psi_{N;0}\vert \hat\psi_{\sigma}^{\dag}({\Bf r})
\hat\psi_{\sigma}({\Bf r}')\vert\Psi_{N;0}\rangle
\equiv \Gamma^{(1)}({\Bf r}\sigma;{\Bf r}'\sigma).
\end{equation}

\vspace{0.2cm}
(ii) Consequently, for the partial number density of particles with 
spin $\sigma$ one has $n_{\sigma}({\Bf r}) \equiv \Gamma^{(1)}
({\Bf r}\sigma;{\Bf r}\sigma)$; note in passing that from 
Eq.~(\ref{eb2}) we have $\Gamma^{(1)}({\Bf r} \sigma;{\Bf r}'\sigma') 
\equiv 0$ for $\sigma\not=\sigma'$. 

\vspace{0.2cm}
(iii) The van Hove (1954a,b) pair correlation function
\begin{eqnarray}
\label{eb17}
&&g_{\sigma,\sigma'}({\Bf r},{\Bf r}')
{:=} \frac{1}{N (N-1)}\nonumber\\
&&\;\;\;\;\;\;\;\;\;
\times\langle\Psi_{N;0}\vert
\hat\psi_{\sigma}^{\dag}({\Bf r})
\hat\psi_{\sigma'}^{\dag}({\Bf r}')
\hat\psi_{\sigma'}({\Bf r}')
\hat\psi_{\sigma}({\Bf r})
\vert\Psi_{N;0}\rangle\nonumber\\
&&\;\;\;
\equiv \frac{1}{N (N-1)}\, \Gamma^{(2)}({\Bf r}\sigma,
{\Bf r}'\sigma';{\Bf r}\sigma,{\Bf r}'\sigma').
\end{eqnarray}
Following Eq.~(\ref{eb13}), we have 
\begin{equation}
\label{eb18}
\varrho_{\sigma}({\Bf r}',{\Bf r}) 
\equiv \varrho_{\sigma}({\Bf r},{\Bf r}')
\end{equation}
and, following Eq.~(\ref{eb12}),
\begin{equation}
\label{eb19}
g_{\sigma,\sigma'}({\Bf r},{\Bf r}') \equiv 
g_{\sigma',\sigma}({\Bf r}',{\Bf r}). 
\end{equation}
For some applications it is meaningful to consider the following 
normalized van Hove pair correlation function: 
\begin{equation}
\label{eb20}
{\sf g}_{\sigma,\sigma'}({\Bf r},{\Bf r}')
{:=} \Omega^2 g_{\sigma,\sigma'}({\Bf r},{\Bf r}'),
\end{equation}
which in the thermodynamic limit can be written as
\begin{equation}
\label{eb21}
{\sf g}_{\sigma,\sigma'}({\Bf r},{\Bf r}') 
= \frac{1}{n_0^2}\, \Gamma^{(2)}({\Bf r}\sigma,{\Bf r}'\sigma';
{\Bf r}\sigma,{\Bf r}'\sigma'),
\end{equation}
where $n_0$ stands for the {\sl total} concentration of the
particles, defined in Eq.~(\ref{e9}).

For completeness we mention that the expression in Eq.~(\ref{eb8}) 
can be deduced from that in Eq.~(\ref{eb1}) through employing the 
following association between the second-quantized and first-quantized 
operators corresponding to $N$-particle systems;
\begin{eqnarray}
\label{eb22}
\hat\psi_{\sigma}^{\dag}({\Bf r})
\hat\psi_{\sigma'}({\Bf r}')
\rightleftharpoons
\sum_{i=1}^N \delta_{\sigma,\sigma_i} \delta({\Bf r}-{\Bf r}_i)
\nonumber\\
\times {\cal P}^{\rm op}({\Bf r}',{\Bf r}_i)
{\cal S}^{\rm op}(\sigma',\sigma_i),
\end{eqnarray}
where the projection operators ${\cal P}^{\rm op}({\Bf r}',{\Bf r}_i)$,
${\cal S}^{\rm op}(\sigma',\sigma_i)$ replace {\sl all} ${\Bf r}_i$ and 
$\sigma_i$ to their rights by ${\Bf r}'$ and $\sigma'$ respectively.
For illustration, we apply the prescription in Eq.~(\ref{eb22}) to the 
van Hove pair correlation function as defined in Eq.~(\ref{eb17}). To this
end, from Eq.~(\ref{eb17}), making use of the canonical anticommutation 
relations in Eq.~(\ref{e29}), we first deduce that
\begin{eqnarray}
\label{eb23}
N (N-1) g_{\sigma,\sigma'}({\Bf r},{\Bf r}') = 
-\delta_{\sigma,\sigma'} \delta({\Bf r}-{\Bf r}')\,
n_{\sigma}({\Bf r})\nonumber\\
+ \langle \Psi_{N;0}\vert
{\hat n}_{\sigma}({\Bf r}) {\hat n}_{\sigma'}({\Bf r}')
\vert \Psi_{N;0}\rangle,
\end{eqnarray}
where 
\begin{equation}
\label{eb24}
{\hat n}_{\sigma}({\Bf r}) {:=} 
\hat\psi_{\sigma}^{\dag}({\Bf r})
\hat\psi_{\sigma}({\Bf r}).
\end{equation}
From the prescription in Eq.~(\ref{eb22}) it follows that
\begin{eqnarray}
\label{eb25}
&&\langle \Psi_{N;0}\vert
{\hat n}_{\sigma}({\Bf r}){\hat n}_{\sigma'}({\Bf r}')
\vert \Psi_{N;0}\rangle
= \sum_{i=1}^N \sum_{j=1}^N\int {\rm d}x_1\dots {\rm d}x_N\;
\nonumber\\
&&\;\;\;\;\;\;\;\;\;\;\;\;\;\;
\times \Psi_{N;0}^*(x_1,\dots,x_N)\, \delta_{\sigma,\sigma_i}\,
\delta({\Bf r}-{\Bf r}_i) \nonumber\\
&&\;\;\;\;\;\;\;\;\;\;\;\;\;\;
\times \delta_{\sigma',\sigma_j}\,
\delta({\Bf r}'-{\Bf r}_j)\,
\Psi_{N;0}(x_1,\dots,x_N) \nonumber\\
&&\;\;\;
= \delta_{\sigma,\sigma'}\,\delta({\Bf r}-{\Bf r}')\,
n_{\sigma}({\Bf r})
+ N (N-1) \int {\rm d}x_3\dots {\rm d}x_N\;\nonumber\\
&&\;\;\;\;\;\;\;\;\;\;\;\;\;\;
\times\left|\Psi_{N;0}({\Bf r}\sigma,{\Bf r}'\sigma',x_3,\dots,
x_N)\right|^2,
\end{eqnarray}
which in combination with Eq.~(\ref{eb23}) yields the expression
for $g_{\sigma,\sigma'}({\Bf r},{\Bf r}')$ on the right-most side 
of Eq.~(\ref{eb17}) in which $\Gamma^{(2)}$ is replaced by its 
configuration-space integral representation according to Eq.~(\ref{eb8}). 
It should be noted that, in obtaining the RHS of Eq.~(\ref{eb25}), 
we have first 
written 
\begin{equation}
\label{eb26}
\sum_{i=1}^N \sum_{j=1}^N a_{ij} = \sum_{i=1}^N a_{ii} 
+\sum_{i=1}^N \sum_{j=1\atop j\not=i}^N a_{ij}, 
\end{equation}
and subsequently made use of the invariance of $\left|\Psi_{N;0}
(x_1,\dots,x_N)\right|$ under the permutation of its arguments; 
in this way, the sums have been replaced by the number of summands, 
namely
\begin{equation}
\label{eb27}
\sum_{i=1}^N 1 = N,\;\;\;\;\;\;\;
\sum_{i=1}^N \sum_{j=1\atop j\not=i}^N 1 = N (N-1),
\end{equation}
times the value of a single summand. 
 
Finally, for a concise notation, in the main text we employ the 
following auxiliary functions which all involve $\Gamma^{(2)}$:
\begin{eqnarray}
\label{eb28}
&&{\cal A}({\Bf r},{\Bf r}') {:=}
\int {\rm d}^dr_1''\, {\rm d}^dr_2''\;
v({\Bf r}-{\Bf r}_1'') v({\Bf r}'-{\Bf r}_2'')
\nonumber\\ 
&&\;\;\;\;\;\;\;\;\;\;\;\;
\times \sum_{\sigma_1',\sigma_2'}
\Gamma^{(2)}({\Bf r}_1''\sigma_1',{\Bf r}_2''\sigma_2';
{\Bf r}_1''\sigma_1',{\Bf r}_2''\sigma_2'),
\end{eqnarray}
\begin{eqnarray}
\label{eb29}
&&{\cal B}_{\sigma}({\Bf r},{\Bf r}') {:=} 
\int {\rm d}^dr''\;
v({\Bf r}-{\Bf r}'')\nonumber\\
&&\;\;\;\;\;\;\;\;\;\;\;\;\;\;\;\;\;\;\;\;\;\;
\times \sum_{\sigma'} 
\Gamma^{(2)}({\Bf r}'\sigma,{\Bf r}''\sigma';
{\Bf r}''\sigma',{\Bf r}\sigma),
\end{eqnarray}
\begin{eqnarray}
\label{eb30}
&&{\cal C}({\Bf r},{\Bf r}') {:=} \int {\rm d}^dr_1''\,
{\rm d}^dr_2''\; v({\Bf r}-{\Bf r}_1'')
v({\Bf r}'-{\Bf r}_2'') \nonumber\\
&&\;\;\;\;\;\;\;\;\;\;\;\;\;\;\;\;\;\;\;\;\;\;
\times\sum_{\sigma'} \Gamma^{(2)}({\Bf r}_1''\sigma',
{\Bf r}_2''\sigma';{\Bf r}_1''\sigma',
{\Bf r}_2''\sigma').
\end{eqnarray}
One observes that ${\cal A}({\Bf r},{\Bf r}')$ and ${\cal C}({\Bf r},
{\Bf r}')$ are directly related to the van Hove pair correlation 
function (see Eq.~(\ref{eb17}) above), whereas this is the case for 
${\cal B}_{\sigma}({\Bf r},{\Bf r}')$ {\sl only} when ${\Bf r}={\Bf r}'$. 
We point out that calculations of ${\cal A}({\Bf r},{\Bf r}')$, 
${\cal B}_{\sigma}({\Bf r},{\Bf r}')$ and ${\cal C}({\Bf r},{\Bf r}')$ 
corresponding to {\sl extended} systems of fermions interacting 
through a long-range interaction function $v$ have to be preceded by 
careful analyses of the long-distance asymptotic behaviours of the 
integrands on the RHSs of Eqs.~(\ref{eb28}) - (\ref{eb30}). In 
Appendix F we carry out the necessary analyses for the specific case 
of $v\equiv v_c$ in $d=3$. In this context we identify the unbounded 
contributions to these functions and further establish the links 
between these and their counter-contributions that have their origins 
in other functions. For clarity of presentation, unless indicated or 
implied otherwise, in this work we denote the function obtained by 
removing the fundamentally unbounded part of an original function 
(defined in terms of an integral), by means of the symbol of the 
original function complemented by a single prime; thus, for instance, 
by ${\cal A}'({\Bf r},{\Bf r}')$ we denote ${\cal A}({\Bf r},{\Bf r}')$ 
bar its unbounded contribution (see Eqs.~(\ref{ef2}) and (\ref{ef58})). 
Without further specifying the details (which are sufficiently 
clearly indicated in the appropriate places in this paper), we draw 
attention to the fact that, in general, two and three primes attached 
to the symbol of an earlier-defined function similarly carry 
significance according to the notation adopted in this paper (see 
the Contents for a glimpse of such functions).

\hfill $\Box$

\section{The single Slater-determinant approximation (SSDA)}
\label{s45}

Approximating the interacting $N$-particle GS wavefunction 
$\Psi_{N;0}(x_1,\dots, x_N)$ by a SSD, which we denote by 
$\Phi_{N;0}(x_1,\dots,x_N)$, composed of $N$ orthonormal one-particle 
spin-orbitals (in principle, these are the $N$ normalized 
eigenfunctions $\varphi_{\varsigma;\sigma}({\Bf r})$ of 
Eq.~(\ref{e56}) corresponding to the lowest $N$ eigenvalues 
$\varepsilon_{\varsigma;\sigma}^{(0)}$; however, this aspect is of 
no fundamental significance to our considerations in this Appendix), 
results in considerable simplifications in the expressions derived 
and presented in this work. Viewed from the standpoint that gives 
especial prominence to interaction effects as manifested in the GS 
of interacting systems, a SSDA to $\Psi_{N;0}(x_1,\dots,x_N)$ is 
seen to be largely of interest through the fact that it provides 
{\sl insight} into various aspects of often very complicated 
expressions. Interestingly however, some of the results deduced 
within the framework of the SSDA are {\sl qualitatively} correct 
and in some instances are even amenable to being made 
{\sl quantitatively} exact upon making a judicious choice for the 
`non-interacting' many-body Hamiltonian $\wh{H}_0$ of which
$\vert\Phi_{N;0}\rangle$ is the $N$-particle GS. Below we first 
describe the procedure of calculating the $m$-particle correlation 
function $\Gamma^{(m)}(\{x_i\};\{x_i'\})$ within the framework of 
the SSDA, formally defined as follows ({\it cf}. Eq.~(\ref{eb1}))
\begin{eqnarray}
\label{ec1}
&&\Gamma^{(m)}_{\rm s}({\Bf r}_1\sigma_1,\dots,{\Bf r}_m\sigma_m;
{\Bf r}_1'\sigma_1',\dots,{\Bf r}_m'\sigma_m')\nonumber\\
&&\;\;\;\;\;\;\;\;\;\;\;\;\,
{:=} \langle\Phi_{N;0}\vert
\hat\psi_{\sigma_1}^{\dag}({\Bf r}_1)\dots
\hat\psi_{\sigma_m}^{\dag}({\Bf r}_m)\nonumber\\
&&\;\;\;\;\;\;\;\;\;\;\;\;\;\;\;\;\;\;\;\;\;\;\;\;
\times\hat\psi_{\sigma_m'}({\Bf r}_m')\dots
\hat\psi_{\sigma_1'}({\Bf r}_1')\vert\Phi_{N;0}\rangle.
\end{eqnarray}
Employing the expression for $\Gamma^{(m)}_{\rm s}(\{x_i\};\{x_i'\})$ 
in terms of the single-particle Slater-Fock density matrices 
$\{\Gamma_{\rm s}^{(1)}(x_i;x'_j) \}$ ({\it cf}. Eq.~(\ref{eb16}); see 
also Eq.~(\ref{e169})), we subsequently present the expressions within 
the framework of the SSDA of the correlation functions defined in the 
closing part of Appendix B. Before proceeding, we mention that in 
Appendix F (\S~1.c herein) we introduce the set of generalized density 
matrices $\Gamma^{(m)}_{\Bf\xi}(\{x_i\};\{x_i'\})$ in terms of the 
complete set $\{ \vert {\Bf\xi}\rangle\}$ of $N$-particle Slater 
determinants; identifying $\vert {\Bf\xi} ={\bf 0}\rangle$ with 
$\vert\Phi_{N;0}\rangle$, we have 
\begin{equation}
\label{ec2}
\Gamma^{(m)}_{{\Bf\xi}={\bf 0}}(\{x_i\};\{x_i'\}) \equiv 
\Gamma^{(m)}_{\rm s}(\{x_i\};\{x_i'\}). 
\end{equation}
In Appendix F we further encounter the so-called {\sl transition} 
correlation function $\Gamma^{(m)}_{{\Bf\xi},{\Bf\xi}'}(\{x_i\};
\{x_i'\})$ (McWeeny 1992, \S~5.4) defined as follows: 
\begin{eqnarray}
\label{ec3}
&&\Gamma^{(m)}_{{\Bf\xi},{\Bf\xi}'}({\Bf r}_1\sigma_1,
\dots,{\Bf r}_m\sigma_m;{\Bf r}_1'\sigma_1',\dots,
{\Bf r}_m'\sigma_m')\nonumber\\
&&\;\;\;\;\;\;\;\;\;\;\;\;\;\;\;\;\;\;
{:=} \langle {\Bf\xi}\vert
\hat\psi_{\sigma_1}^{\dag}({\Bf r}_1)\dots
\hat\psi_{\sigma_m}^{\dag}({\Bf r}_m)\nonumber\\
&&\;\;\;\;\;\;\;\;\;\;\;\;\;\;\;\;\;\;\;\;\;\;\;\;
\times\hat\psi_{\sigma_m'}({\Bf r}_m')\dots
\hat\psi_{\sigma_1'}({\Bf r}_1')\vert {\Bf\xi}'\rangle.
\end{eqnarray}
For the generalized $m$-point correlation function 
$\Gamma^{(m)}_{\Bf\xi}$ we have
\begin{equation}
\label{ec4}
\Gamma^{(m)}_{\Bf\xi}(\{x_i\};\{x_i'\}) {:=} 
\Gamma^{(m)}_{{\Bf\xi},{\Bf\xi}}(\{x_i\};\{x_i'\}).
\end{equation}  
The prescription that we present below for expressing 
$\Gamma^{(m)}_{\rm s}$ in terms of $\Gamma^{(1)}_{\rm s}$ (see 
Eq.~(\ref{eb16})), identically applies to $\Gamma^{(m)}_{\Bf\xi}$, 
expressing this in terms of $\Gamma_{{\Bf\xi}}^{(1)}$. For the 
general rules concerning determination of $\Gamma^{(m)}_{{\Bf\xi},
{\Bf\xi}'}$, for $m=1,2$, in terms of simpler functions, we refer 
the reader to Eq.~(5.4.14) of the book by McWeeny (1992).
\footnote{\label{f120}
For completeness we indicate the following corrections to the 
mentioned Eq.~(5.4.14): all functions 
$\psi_{R}^{\prime *}({\Bf x}_1)$ and 
$\psi_{S}^{\prime *}({\Bf x}_2)$ 
in this equation are to be replaced by
$\psi_{R'}^{*}({\Bf x}_1)$ and 
$\psi_{S'}^{*}({\Bf x}_2)$ respectively (i.e. the primes 
pertain to the {\sl subscripts} ${\small R}$ and ${\small S}$ 
and {\sl not} to the spin-orbitals $\psi_{R}$ and $\psi_{S}$); 
further, in (ii), $\pi(\kappa \lambda\vert {\Bf x}_1,{\Bf x}_1)$ 
is to be replaced by $\pi(\kappa \lambda\vert {\Bf x}_1,{\Bf x}_2)$. }

Within the framework of the SSDA, $\Gamma^{(m)}(\{x_i\};\{x_i'\})$ 
is fully expressed in terms of $\Gamma^{(1)}_{\rm s}(x;x')$ $\equiv$ 
$\delta_{\sigma,\sigma'}$ $\Gamma^{(1)}_{\rm s}({\Bf r}\sigma;
{\Bf r}'\sigma)$ $\equiv$ $\delta_{\sigma,\sigma'}$ $\varrho_{{\rm s};
\sigma}({\Bf r},{\Bf r}')$ (for example McWeeny 1992, pp. 125-128). 
This is easily demonstrated by employing the Laplace expansion of 
determinants, organized in such a way that the pertinent 
sub-determinants of order $(N-m)\times (N-m)$ do not involve functions 
with arguments $x_1,\dots, x_m$, $x_1',\dots, x_m'$; a subsequent 
use of the orthogonality of the corresponding $(N-m)$-particle Slater 
determinants, gives rise to the aforementioned simplification in the 
expression for $\Gamma^{(m)}_{\rm s}(\{x_i\};\{x_i'\})$; explicitly, 
in this expression, $\Gamma^{(m)}_{\rm s}(\{x_i\};\{x_i'\})$ is the 
determinant of an $m\times m$ matrix whose $(i,j)$th entry is equal to 
$\Gamma^{(1)}_{\rm s}(x_i;x_j') \equiv \delta_{\sigma_i,\sigma_j} 
\varrho_{{\rm s};\sigma_i}({\Bf r}_i,{\Bf r}_j)$.
\footnote{\label{f121}
The validity of this prescription is readily verified by 
{\sl incomplete induction}. }
It is tempting to replace the latter Slater-Fock density matrix by 
its correlated counterpart $\varrho_{\sigma_i}({\Bf r}_i,{\Bf r}_j)$ 
in order to regain, through the latter function, some of the lost 
correlation effects brought about by the SSDA. However, as
$\varrho_{\sigma_i}({\Bf r}_i,{\Bf r}_j)$, in contrast with 
$\varrho_{{\rm s};\sigma_i}({\Bf r}_i,{\Bf r}_j)$, is {\sl not} idempotent 
(i.e. whereas $\varrho_{{\rm s};\sigma} \varrho_{{\rm s};\sigma} 
\equiv \varrho_{{\rm s};\sigma}$, unless $v\equiv 0$, $\varrho_{\sigma} 
\varrho_{\sigma} \not\equiv \varrho_{\sigma}$), such a direct 
replacement gives rise to fundamentally erroneous results, such as 
incorrect values associated with the integrals of $\Gamma^{(m)}
(\{x_i\};\{x_i'\})$ (see, for example, the text following 
Eq.~(\ref{ef11})).  

Finally, we present the expressions for ${\cal A}({\Bf r},{\Bf r}')$,
${\cal B}_{\sigma}({\Bf r},{\Bf r}')$ and ${\cal C}({\Bf r},{\Bf r}')$,
defined in Eqs.~(\ref{eb28}), (\ref{eb29}) and (\ref{eb30}) respectively,
within the SSDA scheme. Using the above prescription, we readily obtain
\begin{eqnarray}
\label{ec5}
&&{\cal A}_{\rm s}({\Bf r},{\Bf r}') =
v_H({\Bf r};[n]) v_H({\Bf r}';[n])\nonumber\\
&&
-\int {\rm d}^dr_1''\, {\rm d}^dr_2''\;
v({\Bf r}-{\Bf r}_1'') v({\Bf r}'-{\Bf r}_2'')
\sum_{\sigma'}
\varrho_{{\rm s};\sigma'}^2({\Bf r}_1'',{\Bf r}_2''),
\end{eqnarray}
\begin{eqnarray}
\label{ec6}
&&{\cal B}_{{\rm s};\sigma}({\Bf r},{\Bf r}') =
-v_H({\Bf r};[n]) \varrho_{{\rm s};\sigma}({\Bf r}',{\Bf r})
\nonumber\\
&&\;\;\;\;\;\;\;
+\int {\rm d}^dr''\; v({\Bf r}-{\Bf r}'')
\varrho_{{\rm s};\sigma}({\Bf r}',{\Bf r}'')
\varrho_{{\rm s};\sigma}({\Bf r}'',{\Bf r}),
\end{eqnarray}
\begin{eqnarray}
\label{ec7}
&&{\cal C}_{\rm s}({\Bf r},{\Bf r}') =
\sum_{\sigma'} v_H({\Bf r};[n_{\sigma'}]) 
v_H({\Bf r}';[n_{\sigma'}])\nonumber\\
&&-\int {\rm d}^dr_1''\, {\rm d}^dr_2''\;
v({\Bf r}-{\Bf r}_1'') v({\Bf r}'-{\Bf r}_2'')
\sum_{\sigma'} 
\varrho_{{\rm s};\sigma'}^2({\Bf r}_1'',{\Bf r}_2'').
\end{eqnarray}
On comparing the expressions in Eqs.~(\ref{ec5}) and
(\ref{ec7}), one readily deduces that
\begin{eqnarray}
\label{ec8}
{\cal C}_{\rm s}({\Bf r},{\Bf r}')
\equiv
{\cal A}_{\rm s}({\Bf r},{\Bf r}')
&+& \sum_{\sigma'} v_H({\Bf r};[n_{\sigma'}])
v_H({\Bf r}';[n_{\sigma'}])\nonumber\\
&-& v_H({\Bf r};[n]) v_H({\Bf r}';[n]).
\end{eqnarray}
Because of the close similarity between ${\cal A}({\Bf r},{\Bf r}')$ 
and ${\cal C}({\Bf r},{\Bf r}')$ ({\it cf}. Eqs.~(\ref{eb28}) and
(\ref{eb30})), and specifically between their SSDA counterparts
as exposed in Eq.~(\ref{ec8}), the regularized expression for
${\cal C}({\Bf r},{\Bf r}')$ appropriate to the case corresponding 
to $v\equiv v_c$ and $d=3$ is readily deduced along the lines of 
Appendix F.
\hfill $\Box$

\section{The time-reversal symmetry and the vanishing of
${\cal J}_{\sigma}(\lowercase{\Bf r},\lowercase{\Bf r}')$}
\label{s46}

In the most complete expression for $G_{\sigma;\infty_4}({\Bf r},
{\Bf r}')$ presented in Eq.~(\ref{e194}), we encounter the GS 
correlation function 
\begin{eqnarray}
\label{ed1}
&&{\cal J}_{\sigma}({\Bf r},{\Bf r}') 
\equiv v({\Bf r}-{\Bf r}')\, \gamma_{\sigma}({\Bf r},{\Bf r}'),\\
\label{ed2}
&&\gamma_{\sigma}({\Bf r},{\Bf r}') {:=}
-\int {\rm d}^dr''\; v({\Bf r}'-{\Bf r}'')
\lim_{{\tilde {\Bf r}}''\to {\Bf r}''} \tau({\Bf r}'')
\nonumber\\
&&\;\;\;\;\;\;\;\;\;
\times
\sum_{\sigma'} \Big\{
\Gamma^{(2)}({\Bf r}'\sigma,{\Bf r}''\sigma';
{\Bf r}\sigma,{\tilde {\Bf r}}''\sigma')
\nonumber\\
&&\;\;\;\;\;\;\;\;\;\;\;\;\;\;\;\;\;\;
-\Gamma^{(2)}({\Bf r}'\sigma,{\tilde {\Bf r}}''\sigma';
{\Bf r}\sigma,{\Bf r}''\sigma')\Big\},
\end{eqnarray}
which is seen to be asymmetric with respect to the exchange of ${\Bf r}$ 
and ${\Bf r}'$, in apparent contradiction with the required property 
$G_{\sigma;\infty_4}({\Bf r}', {\Bf r}) \equiv G_{\sigma;\infty_4}
({\Bf r}, {\Bf r}')$ (see Eq.~(\ref{e178})). Here we demonstrate that 
in fact
\begin{equation}
\label{ed3}
{\cal J}_{\sigma}({\Bf r},{\Bf r}') \equiv 0.
\end{equation}
As we shall see, the proof of this statement is more direct for systems 
with {\sl non}-uniform GSs than for those with uniform GSs. In the 
course of obtaining the result in Eq.~(\ref{ed3}), we encounter a 
direct relationship between ${\cal J}_{\sigma}({\Bf r},{\Bf r}')$ and 
a matrix element of the total paramagnetic flux-density operator 
${\hat {\Bf j}}_{\rm p}({\Bf r})$ with respect to an $(N_{\sigma}-1
+N_{\bar\sigma})$-particle state directly associated with the 
$(N_{\sigma}+N_{\bar\sigma})$-particle GS of $\wh{H}$. This relationship 
thus suggests a link between ${\cal J}_{\sigma}({\Bf r},{\Bf r}')\equiv 
0$ and our assumption with regard to the non-degeneracy, and thus the 
time-reversal symmetry, of the GS of the system under consideration.
\footnote{\label{f122}
In the absence of an external magnetic field, the time-reversal symmetry 
can be broken only spontaneously; such type of broken-symmetry state
can arise only in a manifold of degenerate GSs. }

From the defining expression for $\gamma_{\sigma}({\Bf r},{\Bf r}')$ in 
Eq.~(\ref{ed2}), making use of the symmetry relation in Eq.~(\ref{eb13}), 
it follows that 
\begin{equation}
\label{ed4}
\gamma_{\sigma}({\Bf r},{\Bf r}) \equiv 0,
\end{equation}
which we now show to have important implications for the integrand of 
the ${\Bf r}''$ integral on the RHS of Eq.~(\ref{ed2}). To this end,
we first express $\gamma_{\sigma}({\Bf r},{\Bf r}')$ as defined in
Eq.~(\ref{ed2}) in its most elementary form, namely
\begin{eqnarray}
\label{ed5}
&&\gamma_{\sigma}({\Bf r},{\Bf r}')
\equiv \frac{\hbar^2}{2 m_e}
\langle\Psi_{N;0}\vert \hat\psi_{\sigma}^{\dag}({\Bf r}')
\nonumber\\
&&\;\;\;
\times\sum_{\sigma'}\int {\rm d}^dr''\; v({\Bf r}'-{\Bf r}'')
\Big\{ \big[\nabla^2_{{\Bf r}''} 
\hat\psi_{\sigma'}^{\dag}({\Bf r}'')\big]
\hat\psi_{\sigma'}({\Bf r}'')
\nonumber\\
&&\;\;\;\;\;\;\;\;\;\;\;\;\;
-\hat\psi_{\sigma'}^{\dag}({\Bf r}'')
\big[\nabla^2_{{\Bf r}''}
\hat\psi_{\sigma'}({\Bf r}'')\big]\Big\}\,
\hat\psi_{\sigma}({\Bf r})\vert\Psi_{N;0}\rangle.
\end{eqnarray}
Making use of 
\begin{equation}
\label{ed6}
\big[\nabla^2 f({\Bf r})\big] g({\Bf r}) = 
{\Bf\nabla}\cdot\big(\big[{\Bf\nabla} f({\Bf r})\big] 
g({\Bf r})\big)-\big[{\Bf\nabla} f({\Bf r})\big]\cdot
\big[{\Bf\nabla} g({\Bf r})\big]
\end{equation}
and the Gauss divergence theorem (or its equivalent in $d\not=3$), 
we readily obtain
\begin{eqnarray}
\label{ed7}
&&\gamma_{\sigma}({\Bf r},{\Bf r}') \equiv
i\hbar \int {\rm d}^dr''\;
\big[{\Bf\nabla}_{{\Bf r}''} v({\Bf r}'-{\Bf r}'')\big]
\cdot {\Bf A}_{{\Bf r}''\sigma}({\Bf r},{\Bf r}'),
\nonumber\\
\end{eqnarray}
where
\begin{eqnarray}
\label{ed8}
{\Bf A}_{{\Bf r}''\sigma}({\Bf r},{\Bf r}')
{:=} \langle\Psi_{N;0}\vert \hat\psi_{\sigma}^{\dag}({\Bf r}')
\, {\hat {\Bf j}}_{\rm p}({\Bf r}'')\, \hat\psi_{\sigma}({\Bf r})
\vert\Psi_{N;0}\rangle,
\end{eqnarray}
with ${\hat {\Bf j}}_{\rm p}({\Bf r}) \equiv \sum_{\sigma} 
{\hat {\Bf j}}_{{\rm p};\sigma}({\Bf r})$ the {\sl total} paramagnetic 
particle flux density operator, where
\begin{eqnarray}
\label{ed9}
&&{\hat {\Bf j}}_{{\rm p};\sigma}({\Bf r}) {:=}
\frac{-i\hbar}{2 m_e}\,
\Big\{ \hat\psi_{\sigma}^{\dag}({\Bf r})
\big[{\Bf\nabla}_{{\Bf r}}\hat\psi_{\sigma}({\Bf r})\big]
-\big[{\Bf\nabla}_{\Bf r}\hat\psi_{\sigma}^{\dag}({\Bf r})\big]
\hat\psi_{\sigma}({\Bf r})\Big\}.\nonumber\\
\end{eqnarray}
The expression in Eq.~(\ref{ed8}) is the aforementioned direct 
relationship between ${\cal J}_{\sigma}({\Bf r},{\Bf r}')$ (see 
Eq.~(\ref{ed7})) and the matrix elements of the total paramagnetic 
flux density operator with respect to an $(N_{\sigma}-1
+N_{\bar\sigma})$-particle state directly associated with
the $(N_{\sigma}+N_{\bar\sigma})$-particle GS of $\wh{H}$. 

From Eq.~(\ref{ed8}) it can be verified that
\begin{equation}
\label{ed10}
\Big({\Bf A}_{{\Bf r}''\sigma}({\Bf r},{\Bf r}')\Big)_{\alpha} \equiv 
\Big({\Bf A}_{{\Bf r}''\sigma}({\Bf r}',{\Bf r})\Big)_{\alpha}^*,\;\;
\alpha=x,y,\dots,
\end{equation}
that is ${\Bf A}_{{\Bf r}''\sigma}({\Bf r},{\Bf r}')$ is Hermitian 
(here $\{\alpha\}$ denotes the components of {\sl vector} 
${\Bf A}_{{\Bf r}''\sigma}({\Bf r},{\Bf r}')$ with respect to some 
$d$-dimensional basis). Since $\gamma_{\sigma}({\Bf r},{\Bf r}')$ is 
real valued (see the defining expression in Eq.~(\ref{ed2}) and consider 
our convention set out in the second paragraph of Appendix B), the 
property in Eq.~(\ref{ed10}), which implies ${\Bf A}_{{\Bf r}''\sigma}
({\Bf r},{\Bf r})$ to be real valued, in combination with 
Eq.~(\ref{ed7}) leads to the conclusion that
\begin{equation}
\label{ed11}
{\Bf A}_{{\Bf r}''\sigma}({\Bf r},{\Bf r}) \equiv 0,\;\;\,
\forall {\Bf r}'', {\Bf r}, \sigma;\;\;\;\;
\gamma_{\sigma}({\Bf r},{\Bf r}) \equiv 0,\;\;\,
\forall {\Bf r}, \sigma.
\end{equation}
The second of these results is exactly that directly obtained from 
the defining expression for $\gamma_{\sigma}({\Bf r},{\Bf r}')$ in 
Eq.~(\ref{ed2}) and presented in Eq.~(\ref{ed4}).

The Hermitian property of ${\Bf A}_{{\Bf r}''\sigma}({\Bf r},{\Bf r}')$ 
implies that this function has the following spectral resolution:
\footnote{\label{f123}
We point out that, whereas the spectral content of 
$\big( {\Bf A}_{{\Bf r}''\sigma}({\Bf r},{\Bf r}') \big)_{\alpha}$
is not exhausted solely by a point spectrum (i.e. by a set of
`eigenvalues'; see footnote \protect\ref{f39}), the spectral 
functions $\{ u_{{\Bf r}''\sigma;\varsigma}({\Bf r})\}$ are 
orthonormal and thus complete, as opposed to `over-complete'. }
\begin{eqnarray}
\label{ed12}
\Big({\Bf A}_{{\Bf r}''\sigma}({\Bf r},{\Bf r}')\Big)_{\alpha} 
= \sum_{\varsigma} \lambda_{{\Bf r}''\sigma;\varsigma}^{(\alpha)}\,
u_{{\Bf r}''\sigma;\varsigma}^{(\alpha)}({\Bf r})
u_{{\Bf r}''\sigma;\varsigma}^{(\alpha)*}({\Bf r}'),
\end{eqnarray}
where $\lambda_{{\Bf r}''\sigma;\varsigma}^{(\alpha)}$, a spectral
point of $\Big({\Bf A}_{{\Bf r}''\sigma}({\Bf r},{\Bf r}')\Big)_{\alpha}$,
is real and the spectral functions satisfy $\langle u_{{\Bf r}''\sigma;
\varsigma}^{(\alpha)},u_{{\Bf r}''\sigma;\varsigma'}^{(\alpha)}\rangle 
= \delta_{\varsigma,\varsigma'}$; in principle the latter result applies 
when $\lambda_{{\Bf r}''\sigma;\varsigma}^{(\alpha)} \not=
\lambda_{{\Bf r}''\sigma;\varsigma'}^{(\alpha)}$, however, in the case 
of degeneracy, the spectral functions can be made orthogonal through 
the Gram-Schmidt orthogonalization procedure. The representation in 
Eq.~(\ref{ed12}) makes evident that for {\sl non-uniform} states,
\begin{equation}
\label{ed13}
\Big({\Bf A}_{{\Bf r}''\sigma}({\Bf r},{\Bf r})\Big)_{\alpha} 
\equiv 0,\; \forall {\Bf r}\;
\iff \lambda_{{\Bf r}''\sigma;\varsigma}^{(\alpha)} \equiv 0,\;
\forall \varsigma.
\end{equation}
In other words, for {\sl non-uniform} GSs, ${\Bf A}_{{\Bf r}''\sigma}
({\Bf r},{\Bf r}) \equiv {\bf 0}$, $\forall {\Bf r}$, suffices to 
establish that ${\Bf A}_{{\Bf r}''\sigma}({\Bf r},{\Bf r}') \equiv 
{\bf 0}$, $\forall {\Bf r}, {\Bf r}'$. The essential role played by 
{\sl non}-uniformity of the GS is that for these states the 
corresponding ${\Bf A}_{{\Bf r}''\sigma}({\Bf r},{\Bf r}')$ is 
{\sl not} a function of ${\Bf r}-{\Bf r}'$, so that ${\Bf A}_{{\Bf r}''
\sigma}({\Bf r},{\Bf r})$ non-trivially depends on the {\sl continuous} 
variable ${\Bf r}$. Consequently, whereas it is possible that for 
{\sl a given} ${\Bf r}$ (or for a {\sl finite} number of distinct 
${\Bf r}$) the set of $\lambda_{{\Bf r}''\sigma;\varsigma}^{(\alpha)}$ 
can balance out the contributions of {\sl all} $\vert u_{{\Bf r}''\sigma;
\varsigma}^{(\alpha)}({\Bf r})\vert^2$, resulting in a vanishing total 
contribution $\big( {\Bf A}_{{\Bf r}''\sigma}({\Bf r},{\Bf r})
\big)_{\alpha}$ (see Eq.~(\ref{ed12})), the ${\Bf r}$ {\sl independence} 
of $\big\{\lambda_{{\Bf r}''\sigma;\varsigma}^{(\alpha)}\big\}$, on the 
one hand, and the non-trivial dependence of $\big\{\vert u_{{\Bf r}''\sigma;
\varsigma}^{(\alpha)}({\Bf r})\vert^2\big\}$, on ${\Bf r}$ on the other
hand (see further), do not allow for any other possibility but that 
presented in Eq.~(\ref{ed13}). For clarity, it is here essential to 
realize that, even though ${\Bf r}''$ is also a continuous variable, the 
dependence on ${\Bf r}''$ of both $\big\{\lambda_{{\Bf r}''\sigma;
\varsigma}^{(\alpha)} \big\}$ and $\big\{\vert u_{{\Bf r}''\sigma;
\varsigma}^{(\alpha)}({\Bf r})\vert^2 \big\}$ implies that the result 
in Eq.~(\ref{ed13}) is {\sl not} necessarily true for uniform systems 
where $\big\{\vert u_{{\Bf r}''\sigma;\varsigma}^{(\alpha)}({\Bf r})
\vert^2\big\}$ is {\sl not} capable of being varied (through 
variation of the continuous variable ${\Bf r}''$) independently 
from $\big\{\lambda_{{\Bf r}''\sigma;\varsigma}^{(\alpha)} \big\}$.

Now we demonstrate that the result in Eq.~(\ref{ed13}) holds true also 
for systems with uniform and isotropic GSs (the assumption concerning
the isotropy of the GS is {\sl not} essential to our arguments that 
follow; its relaxation, however, necessitates introduction of some 
additional symbols which we hereby avoid for simplicity but without loss 
of generality). To this end, we first point out that, for such systems, 
$\gamma_{\sigma}({\Bf r},{\Bf r}') \equiv \gamma_{\sigma}^{\rm h}
(\|{\Bf r}-{\Bf r}'\|)$ and ${\Bf A}_{{\Bf r}''\sigma}({\Bf r},{\Bf r}') 
\equiv {\Bf A}_{{\Bf r}''\sigma}^{\rm h}(\|{\Bf r}-{\Bf r}'\|)$ (see 
Eq.~(\ref{ef9})), where on account of the result in Eq.~(\ref{ed11}) 
we have $\gamma_{\sigma}^{\rm h}(0) = 0$ and ${\Bf A}_{{\Bf r}''
\sigma}^{\rm h}(0) \equiv 0$. Thus for uniform systems, following a 
shift transformation in the variable of integration and application 
of the result ${\Bf\nabla}_{{\Bf r}''} v({\Bf r}-{\Bf r}'') \equiv 
-{\Bf\nabla}_{\Bf r} v({\Bf r}-{\Bf r}'')$, Eq.~(\ref{ed7}) can be 
written as 
\begin{eqnarray}
\label{ed14}
\gamma_{\sigma}^{\rm h}(\|\delta {\Bf r}\|) =
\frac{\hbar}{i} {\Bf\nabla}_{\Bf r} \cdot \int {\rm d}^dr''\; 
v({\Bf r}-{\Bf r}'')&& 
{\Bf A}_{{\Bf r}''+\delta {\Bf r}\sigma}^{\rm h}(\|\delta {\Bf r}\|),
\nonumber\\
&&\;\;\;\;\;\;\;\;\;\; \forall {\Bf r},
\end{eqnarray}
where $\delta {\Bf r} {:=} {\Bf r}' - {\Bf r}$ is an independent
variable. The significance of this result lies in the fact that its 
LHS is independent of ${\Bf r}$ whereas its RHS explicitly depends on 
${\Bf r}$. For cases where $v({\Bf r}-{\Bf r}'')$ is short range, it 
is evident from Eq.~(\ref{ed14}) that, for sufficiently large $\|{\Bf r}\|$, 
to leading order in $1/\|{\Bf r}\|$ the integral on the RHS of 
Eq.~(\ref{ed14}) is proportional to ${\Bf\nabla}_{\Bf r}\cdot 
{\Bf A}_{{\Bf r}+\delta {\Bf r}\sigma}^{\rm h}(\|\delta{\Bf r}\|)$, 
which in combination with the independence of the LHS results in the 
conclusion that   
\begin{equation}
\label{ed15}
\gamma_{\sigma}^{\rm h}(r) \equiv 0,\;\;\; \forall r.
\end{equation}
For the long-range Coulomb interaction in $d=3$, using the multi-pole 
expansion of this function (see Eq.~(\ref{ef20})), one observes that 
the RHS of Eq.~(\ref{ed14}) is a power-low decaying function for 
$\|{\Bf r}\|\to\infty$, which by the same reasoning as above, results 
in the conclusion presented in Eq.~(\ref{ed15}). We thus arrive at
the conclusion that the expression in Eq.~(\ref{ed13}) applies to all 
systems, irrespective of whether the corresponding GSs are uniform or 
otherwise.
\hfill $\Box$

\section{Symmetry of some correlation functions}
\label{s47}

In the main text we encounter some two-point correlation functions,
here denoted by, for example, $f({\Bf r},{\Bf r}')$, which belong to 
one of the following categories: 

\vspace{0.2cm}
(i) explicitly symmetric, that is those functions that {\sl directly} 
transform into themselves upon effecting the exchange ${\Bf r} 
\rightleftharpoons {\Bf r}'$; 

\vspace{0.2cm}
(ii) asymmetric functions (such as $f({\Bf r},{\Bf r}')$) that 
occur in pairs, $f({\Bf r},{\Bf r}') + f({\Bf r}',{\Bf r})$, one 
transforming into another upon applying ${\Bf r}\rightleftharpoons 
{\Bf r}'$; 

\vspace{0.2cm}
(iii) explicitly asymmetric functions that, however, can be shown to 
be symmetric upon some algebraic manipulation; 

\vspace{0.2cm}
(iv) functions (such as $f_1({\Bf r},{\Bf r}')$) that are non-symmetric 
but that can be shown to lead to symmetric functions in combination 
with other equally non-symmetric functions (such as $f_2({\Bf r},
{\Bf r}')$, with $f_1({\Bf r},{\Bf r}') + f_2({\Bf r},{\Bf r}')$ 
symmetric); 

\vspace{0.2cm}
(v) functions that are explicitly asymmetric but detailed 
considerations reveal these to be identically vanishing. 

\vspace{0.2cm}
\noindent
In Appendix D we deal with the only function of the latter category 
that one encounters in considerations of $\Sigma_{\sigma;\infty_m}
({\Bf r},{\Bf r}')$ with $m$ limited to $0, 1$ and $2$. In this 
Appendix we consider functions from categories (iii) and (iv).

\subsection{Implicitly symmetric functions}
\label{s48}

Consider
\begin{eqnarray}
\label{ee1}
{\cal I}_{\sigma}({\Bf r},{\Bf r}') {:=}
\int {\rm d}^dr''\; v({\Bf r}'-{\Bf r}'') 
\lim_{{\tilde {\Bf r}}''\to {\Bf r}''} \tau({\Bf r}'') 
\nonumber\\ 
\times v({\Bf r}-{\Bf r}'')\,
\sum_{\sigma'} \Gamma^{(2)}({\Bf r}'\sigma,
{\Bf r}''\sigma';{\Bf r}\sigma,{\tilde {\Bf r}}''\sigma').
\end{eqnarray}
Using the definition for $\Gamma^{(2)}$ in Eq.~(\ref{eb1}), the 
expression in Eq.~(\ref{ee1}) is readily brought into the following 
elementary form
\begin{eqnarray}
\label{ee2}
&&{\cal I}_{\sigma}({\Bf r},{\Bf r}')
= - \langle\Psi_{N;0}\vert
\hat\psi_{\sigma}^{\dag}({\Bf r}') \sum_{\sigma'}
\int {\rm d}^dr''\; \Big\{ v({\Bf r}'-{\Bf r}'')\nonumber\\
&&\;\;\times\big[ \tau({\Bf r}'')
v({\Bf r}-{\Bf r}'') \hat\psi_{\sigma'}^{\dag}({\Bf r}'')\big]
\hat\psi_{\sigma'}({\Bf r}'')\Big\}
\hat\psi_{\sigma}({\Bf r}) \vert\Psi_{N;0}\rangle.
\end{eqnarray}
Through applying the Gauss divergence theorem (or its equivalent in
$d\not=3$), one readily obtains
\begin{eqnarray}
\label{ee3}
&&{\cal I}_{\sigma}({\Bf r},{\Bf r}') 
= - \langle\Psi_{N;0}\vert
\hat\psi_{\sigma}^{\dag}({\Bf r}') \sum_{\sigma'}
\int {\rm d}^dr''\; \Big\{ v({\Bf r}-{\Bf r}'') \nonumber\\
&&\;\;\times \hat\psi_{\sigma'}^{\dag}({\Bf r}'')
\big[ \tau({\Bf r}'')
v({\Bf r}'-{\Bf r}'') \hat\psi_{\sigma'}({\Bf r}'')\big]
\Big\} \hat\psi_{\sigma}({\Bf r})
\vert\Psi_{N;0}\rangle \nonumber\\
&&\;\;\;\;\;\;\;\;\;\;\;\;\;\;\,
\equiv {\cal I}_{\sigma}^*({\Bf r}',{\Bf r}). 
\end{eqnarray} 
Since ${\cal I}_{\sigma}({\Bf r},{\Bf r}')$ is real-valued (see
the second paragraph in Appendix B), the result in Eq.~(\ref{ee3}) 
demonstrates the symmetry of ${\cal I}_{\sigma}({\Bf r},{\Bf r}')$ 
with respect to ${\Bf r}\rightleftharpoons {\Bf r}'$.

In a similar fashion as above, one can also deduce that (here
${\cal I}_{\sigma}$ is merely a generic symbol for functions
defined in terms of integrals) 
\begin{eqnarray}
\label{ee4}
{\cal I}_{\sigma}({\Bf r},{\Bf r}') {:=}
\int {\rm d}^dr''\; v({\Bf r}'-{\Bf r}'') 
\lim_{{\tilde {\Bf r}}''\to {\Bf r}''} \tau({\Bf r}'') 
\nonumber\\ 
\times v({\Bf r}-{\Bf r}'')\,
\sum_{\sigma'} \Gamma^{(2)}({\Bf r}'\sigma,
{\tilde {\Bf r}}''\sigma';{\Bf r}\sigma,{\Bf r}''\sigma')
\end{eqnarray}
is symmetric, that is ${\cal I}_{\sigma}({\Bf r}',{\Bf r}) \equiv 
{\cal I}_{\sigma}({\Bf r},{\Bf r}')$.
\hfill $\Box$

\subsection{Asymmetric functions
${\cal B}_{\sigma}({\Bf r},{\Bf r}')$, 
${\cal G}_{\sigma}({\Bf r},{\Bf r}')$ 
and their symmetric combinations
${\cal D}_{\sigma}({\Bf r},{\Bf r}')$ and
${\cal F}_{\sigma}({\Bf r},{\Bf r}')$} 
\label{s49}

The function ${\cal B}_{\sigma}({\Bf r},{\Bf r}')$ as defined in
Eq.~(\ref{eb29}) is evidently asymmetric. Here we demonstrate that 
\begin{equation}
\label{ee5}
{\cal D}_{\sigma}({\Bf r},{\Bf r}') {:=} h_0({\Bf r})
\varrho_{\sigma}({\Bf r}',{\Bf r}) - {\cal B}_{\sigma}
({\Bf r},{\Bf r}')
\end{equation}
is symmetric. To this end we consider the equation of motion for 
the annihilation field operator $\hat\psi_{\sigma}({\Bf r} t)$ in 
the Heisenberg picture (Fetter and Walecka 1971, p.~59)
(see footnote \ref{f44})
\begin{eqnarray}
\label{ee6}
\big[i\hbar \frac{\rm d}{{\rm d} t} - h_0({\Bf r})\big]
\hat\psi_{\sigma}({\Bf r} t) &=&
\sum_{\sigma'} \int {\rm d}^dr''\;
v({\Bf r}-{\Bf r}'') \nonumber\\
& &\times\hat\psi_{\sigma'}^{\dag}({\Bf r}'' t)
\hat\psi_{\sigma'}({\Bf r}'' t)
\hat\psi_{\sigma}({\Bf r} t),
\end{eqnarray}
where we have employed the Hamiltonian $\wh{H}$ in Eqs.~(\ref{e1}) 
and (\ref{e2})and the equal-time anticommutation relations for 
the Heisenberg-picture field operators, which are identical with 
those in Eq.~(\ref{e29}) which concern the Schr\"odinger-picture
field-operators. Multiplying both sides of Eq.~(\ref{ee6}) from the 
left by $\hat\psi_{\sigma}^{\dag}({\Bf r}' t')$, with $t' > t$, taking 
the expectation value of the resulting equation with respect to 
$\vert\Psi_{N;0}\rangle$, from the definition of the single-particle 
GF $G_{\sigma}({\Bf r} t, {\Bf r}' t')$, namely
\begin{equation}
\label{ee7}
G_{\sigma}({\Bf r} t, {\Bf r}' t')
{:=} -i\,\langle\Psi_{N;0}\vert {\rm T}
\big\{ \hat\psi_{\sigma}({\Bf r}t) 
\hat\psi_{\sigma}^{\dag}({\Bf r}'t') \big\}\vert
\Psi_{N;0}\rangle,
\end{equation}
where ${\rm T}$ stands for the fermion time-ordering operator 
(Fetter and Walecka 1971, p.~65), we obtain
\begin{eqnarray}
\label{ee8}
&&-i\big[i\hbar \frac{\rm d}{{\rm d} t} - h_0({\Bf r})\big]
G_{\sigma}({\Bf r}t, {\Bf r}'t')
= \int {\rm d}^dr''\; v({\Bf r}-{\Bf r}'')\nonumber\\
&&\;\;\;\;
\times \sum_{\sigma'}\, \langle\Psi_{N;0}\vert
\hat\psi_{\sigma}^{\dag}({\Bf r}'t')
\hat\psi_{\sigma'}^{\dag}({\Bf r}''t)
\hat\psi_{\sigma'}({\Bf r}''t)
\hat\psi_{\sigma}({\Bf r}t)\vert\Psi_{N;0}\rangle,\nonumber\\
&&\;\;\;\;\;\;\;\;\;\;\;\;\;\;\;\;\;\;\;\;\;\;\;\;\;\; 
\;\;\;\;\;\;\;\;\;\;\;\;\;\;\;\;\;\;\;\;\;\;\;\;\;\; 
\;\;\;\;\;\;\;\;\;\;\;\;\;\;\;\;\; t' > t.
\end{eqnarray} 
Now we employ the Fourier representation
\begin{equation}
\label{ee9}
G_{\sigma}({\Bf r}t,{\Bf r}'t')
= \int_{-\infty}^{\infty}
\frac{{\rm d}\varepsilon}{2\pi\hbar}\,
{\rm e}^{-i\varepsilon (t-t')/\hbar}\,
G_{\sigma}({\Bf r},{\Bf r}';\varepsilon),
\end{equation}
from which we obtain (below $\eta\downarrow 0$, corresponds to
$t'\downarrow t$)
\begin{eqnarray}
\label{ee10}
i\hbar \frac{\rm d}{{\rm d} t}
\left. G_{\sigma}({\Bf r}t, {\Bf r}'t')\right|_{t'\downarrow t}
&=& \frac{i}{\hbar} \int_{-\infty}^{\infty}
\frac{{\rm d}\varepsilon}{2\pi i}\;
{\rm e}^{i\varepsilon\eta/\hbar}\,
\varepsilon\, G_{\sigma}({\Bf r},{\Bf r}';\varepsilon),\nonumber\\ \\
\label{ee11}
-i\left. G_{\sigma}({\Bf r}t, {\Bf r}'t')\right|_{t'\downarrow t}
&=& \frac{1}{\hbar} \int_{-\infty}^{\infty} \frac{{\rm d}
\varepsilon}{2\pi i}\;
{\rm e}^{i\varepsilon\eta/\hbar}\,
G_{\sigma}({\Bf r},{\Bf r}';\varepsilon)\nonumber\\
&\equiv& \varrho_{\sigma}({\Bf r}',{\Bf r}).
\end{eqnarray}
Upon taking the limit $t'\downarrow t$ on both sides of Eq.~(\ref{ee8})
and making use of the results in Eqs.~(\ref{ee10}) and (\ref{ee11}), we 
obtain (see Eq.~(\ref{eb29}))
\begin{eqnarray}
\label{ee12}
&&\frac{1}{\hbar} \int_{-\infty}^{\infty}
\frac{{\rm d}\varepsilon}{2\pi i}\; 
{\rm e}^{i\varepsilon \eta/\hbar}
\big[ \varepsilon - h_0({\Bf r}) \big]
G_{\sigma}({\Bf r},{\Bf r}';\varepsilon)\nonumber\\
&&\;\;\;\;\;\;
= \int {\rm d}^dr''\;
v({\Bf r}-{\Bf r}'') \sum_{\sigma'}
\Gamma^{(2)}({\Bf r}'\sigma,{\Bf r}''\sigma';
{\Bf r}\sigma,{\Bf r}''\sigma')\nonumber\\
&&\;\;\;\;\;\;
\equiv -{\cal B}_{\sigma}({\Bf r},{\Bf r}'),
\end{eqnarray}
from which, making use of Eq.~(\ref{ee11}), we deduce ({\it cf}.
Eq.~(\ref{ee5}))
\begin{eqnarray}
\label{ee13}
&&h_0({\Bf r}) \varrho_{\sigma}({\Bf r}',{\Bf r})
- {\cal B}_{\sigma}({\Bf r},{\Bf r}')
\equiv {\cal D}_{\sigma}({\Bf r},{\Bf r}')\nonumber\\
&&\;\;\;\;\;\;\;\;\;
= \frac{1}{\hbar} \int_{-\infty}^{\infty}
\frac{{\rm d}\varepsilon}{2\pi i}\;
{\rm e}^{i\varepsilon \eta/\hbar}\, \varepsilon\,
G_{\sigma}({\Bf r},{\Bf r}';\varepsilon).
\end{eqnarray}
Since $G_{\sigma}({\Bf r},{\Bf r}';\varepsilon) \equiv G_{\sigma}
({\Bf r}',{\Bf r};\varepsilon)$ (for a detailed demonstration see 
Farid (1999a)), Eq.~(\ref{ee13}) establishes that indeed
\begin{equation}
\label{ee14}
{\cal D}_{\sigma}({\Bf r},{\Bf r}') \equiv 
{\cal D}_{\sigma}({\Bf r}',{\Bf r}). 
\end{equation}
For completeness, we mention that, through employing the Lehmann 
representation in Eq.~(\ref{e17}) and application of the residue 
theorem, from the right-most expression on the RHS of Eq.~(\ref{ee13}) 
one obtains
\begin{equation}
\label{ee15}
{\cal D}_{\sigma}({\Bf r},{\Bf r}')
\equiv \sum_s^{<} \varepsilon_{s;\sigma}
f_{s;\sigma}({\Bf r}) f_{s;\sigma}^*({\Bf r}'),
\end{equation} 
where $\sum_s^{<}$ denotes the sum over {\sl all} $s$ for which 
$\varepsilon_{s;\sigma} < \mu$ holds (see Eq.~(\ref{e168}) and 
the succeeding text). We note in passing that, from Eq.~(\ref{ee5}), 
making use of Eq.~(\ref{e43}), one readily obtains
\begin{eqnarray}
\label{ee16}
u({\Bf r}) &=& \frac{1}{n({\Bf r})}\, \sum_{\sigma}
\Big\{ {\cal B}_{\sigma}({\Bf r},{\Bf r})
+ {\cal D}_{\sigma}({\Bf r},{\Bf r}) \nonumber\\
& &\;\;\;\;\;\;\;\;\;\;\;\;\;\;\;\;\;
-\lim_{{\Bf r}'\to {\Bf r}} \tau({\Bf r})
\varrho_{\sigma}({\Bf r}',{\Bf r}) \Big\}.
\end{eqnarray}
This expression is of conceptual relevance in exposing the one-to-one 
correspondence between $u({\Bf r})$ and $n({\Bf r})$ corresponding 
to non-degenerate GSs of many-particle systems, as expected through
a theorem due to Hohenberg and Kohn (1964).
\footnote{\label{f124}
According to this theorem, $u({\Bf r})$ is {\sl up to a constant}
uniquely determined by $n({\Bf r})$; this aspect is not reflected
in the expression in Eq.~(\protect\ref{ee16}) (i.e. there is {\sl no}
arbitrariness in this expression) because in our considerations we 
have assumed the GS wavefunction to be real-valued, whereby we have 
fixed the global gauge in the problem under consideration (see the 
second paragraph in Appendix B). }
It is important to realize that, for the expression in Eq.~(\ref{ee16}) 
not to reduce into an {\sl identity}, it is important that 
${\cal B}_{\sigma}$ herein be calculated from the defining expression 
in Eq.~(\ref{eb29}) and ${\cal D}_{\sigma}$ from the right-most 
expression on the RHS of Eq.~(\ref{ee13}). The result in Eq.~(\ref{ee16}) 
may be employed for the purpose of verifying the correctness and 
establishing the accuracy of the (numerically) calculated functions 
$n$, $\varrho_{\sigma}$, ${\cal B}_{\sigma}$ and ${\cal D}_{\sigma}$ 
in applications of the formalism described in this paper.

It is sometimes advantageous to make the symmetry property in
Eq.~(\ref{ee14}) explicit by means of symmetrizing the RHS of 
Eq.~(\ref{ee5}), thus obtaining
\begin{eqnarray}
\label{ee17}
{\cal D}_{\sigma}({\Bf r},{\Bf r}') &=& \frac{1}{2} 
\big[h_0({\Bf r}) + h_0({\Bf r}')\big]\,
\varrho_{\sigma}({\Bf r}',{\Bf r})\nonumber\\
&-&\frac{1}{2} \big[ {\cal B}_{\sigma}({\Bf r},{\Bf r}')
+ {\cal B}_{\sigma}({\Bf r}',{\Bf r})\big].
\end{eqnarray} 

A second non-symmetric function of ${\Bf r}$ and ${\Bf r}'$
encountered in the main text is the following
\begin{eqnarray}
\label{ee18}
&&{\cal G}_{\sigma}({\Bf r},{\Bf r}') {:=}
\int {\rm d}^dr''\; v({\Bf r}-{\Bf r}'')
v({\Bf r}-{\Bf r}'')\nonumber\\
&&\;\;\;\;\;\;\;\;\;\;\;\;\;\;\;\;\;\;\;\;\;\;
\times\sum_{\sigma'} \Gamma^{(2)}({\Bf r}'\sigma,
{\Bf r}''\sigma';{\Bf r}\sigma,{\Bf r}''\sigma')\nonumber\\
&&\;
+\int {\rm d}^dr_1'' {\rm d}^dr_2''\;
v({\Bf r}-{\Bf r}_1'') v({\Bf r}-{\Bf r}_2'')\nonumber\\
&&\;\;\;\;\;\;\;\;
\times\sum_{\sigma_1',\sigma_2'}
\Gamma^{(3)}({\Bf r}'\sigma,{\Bf r}_1''\sigma_1',
{\Bf r}_2''\sigma_2';{\Bf r}\sigma,{\Bf r}_1''\sigma_1',
{\Bf r}_2''\sigma_2').
\end{eqnarray}
We now determine the counter-terms which in combination with this 
function give rise to a {\sl symmetric} function. To this end we 
proceed from the Heisenberg equation of motion in Eq.~(\ref{ee6}) 
above. Multiplying both sides of this equation from left by 
$[i\hbar {\rm d}/{\rm d} t - h_0({\Bf r})] 
\hat\psi_{\sigma}^{\dag}({\Bf r}'t')$,
replacing $[i\hbar {\rm d}/{\rm d} t - h_0({\Bf r})] 
\hat\psi_{\sigma}({\Bf r}t)$ on the RHS of the thus-obtained
equation by the RHS of Eq.~(\ref{ee6}) followed by a normal-ordering 
of the equal-time Heisenberg field operators, we obtain
\begin{eqnarray}
\label{ee19}
&&\big[i\hbar\frac{\rm d}{{\rm d} t} - h_0({\Bf r})\big]^2
\hat\psi_{\sigma}^{\dag}({\Bf r}'t')
\hat\psi_{\sigma}({\Bf r}t)
\nonumber\\
&&\;\;\;\;
= \int {\rm d}^dr''\; v({\Bf r}-{\Bf r}'')
v({\Bf r}-{\Bf r}'')\nonumber\\
&&\;\;\;\;\;\;\;\;\;\;\;\;
\times\sum_{\sigma'}  
\hat\psi_{\sigma}^{\dag}({\Bf r}'t') 
\hat\psi_{\sigma'}^{\dag}({\Bf r}''t) 
\hat\psi_{\sigma'}({\Bf r}''t) 
\hat\psi_{\sigma}({\Bf r}t)\nonumber\\ 
&&\;\;\;\;
+\int {\rm d}^dr_1'' {\rm d}^dr_2''\;
v({\Bf r}-{\Bf r}_1'') v({\Bf r}-{\Bf r}_2'')\nonumber\\
&&\;\;\;\;\;\;\;\;\;\;\;\;
\times \sum_{\sigma_1',\sigma_2'}
\hat\psi_{\sigma}^{\dag}({\Bf r}'t')
\hat\psi_{\sigma_1'}^{\dag}({\Bf r}_1''t)
\hat\psi_{\sigma_2'}^{\dag}({\Bf r}_2''t)\nonumber\\
&&\;\;\;\;\;\;\;\;\;\;\;\;\;\;\;\;\;\;\;\;
\times \hat\psi_{\sigma_2'}({\Bf r}_2''t)
\hat\psi_{\sigma}({\Bf r}_1''t) 
\hat\psi_{\sigma}({\Bf r}t).
\end{eqnarray}
Along similar lines as above, that is by taking the expectation
value of both sides of Eq.~(\ref{ee19}) with respect to
$\vert\Psi_{N;0}\rangle$ and subsequently taking the limit
$t'\downarrow t$, we arrive at
\begin{eqnarray}
\label{ee20}
{\cal G}_{\sigma}({\Bf r},{\Bf r}')
&=& {\cal F}_{\sigma}({\Bf r},{\Bf r}')
- 2 h_0({\Bf r}) {\cal D}_{\sigma}({\Bf r},{\Bf r}')
\nonumber\\
&+& h_0({\Bf r}) h_0({\Bf r}) \varrho_{\sigma}({\Bf r}',{\Bf r}),
\end{eqnarray} 
where (below $\eta\downarrow 0$)
\begin{eqnarray}
\label{ee21}
{\cal F}_{\sigma}({\Bf r},{\Bf r}')
&{:=}& \frac{1}{\hbar} \int_{-\infty}^{\infty}
\frac{{\rm d}\varepsilon}{2\pi i}\;
{\rm e}^{i\varepsilon\eta/\hbar}\,
\varepsilon^2\, G_{\sigma}({\Bf r},{\Bf r}';\varepsilon)\nonumber\\
&\equiv& -i \left(i\hbar\frac{\rm d}{{\rm d} t}\right)^2\,
\left. G_{\sigma}({\Bf r}t,{\Bf r}'t')\right|_{t'\downarrow t}.
\end{eqnarray}
Here again $G_{\sigma}({\Bf r}',{\Bf r};\varepsilon) \equiv G_{\sigma}
({\Bf r},{\Bf r}';\varepsilon)$ (Farid 1999a) implies that 
${\cal F}_{\sigma}({\Bf r}',{\Bf r}) \equiv {\cal F}_{\sigma}({\Bf r},
{\Bf r}')$. From Eq.~(\ref{ee20}) we have the combination of functions 
that together with the asymmetric function ${\cal G}_{\sigma}({\Bf r},
{\Bf r}')$ result in a symmetric function, namely ${\cal F}_{\sigma}
({\Bf r},{\Bf r}')$. Similar to the expression for ${\cal D}_{\sigma}
({\Bf r},{\Bf r}')$ in Eq.~(\ref{ee17}), the following symmetrized 
expression brings out the true symmetry of ${\cal F}_{\sigma}({\Bf r},
{\Bf r}')$ and may be employed in applications
\begin{eqnarray}
\label{ee22}
&&{\cal F}_{\sigma}({\Bf r},{\Bf r}') 
= \frac{1}{2} \big[ {\cal G}_{\sigma}({\Bf r},{\Bf r}')
+ {\cal G}_{\sigma}({\Bf r}',{\Bf r})\big] \nonumber\\
&&\;\;\;
+\big[ h_0({\Bf r}) + h_0({\Bf r}')\big] {\cal D}_{\sigma}
({\Bf r},{\Bf r}')\nonumber\\
&&\;\;\;
-\frac{1}{2} \big[ h_0({\Bf r}) h_0({\Bf r})
+ h_0({\Bf r}') h_0({\Bf r}')\big] 
\varrho_{\sigma}({\Bf r}',{\Bf r})\nonumber\\
&&\equiv \frac{1}{2} \big[ {\cal G}_{\sigma}({\Bf r},{\Bf r}')
+ {\cal G}_{\sigma}({\Bf r}',{\Bf r})\big]
+ h_0({\Bf r}) h_0({\Bf r}') \varrho_{\sigma}({\Bf r}',{\Bf r})
\nonumber\\
&&\;
-\frac{1}{2} \big[ h_0({\Bf r}) + h_0({\Bf r}')\big]\,
\big[ {\cal B}_{\sigma}({\Bf r},{\Bf r}')
+ {\cal B}_{\sigma}({\Bf r}',{\Bf r})\big].
\end{eqnarray} 
\hfill $\Box$

\section{Regularization of some correlation functions pertaining
to Coulomb-interacting fermion systems}
\label{s50}

Some essential integrals in our present work involving the correlation 
functions $\Gamma^{(m)}(\{x_i\};\{x_i'\})$, $m=2,3$, and the 
long-range Coulomb potential $v_c$ in $d=3$, are {\sl not} 
well-defined in the thermodynamic limit. In this Appendix we deduce 
and present the regularized forms of these integrals which can be 
directly employed and numerically evaluated.

\subsection{${\cal A}({\Bf r},{\Bf r}')$ and its 
regularized form; ${\cal A}'({\Bf r},{\Bf r}')$}
\label{s51}

\subsubsection{Arbitrary systems}
\label{s52}

Consider ${\cal A}({\Bf r},{\Bf r}')$ as defined in Eq.~(\ref{eb28}). 
Making use of the anticommutation relations in Eq.~(\ref{e29}), we 
readily obtain
\begin{eqnarray}
\label{ef1}
&&\Gamma^{(2)}({\Bf r}_1''\sigma_1',{\Bf r}_2''\sigma_2';
{\Bf r}_1''\sigma_1',{\Bf r}_2''\sigma_2')\nonumber\\
&&\;\; = -\delta_{\sigma_1',\sigma_2'}\,
\delta({\Bf r}_1''-{\Bf r}_2'')\, 
n_{\sigma_1'}({\Bf r}_1'') + n_{\sigma_1'}({\Bf r}_1'') 
n_{\sigma_2'}({\Bf r}_2'') \nonumber\\
&&\;\;\;\;\;\;
+\langle\Psi_{N;0}\vert
\big[{\hat n}_{\sigma_1'}({\Bf r}_1'') -
n_{\sigma_1'}({\Bf r}_1'')\big]\,\nonumber\\
&&\;\;\;\;\;\;\;\;\;\;\;\;\;\;\;\;\,
\times \big[{\hat n}_{\sigma_2'}({\Bf r}_2'') -
n_{\sigma_2'}({\Bf r}_2'')\big]\vert\Psi_{N;0}\rangle.
\end{eqnarray}
Substitution of this expression in the RHS of Eq.~(\ref{eb28}) 
yields (see text following Eq.~(\ref{e15}) in \S~II.A concerning our 
convention with regard to $v$ and $v'$)
\begin{eqnarray}
\label{ef2}
&&{\cal A}({\Bf r},{\Bf r}')
= v_H({\Bf r};[n])\, v_H({\Bf r}';[n]) \nonumber\\
&&\;\;
-\int {\rm d}^dr''\; v({\Bf r}-{\Bf r}'')
v({\Bf r}'-{\Bf r}'')\, n({\Bf r}'')
+{\cal A}'({\Bf r},{\Bf r}'), 
\end{eqnarray}
where 
\begin{equation}
\label{ef3}
{\cal A}'({\Bf r},{\Bf r}') {:=}
\int {\rm d}^dr_1'' {\rm d}^dr_2''\;
v({\Bf r}-{\Bf r}_1'') 
v({\Bf r}'-{\Bf r}_2'')\,
{\cal U}({\Bf r}_1'',{\Bf r}_2''),
\end{equation}
in which 
\begin{eqnarray}
\label{ef4}
&&{\cal U}({\Bf r}_1'',{\Bf r}_2'') {:=}
\sum_{\sigma_1',\sigma_2'}
\langle\Psi_{N;0}\vert
\big[{\hat n}_{\sigma_1'}({\Bf r}_1'') -
n_{\sigma_1'}({\Bf r}_1'')\big]\,\nonumber\\
&&\;\;\;\;\;\;\;\;\;\;\;\;\;\;\;\;\;\;\;\;\;\;\;\;\,
\times \big[{\hat n}_{\sigma_2'}({\Bf r}_2'') -
n_{\sigma_2'}({\Bf r}_2'')\big]\vert\Psi_{N;0}\rangle.
\end{eqnarray}

In $d=3$ and for $v\equiv v_c$, the Coulomb potential, neither
\begin{eqnarray}
v_H({\Bf r};[n_0])\, v_H({\Bf r}';[n_0])\nonumber
\end{eqnarray}
nor 
\begin{eqnarray}
-n_0 \int {\rm d}^3r''\; v({\Bf r}-{\Bf r}'') 
v({\Bf r}'-{\Bf r}'')\nonumber
\end{eqnarray}
exists for $\kappa\downarrow 0$; here $n_0$ stands for the total 
concentration of fermions defined in Eq.~(\ref{e9}). Further, as 
the former contribution depends {\sl quadratically} upon $n_0$ while 
the latter {\sl linearly} upon $n_0$, it is evident that these 
contributions cannot cancel so that for the case under consideration, 
either ${\cal A}({\Bf r},{\Bf r}')$ is unbounded or otherwise 
${\cal A}'({\Bf r},{\Bf r}')$ must contain contributions that remove 
the indicated unbounded contributions. Before presenting details, 
demonstrating that ${\cal A}'({\Bf r},{\Bf r}')$ {\sl is} well-defined 
(i.e., it is {\sl not} unbounded), we point out that the local part of 
$\Sigma_{\sigma; \infty_1}({\Bf r},{\Bf r}')$, that is $\Sigma_{\sigma;
\infty_1}^{\rm l}({\Bf r},{\Bf r}')$ as presented in Eq.~(\ref{e187}), 
is {\sl exactly} equal to $\hbar^{-1}{\cal A}'({\Bf r},{\Bf r})\, 
\delta({\Bf r}-{\Bf r'})$ so that the possible unboundedness of 
${\cal A}'({\Bf r},{\Bf r})$ would amount to the result that in the 
AS of $\wt{\Sigma}_{\sigma}({\Bf r},{\Bf r}';z)$ for 
$\vert z\vert\to \infty$, the leading asymptotic term following the 
constant $\Sigma_{\sigma;\infty_0}({\Bf r},{\Bf r}')$, would not be 
of the form $\Sigma_{\sigma;\infty_1}({\Bf r},{\Bf r}')/z$, but that 
this would be preceded by a more dominant term (see \S~II.B). In what 
follows we shall in the main focus on the case corresponding to $d=3$ 
and $v\equiv v_c$.

We now demonstrate that ${\cal A}'({\Bf r},{\Bf r}')$ is {\sl bounded} 
(in contrast, for $v\equiv v_c$ in $d=2$, ${\cal A}'({\Bf r},{\Bf r}')$ 
turns out to be unbounded; B. Farid, 2001, unpublished). To this end, 
we first point out that 
\begin{equation}
\label{ef5}
\int {\rm d}^dr_1''\; {\cal U}({\Bf r}_1'',{\Bf r}_2'') \equiv 
\int {\rm d}^dr_2''\; {\cal U}({\Bf r}_1'',{\Bf r}_2'') \equiv 0,
\end{equation}
which follow from the combination of the following facts. 

\vspace{0.2cm}
(i) $\sum_{\sigma'}\int {\rm d}^dr'\; {\hat n}_{\sigma'} ({\Bf r}') 
= \wh{N}$, the {\sl total} number operator. 

\vspace{0.2cm}
(ii) $\vert\Psi_{N;0}\rangle$ is an eigenstate of $\wh{N}$. 

\vspace{0.2cm}
(iii) $\sum_{\sigma'} \int {\rm d}^dr'\; n_{\sigma'}({\Bf r}') = N$. 

\vspace{0.2cm}
\noindent
For large values of $\|{\Bf r}\|$ and $\|{\Bf r}'\|$, employing the 
asymptotic results $v_c({\Bf r}-{\Bf r}'') \sim v_c({\Bf r}) \propto 
1/\|{\Bf r}\|$ and $v_c({\Bf r}'-{\Bf r}'') \sim v_c({\Bf r}') 
\propto 1/\|{\Bf r}'\|$ in the integrand on the RHS of Eq.~(\ref{ef3}), 
we arrive at the conclusion that ${\cal A}'({\Bf r},{\Bf r}') \sim 0$ 
to orders $1/\|{\Bf r}\|$ and $1/\|{\Bf r}'\|$, as $\|{\Bf r}\|, 
\|{\Bf r}'\| \to\infty$. This result in particular implies that 
${\cal A}'({\Bf r},{\Bf r}')$ {\sl cannot} involve a {\sl constant} 
unbounded contribution under the conditions considered in this 
Appendix. In contrast, $v_H({\Bf r};[n_0])$ and $v_H({\Bf r}';[n_0])$ 
diverge for {\sl all} ${\Bf r}$ and ${\Bf r}'$ as $\kappa\downarrow 0$ 
(see Eq.~(\ref{e15})). We point out that for uniform and isotropic 
systems, ${\cal A}'({\Bf r},{\Bf r}')$ is a function of 
$\|{\Bf r}-{\Bf r}'\|$, so that in these systems ${\cal A}'({\Bf r},
{\Bf r}') \sim 0$ to order $1/\|{\Bf r}-{\Bf r}'\|$ for $\|{\Bf r}
-{\Bf r}'\| \to \infty$. It is important to take this fact into account 
when considering uniform GSs, since in these the conditions $\|{\Bf r}\|, 
\|{\Bf r}'\| \to \infty$ do {\sl not} necessarily imply $\|{\Bf r}
-{\Bf r}'\|\to\infty$.

Now, making use of Eq.~(\ref{e16}), the application of 
$\nabla_{{\Bf r}}^2$ and $\nabla_{{\Bf r}'}^2$ from the left to 
${\cal A}'({\Bf r},{\Bf r}')$ in Eq.~(\ref{ef3}) yields
\begin{eqnarray}
\label{ef6}
&&\nabla_{{\Bf r}}^2 \nabla_{{\Bf r}'}^2\,
{\cal A}'({\Bf r},{\Bf r}') = 
\big(-e^2/\epsilon_0\big)^2\, 
{\cal U}({\Bf r},{\Bf r'})\nonumber\\
&&\;\;\;\;\;\;
\equiv\big(-e^2/\epsilon_0\big)^2\, 
\Big( \sum_{\sigma,\sigma'}
\Gamma^{(2)}({\Bf r}\sigma,{\Bf r}'\sigma';
{\Bf r}\sigma,{\Bf r}'\sigma') \nonumber\\
&&\;\;\;\;\;\;\;\;\;\;\;\;\;\;\;\;\;\;\;\;\;\;\;\;\;\;\;\;\;
- n({\Bf r}) n({\Bf r}')
+ n({\Bf r})\, \delta({\Bf r}-{\Bf r}')
\Big),
\end{eqnarray}
where we have made use of Eqs.~(\ref{ef1}) and (\ref{ef4}). From 
the definition of $\Gamma^{(2)}$ in Eq.~(\ref{eb1}) it is evident 
that the first contribution enclosed by large parentheses on the RHS 
of Eq.~(\ref{ef6}), similar to the second contribution, is bounded 
for {\sl all} 
\footnote{\label{f125}
Note that $\Gamma^{(2)}({\Bf r}\sigma,{\Bf r}'\sigma';{\Bf r}\sigma,
{\Bf r}'\sigma') \equiv n_0^2\, {\sf g}_{\sigma,\sigma'}({\Bf r},
{\Bf r}')$ (see Eq.~(\protect\ref{eb21})). }
${\Bf r}$ and ${\Bf r}'$ so that $\nabla_{{\Bf r}}^2 
\nabla_{{\Bf r}'}^2\,{\cal A}'({\Bf r},{\Bf r}')$ is bounded 
everywhere {\sl except} at ${\Bf r}={\Bf r}'$. Thus, for 
${\Bf r}\approx {\Bf r}'$ the {\sl most dominant} contribution to
${\cal A}'({\Bf r},{\Bf r}')$ is obtained by the approximate 
differential equation ({\it cf}. Eq.~(\ref{ef6}))
\begin{equation}
\label{ef7}
\nabla_{{\Bf r}}^2 \nabla_{{\Bf r}'}^2\,
{\cal A}'({\Bf r},{\Bf r}') \approx
\big(e^2/\epsilon_0\big)^2\, n({\Bf r})\, 
\delta({\Bf r}-{\Bf r}').
\end{equation} 
It is readily verified that this equation has the following 
solution
\begin{eqnarray}
\label{ef8}
&&{\cal A}'({\Bf r},{\Bf r}') \approx {\cal A}_0' 
-\frac{n_0 e^4}{8\pi \epsilon_0^2}\,\|{\Bf r}-{\Bf r}'\| 
\nonumber\\
&&\;\;\;\;\;\;\;
+\int {\rm d}^3r''\;
{\rm e}^{-\kappa \|{\Bf r}-{\Bf r}''\|}\,
{\rm e}^{-\kappa \|{\Bf r}'-{\Bf r}''\|}\,\nonumber\\
&&\;\;\;\;\;\;\;\;\;\;\;\;\;\;\;
\times v_c({\Bf r}-{\Bf r}'') v_c({\Bf r}'-{\Bf r}'')\,
n'({\Bf r}''),
\end{eqnarray}
which is demonstrably {\sl bounded} (for $n'({\Bf r}'')$ see
Eq.~(\ref{e12}); note that for uniform GSs, $n'\equiv 0$). Here 
${\cal A}_0'$ is a constant. We point out that similar to the 
{\sl exact} ${\cal A}'({\Bf r},{\Bf r}')$, this approximate 
solution (which has a `cusp'-like behaviour) is symmetric with 
respect to the exchange ${\Bf r} \rightleftharpoons {\Bf r}'$.

\subsubsection{Uniform and isotropic systems}
\label{s53}

We now complement our above general considerations by directly dealing 
with ${\cal A}'({\Bf r},{\Bf r}')$ pertaining to uniform and isotropic 
systems, giving especial attention to ${\cal A}'({\Bf r},{\Bf r}')$
within the framework of the SSDA (see Appendix C). As we shall 
see, these considerations directly lead us to deduce the functional 
form of ${\cal A}^{\prime\rm h}(\|{\Bf r}\|=0)$ pertaining to the fully 
{\sl interacting} system. For clarity, here as elsewhere in this paper 
we employ the notation
\begin{equation}
\label{ef9}
F^{\rm h}(\|{\Bf r}-{\Bf r}'\|) \equiv F({\Bf r},{\Bf r}')
\end{equation}
when dealing with two-point functions pertaining to uniform and 
isotropic GSs. The following analyses will in addition demonstrate 
the hazards of inappropriately neglecting the soft cut-off function 
$\exp(-\kappa\|{\Bf r}-{\Bf r}'\|)$ with which $v({\Bf r}-{\Bf r}')$ 
has to be multiplied when $v\equiv v_c$ ({\it cf}. Eq.~(\ref{e13})). 
We note that ${\cal A}^{\prime\rm h}(0)$ is the same quantity as 
${\cal A}_0'$ in Eq.~(\ref{ef8}) (recall that here $n'\equiv 0$) and 
that, for uniform GSs, it {\sl fully} determines the {\sl local} part 
of $\Sigma_{\sigma;\infty_1}({\Bf r},{\Bf r}')$ ({\it cf}. 
Eq.~(\ref{e187})); for {\sl non}-uniform systems, on the other hand, 
$\Sigma_{\sigma;\infty_1}^{\rm l}({\Bf r},{\Bf r}')$ is fully determined 
by the ${\Bf r}$-dependent function ${\cal A}'({\Bf r},{\Bf r})$. 

Making use of Eq.~(\ref{eb1}), while employing (see Appendix C)
\begin{eqnarray}
\label{ef10}
\Gamma^{(2)}_{\rm s}({\Bf r}\sigma,{\Bf r}'\sigma';
{\Bf r}\sigma,{\Bf r}'\sigma') &=& 
n_{{\rm s};\sigma}({\Bf r}) 
n_{{\rm s};\sigma'}({\Bf r}') \nonumber\\
&-&\delta_{\sigma,\sigma'} 
\varrho_{{\rm s};\sigma}^2({\Bf r}',{\Bf r}), 
\end{eqnarray}
from Eq.~(\ref{ef4}) we obtain 
\begin{eqnarray}
\label{ef11}
{\cal U}_{\rm s}({\Bf r}_1'',{\Bf r}_2'')
= n_{\rm s}({\Bf r}_1'')\, \delta({\Bf r}_1''-{\Bf r}_2'')
-\sum_{\sigma'} 
\varrho_{{\rm s};\sigma'}^2({\Bf r}_2'',{\Bf r}_1'').
\end{eqnarray}
Here as elsewhere in the present work, $\varrho_{{\rm s};\sigma}$ 
denotes the single-particle Slater-Fock density matrix and $n_{\rm s}$ 
the {\sl total} number density pertaining to the SSDA of the GS of 
the system; the use in Eq.~(\ref{ef11}) of the interacting density 
matrix $\varrho_{\sigma}$ (rather than $\varrho_{{\rm s};\sigma}$) 
gives rise to violation of the exact results in Eq.~(\ref{ef5}) 
(irrespective of whether in Eq.~(\ref{ef11}) $n_{\rm s}$ is 
maintained {\sl or} replaced by its exact counterpart $n$), this 
owing to the {\sl non}-idempotency of the interacting density matrix 
(see Appendix C). 

For later reference, we re-write the result in Eq.~(\ref{ef4}) in 
the following appealing form ({\it cf}. Eq.~(\ref{ef11}))
\begin{equation}
\label{ef12}
{\cal U}({\Bf r}_1'',{\Bf r}_2'') = n({\Bf r}_1'')\,
\delta({\Bf r}_1''-{\Bf r}_2'')
- \rho({\Bf r}_2'',{\Bf r}_1''),
\end{equation}
where
\begin{eqnarray}
\label{ef13}
&&\rho({\Bf r}_2'',{\Bf r}_1'') {:=}
n({\Bf r}_1'') n({\Bf r}_2'')\nonumber\\
&&\;\;\;\;\;\;\;\;\;\;
-\sum_{\sigma_1',\sigma_2'}
\Gamma^{(2)}({\Bf r}_1''\sigma_1',{\Bf r}_2''\sigma_2';
{\Bf r}_1''\sigma_1',{\Bf r}_2''\sigma_2').
\end{eqnarray}
Evidently, similar to $\sum_{\sigma'} \varrho_{{\rm s};\sigma'}^2
({\Bf r}_2'',{\Bf r}_1'') \equiv \rho_{\rm s}({\Bf r}_2'',{\Bf r}_1'')$ 
on the RHS of Eq.~(\ref{ef11}), $\rho({\Bf r}_2'',{\Bf r}_1'')$ is 
symmetric with respect to the operation ${\Bf r}_1'' \rightleftharpoons 
{\Bf r}_2''$. Further, making use of the contraction formula in 
Eq.~(\ref{eb11}), one can readily show that
\begin{equation}
\label{ef14}
\int {\rm d}^dr_2''\; \rho({\Bf r}_2'',{\Bf r}_1'')=n({\Bf r}_1''),
\end{equation}
which guarantees satisfaction of the conditions in Eq.~(\ref{ef5}).
It is interesting to note that, owing to the idempotency of
$\varrho_{{\rm s};\sigma}({\Bf r}_2'',{\Bf r}_1'')$ ({\it cf}.
Eq.~(\ref{e166})), 
we similarly have
\begin{equation}
\label{ef15}
\int {\rm d}^dr_2''\; \sum_{\sigma'} 
\varrho_{{\rm s};\sigma'}^2({\Bf r}_2'',{\Bf r}_1'')
=n_{\rm s}({\Bf r}_1'').
\end{equation}
For uniform GSs, $n_{\rm s}$ coincides with $n_0$, the total 
concentration of particles (see Eq.~(\ref{e9})), and is therefore 
independent of interaction. For non-uniform GSs, on the other hand, 
in general $n_{\rm s}({\Bf r})\not\equiv n({\Bf r})$, however under the 
conditions that we have specified in \S~III.C (see the paragraph that 
includes Eqs.~(\ref{e54}) and (\ref{e55})), it {\sl is} in general 
possible to choose the `non-interacting' Hamiltonian $\wh{H}_0$ such 
that the corresponding GS number density $n_{\rm s}({\Bf r})$ 
identically coincides with $n({\Bf r})$. By choosing the Slater
determinant employed within the framework of the SSDA to be the GS 
of such an $\wh{H}_0$, not only do the first terms on the RHSs of 
Eqs.~(\ref{ef11}) and (\ref{ef12}) become identical, but also the 
integrals of the second terms yield identical results, as evidenced 
by Eqs.~(\ref{ef14}) and (\ref{ef15}) (see footnote \ref{f7} and the 
associated text in \S~I.B). Finally, for uniform GSs where 
$n({\Bf r}_1'') n({\Bf r}_2'') = n_0^2$ (see Eq.~(\ref{e12})), 
$\rho({\Bf r}_2'',{\Bf r}_1'')$ in Eq.~(\ref{ef13}) correctly 
reduces to a function of ${\Bf r}_1''-{\Bf r}_2''$. In what follows, 
we concentrate on systems with uniform {\sl and} isotropic GSs and 
consequently where appropriate, denote $\varrho_{{\rm s};\sigma'}
({\Bf r}_2'',{\Bf r}_1'')$ and $\rho({\Bf r}_2'',{\Bf r}_1'')$ by 
$\varrho_{{\rm s};\sigma'}^{\rm h}(\|{\Bf r}_2''-{\Bf r}_1''\|)$ and 
$\rho^{\rm h}(\|{\Bf r}_2''-{\Bf r}_1''\|)$, respectively (see 
Eq.~(\ref{ef9}) above).

Substitution of the RHS of Eq.~(\ref{ef12}) into that of 
Eq.~(\ref{ef3}), followed by two consequent transformations of 
variables, yields
\begin{eqnarray}
\label{ef16}
&&{\cal A}^{\prime\rm h}(\|{\tilde {\Bf r}}\|)
= \int {\rm d}^3r''\;
v_c({\Bf r}'') \,
\Big[ n_0\,\exp(-\kappa \|{\tilde {\Bf r}}-{\Bf r}''\|)\nonumber\\
&&\;\;\;\;\;\;\;\;\;\;\;\;\;\;\;\;\;\;\;\;\;\;\;
\times v_c({\tilde {\Bf r}} - {\Bf r}'')
- {\cal I}_{\kappa}({\tilde 
{\Bf r}} - {\Bf r}'') \Big],
\end{eqnarray}
where ${\tilde {\Bf r}} {:=} {\Bf r}-{\Bf r}'$ and
\begin{equation}
\label{ef17}
{\cal I}_{\kappa}({\Bf r}) {:=}
\int {\rm d}^3r''\;
\exp(-\kappa \|{\Bf r}-{\Bf r}''\|)\,
v_c({\Bf r}-{\Bf r}'')\,
\rho^{\rm h}(\|{\Bf r}''\|).
\end{equation}
The expression for ${\cal A}^{\prime\rm h}_{\rm s}(\|{\tilde {\Bf r}}\|)$ 
is directly deduced from that in Eq.~(\ref{ef16}) through replacing 
${\cal I}_{\kappa}({\tilde {\Bf r}} - {\Bf r}'')$ herein by 
${\cal I}_{{\rm s};\kappa}({\tilde {\Bf r}} - {\Bf r}'')$, where
\begin{equation}
\label{ef18}
{\cal I}_{{\rm s};\kappa}({\Bf r}) \equiv
\left. {\cal I}_{\kappa}({\Bf r})
\right|_{\rho^{\rm h} \rightharpoonup \sum_{\sigma'} 
(\varrho_{{\rm s};\sigma'}^{\rm h})^2}.
\end{equation}

In order to investigate the behaviour of 
${\cal A}^{\prime\rm h}_{\rm s}(\|{\tilde {\Bf r}}\|)$, we first 
consider ${\cal I}_{{\rm s};\kappa}({\Bf r})$ in Eq.~(\ref{ef18}),
with ${\cal I}_{\kappa}({\Bf r})$ herein defined according to
Eq.~(\ref{ef17}). Since $\varrho_{{\rm s};\sigma'}^{\rm h}(\|{\Bf r}''\|)$ 
is bounded everywhere, the possibility that ${\cal I}_{{\rm s};\kappa}
({\Bf r})$ be unbounded solely depends on the possibility of a slow 
decay of the integrand in the defining expression for ${\cal I}_{{\rm s};
\kappa}({\Bf r})$ as $\|{\Bf r}''\|\to\infty$. Later in this Appendix we 
shall see that ({\it cf}. Eq.~(\ref{ef24}) below; see also Appendix J) 
\begin{eqnarray}
\label{ef19}
\varrho_{{\rm s};\sigma}^{\rm h}(\|{\Bf r}''\|)
\sim \frac{-k_{F;\sigma}}{2\pi^2}\,
\frac{\cos(k_{F;\sigma} \|{\Bf r}''\|)}{\|{\Bf r}''\|^2},\;\;\;
\|{\Bf r}''\| \to \infty,
\end{eqnarray}
where $k_{F;\sigma}$ stands for the Fermi wavenumber of the fermions 
with spin $\sigma$ (here we are assuming that $N_{\sigma}\not=0$). 
The result in Eq.~(\ref{ef19}) applies also to the density matrix 
pertaining to the fully {\sl interacting} uniform and isotropic GS, 
with the provision that the RHS of Eq.~(\ref{ef19}) be multiplied by 
$Z_{F;\sigma}$ (Farid 2000a --- see also Appendix J), the amount of 
jump in the momentum distribution function ${\sf n}_{\sigma}({\Bf k})$ 
at $k_{F;\sigma}$ (for definition see Eq.~(\ref{ej2}) below); here we 
make the implicit assumption that the interacting GS is uniform and 
metallic, and moreover the corresponding Fermi seas are spherical 
in shape. For completeness, we mention the important fact that in 
cases where $Z_{F;\sigma}=0$ (for {\sl some} $\sigma$), the 
leading-order term in the large-$\|{\Bf r}\|$ AS for 
$\varrho_{\sigma}^{\rm h}(\|{\Bf r}\|)$ is fundamentally different 
in comparison with that on the RHS of Eq.~(\ref{ef19}); in the case 
of $Z_{F;\sigma}=0$, $\varrho_{\sigma}^{\rm h}(\|{\tilde {\Bf r}}\|)$ 
to leading order decays in magnitude like $1/\|{\tilde {\Bf r}}\|^{2
+\alpha}$ (when $d=3$), where $\alpha > 0$ is referred to as the 
{\sl anomalous exponent}. We note in passing that the property 
$Z_{F;\sigma} < 1$ reflects the {\sl non}-idempotency of 
$\varrho_{\sigma}^{\rm h}(\|{\tilde {\Bf r}}\|)$ which in turn is a 
manifestation of the overcompleteness of the set of the Lehmann 
amplitudes (see Appendix A).
 
From the result in Eq.~(\ref{ef19}), it follows that for $\|{\Bf r}''\|
\to\infty$, barring the exponential cut-off function, the integrand 
on the RHS of Eq.~(\ref{ef17}), with $\rho^{\rm h}$ herein replaced by 
$\sum_{\sigma}(\varrho_{{\rm s};\sigma})^2$ (see Eq.~(\ref{ef18})), 
{\sl decays} like $1/\|{\Bf r}''\|^5$ which is sufficient for the 
existence of ${\cal I}_{{\rm s};\kappa}({\Bf r})$ with $\kappa=0$: 
with ${\rm d}^3r'' = {\rm d}\varphi\, {\rm d}\theta\, {\rm d}r''\, 
{r''}^2 \sin(\theta)$ in the spherical-polar coordinates system, it 
is seen that the integrand in the radial direction {\sl decays} like 
$1/{r''}^3$ for $r''\to\infty$; consequently, ${\cal I}_{{\rm s};
\kappa=0}({\Bf r}) \equiv \lim_{\kappa\downarrow 0} {\cal I}_{{\rm s};
\kappa}({\Bf r})$. As we shall see, this fact notwithstanding, 
identifying $\kappa$ in Eq.~(\ref{ef16}) with zero {\sl prior to} 
evaluating ${\cal A}^{\prime\rm h}_{\rm s}(\|{\Bf r}\|=0)$, gives 
rise to an erroneous outcome in the case at hand.
 
We now proceed with considering ${\cal A}^{\prime\rm h}_{\rm s}
(\|{\tilde {\Bf r}}\|)$ by investigating the behaviour of the integrand 
of the ${\Bf r}''$ integral on the RHS of Eq.~(\ref{ef16}). Since,
following our above discussion, ${\cal I}_{{\rm s};\kappa}({\tilde 
{\Bf r}}-{\Bf r}'')$ is bounded for {\sl all} ${\Bf r}''$, and since 
the singularities of $v_c({\Bf r}'')$ and $v_c({\tilde {\Bf r}}
-{\Bf r}'')$ at ${\Bf r}''={\Bf 0}$ and ${\Bf r}''={\tilde {\Bf r}}$ 
respectively are integrable, the boundedness or otherwise of 
${\cal A}^{\prime\rm h}_{\rm s}(\|{\tilde {\Bf r}}\|)$ is entirely 
determined by the behaviour of the integrand of the ${\Bf r}''$ integral 
for $\|{\Bf r}''\|\to \infty$. In order to establish this behaviour, 
we need to determine that of ${\cal I}_{{\rm s};\kappa}({\tilde {\Bf r}}
-{\Bf r}'')$ in the asymptotic region $\|{\Bf r}''\|\to \infty$, which 
for finite $\|{\tilde {\Bf r}}\|$ corresponds to $\|{\tilde {\Bf r}}
-{\Bf r}''\| \to \infty$. To this end, we employ the  multipole 
expansion (for example Jackson (1975) or Morse and Feshbach (1953)) 
(below $\wh{\Bf r} {:=} {\Bf r}/\|{\Bf r}\|$) 
\begin{equation}
\label{ef20}
v_c({\Bf r}-{\Bf r}'') = v_c({\Bf r}) +\frac{ v_c({\Bf r})\,
\wh{{\Bf r}}\cdot {\Bf r}''}{\|{\Bf r}\|}\, 
+ {\cal O}\big(1/\|{\Bf r}\|^3\big)
\end{equation}
in the regime of $\|{\Bf r}\|\to\infty$ (which for finite values of 
$\|{\Bf r}''\|$, corresponds to $\|{\Bf r}-{\Bf r}''\|\to \infty$).
From this, making use of Eq.~(\ref{ef14}), for ${\cal I}_{\kappa}
({\Bf r})$ in Eq.~(\ref{ef17}) we readily obtain the following
leading-order result:
\begin{equation}
\label{ef21}
{\cal I}_{\kappa}({\Bf r}) \sim
\exp(-\kappa \|{\Bf r}\|)\, n_0\, v_c({\Bf r}), \;\; 
\|{\Bf r}\| \to \infty;
\end{equation}
since $n_0$ is independent of interaction, the RHS of Eq.~(\ref{ef21}) 
equally applies to ${\cal I}_{{\rm s};\kappa}({\Bf r})$. In light
of Eq.~(\ref{ef21}), for $\|{\Bf r}''\| \to\infty$ (or 
$\|{\tilde {\Bf r}}-{\Bf r}''\|\to\infty$ when $\|{\tilde {\Bf r}}\| 
< \infty$), the {\sl leading} term in the AS of the integrand of the 
${\Bf r}''$ integral on the RHS of Eq.~(\ref{ef16}) is determined by 
the {\sl next-to-leading} term in the AS of ${\cal I}_{\kappa}
({\tilde {\Bf r}}-{\Bf r}'')$. Due to the isotropy of $\rho^{\rm h}
(\|{\Bf r}''\|)$, 
\begin{equation}
\label{ef22}
\int {\rm d}^3r''\; {\rm e}^{-\kappa \|{\Bf r}-{\Bf r}''\|}\, 
{\hat {\Bf r}}\cdot {\Bf r}''\, \rho^{\rm h}(\|{\Bf r}''\|) 
\equiv 0,\;\;\;
\kappa \downarrow 0,
\end{equation}
so that the term in the indicated large-$\|{\Bf r}''\|$ AS for
${\cal I}_{\kappa}({\tilde {\Bf r}}-{\Bf r}'')$, which according to 
Eq.~(\ref{ef20}) would decay like $v_c({\tilde {\Bf r}}
-{\Bf r}'')/\|{\tilde {\Bf r}}-{\Bf r}''\|$, is vanishing.
The absence of a terms decaying like 
\begin{eqnarray}
\frac{v_c({\tilde {\Bf r}}-{\Bf r}'')}{\|{\tilde {\Bf r}}-{\Bf r}''\|}
\propto \frac{1}{\|{\tilde {\Bf r}}-{\Bf r}''\|^2}
\nonumber
\end{eqnarray}
in the large-$\|{\Bf r}''\|$ AS of ${\cal I}_{\kappa}
({\tilde {\Bf r}}-{\Bf r}'')$ (or ${\cal I}_{{\rm s};\kappa}
({\tilde {\Bf r}}-{\Bf r}'')$) implies that the term enclosed 
by the square brackets on the RHS of Eq.~(\ref{ef16}) decays 
{\sl more rapidly} 
\footnote{\label{f126}
A note of caution is in place here; contrary to the considerations 
related to the AS expansions of functions of $\varepsilon$ (or of $z$), 
for $\vert\varepsilon\vert \to\infty$ (or $\vert z\vert\to \infty$) 
considered in this paper, here the Poincar\'e definition of AS (see 
\S~II.B) is restrictive (we have some circumstantial evidence 
indicating this to be the case). See footnote \protect\ref{f28}. }
than $1/\|{\tilde {\Bf r}}-{\Bf r}''\|^2$. As we shall see shortly, 
this aspect, which is of vital consequence for the boundedness of 
${\cal A}^{\prime\rm h}_{\rm s}(0)$ in the limit $\kappa\downarrow 0$, 
is {\sl not} accounted for by the expression for ${\cal I}_{{\rm s};
\kappa=0}({\tilde {\Bf r}}-{\Bf r}'')$ (or ${\cal I}_{\kappa=0}
({\tilde {\Bf r}}-{\Bf r}'')$). It should be mentioned that, through 
identifying $\kappa$ with zero {\sl prior to} carrying out the 
${\Bf r}''$ integral on the LHS of Eq.~(\ref{ef22}), this integral 
would become ill-defined, rendering the result in Eq.~(\ref{ef22}) only 
conditionally valid; to deduce Eq.~(\ref{ef22}) (while $\kappa=0$), 
it is required that, in evaluating the ${\Bf r}''$ integral in terms 
of the spherical polar coordinates of ${\Bf r}''$, that is 
$(r'',\theta,\varphi)$, the integration with respect to $\varphi \in 
[0,2\pi)$ be carried out {\sl prior to} that with respect to $r'' \in 
[0,R)$, $R=\infty$; with $\kappa=0$, in view of the asymptotic result 
to be presented in Eqs.~(\ref{ef52}) and (\ref{ef86}) below, the 
latter integral can be shown to diverge like $\sum_{\sigma'} 
\ln(k_{F;\sigma'}\, R)$ as $R\to\infty$ (see \S~II.A and the text 
following Eq.~(\ref{e10})).

Since, in the limit $\kappa\downarrow 0$, the contribution corresponding 
to the third term in the multipole expansion of $v_c({\Bf r}-{\Bf r}'')$ 
gives rise to an unbounded ${\Bf r}''$ integral on the RHS of 
Eq.~(\ref{ef17}), we deduce that the {\sl next-to-leading} term 
in the large-$\|{\Bf r}\|$ AS for ${\cal I}_{\kappa}({\Bf r})$ (or 
${\cal I}_{{\rm s};\kappa}({\Bf r})$) {\sl cannot} be one decaying 
like $1/\|{\Bf r}\|^3$, but a term decaying more slowly; however, it 
decays more rapidly than (see footnote \ref{f126}) $1/\|{\Bf r}\|^2$. 
This implies that ${\cal A}^{\prime\rm h}(\|{\tilde {\Bf r}}\|)$ 
(or ${\cal A}^{\prime\rm h}_{\rm s}(\|{\tilde {\Bf r}}\|)$) in 
Eq.~(\ref{ef16}) is indeed a bounded function for {\sl all} 
${\tilde {\Bf r}}$, in full conformity with our finding based on 
general considerations (see Eq.~(\ref{ef8}) above and the text following 
it; see also Eq.~(\ref{ef50}) below and the succeeding text as
well as footnote \ref{f129}). 

We now consider the crucial role played by a non-vanishing $\kappa$ 
in rendering the expression in Eq.~(\ref{ef16}) well defined. To this 
end we first recall that, for non-interacting uniform systems, we have 
(see Eqs.~(\ref{e46}) and (\ref{e47}))
\begin{equation}
\label{ef23}
f_{s;\sigma}({\Bf r}) \Longrightarrow 
\phi_{\Bf k}({\Bf r}) \equiv \frac{1}{\Omega^{1/2}} 
\exp(i {\Bf k}\cdot {\Bf r}), 
\end{equation}
so that for uniform and isotropic GSs (the attribute `isotropic'
is reflected in the condition $\|{\Bf k}\| \le k_{F;\sigma}$ invoked
below) (see Eq.~(\ref{e169})) (March, {\sl et al.} 1967, 
pp.~15 and 16),
\begin{eqnarray}
\label{ef24}
\varrho_{{\rm s};\sigma}^{\rm h}(\|{\Bf r}-{\Bf r}'\|)
&=& \frac{1}{\Omega} \sum_{\|{\Bf k}\| \le k_{F;\sigma}}
\exp(-i {\Bf k}\cdot [{\Bf r}-{\Bf r}'])\nonumber\\
&=& \frac{1}{(2\pi)^3} \int_{\|{\Bf k}\| \le k_{F;\sigma}}
{\rm d}^3k\; \exp(-i {\Bf k}\cdot [{\Bf r}-{\Bf r}'])
\nonumber\\
&=& \frac{k_{F;\sigma}^2}{2\pi^2} \,
\frac{j_1\big(k_{F;\sigma} 
\|{\Bf r}-{\Bf r}'\|\big)}{\|{\Bf r}-{\Bf r}'\|},
\end{eqnarray}
where $j_1(x)$ stands for the spherical Bessel function of the 
first order, defined as (Abramowitz and Stegun 1972, p.~438) 
\begin{equation}
\label{ef25}
j_1(x) {:=} \frac{\sin(x) - x \cos(x)}{x^2}.
\end{equation}
The asymptotic result in Eq.~(\ref{ef19}) is trivially deduced from 
the expression in Eq.~(\ref{ef24}) through making use of $j_1(x) \sim 
-\cos(x)/x$ for $\vert x\vert \to \infty$. 

From Eqs.~(\ref{ef11}) and (\ref{ef24}) and employing $n_{0;\sigma}
= k_{F;\sigma}^3/[6\pi^2]$, we obtain
\begin{eqnarray}
\label{ef26}
{\cal U}_{\rm s}({\Bf r},{\Bf r}')
&=& n_0 \delta({\Bf r}-{\Bf r}')
-\sum_{\sigma'}\frac{9 n_{0;\sigma'}^2 }{k_{F;\sigma'}^2}\,
\frac{j_1^2(k_{F;\sigma'} \|{\Bf r}-{\Bf r}'\|)}{\|{\Bf r}
-{\Bf r}'\|^2}.\nonumber\\
\end{eqnarray}
From now onwards we assume $n_{0;\sigma} = n_{0;{\bar\sigma}} = n_0/2$ 
(corresponding to uniform and isotropic systems of spin-$1/2$ fermions, 
i.e. ${\sf s}=1/2$, in the paramagnetic state) and thus $k_{F;\sigma}
=k_{F;{\bar\sigma}}= k_F =(3\pi^2 n_0)^{1/3}$.

Employing the Fourier representation of $v_c({\Bf r}-{\Bf r}')$ (see 
Eq.~(\ref{eg2})), from Eqs.~(\ref{ef17}), (\ref{ef18}) and (\ref{ef24})
we obtain
\begin{eqnarray}
\label{ef27}
&&{\cal I}_{{\rm s};\kappa=0}({\tilde {\Bf r}}-{\Bf r}'')
= \frac{9 n_0^2}{2 k_F^2} \int {\rm d}^3r_1''\;
v_c({\tilde {\Bf r}}-{\Bf r}_1'') 
\nonumber\\
&&\;\;\;\;\;\;\;\;\;\;\;\;\;\;\;\;\;\;\;\;\;\;\;\;\;\;\;\;
\times\frac{j_1^2(k_F \|{\Bf r}''-{\Bf r}_1''\|)}{\|{\Bf r}''
-{\Bf r}_1''\|^2}\nonumber\\
&&\;\;\;\;
= \frac{9 n_0^2}{2 k_F^2} \int \frac{{\rm d}^3q}{(2\pi)^3}\;
\frac{e^2/\epsilon_0}{\|{\Bf q}\|^2}\,
{\rm e}^{i {\Bf q}\cdot ({\tilde {\Bf r}}-{\Bf r}'')}\,
\nonumber\\
&&\;\;\;\;\;\;\;\;\;\;
\times \frac{4\pi}{\|{\Bf q}\|}\,
\int_0^{\infty} {\rm d}r_1''\;
\frac{j_1^2(k_F r_1'')\, \sin(\|{\Bf q}\| r_1'')}{r_1''}.
\end{eqnarray}
With $\|{\Bf q}\| > 0$, we have the standard result
\begin{eqnarray}
\label{ef28}
&&\int_0^{\infty} {\rm d}r_1''\;
\frac{j_1^2(k_F r_1'')\, \sin(\|{\Bf q}\| r_1'')}{r_1''}
= \frac{\pi\|{\Bf q}\|}{96 k_F^4}\,
(2 k_F - \|{\Bf q}\|)^2\nonumber\\
&&\;\;\;\;\;\;\;\;\;\;\;\;\;\;\;\;\;\;\;\;\;\;\;\;
\times (4 k_F + \|{\Bf q}\|)\,
\Theta(2 k_F - \|{\Bf q}\|),
\end{eqnarray}
substitution of which into the RHS of Eq.~(\ref{ef27}) yields
\footnote{\label{f127}
Recall that $9 n_0/[2 k_F^2] = 3 k_F/[2\pi^2]$. }
\begin{eqnarray}
\label{ef29}
&&{\cal I}_{{\rm s};\kappa=0}({\tilde {\Bf r}}-{\Bf r}'') = 
\frac{n_0}{4\pi k_F^3}\, 
v_c({\tilde {\Bf r}}-{\Bf r}'')\,
\frac{1}{\|{\tilde {\Bf r}}-{\Bf r}''\|^3}
\nonumber\\
&&\;\;\;\;\;
\times \Big\{-1 -6 k_F^2 \|{\tilde {\Bf r}}-{\Bf r}''\|^2
\nonumber\\
&&\;\;\;\;\;\;\;\;\;\;\;
+\big(1 + 4 k_F^2 \|{\tilde {\Bf r}}-{\Bf r}''\|^2\big)\,
\cos(2 k_F \|{\tilde {\Bf r}}-{\Bf r}''\|)\nonumber\\
&&\;\;\;\;\;\;\;\;\;\;\;
+2 k_F \|{\tilde {\Bf r}}-{\Bf r}''\| 
\sin(2 k_F \|{\tilde {\Bf r}}-{\Bf r}''\|)\,
\nonumber\\
&&\;\;\;\;\;\;\;\;\;\;\;
+8 k_F^3 \|{\tilde {\Bf r}}-{\Bf r}''\|^3\,
{\rm Si}(2 k_F \|{\tilde {\Bf r}}-{\Bf r}''\|)\Big\},
\end{eqnarray}
where ${\rm Si}(x)$ stands for the sine-integral function (Abramowitz 
and Stegun 1972, p.~231). The linearity with respect to $\|{\Bf q}\|$
of the expression on the RHS of Eq.~(\ref{ef28}) for small values of 
$\|{\Bf q}\|$ (to leading order, one has $\pi\|{\Bf q}\|/[6 k_F]$), as 
well as the infinitely sharp cut-off $\Theta(2 k_F - \|{\Bf q}\|)$ in 
this expression, are consequences of identifying $\kappa$ with zero. 
Below we shall derive the exact expression for 
${\cal A}^{\prime\rm h}_{\rm s}(0)$ but, before doing so, consider in 
some detail the undesired consequence to which a premature 
identification of $\kappa$ with zero gives rise.

Making use of the asymptotic results ${\rm Si}(z) \sim z - z^3/18 
+ \dots$, for $\vert z\vert\to 0$ and 
\footnote{\label{f128}
The {\sl transcendental} functions $\cos(z)$ and $\sin(z)$ encountered 
in the large-$\vert z\vert$ AS of ${\rm Si}(z)$ are partly responsible 
for rendering the Poincar\'e definition of AS (see \S~II.B) inadequate 
in regard to the ensuing functions of spatial coordinates (see footnote 
\protect\ref{f126}). } 
${\rm Si}(z) \sim \pi/2 - 
\cos(z)/z - \sin(z)/z^2 + \dots$, for $\vert z\vert\to \infty$ 
($\vert {\rm arg}(z) \vert < \pi$) (see Abramowitz and Stegun 1972, 
pp. 232 and 233), we readily obtain
\begin{eqnarray}
\label{ef30}
{\cal I}_{{\rm s};\kappa=0}({\tilde {\Bf r}}-{\Bf r}'')
&\sim& \frac{e^2}{4\pi\epsilon_0}\, \frac{3 k_F n_0}{2\pi^2},\;\;\;\;
\|{\tilde {\Bf r}}-{\Bf r}''\|\to 0;\\
\label{ef31}
{\cal I}_{{\rm s};\kappa=0}({\tilde {\Bf r}}-{\Bf r}'')
&\sim& n_0 \Big(1  
-\frac{3/[2\pi k_F]}{\|{\tilde {\Bf r}}-{\Bf r}''\|} 
+ \dots\Big)\, v_c({\tilde {\Bf r}}-{\Bf r}''),\nonumber\\
& &\;\;\;\;\;\;\;\;\;\;\;\;\;\;\;\;\;\;\;\;\;
\|{\tilde {\Bf r}}-{\Bf r}''\|\to \infty.
\end{eqnarray}
Compare the leading term on the RHS of Eq.~(\ref{ef31}) with RHS of 
Eq.~(\ref{ef21}), with $\kappa$ herein identified with zero. From 
Eq.~(\ref{ef16}) and the results in Eqs.~(\ref{ef30}) and (\ref{ef31}), 
we observe that for $\|{\tilde {\Bf r}}-{\Bf r}''\| \to 0$ the 
contribution of the terms enclosed by square brackets on the RHS of 
Eq.~(\ref{ef16}) is asymptotically equal to $n_0 v_c({\tilde {\Bf r}}
-{\Bf r}'') \propto 1/\|{\tilde {\Bf r}}-{\Bf r}''\|$, whereas for 
$\|{\tilde {\Bf r}}-{\Bf r}''\| \to \infty$, the indicated contribution 
is asymptotically equal to $(3 n_0/[2\pi k_F]) v_c({\tilde {\Bf r}}
-{\Bf r}'')/\|{\tilde {\Bf r}}-{\Bf r}''\| \propto 1/\|
{\tilde {\Bf r}}-{\Bf r}''\|^2$. This result contradicts our earlier 
conclusion (see text following Eq.~(\ref{ef22}) above), namely that, 
for $\|{\tilde {\Bf r}} -{\Bf r}''\| \to \infty$, the function 
enclosed inside the curly braces on the RHS of Eq.~(\ref{ef16}) decays 
{\sl more rapidly} than $1/\|{\tilde {\Bf r}}-{\Bf r}''\|^2$, that is the 
coefficient of $1/\|{\tilde {\Bf r}}-{\Bf r}''\|^2$ must be identically 
vanishing.

Now we proceed with the determination of ${\cal A}_{\rm s}^{\prime\rm h}
(0)$. In doing so, we concentrate on the {\sl exact} 
${\cal A}^{\prime\rm h}(0)$ and only in the final stage explicitly 
deal with this function within the framework of the SSDA. To this end, 
we first introduce the pair of Fourier transforms (below $r {:=} 
\|{\Bf r}\|$ and $q {:=} \|{\Bf q}\|$)
\begin{eqnarray}
\label{ef32}
\rho^{\rm h}(r) &=& \int \frac{{\rm d}^3q}{(2\pi)^3}\;
{\rm e}^{i {\Bf q}\cdot {\Bf r}}\, {\bar\rho}^{\rm h}(q), \\
\label{ef33}
{\bar\rho}^{\rm h}(q) &=& \int {\rm d}^3r\; 
{\rm e}^{-i {\Bf q}\cdot {\Bf r}}\, \rho^{\rm h}(r),
\end{eqnarray}
from which and Eq.~(\ref{ef17}) we readily obtain
\begin{equation}
\label{ef34}
{\cal I}_{\kappa}({\Bf r})
= \frac{2}{\pi}\, v_c({\Bf r})\,
\int_0^{\infty} {\rm d}q\; 
\frac{q}{q^2 + \kappa^2}\,
{\bar\rho}^{\rm h}(q)\, \sin(q r).
\end{equation}
We note that ${\bar\rho}^{\rm h}(0) = n_0$ (see Eq.~(\ref{ef14})) 
and that
\begin{equation}
\label{ef35}
\int_0^{\infty} {\rm d}x\, \frac{x}{x^2+\kappa^2}\,
\sin(a x) = \frac{\pi}{2} {\rm sgn}(a)\,
{\rm e}^{-\kappa a},
\end{equation}
so that ({\it cf}. Eq.~(\ref{ef21}))
\begin{equation}
\label{ef36}
\frac{2}{\pi}\, v_c({\Bf r})\,
\int_0^{\infty} {\rm d}q\;
\frac{q}{q^2+\kappa^2}\, {\bar\rho}^{\rm h}(0)\,
\sin(q r) = 
{\rm e}^{-\kappa r}\, n_0\, v_c({\Bf r}). 
\end{equation}
From this and Eqs.~(\ref{ef16}) and (\ref{ef34}) we obtain
\begin{eqnarray}
\label{ef37}
{\cal A}^{\prime\rm h}(0) &=&
-\frac{2}{\pi} \int {\rm d}^3r\; v_c^2({\Bf r})\,
{\rm e}^{-\kappa r}\, {\cal J}_{\kappa}(r)\nonumber\\
&=& -8 \left(\frac{e^2}{4\pi\epsilon_0}\right)^2 
\int_0^{\infty} {\rm d}r\;
{\rm e}^{-\kappa r}\, {\cal J}_{\kappa}(r),\;\; 
\kappa \downarrow 0,
\end{eqnarray}
where
\begin{equation}
\label{ef38}
{\cal J}_{\kappa}(r) {:=}
\int_0^{\infty} {\rm d}q\;
\frac{q\, [{\bar\rho}^{\rm h}(q) - 
{\bar\rho}^{\rm h}(0)]}{q^2 + \kappa^2}\, \sin(q r).
\end{equation}
It can be shown that
\begin{equation}
\label{ef38a}
\left. {\cal J}_{0}(r)\right|_{\rm Eq.~(\ref{ef38})}
\sim \frac{{\bar\rho}^{\rm h\prime}(0)}{r},
\;\;\;\mbox{\rm for}\;\;\; r \to \infty,
\end{equation}
where ${\bar\rho}^{\rm h\prime}(0) \equiv 
{\rm d} {\bar\rho}^{\rm h}(q)/{\rm d}q\vert_{q\downarrow 0}$
(concerning the existence of this quantity, see Eqs.~(\ref{ef43}) 
and (\ref{ef51}) below and the subsequent considerations). It follows 
that ${\cal A}^{\prime\rm h}(0)$ according to Eq.~(\ref{ef37}), with 
${\cal J}_{\kappa}(r)$ according to Eq.~(\ref{ef38}), diverges like 
$\ln(\kappa/k_F)$ for $\kappa\downarrow 0$. This divergence has the 
same origin as discussed above. Below, we construct an expression 
for ${\cal J}_{\kappa}(r)$ which gives rise to the expected finite 
value for ${\cal A}^{\prime\rm h}(0)$ for $\kappa\downarrow 0$.

From Eq.~(\ref{ef33}) we readily deduce
\begin{equation}
\label{ef39}
{\bar\rho}^{\rm h}(q) - {\bar\rho}^{\rm h}(0) 
= 4\pi \int_0^{\infty} {\rm d}r\,
r^2\, \rho^{\rm h}(r)\,
\Big[ \frac{\sin(q r)}{q r} - 1\Big].
\end{equation}
By assuming 
\footnote{\label{f129}
This assumption coincides with fact insofar as $\rho_{\rm s}^{\rm h}
(r)$ as well as $\rho^{\rm h}(r)$ pertaining to weakly interacting
systems are concerned. As our considerations following 
Eq.~(\protect\ref{ef51}) below show, however, with $\rho^{\rm h}(r)$ 
decaying like $1/r^{\alpha}$ for $r\to\infty$, although it is {\sl not} 
excluded that $\alpha > 4$, it is strictly excluded that $\alpha < 4$; 
the latter possibility implies divergence of ${\cal A}^{\prime\rm h}(0)$ 
(see Eqs.~(\protect\ref{ef50}), (\protect\ref{ef50a}),
(\protect\ref{ef50b}), (\protect\ref{ef51}) and (\protect\ref{ef48f}) 
below) and therefore instability of the GS of the system under 
consideration. Recall that in states with off-diagonal long-range 
order (ODLRO) (for definition and some details see, for example, Reichl 
(1980, pp. 202-205)), including those with {\sl algebraic} ODLRO (here 
corresponding to cases where $\alpha < 4$; recall that here we are 
explicitly considering $d=3$), the {\sl normal} GS assumed here, 
is indeed not the GS and therefore unstable. In this context, it should 
be noted that according to Kohn and Luttinger (1965), below a finite, 
but extremely small, transition temperature, the true GS of a uniform 
and isotropic system of fermions is {\sl not} normal, but 
superconducting. It would therefore appear (since we consider the 
absolute temperature to be zero) that, by using the {\sl true} GS of 
the system under consideration, we should expect 
${\cal A}^{\prime\rm h}(0)$ to be unbounded. Use of such a GS, 
however, is {\sl not} justified within the framework of our present 
work, where by assumption the GS of the system is {\sl normal}; 
in order to be able to deal with systems with, say, superconducting 
GSs, we should have utilized the Nambu (1960) formalism (Fetter and 
Walecka 1971, p.~443), thus taking account of the possibility of a 
non-vanishing {\sl anomalous} GF, signifying ODLRO in the system. }
that $\rho^{\rm h}(r)$, similar to $\sum_{\sigma'} 
\big(\varrho_{{\rm s};\sigma'}^{\rm h}(r)\big)^2 \equiv 
\rho_{\rm s}^{\rm h}(r)$, decays like $1/r^4$ for $r\to \infty$ (see 
Eqs.~(\ref{ef52}) and (\ref{ef86}) below), we have 
\begin{equation}
\label{ef40}
\vert \int_0^{\infty} {\rm d}r\; r^{\alpha}\, \rho^{\rm h}(r)
\vert < \infty \;\;\;\mbox{\rm for}\;\;\; -1 < \alpha < 3.
\end{equation}
Consequently, from Eq.~(\ref{ef39}) we directly deduce that
${\bar\rho}^{\rm h}(q)$ is {\sl continuous} at $q=0$, that is
\begin{equation}
\label{ef41}
{\bar\rho}^{\rm h}(q) - {\bar\rho}^{\rm h}(0) = o(1) 
\end{equation}
for $q \in [0,Q]$ with $Q$ some {\sl positive} constant. This is 
clarified by the observation that on decreasing $q$ towards zero, 
the minimum value of $r$ beyond which $[\sin(q r)/(q r) - 1]$ 
appreciably deviates from zero increases (by an amount proportional 
to $1/q$), which, in view of the fact that the integral in 
Eq.~(\ref{ef40}) is bounded for $\alpha=2$, through Eq.~(\ref{ef39}) 
results in Eq.~(\ref{ef40}). The condition in Eq.~(\ref{ef40}) is 
weaker than that in which $o(1)$ on the RHS of Eq.~(\ref{ef41}) is 
replaced by ${\cal O}(q)$, implying ${\bar\rho}^{\rm h}(q)$ to be 
a (right) {\sl differentiable} function of $q$ in $[0,Q]$ (see text 
following Eq.~(\ref{ef43}) below). By employing $\sin(x)/x \sim 1 - 
x^2/6$, for $x\to 0$, in the RHS of Eq.~(\ref{ef39}), we further 
deduce that $o(1)$ on the RHS of Eq.~(\ref{ef41}) {\sl cannot} be equal 
to ${\cal O}(q^2)$ but to some function which for $q\downarrow 0$ 
is more dominant than ${\cal O}(q^2)$ (but, by continuity, less 
dominant than $o(1)$; see above). This follows from the fact that 
the integral in Eq.~(\ref{ef40}) is divergent for $\alpha=4$ (see the 
considerations in \S~II.B). Our calculations, to be presented below, 
combined with our above observations, lead us to the conclusion that 
in addition to being {\sl continuous}, ${\bar\rho}^{\rm h}(q)$ is a 
{\sl continuously differentiable} function of $q$ for $q \in [0,Q]$ 
with $Q > 0$, i.e.
\begin{eqnarray}
\label{ef42}
{\bar\rho}^{\rm h}(q) &=& {\bar\rho}^{\rm h}(0)
+{\bar\rho}^{\rm h\prime}(0)\, q + o(q),\nonumber\\
& &\;\;\;\mbox{\rm where {\sl possibly}}\;\; q^2/o(q) \to 0 \;\;
\mbox{\rm as}\;\, q \downarrow 0;
\end{eqnarray}
for ${\bar\rho}^{\rm h}_{\rm s}(q)$, the above $o(q)$ is equal to 
${\cal O}(q^3)$, however. For later reference, from Eq.~(\ref{ef39}) 
we have
\begin{eqnarray}
\label{ef43}
&&{\bar\rho}^{\rm h\prime}(q) {:=}
\frac{{\rm d} {\bar\rho}^{\rm h}(q)}{{\rm d} q} \nonumber\\
&&\;\;
=\frac{4\pi}{q}\,
\int_0^{\infty} {\rm d}r\; r^2\, \rho^{\rm h}(r)\,
\Big[\cos(q r) - \frac{\sin(q r)}{q r} \Big].
\end{eqnarray}
We point out that since by definition $q\ge 0$, in Eq.~(\ref{ef42})
as well as below, ${\bar\rho}^{\rm h\prime}(0)$ should be considered
as denoting the {\sl right}-derivative of ${\bar\rho}^{\rm h}(q)$ with
respect to $q$ at $q=0$. Later we investigate the condition for
the boundedness of ${\bar\rho}^{\rm h\prime}(0)$ is some detail
(see the paragraph containing Eq.~(\ref{ef51}) below).

From the expression in Eq.~(\ref{ef39}), through repeated application
of integration by parts (compare with a similar procedure adopted in
Appendix J) and making use of the Riemann-Lebesgue lemma (Whittaker 
and Watson 1927, p.~172) we deduce (below $\rho^{\rm h\prime}(0) 
{:=} {\rm d}\rho^{\rm h}(r)/{\rm d}r\vert_{r\downarrow 0}$; we note
that $\rho^{\rm h\prime}_{\rm s}(0)=0$)
\begin{equation}
\label{ef44}
{\bar\rho}^{\rm h}(q) \sim -8\pi\, \frac{\rho^{\rm h\prime}(0)}{q^4}
\;\;\; \mbox{\rm for}\;\;\; q\to \infty,
\end{equation}
so that through integration by parts, from Eq.~(\ref{ef38}) we obtain
\begin{equation}
\label{ef45}
{\cal J}_{\kappa}(r)
= \frac{1}{r}\,
\int_0^{\infty} {\rm d}q\;
\Big[ \frac{\rm d}{{\rm d} q}\,
\frac{q\, [{\bar\rho}^{\rm h}(q)-{\bar\rho}^{\rm h}(0)]}{q^2 
+ \kappa^2} \Big]\, \cos(q r),
\end{equation}
to which no boundary terms have contributed, for 
\begin{equation}
\label{ef46}
\left. \frac{-1}{r}\,\cos(q r)\,  
\frac{q\, [{\bar\rho}^{\rm h}(q) - 
{\bar\rho}^{\rm h}(0)]}{q^2 + \kappa^2}\right|_{q=0}^{\infty} = 0.
\end{equation}
It is here that a finite $\kappa$, even though infinitesimally small, 
manifests its essential role; from Eq.~(\ref{ef46}) we observe that 
with $\kappa=0$ (as opposed to $\kappa\downarrow 0$), the LHS of 
Eq.~(\ref{ef46}) would have been equal to ${\cal O}(1/r)$ (to be
precise, in view of Eq.~(\ref{ef42}), equal to ${\bar\rho}^{\rm h\prime}
(0)/r$) arising from the $q=0$ side of the integration boundary, rather 
than zero as in Eq.~(\ref{ef46}). This non-vanishing contribution of 
order $1/r$ is directly related to the second term inside the large
parentheses on the RHS of Eq.~(\ref{ef31}) resulting from a premature 
identification of $\kappa$ with zero; in this connection note that 
$-n_0 \times 3/[2\pi k_F]$ encountered on the RHS of Eq.~(\ref{ef31}) 
is exactly equal to $(2/\pi) \times {\bar\rho}^{\rm h\prime}_{\rm s}(0)$,
where ${\bar\rho}^{\rm h\prime}_{\rm s}(0) = -k_F^2/[4\pi^2] \equiv 
-3 n_0/[4 k_F]$ and where $2/\pi$ originates from the RHS of 
Eq.~(\ref{ef34}).

Making use of $[{\bar\rho}^{\rm h}(q)-{\bar\rho}^{\rm h}(0)]/q =
{\bar\rho}^{\rm h\prime}(0) + o(1)$ and ${\bar\rho}^{\rm h\prime}(q)
-{\bar\rho}^{\rm h\prime}(0) = o(1)$, for $q\to 0$, both of which 
follow from Eq.~(\ref{ef42}), we have 
\begin{eqnarray}
\label{ef47}
\left. \frac{\rm d}{{\rm d} q}\,
\frac{q\, [{\bar\rho}^{\rm h}(q)
-{\bar\rho}^{\rm h}(0)]}{q^2 + \kappa^2}\right|_{\kappa=0}
&\equiv& 
\frac{ {\bar\rho}^{\rm h\prime}(q) - 
[{\bar\rho}^{\rm h}(q)-{\bar\rho}^{\rm h}(0)]/q}{q} \nonumber\\
&=& o(1/q)
\;\;\;\mbox{\rm for}\;\;\; q\to 0.
\end{eqnarray}
From this we conclude that for $\kappa=0$, the integrand of the 
$q$ integral on the RHS of Eq.~(\ref{ef45}) is integrable in the 
neighbourhood of $q=0$. Further, from Eq.~(\ref{ef44}) we have (for 
considerations related to derivatives of AS and AS of derivatives 
see, for example, Lauwerier (1977)) ${\bar\rho}^{\rm h\prime}(q) 
\sim 32\pi\rho^{\rm h\prime}(0)/q^5$ for $q\to\infty$ so that, for 
$q\to\infty$ the integrand of the integral on the RHS of 
Eq.~(\ref{ef45}), as enclosed by square brackets, behaves like 
${\bar\rho}^{\rm h}(0)/q^2$. We thus conclude that 
${\cal J}_{\kappa}(r)$ according to Eq.~(\ref{ef45}) is a well-defined 
function of $r$ for $\kappa=0$. 

On account of the Riemann-Lebesgue lemma (Whittaker and Watson 1927, 
p.~172) we have
\begin{equation}
\label{ef48}
\int_0^{\infty} {\rm d}q\;
\Big[ \frac{\rm d}{{\rm d} q}
\frac{q\, [{\bar\rho}^{\rm h}(q)-{\bar\rho}^{\rm h}(0)]}{q^2 + \kappa^2} 
\Big] \cos(q r) = o(1)\;\,
\mbox{\rm for}\;\, r\to\infty,
\end{equation}
so that ${\cal J}_{\kappa}(r)$ as presented in Eq.~(\ref{ef45}) 
decays {\sl faster} than $1/r$ for $r\to\infty$. The behaviour 
${\cal J}_{\kappa}(r) = o(1/r)$ (to be contrasted with ${\cal O}(1/r)$ 
discussed above) for $r\to\infty$, exactly reflects the property that 
we have earlier deduced through employing the multipole expansion in 
Eq.~(\ref{ef20}) in our investigation of the large-$\|{\Bf r}\|$ 
asymptotic behaviour of ${\cal I}_{\kappa}({\Bf r})$, culminating in 
Eq.~(\ref{ef22}) and the subsequent conclusions. In spite of this
desired property of the ${\cal J}_{\kappa}(r)$ in Eq.~(\ref{ef45}) 
as $r\to\infty$, for $\kappa=0$ this function can be shown to behave 
like ({\it cf}. Eq.~(\ref{ef38a}) above)
\begin{equation}
\label{ef48a}
\left. {\cal J}_{0}(r)\right|_{\rm Eq.~(\ref{ef45})}
\sim -\frac{{\bar\rho}^{\rm h\prime}(0)}{r},\;\;\;
\mbox{\rm for}\;\;\; r\to 0.
\end{equation}
It follows that ${\cal A}^{\prime\rm h}(0)$ according to
Eq.~(\ref{ef37}) with ${\cal J}_{\kappa}(r)$ as presented in 
Eq.~(\ref{ef45}) also diverges like $\ln(\kappa/k_F)$ for 
$\kappa\downarrow 0$. Thus ${\cal J}_{\kappa}(r)$ neither according 
to Eq.~(\ref{ef38}) nor according to Eq.~(\ref{ef45}) is the 
appropriate function for calculating ${\cal A}^{\prime\rm h}(0)$. 

The equivalence of the expressions in Eqs.~(\ref{ef38}) and (\ref{ef45}) 
for $\kappa > 0$ implies that for $\kappa > 0$ we can define 
${\cal J}_{\kappa}(r)$ as follows
\begin{eqnarray}
\label{ef48b}
&&{\cal J}_{\kappa}(r) {:=} \frac{1}{2}
\left\{
\int_0^{\infty} {\rm d}q\;
\frac{q\, [{\bar\rho}^{\rm h}(q) -
{\bar\rho}^{\rm h}(0)]}{q^2 + \kappa^2}\, \sin(q r) \right. \nonumber\\
&&\;\;\;\;\; \left.
+\frac{1}{r}\,
\int_0^{\infty} {\rm d}q\;
\Big[ \frac{\rm d}{{\rm d} q}\,
\frac{q\, [{\bar\rho}^{\rm h}(q)-{\bar\rho}^{\rm h}(0)]}{q^2
+ \kappa^2} \Big]\, \cos(q r) \right\}.
\end{eqnarray}
From Eqs.~(\ref{ef38a}) and (\ref{ef48a}) one observes that
${\cal A}^{\prime\rm h}(0)$ as calculated in terms of
${\cal J}_{\kappa}(r)$ in Eq.~(\ref{ef48b}) (see Eq.~(\ref{ef37}) above)
has a finite limit for $\kappa\downarrow 0$, as in this limit the 
logarithmically divergent contributions to ${\cal A}^{\prime\rm h}(0)$ 
identically cancel. This cancellation can be explicitly effected
through introducing the transformation $k_F r \rightharpoonup 1/[k_F r]$ 
in the $r$ integral on the RHS of Eq.~(\ref{ef37}) as applied to one 
of the two functions of $r$ on the RHS of Eq.~(\ref{ef48b}). Following 
this, we shall be able to set $\kappa$ equal to zero {\sl prior} to 
evaluating the resulting total $r$ integral. 
 
To proceed, we note that in $d$ spatial dimensions, 
${\bar\rho}^{\rm h}(q)$ has the dimension m$^{-d}$ (here, $d=3$), that
is inverse meter to the power $d$. Consequently, $[{\bar\rho}^{\rm h}(q)
-{\bar\rho}^{\rm h}(0)]/q$ has the dimension m$^{1-d}$ so that 
$k_F^{1-d} [{\bar\rho}^{\rm h}(q)-{\bar\rho}^{\rm h}(0)]/q$ is 
dimensionless. Since the problem at hand has only one length scale, 
namely $1/k_F$, it follows that 
\begin{equation}
\label{ef48c}
\frac{[{\bar\rho}^{\rm h}(q)-{\bar\rho}^{\rm h}(0)]}{q}
= k_F^{d-1}\, {\sf f}(q/k_F),
\end{equation}
where ${\sf f}(x)$ stands for a dimensionless function of the 
dimensionless variable $x$ that has {\sl no} explicit dependence 
on $k_F$. Consider
\begin{eqnarray}
\label{ef48d}
{\sf f}(x) \equiv
\left\{ \begin{array}{ll}
\displaystyle
{\sf f}^{<}(x), &\;\; x \le 2, \\ \\
\displaystyle
{\sf f}^{>}(x), &\;\; x \ge 2,
\end{array} \right.
\end{eqnarray}
where we have explicitly assumed that ${\sf f}(x)$ is continuous
at $x=2$. For $d=3$ and within the framework of the SSDA (whence 
the subscript `s' below) we have (see Eq.~(\ref{ef28}) above)
\begin{equation}
\label{ef48e}
{\sf f}_{\rm s}^{<}(x) \equiv \frac{-1}{4\pi^2}
+\frac{x^2}{48 \pi^2},\;\;\;\;\;
{\sf f}_{\rm s}^{>}(x) \equiv \frac{-1}{3\pi^2 x}.
\end{equation}
It is seen that indeed ${\sf f}_{\rm s}^{<}(2) =
{\sf f}_{\rm s}^{>}(2)$; as a matter of fact, we have also
${\rm d} {\sf f}_{\rm s}^{<}(x)/{\rm d} x\vert_{x=2} =
{\rm d} {\sf f}_{\rm s}^{>}(x)/{\rm d} x\vert_{x=2}$; however,
${\rm d}^m {\sf f}_{\rm s}^{<}(x)/{\rm d} x^m\vert_{x=2} \not=
{\rm d}^m {\sf f}_{\rm s}^{>}(x)/{\rm d} x^m\vert_{x=2}$ for $m\ge 2$.
From Eqs.~(\ref{ef48c}) and (\ref{ef48d}) we deduce
\begin{equation}
\label{ef48f}
{\bar\rho}^{\rm h\prime}(0) =
\lim_{q\downarrow 0} 
\frac{{\bar\rho}^{\rm h}(q)-{\bar\rho}^{\rm h}(0)}{q}
= k_F^{d-1}\, {\sf f}(0) \equiv 
k_F^{d-1}\, {\sf f}^{<}(0). 
\end{equation}
Further, from Eq.~(\ref{ef48c}) we have 
\begin{equation}
\label{ef48g}
\frac{\rm d}{{\rm d} q}\,
\frac{{\bar\rho}^{\rm h}(q)-{\bar\rho}^{\rm h}(0)}{q}
\equiv k_F^{d-2}\, {\sf f}'(q/k_F),
\end{equation}
where ${\sf f}'(x) \equiv {\rm d} {\sf f}(x)/{\rm d}x$. 

From the expressions in Eqs.~(\ref{ef48c}) and (\ref{ef48g}),
making use of the expression in Eq.~(\ref{ef48b}), we obtain
\begin{eqnarray}
\label{ef48h}
&&\int_0^{\infty} {\rm d}r\; {\rm e}^{-\kappa r}\,
{\cal J}_{0}(r) = \frac{k_F^2}{2} \,
\int_0^{\infty} {\rm d}y\;
{\rm e}^{-(\kappa/k_F) y}\, \nonumber\\
&&\;\;\;\;\;\;\;\;\;\;\;\;\;\;\;\;\;\;
\times \Big\{ \frac{1}{y} \int_0^{\infty} {\rm d}x\;
{\sf f}(x/y) \sin(x) \nonumber\\
&&\;\;\;\;\;\;\;\;\;\;\;\;\;\;\;\;\;\;\;\;
+ \frac{1}{y^2} \int_0^{\infty} {\rm d}x\;
{\sf f}'(x/y) \cos(x) \Big\}.
\end{eqnarray}
In view of the result in Eq.~(\ref{ef38a}) and with reference
to our above considerations, we apply the transformation $y 
\rightharpoonup 1/y$ in the $y$ integral corresponding to the 
first term enclosed by the curly brackets on the RHS of 
Eq.~(\ref{ef48h}), upon which we can identify $\kappa$ on the 
RHS of Eq.~(\ref{ef48h}) with zero; thus 
\begin{eqnarray}
\label{ef48i}
&&\int_0^{\infty} {\rm d}r\; {\rm e}^{-\kappa r}\,
{\cal J}_{0}(r) = \frac{k_F^2}{2} \,
\int_0^{\infty} \frac{{\rm d}y}{y}\; \nonumber\\
&&\;\;\;
\times \int_0^{\infty} {\rm d}x\;
\Big\{ {\sf f}(x y) \sin(x) + \big[\frac{\rm d}{{\rm d} x}
{\sf f}(x/y)] \cos(x) \Big\} \nonumber\\
&&\equiv -\frac{k_F^2}{2} \, 
\int_0^{\infty} \frac{{\rm d}y}{y}
\Big\{ {\sf f}(0) - \int_0^{\infty} {\rm d}x\;
\big[ {\sf f}(x y) + {\sf f}(x/y)\big]\sin(x) \Big\}, \nonumber\\
\end{eqnarray}
where in arriving at the last expression we have assumed ${\sf f}(x)$
to be continuous for $x > 0$ (specifically at $x=2$; see text
following Eq.~(\ref{ef48d}) above) and applied integration by parts.
From the result in Eq.~(\ref{ef48i}) and the defining expression for
${\cal A}^{\prime\rm h}(0)$ in Eq.~(\ref{ef37}) above we obtain 
\begin{equation}
\label{ef50}
{\cal A}^{\prime\rm h}(0) =
\frac{2 {\sf a} }{\pi}
\left(\frac{e^2}{4\pi\epsilon_0}\right)^2\, k_F^2,
\end{equation}
where 
\begin{equation}
\label{ef50a}
{\sf a} {:=} \frac{4}{\pi} \int_0^{1} {\rm d}y\;
{\sf Y}(y),
\end{equation}
in which
\begin{eqnarray}
\label{ef50b}
&&{\sf Y}(y) {:=} 
\frac{\pi^2}{y}
\Big\{ {\sf f}(0)
-\int_0^{\infty} {\rm d}x\;
\big[ {\sf f}(x y) + {\sf f}(x/y)\big]\sin(x) \Big\} \nonumber\\
&&\;\;\;
\equiv \frac{-\pi^2}{y} \Big\{
\int_0^{2/y} {\rm d}x\; \big[{\sf f}^{<}(x y)
-{\sf f}^{<}(0)\big] \sin(x) \nonumber\\
&&\;\;\;\;\;
+\int_{2/y}^{\infty} {\rm d}x\; {\sf f}^{>}(x y) \sin(x)
+\int_0^{2 y} {\rm d}x\; {\sf f}^{<}(x/y) \sin(x)\nonumber\\
&&\;\;\;\;\;
+\int_{2 y}^{\infty} {\rm d}x\;
{\sf f}^{>}(x/y) \sin(x) - {\sf f}^{<}(0) \cos(2/y) \Big\},
\end{eqnarray}
where we have employed the identity 
\begin{eqnarray}
1 = \cos(2/y) + \int_0^{2/y} {\rm d}x\; \sin(x). \nonumber
\end{eqnarray}
In arriving at the result in Eq.~(\ref{ef50a}) we have made
use of the property
\begin{equation}
\label{ef50c}
{\sf Y}(1/y) = y^2\, {\sf Y}(y),
\end{equation}
which is manifest from the first expression on the RHS of
Eq.~(\ref{ef50b}), and consequently employed $\int_0^{\infty} 
{\rm d}y\; {\sf Y}(y) = 2 \int_0^1 {\rm d}y\; {\sf Y}(y)$. In 
the light of Eqs.~(\ref{ef50}) and (\ref{ef50a}) and our above 
considerations, ${\cal A}^{\prime\rm h}(0)$ is finite provided 
${\sf Y}(y) \sim o(1/y)$ for $y\downarrow 0$. Making use of the
result in Eq.~(\ref{ef48e}), for ${\sf Y}(y)$ within the framework 
of the SSDA we obtain ($d=3$)
\begin{eqnarray}
\label{ef50d}
&&{\sf Y}_{\rm s}(y) = \frac{1-\cos(2/y)}{4 y} \nonumber\\
&&\;\;\;\;\;
+ \frac{1}{24} \Big( y - (y - 2/y) \cos(2/y) - 2\sin(2/y) \Big)
\nonumber\\
&&\;\;\;\;\; 
+\frac{\pi/2 - {\rm Si}(2/y)}{3 y^2}
+ \frac{1 - \cos(2 y) - 2 y \sin(2 y)}{24 y^3} \nonumber\\
&&\;\;\;\;\; 
+\frac{1}{3} \Big( \frac{\pi}{2} - {\rm Si}(2 y)
- \frac{\cos(2 y)}{2 y} \Big),
\end{eqnarray}
where ${\rm Si}(z)$ stands for the sine-integral function (Abramowitz 
and Stegun 1972, pp. 231 and 232). From the expression in 
Eq.~(\ref{ef50d}) we readily deduce that
\begin{equation}
\label{ef50e}
{\sf Y}_{\rm s}(y) \sim \frac{\pi}{6}
- \frac{5 + 3 \cos(2/y)}{24}\, y,\;\;\;
\mbox{\rm for}\;\;\; y \to 0.
\end{equation} 
It follows that ${\sf a}_{\rm s}$ (the value of ${\sf a}$ according 
to the SSDA) is indeed bounded. By means of numerical integration, 
we have obtained 
\begin{equation}
\label{ef50f}
{\sf a}_{\rm s} = 0.563~523~995~\dots,
\end{equation}
which is very close to the contribution of the RHS of Eq.~(\ref{ef50e}) 
to ${\sf a}_{\rm s}$, for which we have
\begin{eqnarray}
\frac{4}{\pi} \int_0^1 {\rm d}y\;
\Big( \frac{\pi}{6} - \frac{5 + 3\cos(2/y)}{24}\, y \Big)
= 0.577~233~661~\dots. \nonumber
\end{eqnarray}
For ${\cal A}^{\prime\rm h}_{\rm s}(0)$ we therefore have 
\footnote{\label{f129a}
The author should like to thank Professor Roland Zimmermann
whose expression of concern with regard to the possibility of
unboundedness of ${\cal A}^{\prime\rm h}_{\rm s}(0)$ led the 
author to detect an error in the original derivation of the result 
in Eq.~(\protect\ref{ef55}), causing ${\sf a}_{\rm s}$ herein to 
be incorrectly identified with unity; the same error had caused
the constant ${\sf a}$ on the RHS of Eq.~(\protect\ref{ef50}) to 
be different from that in Eq.~(\protect\ref{ef50a}) above. }
({\it cf}. Eq.~(\ref{ef50}))
\begin{equation}
\label{ef55}
{\cal A}^{\prime\rm h}_{\rm s}(0) = \frac{2 {\sf a}_{\rm s} }{\pi}\,
\left(\frac{e^2}{4\pi\epsilon_0}\right)^2\, k_F^2.
\end{equation}
With reference to our considerations in \S~III.E, we point out that
(see footnote \ref{f83})
\footnote{\label{f130}
$(2 {\sf a}_{\rm s}/\pi)\, (9\pi/4)^{2/3} 
\approx 2.34\, {\sf a}_{\rm s} \approx 1.32$.}
\begin{equation}
\label{ef56}
{\BAr{\cal A}}^{\prime\rm h}_{\rm s}(0) \equiv
\frac{ {\bar {\cal A}}^{\prime\rm h}_{\rm s}(0)}{e_0^2}
= \frac{2 {\sf a}_{\rm s} }{\pi}\, 
\left(\frac{9\pi}{4}\right)^{2/3}\, r_s^2,
\end{equation}
so that in view of Eq.~(\ref{e108}), $(2 {\sf a}_{\rm s}/\pi) 
(9\pi/4)^{2/3}$ on the RHS of Eq.~(\ref{ef56}) amounts to a 
contribution to ${\cal S}_{\sigma;\infty_1}^{(2)}
({\bar k})\vert_{\rm s}$. It is interesting to compare the RHS 
of Eq.~(\ref{ef56}) with that of Eq.~(\ref{e130}), taking into 
account that for small values of ${\bar k}/{\bar k}_F$, 
${\sf F}({\bar k}/{\bar k}_F) \approx 1$. 

Above we have encountered ${\bar\rho}^{\rm h\prime}(0) \equiv 
{\rm d} {\bar\rho}^{\rm h}(q)/{\rm q}\vert_{q\downarrow 0}$ 
for a number of times. Here we consider this quantity in some
detail. Note that an unbounded ${\bar\rho}^{\rm h\prime}(0)
\equiv k_F^{1-d} {\sf f}(0)$ (see Eq.~(\ref{ef48f}) above)
has through Eq.~(\ref{e187}) a far-reaching consequence for the
behaviour of the SE operator and therefore the stability of the 
{\sl normal} GS assumed here (see footnote \ref{f129}). Through 
the variable transformation $q r\rightharpoonup r$, for $q\not=0$, 
in the integral on the RHS of Eq.~(\ref{ef43}), we obtain
\begin{equation}
\label{ef51}
{\bar\rho}^{\rm h\prime}(0) = \lim_{q\downarrow 0}\, \frac{4\pi}{q^4}\,
\int_0^{\infty} {\rm d}r\; r^2\, \rho^{\rm h}(r/q)\,
\Big[ \cos(r) - \frac{\sin(r)}{r} \Big].
\end{equation}
From this expression it is seen that for ${\bar\rho}^{\rm h\prime}(0)$
to be bounded, it is required that for $q\downarrow 0$ the integral on 
the RHS approach zero like $q^{\alpha}$, with $\alpha \ge 4$: $\alpha=4$ 
gives rise to a finite ${\bar\rho}^{\rm h\prime}(0)$, whereas 
$\alpha > 4$ yields a vanishing ${\bar\rho}^{\rm h\prime}(0)$; the 
possibility of $\alpha < 4$ corresponds to an unbounded 
${\bar\rho}^{\rm h\prime}(0)$ which contradicts the result in 
Eq.~(\ref{ef42}). Since for any {\sl finite} value of $q$, the 
integrand of the $r$ integral on the RHS of Eq.~(\ref{ef51}) behaves 
like $r^4$ for $(r/q)\downarrow 0$, the {\sl leading} term in the AS 
of this integral for $q\downarrow 0$ is determined by the behaviour 
of $\rho^{\rm h}(r)$ for $r \to\infty$. This implies that 
${\bar\rho}^{\rm h\prime}(0)$ is fully determined by the leading 
term in the large-$r$ AS of $\rho^{\rm h}(r)$. It is here that 
the knowledge of this term proves most useful. Within the SSDA, 
making use of Eq.~(\ref{ef19}) we have
\begin{equation}
\label{ef52}
\rho_{\rm s}^{\rm h}(r/q) \equiv 
\sum_{\sigma'} \big(\varrho_{{\rm s};\sigma'}^{\rm h}(r/q)\big)^2
\sim \frac{k_F^2}{2\pi^4}\,
q^4\, \frac{\cos^2(k_F\, r/q)}{r^4},
\end{equation}
from which and Eq.~(\ref{ef51}), making use of the standard integral
\begin{equation} 
\label{ef53}
\int_0^{\infty} {\rm d}r\; \frac{\cos^2(\alpha r)}{r^2}
\Big[ \cos(r) - \frac{\sin(r)}{r} \Big] = -\frac{\pi}{8},
\;\;\; \alpha\not=0,
\end{equation}
we obtain
\begin{equation}
\label{ef54}
{\bar\rho}^{\rm h\prime}_{\rm s}(0) = -\frac{k_F^2}{4\pi^2},
\end{equation}
which is exactly the result obtained from Eqs.~(\ref{ef48e})
and (\ref{ef48g}) above for $d=3$.

Below, in \S~F.1.c, we consider the behaviour of the {\sl exact} 
$\rho^{\rm h}(r)$ in some detail. Here we deduce an approximate 
expression for this function in the regime of weak interaction. 
From our above considerations, making use of Eq.~(\ref{ef48f}) 
above and the approximate expression in Eq.~(\ref{ef86}) below, 
we obtain
\begin{equation}
\label{ef57}
{\sf f}(0) \approx 
\Big[ Z_F^2 + \vert C_{\bf 0}\vert^2 (1-Z_F)^2\Big]\,
{\sf f}_{\rm s}(0),
\end{equation} 
where $C_{\bf 0}$ denotes the overlap of the {\sl exact} $N$-particle 
GS of the system with the $N$-particle GS of the non-interacting 
system (see Eq.~(\ref{ef66}) below). From Eq.~(\ref{ef48e}) above
we have ${\sf f}_{\rm s}(0) = -1/[4\pi^2]$.

Since in practice one calculates $\Gamma^{(m)}$, $m=1,2,\dots$, rather 
than, say, ${\cal U}({\Bf r}_1'',{\Bf r}_2'')$ as defined in 
Eq.~(\ref{ef4}), the expression for ${\cal A}'({\Bf r},{\Bf r}')$ in 
Eq.~(\ref{ef3}) is not useful. To obtain a practically useful expression, 
we make use of Eq.~(\ref{ef2}) which in fact {\sl defines} ${\cal A}'
({\Bf r},{\Bf r}')$. Through some rearrangement of terms, we obtain the 
following result
\begin{eqnarray}
\label{ef58}
&&{\cal A}'({\Bf r},{\Bf r}')
= \int {\rm d}^dr_1''\;
v({\Bf r}-{\Bf r}_1'')
\Big\{ v({\Bf r}'-{\Bf r}_1'') n({\Bf r}_1'')
\nonumber\\
&&\;
+\int {\rm d}^dr_2''\; v({\Bf r}'-{\Bf r}_2'')
\sum_{\sigma_1',\sigma_2'}
\Big[ \Gamma^{(2)}({\Bf r}_1''\sigma_1',{\Bf r}_2''\sigma_2';
{\Bf r}_1''\sigma_1',{\Bf r}_2''\sigma_2')\nonumber\\ 
&&\;\;\;\;\;\;\;\;\;\;\;\;\;\;\;\;\;\;\;\;\;\;\;\;\;
\;\;\;\;\;\;\;\;\;\;\;\;\;\;\;\;\;\;\;\;\,
-n_{\sigma_1'}({\Bf r}_1'') 
n_{\sigma_2'}({\Bf r}_2'')\Big] \Big\}.
\end{eqnarray}
We emphasize that in evaluating the ${\Bf r}_1''$ integral it is 
important (it is even vital, when $v\equiv v_c$) not to separate 
the terms enclosed by the curly brackets, since cancellation among 
these terms plays a crucial role in rendering the integrand 
sufficiently rapidly decaying for large values of $\|{\Bf r}_1''\|$. 

Finally, from Eq.~(\ref{ef58}) for ${\cal A}'({\Bf r},{\Bf r}')$ 
within the framework of the SSDA (see Appendix C) we obtain
\begin{eqnarray}
\label{ef59}
&&{\cal A}'_{\rm s}({\Bf r},{\Bf r}')
= \int {\rm d}^dr_1''\;
v({\Bf r}-{\Bf r}_1'')
\Big\{ v({\Bf r}'-{\Bf r}_1'') n({\Bf r}_1'')
\nonumber\\
&&\;\;\;\;\;\;\;\;
-\int {\rm d}^dr_2''\; v({\Bf r}'-{\Bf r}_2'')
\sum_{\sigma'} \varrho_{{\rm s};\sigma'}^2({\Bf r}_1'',
{\Bf r}_2'') \Big\}.
\end{eqnarray}
We have already dealt with this expression (see Eqs.~(\ref{ef16})
and (\ref{ef18}) above) while considering uniform and isotropic GSs.

\subsubsection{On the behaviour of $\rho({\Bf r},{\Bf r}')$ 
pertaining to uniform and isotropic ground states}
\label{s54}

In \S~1 of this Appendix, we introduced the function $\rho({\Bf r}_1'',
{\Bf r}_2'')$ (see Eq.~(\ref{ef13})) which according to the SSDA is 
equal to $\sum_{\sigma'} \varrho^2_{{\rm s};\sigma'}({\Bf r}_1'',
{\Bf r}_2'')$. Here we analyse $\rho({\Bf r}_1'',{\Bf r}_2'')$ 
pertaining to uniform GSs is some detail and deduce the behaviour of 
the leading-order term in the AS of this function for 
$\|{\Bf r}_1''-{\Bf r}_2''\|\to\infty$.

Making use of Eq.~(\ref{ea41}), from Eq.~(\ref{eb1}) we obtain
\begin{eqnarray}
\label{ef60}
&&\Gamma^{(2)}({\Bf r}_1''\sigma_1',{\Bf r}_2''\sigma_2';
{\Bf r}_1''\sigma_1',{\Bf r}_2''\sigma_2')
= \frac{-1}{\Omega^2} 
\sum_{{\Bf k}_1,{\Bf k}_2,{\Bf k}_3}\nonumber\\
&&\;\;\;\;\;\;
\times \langle\Psi_{N;0}\vert
{\sf\hat a}_{{\Bf k}_1;\sigma_1'}^{\dag}
{\sf\hat a}_{{\Bf k}_2;\sigma_2'}^{\dag}
{\sf\hat a}_{{\Bf k}_3;\sigma_1'}
{\sf\hat a}_{{\Bf k}_1+{\Bf k}_2-{\Bf k}_3;\sigma_2'}
\vert\Psi_{N;0}\rangle\nonumber\\
&&\;\;\;\;\;\;\;\;\;\;\;\;
\times {\rm e}^{-i({\Bf k}_1-{\Bf k}_3)\cdot 
({\Bf r}_1''-{\Bf r}_2'')},
\end{eqnarray}
where we have made use of
\begin{equation}
\label{ef61}
\langle\Psi_{N;0}\vert
{\sf\hat a}_{{\Bf k}_1;\sigma_1'}^{\dag}
{\sf\hat a}_{{\Bf k}_2;\sigma_2'}^{\dag}
{\sf\hat a}_{{\Bf k}_3;\sigma_1'}
{\sf\hat a}_{{\Bf k}_4;\sigma_2'}
\vert\Psi_{N;0}\rangle
\propto \delta_{{\Bf k}_1+{\Bf k}_2,{\Bf k}_3+{\Bf k}_4}.
\end{equation}

Let now $\{ \vert {\Bf\xi}\rangle \}$ be the complete set of
normalized $N$-particle Slater determinants, composed of 
single-particle states $\{ \vert {\Bf k}\rangle \}$, 
$\langle {\Bf r}\vert {\Bf k}\rangle =\Omega^{-1/2} \exp(i {\Bf k}
\cdot {\Bf r})$, corresponding to the {\sl same} set $\{N_{\sigma}\}$ 
of partial number-operators eigenvalues as that corresponding to the 
interacting GS $\vert\Psi_{N;0}\rangle$. We identify ${\Bf\xi}$ with 
a vector of $N$ components, comprised of $2{\sf s}+1$ sub-vectors 
${\Bf\xi}_i$ of $N_{\sigma_i}$ components, where $i=1,\dots, 
2{\sf s}+1$. That is
\begin{equation}
\label{ef62}
{\Bf\xi} \equiv {\Bf\xi}_{\sigma_1},\dots,
{\Bf\xi}_{\sigma_{2{\sf s}+1}}.
\end{equation}
For $N_{\sigma_i}\not=0$, each component of ${\Bf\xi}_{\sigma_i}$
is equal to a non-vanishing wave-vector ${\Bf k}$ and, by antisymmetry, 
{\sl no} two vectors (provided that $N_{\sigma_i} \ge 2$) in 
${\Bf\xi}_{\sigma_i}$ are identical. Further, by the 
{\sl indistinguishability} of the particles, any vector
${\tilde {\Bf\xi}}_{\sigma_i}$ whose wave-vector components are 
by permutation related to those of ${\Bf\xi}_{\sigma_i}$ are 
identified as representing the {\sl same} vector, that is 
${\Bf\xi}_{\sigma_i}$.
\footnote{\label{f131}
When {\sl all} $N_{\sigma_i}$ except one are vanishing,
$\vert {\Bf\xi}\rangle$ describes a fully spin-polarized state.} 
Below, we denote the ${\Bf\xi}$ associated with the Slater 
determinant of lowest (kinetic) energy by ${\Bf\xi}={\bf 0}$ so that 
(see Appendix C)
\begin{equation}
\label{ef63}
\vert {\bf 0}\rangle {:=}
\vert {\Bf\xi}={\bf 0}\rangle
\equiv \vert\Phi_{N;0}\rangle;
\end{equation}
the symbol $\vert {\bf 0}\rangle$ should not be confused with the 
single-particle state $\vert {\Bf k}\rangle$ for ${\Bf k}={\bf 0}$.

Expanding $\vert \Psi_{N;0}\rangle$ with respect to the complete 
set $\{ \vert {\Bf\xi}\rangle \}$, we have
\begin{eqnarray}
\label{ef64}
&&\vert\Psi_{N;0}\rangle =
\sum_{\Bf\xi} C_{\Bf\xi}\, \vert {\Bf\xi}\rangle\nonumber\\
&&\;\;\;
\equiv\sum_{{\Bf\xi}_{\sigma_1},\dots,
{\Bf\xi}_{\sigma_{2{\sf s}+1}} }
C_{{\Bf\xi}_{\sigma_1},\dots,
{\Bf\xi}_{\sigma_{2{\sf s}+1}} }\, 
\vert {\Bf\xi}_{\sigma_1},\dots,
{\Bf\xi}_{\sigma_{2{\sf s}+1}}\rangle;
\end{eqnarray}
completeness of $\{ \vert {\Bf\xi}\rangle\}$ together with the 
normalization to unity of both $\vert\Psi_{N;0}\rangle$ and 
$\{\vert {\Bf\xi}\rangle\}$ imply that
\begin{equation}
\label{ef65}
\sum_{\Bf\xi} \vert C_{\Bf\xi}\vert^2 
\equiv \sum_{{\Bf\xi}_{\sigma_1},\dots,
{\Bf\xi}_{\sigma_{2{\sf s}+1}} }
\vert C_{{\Bf\xi}_{\sigma_1},\dots,
{\Bf\xi}_{\sigma_{2{\sf s}+1}} }\vert^2 = 1
\end{equation}
and
\begin{equation}
\label{ef66}
C_{\Bf\xi} = \langle {\Bf\xi}\vert \Psi_{N;0}\rangle.
\end{equation} 

Following the above considerations, we can now write
\begin{eqnarray}
\label{ef67}
&&\langle \Psi_{N;0}\vert 
{\sf\hat a}_{{\Bf k}_1;\sigma_1'}^{\dag}
{\sf\hat a}_{{\Bf k}_2;\sigma_2'}^{\dag}
{\sf\hat a}_{{\Bf k}_3;\sigma_1'}
{\sf\hat a}_{{\Bf k}_1+{\Bf k}_2-{\Bf k}_3;\sigma_2'} 
\vert \Psi_{N;0}\rangle \nonumber\\
&&\;\;\;\;\;
= \sum_{\Bf\xi} \vert C_{\Bf\xi}\vert^2\,
\langle {\Bf\xi}\vert 
{\sf\hat a}_{{\Bf k}_1;\sigma_1'}^{\dag}
{\sf\hat a}_{{\Bf k}_2;\sigma_2'}^{\dag}
{\sf\hat a}_{{\Bf k}_3;\sigma_1'}
{\sf\hat a}_{{\Bf k}_1+{\Bf k}_2-{\Bf k}_3;\sigma_2'} 
\vert {\Bf\xi}\rangle\nonumber\\ 
&&\;\;\;\;\;
+ \delta\langle\Psi_{N;0}\vert 
{\sf\hat a}_{{\Bf k}_1;\sigma_1'}^{\dag}
{\sf\hat a}_{{\Bf k}_2;\sigma_2'}^{\dag}
{\sf\hat a}_{{\Bf k}_3;\sigma_1'}
{\sf\hat a}_{{\Bf k}_1+{\Bf k}_2-{\Bf k}_3;\sigma_2'} 
\vert\Psi_{N;0}\rangle,\nonumber\\
\end{eqnarray}
where
\begin{eqnarray}
\label{ef68}
&&\delta \langle \Psi_{N;0}\vert 
{\sf\hat a}_{{\Bf k}_1;\sigma_1'}^{\dag}
{\sf\hat a}_{{\Bf k}_2;\sigma_2'}^{\dag}
{\sf\hat a}_{{\Bf k}_3;\sigma_1'}
{\sf\hat a}_{{\Bf k}_1+{\Bf k}_2-{\Bf k}_3;\sigma_2'} 
\vert \Psi_{N;0}\rangle\nonumber\\ 
&&\;
{:=}\sum_{{\Bf\xi},{\Bf\xi}'\atop ({\Bf\xi}\not={\Bf\xi}')}\,
C_{\Bf\xi}^*\, C_{{\Bf\xi}'}\,
\langle {\Bf\xi}\vert
{\sf\hat a}_{{\Bf k}_1;\sigma_1'}^{\dag}
{\sf\hat a}_{{\Bf k}_2;\sigma_2'}^{\dag}
{\sf\hat a}_{{\Bf k}_3;\sigma_1'}
{\sf\hat a}_{{\Bf k}_1+{\Bf k}_2-{\Bf k}_3;\sigma_2'}
\vert {\Bf\xi}'\rangle.\nonumber\\
\end{eqnarray}

For the reason that will become evident shortly, it is advantageous 
to express the matrix element in the first term on the RHS of 
Eq.~(\ref{ef67}) in terms of the coordinate representation of 
$\Gamma^{(2)}_{\Bf\xi}$ defined in Eq.~(\ref{ec4}); making use of 
the expression in Eq.~(\ref{ea40}), one readily verifies that
\begin{eqnarray}
\label{ef69}
&&\langle {\Bf\xi}\vert
{\sf\hat a}_{{\Bf k}_1;\sigma_1'}^{\dag}
{\sf\hat a}_{{\Bf k}_2;\sigma_2'}^{\dag}
{\sf\hat a}_{{\Bf k}_3;\sigma_1'}
{\sf\hat a}_{{\Bf k}_1+{\Bf k}_2-{\Bf k}_3;\sigma_2'}
\vert {\Bf\xi}\rangle\nonumber\\
&&\;\;
=\frac{-1}{\Omega^2} \int \Pi_{j=1}^4 {\rm d}^dr_j\,
\Gamma^{(2)}_{\Bf\xi}({\Bf r}_1\sigma_1',
{\Bf r}_2\sigma_2';{\Bf r}_3\sigma_1',{\Bf r}_4\sigma_2')
\nonumber\\
&&\;\;\;\;\;\;\;\;\times
{\rm e}^{i {\Bf k}_1\cdot {\Bf r}_1}\,
{\rm e}^{i {\Bf k}_2\cdot {\Bf r}_2}\,
{\rm e}^{-i {\Bf k}_3\cdot {\Bf r}_3}\,
{\rm e}^{-i ({\Bf k}_1+{\Bf k}_2-{\Bf k}_3)\cdot {\Bf r}_4}.
\end{eqnarray}
Since $\vert {\Bf\xi}\rangle$ is a single Slater determinant, for 
${\Bf\xi}={\bf 0}$, $\Gamma^{(2)}_{\Bf\xi}$ exactly coincides with 
$\Gamma^{(2)}_{\rm s}$ which we have defined in Appendix C. Denoting 
the generalized Slater-Fock density matrix corresponding to 
$\vert {\Bf\xi}\rangle$, by $\varrho_{{\Bf\xi};\sigma}({\Bf r},
{\Bf r}')$, defined according to ({\it cf}. Eq.~(\ref{ef24}) above)
\begin{eqnarray}
\label{ef70}
\varrho_{{\Bf\xi};\sigma}({\Bf r},{\Bf r}') &{:=}&
\langle {\Bf\xi}\vert 
\hat\psi_{\sigma}^{\dag}({\Bf r})
\hat\psi_{\sigma}({\Bf r}')
\vert {\Bf\xi}\rangle\nonumber\\
&=&\frac{1}{\Omega} \sum_{{\Bf k}\in {\Bf\xi}_{\sigma}}
\, \exp(-i {\Bf k}\cdot [{\Bf r}-{\Bf r}']),
\end{eqnarray}
where by ${\Bf k}\in {\Bf\xi}_{\sigma}$ we specify that ${\Bf k}$ 
is one of the $N_{\sigma}$ {\sl non-vanishing} components of vector 
${\Bf\xi}_{\sigma}$, we have (see Appendix C, the paragraph following
that containing Eq.~(\ref{ec4}))
\begin{eqnarray}
\label{ef71}
&&\Gamma^{(2)}_{\Bf\xi}({\Bf r}_1\sigma_1',
{\Bf r}_2\sigma_2';{\Bf r}_3\sigma_1',{\Bf r}_4\sigma_2')
=\varrho_{{\Bf\xi};\sigma_1'}({\Bf r}_1,{\Bf r}_3)\,
\varrho_{{\Bf\xi};\sigma_2'}({\Bf r}_2,{\Bf r}_4) 
\nonumber\\
&&\;\;\;\;\;\;\;\;\;\;\;\;\;\;\;\;\;\;\;\;\;
-\delta_{\sigma_1',\sigma_2'}\,
\varrho_{{\Bf\xi};\sigma_1'}({\Bf r}_1,{\Bf r}_4)\,
\varrho_{{\Bf\xi};\sigma_1'}({\Bf r}_2,{\Bf r}_3).
\end{eqnarray}
This simplified result clarifies our use of the representation in 
Eq.~(\ref{ef69}). We point out that, similar to $\varrho_{{\rm s};
\sigma}$, $\varrho_{{\Bf\xi};\sigma}$ is {\sl idempotent}, that is
({\it cf}. Eq.~(\ref{e166}))
\begin{equation}
\label{ef72}
\int {\rm d}^dr''\;\varrho_{{\Bf\xi};\sigma}({\Bf r},{\Bf r}'')
\varrho_{{\Bf\xi};\sigma}({\Bf r}'',{\Bf r}') =
\varrho_{{\Bf\xi};\sigma}({\Bf r},{\Bf r}').
\end{equation}
The validity of this expression is readily verified through 
employing the second expression on the RHS of Eq.~(\ref{ef70}) for 
$\varrho_{{\Bf\xi};\sigma}$ on both sides of Eq.~(\ref{ef72}).

Substituting the RHS of Eq.~(\ref{ef71}) into that of Eq.~(\ref{ef69}), 
from Eqs.~(\ref{ef60}) and (\ref{ef67}), upon exchange of orders of
wave-vector summations and spatial integrations, we obtain
\begin{eqnarray}
\label{ef73}
&&\Gamma^{(2)}({\Bf r}_1''\sigma_1',{\Bf r}_2''\sigma_2';
{\Bf r}_1''\sigma_1',{\Bf r}_2''\sigma_2')
= -\delta_{\sigma_1',\sigma_2'}\,
\sum_{\Bf\xi} \vert C_{\Bf\xi}\vert^2\,\nonumber\\
&&\;\;\;\;\;\;\;\;\;\;\;\;\;\times
\varrho_{{\Bf\xi};\sigma_1'}^{\rm h\neg i}
({\Bf r}_1''-{\Bf r}_2'')\,
\varrho_{{\Bf\xi};\sigma_1'}^{\rm h\neg i}
(-{\Bf r}_1''+{\Bf r}_2'')\nonumber\\
&&\;\;\;\;
+\sum_{\Bf\xi} \vert C_{\Bf\xi}\vert^2\,
\varrho_{{\Bf\xi},\sigma_1'}^{\rm h\neg i}({\bf 0})
\varrho_{{\Bf\xi},\sigma_2'}^{\rm h\neg i}({\bf 0})\nonumber\\
&&\;\;\;\;
+\delta \Gamma^{(2)}({\Bf r}_1''\sigma_1',{\Bf r}_2''\sigma_2';
{\Bf r}_1''\sigma_1',{\Bf r}_2''\sigma_2'),
\end{eqnarray}
where we have introduced ({\it cf}. Eq.~(\ref{ef9}))
\begin{equation}
\label{ef74}
\varrho_{{\Bf\xi};\sigma}^{\rm h\neg i}({\Bf r}-{\Bf r}')
\equiv \varrho_{{\Bf\xi};\sigma}({\Bf r},{\Bf r}')
\end{equation}
and where $\delta\Gamma^{(2)}$ is the contribution due to the second 
term on the RHS of Eq.~(\ref{ef67}). We note that, with the exception 
of the case where ${\Bf\xi}={\bf 0}$, $\varrho_{{\Bf\xi};
\sigma}^{\rm h\neg i}({\Bf r})$ is {\sl not} isotropic. This we have 
signified by means of the superscript `${\rm \neg i}$'; thus although 
the function in question is invariant under a simultaneous continuous 
translation of ${\Bf r}$ and ${\Bf r}'$ (it is `homogeneous', as signified 
by the superscript `h'), it is {\sl not} necessarily a function of 
$\|{\Bf r}-{\Bf r}'\|$ (i.e. not necessarily `isotropic'). It can, 
however, be shown that by time-reversal symmetry, $\vert C_{\Bf\xi}\vert^2 
= \vert C_{-{\Bf\xi}}\vert^2$, where $-{\Bf\xi}$, the `time-reversed' 
counterpart of ${\Bf\xi}$, denotes the vector obtained from ${\Bf\xi}$ 
through multiplying {\sl all} component wave-vectors herein by $-1$. 
Consequently, the RHS of Eq.~(\ref{ef73}) is indeed an isotropic 
function of ${\Bf r}_1''-{\Bf r}_2''$. For completeness, we mention 
that, in arriving at the expression in Eq.~(\ref{ef73}), we have 
replaced $\varrho_{{\Bf\xi};\sigma_1'}({\Bf r}_1,{\Bf r}_1-{\Bf r}_1''
+{\Bf r}_2'')$, $\varrho_{{\Bf\xi};\sigma_1'}({\Bf r}_1-{\Bf r}_1''
+{\Bf r}_2'',{\Bf r}_1)$, $\varrho_{{\Bf\xi};\sigma_1'}({\Bf r}_1,
{\Bf r}_1)$, and $\varrho_{{\Bf\xi};\sigma_2'}({\Bf r}_1-{\Bf r}_1''
+{\Bf r}_2'',{\Bf r}_1-{\Bf r}_1''+{\Bf r}_2'')$ by the equivalent 
functions $\varrho_{{\Bf\xi};\sigma_1'}^{\rm h\neg i}({\Bf r}_1'
-{\Bf r}_2'')$, $\varrho_{{\Bf\xi};\sigma_1'}^{\rm h\neg i}(-{\Bf r}_1'
+{\Bf r}_2'')$, $\varrho_{{\Bf\xi};\sigma_1'}^{\rm h\neg i}({\bf 0})$ 
and $\varrho_{{\Bf\xi};\sigma_2'}^{\rm h\neg i}({\bf 0})$ respectively.

In view of the fact that 
\begin{equation}
\label{ef75}
{\sf n}_{{\Bf\xi};\sigma}({\Bf k}) {:=}
\langle {\Bf\xi}\vert 
{\sf\hat a}_{{\Bf k};\sigma}^{\dag}
{\sf\hat a}_{{\Bf k};\sigma}
\vert {\Bf\xi}\rangle
\end{equation}
is the Fourier transform with respect to ${\Bf r}$ of
\footnote{\label{f132}
This holds equally true for ${\sf n}_{\sigma}(k)$, the momentum 
distribution function pertaining to uniform and isotropic interacting 
GSs defined in Eq.~(\protect\ref{ej2}), which is the Fourier transform 
with respect to ${\Bf r}$ of $\varrho_{\sigma}^{\rm h}(\|{\Bf r}\|)$ 
(see Eq.~(\protect\ref{ej3})). } 
$\varrho_{{\Bf\xi};\sigma}^{\rm h\neg i}({\Bf r})$, by introducing 
\begin{equation}
\label{ef76}
\delta \varrho_{{\Bf\xi};\sigma}({\Bf r},{\Bf r}')
{:=} \varrho_{{\Bf\xi};\sigma}({\Bf r},{\Bf r}')
-\varrho_{\sigma}({\Bf r},{\Bf r}'),
\end{equation} 
we have the following associated expression
\begin{equation}
\label{ef77}
\delta {\sf n}_{{\Bf\xi};\sigma}({\Bf k})
{:=} {\sf n}_{{\Bf\xi};\sigma}({\Bf k}) - {\sf n}_{\sigma}(k);
\end{equation}
note that, unless ${\Bf\xi} = {\bf 0}$, this function depends on both
the magnitude and direction of ${\Bf k}$ (see our remarks following 
Eq.~(\ref{ef74}) above). Since the momentum-distribution functions
${\sf n}_{{\Bf\xi};\sigma}({\Bf k})$, $\forall {\Bf\xi}$, are defined 
with respect to the $N$-particle states corresponding to the same set of
values $\{N_{\sigma}\}$ as corresponding to the GS of the interacting
system, from Eq.~(\ref{ej3}) (see also Eq.~(\ref{eh10})) we immediately 
deduce the useful result
\begin{equation}
\label{ef78}
\sum_{\Bf k} \delta {\sf n}_{{\Bf\xi};\sigma}({\Bf k}) = 0\;\;
\iff\;\;
\delta \varrho_{{\Bf\xi};\sigma}^{\rm h\neg i}({\Bf r}={\bf 0}) = 0.
\end{equation}
Further, through multiplying both sides of Eq.~(\ref{ef77}) by 
$\vert C_{\Bf\xi}\vert^2$ (note that $\langle {\Bf\xi}\vert
{\sf\hat a}_{{\Bf k};\sigma}^{\dag} {\sf\hat a}_{{\Bf k};\sigma}
\vert {\Bf\xi}'\rangle \propto \delta_{{\Bf\xi},{\Bf\xi}'}$) and 
summing over all ${\Bf\xi}$, from Eq.~(\ref{ef65}) we obtain the 
following equally useful result:
\begin{equation}
\label{ef79}
\sum_{\Bf\xi} \vert C_{\Bf\xi}\vert^2\,
\delta {\sf n}_{{\Bf\xi};\sigma}({\Bf k}) = 0,
\end{equation}
which is equivalent with ({\it cf}. Eq.~(\ref{ej3}))
\begin{equation}
\label{ef80}
\sum_{\Bf\xi} \vert C_{\Bf\xi}\vert^2\,
\delta \varrho_{{\Bf\xi};\sigma}({\Bf r},{\Bf r}') \equiv 0.
\end{equation}
Consequently, making use of Eq.~(\ref{ef76}) we obtain
\begin{eqnarray}
\label{ef81}
&&\sum_{\Bf\xi} \vert C_{\Bf\xi}\vert^2\,
\varrho_{{\Bf\xi};\sigma_1'}({\Bf r},{\Bf r}')\,
\varrho_{{\Bf\xi};\sigma_2'}({\Bf r}',{\Bf r}) \equiv 
\varrho_{\sigma_1'}({\Bf r},{\Bf r}') 
\varrho_{\sigma_2'}({\Bf r}',{\Bf r}) 
\nonumber\\
&&\;\;\;\;\;\;\;\;\;\;\;\;\;\;\;
+\sum_{\Bf\xi} \vert C_{\Bf\xi}\vert^2 
\delta \varrho_{{\Bf\xi};\sigma_1'}({\Bf r},{\Bf r}')\,
\delta \varrho_{{\Bf\xi};\sigma_2'}({\Bf r}',{\Bf r}).
\end{eqnarray}
In view of the second of the results in Eq.~(\ref{ef78}), as well 
as the expression in Eq.~(\ref{ef65}), identification of ${\Bf r}'$ 
with ${\Bf r}$ in Eq.~(\ref{ef81}) leads to an {\sl identity},
reflecting the equality of the sets of particles concentrations,  
$\{ n_{0;\sigma}\}$, associated with all the $N$-particle states 
involved (see the paragraph directly following Eq.~(\ref{ef61}) 
above). Making use of the result in Eq.~(\ref{ef81}), we can 
re-write Eq.~(\ref{ef73}) as follows 
\begin{eqnarray}
\label{ef82}
&&\Gamma^{(2)}({\Bf r}_1''\sigma_1',{\Bf r}_2''\sigma_2';
{\Bf r}_1''\sigma_1',{\Bf r}_2''\sigma_2') 
= n_{0;\sigma_1'} n_{0;\sigma_2'} \nonumber\\
&&\;\;\;
-\delta_{\sigma_1',\sigma_2'}\, \Big(
\varrho_{\sigma_1'}^2({\Bf r}_1'',{\Bf r}_2'')
+\sum_{\Bf\xi} \vert C_{\Bf\xi}\vert^2\,
\delta \varrho_{{\Bf\xi};\sigma_1'}^{\rm h\neg i}
({\Bf r}_1''-{\Bf r}_2'')\,
\nonumber\\
&&\;\;\;\;\;\;\;\;\;\;\;\;\;\;\;\;\;\;\;\;\;\;\;\;\;\;\;\;\;\;
\;\;\;\;\;\;\;\;\times
\delta \varrho_{{\Bf\xi};\sigma_1'}^{\rm h\neg i}({\Bf r}_2''
-{\Bf r}_1'')\Big)\nonumber\\
&&\;\;\;
+\delta \Gamma^{(2)}({\Bf r}_1''\sigma_1',{\Bf r}_2''\sigma_2';
{\Bf r}_1''\sigma_1',{\Bf r}_2''\sigma_2').
\end{eqnarray}
It is interesting to note that neglecting $\delta\Gamma^{(2)}$, the 
off-diagonal elements of $\Gamma^{(2)}({\Bf r}_1''\sigma_1',
{\Bf r}_2''\sigma_2';{\Bf r}_1''\sigma_1',{\Bf r}_2''\sigma_2')$ 
in the spin space are solely determined by $\{ n_{0;\sigma}\}$. 
Making use of the expression in Eq.~(\ref{ef82}), from 
Eq.~(\ref{ef13}) we finally obtain
\begin{eqnarray}
\label{ef83}
&&\rho({\Bf r}_1'',{\Bf r}_2'') = \sum_{\sigma'} 
\varrho_{\sigma'}^2({\Bf r}_1'',{\Bf r}_2'')\nonumber\\
&&\;\;
+\sum_{\Bf\xi} \vert C_{\Bf\xi}\vert^2\,
\sum_{\sigma'}
\delta \varrho_{{\Bf\xi};\sigma'}^{\rm h\neg i}({\Bf r}_1''-{\Bf r}_2'')\,
\delta \varrho_{{\Bf\xi};\sigma'}^{\rm h\neg i}({\Bf r}_2''-{\Bf r}_1'')
\nonumber\\
&&\;\;
+\delta \rho({\Bf r}_1'',{\Bf r}_2''),
\end{eqnarray}
which is our desired result. Here $\delta \rho({\Bf r}_1'',{\Bf r}_2'')$
is the contribution due to $\delta\Gamma^{(2)}$ on the RHS of
Eq.~(\ref{ef82}).

The expression in Eq.~(\ref{ef83}) provides us with some {\sl insight} 
concerning the leading asymptotic behaviour of $\rho({\Bf r}_1'',
{\Bf r}_2'')$, for $\|{\Bf r}_1''-{\Bf r}_2''\|\to\infty$, at least 
in the case of weakly interacting systems. In such systems, by 
appealing to properties (see Eq.~(\ref{ef65}) above)
\begin{equation}
\label{ef84}
\vert C_{\bf 0} \vert \approx 1\;\;\;\mbox{\rm and thus}\;\;\;
\vert C_{\Bf\xi} \vert \approx 0\;\;\mbox{\rm for}\;\;
{\Bf\xi}\not= {\bf 0}, 
\end{equation}
we restrict the sum over ${\Bf\xi}$ on the RHS of Eq.~(\ref{ef83})
to the term corresponding to ${\Bf\xi} ={\bf 0}$, which upon using 
Eq.~(\ref{ef84}) results in
\begin{eqnarray}
\label{ef85}
&&\rho({\Bf r}_1'',{\Bf r}_2'')
\approx \sum_{\sigma'} \Big\{
\big(1 +\vert C_{\bf 0}\vert^2\big)\,
\big(\varrho_{\sigma'}^{\rm h}(\|{\Bf r}_1''-{\Bf r}_2''\|)\big)^2
\nonumber\\
&&\;\;\;\;\;
+\vert C_{\bf 0}\vert^2\,
\varrho_{0;\sigma'}^{\rm h}(\|{\Bf r}_1''-{\Bf r}_2''\|)\nonumber\\
&&\;\;\;\;\;
- 2\vert C_{\bf 0}\vert^2\,
\varrho_{0;\sigma'}^{\rm h}(\|{\Bf r}_1''-{\Bf r}_2''\|)
\big(\varrho_{\sigma'}^{\rm h}(\|{\Bf r}_1''
-{\Bf r}_2''\|)\big)^2 \Big\}.
\end{eqnarray}
We note that $\delta\rho({\Bf r}_1'',{\Bf r}_2'')$ on the RHS of 
Eq.~(\ref{ef83}) has {\sl not} been incorporated in the expression 
on the RHS of Eq.~(\ref{ef85}), following the fact that the sum on 
the RHS of Eq.~(\ref{ef68}) involves $C_{\Bf\xi}^* C_{{\Bf\xi}'}$ 
with ${\Bf\xi}\not={\Bf\xi}'$ so that {\sl at least} one of 
$C_{\Bf\xi}$ and $C_{{\Bf\xi}'}$ is {\sl not} equal to $C_{\bf 0}$.

From Eq.~(\ref{ef85}) and the result in Eq.~(\ref{ej6}) we obtain 
the following {\sl approximate} leading-order asymptotic result 
\begin{eqnarray}
\label{ef86}
&&\rho^{\rm h}(r) \sim \sum_{\sigma}
\frac{k_{F;\sigma}^2}{4\pi^4}\,
\big[ Z_{F;\sigma}^2 + \vert C_{\bf 0}\vert^2 
(1 - Z_{F;\sigma})^2 \big]\,\nonumber\\
&&\;\;\;\;\;\;\;\;\;\;\;\;\;\;\;\;\;\;\;\;\;\;\;\;\;\;
\;\;\;\;\;\;\;\;\;
\times\frac{\cos^2(k_{F;\sigma}\, r)}{r^4},\;\; r\to\infty,
\end{eqnarray}
where ({\it cf}. Eq.~(\ref{ef9}))
\begin{equation}
\label{ef87}
\rho^{\rm h}(r) \equiv \rho^{\rm h}(\|{\Bf r}\|) 
{:=} \rho({\Bf r},{\bf 0}).
\end{equation}
For weakly-interacting fermions, $Z_{F;\sigma}$ is close to unity, so 
that $\vert C_{\bf 0}\vert^2 (1-Z_{F;\sigma})^2$ in Eq.~(\ref{ef86}) 
can be neglected in comparison with $Z_{F;\sigma}^2$, in which case 
$\rho^{\rm h}(r)$ at large $r$ is seen not to depend {\sl explicitly} 
on $C_{\bf 0}$.

\subsubsection{On the van Hove pair correlation function 
${\sf g}_{\sigma,\sigma'}({\Bf r},{\Bf r}')$}
\label{s55}

From Eqs.~(\ref{eb21}) and (\ref{ef82}), for the normalized van 
Hove pair correlation function pertaining to a uniform and 
isotropic system we have
\begin{eqnarray}
\label{ef88}
&&{\sf g}_{\sigma,\sigma'}({\Bf r},{\Bf r}')
= \frac{n_{0;\sigma} n_{0;\sigma'}}{n_0^2} 
-\frac{1}{n_0^2} \Big\{ 
\varrho_{\sigma}^2({\Bf r},{\Bf r}') \nonumber\\
&&\;\;\;\;\; 
+\sum_{\Bf\xi} \vert C_{\Bf\xi}\vert^2\,
\delta \varrho_{{\Bf\xi};\sigma}^{\rm h\neg i}({\Bf r}-{\Bf r}')\,
\delta \varrho_{{\Bf\xi};\sigma}^{\rm h\neg i}({\Bf r}'-{\Bf r})\Big\}
\, \delta_{\sigma,\sigma'}\nonumber\\
&&\;\;\;\;\;
+\delta {\sf g}_{\sigma,\sigma'}({\Bf r},{\Bf r}'),
\end{eqnarray}
where $\delta {\sf g}_{\sigma,\sigma'}({\Bf r},{\Bf r}')$ denotes
the contribution associated with $\delta\Gamma^{(2)}$ on the RHS 
of Eq.~(\ref{ef82}). With (see Eq.~(\ref{ef9}) above)
\begin{equation}
\label{ef89}
{\sf g}^{\rm h}_{\sigma,\sigma'}(r) \equiv 
{\sf g}^{\rm h}_{\sigma,\sigma'}(\|{\Bf r}\|) {:=} 
{\sf g}_{\sigma,\sigma'}({\Bf r},{\bf 0}),
\end{equation}
under the assumption of the validity of the assertions in
Eq.~(\ref{ef84}), from Eq.~(\ref{ef88}) we readily obtain
\begin{eqnarray}
\label{ef90}
&&{\sf g}^{\rm h}_{\sigma,\sigma'}(r) \sim
\frac{n_{0;\sigma} n_{0;\sigma'}}{n_0^2}
- \frac{k_{F;\sigma}^2\, \big[Z_{F;\sigma}^2 + 
\vert C_{\bf 0}\vert^2 
(1 - Z_{F;\sigma})^2\big]}{4 \pi^4\, n_0^2}\,\nonumber\\
&&\;\;\;\;\;\;\;\;\;\;\;\;\;\;\;\;\;\;\;\;\;\;\;\;\;\;
\;\;\;
\times\frac{\cos^2(k_{F;\sigma}\, r)}{r^4}\,
\delta_{\sigma,\sigma'},\;\; r\to \infty,
\end{eqnarray}
where we have made use of the asymptotic results in
Eqs.~(\ref{ef19}) and (\ref{ej6}). In practice one often
deals with the following function
\begin{equation}
\label{ef91}
{\sf g}^{\rm h}(r) {:=} 
\sum_{\sigma,\sigma'} {\sf g}^{\rm h}_{\sigma,\sigma'}(r),
\end{equation}
for which from Eq.~(\ref{ef90}) one readily obtains the following
result
\begin{eqnarray}
\label{ef92}
&&{\sf g}^{\rm h}(r) \sim 1 - \frac{1}{n_0^2}
\sum_{\sigma}
\frac{k_{F;\sigma}^2}{4\pi^4}\,
\big[ Z_{F;\sigma}^2 + \vert C_{\bf 0}\vert^2
(1 - Z_{F;\sigma})^2 \big]\,\nonumber\\
&&\;\;\;\;\;\;\;\;\;\;\;\;\;\;\;\;\;\;\;\;\;\;\;\;\;\;
\;\;\;\;\;\;\;\;\;
\times\frac{\cos^2(k_{F;\sigma}\, r)}{r^4},\;\; r\to\infty.
\end{eqnarray}
Note that the second term on the RHS of Eq.~(\ref{ef92}) is equal
to $1/n_0^2$ times the RHS of Eq.~(\ref{ef86}). For uniform system
of spin-$1/2$ fermions in the paramagnetic phase, where
$k_{F;\sigma}=k_F$ and $Z_{F;\sigma}=Z_F$, Eq.~(\ref{ef92}) takes
on the following simple form
\begin{equation}
\label{ef93}
{\sf g}^{\rm h}(r) \sim 
1 - \frac{9 \big[Z_F^2 + \vert C_{\bf 0}\vert^2
(1 - Z_F)^2 \big]}{2 k_F^4}\, 
\frac{\cos^2(k_F\, r)}{r^4},\, r\to\infty.
\end{equation}
The result in Eq.~(\ref{ef93}) is interesting in that it exposes that 
at least for weakly interacting systems, for which the conditions in 
Eq.~(\ref{ef84}) apply, $ \rho^{\rm h}(r) \sim n_0^2 \big(1 - 
{\sf g}^{\rm h}(r)\big)$ for large values of $r$. Note in passing that 
the appropriate expressions for ${\sf g}^{\rm h}_{\sigma,\sigma'}(r)$ 
and ${\sf g}^{\rm h}(r)$ within the framework of the SSDA are readily 
obtained through identifying with unity both $\vert C_{\bf 0}\vert^2$ 
and $Z_{F;\sigma}$ in Eqs.~(\ref{ef90}) and (\ref{ef92}). It is important 
to realize that, although the approximate result in for instance 
Eq.~(\ref{ef93}) relies on the conditions in Eq.~(\ref{ef84}), 
nonetheless it is far superior to the straightforward approximation in 
which the series on the RHS of Eq.~(\ref{ef64}) is restricted to its 
first term, namely $C_{\bf 0} \vert {\bf 0}\rangle$; such an 
approximation would have resulted in $\vert C_{\bf 0}\vert^2$ as 
being the first term on RHS of Eq.~(\ref{ef92}), rather than the
present {\sl exact} result $1$. Although this may not seem an important 
aspect in the first glance, the fact that $\vert C_{\bf 0}\vert^2 
\not= 1$ (unless the system under consideration is fully 
non-interacting), ${\sf g}^{\rm h}(r) \sim \vert C_{\bf 0}\vert^2$ 
for $r\to\infty$ amounts to a fundamental error as regards the charge 
conservation in the system.

\subsection{${\cal B}_{\sigma}({\Bf r},{\Bf r}')$ and its 
regularized forms ${\cal B}_{\sigma}'({\Bf r},{\Bf r}')$
and ${\cal B}_{\sigma}''({\Bf r},{\Bf r}')$}
\label{s56}

Here we deal with ${\cal B}_{\sigma}({\Bf r},{\Bf r}')$; however, 
contrary to the case of ${\cal A}({\Bf r},{\Bf r}')$ we shall not 
go into the details which are very similar to those underlying 
regularization of ${\cal A}({\Bf r},{\Bf r}')$.

Making use of the anticommutation relations in Eq.~(\ref{e29})
and the defining expression for $\Gamma^{(2)}$ in Eq.~(\ref{eb1}),
we obtain
\begin{eqnarray}
\label{ef94}
&&\Gamma^{(2)}({\Bf r}'\sigma,{\Bf r}''\sigma';
{\Bf r}''\sigma',{\Bf r}\sigma)
=\delta_{\sigma,\sigma'} \delta({\Bf r}'-{\Bf r}'')
\, \varrho_{\sigma}({\Bf r}',{\Bf r})\nonumber\\
&&\;\;\;
-n_{\sigma'}({\Bf r}'') 
\varrho_{\sigma}({\Bf r}',{\Bf r})\nonumber\\
&&\;\;\;
-\langle\Psi_{N;0}\vert\big[{\hat n}_{\sigma'}({\Bf r}'')
-n_{\sigma'}({\Bf r}'')\big] 
{\hat\varrho}_{\sigma}({\Bf r}',{\Bf r})
\vert\Psi_{N;0}\rangle.
\end{eqnarray}
Replacing ${\hat\varrho}_{\sigma}({\Bf r}',{\Bf r}) {:=}
\hat\psi_{\sigma}^{\dag}({\Bf r}') \hat\psi_{\sigma}({\Bf r})$ on 
the RHS of Eq.~(\ref{ef94}) by $\big[{\hat\varrho}_{\sigma}({\Bf r}',
{\Bf r}) -\varrho_{\sigma}({\Bf r}',{\Bf r})\big]$ does {\sl not} 
lead to any quantitative change; however, it renders ${\cal B}_{\sigma}
({\Bf r},{\Bf r})$ a similar {\sl appearance} as ${\cal A}({\Bf r},
{\Bf r})$ in Eq.~(\ref{ef2}). This observation is relevant in that it 
shows that all our earlier considerations in this Appendix with regard 
to ${\cal A}'({\Bf r},{\Bf r})$ corresponding to uniform and isotropic 
GSs directly apply to the regularized counterpart of ${\cal B}_{\sigma}
({\Bf r},{\Bf r})$ corresponding to these systems.

From Eqs.~(\ref{eb29}) and (\ref{ef94}) we obtain
\begin{eqnarray}
\label{ef95}
{\cal B}_{\sigma}({\Bf r},{\Bf r}') 
&=& v({\Bf r}-{\Bf r}') \varrho_{\sigma}({\Bf r}',{\Bf r})
\nonumber\\
&-&v_H({\Bf r};[n]) \varrho_{\sigma}({\Bf r}',{\Bf r})
+{\cal B}'_{\sigma}({\Bf r},{\Bf r}')
\end{eqnarray}
where
\begin{eqnarray}
\label{ef96}
&&{\cal B}'_{\sigma}({\Bf r},{\Bf r}') {:=} 
-\int {\rm d}^dr''\; v({\Bf r}-{\Bf r}'') \nonumber\\
&&\;\;\;
\times\sum_{\sigma'}
\langle\Psi_{N;0}\vert\big[{\hat n}_{\sigma'}({\Bf r}'')
-n_{\sigma'}({\Bf r}'')\big]
{\hat\varrho}_{\sigma}({\Bf r}',{\Bf r})
\vert\Psi_{N;0}\rangle.
\end{eqnarray}
Since the integral with respect to ${\Bf r}''$ of the GS expectation 
value on the RHS of Eq.~(\ref{ef96}) is identically vanishing (compare 
with Eq.~(\ref{ef5}) above), it follows that, for a constant $v$, 
${\cal B}'_{\sigma}({\Bf r},{\Bf r}')\equiv 0$. Consequently, when 
$v\equiv v_c$, ${\cal B}'_{\sigma}({\Bf r},{\Bf r}')$ decreases 
{\sl faster} than $1/\|{\Bf r}\|$ for $\|{\Bf r}\|\to \infty$. This 
is established through invoking on the RHS of Eq.~(\ref{ef96}) the 
multipole expansion for $v_c({\Bf r}-{\Bf r}'')$ as presented in 
Eq.~(\ref{ef20}). 

In practice, where one calculates $\Gamma^{(m)}$, $m=1,2,\dots$,
evaluation of ${\cal B}'_{\sigma}({\Bf r},{\Bf r}')$ according
to the expression in Eq.~(\ref{ef96}) is not useful. A useful 
expression is obtained by employing Eq.~(\ref{ef95}) as the 
{\sl defining} expression for ${\cal B}'_{\sigma}({\Bf r},{\Bf r}')$. 
One readily obtains
\begin{eqnarray}
\label{ef97}
&&{\cal B}'_{\sigma}({\Bf r},{\Bf r}')
= - v({\Bf r}-{\Bf r}') \varrho_{\sigma}({\Bf r}',{\Bf r})
\nonumber\\
&&\;\;\;
-\int {\rm d}^dr''\; v({\Bf r}-{\Bf r}'')
\sum_{\sigma'} \Big\{
\Gamma^{(2)}({\Bf r}\sigma,{\Bf r}''\sigma';
{\Bf r}'\sigma,{\Bf r}''\sigma')\nonumber\\
&&\;\;\;\;\;\;\;\;\;\;\;\;\;\;\;\;\;\;\;\;\;\;\;\;\;\;\;\;\;\;\;
\;\;\;\;\;\;\;\;\;\;\,
-n_{\sigma'}({\Bf r}'') \varrho_{\sigma}({\Bf r}',{\Bf r})
\Big\}.
\end{eqnarray}
In order to facilitate notation in the main text, we introduce
\begin{eqnarray}
\label{ef98}
{\cal B}''_{\sigma}({\Bf r},{\Bf r}')
{:=} {\cal B}'_{\sigma}({\Bf r},{\Bf r}')
+ v({\Bf r}-{\Bf r}') \varrho_{\sigma}({\Bf r}',{\Bf r}).
\end{eqnarray}

For completeness, within the framework of the SSDA we have (see 
Eq.~(\ref{ec6}))
\begin{eqnarray}
\label{ef99}
&&\left. {\cal B}'_{\sigma}({\Bf r},{\Bf r}')\right|_{\rm s}
=-v({\Bf r}-{\Bf r}') \varrho_{{\rm s};\sigma}({\Bf r}',{\Bf r})
\nonumber\\ 
&&\;\;\;\;\;\;
+\int {\rm d}^dr''\; v({\Bf r}-{\Bf r}'')
\varrho_{{\rm s};\sigma}({\Bf r}',{\Bf r}'') 
\varrho_{{\rm s};\sigma}({\Bf r}'',{\Bf r}).
\end{eqnarray}
For $v\equiv v_c$, owing to the idempotency of the Slater-Fock 
density matrices, the leading term in the AS of the 
integral on the RHS of Eq.~(\ref{ef99}) for $\|{\Bf r}\|\to \infty$ 
is equal to 
\footnote{\label{f133}
The LHS of this relation is deduced by employing the multipole 
expansion in Eq.~(\protect\ref{ef20}). }
$v_c({\Bf r}) \varrho_{{\rm s};\sigma}({\Bf r}',{\Bf r}) \sim 
v_c({\Bf r}-{\Bf r}') \varrho_{{\rm s};\sigma}({\Bf r}',{\Bf r})$, 
where we have assumed $\|{\Bf r}'\|$ to be finite. This contribution 
cancels the first term on the RHS of Eq.~(\ref{ef99}), implying that, 
similar to the {\sl exact} ${\cal B}'_{\sigma}({\Bf r},{\Bf r}')$, 
the approximation in Eq.~(\ref{ef99}) decays {\sl faster} than 
$1/\|{\Bf r}\|$ for $\|{\Bf r}\|\to \infty$.

\subsection{${\cal G}_{\sigma}({\Bf r},{\Bf r}')$ and its 
regularized forms ${\cal G}_{\sigma}'({\Bf r},{\Bf r}')$ and
${\cal G}_{\sigma}''({\Bf r},{\Bf r}')$}
\label{s57}

Making use of the anticommutation relations in Eq.~(\ref{e29}),
from the defining equation for $\Gamma^{(3)}$ in Eq.~(\ref{eb1})
we obtain
\begin{eqnarray}
\label{ef100}
&&\Gamma^{(3)}({\Bf r}'\sigma,{\Bf r}_1''\sigma_1',
{\Bf r}_2''\sigma_2';{\Bf r}\sigma,{\Bf r}_1''\sigma_1',
{\Bf r}_2''\sigma_2')\nonumber\\
&&\;\;
= -\delta_{\sigma_1',\sigma_2'}
\delta({\Bf r}_1''-{\Bf r}_2'')
\Gamma^{(2)}({\Bf r}'\sigma,{\Bf r}_1''\sigma_1';
{\Bf r}\sigma,{\Bf r}_1''\sigma_1')\nonumber\\
&&\;\;\;\;\;\;
+\delta_{\sigma,\sigma_1'} \delta_{\sigma,\sigma_2'} 
\delta({\Bf r}'-{\Bf r}_1'')
\delta({\Bf r}-{\Bf r}_2'')
\varrho_{\sigma}({\Bf r}',{\Bf r})\nonumber\\
&&\;\;\;\;\;\;
-\delta_{\sigma,\sigma_1'} \delta({\Bf r}'-{\Bf r}_1'')\,
\langle\Psi_{N;0}\vert 
\hat\varrho_{\sigma}({\Bf r}',{\Bf r})
{\hat n}_{\sigma_2'}({\Bf r}_2'')
\vert\Psi_{N;0}\rangle\nonumber\\
&&\;\;\;\;\;\;
-\delta_{\sigma,\sigma_2'} \delta({\Bf r}'-{\Bf r}_2'')\,
\langle\Psi_{N;0}\vert
{\hat n}_{\sigma_1'}({\Bf r}_1'') 
\hat\varrho_{\sigma}({\Bf r}',{\Bf r})
\vert\Psi_{N;0}\rangle\nonumber\\
&&\;\;\;\;\;\;
+\langle\Psi_{N;0}\vert 
{\hat n}_{\sigma_1'}({\Bf r}_1'')
{\hat n}_{\sigma_2'}({\Bf r}_2'')
\hat\varrho_{\sigma}({\Bf r}',{\Bf r})
\vert\Psi_{N;0}\rangle.
\end{eqnarray}
Employing this result, Eq.~(\ref{ee18}) can be re-written as
\begin{eqnarray}
\label{ef101}
&&{\cal G}_{\sigma}({\Bf r},{\Bf r}')
= v^2({\Bf r}-{\Bf r}') \varrho_{\sigma}({\Bf r}',{\Bf r})
+2 v({\Bf r}-{\Bf r}') {\cal B}'_{\sigma}({\Bf r},{\Bf r}')
\nonumber\\
&&\;\;\;\;
-2 v({\Bf r}-{\Bf r}') v_H({\Bf r};[n]) 
\varrho_{\sigma}({\Bf r}',{\Bf r})
-2 v_H({\Bf r};[n]) {\cal B}'_{\sigma}({\Bf r},{\Bf r}')
\nonumber\\
&&\;\;\;\;
+ v_H^2({\Bf r};[n]) \varrho_{\sigma}({\Bf r}',{\Bf r})
+ {\cal G}'_{\sigma}({\Bf r},{\Bf r}'),
\end{eqnarray}
where
\begin{eqnarray}
\label{ef102}
&&{\cal G}'_{\sigma}({\Bf r},{\Bf r}')
{:=} \int {\rm d}^dr_1'' {\rm d}^dr_2''\;
v({\Bf r}-{\Bf r}_1'') v({\Bf r}-{\Bf r}_2'')\nonumber\\
&&\;\;\;
\times \sum_{\sigma_1',\sigma_2'}
\langle\Psi_{N;0}\vert \big[
{\hat n}_{\sigma_1'}({\Bf r}_1'') 
-n_{\sigma_1'}({\Bf r}_1'')\big]\nonumber\\
&&\;\;\;\;\;\;\;\;\;\;\;\;\;\;
\times \big[ {\hat n}_{\sigma_2'}({\Bf r}_2'')
-n_{\sigma_2''}({\Bf r}_2'')\big] \hat\varrho_{\sigma}
({\Bf r}',{\Bf r})\vert\Psi_{N;0}\rangle.
\end{eqnarray}
Since $\big[ {\hat n}_{\sigma}({\Bf r}),{\hat n}_{\sigma'}({\Bf r}') 
\big]_{-} =0$, $\big[{\hat n}_{\sigma_2'}({\Bf r}_2'') - n_{\sigma_2'}
({\Bf r}_2'') \big]$ in the RHS of Eq.~(\ref{ef102}) can be transposed 
to the left of $\big[ {\hat n}_{\sigma_1'}({\Bf r}_1'') - n_{\sigma_1'}
({\Bf r}_1'')\big]$. It follows that the integrals with respect to both 
${\Bf r}_1''$ {\sl and} ${\Bf r}_2''$ of $\sum_{\sigma_1',\sigma_2'} 
\langle\Psi_{N;0}\vert \dots\vert\Psi_{N;0}\rangle$ on the RHS of 
Eq.~(\ref{e102}) are identically vanishing ({\it cf}. Eq.~(\ref{ef5}); 
see also text following Eq.~(\ref{ef96}) above). Thus, in the case of 
$d=3$ and $v\equiv v_c$, ${\cal G}'_{\sigma}({\Bf r},{\Bf r}')$ decays 
more rapidly than $1/\|{\Bf r}\|^2$ for $\|{\Bf r}\|\to\infty$.

In the case where $d=3$ and $v\equiv v_c$ and the system under 
consideration is in the thermodynamic limit, each $v_c({\Bf r}
-{\Bf r}')$ in the above expressions is accompanied by the appropriate 
smooth cut-off function $\exp(-\kappa \|{\Bf r}-{\Bf r}'\|)$, with 
$\kappa\downarrow 0$ (see Eq.~(\ref{e13}) and the text following
Eq.~(\ref{e15})). Here, as in other cases, the limit $\kappa\downarrow 
0$ is taken {\sl following} evaluation of the pertinent integrals over 
the space occupied by the system with volume $\Omega\to\infty$ (see 
Eq.~(\ref{e10})). Since under these conditions, ${\cal B}'_{\sigma}
({\Bf r},{\Bf r}')$ as well as ${\cal G}'_{\sigma}({\Bf r},{\Bf r}')$ 
are bounded, from Eq.~(\ref{ef101}) and the {\sl unboundedness} of 
$v_H({\Bf r};[n])$ for $\kappa\downarrow 0$, it follows that, for 
$v\equiv v_c$ in $d=3$, ${\cal G}_{\sigma}({\Bf r},{\Bf r}')$ is 
unbounded in the thermodynamic limit. As can be verified from 
Eq.~(\ref{e199}) (see also Eqs.~(\ref{e210}) - (\ref{e212}) and 
(\ref{e214})), $\Sigma_{\sigma;\infty_2}({\Bf r},{\Bf r}')$ does 
{\sl not} depend on ${\cal G}_{\sigma}({\Bf r},{\Bf r}')$, but only 
on ${\cal G}'_{\sigma}({\Bf r},{\Bf r}')$, nor does $\Sigma_{\sigma;
\infty_2}({\Bf r},{\Bf r}')$ depend on $v_H({\Bf r};[n])$ but on 
$v_H({\Bf r};[n'])$, which is bounded in the thermodynamic limit.

The expression for ${\cal G}'_{\sigma}({\Bf r},{\Bf r}')$ in 
Eq.~(\ref{ef102}) is not useful for the purpose of actual 
calculations where the correlation functions $\Gamma^{(m)}$
are the basic functions. Taking Eq.~(\ref{ef101}) as the
defining equation for ${\cal G}'_{\sigma}({\Bf r},{\Bf r}')$,
some elementary algebra yields
\begin{eqnarray}
\label{ef103}
&&{\cal G}'_{\sigma}({\Bf r},{\Bf r}')
= \int {\rm d}^dr_1''\; v({\Bf r}-{\Bf r}_1'') 
\Big( v({\Bf r}-{\Bf r}_1'') \nonumber\\
&&\;\; \times \sum_{\sigma'} \Big[
\Gamma^{(2)}({\Bf r}'\sigma,{\Bf r}_1''\sigma';
{\Bf r}\sigma,{\Bf r}_1''\sigma_1') 
-n_{\sigma'}({\Bf r}_1'') 
\varrho_{\sigma}({\Bf r}',{\Bf r})\Big]\nonumber\\
&&\;\;\;\;
+\int {\rm d}^dr_2''\; v({\Bf r}-{\Bf r}_2'')
\nonumber\\
&&\;\;\;\;\;\;\;\;\;\;\;\;
\times
\sum_{\sigma_1',\sigma_2'} \Big[
\Gamma^{(3)}({\Bf r}'\sigma,{\Bf r}_1''\sigma_1',
{\Bf r}_2''\sigma_2'';{\Bf r}\sigma,
{\Bf r}_1''\sigma_1',{\Bf r}_2''\sigma_2')\nonumber\\
&&\;\;\;\;\;\;\;\;\;\;\;\;\;\;\;\;\;\;\;\;\;\;\;\;\;\;\;\;
-n_{\sigma_1'}({\Bf r}_1'') n_{\sigma_2'}({\Bf r}_2'')
\varrho_{\sigma}({\Bf r}',{\Bf r})\Big]\nonumber\\
&&\;\;\;\;
+\sum_{\sigma'} n_{\sigma'}({\Bf r}_1'')
\Big[ \big\{ 2 v({\Bf r}-{\Bf r}') + v({\Bf r}-{\Bf r}_1'')
\big\} \varrho_{\sigma}({\Bf r}',{\Bf r}) \nonumber\\
&&\;\;\;\;\;\;\;\;\;\;\;\;\;\;\;\;\;\;\;\;\;\;\;\;\;\;\;
+2 {\cal B}'_{\sigma}({\Bf r},{\Bf r}')\Big] \Big)
\nonumber\\
&&\;\; -v^2({\Bf r}-{\Bf r}') 
\varrho_{\sigma}({\Bf r}',{\Bf r})
-2 v({\Bf r}-{\Bf r}') {\cal B}'_{\sigma}({\Bf r},{\Bf r}').
\end{eqnarray}
In order to facilitate notation in the main text, noting from
Eq.~(\ref{ef98}) that $-v^2 \varrho_{\sigma} - 2 v {\cal B}'_{\sigma}
= v^2 \varrho_{\sigma} - 2 v {\cal B}''_{\sigma}$,
we introduce
\begin{eqnarray}
\label{ef104}
{\cal G}''_{\sigma}({\Bf r},{\Bf r}') {:=} 
{\cal G}'_{\sigma}({\Bf r},{\Bf r}') - 
v^2({\Bf r}-{\Bf r}') \varrho_{\sigma}({\Bf r}',{\Bf r}).
\end{eqnarray}

Finally, within the framework of the SSDA (see Appendix C) we have
\begin{eqnarray}
\label{ef105}
&&\left. {\cal G}'_{\sigma}({\Bf r},{\Bf r}')\right|_{\rm s}
= {\cal A}'_{\rm s}({\Bf r},{\Bf r}) 
\varrho_{{\rm s};\sigma}({\Bf r}',{\Bf r})\nonumber\\
&&\;\;\;
- \int {\rm d}^dr''\; v^2({\Bf r}-{\Bf r}'') 
\varrho_{{\rm s};\sigma}({\Bf r}',{\Bf r}'')
\varrho_{{\rm s};\sigma}({\Bf r}'',{\Bf r}) \nonumber\\
&&\;\;\;
+2 \int {\rm d}^dr_1'' {\rm d}^dr_2''\;
v({\Bf r}-{\Bf r}_1'') v({\Bf r}-{\Bf r}_2'')\nonumber\\
&&\;\;\;\;\;\;\;\;\;\;\;\;\;\;\;\;\;\;\;\; \times
\varrho_{{\rm s};\sigma}({\Bf r}',{\Bf r}_1'')
\varrho_{{\rm s};\sigma}({\Bf r}_1'',{\Bf r}_2'')
\varrho_{{\rm s};\sigma}({\Bf r}_2'',{\Bf r})\nonumber\\
&&\;\;\;
-v^2({\Bf r}-{\Bf r}') \varrho_{{\rm s};\sigma}({\Bf r}',{\Bf r})
-2 v({\Bf r}-{\Bf r}') 
{\cal B}'_{{\rm s};\sigma}({\Bf r},{\Bf r}').
\end{eqnarray}

\subsection{${\cal K}_{\sigma}({\Bf r},{\Bf r}')$ and its 
regularized forms ${\cal K}_{\sigma}'({\Bf r},{\Bf r}')$,
${\cal K}_{\sigma}''({\Bf r},{\Bf r}')$ and
${\cal K}_{\sigma}'''({\Bf r},{\Bf r}')$}
\label{s58}

In the expression for $\Sigma_{\sigma;\infty_2}({\Bf r},{\Bf r}')$ in
Eq.~(\ref{e199}) we encounter 
\begin{eqnarray}
\label{ef106}
&&{\cal K}_{\sigma}({\Bf r},{\Bf r}')
{:=} \int {\rm d}^dr''\; v({\Bf r}-{\Bf r}'')
v({\Bf r}'-{\Bf r}'') \nonumber\\ 
&&\;\;\;\;\;\;\;\;\;
\times\sum_{\sigma'} \Gamma^{(2)}({\Bf r}'\sigma,
{\Bf r}''\sigma';{\Bf r}\sigma,{\Bf r}''\sigma')\nonumber\\
&&\;\;\;
+\int {\rm d}^dr_1'' {\rm d}^dr_2''\;
v({\Bf r}-{\Bf r}_1'') v({\Bf r}'-{\Bf r}_2'')
\nonumber\\
&&\;\;\;\;\;\;\;\;
\times \sum_{\sigma_1',\sigma_2'}
\Gamma^{(3)}({\Bf r}'\sigma,{\Bf r}_1''\sigma_1',
{\Bf r}_2''\sigma_2';
{\Bf r}\sigma,{\Bf r}_1''\sigma_1',{\Bf r}_2''\sigma_2'),
\end{eqnarray}
which, like ${\cal A}'({\Bf r},{\Bf r}')$ but unlike 
${\cal B}'_{\sigma}({\Bf r},{\Bf r}')$ and ${\cal G}'_{\sigma}
({\Bf r},{\Bf r}')$, {\sl is} symmetric with respect to ${\Bf r} 
\rightleftharpoons {\Bf r}'$. Making use of the expression in 
Eq.~(\ref{ef100}) together with the commutation relation 
\begin{eqnarray}
\label{ef107}
\big[{\hat n}_{\sigma'}({\Bf r}''), 
\hat\varrho_{\sigma}({\Bf r}',{\Bf r}) \big]_{-} &=& 
\delta_{\sigma,\sigma'} \big\{ \delta({\Bf r}'-{\Bf r}'')
- \delta({\Bf r}-{\Bf r}'')\big\} \nonumber\\
& &\;\;\;\;\;\;\;\;\;\;\;\;\;\;\;\;\;\;\;\;
\times \hat\varrho_{\sigma}({\Bf r}',{\Bf r}),
\end{eqnarray}
we obtain
\begin{eqnarray}
\label{ef108}
&&{\cal K}_{\sigma}({\Bf r},{\Bf r}')
= \Big\{ v^2({\Bf r}-{\Bf r}') 
+v_H({\Bf r};[n]) v_H({\Bf r}';[n])
\nonumber\\
&&\;\;\;\;\;\;\;\;\;\;\;\;\;\;\;\;
-v({\Bf r}-{\Bf r}') \big[ v_H({\Bf r};[n])
+ v_H({\Bf r}';[n]) \big] \Big\}
\varrho_{\sigma}({\Bf r}',{\Bf r})\nonumber\\
&&\;\; +v({\Bf r}-{\Bf r}') 
\big[{\cal B}'_{\sigma}({\Bf r},{\Bf r}') +
{\cal B}'_{\sigma}({\Bf r}',{\Bf r})\big]\nonumber\\
&&\;\;
-v_H({\Bf r};[n]) {\cal B}'_{\sigma}({\Bf r}',{\Bf r})
-v_H({\Bf r}';[n]) {\cal B}'_{\sigma}({\Bf r},{\Bf r}')
+ {\cal K}'_{\sigma}({\Bf r},{\Bf r}'),\nonumber\\
\end{eqnarray}
where
\begin{eqnarray}
\label{ef109}
&&{\cal K}'_{\sigma}({\Bf r},{\Bf r}')
{:=} \int {\rm d}^dr_1'' {\rm d}^dr_2''\;
v({\Bf r}-{\Bf r}_1'') v({\Bf r}'-{\Bf r}_2'')\nonumber\\
&&\;\;\; \times \sum_{\sigma_1',\sigma_2'}
\langle\Psi_{N;0}\vert
\big[{\hat n}_{\sigma_1'}({\Bf r}_1'')
-n_{\sigma_1'}({\Bf r}_1'')\big]
\hat\varrho_{\sigma}({\Bf r}',{\Bf r})\nonumber\\
&&\;\;\;\;\;\;\;\;\;\;\;\;\;\;\;\;\;\;\;\;\;
\times \big[{\hat n}_{\sigma_2'}({\Bf r}_2'')
-n_{\sigma_2'}({\Bf r}_2'')\big]
\vert\Psi_{N;0}\rangle.
\end{eqnarray}
It is easily verified that the integrals with respect to ${\Bf r}_1''$ 
and ${\Bf r}_2''$ of $\sum_{\sigma_1',\sigma_2'} \langle\Psi_{N;0}
\vert\dots \vert\Psi_{N;0}\rangle$ on the RHS of Eq.~(\ref{ef109}) are 
identically vanishing (compare with functions of similar characteristic 
property encountered in dealing with ${\cal A}'({\Bf r},{\Bf r}')$, 
${\cal B}'_{\sigma}({\Bf r},{\Bf r}')$ and ${\cal G}'_{\sigma}
({\Bf r},{\Bf r}')$).

For uniform and isotropic GSs, ${\cal K}'_{\sigma}({\Bf r},{\Bf r}')$ 
is a function of $\|{\Bf r}-{\Bf r}'\|$ and for the reason that the 
integrals with respect to ${\Bf r}_1''$ and ${\Bf r}_2''$ of 
$\sum_{\sigma_1',\sigma_2'} \langle\Psi_{N;0}\vert \dots
\vert\Psi_{N;0}\rangle$ on the RHS of Eq.~(\ref{ef109}) are vanishing, 
it decays {\sl more rapidly} than $1/\|{\Bf r}-{\Bf r}'\|$ for 
$\|{\Bf r}-{\Bf r}'\|\to\infty$. For non-uniform systems, 
${\cal K}'_{\sigma}({\Bf r},{\Bf r}')$ decreases {\sl more rapidly} 
than $1/\|{\Bf r}\|$ and $1/\|{\Bf r}'\|$ for $\|{\Bf r}\|, 
\|{\Bf r}'\| \to \infty$. In arriving at these conclusions we have 
relied upon the multipole series expansion in Eq.~(\ref{ef20}) above. 

Similar to the other functions that we have dealt with thus far in 
this Appendix, ${\cal K}'_{\sigma}({\Bf r},{\Bf r}')$ according to 
its defining expression in Eq.~(\ref{ef109}) is {\sl not} amenable to 
direct numerical calculation. In contrast, the following expression, 
which one obtains through employing Eq.~(\ref{ef108}) as the defining 
expression for ${\cal K}'_{\sigma}({\Bf r},{\Bf r}')$ followed by 
judiciously rearranging terms, is readily verified to be suitable 
for direct numerical evaluation:
\begin{eqnarray}
\label{ef110}
&&{\cal K}'_{\sigma}({\Bf r},{\Bf r}') =
v^2({\Bf r}-{\Bf r}') \varrho_{\sigma}({\Bf r}',{\Bf r})
\nonumber\\
&&\;\;\;
-v({\Bf r}-{\Bf r}') \big[
{\cal B}''_{\sigma}({\Bf r},{\Bf r}') + 
{\cal B}''_{\sigma}({\Bf r}',{\Bf r})\big]\nonumber\\
&&\;\;\;
+\int {\rm d}^dr_1''\; v({\Bf r}-{\Bf r}_1'') \Big(
v({\Bf r}'-{\Bf r}_1'') \nonumber\\
&&\;\;\;\;\;\;\;\;\;\;\;\;\;\;\;\;
\times \sum_{\sigma_1'} \Gamma^{(2)}({\Bf r}'\sigma,
{\Bf r}_1''\sigma_1';{\Bf r}\sigma,{\Bf r}_1''\sigma_1')
\nonumber\\
&&\;\;\;\;\;\;\;
+\int {\rm d}^dr_2''\; v({\Bf r}'-{\Bf r}_2'') 
\sum_{\sigma_1',\sigma_2'}
\nonumber\\
&&\;\;\;\;\;\;\;\;\;\;\;
\times\Big[ \Gamma^{(3)}({\Bf r}'\sigma,{\Bf r}_1''\sigma_1',
{\Bf r}_2''\sigma_2';{\Bf r}\sigma,{\Bf r}_1''\sigma_1',
{\Bf r}_2''\sigma_2')\nonumber\\
&&\;\;\;\;\;\;\;\;\;\;\;\;\;\;\;\;\;
- n_{\sigma_1'}({\Bf r}_1'')  
\Gamma^{(2)}({\Bf r}\sigma,{\Bf r}_2''\sigma_2'';
{\Bf r}'\sigma,{\Bf r}_2''\sigma_2'') \nonumber\\
&&\;\;\;\;\;\;\;\;\;\;\;\;\;\;\;\;\;
-n_{\sigma_2'}({\Bf r}_2'') 
\Gamma^{(2)}({\Bf r}'\sigma,{\Bf r}_1''\sigma_1'';
{\Bf r}\sigma,{\Bf r}_1''\sigma_1'') \nonumber\\
&&\;\;\;\;\;\;\;\;\;\;\;\;\;\;\;\;\;
+n_{\sigma_1'}({\Bf r}_1'')
n_{\sigma_2'}({\Bf r}_2'') 
\varrho_{\sigma}({\Bf r}',{\Bf r})\Big] \Big).
\end{eqnarray} 
It can be verified that the ${\Bf r}_1''$ and ${\Bf r}_2''$ integrals 
on the RHS of Eq.~(\ref{ef110}) are bounded for $d=3$ and $v\equiv v_c$. 
This aspect crucially depends on the specific way in which various 
terms are combined in Eq.~(\ref{ef110}), signified by means of large
parentheses and square brackets. It is therefore crucial that in particular 
in the case of $d=3$ and $v\equiv v_c$, the structure of the expression 
in Eq.~(\ref{ef110}) is adhered to as closely as possible.

We introduce
\begin{equation}
\label{ef111}
{\cal K}''_{\sigma}({\Bf r},{\Bf r}')
{:=} {\cal K}'_{\sigma}({\Bf r},{\Bf r}')
-v^2({\Bf r}-{\Bf r}') 
\varrho_{\sigma}({\Bf r}',{\Bf r}). 
\end{equation}
It is further useful to define
\begin{equation}
\label{ef112}
{\cal K}'''_{\sigma}({\Bf r},{\Bf r}')
{:=} {\cal K}''_{\sigma}({\Bf r},{\Bf r}')
+ v({\Bf r}-{\Bf r}') \big[
{\cal B}''_{\sigma}({\Bf r},{\Bf r}') + 
{\cal B}''_{\sigma}({\Bf r}',{\Bf r})\big].
\end{equation}
This enables us to take full account of a contribution to 
$\wt{\Sigma}_{\sigma;\infty_2}({\Bf r},{\Bf r}'\vert z)$, namely 
$\Sigma_{\sigma;\infty_2}^{\rm s_b}({\Bf r},{\Bf r}')$ (see 
Eqs.~(\ref{e210}) and (\ref{e212})), that is bounded almost 
everywhere; however, it is {\sl not} integrable (see criteria 
(A)-(C) in \S~II.B).

Finally, within the framework of the SSDA (see Appendix C) we have
\begin{eqnarray}
\label{ef113}
&&\left. {\cal K}'_{\sigma}({\Bf r},{\Bf r}')\right|_{\rm s} = 
v^2({\Bf r}-{\Bf r}') \varrho_{{\rm s};\sigma}({\Bf r}',{\Bf r})
\nonumber\\
&&\;\;\;
-v({\Bf r}-{\Bf r}') \big[
{\cal B}''_{{\rm s};\sigma}({\Bf r},{\Bf r}') + 
{\cal B}''_{{\rm s};\sigma}({\Bf r}',{\Bf r})\big]\nonumber\\
&&\;\;\;
+\int {\rm d}^dr_1''\; v({\Bf r}-{\Bf r}_1'')
\Big( v({\Bf r}'-{\Bf r}_1'') \sum_{\sigma'} 
n_{\sigma'}({\Bf r}_1'')\nonumber\\
&&\;\;\;\;\;\;\;
-\int {\rm d}^dr_2''\; v({\Bf r}'-{\Bf r}_2'')
\sum_{\sigma'} \varrho_{{\rm s};\sigma'}^2
({\Bf r}_1'',{\Bf r}_2'')\Big) \varrho_{{\rm s};\sigma}
({\Bf r}',{\Bf r})\nonumber\\
&&\;\;\;
-\int {\rm d}^dr''\; v({\Bf r}-{\Bf r}'')
v({\Bf r}'-{\Bf r}'') 
\varrho_{{\rm s};\sigma}({\Bf r}',{\Bf r}'')
\varrho_{{\rm s};\sigma}({\Bf r}'',{\Bf r})\nonumber\\
&&\;\;\;
+2\int {\rm d}^dr_1'' {\rm d}^dr_2''\;
v({\Bf r}-{\Bf r}_1'') v({\Bf r}'-{\Bf r}_2'')
\varrho_{{\rm s};\sigma}({\Bf r}',{\Bf r}_1'')\nonumber\\
&&\;\;\;
\;\;\;\;\;\;\;\;\;\;\;\;\;\;\;\;\;\;\;\;\;\;\;\;\;\;\;\;\;\;
\times \varrho_{{\rm s};\sigma}({\Bf r}_1'',{\Bf r}_2'')
\varrho_{{\rm s};\sigma}({\Bf r}_2'',{\Bf r}).
\end{eqnarray}

\subsection{${\cal L}({\Bf r})$ and its regularized form}
\label{s59}

\subsubsection{Basic considerations; ${\cal L}'({\Bf r})$,
${\cal L}''({\Bf r})$, ${\cal M}({\Bf r})$ and
$\wt{\sf M}({\Bf r};z)$}
\label{s60}

In the expression for $\Sigma_{\sigma;\infty_2}({\Bf r},{\Bf r}')$ 
in Eq.~(\ref{e199}) we encounter 
\begin{eqnarray}
\label{ef114}
&&{\cal L}({\Bf r}) {:=}
\int {\rm d}^dr_1'' {\rm d}^dr_2'' {\rm d}^dr_3''\;
v({\Bf r}-{\Bf r}_1'') v({\Bf r}-{\Bf r}_2'')
v({\Bf r}-{\Bf r}_3'')\nonumber\\
&&\;\;\;
\times
\sum_{\sigma_1',\sigma_2',\sigma_3'}
\Gamma^{(3)}({\Bf r}_1''\sigma_1',
{\Bf r}_2''\sigma_2',{\Bf r}_3''\sigma_3';
{\Bf r}_1''\sigma_1',{\Bf r}_2''\sigma_2',
{\Bf r}_3''\sigma_3')\nonumber\\
\end{eqnarray}
which, making use of the definition for $\Gamma^{(3)}$ in 
Eq.~(\ref{eb1}) and the anticommutation relations in Eq.~(\ref{e29}),
can be put into the following form:
\begin{eqnarray}
\label{ef115}
&&{\cal L}({\Bf r})
= 2 \int {\rm d}^dr''\; 
v^3({\Bf r}-{\Bf r}'') n({\Bf r}'') \nonumber\\
&&\;\;\;
-3 v_H({\Bf r};[n]) \int {\rm d}^dr''\;
v^2({\Bf r}-{\Bf r}'') n({\Bf r}'') \nonumber\\
&&\;\;\;
-3 \int {\rm d}^dr_1'' {\rm d}^dr_2''\;
v^2({\Bf r}-{\Bf r}_1'') v({\Bf r}-{\Bf r}_2'')\,
{\cal U}({\Bf r}_1'',{\Bf r}_2'')\nonumber\\
&&\;\;\;
+3 v_H({\Bf r};[n]) {\cal A}'({\Bf r},{\Bf r})
+ v_H^3({\Bf r};[n]) + {\cal L'}({\Bf r}),
\end{eqnarray}
where ${\cal U}({\Bf r}_1'',{\Bf r}_2'')$ and ${\cal A}'({\Bf r},
{\Bf r})$ are defined in Eqs.~(\ref{ef4}) and (\ref{ef3}) 
respectively, and
\begin{eqnarray}
\label{ef116}
&&{\cal L'}({\Bf r}) {:=}
\int {\rm d}^dr_1'' {\rm d}^dr_2'' {\rm d}^dr_3''\;
v({\Bf r}-{\Bf r}_1'') v({\Bf r}-{\Bf r}_2'')
v({\Bf r}-{\Bf r}_3'') \nonumber\\
&&\;\;\;\;\;\;\;\;\;
\times \sum_{\sigma_1',\sigma_2',\sigma_3'}
\langle\Psi_{N;0}\vert 
\big[ {\hat n}_{\sigma_1'}({\Bf r}_1'')
-n_{\sigma_1'}({\Bf r}_1'')\big]\nonumber\\
&&\;\;\;\;\;\;\;\;\;
\;\;\;\;\;\;\;\;\;\;\;\;\;\;\;\;\;\;\;\;\;
\times \big[ {\hat n}_{\sigma_2'}({\Bf r}_2'')
-n_{\sigma_2'}({\Bf r}_2'')\big]\nonumber\\
&&\;\;\;\;\;\;\;\;\;
\;\;\;\;\;\;\;\;\;\;\;\;\;\;\;\;\;\;\;\;\;
\times \big[ {\hat n}_{\sigma_3'}({\Bf r}_3'')
-n_{\sigma_3'}({\Bf r}_3'')\big]
\vert\Psi_{N;0}\rangle.
\end{eqnarray}
The similarity of this to other (singly-primed) functions, such as 
${\cal A}'({\Bf r},{\Bf r}')$, ${\cal B}'_{\sigma}({\Bf r},{\Bf r}')$ 
in Eqs.~(\ref{ef3}) and (\ref{ef96}) respectively, is suggestive of 
${\cal L}'({\Bf r})$ being bounded for $v\equiv v_c$ in $d=3$. For 
the reasons that we shall present shortly, this is {\sl not} the case 
however. For now, we point out that ${\cal L'}({\Bf r})$ similar to 
${\cal A}'({\Bf r},{\Bf r}')$ and ${\cal B}_{\sigma}'({\Bf r},{\Bf r}')$ 
does {\sl not} suffer from the consequences of the long range of the 
Coulomb potential; it rather suffers from the singular behaviour of 
$v_c({\Bf r}-{\Bf r}')$ for $\|{\Bf r}-{\Bf r}'\|\to 0$ (this is
similar to the situation with regard to ${\cal A}'({\Bf r},{\Bf r}')$
in $d=2$ (B. Farid, 2001, unpublished)). It is not difficult to demonstrate 
that this behaviour of $v_c({\Bf r}-{\Bf r}')$ is of considerable 
consequence not only to $\Sigma_{\sigma;\infty_2}({\Bf r},{\Bf r}')$ 
but also to $\Sigma_{\sigma;\infty_m}({\Bf r},{\Bf r}')$ for {\sl all} 
$m\ge 2$ (here $m=2$ is specific to $d=3$).

Before discussing the behaviour of ${\cal L'}({\Bf r})$ for the 
specific case corresponding to $v\equiv v_c$ in $d=3$, on employing 
Eq.~(\ref{ef115}) as the defining expression for ${\cal L'}({\Bf r})$ 
we obtain the following expression which is suitable for direct 
calculation in terms of $\Gamma^{(m)}$:
\begin{equation}
\label{ef117}
{\cal L}'({\Bf r}) \equiv {\cal M}({\Bf r}) + {\cal L''}({\Bf r}),
\end{equation}
where (see Eq.~(\ref{e214}))
\begin{equation}
\label{ef118}
{\cal M}({\Bf r}) {:=} 
\int {\rm d}^dr_1''\; v^2({\Bf r}-{\Bf r}_1'') 
\Big\{ v({\Bf r}-{\Bf r}_1'') n({\Bf r}_1'')
+\Lambda({\Bf r}_1'',{\Bf r})\Big\},
\end{equation}
in which
\begin{equation}
\label{ef119}
\Lambda({\Bf r}_1'',{\Bf r}) {:=} -\int 
{\rm d}^dr_2''\; v({\Bf r}-{\Bf r}_2'')\,
\rho({\Bf r}_2'',{\Bf r}_1''),
\end{equation}
with $\rho({\Bf r}_2'',{\Bf r}_1'')$ as defined in Eq.~(\ref{ef13}),
and
\begin{eqnarray}
\label{ef120}
&&{\cal L''}({\Bf r}) {:=} 
-2\int {\rm d}^dr_1'' {\rm d}^dr_2''\;
v({\Bf r}-{\Bf r}_1'') v({\Bf r}-{\Bf r}_2'')\nonumber\\ 
&&\;
\times \Big\{ v({\Bf r}-{\Bf r}_1'')\,
\rho({\Bf r}_2'',{\Bf r}_1'') \nonumber\\
&&\;\;\;\;\;
-\frac{1}{2} \int {\rm d}^dr_3''\; v({\Bf r}-{\Bf r}_3'') \,
\Big[ 3 \rho({\Bf r}_2'',{\Bf r}_1'')\, n({\Bf r}_3'')
\nonumber\\
&&\;\;\;\;\;\;\,
+ \sum_{\sigma_1',\sigma_2',\sigma_3'} \Big(
\Gamma^{(3)}({\Bf r}_1''\sigma_1',{\Bf r}_2''\sigma_2',
{\Bf r}_3''\sigma_3';{\Bf r}_1''\sigma_1',
{\Bf r}_2''\sigma_2',{\Bf r}_3''\sigma_3')\nonumber\\
&&\;\;\;\;\;\;\;\;\;\;\;\;\;\;\;\;\;\;\;\;\;\;\;\;\;\;\;
\;\;\;\;\;\;\;\;\;\;\;\;\;\;
-n_{\sigma_1'}({\Bf r}_1'') n_{\sigma_2'}({\Bf r}_2'')
n_{\sigma_3'}({\Bf r}_3'')\Big) \Big]\Big\}.
\nonumber\\
\end{eqnarray}
For uniform GSs, ${\cal L}({\Bf r})$ and its constituent parts
${\cal L}'({\Bf r})$, ${\cal M}({\Bf r})$ and ${\cal L}''({\Bf r})$ 
do {\sl not} depend on ${\Bf r}$. This can be readily verified by 
means of shift transformations of the integration variables in the 
defining expressions of these functions. 

Within the framework of the SSDA (see Appendix C), we have ({\it cf}. 
Eq.~(\ref{ef117}))
\begin{equation}
\label{ef121}
{\cal L}_{\rm s}'({\Bf r}) \equiv
{\cal M}_{\rm s}({\Bf r}) +
{\cal L}_{\rm s}''({\Bf r}),
\end{equation}
where
\begin{eqnarray}
\label{ef122}
{\cal M}_{\rm s}({\Bf r}) &\equiv&
\int {\rm d}^dr_1''\; v^2({\Bf r}-{\Bf r}_1'')
\Big\{ v({\Bf r}-{\Bf r}_1'') n_{\rm s}({\Bf r}_1'') \nonumber\\
& &\;\;\;\;\;\;\;\;\;\;\;\;\;\;\;\;\;\;\;\;\;\;\;\;\;\;\;\;\;\;
\;\;\;\;\;\;\;\;\;\;\;
+\Lambda_{\rm s}({\Bf r}_1'',{\Bf r})\Big\},
\end{eqnarray}
in which ({\it cf}. Eqs.~(\ref{ef119}) and (\ref{ef120}))
\begin{equation}
\label{ef123}
\Lambda_{\rm s}({\Bf r}_1'',{\Bf r}) \equiv
-\int {\rm d}^dr_2''\;
v({\Bf r}-{\Bf r}_2'')\, \rho_{\rm s}({\Bf r}_2'',{\Bf r}_1''),
\end{equation}
and
\begin{eqnarray}
\label{ef124}
&&{\cal L}_{\rm s}''({\Bf r}) \equiv
-2\int {\rm d}^dr_1'' {\rm d}^dr_2''\;
v({\Bf r}-{\Bf r}_1'') v({\Bf r}-{\Bf r}_2'')\nonumber\\
&&\;\;\;\;\;\;\;\;\;\;
\times \Big\{ v({\Bf r}-{\Bf r}_1'') 
\rho_{\rm s}({\Bf r}_1'',{\Bf r}_2'') \nonumber\\
&&\;\;\;\;\;\;\;\;\;\;\;\;\;\;\;
-\int {\rm d}^dr_3''\; v({\Bf r}-{\Bf r}_3'')\,
\sum_{\sigma'} \varrho_{\rm s;\sigma'}({\Bf r}_1'',{\Bf r}_2'') 
\nonumber\\
&&\;\;\;\;\;\;\;\;\;\;\;\;\;\;\;\;\;\;\;\;\;\;\;\;\;\;\;\;\;\;
\times \varrho_{{\rm s};\sigma'}({\Bf r}_2'',{\Bf r}_3'') 
\varrho_{{\rm s};\sigma'}({\Bf r}_3'',{\Bf r}_1'') \Big\},
\end{eqnarray}
where $\rho_{\rm s}({\Bf r}_1'',{\Bf r}_2'') \equiv \sum_{\sigma'} 
\varrho_{{\rm s};\sigma'}^2({\Bf r}_1'',{\Bf r}_2'')$ (see
Eq.~(\ref{ef13}) above and the subsequent text).

It can be shown that ${\cal L}''({\Bf r})$ is a well-defined and 
bounded function of ${\Bf r}$, a fact that can be readily verified 
through examining ${\cal L}_{\rm s}''({\Bf r})$, taking into account 
the simplifying aspect associated with the idempotency of 
$\varrho_{{\rm s};\sigma}$.

The analysis of the exact ${\cal M}({\Bf r})$ is facilitated by first 
considering ${\cal M}_{\rm s}({\Bf r})$ in Eq.~(\ref{ef122}). To this 
end we first point out that in the case of $v\equiv v_c$ in $d=3$, 
\footnote{\label{f134}
The result in Eq.~(\protect\ref{ef125}) is easiest deduced by considering 
the fact that, since $\sum_{\sigma'}\varrho_{{\rm s};\sigma'}^2
({\Bf r}_2'',{\Bf r}_1'') \equiv \rho_{\rm s}({\Bf r}_2'',{\Bf r}_1'')$ 
is a strongly decaying function for $\|{\Bf r}_1''-{\Bf r}_2''\|\to
\infty$ (see, for example, Eq.~(\protect\ref{ef52})), from which it 
follows that for $\|{\Bf r}_1''\| \to\infty$ the most significant 
contribution to the integral on the RHS of Eq.~(\protect\ref{ef123}) 
originates from the ${\Bf r}_2''$ integration in a close neighbourhood 
of ${\Bf r}_1''$. Consequently, replacing $v_c({\Bf r}-{\Bf r}_2'')$ 
by $v_c({\Bf r}-{\Bf r}_1'')$, owing to the idempotency of 
$\varrho_{{\rm s};\sigma'}({\Bf r}_1'',{\Bf r}_2'')$ the subsequent 
integration with respect to ${\Bf r}_2''$ of $\sum_{\sigma'} 
\varrho_{{\rm s};\sigma'}^2({\Bf r}_1'',{\Bf r}_2'')$ yields 
$\sum_{\sigma'} n_{{\rm s};\sigma'}({\Bf r}_1'') \equiv n_{\rm s}
({\Bf r}_1'')$. Thus one arrives at the RHS of Eq.~(\protect\ref{ef125}). 
From Eqs.~(\protect\ref{ef14}) and (\protect\ref{ef119}), making use 
of the multipole expansion in Eq.~(\protect\ref{ef20}), one readily 
deduces that similarly one has $\Lambda({\Bf r}_1'',{\Bf r}) \sim 
-v_c({\Bf r}-{\Bf r}_1'')\, n({\Bf r}_1'')$ for $\|{\Bf r}_1''\|
\to\infty$. }
\begin{equation}
\label{ef125}
\Lambda_{\rm s}({\Bf r}_1'',{\Bf r}) 
\sim -v_c({\Bf r}-{\Bf r}_1'')\, n_{\rm s}({\Bf r}_1''),
\;\; \|{\Bf r}_1''\|\to\infty, 
\end{equation}
so that for $\|{\Bf r}_1''\| \to \infty$ the integrand of the
${\Bf r}_1''$ integral on the RHS of Eq.~(\ref{ef118}) decays 
{\sl more rapidly} than $1/\|{\Bf r}_1''\|^3$. Consequently, the 
${\Bf r}_1''$ integral is {\sl not} infrared divergent. If there
was no $v_c({\Bf r}-{\Bf r}_1'') n_{\rm s}({\Bf r}_1'')$ on 
the RHS of Eq.~(\ref{ef125}), the integrand of the ${\Bf r}_1''$ 
integral on the RHS of Eq.~(\ref{ef122}) would have decayed like 
$1/\|{\Bf r}_1''\|^3$ and consequently this integral would have 
been infrared divergent. Thus a contribution of the form 
$\int {\rm d}^3r_1''\; v_c^3({\Bf r}-{\Bf r}_1'')\, n_{\rm s}
({\Bf r}_1'')$ to ${\cal M}_{\rm s}({\Bf r})$ seems necessary. On 
the other hand, it is evident that the singularity of $v_c^3({\Bf r}
-{\Bf r}_1'')$ at ${\Bf r}_1''={\Bf r}$ is {\sl not} integrable in 
$d=3$. Given the fact that $\Lambda_{\rm s}({\Bf r}_1'',{\Bf r})$ 
is bounded for ${\Bf r}_1''={\Bf r}$ and that the singularity of 
$v_c^2({\Bf r}-{\Bf r}_1'')$ at ${\Bf r}_1''={\Bf r}$ {\sl is} 
integrable, we observe that, in the case where $v\equiv v_c$ and 
$d=3$, the ${\Bf r}_1''$ integral on the RHS of Eq.~(\ref{ef122}) 
is {\sl unbounded}, that is it is so-called ultraviolet divergent. An 
analysis along the same line (see footnote \ref{f134}) demonstrates 
the fact that the exact ${\cal M}({\Bf r})$ is similar to 
${\cal M}_{\rm s}({\Bf r})$ unbounded for $v\equiv v_c$ and $d=3$. 
Thus, whereas, in common with other `primed' functions that we have 
considered in this Appendix, ${\cal L}'({\Bf r})$ through ${\cal M}
({\Bf r})$ (see Eq.~(\ref{ef117})) is infrared regular (in the case 
$v\equiv v_c$ and $d=3$), it nevertheless is {\sl unbounded} owing 
to the ultraviolet-divergent integral $\int {\rm d}^3r_1''\; 
v_c^3({\Bf r}-{\Bf r}_1'') n({\Bf r}_1'')$. 

In the light of our above discussion, from the considerations in 
\S~II.B it follows that $\hbar^{-1} {\cal M}({\Bf r})\, \delta({\Bf r}
-{\Bf r}')$ must have local counterparts pertaining to $\Sigma_{\sigma;
\infty_m}({\Bf r},{\Bf r}')$ with $m > 2$, described by 
ultraviolet-divergent integrals. Indeed, it can be shown that 
$\Sigma_{\sigma;\infty_m}({\Bf r},{\Bf r}')$ involves the constituent 
local term $\hbar^{-1} \int {\rm d}^dr_1''\; v^m({\Bf r}-{\Bf r}_1'')\, 
n({\Bf r}_1'')$ for $m \ge 2$. In fact, with reference to the 
expressions for $\Sigma_{\sigma;\infty_0}({\Bf r},{\Bf r}')$ and 
$\Sigma_{\sigma;\infty_1}({\Bf r},{\Bf r}')$ in Eqs.~(\ref{e174}) and 
(\ref{e185}) respectively, it is readily verified that indeed 
\begin{eqnarray}
\frac{1}{\hbar} 
&&v_H({\Bf r};[n])\, \delta({\Bf r}-{\Bf r}') \nonumber\\
&&\;\;\;\;\;\;\;\;\;\;\;\;\;\;\;\;
\equiv \frac{1}{\hbar} 
\int {\rm d}^dr_1''\, v({\Bf r}-{\Bf r}_1'')\, n({\Bf r}_1'') 
\delta({\Bf r}-{\Bf r}') \nonumber
\end{eqnarray}
is a component part of the former function and that
\begin{eqnarray}
\frac{1}{\hbar} \int {\rm d}^dr_1''\, v^2({\Bf r} -{\Bf r}_1'')\, 
n({\Bf r}_1'')\, \delta({\Bf r}-{\Bf r}')\nonumber
\end{eqnarray}
is a component part of the latter function. We thus arrive at the 
conclusion that (see text following Eq.~(\ref{e111}))
\begin{eqnarray}
\label{ef126}
&&\wt{\sf M}({\Bf r};z) {:=}
\frac{1}{z^2} \int {\rm d}^dr''\;
v^2({\Bf r}-{\Bf r}'') \nonumber\\
& &\;\;\;\;\;\;\;\;\;\;\;\;\;\;\;\;\;\;
\times \Big(\frac{v({\Bf r}-{\Bf r}'') n({\Bf r}'')}
{1 - v({\Bf r}-{\Bf r}'')/z}+\Lambda({\Bf r}'',{\Bf r})\Big)
\end{eqnarray}
is the appropriate function associated with the regularization of 
${\cal M}({\Bf r})$. For $v\equiv v_c$ in $d=3$, it is seen that, 
since for $\|{\Bf r}-{\Bf r}''\|\to\infty$ (which for finite values 
of $\|{\Bf r}\|$ implies the condition $\|{\Bf r}''\| \to\infty$) the 
first function enclosed by the large parentheses on the RHS of 
Eq.~(\ref{ef126}) approaches $v_c({\Bf r}-{\Bf r}'')\, n({\Bf r}'')$, 
$\wt{\sf M}({\Bf r};z)$ similar to ${\cal M}({\Bf r})$ is free from 
an infrared divergence; on the other hand, as it is evident, in 
contrast with ${\cal M}({\Bf r})$, $\wt{\sf M}({\Bf r};z)$ is {\sl not} 
ultraviolet divergent. In the next Section (\S~F.5.b) we explicitly 
consider the case corresponding to $d=3$ and $v\equiv v_c$ and calculate 
the leading term in the large-$\vert z\vert$ AS for $\wt{\sf M}
({\Bf r};z)$, that is $\wt{\sf M}_{\infty_2}({\Bf r} \vert z)/z^2$, 
with $\wt{\sf M}_{\infty_2}({\Bf r}\vert z) \equiv 
{\sf M}^{\rm r}_{\infty_2}({\Bf r}) + \wt{\sf M}_{\infty_2}^{\rm s}
({\Bf r}\| z)$ (for our notational conventions see \S~III.E.2; see
also footnote \ref{f31}); here ${\sf M}_{\infty_2}^{\rm r}({\Bf r})$ 
is a regular function, independent of $z$, and 
$\wt{\sf M}_{\infty_2}^{\rm s}({\Bf r}\| z)$ in contrast diverges 
logarithmically as $\vert z\vert\to\infty$ (see Eqs.~(\ref{ef144}) 
and (\ref{ef145}) below). In \S~II.B we have in detail discussed the 
principles underlying the appearance of transcendental functions of $z$ 
in the regularized coefficients corresponding to the large-$\vert z\vert$ 
AS for $\wt{\Sigma}_{\sigma}({\Bf r},{\Bf r}';z)$ and therefore will 
not elaborate on these here.

\subsubsection{The large-$\vert z\vert$ AS for
$\wt{\sf M}({\Bf r};z)$; $\wt{\sf M}_{\infty_2}({\Bf r}\vert z)$,
${\sf M}_{\infty_2}^{\rm r}({\Bf r})$ and 
$\wt{\sf M}_{\infty_2}^{\rm s}({\Bf r}\| z)$}
\label{s61}

Here we consider the case where $d=3$ and $v\equiv v_c$. By shifting 
the origin of integration and some evident algebraic manipulations, 
from Eq.~(\ref{ef126}) we deduce
\begin{eqnarray}
\label{ef127}
&&\wt{\sf M}({\Bf r};z)
\equiv \frac{1}{z^2} \Big[
\int_{\Omega_R} {\rm d}^3r''\; v_c^2({\Bf r}'')
\Lambda({\Bf r}-{\Bf r}'',{\Bf r})\nonumber\\
&&\;\;
+ \int_{\Omega\setminus \Omega_R} {\rm d}^3r''\;
v_c^2({\Bf r}'')\, 
\Big\{ v_c({\Bf r}'') n({\Bf r}-{\Bf r}'')
+\Lambda({\Bf r}-{\Bf r}'',{\Bf r})\Big\}\nonumber\\
&&\;\;
+\int_{\Omega_R} {\rm d}^3r''\;
\frac{v_c^3({\Bf r}'') n({\Bf r}-{\Bf r}'')}
{1 -v_c({\Bf r}'')/z} \nonumber\\
&&\;\;
+ \int_{\Omega\setminus \Omega_R} {\rm d}^3r''\;
v_c^3({\Bf r}'') 
\Big\{ \frac{1}{1- v_c({\Bf r}'')/z} - 1\Big\}\,
n({\Bf r}-{\Bf r}'') \Big],\nonumber\\
\end{eqnarray}
where $\Omega$ stands for the (macroscopic) volume of the system 
and $\Omega_R$ a spherical volume with radius $R$ centred at the 
origin; $\int_{\Omega_R}$ stands for spatial integration {\sl inside} 
$\Omega_R$ and $\int_{\Omega\setminus\Omega_R}$ for that {\sl outside} 
$\Omega_R$ but inside $\Omega$. In the following we assume that
\begin{equation}
\label{ef128}
R \gg \frac{e^2}{4\pi\epsilon_0 \vert z\vert}.
\end{equation}
We should emphasize that, for our purpose, {\sl any} $R > 
e^2/(4\pi\epsilon_0 \vert z\vert)$ would suffice; however, the 
condition in Eq.~(\ref{ef128}) has the advantage of reducing the 
complexity of the calculations to be encountered below (see text 
following Eq.~(\ref{ef138}) below).

We proceed by considering the third contribution to the RHS of 
Eq.~(\ref{ef127}), namely
\begin{eqnarray}
\label{ef129}
\wt{\sf S}({\Bf r},R;z)
{:=} \frac{1}{z^2} \int_{\Omega_R} {\rm d}^3r''\;
\frac{v_c^3({\Bf r}'') n({\Bf r}-{\Bf r}'')}
{1-v_c({\Bf r}'')/z}.
\end{eqnarray}
We assume the system under consideration to be in the thermodynamic 
limit and consequently adopt the decomposition of the total number 
density $n({\Bf r})$ in terms of $n_0$ and $n'({\Bf r})$ as presented 
in Eq.~(\ref{e12}). In order to gain insight into the behaviour to 
be expected from $\wt{\sf S}({\Bf r},R;z)$, we first deal with 
the case where $n\equiv n_0$. In this case, the integrand in the 
expression on the RHS of Eq.~(\ref{ef129}) is fully isotropic so that, 
by employing the spherical polar coordinate system, we readily obtain 
({\it cf}. Eq.~(\ref{ef140}) below)
\begin{eqnarray}
\label{ef130}
&&\left.\wt{\sf S}({\Bf r},R;z)\right|_{n=n_0}
= \frac{4\pi}{z^2} 
\left(\frac{e^2}{4\pi\epsilon_0}\right)^3\, n_0\nonumber\\
&&\;\;\;\;\;\;\;\;\;\;\;\;\;\;\;\;\;\;\;\;\;\;
\times \Big\{ \ln\Big(R/{\sf a}_0 
- \frac{e^2}{4\pi\epsilon_0 {\sf a}_0 z}\Big)
-\ln\Big(\frac{-e^2}{4\pi\epsilon_0 {\sf a}_0 z}\Big)\Big\} 
\nonumber\\
&&\;\;\;\;
= 4\pi \left(\frac{e^2}{4\pi\epsilon_0}\right)^3\, n_0\,
\Big\{ \frac{\ln(-z/\varepsilon_0)}{z^2}
+ \frac{\ln(\varepsilon_0/e_R)}{z^2} \Big\}\nonumber\\
&&\;\;\;\;\;\;\;\;\;\;\;\;\;\;\;\;\;\;\;\;\;\;\;\;\;\;
\;\;\;\;\;\;\;\;\;\;\;\;\;\;\;\;\;\;\;\;\;\;\;\;\;\;\,
+ {\cal O}\big(\frac{1}{z^3}\big),
\end{eqnarray}
where $\varepsilon_0$ stands for a positive constant energy,
${\sf a}_0$ for a positive constant length and
\begin{equation}
\label{ef131}
e_R {:=} \frac{e^2}{4\pi\epsilon_0 R}.
\end{equation}
Our choice in Eq.~(\ref{ef128}) implies that $\vert z\vert/e_R
\gg 1$. In Eq.~(\ref{ef130}), $\ln$ stands for the principal 
branch of the logarithm function, that is ${\rm Im}[\ln(z)] 
\in (-\pi,\pi)$. 

Owing to the {\sl linearity} of $\wt{\sf S}({\Bf r},R;z)$ with respect
to $n({\Bf r})$ and with reference to the result in Eq.~(\ref{ef130}), 
it remains only to determine $\wt{\sf S}({\Bf r},R;z)\vert_{n=n'}$. To 
this end we employ the following Fourier integral representation
\begin{equation}
\label{ef132}
n'({\Bf r}) = \int \frac{{\rm d}^3 q}{(2\pi)^3}\,
{\bar n}'({\Bf q})\, \exp(i {\Bf q}\cdot {\Bf r})
\end{equation}
which in cases corresponding to periodic solids, can be made into a 
discrete Fourier series by considering ${\bar n}'({\Bf q})= (2\pi)^3 
\sum_{\small\Bf G} {\bar n}'_{\small\Bf G}\, \delta({\Bf q}-{\Bf G})$, 
where the sum is over the set of reciprocal-lattice vectors corresponding 
to the underlying lattice. We point out that since $n'({\Bf r})$ has 
by definition a zero average (see Eq.~(\ref{e12})), ${\bar n}'({\Bf q}) 
\to 0$ for $\|{\Bf q}\| \to 0$; in the case of periodic solids, 
${\bar n}'_{{\small\Bf G}={\Bf 0}}=0$. Further, the analytic property 
of $n'({\Bf r})$ in real space implies certain behaviour for 
${\bar n}'({\Bf q})$ as $\|{\Bf q}\|\to \infty$; as we shall discuss 
in some detail below (see also Appendix K), this behaviour is of 
important consequence to that of $\wt{\sf S}({\Bf r},R;z)\vert_{n=n'}$ 
for large $\vert z\vert$.

From Eqs.~(\ref{ef129}) and (\ref{ef132}), performing the ${\Bf r}''$
integral in terms of the spherical polar coordinates of ${\Bf r}''$,
with ${\Bf q}$ the polar axis, we readily obtain
\begin{eqnarray}
\label{ef133}
&&\left. \wt{\sf S}({\Bf r},R;z)\right|_{n=n'}
=\frac{4\pi}{z^2} \left(\frac{e^2}{4\pi\epsilon_0}\right)^3
\nonumber\\
&&\;\;\;\;\;\;\;
\times \int \frac{{\rm d}^3 q}{(2\pi)^3} \,
\frac{{\bar n}'({\Bf q})\, 
{\tilde g}(\|{\Bf q}\|,R;z)}{\|{\Bf q}\|} \,
\exp(i {\Bf q}\cdot {\Bf r}),
\end{eqnarray} 
where
\footnote{\label{f135}
The condition in Eq.~(\protect\ref{ef128}) is equivalent to
$R \gg \vert\zeta\vert$. }
\begin{equation}
\label{ef134}
{\tilde g}(\|{\Bf q}\|,R;z)
{:=} \int_0^{R} {\rm d}r\;
\frac{\sin(\|{\Bf q}\| r)}{r (r - \zeta)},\;\;\;
\zeta {:=} \frac{e^2}{4\pi\epsilon_0 z}.
\end{equation}
Following our discussions in \S~II.B, on general grounds (i.e.
without recourse to the expressions in Eqs.~(\ref{ef133}) and 
(\ref{ef134})) we expect the leading asymptotic contribution to 
$\wt{\sf S}({\Bf r},R;z) \vert_{n=n'}$ for $\vert z\vert\to \infty$ 
(i.e. $\zeta\to 0$) to be more dominant than $1/z^2$ but less so than 
$1/z$. This aspect is evident from the expression for ${\tilde g}
(\|{\Bf q}\|,R;z)$ in Eq.~(\ref{ef134}) where, for $\zeta\to 0$ or 
$\vert z\vert\to\infty$, the denominator of the integrand approaches 
$r^2$, which in combination with $\sin(\|{\Bf q}\| r)$ in the numerator 
gives rise to an integrand that diverges like $1/r$ for $r\to 0$, 
implying a logarithmically-divergent integral as $\zeta\to 0$. This 
behaviour is explicit in the expression for $\wt{\sf S}({\Bf r},R;
z)\vert_{n=n_0}$ in Eq.~(\ref{ef130}) above.

In order to obtain the expressions for the terms in the 
large-$\vert z\vert$ AS for ${\tilde g}(\|{\Bf q}\|,R;z)$, we first 
invoke the following decomposition:
\begin{equation}
\label{ef135}
\frac{1}{r (r - \zeta)} = -\frac{1}{\zeta}\,
\Big(\frac{1}{r} - \frac{1}{r - \zeta}\Big).
\end{equation}
In view of the expected behaviour of ${\tilde g}(\|{\Bf q}\|,R;z)$
for $\vert z\vert\to \infty$, the contribution of $1/r$ in this 
fractional expansion to ${\tilde g}(\|{\Bf q}\|,R;z)$ does {\sl not} 
need to be considered, since $(-1/\zeta) \int_0^{R} {\rm d}r\; 
\sin(\|{\Bf q}\| r)/r$ scales like $z$ and therefore must be 
cancelled by a counter contribution in the large-$\vert z\vert$ 
AS for ${\tilde g}(\|{\Bf q}\|,R;z)$ originating from $(-1/\zeta) 
(-1)/(r-\zeta)$. Indeed, in the formal series expansion for 
$-1/(r-\zeta) \equiv (-1/r)/(1-\zeta/r)$ around $\zeta/r=0$, the 
leading term exactly cancels the $1/r$ enclosed by large parentheses 
on the RHS of Eq.~(\ref{ef135}). 
\footnote{\label{f136}
Through Eq.~(\protect\ref{ef135}), the next-to-leading term in the 
above-mentioned expansion corresponds to a contribution of the form 
$1/r^2$ to $1/(r [r-\zeta])$, which, upon substitution in the RHS of 
Eq.~(\protect\ref{ef134}), gives rise to a divergent integral. This 
divergence signals the fact (see \S~II.B) that the leading term in the 
large-$\vert z\vert$ AS of ${\tilde g}(\|{\Bf q}\|,R;z)$ is {\sl not} 
a constant but a more dominant contribution. Our calculations 
indeed show (see Eqs.~(\protect\ref{ef130}) and (\protect\ref{ef140})) 
that this contribution scales like $\ln(-z/\varepsilon_0)$, where 
$\varepsilon_0$ stands for a positive constant energy. }
Thus, defining
\begin{equation}
\label{ef136}
{\tilde {\cal I}}(\zeta) {:=} \int_0^{R} {\rm d}r\;
\frac{\sin(\|{\Bf q}\| r)}{r-\zeta},
\end{equation}
from the closed expression for $\wt{\cal I}(\zeta)$ we shall be able 
to deduce the leading term in the AS for ${\tilde g}(\|{\Bf q}\|,R;z)$ 
corresponding to $\vert z\vert \to \infty$; we shall specify the 
underlying procedure for this shortly. Note the simplification achieved 
through replacing the need for the evaluation of the integral on the 
RHS of Eq.~(\ref{ef134}) by the need to evaluate the simpler integral 
on the RHS of Eq.~(\ref{ef136}). 

Following the considerations in Appendix I (use the substitution $\zeta' 
\rightharpoonup -\zeta$ in this Appendix), assuming ${\rm Re}(z) \times 
{\rm Im}(z) > 0$ (i.e. $z$ is located either in the first or in the 
third quadrant of the complex $z$ plane; for some relevant details 
see later), we obtain (below $q\equiv \|{\Bf q}\|$)
\begin{eqnarray}
\label{ef137}
{\tilde {\cal I}}(\zeta) &=& -\frac{1}{2} \sin(q \zeta)
\Big[ {\rm Ei}(i q\zeta) + {\rm Ei}(-i q\zeta)\Big]\nonumber\\
& &-\frac{i}{2} \cos(q \zeta) 
\Big[ {\rm Ei}(i q\zeta) - {\rm Ei}(-i q\zeta)\Big]\nonumber\\
& &+\pi \exp(-i q\zeta)\,\Theta\big({\rm Re}(z)\big)\nonumber\\
& &+{\rm Ci}\big(q (R-\zeta)\big) \sin(q\zeta)\nonumber\\
& &-\Big[ \frac{\pi}{2} - {\rm Si}\big(q (R-\zeta)\big) \Big]
\cos(q\zeta),
\end{eqnarray}
where ${\rm Ei}$ stands for the exponential-integral function (see 
Appendix I), ${\rm Si}$ for the sine-integral function and ${\rm Ci}$ 
for the cosine-integral function (Abramowitz and Stegun 1972, pp. 231 
and 232). For completeness, in arriving at the expression in 
Eq.~(\ref{ef137}), we have employed the decomposition $\int_0^{R} 
{\rm d}r\, (\dots) = \int_0^{\infty} {\rm d}r\, (\dots) 
-\int_{R}^{\infty} {\rm d}r\, (\dots)$, the contribution of the latter 
integral to ${\tilde {\cal I}}(\zeta)$ being the last two lines on 
the RHS of Eq.~(\ref{ef137}), that is those involving ${\rm Ci}$ and 
${\rm Si}$. Further, we point out that, as a consequence of the inequality 
in Eq.~(\ref{ef128}), evaluation of the integral with respect to $r$ of 
$\sin(\|{\Bf q}\| r)/(r-\zeta)$ over $[R,\infty)$ does {\sl not} require 
the sophistication of the treatment necessary for the evaluation of the 
integral over $[0,R]$ (or $[0,\infty)$; see Appendix I) since, in 
contrast with the latter case, in the former the integral is an analytic 
functions of $\zeta$ in a finite neighbourhood of $\zeta=0$; this is 
evident from the last two contributions on the RHS of Eq.~(\ref{ef137}) 
where we observe ${\rm Ci}\big(q (R-\zeta)\big)$ and ${\rm Si}\big(q
(R-\zeta)\big)$ (see footnote \ref{f135}).

Concerning ${\rm Re}(z) \times {\rm Im}(z) > 0$, we have invoked this 
condition for the reason that it enables one to deduce {\sl directly} 
the `physical' ${\tilde {\cal I}}(\zeta)$ for real energies, that 
is ${\cal I}(e^2/[4\pi \epsilon_0 \varepsilon])$; depending on whether 
$\varepsilon \to +\infty$ or $\varepsilon \to -\infty$, the `physical' 
${\tilde {\cal I}}(\zeta)$ is obtained through approaching the real 
$\varepsilon$ axis from the upper half-plane and the lower half-plane 
respectively of the complex $z$ plane ({\it cf}. Eqs.~(\ref{e25}), 
(\ref{e65}) and (\ref{e66})). We should, however, emphasize that 
knowledge of ${\tilde {\cal I}}(e^2/[4\pi\epsilon_0 z])$ in {\sl any} 
open region of the $z$ plane suffices for calculation of 
${\tilde {\cal I}}(e^2/[4\pi\epsilon_0 z])$ in any other open region 
of the complex $z$ plane through the application of the process of 
analytic continuation. 

Let now ${\sf C}_{\zeta;m}\big[{\tilde f}(\zeta)\big]$ denote the 
coefficient of $\zeta^m$ in the {\sl regularized} AS of ${\tilde f}
(\zeta)$, with respect to the asymptotic sequence $\{1,\zeta,\zeta^2, 
\dots\}$, for $\zeta \to 0$; by definition, since we have to do with 
{\sl regularized} AS, ${\sf C}_{\zeta;m}\big[{\tilde f}(\zeta)\big]$ 
can involve transcendental functions of $\zeta$ (see \S~II.B). Our above 
considerations have made evident that the leading asymptotic 
contribution to $\wt{\sf S}({\Bf r},R;z)\vert_{n=n'}$ for 
$\vert z\vert\to \infty$ is determined by 
${\sf C}_{\zeta;1}\big[\wt{\cal I}(\zeta)\big]$. Making use of the 
expression in Eq.~(\ref{ef137}) and the asymptotic expression in 
Eq.~(\ref{ei4}), we obtain
\begin{eqnarray}
\label{ef138}
&&{\sf C}_{\zeta;1}\big[{\tilde {\cal I}}(\zeta)\big]
= q \Big( (\gamma-1) + \frac{1}{2} 
\big[ \ln(i q\zeta) + \ln(-i q\zeta)\big]\nonumber\\
&&\;\;\;\;\;\;\;\;\;\;
+\pi i \Theta\big({\rm Re}(z)\big)
-{\rm Ci}(qR) + \frac{\sin(q R)}{q R}
 \Big)\nonumber\\
&&\;\;\;
=\|{\Bf q}\| \Big[ (\gamma -1) 
+\ln(\|{\Bf q}\| R)-\ln(-z/e_R)
\nonumber\\
&&\;\;\;\;\;\;\;\;\;\;\;\;\;\;\;
-{\rm Ci}(\|{\Bf q}\| R) + \frac{\sin(\|{\Bf q}\| R)}
{\|{\Bf q}\| R} \Big],
\end{eqnarray}
where $\gamma = 0.57721~566\dots$ stands for the Euler constant. In 
arriving at the last expression on the RHS of Eq.~(\ref{ef138}) we have 
made use of the fact that $\big[\ln(i q\zeta) + \ln(-i q\zeta)\big]/2 
= \ln\vert q\zeta\vert - i \varphi$, where $\varphi = {\rm arg}(z)$ 
when $0\le {\rm arg}(z) < \pi/2$ and $\varphi = \pi + {\rm arg}(z)$ 
when $-\pi < {\rm arg}(z) \le -\pi/2$, as well as $\ln\vert z\vert + 
i \big(\pi + {\rm arg}(z)\big) -2\pi i \Theta\big({\rm Re}(z)\big)\equiv 
\ln(-z)$ for $0 < {\rm arg}(z) < \pi/2$ and $-\pi < {\rm arg}(z) < 
-\pi/2$; we have in addition invoked the condition $R \gg 
\vert\zeta\vert$ (see Eq.~(\ref{ef128}) above and footnote \ref{f135}), 
permitting us to disregard the branch cuts of ${\rm Si}$ and ${\rm Ci}$, 
an aspect reflected in the following result ({\it cf}. Eq.~(\ref{ef138}))
\begin{eqnarray}
\label{ef139}
&&{\sf C}_{\zeta;1}\Big[
{\rm Ci}\big(q (R - \zeta)\big) \sin(q\zeta)
+\big\{\frac{\pi}{2} - {\rm Si}\big(q (R-\zeta)\big)\big\}
\cos(q\zeta) \Big]\nonumber\\
&&\;\;\;\;\;\;\;\;\;\;\;\;\;\;\;\;\;\;\;\;\;\;\;\;\;\;
\;\;\;\;\;\;\;\;\;\;\;
= {\rm Ci}(q R) q - \sin(q R)/R.
\end{eqnarray}  

Combining the above results, we eventually obtain ({\it cf}. 
Eq.~(\ref{ef130}))
\begin{eqnarray}
\label{ef140}
&&\left. \wt{\sf S}({\Bf r},R;z)\right|_{n=n'} 
\sim 4\pi \left(\frac{e^2}{4\pi\epsilon_0} \right)^3 
n'({\Bf r})\, \frac{\ln(-z/\varepsilon_0)}{z^2}\nonumber\\
&&+4\pi \left(\frac{e^2}{4\pi\epsilon_0} \right)^3 
\Big\{ (1-\gamma)\, n'({\Bf r}) 
+\int \frac{{\rm d}^3q}{(2\pi)^3}\; {\bar n}'({\Bf q})\,
\nonumber\\
&&\;\;\;\;\;\;
\times \Big[ {\rm Ci}(\|{\Bf q}\| R) - \ln(\|{\Bf q}\| R)
-\frac{\sin(\|{\Bf q}\| R)}{\|{\Bf q}\| R}\Big]
\,\exp(i {\Bf q}\cdot {\Bf r})\nonumber\\
&&\;\;\;\;\;\;\;\;\;\;\;\;\;\;\;\;\;\;\;\;\;\;\;\;\;\;\,
+ n'({\Bf r}) \ln\left(\frac{\varepsilon_0}{e_R}\right)
\Big\} \frac{1}{z^2},
\;\vert z\vert\to \infty.
\end{eqnarray}
In order to verify the correctness of this result, let us assume that 
${\bar n}'({\Bf q})$ in Eq.~(\ref{ef140}) were the Fourier transform 
of the {\sl total} $n({\Bf r})$ rather than that of $n'({\Bf r})$. 
Assuming $n({\Bf r}) \equiv n_0$, we would have ${\bar n}'({\Bf q}) 
= (2\pi)^3 n_0 \delta({\Bf q})$. Making use of $\lim_{z=0} \sin(z)/z 
= 1$ and ${\rm Ci}(z) \sim \gamma + \ln(z) - z^2/4 +\dots$ for 
$z\to 0$ (see Abramowitz and Stegun 1972, p.~232), the ${\Bf q}$ 
integral on the RHS of Eq.~(\ref{ef140}) would identically cancel 
$(1-\gamma) n'({\Bf r}) \equiv (1-\gamma) n_0$, thus correctly 
reducing the result in Eq.~(\ref{ef140}) into that in Eq.~(\ref{ef130}). 
From Eqs.~(\ref{ef130}) and (\ref{ef140}) we also observe that, whereas 
the full total number density $n({\Bf r})$ contributes to the 
`coefficient' of the $\ln(-z/e_R)/z^2$ term in the large-$\vert z\vert$ 
AS of $\wt{\sf S}({\Bf r},R;z)$, it is only the `fluctuating' part of 
$n({\Bf r})$, namely $n'({\Bf r})\equiv n({\Bf r})-n_0$ (see 
Eq.~(\ref{e12})), that contributes to the `coefficient' of the 
$1/z^2$ term in this series. 

In order to establish the correctness of the asymptotic result in 
Eq.~(\ref{ef140}), it remains to consider the existence of the 
${\Bf q}$ integral on the RHS of Eq.~(\ref{ef140}). This aspect is 
of utmost relevance since boundedness of this integral is not an 
{\sl a priori} necessity; with reference to our considerations 
in \S~II.B, it is evident that unboundedness of this integral
would signal the fact that, in a finite-order AS for $\wt{\sf S}
({\Bf r},R;z)$ corresponding to $\vert z\vert\to\infty$, the 
asymptotic term scaling like $1/z^2$ would be preceded by a more 
dominant term different from and in addition to 
$\ln(-z/\varepsilon_0)/z^2$, which is not {\sl a priori} ruled 
out. The likelihood of this possibility should be the more
appreciated by realizing the fact that in determining 
${\sf C}_{\zeta;1}[\dots]$ in Eq.~(\ref{ef138}) we have made use 
of AS of {\sl transcendental} functions $\sin(\zeta')$ and 
$\cos(\zeta')$ (which are bounded for ${\rm Im}(\zeta')=0$), with 
$\zeta' = q\zeta$, for $\zeta' \to 0$. 
\footnote{\label{f137}
See footnotes \protect\ref{f28} and \protect\ref{f126} concerning the 
limitations of the Poincar\'e definition of AS. }

For establishing the existence or otherwise of the integral on the 
RHS of Eq.~(\ref{ef140}), we first note that ${\rm Ci}(x) \sim \sin(x)/x$ 
for $\vert x\vert\to \infty$ (see Abramowitz and Stegun 1972, pp.~232 
and 233), implying that 
\begin{eqnarray}
\label{ef141}
\Big({\rm Ci}(\|{\Bf q}\| R) -\ln(\|{\Bf q}\| R) 
-\frac{\sin(\|{\Bf q}\|R)}{\|{\Bf q}\| R}\Big)
&\sim& -\ln(\|{\Bf q}\| R)\nonumber\\
\mbox{\rm for}\;\,\|{\Bf q}\| &\to& \infty. 
\end{eqnarray}
In Appendix K we consider the `smoothness' properties of $n({\Bf r})$. 
Here we arrive at the well-established fact that $n({\Bf r})$ possesses 
cusps at the positions $\{{\Bf R}_j\}$ of atomic nuclei when the bare 
external potential $u({\Bf r})$ diverges like the Coulomb potential
$v_c({\Bf r}-{\Bf R}_j)$ for ${\Bf r}$ approaching ${\Bf R}_j$ (see 
Eq.~(\ref{ek6})). In such cases, which we consider as corresponding 
to the most singular of physical external potentials, $\nabla_{\Bf r}^2 
n({\Bf r})$ diverges like $1/\|{\Bf r}-{\Bf R}_j\|$ for ${\Bf r}\to 
{\Bf R}_j$, $\forall j$ (see Eq.~(\ref{ek16})). This implies that, 
neglecting logarithmic corrections, ${\bar n}'({\Bf q})$ {\sl cannot} 
decrease {\sl more slowly} than $1/\|{\Bf q}\|^4$ for $\|{\Bf q}\|\to 
\infty$. Such asymptotic behaviour is sufficiently strongly decaying for 
countering the divergent behaviour of $\ln(\|{\Bf q}\| R)$ for 
$\|{\Bf q}\| \to\infty$ (see Eq.~(\ref{ef141}) above) and securing 
convergence of the ${\Bf q}$ integral on the RHS of Eq.~(\ref{ef140}). 
We point out that the integrand of this integral is regular over the 
entire ${\Bf q}$ space (for the behaviour of this integrand as 
$\|{\Bf q}\| \to 0$, see text following Eq.~(\ref{ef140}) above).

The {\sl physical} function ${\sf S}({\Bf r},R;\varepsilon)$ is obtained 
from $\wt{\sf S}({\Bf r},R;z)$ through the substitution $z\rightharpoonup 
\varepsilon \pm i\eta$, $\eta\downarrow 0$, for $\varepsilon \to 
\pm\infty$. With ${\rm Im}\{\ln(-[\varepsilon +i\eta])\} = -\pi$ for 
$\varepsilon >0$ and ${\rm Im}\{\ln(-[\varepsilon-i\eta])\} = 0$ for 
$\varepsilon <0$, we observe that, when $z=\varepsilon+i\eta$ with 
$\varepsilon > 0$, the imaginary parts of the RHSs of Eqs.~(\ref{ef130}) 
and (\ref{ef140}) are non-vanishing, decreasing like $1/\varepsilon^2$ 
for increasing $\varepsilon$, in contrast with the case corresponding 
to $z=\varepsilon-i\eta$ with $\varepsilon < 0$ where the RHSs of 
Eqs.~(\ref{ef130}) and (\ref{ef140}) are purely real valued. This can 
be easily understood by the fact that the positive definiteness of 
the Coulomb potential $v_c({\Bf r}'')$ implies that for $z=\varepsilon
-i\eta$ and $\varepsilon < 0$ the denominator of the integrand on the 
RHS of Eq.~(\ref{ef129}) is never vanishing; for $z=\varepsilon+i\eta$ 
and $\varepsilon > 0$, on the other hand, this denominator clearly 
changes sign in some vicinity of ${\Bf r}''= {\bf 0}$. 

By writing (in analogy with $\wt{\Sigma}_{\sigma}({\Bf r},{\Bf r}';z)$ 
in Eq.~(\ref{e112}))
\begin{equation}
\label{ef142}
\wt{\sf M}({\Bf r};z) \sim 
\frac{\wt{\sf M}_{\infty_2}({\Bf r}\vert z)}{z^2} + \dots,\;\;\;
\vert z\vert\to\infty,
\end{equation}
from Eqs.~(\ref{ef127}), (\ref{ef130}) and (\ref{ef140}), employing the 
convention described in \S~III.E.2, we have ({\it cf}. Eq.~(\ref{e110}))
\begin{equation}
\label{ef143}
\wt{\sf M}_{\infty_2}({\Bf r}\vert z)
\equiv {\sf M}^{\rm r}_{\infty_2}({\Bf r})
+ \wt{\sf M}^{\rm s}_{\infty_2}({\Bf r}\| z),
\end{equation}
where (see Eq.~(\ref{e213}))
\begin{eqnarray}
\label{ef144}
&&{\sf M}^{\rm r}_{\infty_2}({\Bf r}) \equiv
\int_{\Omega_R} {\rm d}^3r''\; v_c^2({\Bf r}'') 
\Lambda({\Bf r}-{\Bf r}'',{\Bf r}) \nonumber\\
&&\;\;
+\int_{\Omega\setminus\Omega_R} {\rm d}^3r''\;
v_c^2({\Bf r}'')\,\Big\{ v_c({\Bf r}'') n({\Bf r}-{\Bf r}'') 
+\Lambda({\Bf r}-{\Bf r}'',{\Bf r})\Big\}\nonumber\\
&&\;\;
+4\pi \left(\frac{e^2}{4\pi\epsilon_0} \right)^3 
\Big\{ (1-\gamma)\, n'({\Bf r}) 
+\int \frac{{\rm d}^3q}{(2\pi)^3}\; {\bar n}'({\Bf q})\,
\nonumber\\
&&\;\;\;\;\;\;
\times\Big[ {\rm Ci}(\|{\Bf q}\| R) - \ln(\|{\Bf q}\| R)
-\frac{\sin(\|{\Bf q}\| R)}{\|{\Bf q}\| R}\Big]
\,\exp(i {\Bf q}\cdot {\Bf r})\nonumber\\
&&\;\;\;\;\;\;\;\;\;\;\;\;\;\;\;\;\;\;\;\;\;\;\;\;\;\;\;\;\;\;\;\;
\;\;\;\;\;\;\;\;\;\;\;\;\,
+n({\Bf r})\, \ln\left(\frac{\varepsilon_0}{e_R}\right)\Big\},
\end{eqnarray}
and (see Eq.~(\ref{e213}))
\begin{equation}
\label{ef145}
\wt{\sf M}^{\rm s}_{\infty_2}({\Bf r}\| z) \equiv
4\pi \left(\frac{e^2}{4\pi\epsilon_0} \right)^3 \,
n({\Bf r}) \, \ln\left(\frac{-z}{\varepsilon_0}\right).
\end{equation}
It is important to note that the leading-order contribution to 
$\wt{\sf M}({\Bf r};z)$ of the last term on the RHS of Eq.~(\ref{ef127}) 
scales like $1/z^3$ for $\vert z\vert\to \infty$ when $R >
e^2/(4\pi\epsilon_0 \vert z\vert)$ so that, on account of our 
implicit assumption in Eq.~(\ref{ef128}), there is no contribution to 
$\wt{\sf M}_{\infty_2}({\Bf r}\vert z)$ in Eq.~(\ref{ef143}) arising 
from the above-mentioned term on the RHS of Eq.~(\ref{ef127}). 

Finally, in \S~III.I.2 we require the leading-order imaginary part of 
$\wt{\sf M}({\Bf r};z)$ for $z =\varepsilon \pm i \eta$, $\eta\downarrow 
0$, as $\varepsilon \to \pm \infty$; with reference to Eq.~(\ref{ef142}), 
from Eqs.~(\ref{ef143}) - (\ref{ef145}) we readily obtain ({\it cf}. 
Eq.~(\ref{eg17}))
\footnote{\label{f138}
Owing to $\vert\varepsilon\vert\to\infty$, in \S~III.I.2 we replace 
$\Theta(\varepsilon)$ by $\Theta(\varepsilon-\mu)$ where $\mu$ stands 
for the `chemical potential' introduced in Eq.~(\protect\ref{e22}). }
\begin{equation}
\label{ef146}
{\rm Im}[{\sf M}({\Bf r};\varepsilon)]\sim
-4\pi^2 \left(\frac{e^2}{4\pi\epsilon_0}\right)^3
n({\Bf r})\, \frac{\Theta(\varepsilon)}{\varepsilon^2},
\;\;\; \vert\varepsilon\vert\to \infty.
\end{equation}
We note in passing that by introducing the {\sl local} Wigner-Seitz 
radius, namely ({\it cf}.  Eq.~(\ref{e93}))
\footnote{\label{f139}
The {\sl total} GS number density is positive everywhere.}
\begin{equation}
\label{ef147}
r_0({\Bf r}) {:=} \left(\frac{3}{4\pi n({\Bf r})}\right)^{1/3},
\end{equation}
the coefficient of $\ln(-z/\varepsilon_0)$ in Eq.~(\ref{ef145}), that 
is $3\big(e^2/[4\pi\epsilon_0 r_0({\Bf r})]\big)^3$, is seen to be
equal to three times the third power of the Coulomb repulsion energy 
due to two particles of equal charge (either $-e$ or $+e$) at distance 
$r_0$ whose value depends on ${\Bf r}$. 
\hfill $\Box$

\section{Regularization of ${\cal T}_{\sigma,\bar\sigma}
(\lowercase{\Bf r})$ and the large-$\vert\lowercase{z}\vert$ 
asymptotic series for 
$\wt{\sf T}_{\sigma,\bar\sigma}(\lowercase{\Bf r};\lowercase{z})$
(${\sf T}^{\lowercase{\rm r}}_{\sigma,\bar\sigma;\infty_2}
(\lowercase{\Bf r})$, 
${\sf T}^{\lowercase{\rm s_b}}_{\sigma,\bar\sigma;\infty_2}
(\lowercase{\Bf r})$ and 
$\wt{\sf T}^{\lowercase{\rm s}}_{\sigma,\bar\sigma;\infty_2}
(\lowercase{\Bf r}\| \lowercase{z})$)}
\label{s62}

In the expression for the {\sl exact} $\Sigma_{\sigma;\infty_2}({\Bf r},
{\Bf r}')$ corresponding to $v\equiv v_c$ in $d=3$ we encounter the 
local contribution $\hbar^{-1} {\cal T}_{\sigma,\bar\sigma}({\Bf r})\,
\delta({\Bf r}-{\Bf r}')$ with ${\cal T}_{\sigma,\bar\sigma}({\Bf r})$ 
as presented in Eq.~(\ref{e209}). For the reasons presented in 
\S~III.H.2, ${\cal T}_{\sigma,\bar\sigma}({\Bf r})$ is unbounded. In 
the expression for $\Sigma_{\sigma;\infty_2}^{(1)}({\Bf r},{\Bf r}')$ 
corresponding to $v\equiv v_c$ in $d=3$ we encounter an equally 
unbounded local contribution, denoted by $\hbar^{-1} {\cal T}^{(1)}
({\Bf r})\,\delta({\Bf r}-{\Bf r}')$ (see Eq.~(\ref{e278})). Although 
${\cal T}_{\sigma,\bar\sigma}({\Bf r})$ and ${\cal T}^{(1)}({\Bf r})$ 
have different functional forms (among others, whereas 
${\cal T}_{\sigma,\bar\sigma}({\Bf r})$ is identically vanishing when 
$n_{\sigma}({\Bf r}) \equiv n_{\bar\sigma}({\Bf r})$, ${\cal T}^{(1)}
({\Bf r})$ is {\sl never} identically vanishing), they are of the same 
origin and their unboundedness is rooted in the fact that, up to a 
multiplicative constant, the Coulomb potential $v_c$, viewed as an 
operator, and the single-particle kinetic-energy operator $\tau$ are 
each other's inverses (see Eq.~(\ref{e205})). In this Appendix we shall 
in the main consider regularization of ${\cal T}_{\sigma,\bar\sigma}
({\Bf r})$, effected through a summation over an infinite number of 
unbounded contributions pertaining to $\Sigma_{\sigma;\infty_m}({\Bf r},
{\Bf r}')$, with $m \ge 2$; we denote the regularized extension of 
${\cal T}_{\sigma,\bar\sigma}({\Bf r})$ by $\wt{\sf T}_{\sigma,
\bar\sigma}({\Bf r};z)$ (for our notational conventions see \S~III.E.2; 
see specifically the text following Eq.~(\ref{e111})). The regularized 
extension of ${\cal T}^{(1)}({\Bf r})$, denoted by $\wt{\sf T}^{(1)}
({\Bf r};z)$, is directly deduced from $\wt{\sf T}_{\sigma,\bar\sigma}
({\Bf r};z)$ through the substitution $(n_{\bar\sigma}-n_{\sigma}) 
\rightharpoonup n$ that transforms ${\cal T}_{\sigma,\bar\sigma}
({\Bf r})$ into ${\cal T}^{(1)}({\Bf r})$ ({\it cf}. Eqs.~(\ref{e209}) 
and (\ref{e278})). 

The functional similarities between ${\cal T}^{(1)}({\Bf r})$ and 
${\cal T}_{\sigma,\bar\sigma}({\Bf r})$ and our calculation of 
$\Sigma_{\sigma;\infty_3}^{(1)}({\Bf r},{\Bf r}')$ in \S~IV.C, 
confirm that the following regularizing series, deduced solely on 
the basis of the expression for $\Sigma_{\sigma;\infty_2}({\Bf r},
{\Bf r}')$, correctly produces the pertinent unbounded term pertaining 
to $\Sigma_{\sigma;\infty_3}^{(1)}({\Bf r},{\Bf r}')$ upon the
substitution $(n_{\bar\sigma} - n_{\sigma}) \rightharpoonup n$:
\begin{eqnarray}
\label{eg1}
&&\wt{\sf T}_{\sigma,\bar\sigma}({\Bf r};z) {:=} \frac{1}{z} 
\sum_{m=1}^{\infty} \int {\rm d}^3r''\; 
v_c({\Bf r}-{\Bf r}'')
\big[ n_{\bar\sigma}({\Bf r}'') -
n_{\sigma}({\Bf r}'')\big] \nonumber\\
&&\;\;\;\;\;\;\;\;\;\;\;\;\;\;\;\;\;\;\;\;\;
\;\;\;\;\;\;\;\;\;\;\;
\times \big[\frac{1}{z^m} \tau^m({\Bf r}'') 
v_c({\Bf r}-{\Bf r}'')\big]
\nonumber\\
&&\;\;\;\;\;\;\;\;
=\frac{1}{z^2} \int {\rm d}^3r''\; v_c({\Bf r}-{\Bf r}'')
\big[ n_{\bar\sigma}({\Bf r}'') -
n_{\sigma}({\Bf r}'')\big]\nonumber\\
&&\;\;\;\;\;\;\;\;\;\;\;\;\;
\;\;\;\;\;\;\;\;
\times\tau({\Bf r}'') \big(1 - \frac{1}{z} 
\tau({\Bf r}'')\big)^{-1} v_c({\Bf r}-{\Bf r}''),
\end{eqnarray}
where the second expression on the RHS is obtained through the formal 
summation of the geometric series in the first expression (see text 
following Eq.~(\ref{e111})).

In order to determine a finite-order large-$\vert z\vert$ AS for 
$\wt{\sf T}_{\sigma,\bar\sigma}({\Bf r};z)$, we first employ the 
following Fourier representation of the Coulomb potential:
\begin{eqnarray}
\label{eg2}
v_c({\Bf r}-{\Bf r}'') = \int \frac{{\rm d}^3q}{(2\pi)^3}\;
\frac{e^2/\epsilon_0}{\|{\Bf q}\|^2}\,
\exp(i {\Bf q}\cdot [{\Bf r}-{\Bf r}'']), 
\end{eqnarray}
from which it readily follows that
\begin{eqnarray}
\label{eg3}
&&\tau({\Bf r}'') \big(1 -\frac{1}{z} \tau({\Bf r}'')\big)^{-1}
v_c({\Bf r}-{\Bf r}'')\nonumber\\
&&\;\;\;\;\;\;\;\;\;\;\;\;\;\;
= \frac{e^2 \hbar^2}{2 m_e \epsilon_0}
\int \frac{{\rm d}^3q}{(2\pi)^3} \;
\frac{\exp(i {\Bf q}\cdot [{\Bf r}-{\Bf r}''])}{1
-\hbar^2 \|{\Bf q}\|^2/[2 m_e z]}.
\end{eqnarray}
Using the spherical polar coordinates for ${\Bf q}$, with 
${\Bf r}-{\Bf r}''$ the polar axis, the integral on the RHS of 
Eq.~(\ref{eg3}) reduces to $(\hbar^2/[\pi m_e]) v_c({\Bf r}
-{\Bf r}'')$ times the one-dimensional integral $\int_0^{\infty} 
{\rm d}q\, q \sin(a q)/(1 - \zeta q^2)$, where $a {:=} \|{\Bf r} 
-{\Bf r}''\|$ and $\zeta {:=} \hbar^2/(2 m_e z)$. Making use of 
the fact that the integrand of this integral is an {\sl even} 
function of $q$, so that $\int_0^{\infty} {\rm d}q\, (\dots) = 
\frac{1}{2} \int_{-\infty}^{\infty} {\rm d}q\, (\dots)$, applying 
the residue theorem to two integrals that arise from use of 
$\sin(a q) = \frac{1}{2 i} \{ \exp(i a q) - \exp(-i a q)\}$ in the 
latter integral, we readily obtain
\begin{eqnarray}
\label{eg4}
&&\tau({\Bf r}'') \big(1 -\frac{1}{z}\tau({\Bf r}'')\big)^{-1}
v_c({\Bf r}-{\Bf r}'')
= -z\, v_c({\Bf r}-{\Bf r}'') \nonumber\\
&&\;\;\;\;\;\;\;\;\;\;\;\;\;\;\;\;\;\;\;
\times \exp\big(-\sqrt{2 m_e}\, (-z)^{1/2} 
\|{\Bf r}-{\Bf r}''\| /\hbar\big),
\end{eqnarray}
where $z^{1/2}$ stands for the principal branch of the 
square-root function, so that
\begin{eqnarray}
\label{eg5}
(-z)^{1/2} = \mp i z^{1/2},
\;\;\,
\left\{ \begin{array}{rr}
0 < {\rm arg}(z) < \pi, \\
-\pi < {\rm arg}(z) < 0.
\end{array}\right.
\end{eqnarray}
Substituting the RHS of Eq.~(\ref{eg4}) into that of Eq.~(\ref{eg1}), 
transforming the variable of integration (so that ${\Bf r}''$ in the
arguments of $n_{\bar\sigma}({\Bf r}'')$ and $n_{\sigma}({\Bf r}'')$ 
is changed into ${\Bf r}-{\Bf r}''$), followed by employing the Fourier
integral representation ({\it cf.} Eq.~(\ref{ef132}))
\begin{eqnarray}
\label{eg6}
n_{\sigma}({\Bf r}) = \int \frac{{\rm d}^3q}{(2\pi)^3}\;
{\bar n}_{\sigma}({\Bf q})\, \exp(i {\Bf q}\cdot {\Bf r}),
\end{eqnarray}
the resulting integration with respect to ${\Bf r}''$ is easily carried 
out in terms of the spherical polar coordinates of ${\Bf r}''$; making 
use of the standard result (see Abramowitz and Stegun 1972, p.~1028)
\begin{eqnarray}
\int_0^{\infty} {\rm d}x\; {\rm e}^{-s x}\, 
\frac{\sin(\alpha x)}{x} = \tan^{-1}\left(\frac{\alpha}{s}\right),\;\;
\left\{ \begin{array}{rr}
{\rm Re}(s) > 0, \nonumber\\
{\rm Im}(\alpha)=0, 
\end{array}\right.
\end{eqnarray}
we obtain
\begin{eqnarray}
\label{eg7}
&&\wt{\sf T}_{\sigma,\bar\sigma}({\Bf r};z) = \frac{-4\pi}{z}
\left(\frac{e^2}{4\pi\epsilon_0}\right)^2
\int \frac{{\rm d}^3q}{(2\pi)^3} \;
\frac{{\bar n}_{\bar\sigma}({\Bf q})
-{\bar n}_{\sigma}({\Bf q})}{\|{\Bf q}\|}\,
\nonumber\\
&&\;\;\;\;\;\;\;\;\;\times
\tan^{-1}\Big(\frac{\|{\Bf q}\|}{\sqrt{2 m_e}\,
(-z)^{1/2}/\hbar } \Big)\, \exp(i {\Bf q}\cdot {\Bf r}), 
\end{eqnarray}
where $\tan^{-1}(z)$ stands for the principal branch of the inverse
of $\tan(z)$. 

Since $\tan^{-1}(z) = \sum_{m=0}^{\infty} (-1)^m z^{2m+1}/(2m+1)$ is 
uniformly convergent for $\vert z\vert \leq 1$, $z^2\not=-1$, if follows 
that in cases where ${\bar n}_{\bar\sigma}({\Bf q}) - {\bar n}_{\sigma}
({\Bf q}) \equiv 0$ for $\|{\Bf q}\| > Q$, with $Q$ some {\sl finite} 
wavenumber, $(-z)^{1/2}\, \wt{\sf T}_{\sigma,\bar\sigma}({\Bf r};z)$ has 
a uniformly convergent series in powers of $1/z$ for $\vert z\vert > 
\hbar^2 Q^2/[2 m_e]$.

In order to obtain an AS for $\wt{\sf T}_{\sigma,\bar\sigma}
({\Bf r};z)$, $\vert z\vert\to \infty$, suitable for a general 
case, we replace ${\bar n}_{\bar\sigma}({\Bf q})-{\bar n}_{\sigma}
({\Bf q})$ on the RHS of Eq.~(\ref{eg7}) by $\int {\rm d}^3r\, 
[n_{\bar\sigma}({\Bf r})-n_{\sigma}({\Bf r})] \exp(-i {\Bf q} \cdot 
{\Bf r})$ and exchange the order of integrations. Through formally 
replacing $\tan^{-1}(\dots)$ in this expression by the 
above-presented series, we obtain a series involving integrals of 
the form $\int {\rm d}^3q\, \|{\Bf q}\|^{2m} \exp(i {\Bf q}\cdot 
[{\Bf r}-{\Bf r}'])$, which is equal to $(-1)^m (2\pi)^3\, 
\nabla_{{\Bf r}}^{2m} \delta ({\Bf r}-{\Bf r}')$. Upon exchanging 
the order of differentiation and integration we obtain the following 
formal expression:
\begin{eqnarray}
\label{eg8}
&&\wt{\sf T}_{\sigma,\bar\sigma}({\Bf r};z) = 
\frac{-4\pi\hbar}{\sqrt{2 m_e}}
\left(\frac{e^2}{4\pi\epsilon_0}\right)^2\,
\frac{1}{(-z)^{1/2} z} \nonumber\\
&&\;\;\;\;\;\;\;\;\;\;
\times \sum_{m=0}^{\infty}
\frac{1}{2 m + 1}\, \frac{1}{z^m} \tau^m({\Bf r}) 
\big[ n_{\bar\sigma}({\Bf r})-n_{\sigma}({\Bf r})\big].
\end{eqnarray}
With $\sum_{m=0}^{\infty} z^m/(2m+1) = z^{-1/2} \tanh^{-1}(z^{1/2})$, 
for $\vert z\vert < 1$, the result in Eq.~(\ref{eg8}) can be written as
\begin{eqnarray}
\label{eg9}
&&\wt{\sf T}_{\sigma,\bar\sigma}({\Bf r};z) = 
\frac{-4\pi\hbar}{\sqrt{2 m_e}}
\left(\frac{e^2}{4\pi\epsilon_0}\right)^2
\frac{1}{(-z)^{1/2} z} \nonumber\\
&&\;\;\;\;\;
\times \frac{\tanh^{-1}\big(\tau^{1/2}({\Bf r})/z^{1/2}\big)}
{\tau^{1/2}({\Bf r})/z^{1/2} }\,
\big[ n_{\bar\sigma}({\Bf r})-n_{\sigma}({\Bf r})\big],
\end{eqnarray}
which in essence is an abstract representation of the expression in 
Eq.~(\ref{eg8}); the expressions in Eqs.~(\ref{eg7}) and (\ref{eg9}) 
are more general than the formal expression in Eq.~(\ref{eg8}), for
with $M$ the smallest finite integer for which $\tau^{M}({\Bf r}) 
[n_{\bar\sigma}({\Bf r})-n_{\sigma}({\Bf r})]$ is non-integrably 
unbounded for {\sl some} ${\Bf r}$ (see criterion (B) in \S~II.B), 
the terms on the RHS of Eq.~(\ref{eg8}) corresponding to $m \ge M$ are 
{\sl not} directly meaningful. The existence of such an $M$ implies that 
\begin{eqnarray}
\label{eg10}
&&\wt{\sf T}_{\sigma,\bar\sigma}({\Bf r};z)
\sim \frac{-4\pi\hbar}{\sqrt{2 m_e}}
\left(\frac{e^2}{4\pi\epsilon_0}\right)^2
\frac{1}{(-z)^{1/2}} \Big(
\frac{1}{z} + \dots \nonumber\\
&&\;\;\;\;\;\;\;
+ \frac{1/[2M-1]}{z^M}\, 
\tau^{M-1}({\Bf r}) \Big) \, 
\big[ n_{\bar\sigma}({\Bf r})-
n_{\sigma}({\Bf r})\big]
\end{eqnarray}
constitutes the $M$ leading terms in the AS of $\wt{\sf T}_{\sigma,
\bar\sigma}({\Bf r};z)$ for $\vert z\vert\to\infty$ which directly 
can be Fourier transformed, thus yielding the complete $M$ leading 
terms in the AS pertaining to the Fourier transform with respect to 
${\Bf r}$ of $(-z)^{1/2} \wt{\sf T}_{\sigma,\bar\sigma}({\Bf r};z)$ 
in terms of the asymptotic sequence $\{1,1/z,\dots\}$; the 
non-integrability of $\tau^{M}({\Bf r}) [n_{\bar\sigma}({\Bf r})
-n_{\sigma}({\Bf r})]$ in a neighbourhood of some ${\Bf r}$ implies 
that the term immediately subsequent to that decaying like 
$1/\vert z\vert^M$ in the large-$\vert z\vert$ AS of the Fourier 
transform of $(-z)^{1/2} \wt{\sf T}_{\sigma,\bar\sigma}({\Bf r};z)$, 
does {\sl not} decay like $1/\vert z\vert^{M+1}$, but more slowly (see 
\S~II.B). In Appendix K we deduce and present a direct relationship 
between the behaviour of $\tau({\Bf r}) n_{\sigma}({\Bf r})$ and that 
of the local external potential $u({\Bf r})$, in $d=3$ (see in particular 
Eq.~(\ref{ek16})), from which it follows that for $u({\Bf r})$ diverging 
as ${\Bf r} \to {\Bf R}_j$, with ${\Bf R}_j$ the position vector of an 
ionic nucleus, $\tau({\Bf r}) n_{\sigma}({\Bf r})$ also diverges, like 
$u({\Bf r})$, as ${\Bf r}$ approaches ${\Bf R}_j$, $\forall j$. Since 
the singularity of $v_c({\Bf r}-{\Bf r}')$ at ${\Bf r}={\Bf r}'$ is 
integrable, it follows that provided $n_{\bar\sigma}({\Bf r}) 
\not\equiv n_{\sigma}({\Bf r})$, for $u({\Bf r}) \propto v_c({\Bf r}
-{\Bf R}_j)$ as ${\Bf r}\to {\Bf R}_j$ ({\it cf}. Eq.~(\ref{ek6})), 
$\tau({\Bf r}) [n_{\bar\sigma}({\Bf r})-n_{\sigma}({\Bf r})]$ is 
integrably divergent for ${\Bf r} \to {\Bf R}_j$, $\forall j$. 
Consequently, the term subsequent to that {\sl decaying} like 
$1/\vert z\vert^{3/2}$ in the large-$\vert z\vert$ AS of the Fourier 
transform of $\wt{\sf T}_{\sigma,\bar\sigma}({\Bf r};z)$ with respect 
to ${\Bf r}$ (evidently $[n_{\bar\sigma}({\Bf r})-n_{\sigma}({\Bf r})]$ 
is integrable with respect to ${\Bf r}$) is {\sl not} one decaying 
more slowly than or as slow as $1/\vert z\vert^2$, but more quickly (to 
be explicit, as quickly as $1/\vert z\vert^{5/2}$). Since in the present 
paper we do {\sl not} consider contributions to the large-$\vert z\vert$ 
AS of $\wt{\Sigma}_{\sigma}({\Bf r},{\Bf r}';z)$ decaying like 
$1/\vert z\vert^m$, with $m > 2$, we conclude that from the perspective 
of our considerations in this paper, the following asymptotic expression, 
which can be directly Fourier transformed, is complete ({\it cf}. 
Eq.~(\ref{e112}))
\begin{equation}
\label{eg11}
\wt{\sf T}_{\sigma,\bar\sigma}({\Bf r};z)
\sim \frac{\wt{\sf T}_{\sigma,\bar\sigma;\infty_2}
({\Bf r}\vert z)}{z^2},\;\;\; \vert z\vert \to\infty,
\end{equation}
where ({\it cf}. Eq.~(\ref{e110}))
\begin{equation}
\label{eg12}
\wt{\sf T}_{\sigma,\bar\sigma;\infty_2}({\Bf r}\vert z)
= {\sf T}_{\sigma,\bar\sigma;\infty_2}^{\rm r}({\Bf r}) +
{\sf T}_{\sigma,\bar\sigma;\infty_2}^{\rm s_b}({\Bf r}) +
\wt{\sf T}_{\sigma,\bar\sigma;\infty_2}^{\rm s}({\Bf r}\| z),
\end{equation}
in which (see footnote \ref{f31})
\begin{eqnarray}
\label{eg13}
&&{\sf T}_{\sigma,\bar\sigma;\infty_2}^{\rm r}({\Bf r}) \equiv 0,\\
\label{eg14}
&&{\sf T}_{\sigma,\bar\sigma;\infty_2}^{\rm s_b}({\Bf r}) \equiv 0,\\
\label{eg15}
&&\wt{\sf T}_{\sigma,\bar\sigma;\infty_2}^{\rm s}({\Bf r}\| z)
\equiv \frac{4\pi\hbar}{\sqrt{2 m_e}}
\left(\frac{e^2}{4\pi\epsilon_0}\right)^2
(-z)^{1/2}\, \nonumber\\
&&\;\;\;\;\;\;\;\;\;\;\;\;\;\;\;\;\;\;\;\;\;\;\;\;\;\;\;\;\;
\;\;\;\;\;\;\;\;
\times \big[ n_{\bar\sigma}({\Bf r})-
n_{\sigma}({\Bf r})\big].
\end{eqnarray}
We have introduced these auxiliary functions in order to conform 
with the notational convention adopted in this paper (see
\S~III.E.2). 

As is apparent from Eq.~(\ref{eg15}), $\wt{\sf T}^{\rm s}_{\sigma,
\bar\sigma;\infty_2}({\Bf r}\| z)$ (and thus $\wt{\sf T}_{\sigma,
\bar\sigma;\infty_2}({\Bf r}\vert z)$) approaches a {\sl purely real} 
value for ${\rm arg}(z) \to \pm \pi$, while it approaches a {\sl purely 
imaginary} value for ${\rm arg}(z) \to 0$ ({\it cf}. Eq.~(\ref{eg17}) 
below). These observations can be clarified by considering the fact that 
$\tau({\Bf r})$ is a positive-definite operator, so that for a real and 
negative $\varepsilon$, $\big(1-\tau({\Bf r})/\varepsilon\big)$ does 
{\sl not} vanish and consequently, for $z=\varepsilon \pm i\eta$, with 
$\eta\downarrow 0$, the integral on the RHS of Eq.~(\ref{eg1}) 
{\sl cannot} give rise to an imaginary contribution; for a real and 
positive $\varepsilon$, on the other hand, $\big(1
-\tau({\Bf r})/\varepsilon\big)$ does vanish (the spectrum of 
$\tau({\Bf r})$ is unbounded from above), thus in this case giving 
rise to a $\wt{\sf T}^{\rm s}_{\sigma,\bar\sigma;\infty_2}({\Bf r}
\| z)$, and thus a $\wt{\sf T}_{\sigma,\bar\sigma;\infty_2}({\Bf r}
\vert z)$, that involves an imaginary contribution for $z=\varepsilon 
\pm i\eta$, $\eta\downarrow 0$. 

Following Eq.~(\ref{eg7}) above, we have indicated that, in cases 
where ${\bar n}_{\sigma'}({\Bf q})\equiv 0$ for $\|{\Bf q}\| > Q$, 
with $\sigma'\in \{\sigma,\bar\sigma\}$, $n_{\sigma'}({\Bf r})$ is 
differentiable with respect to ${\Bf r}$ to any arbitrary finite 
order so that, for number densities of this type, $(-z)^{1/2} 
\wt{\sf T}_{\sigma,\bar\sigma}({\Bf r};z)$ can be described in 
terms of a uniformly convergent series in powers of $1/z$ for 
$\vert z\vert > \hbar^2 Q^2/[2 m_e]$. In the most idealized case of 
uniform GSs, corresponding to ${\bar n}_{\sigma'}({\Bf q}) = (2\pi)^3 
n_{0;\sigma'}\, \delta({\Bf q})$, we have
\begin{eqnarray}
\label{eg16}
\wt{\sf T}_{\sigma,\bar\sigma}({\Bf r};z)
&\equiv& \frac{\wt{\sf T}_{\sigma,\bar\sigma;\infty_2}
({\Bf r}\vert z)}{z^2}
\nonumber\\
&=& \frac{4\pi\hbar}{\sqrt{2 m_e}}
\left(\frac{e^2}{4\pi\epsilon_0}\right)^2 (-z)^{1/2}\,
\frac{n_{0;\bar\sigma} - n_{0;\sigma}}{z^2},
\end{eqnarray}
which in the case of spin-$1/2$ fermions, is identically vanishing 
in the paramagnetic phase. With reference to Eq.~(\ref{e65}), for 
$z=\varepsilon\pm i\eta$, $\eta\downarrow 0$, as $\varepsilon\,
\Ieq{<}{>}\, \mu$, from Eq.~(\ref{eg16}) we readily obtain
\footnote{\label{f140}
The result in this equation (as well as that in Eq.~(\protect\ref{eg16})) 
is {\sl exact} for {\sl all} $\varepsilon$; however, unless 
$\vert\varepsilon\vert\to\infty$, {\sl no} physical significance can be 
ascribed to function ${\sf T}_{\sigma,\bar\sigma}({\Bf r};\varepsilon)$ 
nor to its divergence at $\varepsilon=0$. }
\begin{eqnarray}
\label{eg17}
{\sf T}_{\sigma,\bar\sigma}({\Bf r};\varepsilon)
&=& \frac{4\pi\hbar}{\sqrt{2 m_e}}\,
\left(\frac{e^2}{4\pi\epsilon_0}\right)^2
[ n_{0;\bar\sigma} - n_{0;\sigma} ]\,\nonumber\\
& &\;\;\;\;\;\;\;\;\;\;\;\;\;\;\;\;\;
\times \frac{\Theta(\mu-\varepsilon) - i 
\Theta(\varepsilon-\mu)}{\vert\varepsilon\vert^{3/2}}.
\end{eqnarray}

In view of our considerations in \S~III.E.4, we mention that the 
double Fourier transform of $\wt{\sf T}_{\sigma,\bar\sigma;\infty_2}
({\Bf r}\vert z)\, \delta({\Bf r}-{\Bf r}')$ (see Eq.~(\ref{eh12})), 
which we denote by $\wt{\Ol{\sf T}}_{\sigma,\bar\sigma;\infty_2}
({\Bf q},{\Bf q}'\vert z)$, has the following form in the case of 
uniform GSs: 
\begin{eqnarray}
\label{eg18}
\wt{\Ol{\sf T}}_{\sigma,\bar\sigma;\infty_2}
({\Bf q},{\Bf q}'\vert z)
&=& \frac{4\pi\hbar}{\sqrt{2 m_e}}
\left(\frac{e^2}{4\pi\epsilon_0}\right)^2\, (-z)^{1/2}\,\nonumber\\
& &\;\;\;\;\;\;\;\;\;\;\;\;
\times \big[ n_{0;\bar\sigma}-n_{0;\sigma}\big]\,
\delta_{{\Bf q},{\Bf q}'}.
\end{eqnarray}
Making use of the conventions in \S~III.E.2, for (see footnote 
\ref{f83})
\begin{equation}
\label{eg19}
\wt{\Ol{\ol{\sf T}}}_{\sigma,\bar\sigma;\infty_2}
({\bar {\Bf q}},{\bar {\Bf q}}'\vert {\bar z}) 
{:=} \frac{1}{e_0^3}\,
\wt{\Ol{\sf T}}_{\sigma,\bar\sigma;\infty_2}
({\Bf q},{\Bf q}'\vert z)
\end{equation}
we therefore have
\begin{eqnarray}
\label{eg20}
&&\wt{\Ol{\ol{\sf T}}}_{\sigma,\bar\sigma;\infty_2}
({\bar {\Bf q}},{\bar {\Bf q}}'\vert {\bar z}) =
\frac{3}{\sqrt{2}}\, r_s^2\, (-{\bar z})^{1/2}\,\nonumber\\
&&\;\;\;\;\;\;\;\;\;\;\;\;\;\;\;\;\;\;\;\;\;\;\;\;\;\;
\times \big[n_{0;\bar\sigma}/n_0-n_{0;\sigma}/n_0\big]\,
\delta_{{\bar {\Bf q}},{\bar {\Bf q}}'}.
\end{eqnarray}
We employ this result in \S~III.E.3, 4 (see Eqs.~(\ref{e114})
and (\ref{e129})).
\hfill $\Box$

\section{Regularization of the momentum representation of
$\Sigma_{\sigma;\infty_2}^{\lowercase{\rm s_b}}$
and the double Fourier transform of 
$\wt{\Sigma}_{\sigma;\infty_2}^{\lowercase{\rm s_b}}
(\lowercase{\Bf r},\lowercase{\Bf r}';\lowercase{z})$}
\label{s63}

The term $-\hbar^{-1} v^3({\Bf r}-{\Bf r}') \varrho_{\sigma}({\Bf r}',
{\Bf r})$ as contributing to $\Sigma_{\sigma;\infty_2}({\Bf r},{\Bf r'})$ 
(see Eqs.~(\ref{e199}) and (\ref{e212})), although perfectly well defined, 
is in the case where $v\equiv v_c$ in $d=3$ non-integrable (see criterion 
(B) in \S~II.B) with respect to either ${\Bf r}$ or ${\Bf r}'$ so 
that, for instance, it does {\sl not} possess a Fourier representation. 
With reference to the expressions for $\Sigma_{\sigma;\infty_m}({\Bf r},
{\Bf r}')$ corresponding to $m=0$ and $m=1$ in Eqs.~(\ref{e173}) and 
(\ref{e185}) respectively, we observe that $- \hbar^{-1} v_c^3({\Bf r}
-{\Bf r}') \varrho_{\sigma}({\Bf r}',{\Bf r})$ is the third term (and 
the first non-integrable term with respect to ${\Bf r}'$ in a neighbourhood 
of ${\Bf r}$) in the formal geometric series of the following function:
\begin{eqnarray}
\label{eh1}
\wt{\sf S}_{\sigma}({\Bf r},{\Bf r}';z)
{:=} \frac{-1}{\hbar} \frac{v_c({\Bf r}-{\Bf r}') 
\varrho_{\sigma}({\Bf r}',{\Bf r})}
{1-v_c({\Bf r}-{\Bf r}')/z}.
\end{eqnarray}

In \S~II.B we discuss in some detail the fact that non-integrable
contributions of the type $- \hbar^{-1} v_c^3({\Bf r}-{\Bf r}')\,
\varrho_{\sigma}({\Bf r}',{\Bf r})$ in the AS of $\wt{\Sigma}_{\sigma}
({\Bf r},{\Bf r}';z)$ for $\vert z\vert\to \infty$ do {\sl not} 
invalidate this as being an AS; however, such series {\sl cannot} be 
directly used in order to obtain a large-$\vert z\vert$ AS pertaining 
to the momentum representation of the SE {\sl operator} 
$\wt{\Sigma}_{\sigma}(z)$. In the particular case at hand, the latter 
series is to be deduced by first determining the double Fourier 
transform of $\wt{\sf S}_{\sigma}({\Bf r},{\Bf r}';z)$ or a related 
function. These are the tasks that we undertake to perform in this 
Appendix.

With reference to our notational convention that we have introduced
in \S~III.E.2, and in view of the fact that the first two terms in 
the geometric series expansion of $\wt{\sf S}_{\sigma}({\Bf r},
{\Bf r}';z)$ in Eq.~(\ref{eh1}) are integrable functions contributing 
to $\Sigma_{\sigma;\infty_0}({\Bf r},{\Bf r}')$ and $\Sigma_{\sigma;
\infty_1}({\Bf r},{\Bf r}')/z$ respectively (see Eqs.~(\ref{e174}) and 
(\ref{e185})), we have (see text following Eq.~(\ref{e111})) 
\begin{eqnarray}
\label{eh2}
\hbar\wt{\Sigma}_{\sigma;\infty_2}^{\rm s_b}({\Bf r},{\Bf r}'; z) 
\equiv -\frac{1}{z^2} \frac{v_c^3({\Bf r}-{\Bf r}')
\varrho_{\sigma}({\Bf r}',{\Bf r})}
{1-v_c({\Bf r}-{\Bf r}')/z},
\end{eqnarray}
which is evidently integrable with respect to ${\Bf r}'$ in any 
finite neighbourhood of ${\Bf r}'={\Bf r}$ (see criterion (B)
in \S~II.B). It is readily verified that the RHS of Eq.~(\ref{eh2}) 
also satisfies criterion (C) in \S~II.B.

Below we evaluate the first two leading terms in the large-$\vert z\vert$ 
AS of the double Fourier transform of $\hbar\wt{\Sigma}_{\sigma;
\infty_2}^{\rm s_b}({\Bf r},{\Bf r}';z)$ with respect to ${\Bf r}$ and 
${\Bf r}'$. These are the only terms of direct relevance to our 
considerations in this paper; in particular the expressions for these 
terms specialized to uniform and isotropic GSs are of direct use in 
\S~III.E.

In view of the specific form of the function on the RHS of 
Eq.~(\ref{eh2}), it is convenient to introduce the `relative' and 
`centre of mass' coordinates
\begin{equation}
\label{eh3}
{\Bf\rho} \equiv {\Bf r}-{\Bf r'},\;\;\;
{\Bf\zeta} \equiv \frac{1}{2} ({\Bf r} + {\Bf r}')
\end{equation}
and define (the `Wigner-transformed' functions)
\begin{eqnarray}
\label{eh4}
&&\ul{\varrho}_{\,\sigma}({\Bf\zeta},{\Bf\rho}) {:=}
\varrho_{\sigma}({\Bf r}',{\Bf r}),\\
\label{eh5}
&&\wt{\ul{\Sigma}}_{\,\sigma;\infty_2}^{\;\rm s_b}
({\Bf\rho},{\Bf\zeta};z)
{:=} \wt{\Sigma}_{\sigma;\infty_2}^{\rm s_b}
({\Bf r},{\Bf r}';z).
\end{eqnarray}
With
\begin{equation}
\label{eh6}
\ul{\ol{\varrho}}_{\,\sigma}({\Bf\zeta};{\Bf q})
{:=} \int {\rm d}^3\rho\;\,
\ul{\varrho}_{\,\sigma}({\Bf\zeta},{\Bf\rho})\,
{\rm e}^{-i {\Bf q}\cdot {\Bf\rho}},
\end{equation}
we obtain
\begin{eqnarray}
\label{eh7}
&&\hbar\,\ul{\wt{\ol{\Sigma}}}_{\,\sigma;\infty_2}^{\;\rm s_b}
({\Bf q};{\Bf\zeta};z)
{:=} \int {\rm d}^3\rho\;\hbar\,
\ul{\wt{\Sigma}}_{\,\sigma}^{\;\rm s_b}({\Bf\rho},{\Bf\zeta}\vert z)\,
{\rm e}^{-i {\Bf q}\cdot {\Bf\rho}}\nonumber\\
&&\;
=\frac{-1}{z^2} \int \frac{{\rm d}^3q'}{(2\pi)^3}\;
\,\ul{\ol{\varrho}}_{\,\sigma}({\Bf\zeta};{\Bf q}')
\int {\rm d}^3\rho\; 
\frac{v_c^3({\Bf\rho})}{1-v_c({\Bf\rho})/z}\,
{\rm e}^{-i ({\Bf q}-{\Bf q}')\cdot {\Bf\rho}}\nonumber\\
&&\;
= \frac{-4\pi}{z^2}
\left(\frac{e^2}{4\pi\epsilon_0}\right)^3
\int \frac{{\rm d}^3q'}{(2\pi)^3}\;
\frac{\,\ul{\ol{\varrho}}_{\,\sigma}({\Bf\zeta};{\Bf q}')}
{\|{\Bf q}-{\Bf q}'\|}\,
{\tilde g}(\|{\Bf q}-{\Bf q}'\|,\infty;z),\nonumber\\
\end{eqnarray}
where ${\tilde g}(\|{\Bf q}-{\Bf q}'\|,\infty;z) \equiv 
\lim_{R\to\infty} {\tilde g}(\|{\Bf q}-{\Bf q}'\|,R;z)$,
with ${\tilde g}(\|{\Bf q}\|,R;z)$ defined in Eq.~(\ref{ef134}). 
Following the same procedure as in Appendix F (see \S~F.5.b),
we obtain ({\it cf}. Eq.~(\ref{ef138}))
\begin{eqnarray}
\label{eh8}
{\tilde g}(q,\infty;z)
\sim - q \Big\{ (\gamma -1) &+& \ln\Big(\frac{e^2 q}{4\pi
\epsilon_0 \varepsilon_0}\Big) - 
\ln\Big(\frac{-z}{\varepsilon_0}\Big)\Big\},
\nonumber\\
& &\;\;\;\;\;\;\;\;\;\;\;\;\;\;\;
\vert z\vert \to\infty,
\end{eqnarray}
where $\varepsilon_0$ is an arbitrary positive constant energy which 
replaces the $e_R$ in Appendix F. From Eqs.~(\ref{eh7}) and (\ref{eh8}) 
we finally obtain
\begin{eqnarray}
\label{eh9}
&&\hbar\,\ul{\wt{\ol{\Sigma}}}_{\,\sigma;\infty_2}^{\;\rm s_b}
({\Bf q};{\Bf\zeta};z) 
\sim -4\pi \left(\frac{e^2}{4\pi\epsilon_0}\right)^3\nonumber\\
&&\;\;\;\times 
\Big\{ \,\ul{\varrho}_{\,\sigma}({\Bf\zeta},{\Bf\rho}={\bf 0})\,
\frac{\ln\big(-z/\varepsilon_0\big)}{z^2}\nonumber\\
&&\;\;\;\;\;\;\;\;\;
-\Big[ (\gamma-1)\; \ul{\varrho}_{\,\sigma}({\Bf\zeta},{\Bf\rho}={\bf 0})
+\int \frac{{\rm d}^3q'}{(2\pi)^3}\;
\,\ul{\ol{\varrho}}_{\,\sigma}({\Bf\zeta};{\Bf q}')\,\nonumber\\
&&\;\;\;\;\;\;\;\;\;\;\;\;\;\;\;\;\;\;\;\;\;\;
\times\ln\left(\frac{\|{\Bf q}-{\Bf q}'\| 
e^2}{4\pi\epsilon_0\,\varepsilon_0}\right)\Big] 
\frac{1}{z^2}\Big\}, 
\;\;\vert z\vert\to\infty.
\end{eqnarray}
We point out that in the case of {\sl non}-uniform GSs, 
$\ul{\varrho}_{\,\sigma}({\Bf\zeta},{\Bf\rho}={\bf 0}) \not\equiv 
n_{\sigma}({\Bf r})$; however, in the case of systems with uniform and 
isotropic GSs for which $\varrho_{\sigma}({\Bf r}',{\Bf r})$ is a 
function of $\|{\Bf r}-{\Bf r}'\|$, 
\begin{equation}
\label{eh10}
\ul{\varrho}_{\,\sigma}({\Bf\zeta},{\Bf\rho}={\bf 0}) 
\equiv n_{0;\sigma}. 
\end{equation}
From this result and Eq.~(\ref{eh9}), making use of Eq.~(\ref{eh17})
to be presented below, we arrive at the following {\sl exact} 
leading-order contribution for homogeneous systems of fermions with
uniform and isotropic GSs:
\begin{eqnarray}
\label{eh11}
&&\wt{\Ol{\Sigma}}_{\sigma;\infty_2}^{\rm s_b}({\Bf q},{\Bf q}';z) 
\sim -\frac{4\pi}{\hbar}\, \left(\frac{e^2}{4\pi\epsilon_0}\right)^3 
n_{0;\sigma} \frac{\ln\big(-z/\varepsilon_0\big)}{z^2}\,
\delta_{{\Bf q},{\Bf q}'},\nonumber\\
&&\;\;\;\;\;\;\;\;\;\;\;\;\;\;\;\;\;\;\;\;\;\;\;\;\;\;\;\;\;
\;\;\;\;\;\;\;\;\;\;\;\;\;\;\;\;\;\;\;\;\;\;\;\;\;\;\;\;\;\;\;
\vert z\vert\to\infty.
\end{eqnarray}
We point out that, according to our conventions introduced in \S~III.E.2, 
the RHS of Eq.~(\ref{eh11}) is simply the leading contribution to 
$\wt{\Ol{\Sigma}}_{\sigma;\infty_2}^{\rm s_b}({\Bf q},{\Bf q}'\| z)/z^2$ 
(see footnote \ref{f72}); with reference to our convention in this paper 
(see footnote \ref{f31}), we note that $\wt{\Ol{\Sigma}}_{\sigma;
\infty_m}^{\rm s_b}({\Bf q},{\Bf q}'\| z)$ {\sl can} involve a 
$z$-independent contribution, which from Eq.~(\ref{eh9}) can be seen to 
be indeed the case for the $\wt{\Ol{\Sigma}}_{\sigma;\infty_2}^{\rm s_b}
({\Bf q},{\Bf q}'\| z)$ considered here (see Eq.~(\ref{eh18}) below). 
When appropriate, we denote the $z$-independent contribution to 
$\wt{\Ol{\Sigma}}_{\sigma;\infty_m}^{\rm s_b}({\Bf q},{\Bf q}'\| z)$ by 
$\Ol{\Sigma}_{\sigma;\infty_m}^{\rm s_b}({\Bf q},{\Bf q}')$ (see 
Eq.~(\ref{e113}) and footnote \ref{f72}).

For completeness we note that the double Fourier transform
\footnote{\label{f141}
Our convention, involving $1/\Omega$, implies use of box boundary 
condition.}
\begin{equation}
\label{eh12}
\BAr{f}({\Bf q},{\Bf q}') {:=}\frac{1}{\Omega}
\int {\rm d}^3r {\rm d}^3r'\;
{\rm e}^{-i {\Bf q}\cdot {\Bf r}}\,
f({\Bf r},{\Bf r}')\, {\rm e}^{i {\Bf q}'\cdot {\Bf r}'}
\end{equation}
can be expressed in terms of
\begin{equation}
\label{eh13}
\ul{f}\,({\Bf\rho},{\Bf\zeta}) {:=} f({\Bf r},{\Bf r}')
\end{equation}
as follows
\begin{equation}
\label{eh14}
\BAr{f}({\Bf q},{\Bf q}') =\frac{1}{\Omega}
\int {\rm d}^3\rho\, {\rm d}^3\zeta\;
{\rm e}^{-i ({\Bf q}+{\Bf q}')\cdot {\Bf\rho}/2}\,
\ul{f}\,({\Bf\rho},{\Bf\zeta})\,
{\rm e}^{i ({\Bf q}'-{\Bf q})\cdot {\Bf\zeta}}.
\end{equation}
Thus with
\begin{equation}
\label{eh15}
{\Bf Q} {:=} \frac{1}{2} ({\Bf q}'+{\Bf q}),\;\;\;
{\Bf Z} {:=} {\Bf q}'-{\Bf q},
\end{equation}
we have
\begin{equation}
\label{eh16}
\BAr{f}({\Bf q},{\Bf q}')
\equiv\;\Bar{\ul{f}\,}({\Bf Q},{\Bf Z}).
\end{equation}
Making use of this result, the double Fourier transform of 
$\wt{\Sigma}_{\sigma;\infty_2}^{\rm s_b}({\Bf r},{\Bf r}';z)$ is 
obtained from
\begin{equation}
\label{eh17}
\wt{\Ol{\Sigma}}_{\sigma;\infty_2}^{\rm s_b}({\Bf q},{\Bf q}';z)
\equiv \frac{1}{\Omega}\!\int {\rm d}^3\zeta\;
\ul{\wt{\ol{\Sigma}}}_{\,\sigma}^{\;\rm s_b}\big(({\Bf q}+{\Bf q}')/2;
{\Bf\zeta};z\big) \,
{\rm e}^{i ({\Bf q}'-{\Bf q})\cdot {\Bf\zeta}}.
\end{equation}
We point out that, in the case of uniform and isotropic GSs, 
$\,\ul{\wt{\ol{\Sigma}}}_{\,\sigma;\infty_2}^{\;\rm s_b}
({\Bf q};{\Bf\zeta};z)$ is {\sl independent} of ${\Bf\zeta}$ (see 
Eq.~(\ref{eh3}) above); with reference to Eq.~(\ref{eh10}), the RHS of 
Eq.~(\ref{eh9}) is readily verified to obey this expected behaviour. 
The independence with respect to ${\Bf\zeta}$ of 
$\,\ul{\wt{\ol{\Sigma}}}_{\,\sigma;\infty_2}^{\;\rm s_b}({\Bf q};
{\Bf\zeta};z)$ in the case of uniform and isotropic GSs implies that 
$\,\ul{\wt{\Ol{\Sigma}}}_{\,\sigma;\infty_2}^{\;\rm s_b}({\Bf q},
{\Bf q}';z)$ is equal to $\,\ul{\wt{\Ol{\Sigma}}}_{\,\sigma;
\infty_2}^{\;\rm s_b}({\Bf q};{\Bf\zeta};z)$ times $(2\pi)^3 
\delta({\Bf q}-{\Bf q}')/\Omega \equiv \delta_{{\Bf q},{\Bf q}'}$; 
from this, making use of Eqs.~(\ref{eh9}) and (\ref{eh10}), we readily 
obtain ({\it cf}. Eq.~(\ref{eh11}) above)
\begin{eqnarray}
\label{eh18}
&&\wt{\Ol{\Sigma}}_{\sigma;\infty_2}^{\rm s_b}
({\Bf q},{\Bf q}';z) \sim
-\frac{4\pi}{\hbar}\,\left(\frac{e^2}{4\pi\epsilon_0}\right)^3 n_0\,
\Big\{ \frac{n_{0;\sigma}}{n_0}\,
\frac{\ln(-z/\varepsilon_0)}{z^2} \nonumber\\
&&\;\;\;
-\Big[ (\gamma-1) \frac{n_{0;\sigma}}{n_0}
+ \int \frac{{\rm d}^3q''}{(2\pi)^3}\;
\frac{\ol{\varrho}_{\sigma}^{\rm h}(\|{\Bf q}''\|)}{n_0}\,\nonumber\\
&&\;\;\;\;\;\;\;\;\;\;
\times\ln\left(\frac{e^2 \|{\Bf q}-{\Bf q}''\|}{4\pi\epsilon_0
\varepsilon_0}\right)\Big] \frac{1}{z^2} \Big\}\,
\delta_{{\Bf q},{\Bf q}'},\;\;
\vert z\vert\to\infty.
\end{eqnarray}

Considering spin-$1/2$ fermions in the paramagnetic phase, replacing 
$\varrho_{\sigma}({\Bf r}',{\Bf r})$ in Eq.~(\ref{eh2}) by its 
Slater-Fock counterpart as presented in Eq.~(\ref{ef24}), we obtain 
the following expression within the framework of the SSDA ({\it cf}. 
Eq.~(\ref{eh18}) above)
\footnote{\label{f142}
In evaluating the expression in Eq.~(\protect\ref{eh19}) directly 
from Eq.~(\protect\ref{eh18}), one encounters integrals that turn 
out to be extremely unwieldy (with the exception of the case 
corresponding to ${\Bf q}={\bf 0}$). We have obtained the result 
in Eq.~(\protect\ref{eh19}) by determining the asymptotic terms 
{\sl after} evaluation of the Fourier transform of $\wt{\Sigma}_{\sigma;
\infty_2}^{\;\rm s_b}({\Bf r},{\Bf r}';z)$ as presented in 
Eq.~(\protect\ref{eh2}). Although this approach gives rise to somewhat 
lengthy algebra, nonetheless the technique on which our derivations in 
Appendix F.5.b is based made it possible to bypass evaluation of 
integrals that are at least very difficult to express in closed form.}
\begin{eqnarray}
\label{eh19}
&&\left. \wt{\Ol{\Sigma}}_{\sigma;\infty_2}^{\rm s_b}
({\Bf q},{\Bf q}';z)\right|_{\rm s} 
\sim -\frac{4\pi}{\hbar}\,\left(\frac{e^2}{4\pi\epsilon_0}\right)^3 
n_0\, \Big\{ \frac{1}{2}\, \frac{\ln\big(-z/\varepsilon_0\big)}{z^2}
\nonumber\\
&&+\Big( \frac{1}{32} \frac{k_F}{\|{\Bf q}\|}
\Big[ (1-\|{\Bf q}\|/k_F)^3 (3 + \|{\Bf q}\|/k_F)
\ln\big(\frac{e^2 \vert k_F - 
\|{\Bf q}\|\vert}{4\pi\epsilon_0 \varepsilon_0}\big)
\nonumber\\
&&\;\;\;\;\;\;
-(1+\|{\Bf q}\|/k_F)^3 (3 - \|{\Bf q}\|/k_F)
\ln\big(\frac{e^2 \vert k_F + 
\|{\Bf q}\|\vert}{4\pi\epsilon_0 \varepsilon_0}\big) \Big]\nonumber\\
&&\;\;\;\;\;\;
- \big[ (\|{\Bf q}\|/k_F)^2/16 + 
\gamma/2 - 41/48 \big] \Big) \frac{1}{z^2} \Big\}\,
\delta_{{\Bf q},{\Bf q}'},\nonumber\\
&&\;\;\;\;\;\;\;\;\;\;\;\;\;\;\;\;\;\;\;\;\;\;\;\;\;\;\;\;
\;\;\;\;\;\;\;\;\;\;\;\;\;\;\;\;\;\;\;\;\;\;\;\;\;\;\;\;\;
\vert z\vert\to\infty,
\end{eqnarray}
where $\varepsilon_0$ stands for an arbitrary positive constant 
energy. In deducing this result, we have employed $n_{0;\sigma} = 
k_F^3/[6\pi^2]$ for the partial number density of the spin-$1/2$ 
fermions in the paramagnetic state. 

In the calculations of \S~III.I, we require the imaginary part of the 
RHS of Eq.~(\ref{eh18}) for $z = \varepsilon \pm i \eta$, $\eta\downarrow 
0$, as $\varepsilon \to \pm\infty$; with reference to the definition in 
Eq.~(\ref{e65}), from Eq.~(\ref{eh18}) for $\vert\varepsilon\vert\to
\infty$ we readily obtain
\begin{equation}
\label{eh20}
{\rm Re}[\Ol{\Sigma}_{\sigma;\infty_2}^{\rm s_b}
({\Bf q},{\Bf q}';\varepsilon)]\sim 
\frac{-4\pi}{\hbar}\,\left(\frac{e^2}{4\pi\epsilon_0}\right)^3\!\!
n_{0;\sigma}\, \frac{\ln(\vert\varepsilon\vert/\varepsilon_0)}
{\varepsilon^2}\,\delta_{{\Bf q},{\Bf q}'},
\end{equation}
\begin{equation}
\label{eh21}
{\rm Im}[\Ol{\Sigma}_{\sigma;\infty_2}^{\rm s_b}
({\Bf q},{\Bf q}';\varepsilon)] \sim 
\frac{4\pi^2}{\hbar}\,\left(\frac{e^2}{4\pi\epsilon_0}\right)^3\!\!
n_{0;\sigma}\, \frac{\Theta(\varepsilon)}{\varepsilon^2}\,
\delta_{{\Bf q},{\Bf q}'}.
\end{equation}

It is important to point out that the apparent dependence on 
$\varepsilon_0$ of ${\rm Re}[\Ol{\Sigma}_{\sigma;\infty_2}^{\rm s_b}
({\Bf q},{\Bf q}';\varepsilon)]$ in Eq.~(\ref{eh20}) is a consequence 
of {\sl not} presenting the next-to-leading-order term in the AS for 
this function; making use of $(1-\|{\Bf q}\|/k_F)^3 (3+\|{\Bf q}\|/k_F)
-(1+\|{\Bf q}\|/k_F)^3 (3-\|{\Bf q}\|/k_F) = -16\,\|{\Bf q}\|/k_F$, 
as well as $n_{0;\sigma} = k_F^3/[6\pi^2]$ (for uniform and isotropic 
systems of spin-$1/2$ fermions in the paramagnetic GS), from the 
expression on the RHS of Eq.~(\ref{eh19}) we readily deduce that (at 
least) the {\sl value} of $\wt{\Ol{\Sigma}}_{\sigma;\infty_2}^{\rm s_b}
({\Bf q},{\Bf q}';z)\vert_{\rm s}$ in Eq.~(\ref{eh19}) does {\sl not} 
depend on the {\sl value} of $\varepsilon_0$.

Following the conventions in \S~III.E.1, replacing $\varepsilon_0$
by $e_0$ as defined in Eq.~(\ref{e103}), for ({\it cf}. 
Eq.~(\ref{e104}) and see the subsequent text)
\begin{equation}
\label{eh22}
\wt{\Ol{\ol{\Sigma}}}_{\sigma;\infty_2}^{\rm s_b}
({\bar {\Bf q}},{\bar {\Bf q}}';{\bar z}) {:=}
\frac{1}{e_0}\,\hbar
\wt{\Ol{\Sigma}}_{\sigma;\infty_2}^{\rm s_b}
({\Bf q},{\Bf q}';z)
\end{equation}
we obtain (for systems of spin-$1/2$ fermions)
\begin{eqnarray}
\label{eh23}
&&\wt{\Ol{\ol{\Sigma}}}_{\sigma;\infty_2}^{\rm s_b}
({\bar {\Bf q}},{\bar {\Bf q}}';{\bar z}) 
\sim -3\, r_s^3\,
\Big\{ \frac{1}{2}\, \frac{\ln(-{\bar z})}{{\bar z}^2}
-\frac{1}{2}\,\frac{\ln(r_s)}{{\bar z}^2} \nonumber\\
&&\;\;\;
-\Big[ \frac{1}{2} (\gamma-1)
+ \int \frac{{\rm d}^3q''}{(2\pi)^3}\;
\frac{\ol{\varrho}_{\sigma}^{\rm h}(\|{\Bf q}''\|)}{n_0}\,\nonumber\\
&&\;\;\;\;\;\;\;\;\;\;
\times\ln\left(\|{\bar {\Bf q}} - r_0 {\Bf q}''\|\right)\Big] 
\frac{1}{{\bar z}^2} \Big\}\, 
\delta_{{\bar {\Bf q}},{\bar {\Bf q}}'},\;\;
\vert {\bar z}\vert\to\infty,
\end{eqnarray}
where $r_0$ stands for average distance between the particles 
(independent of their spin indices) in the GS defined in 
Eq.~(\ref{e93}). Similarly,
\begin{eqnarray}
\label{eh24}
&&\left. \wt{\Ol{\ol{\Sigma}}}_{\sigma;\infty_2}^{\rm s_b}
({\bar {\Bf q}},{\bar {\Bf q}}';{\bar z})\right|_{\rm s} 
\sim -3\, r_s^3\,
\Big\{ \frac{1}{2}\, \frac{\ln(-{\bar z})}{{\bar z}^2}
-\frac{1}{2}\,\frac{\ln(r_s)}{{\bar z}^2}
\nonumber\\
&&+\Big( \frac{1}{32} \frac{{\bar k}_F}{\|{\bar {\Bf q}}\|}
\Big[ (1-\|{\bar {\Bf q}}\|/{\bar k}_F)^3 
(3 + \|{\bar {\Bf q}}\|/{\bar k}_F)
\ln(\vert {\bar k}_F -\|{\bar {\Bf q}}\|\vert)
\nonumber\\
&&\;\;\;\;\;\;
-(1+\|{\bar {\Bf q}}\|/{\bar k}_F)^3 
(3 - \|{\bar {\Bf q}}\|/{\bar k}_F)
\ln(\vert {\bar k}_F + 
\|{\bar {\Bf q}}\|\vert) \Big]\nonumber\\
&&\;\;\;\;\;\;
-\big[ (\|{\bar {\Bf q}}\|/{\bar k}_F)^2/16 + 
\gamma/2 - 41/48 \big] \Big) \frac{1}{{\bar z}^2} \Big\}\,
\delta_{{\bar {\Bf q}},{\bar {\Bf q}}'},\nonumber\\
&&\;\;\;\;\;\;\;\;\;\;\;\;\;\;\;\;\;\;\;\;\;\;\;\;\;\;\;\;
\;\;\;\;\;\;\;\;\;\;\;\;\;\;\;\;\;\;\;\;\;\;\;\;\;\;\;\;\;
\vert {\bar z}\vert\to\infty.
\end{eqnarray}
For completeness, for $d=3$ and for spin-$1/2$ fermions in the 
paramagnetic phase we have ${\bar k}_F = (9\pi/4)^{1/3}$. Note the 
equality of the first two contributions on the RHS of Eq.~(\ref{eh23}) 
with those on the RHS of Eq.~(\ref{eh24}). Above we have indicated 
that the equality of the first contributions on the RHSs of 
Eqs.~(\ref{eh23}) and (\ref{eh24}) is a direct consequence of the 
relationship in Eq.~(\ref{eh10}). Since, for $r_s\to 0$, 
$\ol{\varrho}_{\sigma}^{\rm h}(\|{\Bf q}''\|) \to \ol{\varrho}_{{\rm s};
\sigma}^{\rm h}(\|{\Bf q}''\|)$, it follows that, as $r_s\to 0$, the 
${\Bf q}''$ integral on the RHS of Eq.~(\ref{eh23}) {\sl cannot} be 
capable of producing a contribution proportional to $\ln(r_s)$ (although 
it may produce a result proportional to $r_s \ln(r_s)$), so that the 
equality of the second contributions on the RHSs of Eqs.~(\ref{eh23}) 
and (\ref{eh24}) {\sl is} exact, independent of the value of $r_s$.
\hfill $\Box$

\section{Two basic integrals}
\label{s64}

In this work we encounter two integrals of the forms
\begin{eqnarray}
\label{ei1}
\left. \begin{array}{ll}
\displaystyle
{\cal I}_1(\zeta';\alpha) {:=}
\int_{\zeta'}^{\infty} {\rm d}z\; \frac{\exp(i\alpha z)}{z} \\ \\
\displaystyle
{\cal I}_2(\zeta';\alpha) {:=}
\int_{\zeta'}^{\infty} {\rm d}z\; \frac{\exp(-i\alpha z)}{z}\\
\end{array} \right\}
\;\;\;\; \alpha > 0,
\end{eqnarray}
where $\zeta'$ is a complex variable. These integrals which enter 
into our calculations through, for instance, the following 
expression
\begin{eqnarray}
\label{ei2}
&&\int_0^{\infty} {\rm d}x\;
\frac{\sin(\alpha x)}{x+\zeta'} 
\equiv \int_{\zeta'}^{\infty} {\rm d}z\;
\frac{\sin(\alpha [z-\zeta'])}{z}\nonumber\\
&&\;\;\;\;\;\;\;\;\;\;\;\;\;
=\frac{1}{2 i} \big\{ 
{\rm e}^{-i\alpha\zeta'}\,
{\cal I}_1(\zeta';\alpha) -
{\rm e}^{i\alpha\zeta'}\,
{\cal I}_2(\zeta';\alpha) \big\},
\end{eqnarray}
can be expressed in terms of the exponential-integral function
(Gradshtyn and Ryzhik 1965, p.~925)
\begin{equation}
\label{ei3}
{\rm Ei}(z) {:=} - \int_{-z}^{\infty} {\rm d}\zeta\; 
\frac{\exp(-\zeta)}{\zeta};
\end{equation}
the contour of integration in the complex $\zeta$ plane is depicted 
in Fig.~1 (for some related considerations see Sansone and Gerretsen 
(1960, pp. 406-412)). This function has two branch points: one at 
$z=0$ and the other at $1/z=0$. The principal branch of ${\rm Ei}(z)$, 
with ${\rm arg}(-z) \in (-\pi,\pi)$, is specified by the requirement 
that ${\rm Ei}(x)$ be real for $x < 0$. In the present work we employ 
{\sl truncated} forms of the following series (Gradshtyn and Ryzhik 
1965, p.~927):
\begin{equation}
\label{ei4}
{\rm Ei}(z) = \gamma + \ln(-z) + \sum_{k=1}^{\infty}
\frac{z^k}{k\, k!}, 
\end{equation} 
in the asymptotic regime $\vert z\vert \to 0$; in our consideration 
we need to employ AS of up to and including $z^5/600$. Here $\gamma 
= 0.577~215~66\dots$ stands for the Euler constant and $\ln$ for the 
principal branch of the logarithm function for which we have $\ln(-z) 
= \ln\vert z\vert + i\, {\rm arg}(-z)$, with ${\rm arg}(-z) \in 
(-\pi,\pi)$. For clarity, the imaginary part of the `principal branch' 
of the logarithm function is equal to the `${\rm arg}$' (or `phase')
of its argument as measured with respect to the positive real axis, 
with `${\rm arg}$' covering the range $(-\pi,\pi)$; the requirement 
concerning measurement of the `${\rm arg}$' with respect to the positive 
real axis is {\sl not} affected by the location of the branch cut 
in the $z$ plane of the functions under consideration; thus 
${\rm arg}(-z)$ is similar to ${\rm arg}(z)$ measured with respect 
to the positive real axis (see Fig.~1), even though the branch cut of 
$\ln(-z)$ is located along the positive real axis, to be contrasted 
with that of $\ln(z)$ which is along the negative real axis. Note in 
passing that $\ln(-z) = \ln(z) \mp i\pi$, ${\rm arg}(z)\, \Ieq<>\, 0$. 

In dealing with ${\cal I}_1(\zeta';\alpha)$ and ${\cal I}_2(\zeta';
\alpha)$, careful account has to be taken of the multi-valued nature 
of ${\rm Ei}(z)$. To appreciate the significance of this, consider 
\begin{equation}
\label{ei5}
{\cal J}_{C}(z) {:=} \int_{C} {\rm d}\zeta\; 
\frac{\exp(-\zeta)}{\zeta},
\end{equation}
where the integration is carried out along the contour $C$ as depicted 
in Fig.~2. As can be seen from this Figure, ${\cal J}_C(z)$ can be 
expressed in terms of ${\rm Ei}(z)$, provided that $C$ is supplemented 
by the contour $\delta^+$, depicted as a broken line. By the Cauchy 
theorem we have
\begin{equation}
\label{ei6}
{\cal J}_C(z) = -2\pi i - {\rm Ei}(z) \equiv -2\pi i 
+ {\cal J}_{C'}(z);
\end{equation} 
the difference $-2\pi i$ between ${\cal J}_C(z)$ and $-{\rm Ei}(z)$ 
is attributable to the branch-cut discontinuity of $\ln(-z)$ in the 
expression on the RHS of Eq.~(\ref{ei4}) for $-z$ along the negative 
real axis.

In order to express ${\cal I}_1(\zeta';\alpha)$ in terms of ${\rm Ei}$, 
we first apply the variable transformation $-i\alpha z \rightharpoonup 
z$, upon which (since we have assumed $\alpha > 0$) $\zeta'$ is rotated 
{\sl clockwise} by $\pi/2$ around the origin while its distance from 
the origin is multiplied by $\alpha$. Since $\zeta'$ is an end-point 
of the contour of integration, the transformation $-i\alpha z 
\rightharpoonup z$ ``drags along'' the contour in the manner shown in 
Figs.~3 and 4. As can be directly seen from Fig.~4, in the case where
${\rm arg}(\zeta')\in (-\pi,-\pi/2)$, the transformation $-i\alpha z 
\rightharpoonup z$ gives rise to $\zeta'$ crossing the negative real 
axis in its change to $-i\alpha\zeta'$. In a manner similar to the 
treatment of ${\cal J}_C(z)$ presented above, we obtain
\begin{eqnarray}
\label{ei7}
&&{\cal I}_1(\zeta';\alpha)
= \left\{ \begin{array}{ll}
-{\rm Ei}(i\alpha \zeta'),\;\; 
&{\rm arg}(\zeta') \in (-\pi/2, \pi)\nonumber \\ \\
-{\rm Ei}(i\alpha\zeta')+2\pi i,\;\;
&{\rm arg}(\zeta') \in(-\pi, -\pi/2).
\end{array} \right.\\
&&
\end{eqnarray}

Along the same lines as presented above, employing, however, the variable 
transformation $i\alpha z \rightharpoonup z$ which for $\alpha > 0$ 
brings about a {\sl counter-clockwise} rotation by $\pi/2$ of $\zeta'$ 
around the origin together with changing the amplitude of $\zeta'$ 
into $\alpha \vert\zeta'\vert$, for ${\cal I}_2(\zeta';\alpha)$ we 
obtain (see Figs.~5 and 6)
\begin{eqnarray}
\label{ei8}
&&{\cal I}_2(\zeta';\alpha)
= \left\{ \begin{array}{ll}
-{\rm Ei}(-i\alpha \zeta'),\;\; 
&{\rm arg}(\zeta') \in (-\pi, \pi/2)\nonumber \\ \\
-{\rm Ei}(-i\alpha\zeta')-2\pi i,\;\;
&{\rm arg}(\zeta') \in(\pi/2, \pi).
\end{array} \right.\\
&&
\end{eqnarray}
\hfill $\Box$

\section{Asymptotic behaviour of the density matrices
pertaining to uniform and isotropic Fermi liquids}
\label{s65}

In several instances in this paper, knowledge of the behaviour of the
single-particle density matrix $\varrho_{\sigma}({\Bf r}',{\Bf r})$,
pertaining to uniform and isotropic GSs, for $\|{\Bf r}-{\Bf r}'\|
\to\infty$ turns out to be of crucial significance. Here we deduce 
the first two leading terms in the large-$r$ AS of 
$\varrho_{\sigma}^{\rm h}(r)$ for systems in $d=3$ ({\it cf}.
Eq.~(\ref{ef9})), with special emphasis on Fermi-liquid metallic states. 

From the definition of the single-particle GF $G_{\sigma}({\Bf r}t,
{\Bf r}'t')$ in Eq.~(\ref{ee7}), it is readily deduced that 
(see Eq.~(\ref{eb16}))
\begin{equation}
\label{ej1}
\varrho_{\sigma}({\Bf r}',{\Bf r}) = -i 
G_{\sigma}({\Bf r}t,{\Bf r}'t^+),
\end{equation}
where $t^+$ is infinitesimally larger than $t$. Specializing to 
uniform and isotropic GSs for which $G_{\sigma}({\Bf r}t,{\Bf r}'t')$ 
is a function of $\|{\Bf r}-{\Bf r}'\|$, with
\begin{equation}
\label{ej2}
{\sf n}_{\sigma}({\Bf k}) {:=}
\langle\Psi_{N;0}\vert 
{\sf\hat a}_{{\Bf k};\sigma}^{\dag}
{\sf\hat a}_{{\Bf k};\sigma} \vert\Psi_{N;0}\rangle
\end{equation}
the GS momentum distribution function ({\it cf}. Eq.~(\ref{ea7})), 
we have
\begin{eqnarray}
\label{ej3}
&&\varrho_{\sigma}^{\rm h}(\|{\Bf r}-{\Bf r}'\|)
=\int \frac{{\rm d}^3k}{(2\pi)^3}\;
{\rm e}^{i {\Bf k}\cdot ({\Bf r}-{\Bf r}')}\,
{\sf n}_{\sigma}(\|{\Bf k}\|)\nonumber\\
&&\;\;\;
= \frac{1}{2\pi^2 \|{\Bf r}-{\Bf r}'\|}\,
\int_0^{\infty} {\rm d}k\; k\, {\sf n}_{\sigma}(k)\,
\sin(k \|{\Bf r}-{\Bf r}'\|), 
\end{eqnarray}
where we have chosen to suppress the subscript `h' in 
${\sf n}_{\sigma}^{\rm h}(k)$ (see Eq.~(\ref{ef9})) that would 
visibly distinguish this {\sl isotropic} function from its more 
general counterpart ${\sf n}_{\sigma}({\Bf k})$; we shall adhere to 
this choice in the remaining part of this Appendix. For completeness, 
according to Yasuhara and Kawazoe (1976), for uniform and isotropic 
electron systems in the paramagnetic phase one has (below ${\sf n}(k) 
{:=} \sum_{\sigma=\uparrow,\downarrow} {\sf n}_{\sigma}(k)$)
\begin{eqnarray}
\label{ej4}
{\sf n}(k) &\sim& \frac{4 m_e^2}{\pi^2 (9\pi)^{4/3} \hbar^4}\,
\left(\frac{e^2}{4\pi\epsilon_0}\right)^2\,
n_0^{-2/3} {\sf g}(0)\, \left(\frac{k_F}{k}\right)^8,
\nonumber\\
& &\;\;\;\;\;\;\;\;\;\;\;\;\;\;\;\;\;\;\;\;\;\;\;\;\;
\;\;\;\;\;\;\;\;
\mbox{\rm for}\;\; k\to\infty,
\end{eqnarray}
where ${\sf g}(0)$ stands for the normalized total pair correlation 
function ${\sf g}(r)$ at $r=0$ (see Appendix F, \S~F.1.d, in particular 
Eq.~(\ref{ef91})); according to Yasuhara (1972), for ${\sf g}(0)$ 
one has the following {\sl approximate} result 
\begin{equation}
\label{ej5}
{\sf g}(0) = \frac{1}{8} \left(\frac{x}{I_1(x)}\right)^2,
\;\;\; x {:=} \frac{4 m_e^{1/2}}{3^{1/6} \pi^{5/6} \hbar}\,
\left(\frac{e^2}{4\pi\epsilon_0}\right)^{1/2}\,
n_0^{-1/6},
\end{equation}
where $I_1(x)$ stands for the modified Bessel function of first kind 
and first order (Abramowitz and Stegun 1972, p.~374). The expression 
on the RHS of Eq.~(\ref{ej4}) shows that the upper boundary of the 
$k$ integral on the RHS of Eq.~(\ref{ej3}) can be reduced from 
$\infty$ to some finite multiple of $k_{F;\sigma}$ without considerably 
affecting the behaviour of $\varrho_{\sigma}^{\rm h}(\|{\Bf r}
-{\Bf r}'\|)$; in fact, as we shall see below, the behaviour of the 
first two leading terms in the AS of this function for $\|{\Bf r}
-{\Bf r}'\|\to\infty$ is fully determined by the behaviour of 
${\sf n}_{\sigma}(k)$ for $k$ in some infinitesimal neighbourhood 
of $k_{F;\sigma}$, the Fermi wavenumber pertaining to fermions with 
spin index $\sigma$. We note that, in the paramagnetic phase, 
${\sf n}(k)$ as calculated within the framework of the random-phase 
approximation (RPA) also decays like $1/k^8$ as $k\to\infty$ (Daniel 
and Vosko 1960). On the other hand, this decay is like $1/k^4$ 
according to the calculations by Belyakov (1961). Although both 
outcomes concern $d=3$, the result by the former workers corresponds 
to $v\equiv v_c$, whereas that by the latter researcher corresponds 
to the extreme limit of a short-range $v$.

Subdividing (in anticipation of the discontinuity of ${\sf n}_{\sigma}
(k)$ at $k=k_{F;\sigma}$) the $k$ integral on the RHS of Eq.~(\ref{ej3}) 
into two, covering the intervals $[0,k_{F;\sigma})$ and $(k_{F;\sigma},
\infty)$, upon twice applying of integration by parts, we obtain
\footnote{\label{f143}
In this Appendix we tacitly assume that, with the exception of $k=
k_{F;\sigma}$, ${\sf n}_{\sigma}(k)$ is {\sl everywhere} continuous 
and, unless explicitly specified otherwise, sufficiently many times 
differentiable. } 
\begin{eqnarray}
\label{ej6}
&&\varrho_{\sigma}^{\rm h}(r)
= -\frac{Z_{F;\sigma}\, k_{F;\sigma}}{2\pi^2}\,
\frac{\cos(k_{F;\sigma}\, r)}{r^2}
+\frac{1}{2\pi^2} \Big\{ Z_{F;\sigma} \nonumber\\
&&\, + k_{F;\sigma} \big[
\left. \frac{\rm d}{{\rm d}k} 
{\sf n}_{\sigma}(k)\right|_{k\uparrow k_{F;\sigma}}\!
-\left. \frac{\rm d}{{\rm d}k} 
{\sf n}_{\sigma}(k)\right|_{k\downarrow k_{F;\sigma}} 
\big]\Big\}\,\frac{\sin(k_{F;\sigma}\, r)}{r^3}\nonumber\\
&&\;\;\;\;\;\;\;\;\;\;\;\,
-\frac{1}{2\pi^2 r^3}\, \wp\!\int_0^{\infty}
{\rm d}k\; \big[ \frac{{\rm d}^2}{{\rm d} k^2}\,
k\, {\sf n}_{\sigma}(k)\big]\, \sin(k\, r),
\end{eqnarray}
where the Cauchy principal-value integral excludes an infinitesimal
interval centred at $k=k_{F;\sigma}$, and (Migdal 1957, Luttinger 1960) 
\begin{equation}
\label{ej7}
Z_{F;\sigma} {:=} {\sf n}_{\sigma}(k_{F;\sigma}-0^+) -
{\sf n}_{\sigma}(k_{F;\sigma}+0^+).
\end{equation}
Provided that the terms enclosed by square brackets (involving the left 
and right derivatives of ${\sf n}_{\sigma}(k)$ at $k=k_{F;\sigma}$) 
as well as the principal-value integral over $[0,\infty)$ exist,
the expression in Eq.~(\ref{ej6}) is an {\sl exact} reformulation 
of that in Eq.~(\ref{ej3}). 
\footnote{\label{f144}
Thus, for instance, by replacing ${\sf n}_{\sigma}(k)$ in 
Eq.~(\protect\ref{ej6}) by ${\sf n}_{{\rm s};\sigma}(k) {:=} 
\Theta(k_{F;\sigma}-k)$ and identifying $Z_{F;\sigma}$ with unity, we 
immediately obtain the exact expression for $\varrho_{{\rm s};
\sigma}^{\rm h}(r)$ in Eq.~(\protect\ref{ef24}). In this connection, 
note that the left and right derivatives of ${\sf n}_{{\rm s};\sigma}
(k)$ with respect to $k$ at $k=k_{F;\sigma}$ are identically vanishing. }
When this is the case, by the Riemann-Lebesgue
lemma (Whittaker and Watson 1927, p.~172) we have
\begin{equation}
\label{ej8}
\wp\!\int_0^{\infty}
{\rm d}k\; \big[ \frac{{\rm d}^2}{{\rm d} k^2}\,
k\, {\sf n}_{\sigma}(k)\big]\, \sin(k\, r)
= o(1)\;\;\; \mbox{\rm for}\;\;\; r\to\infty,
\end{equation}
so that by disregarding the $k$ integral on the RHS of Eq.~(\ref{ej6}), 
we retain an AS of $\varrho_{\sigma}^{\rm h}(r)$ for $r\to\infty$, 
involving the two most leading terms of the complete AS of this function 
(see \S~II.B); compare the first term on the RHS of Eq.~(\ref{ej6}) with 
the RHS of Eq.~(\ref{ef19}) and note that the effect of interaction on the 
leading-order term in the large-$r$ AS for $\varrho_{\sigma}^{\rm h}(r)$ 
is reduction of the magnitude of the corresponding non-interacting term 
by the factor $Z_{F;\sigma}$ (for completeness, compare the leading-order 
term on the RHS of Eq.~(\ref{ej6}) with its counterpart in $d=2$ as 
derived and presented by Farid (2000a)). From the series in Eq.~(\ref{ej6}) 
it is directly evident that the two leading terms in the large-$r$ AS 
for $\varrho_{\sigma}^{\rm h}(r)$ are solely determined by $k_{F;\sigma}$, 
${\sf n}_{\sigma}(k_{F;\sigma}-0^+) - {\sf n}_{\sigma}(k_{F;\sigma}+0^+)$ 
and ${\rm d} {\sf n}_{\sigma}(k)/{\rm d}k\vert_{k\uparrow k_{F;\sigma}} 
-{\rm d} {\sf n}_{\sigma}(k)/{\rm d}k\vert_{k\downarrow k_{F;\sigma}}$. 
Provided that the application of integration by parts to the $k$ 
integral on the RHS of Eq.~(\ref{ej6}) does not lead to any unbounded 
contribution,
\footnote{\label{f145}
We should emphasize that the singular nature of ${\sf n}_{\sigma}
(k)$ at $k=k_{F;\sigma}$ necessitates (and this is signified by the 
principal-value nature of the integral on the RHS of 
Eq.~(\protect\ref{ej6})) that integration by parts be carried out 
separately on the integrals over $[0,k_{F;\sigma}-0^+]$ and 
$[k_{F;\sigma}+0^+,\infty)$. } 
it can be directly verified that the third leading term in the 
large-$r$ AS of $\varrho_{\sigma}^{\rm h}(r)$ is in addition also 
partly determined by ${\rm d} {\sf n}_{\sigma}(k)/{\rm d}k\vert_{k
\downarrow 0}$ which at least in the weak-coupling regime, if 
non-vanishing, should be rather small ({\it cf}. Eq.~(9) in the
paper by Daniel and Vosko (1960)).

Finally, general considerations show (Farid 1999c; see \S~6, and 
in particular the third paragraph)
\footnote{\label{f146}
The considerations by Farid (1999c) are specialized to uniform an 
isotropic systems of spin-less fermions or those in the paramagnetic 
state; however, the pertinent conclusions concerning ${\sf n}(k)$ 
arrived at by Farid (1999c) equally apply to ${\sf n}_{\sigma}(k)$. } 
that, in the case of Fermi liquids, it is in principle possible that 
${\rm d}{\sf n}_{\sigma}(k)/{\rm d}k$ is divergent for both $k\uparrow 
k_{F;\sigma}$ and $k\downarrow k_{F;\sigma}$. Whereas this is not the 
case according to the RPA for ${\sf n}(k)$ as calculated by Daniel 
and Vosko (1960), the ${\sf n}(k)$ due to Belyakov (1961) has the 
following form in the vicinity of $k=k_F$ (see text following 
Eq.~(\ref{ea70}) in Appendix A)
\begin{equation}
\label{ej9}
{\sf n}(k) \sim {\sf n}(k_F \mp 0^+) + {\sf A}\, (k/k_F -1)
\ln\vert k/k_F -1\vert,\;
\begin{array}{l}
k\uparrow k_F \\
k\downarrow k_F
\end{array} 
\end{equation} 
where ${\sf A}$ stands for a (positive) constant. From this expression, 
one deduces that 
\begin{eqnarray}
\frac{{\rm d}{\sf n}(k)}{{\rm d}k} \sim \frac{{\sf A}}{k_F}\, 
\big\{ \ln\vert k/k_F - 1\vert + 1 \big\},\;\;\;
k\to k_F, \nonumber
\end{eqnarray}
and that 
\begin{eqnarray}
\frac{{\rm d}^2{\sf n}(k)}{{\rm d}k^2} \sim 
\frac{{\sf A}/k_F}{k-k_F},\;\;\; 
k\to k_F, \nonumber
\end{eqnarray}
so that not only the (divergent) contributions enclosed by square 
brackets on the RHS of Eq.~(\ref{ej6}) {\sl entirely} cancel, but
also the integral on the RHS of Eq.~(\ref{ej6}) is bounded. Thus 
Eq.~(\ref{ej6}) indeed also applies to such (unconventional) 
Fermi-liquid metallic states as that dealt with by Belyakov (1961).
\hfill $\Box$

\section{On the differentiability property of the ground-state 
partial number densities}
\label{s66}

In \S~III.H.2, we encounter the {\sl unbounded} function
${\cal T}_{\sigma,\bar\sigma}({\Bf r})$ (see Eq.~(\ref{e209})) 
whose regularization, with which we deal in Appendix G, results in 
$\wt{\sf T}_{\sigma,\bar\sigma}({\Bf r};z)$. The leading asymptotic 
term of this function for $\vert z\vert\to\infty$, namely 
$\wt{\sf T}_{\sigma,\bar\sigma;\infty_2}^{\rm s}({\Bf r}\| z)$ 
(see Eqs.~(\ref{eg1}), (\ref{eg11}) and (\ref{eg15})),
\footnote{\label{f147}
For {\sl constant} number densities $\{ n_{\sigma}\}$, 
$\wt{\sf T}_{\sigma,\bar\sigma;\infty_2}^{\rm s}({\Bf r}\| z)/z^2$ 
{\sl identically} coincides with $\wt{\sf T}_{\sigma,\bar\sigma}
({\Bf r};z)$ (see Eq.~(\protect\ref{eg8})). }
contributes to $\wt{\Sigma}_{\sigma;\infty_2}^{\rm s}({\Bf r},{\Bf r}'
\| z)$ presented in Eq.~(\ref{e213}). In Appendix G we discuss the
significance of the value for the smallest integer $M$ for which 
$\tau^M({\Bf r}) [ n_{\bar\sigma}({\Bf r}) - n_{\sigma}({\Bf r})]$ 
becomes non-integrably unbounded for {\sl some} ${\Bf r}$ (see criterion 
(B) in \S~II.B). Here we investigate the differentiability property of 
the partial number densities $\{ n_{\sigma}({\Bf r})\}$ with respect 
to ${\Bf r}$. In doing so, we pay especial attention to the case where 
the ionic potential $u({\Bf r})$ is a linear superposition of 
electrostatic Coulomb potentials (in $d=3$) centred at the ionic 
positions $\{ {\Bf R}_j \}$. For this case we demonstrate that 
$\tau({\Bf r}) n_{\sigma}({\Bf r}) \sim 2 Z_j v_c({\Bf r}-{\Bf R}_j) 
n_{\sigma}({\Bf R}_j)$ for ${\Bf r} \to {\Bf R}_j$ (see Eqs.~(\ref{ek6}) 
and (\ref{ek16}) below) where $Z_j > 0$ stands for the effective 
atomic number 
\footnote{\label{f148}
By `effective' we make provision for cases where $Z_j$ may differ 
from the number of protons constituting the nucleus of the atom at 
${\Bf R}_j$ by the number of electrons that may have formed a tightly 
bound closed shell centred at ${\Bf R}_j$ and as a result have not
been taken account of in determining $N$ in Eq.~(\protect\ref{e20}). }  
of the ion at ${\Bf R}_j$; although divergent like $1/\|{\Bf r}
-{\Bf R}_j\|$ for ${\Bf r} \to {\Bf R}_j$, $\tau({\Bf r}) n_{\sigma}
({\Bf r})$ is integrable. Consequently we deduce that the second term 
in the large-$\vert z\vert$ AS for the Fourier transform with respect
to ${\Bf r}$ of $\wt{\sf T}_{\sigma,\bar\sigma}({\Bf r};z)$ decays 
more rapidly than $1/\vert z\vert^2$ (see Eq.~(\ref{eg10})).

Applying $\tau({\Bf r})$ to both sides of the defining expression 
for $n_{\sigma}({\Bf r})$ (see Eq.~(\ref{e163})),
\begin{equation}
\label{ek1}
n_{\sigma}({\Bf r}) {:=}
\langle\Psi_{N;0}\vert \hat\psi_{\sigma}^{\dag}({\Bf r})
\hat\psi_{\sigma}({\Bf r})\vert\Psi_{N;0}\rangle,
\end{equation}
while making use of chain rule of differentiation and
\begin{eqnarray}
\label{ek2}
&&\langle\Psi_{N;0}\vert 
\big[\tau({\Bf r}) \hat\psi_{\sigma}^{\dag}({\Bf r})\big]
\hat\psi_{\sigma}({\Bf r})\vert\Psi_{N;0}\rangle \nonumber\\
&&\;\;\;\;\;\;
= \big( \langle\Psi_{N;0}\vert \hat\psi_{\sigma}^{\dag}({\Bf r})
\tau({\Bf r}) \hat\psi_{\sigma}({\Bf r})
\vert\Psi_{N;0}\rangle\big)^*,
\end{eqnarray}
we deduce
\begin{eqnarray}
\label{ek3}
&&\tau({\Bf r}) n_{\sigma}({\Bf r}) 
=2 {\rm Re}\big[\langle\Psi_{N;0}\vert 
\hat\psi_{\sigma}^{\dag}({\Bf r}) \tau({\Bf r}) 
\hat\psi_{\sigma}({\Bf r}) \vert\Psi_{N;0} \rangle\big]
\nonumber\\
&&\;\;\;\;\;\;
-\frac{\hbar^2}{m_e}
\langle\Psi_{N;0}\vert 
\big[{\Bf\nabla}_{\Bf r}
\hat\psi_{\sigma}^{\dag}({\Bf r})\big]\cdot 
\big[{\Bf\nabla}_{\Bf r}\hat\psi_{\sigma}({\Bf r})\big] 
\vert\Psi_{N;0}\rangle.
\end{eqnarray}
Making use of (see Eqs.~(\ref{e3}) and (\ref{eb16})) 
\begin{eqnarray}
\label{ek4}
\langle\Psi_{N;0}\vert \hat\psi_{\sigma}^{\dag}({\Bf r})
\tau({\Bf r}) \hat\psi_{\sigma}({\Bf r})
\vert\Psi_{N;0}\rangle = \lim_{{\Bf r}'\to {\Bf r}}
\tau({\Bf r})\, \varrho_{\sigma}({\Bf r}',{\Bf r})
\end{eqnarray} 
and Eq.~(\ref{ee13}), from Eqs.~(\ref{ek3}) and (\ref{e43}) we deduce 
\footnote{\label{f149}
We have $\Gamma^{(2)}({\Bf r}\sigma,{\Bf r}''\sigma';{\Bf r}\sigma,
{\Bf r}''\sigma') \equiv n_0^2\, {\sf g}_{\sigma,\sigma'}({\Bf r},
{\Bf r}'')$ (see Eq.~(\protect\ref{eb21})). For the large-$\|{\Bf r}
-{\Bf r}''\|$ asymptotic behaviour of ${\sf g}_{\sigma,\sigma'}({\Bf r},
{\Bf r}'')$ pertaining to uniform and isotropic systems see Appendix 
F, \S~F.1.d. We further note that the third term on the RHS of 
Eq.~(\protect\ref{ek5}) is identical with $2 {\cal B}_{\sigma}''({\Bf r},
{\Bf r})$ (see Eqs.~(\protect\ref{ef98}) and (\protect\ref{ef97})) and 
the fourth term identical with $2 {\cal D}_{\sigma}({\Bf r},{\Bf r})$ 
(see Eq.~(\protect\ref{ee13})). }
(below $\eta\downarrow 0$)
\begin{eqnarray}
\label{ek5}
&&\tau({\Bf r})\, n_{\sigma}({\Bf r}) 
= -2 u({\Bf r})\, n_{\sigma}({\Bf r}) - 2 v_H({\Bf r};[n])\,
n_{\sigma}({\Bf r}) \nonumber\\
&&- 2 \int {\rm d}^dr''\; v({\Bf r}-{\Bf r}'')
\sum_{\sigma'} \Big\{
\Gamma^{(2)}({\Bf r}\sigma,{\Bf r}''\sigma';
{\Bf r}\sigma,{\Bf r}''\sigma')\nonumber\\
&&\;\;\;\;\;\;\;\;\;\;\;\;\;\;\;\;\;\;\;\;\;\;\;\;\;\;\;\;\;\;
\;\;\;\;\;\;\;\;\;\;\;\;\;\;\;\;\;\;\;\;\;\;\;\,
- n_{\sigma'}({\Bf r}'') n_{\sigma}({\Bf r}) \Big\}\nonumber\\
&&+ \frac{2}{\hbar} \int_{-\infty}^{\infty}
\frac{{\rm d}\varepsilon}{2\pi i}\;
{\rm e}^{i\varepsilon \eta/\hbar}\,
\varepsilon \, G_{\sigma}({\Bf r},{\Bf r};
\varepsilon)\nonumber\\
&&-\frac{\hbar^2}{m_e}\,
\langle\Psi_{N;0}\vert
\big[{\Bf\nabla}_{{\Bf r}}
\hat\psi_{\sigma}^{\dag}({\Bf r})\big]\cdot
\big[{\Bf\nabla}_{{\Bf r}}
\hat\psi_{\sigma}({\Bf r})\big]\vert\Psi_{N;0}\rangle.
\end{eqnarray}
We have arranged the terms in this expression in such a way that it 
is also applicable to $d=3$ and $v\equiv v_c$, where $\varpi_{\kappa}$ 
as arising from $v_H({\Bf r};[n]) \equiv \varpi_{\kappa} + 
v_H({\Bf r};[n'])$ (see Eqs.~(\ref{e15}) and (\ref{e5})) exactly 
cancels that from $u({\Bf r})$ (see Eq.~(\ref{e11})). We therefore 
do {\sl not} need to deal separately with the cases corresponding 
to $v\not\equiv v_c$ and $v\equiv v_c$ in $d=3$. The appearance of 
$u({\Bf r})$ on the RHS of Eq.~(\ref{ek5}) implies that $\tau({\Bf r}) 
n_{\sigma}({\Bf r})$ must be unbounded at ${\Bf r}={\Bf r}_0$ if and 
only if $u({\Bf r})$ is so at ${\Bf r}={\Bf r}_0$. The validity of 
this assertion relies on the consideration that {\sl none} of the 
contributions on the RHS of Eq.~(\ref{ek5}), except the first, 
can be unbounded. 

Below we demonstrate the boundedness of all the contributions on the 
RHS of Eq.~(\ref{ek5}) that follow the first term. For definiteness,
we assume $d=3$ and $v\equiv v_c$ and further that the electron-ion 
interaction is of the following form (for $Z_j$ see above)
\begin{equation}
\label{ek6}
u({\Bf r}) = \sum_{j} -Z_j\, v_c({\Bf r}-{\Bf R}_j).
\end{equation} 
We therefore view the $v_H({\Bf r};[n])$ on the RHS of Eq.~(\ref{ek5}) as 
being replaced by $v_H({\Bf r};[n'])$ and $u({\Bf r})$ as not containing 
$\varpi_{\kappa}$ (see above). One readily verifies that the second, 
third and fourth terms on the RHS of Eq.~(\ref{ek5}) are bounded 
{\sl everywhere}. For completeness, the above-mentioned third and 
fourth terms are identical with $2 {\cal B}_{\sigma}''({\Bf r},
{\Bf r})$ (see Eq.~(\ref{ef98})) and $2 {\cal D}_{\sigma}({\Bf r},
{\Bf r})$ (see Eq.~(\ref{ee5})) respectively. 

In order to investigate the behaviour of the last term on the RHS 
of Eq.~(\ref{ek5}), we need to establish the behaviour of 
${\Bf\nabla}_{{\Bf r}}\hat\psi_{\sigma} ({\Bf r})$. To this end we 
point out that (as mentioned above, here we consider $d=3$) with
\begin{equation}
\label{ek7}
{\sf G}({\Bf r}-{\Bf r}') \equiv 
\frac{-1}{4\pi\, \|{\Bf r}-{\Bf r}'\|},
\end{equation}
\begin{equation}
\label{ek8}
f({\Bf r}) = \int {\rm d}^3r'\; {\sf G}({\Bf r}-{\Bf r}')
\, g({\Bf r}')
\end{equation}
is the solution of
\begin{equation}
\label{ek9}
\nabla^2_{{\Bf r}} f({\Bf r}) = g({\Bf r}).
\end{equation}
Making use of ${\Bf\nabla}_{{\Bf r}} {\sf G}({\Bf r}-{\Bf r}') \equiv
-{\Bf\nabla}_{{\Bf r}'} {\sf G}({\Bf r}-{\Bf r}')$, on application 
of the divergence theorem from Eq.~(\ref{ek8}) we obtain
\begin{equation}
\label{ek10}
{\Bf\nabla}_{{\Bf r}} f({\Bf r})
= \int {\rm d}^3r'\; {\sf G}({\Bf r}-{\Bf r}')\,
{\Bf\nabla}_{{\Bf r}'} g({\Bf r}'). 
\end{equation}
From the Heisenberg equation of motion for the annihilation operator 
$\hat\psi_{\sigma}({\Bf r}t)$ (see Eq.~(\ref{ee6}); see also footnote 
\ref{f44}) we have
\begin{eqnarray}
\label{ek11}
&&\tau({\Bf r}) \hat\psi_{\sigma}({\Bf r})
= i\hbar\left. \frac{\rm d}{{\rm d} t} 
\hat\psi_{\sigma}({\Bf r}t)\right|_{t=0}
- u({\Bf r}) \hat\psi_{\sigma}({\Bf r})
- \wh{U}_{\sigma}({\Bf r} 0),\nonumber\\
\end{eqnarray} 
where
\begin{eqnarray}
\label{ek12}
&&\wh{U}_{\sigma}({\Bf r}t) {:=}
\sum_{\sigma'} \int {\rm d}^dr''\;
v({\Bf r}-{\Bf r}'')\,
\hat\psi_{\sigma'}^{\dag}({\Bf r}''t) 
\hat\psi_{\sigma'}({\Bf r}''t)
\hat\psi_{\sigma}({\Bf r}t),\nonumber\\
\end{eqnarray}
which in the case of $d=3$ and $v\equiv v_c$, is a bounded operator. 
With $\tau({\Bf r})$ as defined in Eq.~(\ref{e3}), from Eq.~(\ref{ek11})
in conjunction with Eq.~(\ref{ek10}) we readily obtain
\begin{eqnarray}
\label{ek13}
{\Bf\nabla}_{\Bf r} \hat\psi_{\sigma}({\Bf r})
&=&\left. \frac{2 m_e}{i \hbar} \frac{\rm d}{{\rm d} t}
\int {\rm d}^3r'\; {\sf G}({\Bf r}-{\Bf r}')\,
{\Bf\nabla}_{{\Bf r}'} 
\hat\psi_{\sigma}({\Bf r}'t)\right|_{t=0} \nonumber\\
&+&\frac{2 m_e}{\hbar^2} \int {\rm d}^3r'\;
{\sf G}({\Bf r}-{\Bf r}') {\Bf\nabla}_{{\Bf r}'}\,
[u({\Bf r}') \hat\psi_{\sigma}({\Bf r}')]\nonumber\\
&+&\frac{2 m_e}{\hbar^2} \int {\rm d}^3r'\;
{\sf G}({\Bf r}-{\Bf r}') {\Bf\nabla}_{{\Bf r}'}\,
\wh{U}_{\sigma}({\Bf r}'0).
\end{eqnarray}

With $u({\Bf r})$ as defined in Eq.~(\ref{ek6}), from Eq.~(\ref{ek13}) 
we observe that the most singular contribution to ${\Bf\nabla}_{\Bf r}
\hat\psi_{\sigma}({\Bf r})$ is due to the following function which 
originates from the second term on the RHS of Eq.~(\ref{ek13}); making 
use of Eqs.~(\ref{ek6}) and (\ref{ek7}), for this contribution we have
\begin{eqnarray}
\label{ek14}
&&\int {\rm d}^3r'\, {\sf G}({\Bf r}-{\Bf r}')
[{\Bf\nabla}_{{\Bf r}'} u({\Bf r}')] \hat\psi_{\sigma}({\Bf r}')
= \frac{-e^2}{(4\pi)^2 \epsilon_0} \sum_j Z_j \nonumber\\
&&\;\;\;\;\;\;\;
\times\int {\rm d}^3r'\; \frac{1}{\|{\Bf r}-{\Bf r}'-{\Bf R}_j\|}\,
\frac{{\Bf r}'}{\|{\Bf r}'\|^3}\,
\hat\psi_{\sigma}({\Bf r}'+{\Bf R}_j).
\end{eqnarray}
The boundedness of this expression is established by demonstrating
that for ${\Bf r}\to {\Bf R}_J$, for some $J$, the contribution of 
the most singular integral on the RHS of Eq.~(\ref{ek14}), namely 
that corresponding to $j=J$, is bounded. This is achieved by equating
${\Bf r}$ with ${\Bf R}_J$ and considering a small spherical volume 
$V_0$ centred around the origin ${\Bf r}'={\bf 0}$, for which holds:
\begin{eqnarray}
\label{ek15}
&&\int_{V_0} {\rm d}^3r'\; \frac{{\Bf r}'}{\|{\Bf r}'\|^4}\,
\hat\psi_{\sigma}({\Bf r}'+{\Bf R}_J)
\approx \hat\psi_{\sigma}({\Bf R}_J)\nonumber\\
&&\;\;\;\;\;\;\;\;\;\;\;\;\;\;\;\;\;\;\;\;\;\;\;\;\;\;\;\;\;\;\;\;\;
\times\int_{V_0} {\rm d}^3r'\; \frac{{\Bf r}'}{\|{\Bf r}'\|^4} 
\equiv {\bf 0}.
\end{eqnarray}
Hereby we have demonstrated that the last term on the RHS of
Eq.~(\ref{ek5}) is bounded.

We have thus shown that $\tau({\Bf r}) n_{\sigma}({\Bf r})$ is 
unbounded {\sl only} there where $u({\Bf r})$ is unbounded, that 
is at the set of points $\{{\Bf R}_j\}$. For ${\Bf r}$ in the
close vicinity of ${\Bf R}_j$, from Eq.~(\ref{ek5}) we therefore
deduce the following asymptotic expressions:
\begin{eqnarray}
\label{ek16}
\tau({\Bf r}) n_{\sigma}({\Bf r}) 
\sim -2 u({\Bf r}) n_{\sigma}({\Bf r})
&\sim& -2 u({\Bf r}) n_{\sigma}({\Bf R}_j),\nonumber\\
& &\;\;\mbox{\rm when}\;\; {\Bf r}\to {\Bf R}_j.
\end{eqnarray}
Expressing ${\Bf r}$ in terms of its spherical polar coordinates 
$(r,\varphi,\theta)$, with the origin being centred at ${\Bf r}
={\Bf R}_j$, the asymptotic differential equation in Eq.~(\ref{ek16}) 
implies that
\begin{equation}
\label{ek17}
\left. \frac{\rm d}{{\rm d} r} 
n_{\sigma}({\Bf r})\right|_{{\Bf r}={\Bf R}_j} =
- \frac{m_e e^2\, Z_j}{2\pi\epsilon_0\hbar^2}\, 
n_{\sigma}({\Bf R}_j), 
\end{equation}
from which we deduce that
\begin{eqnarray}
\label{ek18}
n_{\sigma}({\Bf r}) \sim n_{\sigma}({\Bf R}_j) 
&-&\frac{m_e e^2\, Z_j\, n_{\sigma}({\Bf R}_j)}
{2\pi\epsilon_0\hbar^2}\,
\|{\Bf r}-{\Bf R}_j\|\nonumber\\
& &\;\;\;\;\;\;\;\;\;\;\;\;\;\;\;\;\;\;\;\;
\mbox{\rm when}\;\; {\Bf r}\to {\Bf R}_j.
\end{eqnarray}
The results in Eqs.~(\ref{ek17}) and (\ref{ek18}) coincide with those 
obtained earlier by other workers (Kato 1957, Bingel 1963, Steiner 1963, 
Pack and Brown 1966) and are referred to as the {\sl cusp} conditions 
for $n_{\sigma}({\Bf r})$. Our approach in this Appendix is more 
transparent than those in the latter references. 
\hfill $\Box$

\end{appendix}
 
\vfill
\pagebreak

{\bf\sc\underline{Some frequently-used notation}} 
\label{s67}
\vskip 10pt
\begin{tabular}{lll}
$\sim$ & & $f(x) \sim g(x)$ for $x\to x_0$ (say, $x_0 = \infty$)\\
${}$ & & implies $f(x)/g(x) \to 1$ for $x\to x_0$\\
$o$ & & A Landau's symbol (E.~G.~H. Landau);\\
${}$ & & $f(x) = o\big(g(x)\big)$ for $x\to x_0$ implies \\
${}$ & & $f(x)/g(x) \to 0$ for $x\to x_0$\\
${\cal O}$ & & A Landau's symbol (E.~G.~H. Landau);\\
${}$ & & $f(x) = {\cal O}\big(g(x)\big)$, there exists a constant\\
${}$ & & $C$ such that $\vert f(x)\vert
\leq C \vert g(x)\vert$\\
$a\to b$ & & $a$ approaches $b$\\
$a\rightharpoonup b$ & & $b$ substitutes $a$\\
$a\rightleftharpoons b$ & & $a$ and $b$ are interchanged\\
$\|.\|$ & & Cartesian norm in $d$ dimensions\\
${\rm s}$ & & Used as {\sl subscript}, indicative of SSDA;\\
${}$ & & e.g., $\varrho_{{\rm s};\sigma}$ denotes the SSDA 
of $\varrho_{\sigma}$\\
$s$ & & A compound index, characterizing the\\
${}$ & & single-particle excitations of the \\
${}$ & & {\sl interacting} system; {\it cf}. 
$f_{s;\sigma}({\Bf r})$ \\
${}$ & & (see Eq.~(\ref{e18}))\\
$\varsigma$ & & A compound index, in general characterizing\\ 
${}$ & & the single-particle excitations of the \\
${}$ & & {\sl non}-interacting system; {\it cf}. 
$\varphi_{\varsigma;\sigma}({\Bf r})$ (see\\
${}$ & & Eq.~(\ref{e56}))\\
${\sf s}$ & & Spin of fermions, with multiplicity 
$2 {\sf s}+1$\\
$\sigma$ & & A spin index (one out of $2 {\sf s}+1$ indices)\\
$\bar\sigma$ & & The {\sl set} of $2 {\sf s}$ spin indices 
complimentary\\
${}$ & & to $\sigma$; for ${\sf s}=\frac{1}{2}$, 
$\sigma=\uparrow$ implies $\bar\sigma=\downarrow$\\
$N$ & & $\sum_{\sigma} N_{\sigma}$, total number of fermions 
in the GS\\
$N_{\sigma}$ & & Number of fermions whose spin index is $\sigma$\\
$N_{\bar\sigma}$ & & $N_{\bar\sigma} = N - N_{\sigma}
\equiv \sum_{\sigma'\not=\sigma} N_{\sigma'}$\\
$d$ & & The dimension of the spatial space\\ 
$u$ & & The local external potential\\
$v$ & & The two-body interaction potential\\
$v_c$ & & The two-body Coulomb potential in $d=3$\\
$\varepsilon$ & & Real-valued energy parameter\\
$z$ & & Complex-valued energy parameter; unless\\
${}$ & & otherwise stated, ${\rm Im}(z) \not= 0$\\
$\widetilde{f}(z)$ & & Analytic continuation of 
$f(\varepsilon)$ into the\\
${}$ & & {\sl physical} Riemann sheet;
{\it cf}. $\Sigma_{\sigma}(\varepsilon)$ and 
$\widetilde{\Sigma}_{\sigma}(z)$\\
${\bar {\sf f}}$ & & Fourier transform with respect to ${\Bf r}$
of ${\sf f}({\Bf r})$\\
$\Bar{\sf f}$ & & Normalized ${\bar {\sf f}}$, according to
$\Bar{\sf f} {:=} {\bar {\sf f}}/e_0$ where\\ 
${}$ & & $e_0 {:=} (2/r_s^2)$ Ry (see Eq.~(\ref{e103}))\\
$\Bar{\sf g}$ & & Fourier transform with respect to ${\Bf r}$
and ${\Bf r}'$\\
${}$ & & of ${\sf g}({\Bf r},{\Bf r}')$\\
$\Bar{\bar{\sf g}}$ & & Normalized $\Bar{\sf g}$, according to
$\Bar{\bar{\sf g}} {:=} \Bar{\sf g}/e_0$ where \\
${}$ & & $e_0 {:=} (2/r_s^2)$ Ry (see Eq.~(\ref{e103}))\\
$\,\ul{\sf h}\,$ & & The `Wigner transform' of 
${\sf h}({\Bf r},{\Bf r}')$;\\
${}$ & & $\,\ul{\sf h}\,({\Bf\rho},{\Bf\zeta}) \equiv 
{\sf h}({\Bf r},{\Bf r}')$, where ${\Bf\rho}\equiv 
{\Bf r}-{\Bf r}'$,\\
${}$ & & ${\Bf\zeta} \equiv \frac{1}{2} ({\Bf r}+{\Bf r}')$\\
$F^{\rm h}$ & & $F^{\rm h}(\|{\Bf r}-{\Bf r}'\|) 
\equiv F({\Bf r},{\Bf r}')$ corresponding to\\
${}$ & & homogeneous {\sl and} isotropic GSs; 
$F^{\rm h}\equiv F^{\rm hi}$\\
$F^{\rm h\neg i}$ & & $F^{\rm h\neg i}({\Bf r}-{\Bf r}')
\equiv F({\Bf r},{\Bf r}')$ corresponding to\\
${}$ & & homogeneous but {\sl not} isotropic states\\
\end{tabular}

\vskip 35pt

{\bf\sc\underline{List of some abbreviations}}
\label{s68}
\vskip 10pt
\begin{tabular}{lll}
AS & & Asymptotic series\\
GF & & Green function; if not explicitly specified, \\
${}$ & &the single-particle GF\\
GS & & Ground state\\
LHS & & Left-hand side\\
RHS & & Right-hand side\\
RPA & & Random-phase approximation\\
SE & & Self-energy\\
SSD & & Single-Slater-determinant\\
SSDA & & Single-Slater-determinant approximation\\
\end{tabular}

\vskip 25pt

%
\hrule
\vspace{0.35cm}

\noindent
{\small ${}^{\dagger}${\tt Electronic address: 
bf10000@phy.cam.ac.uk}}\\
{\small ${}^{\ddagger}${\tt Electronic address: 
B.Farid@phys.uu.nl}}\\
\vspace{0.0cm}
\subsection*{References}
\label{s69}
\begin{verse}
{\small

{\sc Abramowitz}, M., and {\sc Stegun}, I.~A. (editors), 1972, 
{\sl Handbook of Mathematical Functions} (New York: Dover
Publications).\\

{\sc Almbladh}, C.-O., and {\sc Hedin}, L., 1983, in
{\sl Handbook on Synchrotron Radiation}, Vol.~1B, edited by
E.~E. Koch (Amsterdam: North-Holland), 607.\\

{\sc Anderson}, P.~W., 1959,
{\it Phys. Rev.} {\bf 115}, 2;
1961, {\it Phys. Rev.}, {\bf 124}, 41;
1987, {\it Science}, {\bf 235}, 1196.\\

{\sc Aryasetiawan}, F., and {\sc Gunnarsson}, O., 1995, 
{\it Phys. Rev. Lett.}, {\bf 74}, 3221.\\

{\sc Aryasetiawan}, F., {\sc Hedin}, L., and {\sc Karlsson},
K., 1996, {\it Phys. Rev. Lett.}, {\bf 77}, 2268.\\

{\sc Aryasetiawan}, F., and {\sc Karlsson}, K., 1996,
{\it Phys. Rev.} B, {\bf 54}, 5353.\\

{\sc Ashcroft}, N.~W., and {\sc Mermin}, N.~D., 1981,
{\sl Solid State Physics} (Philadelphia, Pennsylvania: 
Holt-Saunders).\\

{\sc Auerbach}, A., 1993, in
{\sl Correlated Electron Systems}, edited by V.~J. Emery
(Singapore: World Scientific), 156.\\

{\sc Belyakov}, V.~A., 1961, {\it Soviet Phys. JETP},
{\bf 13}, 850.\\

{\sc Bingel}, W.~A., 1963, {\it Z. Naturfo.} (a), 
{\bf 18a}, 1249.\\

{\sc Brinkman}, W.~F., and {\sc Rice}, T.~M., 1970a,
{\it Phys. Rev.} B, {\bf 2}, 1324;
1970b, {\it ibid.}, {\bf 2}, 4302.\\

{\sc Broer}, L.~J., 1943, {\it Physica}, {\bf 10}, 801.\\

{\sc Ceperley}, D.~M., and {\sc Alder}, B.~J., 1980,
{\it Phys. Rev. Lett.}, {\bf 45}, 566.\\

{\sc Collins}, J.~C., 1984, {\sl Renormalisation} 
(Cambridge University Press).\\

{\sc Copson}, E.~T., 1965, {\sl Asymptotic Expansions}
(Cambridge University Press).\\

{\sc Cornwell}, J.~F., 1984, {\sl Group Theory in Physics}, Vol. I
(London: Academic Press).\\

{\sc Daniel}, E., and {\sc Vosko}, S.~H., 1960, 
{\it Phys. Rev.}, {\bf 120}, 2041.\\

{\sc Debnath}, L., and {\sc Mikusi\'nski}, P., 1990,
{\sl Introduction to Hilbert Spaces with Applications}
(Boston, Massachusetts: Academic Press).\\

{\sc Deisz}, J.~J., {\sc Hess}, D.~W., and {\sc Serene}, J.~W., 
1997, {\it Phys. Rev.} B, {\bf 55}, 2089. \\ 

{\sc Dingle}, R.~B., 1973, {\sl Asymptotic Expansions: Their 
Derivation and Interpretation} (London: Academic Press).\\

{\sc Dreizler}, R.~M., and {\sc Gross}, E.~K.~U., 1990, 
{\sl Density Functional Theory} (Berlin: Springer).\\

{\sc Engel}, G.~E., and {\sc Farid}, B., 1993,
{\it Phys. Rev.} B, {\bf 47}, 15931.\\

{\sc Engel}, G.~E., {\sc Farid}, B., {\sc Nex}, C.~M.~M.,
and {\sc March}, N.~H., 1991, {\it Phys. Rev.} B, {\bf 44},
13356.\\

{\sc Eskes}, H., and {\sc Ole\'s}, A.~M., 1994,
{\it Phys. Rev. Lett.}, {\bf 73}, 1279.\\

{\sc Eskes}, H., {\sc Ole\'s}, A.~M., {\sc Meinders}, M.~B.~J., 
and {\sc Stephan}, W., 1994, {\it Phys. Rev.} B, {\bf 50}, 17980.\\

{\sc Farid}, B., 1997a, {\it Phil. Mag.} B, {\bf 76}, 145;
1997b, {\it Solid St. Commun.}, {\bf 104}, 227;
1998, {\it J. Phys.} C, {\bf 10}, L1;
1999a, {\sl Electron Correlation in the Solid State},
edited by N.~H. March (London: Imperial College Press), chapter 3;
1999b, {\it Phil. Mag. Lett.}, {\bf 79}, 581;
1999c, {\it Phil. Mag.} B, {\bf 79}, 1097;
2000a, {\it Phil. Mag.} B, {\bf 80}, 1599;
2000b, {\it Phil. Mag.} B, {\bf 80}, 1627.\\

{\sc Fetter}, A.~L., and {\sc Walecka}, J.~D., 1971, {\sl Quantum
Theory of Many-Particle Systems} (New York: McGraw-Hill).\\

{\sc Foulkes}, W.~M.~C., {\sc Mitas}, L., {\sc Needs}, R.~J.,
and {\sc Rajagopal}, G., 
2001, {\it Rev. Mod. Phys.}, {\bf 73}, 33.\\ 

{\sc Fulde}, P., 1991, 
{\sl Electron Correlation in Molecules and Solids} 
(Berlin: Springer).\\

{\sc Galitskii}, V.~M., and {\sc Migdal}, A.~B., 1958, 
{\it Soviet Phys. JETP}, {\bf 34}, 96.\\

{\sc Gebhard}, F., 1997, {\sl The Mott Metal-Insulator Transition;
Models and Methods} (Berlin: Springer). \\

{\sc Gelfand}, I.~M., and {\sc Shilov}, G.~E., 1964, 
{\sl Generalized Functions}, Vol. I (London: Academic Press).\\

{\sc Gell-Mann}, M., and {\sc Brueckner}, K., 1957, {\sl Phys. Rev.},
{\bf 106}, 364.\\

{\sc Georges}, A., and {\sc Kotliar}, G., 1992,
{\it Phys. Rev.} B, {\bf 45}, 6479.\\ 

{\sc Georges}, A., {\sc Kotliar}, G., {\sc Krauth}, W.,
and {\sc Rozenberg}, M.~J., 1996,
{\it Rev. Mod. Phys.}, {\bf 68}, 13.\\

{\sc Golub}, G.~H., and {\sc van Loan}, C.~F., 1983, 
{\sl Matrix computations} (London: North Oxford Academic), 
chapter 9.\\

{\sc Gordon}, R.~G., 1968,
{\it J. Math. Phys.}, {\bf 9}, 655.\\

{\sc Gradshtyn}, I.~S., and {\sc Ryzhik}, I.~M., 1965, {\sl Table
of Integrals, Series, and Products} (New York: Academic Press).\\

{\sc Gutzwiller}, M.~C., 
1963, {\it Phys. Rev. Lett.}, {\bf 10}, 159;
1964, {\it Phys. Rev.}, {\bf 134}, A923;
1965, {\it ibid.}, {\bf 137}, A1726.\\

{\sc Harris}, A.~B., and {\sc Lange}, R.~V., 1967, 
{\it Phys. Rev.}, {\bf 157}, 295.\\

{\sc Haydock}, R., 1980, {\sl Solid State Physics}, 
Vol.~{\bf 35}, edited by H. Ehrenreich, F. Seitz, and D. Turnbull 
(New York: Academic), 215.\\

{\sc Hedin}, L., 1965, {\it Phys. Rev.} B, {\bf 139}, A796.\\

{\sc Hedin}, L., and {\sc Lundqvist}, S., 1969, {\sl Solid
State Physics}, Vol.~{\bf 23}, edited by F. Seitz, D. Turnbull, 
and E. Ehrenreich (New York: Academic Press), 1.\\

{\sc Herman}, F., and {\sc March}, N.~H., 1984,
{\it Solid St. Commun.}, {\bf 50}, 725.\\

{\sc Hobson}, E.~W., 1927, {\sl The Theory of Functions of
a Real Variable and the Theory of Fourier's Series}, Vol.~I,
third edition (Cambridge University Press).\\

{\sc Hohenberg}, P., and {\sc Kohn}, W., 1964,
{\it Phys. Rev.}, {\bf 136}, B864.\\

{\sc Hubbard}, J., 
1957, {\it Proc. Roy. Soc.} A, {\bf 240}, 539;
1963, {\it ibid.}, {\bf 277}, 237;
1964, {\it ibid.}, {\bf 281}, 401.\\

{\sc Ince}, E.~L., 1927, {\sl Ordinary Differential Equations}
(London: Longmans, Green).\\

{\sc Izuyama}, T., {\sc Kim}, D.-J., and {\sc Kubo}, R., 1963, 
{\it J. Phys. Soc. Japan}, {\bf 18}, 1025. \\

{\sc Jackson}, J.~D., 1975, {\sl Classical Electrodynamics},
second edition (New York: Wiley).\\

{\sc Kajueter}, H., and {\sc Kotliar}, G., 1996,
{\it Phys. Rev. Lett.}, {\bf 77}, 131.\\

{\sc Kalashnikov}, O.~K., and {\sc Fradkin}, E.~S., 1969,
{\it Soviet Phys. JETP}, {\bf 28}, 317;
1973, {\it Phys. Stat. sol.} (b), {\bf 59}, 9. \\

{\sc Kanamori}, J., 1963,
{\it Prog. Theor. Phys.}, Osaka, {\bf 30}, 275.\\

{\sc Kato}, T., 1957, {\it Commun. pure appl. Math.}, 
{\bf 10}, 151.\\

{\sc Katsnelson}, M.~I., and {\sc Lichtenstein}, A.~I., 2000,
{\it Phys. Rev.} B, {\bf 61}, 8906.\\

{\sc Khinchin}, A.~{\sc Ya.}, 1964, {\sl Continued Fractions}
(University of Chicago Press).\\

{\sc Klein}, A., and {\sc Prange}, R., 1958,
{\it Phys. Rev.}, {\bf 112}, 994.\\

{\sc Kohn}, W., and {\sc Luttinger}, J.~M., 1965, 
{\it Phys. Rev. Lett.}, {\bf 15}, 524.\\

{\sc Kohn}, W., and {\sc Sham}, L.~J., 1965,
{\it Phys. Rev.}, {\bf 140}, A1133.\\

{\sc Kreyszig}, E., 1978, {\sl Introductory Functional
Analysis with Applications} (New York: Wiley).\\

{\sc Kubo}, R., and {\sc Tomita}, K., 1954, 
{\it J. Phys. Soc. Japan}, {\bf 9}, 888.\\

{\sc Landau}, L.~D., 1957, {\it Soviet Phys. JETP},
{\bf 3}, 920.\\

{\sc Landau}, L.~D., and {\sc Lifshitz}, E.~M., 1980, 
{\sl Statistical Physics}, Part I, third revised and enlarged 
edition (Oxford: Pergamon).

{\sc Langer}, J.~S., and {\sc Ambegaokar}, V., 1961, 
{\it Phys. Rev.}, {\bf 121}, 1090.\\

{\sc Lauwerier}, H.~A., 1977, {\sl Asymptotic Analysis},
Part I, second Printing (Amsterdam: Mathematisch Centrum).\\

{\sc Layzer}, A.~J., 1963,
{\it Phys. Rev.}, {\bf 129}, 897.\\

{\sc Lehmann}, H., 1954, {\it Nuovo Cim.}, {\bf 11}, 342.\\

{\sc Lichtenstein}, A.~I., and {\sc Katsnelson}, M.~I., 1998,
{\it Phys. Rev.} B, {\bf 57}, 6884.\\

{\sc Luttinger}, J.~M., 1960, {\it Phys. Rev.}, {\bf 119}, 1153;
1961, {\it ibid.}, {\bf 121}, 942;
1963, {\it J. Math. Phys.}, {\bf 4}, 1154.\\

{\sc Luttinger}, J.~M., and {\sc Ward}, J.~C., 1960, 
{\it Phys. Rev.}, {\bf 118}, 1417.\\

{\sc March}, N.~H., {\sc Young}, W.~H., and
{\sc Sampanthar}, S., 1967, {\sl The Many-Body Problem in
Quantum Mechanics} (Cambridge University Press).\\

{\sc Mattis}, D.~C., and {\sc Lieb}, E.~H., 1965, 
{\it J. Math. Phys.}, {\bf 6}, 304.\\

{\sc Mattuck}, R.~D., 1992, {\sl A Guide to Feynman Diagrams 
in the Many-Body Problem} (New York: Dover Publications).\\

{\sc McWeeny}, R., 1992, {\sl Methods of Molecular Quantum 
Mechanics}, second edition (London: Academic Press).\\

{\sc Metzner}, W., and {\sc Vollhardt}, D., 1989,
{\it Phys. Rev. Lett.}, {\bf 62}, 324.\\

{\sc Migdal}, A.~B., 1957, {\it Soviet Phys. JETP}, {\bf 5}, 333.\\

{\sc Montorsi}, A. (editor), 1992, {\sl The Hubbard Model}, 
reprint volume (Singapore: World Scientific).\\

{\sc Mori}, H., 1965, {\it Prog. Theor. Phys.}, Osaka,
{\bf 33}, 423.\\

{\sc Morse}, P.~M., and {\sc Feshbach}, H., 1953, {\sl Methods of
Theoretical Physics} (New York: McGraw-Hill).\\

{\sc Murray}, J.~D., 1974, {\sl Asymptotic Analysis}
(Oxford University Press).\\

{\sc Nagaoka}, Y., 1965,
{\it Solid St. Commun.}, {\bf 3}, 409;
1966, {\it Phys. Rev.}, {\bf 147}, 392.\\

{\sc Nambu}, Y., 1960, {\it Phys. Rev.}, {\bf 117}, 648.\\

{\sc Negele}, J.~W., and {\sc Orland}, H., 1988,
{\sl Quantum Many-Particle Systems} (Redwood City, California:
Addison-Wesley).\\

{\sc Nolting}, W., 
1972, {\it Z. Phys.}, {\bf 255}, 25;
1977, {\it Phys. Stat. sol.} (b), {\bf 79}, 573;
1978, {\it J. Phys.} C, {\bf 11}, 1427.\\

{\sc Nolting}, W., and {\sc Borgie{\l}}, W., 1989,
{\it Phys. Rev.} B, {\bf 39}, 6962.\\

{\sc Nolting}, W., and {\sc Ole\'s}, A.~M., 
1979, {\it J. Phys.} C, {\bf 13}, 2295;
1980, {\it Phys. Rev.} B, {\bf 22}, 6184;
1981, {\it Phys. Stat. sol.} (b), {\bf 104}, 563;
1987, {\it Physica} A, {\bf 143}, 296. \\ 

{\sc Nozi\`eres}, P., 1964, {\sl Theory of Interacting Fermi
Systems} (Amsterdam: Benjamin).\\

{\sc Ohkawa}, F.~J., 1991,
{\it J. Phys. Soc. Japan}, {\bf 60}, 3218.\\

{\sc Ortiz}, G., {\sc Harris}, M., and {\sc Ballone}, P., 1999,
{\it Phys. Rev. Lett.}, {\bf 82}, 5317.\\

{\sc Pack}, R.~R., and {\sc Brown}, W.~B., 1966, 
{\it J. Chem. Phys.}, {\bf 45}, 556.\\

{\sc Peierls}, R., 1979, {\sl Surprises in Theoretical Physics}
(Princeton University Press).\\

{\sc Perdew}, J.~P., {\sc Parr}, R.~G., {\sc Levy}, M., and 
{\sc Balduz}, J.~L., 1982, {\it Phys. Rev. Lett.}, {\bf 49}, 1691.\\

{\sc Pines}, D., and {\sc Nozi\`eres}, P., 1966, {\sl The Theory
of Quantum Liquids}, Vol. I, {\sl Normal Fermi Liquids} (New York:
Benjamin).\\

{\sc Pryce}, M.~H.~L., and {\sc Stevens}, K.~W.~H., 1951, 
{\it Proc. Phys. Soc.} A, {\bf 63}, 36.\\

{\sc Reichl}, L.~E., 1980, {\sl A Modern Course in Statistical 
Physics} (London: Edward Arnold).\\

{\sc Ruijgrok}, Th.~W., 1962, {\it Physica}, {\bf 28}, 877. \\

{\sc Sansone}, G., and {\sc Gerretsen}, J., 1960,
{\sl Lectures on the Theory of Functions of a Complex
Variable}, I, {\sl Holomorphic Functions} (Groningen: 
P. Noordhoff).\\

{\sc Sartor}, R., and {\sc Mahaux}, C., 1980, {\it Phys. Rev.} C,
{\bf 21}, 1546.\\

{\sc Seitz}, F., 1940, {\sl Modern Theory of Solids}
(New York: McGraw-Hill).\\ 

{\sc Shohat}, J.~A., and {\sc Tamarkin}, J.~D., 1943, 
{\sl The Problem of Moments}, 1970, fourth printing of revised
edition (Providence, Rhode Island: American
Mathematical Society).\\ 

{\sc Slater}, J.~C., 1953,
{\it Rev. Mod. Phys.}, {\bf 25}, 199.\\

{\sc Steiner}, E., 1963, {\it J. Chem. Phys.}, {\bf 39}, 2365.\\

{\sc Stollhoff}, G., 1996, {\it J. Chem. Phys.}, {\bf 105}, 227;
1998, {\it Phys. Rev.} B, {\bf 58}, 9826.\\

{\sc Stollhoff}, G., and {\sc Fulde}, P., 1980, 
{\it J. Chem. Phys.}, {\bf 73}, 4548.\\

{\sc Szabo}, A., and {\sc Ostlund}, N.~S., 1989,
{\sl Modern Quantum Chemistry: Introduction to Advanced
Electronic Structure Theory} (New York: McGraw-Hill).\\

{\sc Titchmarsh}, E.~C., 1939, {\sl The Theory of Functions} 
second edition, 1985, reprinted (Oxford University Press).\\

{\sc van Hove}, L., 1954a, {\it Phys. Rev.}, {\bf 95}, 249;
1954b, {\it ibid.}, {\bf 95}, 1374.\\ 

{\sc van Vleck}, J.~H., 1948, {\it Phys. Rev.}, {\bf 74}, 1168;
1953, {\it Mod. Rev. Phys.}, {\bf 25}, 220.\\

{\sc Voit}, J., 1994, {\it Rep. Prog. Phys.}, {\bf 57}, 977.\\

{\sc Vollhardt}, D., 1993, in
{\sl Correlated Electron Systems}, edited by V.~J. Emery
(Singapore: World Scientific), 57.\\

{\sc von Neumann}, J., 1955, {\sl Mathematical Foundations
of Quantum Mechanics}, 1996, twelfth printing
(Princeton University Press).\\

{\sc Whittaker}, E.~T., and {\sc Watson}, G.~N., 1927, 
{\sl A Course of Modern Analysis}, fourth edition, 1984, 
reprinted (Cambridge University Press).\\

{\sc Wilson}, K.~G., 1973,
{\it Phys. Rev.} D, {\bf 7}, 2911.\\

{\sc Wilson}, K.~G., and {\sc Fisher}, M.~E., 1972,
{\it Phys. Rev. Lett.}, {\bf 28}, 240.\\

{\sc Wohlfarth}, E.~P., 1953,
{\it Rev. Mod. Phys.}, {\bf 25}, 211.\\

{\sc Yasuhara}, H., 1972, {\it Solid St. Commun.}, 
{\bf 11}, 1481.\\

{\sc Yasuhara}, H., and {\sc Kawazoe}, Y., 1976, 
{\it Physica}~A, {\bf 85}, 416.\\

{\sc Zwanzig}, R.~W., 1961, in {\sl Lectures in Theoretical 
Physics}, Vol.~{\bf 3}, edited by W.~E. Brittin, B.~W. Downs,
and J. Downs (New York: Interscience), 106.\\

}
\end{verse}


\vspace{1.0cm}
\hrule
\vspace{0.3cm}

\noindent
{\sc Errata} {\sc concerning the published version of this work} 
[{\it Phil. Mag.} B~{\bf 82}, 1413-1610 (2002)]:\\

--- On the third line of the paragraph on p.~1424, \\
``(Anderson 1959, Gutzwiller 1963, Hubbard 1963, \\
Ruijgrok 1962, Kanamori 1963)'' should be
``(Anderson 1959, Ruijgrok 1962, Gutzwiller 1963, 
Izuyama, {\sl et al.} 1963, Hubbard 1963, Kanamori 1963)''.

--- In equation (106) on page 1456, 
``$\frac{\varepsilon_{k}^{(0)}}{e_0}$'' 
and ``$\frac{1}{2}{\bar k}^2$'' should be separated by ``$\equiv$''.

--- On the LHS of equation (113) on page 1459, ``$\hbar$''
should be suppressed.

--- On the 6th line of equation (199) on page 1483, ``$v^2$''
should be ``$v({\Bf r}-{\Bf r}'')$''.

--- On the 7th line of equation (199) on page 1483, ``$v^2$''
should be ``$v^2({\Bf r}-{\Bf r}'')$''.

--- In equation (223) on page 1494, \\
``${}_2F_1(-\frac{1}{2}-m,1;\frac{3}{2}-m;
\frac{1}{\Delta^2\varepsilon^2})$'' should be \\
``${}_2F_1(\frac{1}{2}-m,1;\frac{3}{2}-m;
\frac{1}{\Delta^2\varepsilon^2})$''.

--- On the LHS of equation (227) on page 1495, \\ 
``$(\varepsilon+i\eta)$'' should be 
``$(\varepsilon\pm i\eta)$''. 

--- On page 1495, 7th line below equation (227), \\
``$(\varepsilon+i\eta)$'' should be 
``$(\varepsilon\pm i\eta)$''.

--- On the LHS of equations (234a) and (234b) on page 1498,
``$\ol{\Sigma}_{\sigma}$'' should be
``$\wt{\ol{\Sigma}}_{\sigma}$''.

--- On page 1523, ``$\varrho$'' on the RHS of the
expression for $\Pi_{\sigma}$ should be ``$\wp$''.

--- On page 1533, last line of footnote 114,
``$\varepsilon_{{\bar{\bF k}};\sigma}$'' should be
``$\varepsilon_{{\bF k};\sigma}$''.

--- On page 1547, on the 3rd line below equation (D10),
``considering'' should be ``consider''. 

--- On page 1572, on the 2nd line below equation (F97),
``$\Gamma^{(2)}({\Bf r}_1''\sigma_1',{\Bf r}_2''\sigma_2';
{\Bf r}_1''\sigma_2',{\Bf r}_2''\sigma_2')$'' should be
``$\Gamma^{(2)}({\Bf r}_1''\sigma_1',{\Bf r}_2''\sigma_2';
{\Bf r}_1''\sigma_1',{\Bf r}_2''\sigma_2')$''. 

--- On page 1587, on the 2nd line above equation (F161),
``(F154)'' should be ``(F157)''. \hfill $\Box$

\vspace{0.3cm}
\hrule

\pagebreak
\begin{figure}[t!]
\label{fi1}
\protect
\centerline{
\psfig{figure=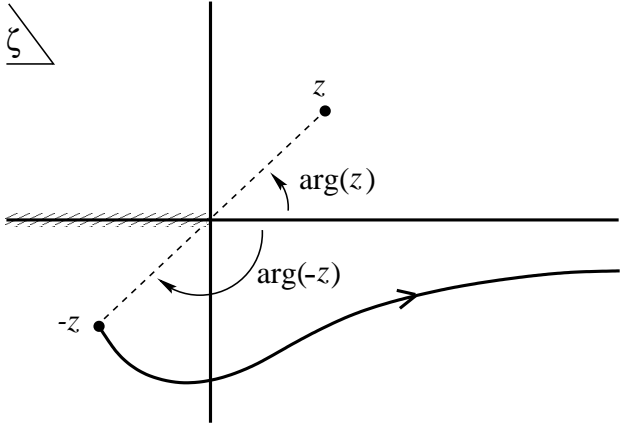,width=3.25in} }
\vskip 20pt
\caption{\rm
The contour of integration in the complex $\zeta$ plane corresponding 
to the integral representation of the exponential-integral function
${\rm Ei}(z)$ as defined in Eq.~(\protect\ref{ei3}). The shaded part 
of the real axis indicates that the branch of the multi-valued function
${\rm Ei}(z)$ (viewed as a function of $-z$) relevant to our 
considerations, that is the `principal' branch, is specified by the 
requirement $-\pi < {\rm arg}(-z) < \pi$. This aspect is directly 
associated with $\ln(-z)$ in Eq.~(\protect\ref{ei4}), with $\ln$ being 
the principle branch of the logarithm function. } 
\end{figure}
%

\begin{figure}[t!]
\label{fi2}
\protect
\centerline{
\psfig{figure=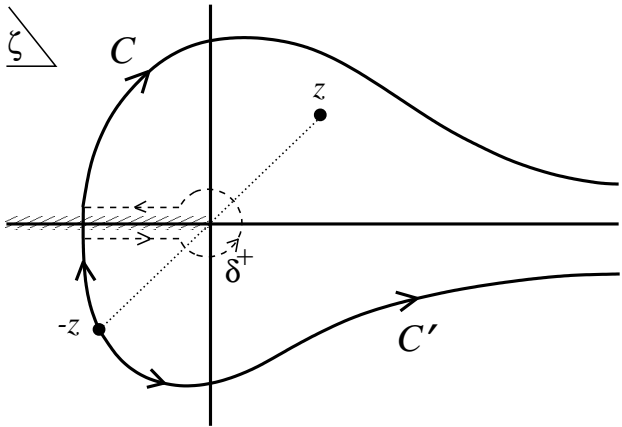,width=3.25in} }
\vskip 20pt
\caption{\rm
The contours $C$ and $C'$ of integration in the complex $\zeta$ plane 
corresponding to ${\cal J}_{C}(z)$, as defined in Eq.~(\protect\ref{ei5}), 
and ${\cal J}_{C'}(z) \equiv {\rm Ei}(z)$, as introduced in 
Eq.~(\protect\ref{ei3}), respectively. The non-vanishing value of the 
integral of $\exp(-\zeta)/\zeta$ along contour $\delta^+$, depicted by 
broken line, establishes that ${\cal J}_{C}(z)$ {\sl cannot} be 
identified with ${\rm Ei}(z)$; see Eq.~(\protect\ref{ei6}). }
\end{figure}
%

\begin{figure}[t!]
\label{fi3}
\protect
\centerline{
\psfig{figure=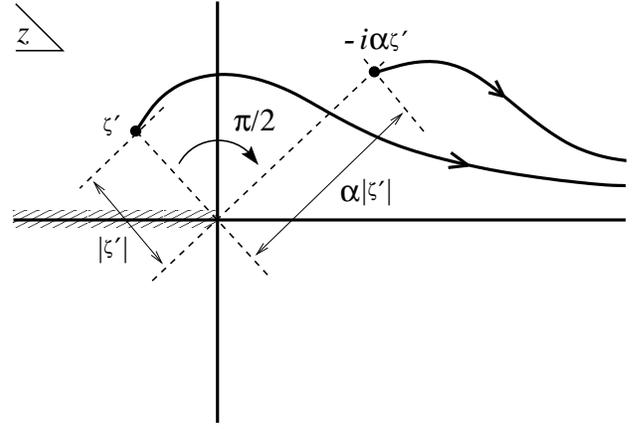,width=3.25in} }
\vskip 20pt
\caption{\rm
In expressing ${\cal I}_1(\zeta';\alpha)$ as defined in 
Eq.~(\protect\ref{ei1}), through a process of transformation of 
the integration variable $z$, in terms of the exponential-integral 
function ${\rm Ei}(i\alpha\zeta')$, account has to be taken of the 
fact that this transformation can give rise to violation of the 
requirement $-\pi < {\rm arg}(-i\alpha\zeta') < \pi$ (see caption 
of Fig.~1). The transformation corresponding to an instance where 
$\zeta'$ is located in the second quadrant of the complex $z$ plane 
and $\alpha$ is a positive constant is shown. }
\end{figure}
%

\begin{figure}[t!]
\label{fi4}
\protect
\centerline{
\psfig{figure=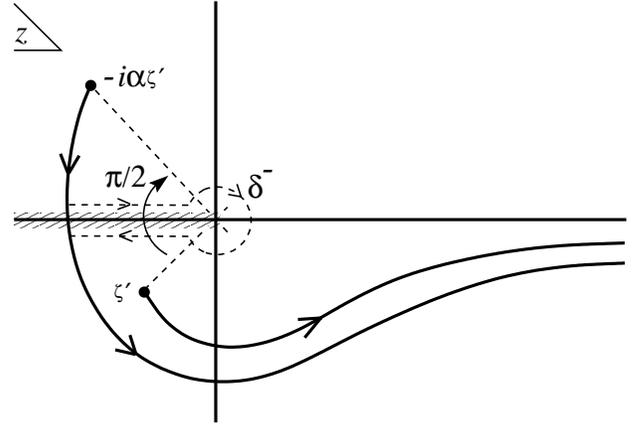,width=3.25in} }
\vskip 20pt
\caption{\rm
Similar to Fig.~3, except that $\zeta'$ is located in the third 
quadrant of the complex $z$ plane. With reference to the caption 
of Fig.~2, it is seen that, for $\alpha > 0$, transformation of 
the integration variable as described in the text does not allow 
direct identification of ${\cal I}_1(\zeta';\alpha)$ with 
$-{\rm Ei}(i\alpha\zeta')$; integration of $\exp(-z)/z$ along
contour $\delta^-$, depicted by means of the broken line, reveals the 
difference between the two functions (see Eq.~(\protect\ref{ei7})). }
\end{figure}
%

\begin{figure}[t!]
\label{fi5}
\protect
\centerline{
\psfig{figure=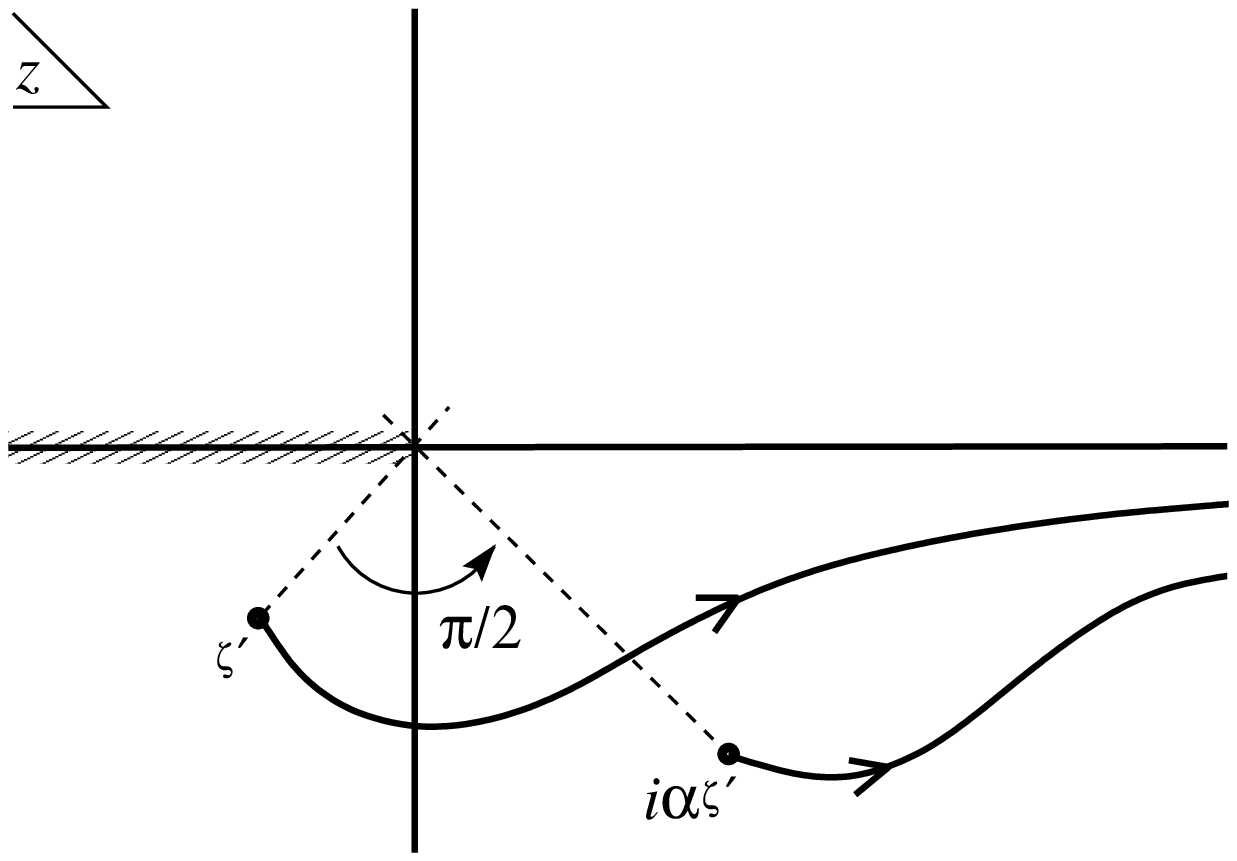,width=3.25in} }
\vskip 20pt
\caption{\rm
Similar to Fig.~4, except that it concerns the transformation of 
the integration variable, that is $z$, in an attempt to express 
${\cal I}_2(\zeta';\alpha)$, with $\alpha > 0$, in terms of 
${\rm Ei}(-i\alpha\zeta')$ (see Eq.~(\protect\ref{ei8})). }
\end{figure}
%

\begin{figure}[t!]
\label{fi6}
\protect
\centerline{
\psfig{figure=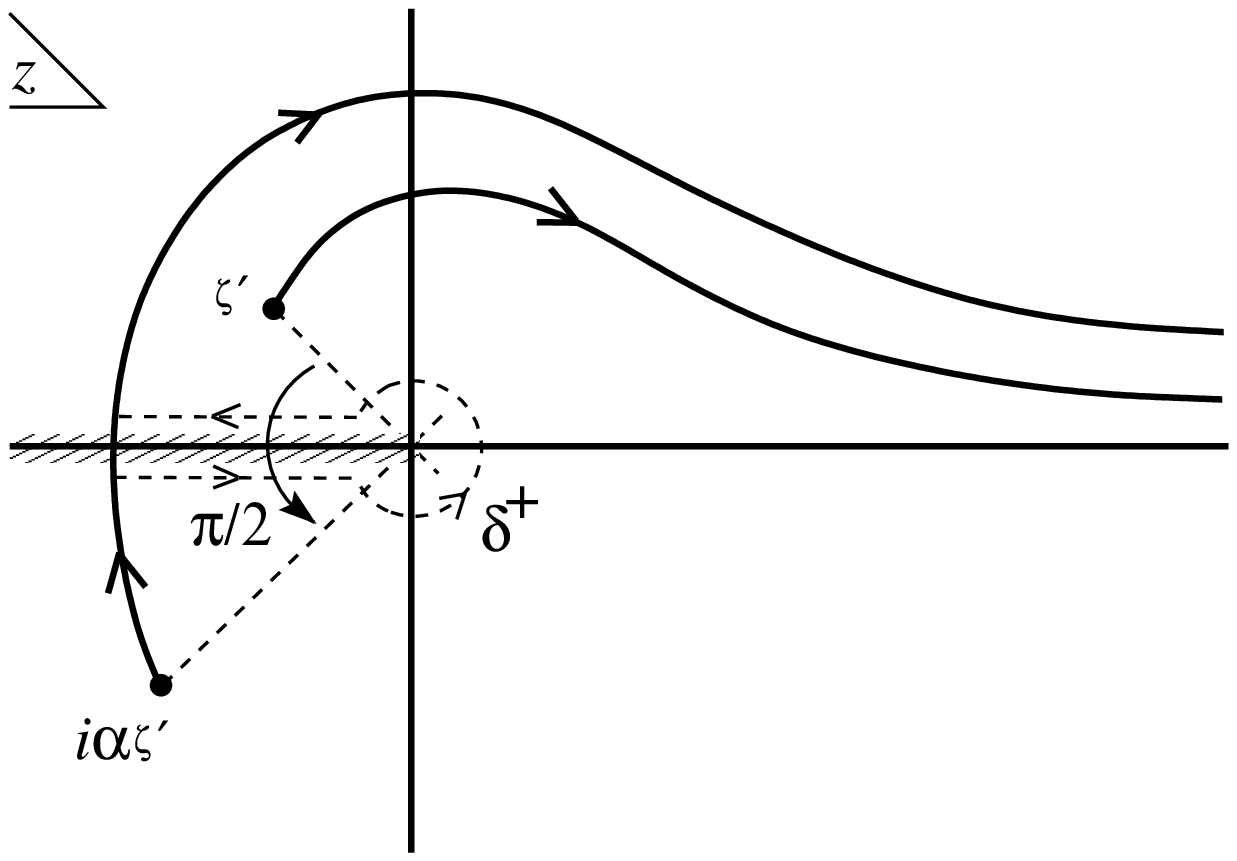,width=3.25in} }
\vskip 20pt
\caption{\rm
Similar to Fig.~3, except that it concerns the transformation of 
the integration variable, that is $z$, in an attempt to express
${\cal I}_2(\zeta';\alpha)$, with $\alpha > 0$, in terms of the 
${\rm Ei}(-i\alpha\zeta')$ (see Eq.~(\protect\ref{ei8})). }
\end{figure}
%

\end{document}